\documentclass[11pt,a4paper]{article}
\usepackage{epsfig}
\usepackage[T1]{fontenc}    
\usepackage{graphics}
\usepackage{graphicx}
\usepackage{pstricks,pst-coil,pst-fill,pst-plot}
\usepackage{cite}
\usepackage[fleqn]{amsmath}    
\usepackage{amssymb}    
\usepackage{amsfonts}   
\usepackage{verbatim}   
\usepackage{mathrsfs}   
\usepackage{dsfont}
\usepackage{euscript}
\usepackage{yfonts}
\usepackage{enumerate}     
\usepackage{amsthm}         
\usepackage{txfonts}
\usepackage{marvosym}
\usepackage{stmaryrd}
\usepackage{vmargin}        
\usepackage{wasysym}		

\usepackage{color}

\setmarginsrb{1.8cm}{2cm}{1.8cm}{2cm}{1cm}{1cm}{1cm}{1.6cm}
 \makeatletter
 \@addtoreset{equation}{section}
 \makeatother

\providecommand{\bysame}{\leavevmode\hbox to3em{\hrulefill}\thinspace}
\providecommand{\MR}{\relax\ifhmode\unskip\space\fi MR }

\providecommand{\href}[2]{#2}

\let\ua=\uparrow
\let\da=\downarrow
\let\tend=\rightarrow

\long\def\symbolfootnote[#1]#2{\begingroup%
\def\thefootnote{\fnsymbol{footnote}}\footnote[#1]{#2}\endgroup}

\newtheorem{theorem}{Theorem}[section]
\newtheorem{prop}[theorem]{Proposition}
\newtheorem*{theorem*}{Theorem}
\newtheorem{cor}[theorem]{Corollary}
\newtheorem{defin}[theorem]{Definition}

\newtheorem{conj}[theorem]{Conjecture}

\newtheorem{lemme}[theorem]{Lemma}
\newtheorem{hypothesis}[theorem]{Hypothesis}
\newtheorem{property}[theorem]{Properties}

\def\Proof{\medskip\noindent {\it Proof --- \ }}

\def\qed{\hfill\rule{2mm}{2mm}}

\newcommand\beq{\begin{equation}}
\newcommand\enq{\end{equation}}
\newcommand\bem{\begin{multline}}
\newcommand\enm{\end{multline}}

\def\beqa{\begin{eqnarray}}
\def\eeqa{\end{eqnarray}}
\def\ba{\begin{array}}
\def\ea{\end{array}}
\def\det{\operatorname{det}}

\newcommand{\f}[2]{{\ensuremath{%
    \mathchoice%
    {\dfrac{#1}{#2}}
    {\dfrac{#1}{#2}}
    {\frac{#1}{#2}}
    {\frac{#1}{#2}}
}}}
\newcommand{\tf}[2]{\ensuremath{#1/#2}}
\newcommand{\pa}[1]{\ensuremath{\left(#1\right)}}

\newcommand{\pac}[1]{\ensuremath{\left[#1\right]}}

\def\a{\alpha}

\def\be{\beta}
\def\ga{\gamma}
\def\Ga{\Gamma}

\def\de{\delta}

\def\De{\Delta}
\def\eps{\epsilon}
\def\veps{\varepsilon}
\def\la{\lambda}
\def\La{\Lambda}

\def\sg{\sigma}
\def\vsg{\varsigma}

\def\Ups{\Upsilon}
\def\ups{\upsilon}
\def\th{\theta}

\def\vth{\vartheta}

\def\Om{\Omega}
\def\om{\omega}
\def\vp{\varphi}

\newcommand{\mc}[1]{\ensuremath{\mathcal{#1}}}
\newcommand{\mf}[1]{\ensuremath{\mathfrak{#1}}}
\newcommand{\msc}[1]{\ensuremath{\mathscr{#1}}}

\newcommand{\bs}[1]{\ensuremath{\boldsymbol{#1}}}

\DeclareFontFamily{OT1}{pzc}{}
\DeclareFontShape{OT1}{pzc}{m}{it}{<-> s * [1.10] pzcmi7t}{}
\DeclareMathAlphabet{\mathpzc}{OT1}{pzc}{m}{it}

\def \i{ \mathrm i}

\newcommand{\ov}[1]{\ensuremath{\overline{#1}}}
\newcommand{\wt}[1]{\ensuremath{\widetilde{#1}}}
\newcommand{\wh}[1]{\ensuremath{\widehat{#1}}}

\newcommand{\Int}[2]{\ensuremath{\int\limits_{#1}^{#2}}}
\newcommand{\Oint}[2]{\ensuremath{\oint\limits_{#1}^{#2}}}
\newcommand{\Fint}[2]{\ensuremath{\fint\limits_{#1}^{#2}}}

\newcommand{\sul}[2]{\ensuremath{\sum\limits_{#1}^{#2}}}
\newcommand{\pl}[2]{\ensuremath{\prod\limits_{#1}^{#2}}}

\newcommand{\R}{\ensuremath{\mathbb{R}}}
\newcommand{\Cx}{\ensuremath{\mathbb{C}}}

\newcommand{\Dp}[1]{\ensuremath{\partial_{#1}}}

\newcommand{\limit}[2]{\ensuremath{\underset{#1 \tend #2}{\longrightarrow} }}

\newcommand{\ex}[1]{\ensuremath{\e{e}^{#1}}}

\newcommand{\op}[1]{ \boldsymbol{ \texttt{#1} } }

\newcommand{\abs}[1]{\ensuremath{\left| #1 \right|}}

\newcommand{\norm}[1]{\ensuremath{  || #1 || }}

\newcommand{\dd}{\mathrm{d}}
\newcommand{\e}[1]{\ensuremath{\mathrm{#1}}}

\newcommand{\intff}[2]{\ensuremath{ [  #1 \,; #2 ] }}
\newcommand{\intfo}[2]{\ensuremath{ [  #1 \,; #2 [ }}
\newcommand{\intof}[2]{\ensuremath{ ]  #1 \,; #2 ] }}
\newcommand{\intoo}[2]{\ensuremath{ ]  #1 \,; #2 [ }}

\newcommand{\intn}[2]{\ensuremath{[\![ \, #1 \,;\, #2 \,]\!]}}

\newcommand{\widesim}[2][1.5]{
  \mathrel{\underset{#2}{\scalebox{#1}[1]{$\sim$}}}
}

\DeclareMathOperator{\Interior}{Int}

\begin{document}

\begin{center}
\begin{LARGE}
{\bf Low-temperature spectrum of the quantum transfer matrix of the XXZ chain in
the massless regime}
\end{LARGE}

\vspace{1cm}

{\large Saskia Faulmann,\footnote{e-mail: saskia.faulmann@uni-wuppertal.de}
Frank G\"{o}hmann\footnote{e-mail: goehmann@uni-wuppertal.de}}
\\[1ex]
Fakult\"at f\"ur Mathematik und Naturwissenschaften, Bergische Universit\"at
Wuppertal, 42097 Wuppertal, Germany.\\[2.5ex]

{\large Karol K. Kozlowski\footnote{e-mail: karol.kozlowski@ens-lyon.fr}}
\\[1ex]
ENSL, CNRS, Laboratoire de physique, F-69342 Lyon, France. \\[2.5ex]

\par 

\end{center}

\vspace{40pt}

\centerline{\bf Abstract} \vspace{1cm}
\parbox{12cm}{\small}
The free energy \textit{per} lattice site of a quantum spin chain in the
thermodynamic limit is determined by a single `dominant' Eigenvalue
of an associated quantum transfer matrix in the infinite Trotter
number limit. For integrable quantum spin chains constructed  from solutions
of the Yang-Baxter equation, the quantum transfer matrix may be taken
as the transfer matrix of an inhomogeneous variant of the underlying
vertex model. Its spectrum can then be studied by Bethe Ansatz methods
and may exhibit universal features such as the emergence of a conformal sub-spectrum
in the low-temperature regime. Access to the full spectrum of the quantum transfer matrix enables
the construction of thermal form factor series representations of the
correlation functions of local operators for the spin chain under
consideration. These are claims, made by physicists, whose rigorous
mathematical justification sets up a long-term research programme.
In this work we implement first steps of this programme with the
example of the XXZ quantum spin chain in the antiferromagnetic massless
parameter regime and in the low-temperature limit. We rigorously
establish the existence and uniqueness of the solutions to a set of
non-linear integral equations, that are equivalent to the Bethe
Ansatz equations for the quantum transfer matrix of this model,
and explicitly characterize the low-temperature form of these solutions.
This allows us to describe that part of the quantum transfer matrix
spectrum that is related to the Bethe Ansatz and that does not collapse
to zero in the infinite Trotter number limit. Within the considered part
of the spectrum we rigorously identify the unique Eigenvalue of largest
modulus and show that those correlations lengths that diverge in the low-temperature
limit are, to the leading order in temperature, in one-to-one correspondence
with the spectrum of the free Boson $c=1$ conformal field theory.  Based on two
conjectures, that are accepted in the physics literature, but that
could so far only be established in the opposite limit of high temperatures,
we prove that the Eigenvalue of largest modulus in the
sub-spectrum we focus on corresponds, in fact, to the dominant Eigenvalue. Its first order
term in temperature is of a universal form conjectured long ago in the
physics literature.
\vspace{20pt}


\tableofcontents

\section{Introduction}

\subsection{The integrable XXZ spin-1/2 chain}
The study of the behaviour of quantum integrable models at finite temperature
was initiated by Yang and Yang \cite{YangYangNLSEThermodynamics} with the
computation of the free energy of a gas of Bosons in one dimension interacting
through a two-body delta-function potential. Their method, nowadays called the
thermodynamic Bethe Ansatz (TBA), was later extended in the independent works
of Gaudin and Takahashi so as to encompass the case of the XXZ spin-1/2 chain
\cite{GaudinTBAXXZMassiveInfiniteSetNLIE,TakahashiTBAforXXZFiniteTinfiniteNbrNLIE}
in the massive antiferromagnetic regime. Subsequently, it was generalised to
many other integrable models, most notably to the XXZ chain in the massless
regime \cite{TakahashiSuzukiFiniteTXXZandStrings} and to the Hubbard model
\cite{TakahashiTBAForHubbartFiniteT}. Extensive expositions of some of the more
important applications of the TBA can be found in
\cite{TakahashiThermodynamics1DSolvModels,TsvelikWiegmannReviewExactResultsMagneticAlloys}.

The TBA approach relies on a number of \textit{ad hoc} assumptions which appear
to be extremely hard to establish on rigorous grounds, at least in the present
state of the art. In fact, the only case when a TBA based description of the thermodynamics
could be established rigorously pertains to the above mentioned Bose gas in one dimension. The closed formula for the free energy,
conjectured in \cite{YangYangNLSEThermodynamics}, was proven in
\cite{DorlasLewisPuleRigorousProofYangYangThermoEqnNLSE} by means of an evaluation
of the large volume behaviour of the model's partition function with the help
of Varadhan's lemma and large deviation estimates for the free fermion model obtained
in \cite{LewisPuleZagrebnovLDPpourMesuresKacEtHardBosons}. Realising this program was possible owing to the particularly
simple form of the Bethe Ansatz equations for the delta function Bose gas. For other, more complicated
models, such as the XXZ chain, there seems to be no hope to make such kind of proof work.

Another drawback of the TBA approach to the thermodynamics is that, generically,
it provides a description of a model's \textit{per}-site free energy and related
thermodynamical quantities in terms of a solution to an infinite set of
coupled non-linear integral equations. This formulation was neither optimal
from the numerical point of view nor was it rigorous. In the late 80s and early
90s, there emerged a new approach to the thermodynamics of quantum integrable
models based on the so-called quantum transfer matrix formalism of Suzuki
\cite{SuzukiArgumentsForInterchangeabilityTrotterAndVolumeLimitInPartFcton}. This was
pioneered in the works of Koma \cite{KomaIntroductionQTM6VertexForThermodynamicsOfXXX,%
KomaIntroductionQTM6VertexForThermodynamicsOfXXZ}, where the characterisation
of the \textit{per} site free energy of the XXZ chain was brought down to the calculation
of the largest Eigenvalue of the associated quantum transfer matrix, a matrix whose
spectrum could, in principle, be studied by the Bethe Ansatz method. Still, even
on heuristic grounds, bringing this problem to an end involved solving, in an efficient
way, the Bethe Ansatz equations and then taking the infinite Trotter number limit
on their level.  The method underwent several technical
improvements  \cite{KlumperNLIEfromQTMDescrThermoRSOSOneUnknownFcton,%
AkustsuSuzukiWadatiModernFormOfQTM4XXZ,TakahashiThermoXXZInfiniteNbrRootsFromQTM}, while
the ground-breaking advance on this problem was achieved independently in
\cite{DestriDeVegaAsymptoticAnalysisCountingFunctionAndFiniteSizeCorrectionsinTBAFirstpaper}
and \cite{KlumperNLIEfromQTMDescrThermoXYZOneUnknownFcton}. Those works reduced
the problem of calculating the \textit{per}-site free energy to the resolution of a
single non-linear integral equation. This turned out to be a highly successful
approach from the numerical point of view and was subsequently extended to
numerous other quantum integrable models
\cite{AufgebauerKlumperTemperleyLiebSpinChainsatFiniteTempe,DamerauKlumperNLIEforsl4UiminSutherlandModel,
JuttnerKlumprSuzukiTauJdModelNLIEandLuttingerLiquidProperties,JuttnerKlumprSuzukiHubberdModelNLIEandLuttingerLiquidProperties,
KlumperPatuThermodynMultiComp1DGAses,SuzukiQTMforXXXatFiniteTempeAnySpin}.
As opposed to the traditional TBA approach, it could also be employed for the explicit calculation of correlation
functions at finite temperature
\cite{GohmannKlumperSeelFinieTemperatureCorrelationFunctionsXXZ}. Still, the validity of
that new method relied on a number of conjectures. The quantum transfer matrix based approach was put on a
rigorous footing only recently for the XXZ chain in the regime of large
enough temperatures \cite{KozGohmannGoomaneeSuzukiRigorousApproachQTMForFreeEnergy}.

The present work \textit{establishes a rigorous solvability theory
for the class of non-linear integral equations} describing the spectrum of the quantum transfer matrix and hence, upon relying on certain conjectures,  the thermodynamics of the XXZ chain
in the so-called massless regime and for temperatures low-enough. Moreover, this is achieved while being able to
access to all the desired information, from the perspective of physical applications, of the solutions. This has several implications:
\begin{enumerate}
\item[i)] We are able to construct  a rich class of solutions to the Bethe Ansatz equations associated with the quantum transfer matrix
at finite but large Trotter numbers and for $T$ small enough.
\item [ii)] We are able to provide an effective description of the infinite Trotter number limit of these solutions.
\item[iii)] We are able to show that the considered solutions to the Bethe Ansatz equations do produce \textit{bona fide} Eigenvectors of the quantum transfer matrix
in that these are non-zero.
\item[iv)] For the considered class of solutions, we explicitly compute, on rigorous grounds, a sub-class of the
sub-dominant Eigenvalues of the quantum transfer matrix. The latter play an important
role in the calculation of correlation functions of local operators within the
thermal form factor approach \cite{KozDugaveGohmannThermaxFormFactorsXXZ,KozGohmannKarbachSuzukiFiniteTDynamicalCorrFcts}.
The considered sub-class contains the family of Eigenvalues which, in the infinite Trotter number limit and in the low-temperature
regime, exactly identify with the spectrum of the $c=1$ free Boson conformal field theory.

\item[v)] Upon relying on certain natural conjectures (commutativity of the Trotter and thermodynamic limit, non-degeneracy of the
maximal Eigenvalue) we are able to provide the low-$T$ behaviour of the XXZ chain's free energy, in particular establishing the connection with the
conformal field theoretic predictions.

\end{enumerate}

The XXZ spin-$1/2$ chain refers to the Hamiltonian operator
\beq
\op{H} \, = \, J \sum_{a=1}^{L} \Big\{ \sigma_a^x \,\sigma_{a+1}^{x} + \sigma_a^y \,\sigma_{a+1}^{y} +  \De \, (\sigma_a^z \,\sigma_{a+1}^{z} + 1)  \Big\} \, - \, \f{h}{2} \sul{a=1}{L} \sigma_a^z  \; .
\label{ecriture hamiltonien XXZ}
\enq
Here  $J>0$ represents the so-called exchange interaction, $\De \in {\mathbb R}$
is the anisotropy parameter, $h>0$ the external magnetic field and $L \in 2\mathbb{N}$
corresponds to the number of sites. $\op{H}$ acts on the Hilbert space
$\mf{h}_{XXZ}=\bigotimes_{a=1}^{L}\mf{h}_a$ with $\mf{h}_a \simeq \Cx^2$.
$\sg^{\a}$ with $\a\in \{x, y, z \}$ are the Pauli matrices. The operator
$\sg_{a}^{\a}$ acts as the Pauli matrix $\sg^{\a}$ on $\mf{h}_a$ and as the
identity on all the other spaces:
\beq
\sg_{a}^{\a} \, = \, \underbrace{ \e{id} \otimes \cdots \otimes \e{id} }_{a-1} \otimes \, \sg^{\a} \otimes \underbrace{ \e{id} \otimes \cdots \otimes \e{id} }_{L-a}  \;.
\enq
Finally, in the expression \eqref{ecriture hamiltonien XXZ}, periodic boundary
conditions, \textit{viz.}\ $\sg_{L+1}^{\a}=\sg_{1}^{\a}$, are understood.

The physical
content of the model depends on the values of the anisotropy $\De$ and of the magnetic
field $h$. At zero temperature and zero magnetic
field\symbolfootnote[1]{There is a long story regarding the rigorous proof of
this simple statement. The result was first argued formally in the
physics literature \cite{BabelondeVegaVialletStringHypothesisWrongXXZ,%
DeVegaWoynarowichFiniteSizeCorrections6VertexNLIEmethod,%
DescloizeauxGaudinExcitationsXXZ+Gap,DescloizeauxPearsonExcitationsXXX,%
DestriLowensteinFirstIntroHKBAEAndArgumentForStringIsWrong,%
VirosztekWoynarovichStudyofExcitedStatesinXXZHigherLevelBAECalculations}.
A proof relevant for the ferromagnetic regime can be found in
\cite{BabbittThomasPlancherelFormulaInfiniteXXX,ThomasGSRepForFerromagneticXXX}.
A proof for the massless regime can be found in \cite{KozProofOfDensityOfBetheRoots}.
The existence of a massive phase when $\De>1$ has not been, to the best of our
knowledge, rigorously established yet.}, the model is ferromagnetic and
has a massive spectrum when $\De \le -1$, while it is antiferromagnetic with a
massless spectrum for $-1 < \De \leq 1$ and with a massive spectrum for $\De>1$.
At finite magnetic field, the boundaries between the different phases
depend on $\De$ and $h$. From now on we shall solely
focus on that part of the massless antiferromagnetic regime for which
$0 < h < 4J (1 + \De)$ and $-1 < \De < 1$. This regime is the most interesting from the point of  view
of physical applications as it corresponds to the range of parameters where
universality arises in the low-temperature limit.
In the parameter regime of interest,
it is convenient to adopt the reparameterisation
\beq
\De = \cos(\zeta) \qquad \e{with} \qquad  \zeta \in \intoo{0}{\pi} \;.
\enq
Here, we exclude the values $0$ and $\pi$, \textit{viz}.\ $\De=\pm1$,
since these are special and demand an independent treatment. Also, for
technical reasons that will be apparent in the following, we shall
assume that $\zeta \not \in \pi \mathbb{Q}$ unless explicitly stated
otherwise.

The \textit{per}-site free energy of the XXZ chain is defined as the limit
\beq
f\, = \; -T \lim_{L\tend + \infty} \Big\{ \f{1}{L} \ln \e{tr}_{\mf{h}_{XXZ}} \Big[ \ex{-\frac{1}{T} \op{H}  } \Big] \Big\} \; .
\label{definition energie libre}
\enq
The limit exists as can be seen from standard handlings of rigorous statistical
mechanics \cite{RuelleRigorousResultsForStatisticalMechanics}. The XXZ spin-$1/2$
chain Hamiltonian is known to be integrable
\cite{McCoyWuFirstProofXXZCommutesWith6VTransferMatrix}.
It is one of the first models the Bethe Ansatz was applied to
\cite{OrbachXXZCBASolution}. The Bethe Ansatz allows one to calculate
explicitly several observables associated with this model.

The quantum transfer matrix approach to the thermodynamics mentioned earlier
may be summed up as follows. By exploiting the integrability of the model,
one may build a one parameter dependent endomorphism called
quantum transfer matrix $\op{t}_{\mf{q}}(\xi)$, \textit{c.f.} Subsection \ref{SousousSection preuve energie libre et identification vp dominante}
for more details. The latter was introduced in full
generality by Klümper \cite{KlumperNLIEfromQTMDescrThermoRSOSOneUnknownFcton}. The
quantum transfer matrix $\op{t}_{\mf{q}}(\xi)$ is built such that it satisfies
\beq
\e{tr}_{\mf{h}_{XXZ}} \Big[ \ex{-\frac{1}{T} \op{H}  } \Big] \; = \; \lim_{N\tend +\infty} \e{tr}_{ \mf{h}_{\mf{q}} } \Big[ \big( \op{t}_{ \mf{q} }(0)\big)^{L} \Big] \;.
\label{ecriture discretisation Trotter pour trace operateur densite}
\enq
We refer to \cite{KozGohmannGoomaneeSuzukiRigorousApproachQTMForFreeEnergy}
for a mathematician-oriented exposition of the construction.

The quantum transfer matrix acts on an auxiliary ``quantum'' space $\mf{h}_{\mf{q}} \, = \,
\bigotimes_{a=1}^{2N} \wt{\mf{h}}_a$, with $\wt{\mf{h}}_a \simeq \Cx^2$. While the dimensionality
of $\mf{h}_{\mf{q}}$ differs from the one of the original Hilbert space
$\mf{h}_{\e{XXZ}}$, the two spaces are built from ``local'' building blocks $\mf{h}_a, \wt{\mf{h}}_a$
which share the same dimensionality. We shall denote the Eigenvalues of the quantum
transfer matrix by $\wh{\La}_k(\xi)$ and set $\wh{\La}_k(0)=\wh{\La}_k$. We choose
the labeling in such a way that $|\wh{\La}_{k - 1}| \ge |\wh{\La}_k|$. It was argued
in \cite{KlumperNLIEfromQTMDescrThermoRSOSOneUnknownFcton} and rigorously established
for $N$ and $T$ large enough in
\cite{KozGohmannGoomaneeSuzukiRigorousApproachQTMForFreeEnergy} that $\op{t}_{\mf{q}}(0)$
has a unique real and positive dominant Eigenvalue $\wh{\La}_{\e{max}} = \wh{\La}_0$ such
that
\beq
     \wh{\La}_{\e{max}} \, >  \, | \wh{\La}_{k} |
\enq
for all $k \ne 0$. It was
further argued in \cite{KlumperNLIEfromQTMDescrThermoRSOSOneUnknownFcton} and rigorously
established for $T$ large enough in
\cite{KozGohmannGoomaneeSuzukiRigorousApproachQTMForFreeEnergy} that one may commute in
\eqref{definition energie libre} the $L \tend +\infty$  with the
$N \tend + \infty$ limit arising from
\eqref{ecriture discretisation Trotter pour trace operateur densite}. This allows
one to compute the $L \tend +\infty$ limit in \eqref{definition energie libre}
explicitly in terms of  $\wh{\La}_{\e{max}}$:
\beq
f\, = \; -T \lim_{N\tend + \infty} \ln \wh{\La}_{\e{max}} \;.
\enq
In this setting, the free energy is thus accessed by taking the infinite
Trotter number limit of $\ln \wh{\La}_{\e{max}}$.

\vspace{4mm}

The quantum transfer matrix $\op{t}_{ \mf{q} }(\xi)$ can be constructed in such
a way that it may be  identified with the transfer matrix of a staggered six-vertex model
\cite{KlumperNLIEfromQTMDescrThermoRSOSOneUnknownFcton}. Hence, at
least some of its Eigenvalues and Eigenvectors can be constructed by solving
the associated system of Bethe Ansatz equations. In full generality, this is a
system of equations for $N^{\prime} \in \{1,\dots, N\}$ unknowns
$\la_1,\dots, \la_{N^{\prime}}$,
\beq
\ex{-\tfrac{h}{T}} (-1)^{ \mf{s} } \f{ \Dp{}^p }{ \Dp{} \xi^p} \Bigg\{ \pl{k=1}{N'} \bigg\{ \f{\sinh( \i\zeta - \xi + \la_k) }{  \sinh( \i \zeta  +  \xi - \la_k) }  \bigg\}
\cdot  \bigg\{ \f{ \sinh\big(   \xi - \tfrac{\aleph}{N} + \i \tfrac{\zeta}{2} \big) \sinh\big(   \i \tfrac{3 \zeta}{2} +  \xi + \tfrac{\aleph}{N}  \big) }
                    {  \sinh\big(   \xi + \tfrac{\aleph}{N} +  \i \tfrac{\zeta}{2} \big) \sinh\big(   \i \tfrac{\zeta}{2} -  \xi + \tfrac{\aleph}{N} \big)  }    \bigg\}^{N} \Bigg\}
\Biggr|_{\xi=\la_a}
\; = \; -\de_{p,0} \; ,
\label{ecriture eqns Bethe Trotter fini}
\enq
with $a=1\dots, N^{\prime}$, $p=0, \dots, k_{\la_a}-1$ and where $k_{\la_a}$ is the multiplicity of $\la_a$. This should be supplemented by the subsidiary condition that the derivative above does not vanish for $p=k_{\la_a}$.
Above, $\aleph$ corresponds to a reparameterisation of the inverse temperature,
\beq
 \aleph=-\i J \f{ \sin (\zeta) }{T } \; ,
\label{definition aleph}
\enq
and
\beq
\mf{s}  = N - N^{\prime}
\label{defintion pseudospin}
\enq
is called the (pseudo) spin.

In order to obtain a joint Eigenvector $\bs{\Psi}\big( \la_1,\dots, \la_{N^{\prime}} \big)$
to the family of mutually commuting quantum transfer matrices $ \op{t}_{ \mf{q} }(\xi)$
from a solution to the Bethe Ansatz equation, one also has to demand that the roots
$\la_1,\dots, \la_{N^{\prime}}$ are admissible, namely that
\begin{itemize}
\item  $\la_{a} \ne\la_{b} \pm \i \zeta \; \e{mod} \; \i \pi$ for any $a,b$;
\item $\la_a\not\in \Big\{ \pm \tfrac{\aleph}{N} - \i\tfrac{\zeta}{2}, \tfrac{\aleph}{N} + \i \tfrac{\zeta}{2} ,  -\tfrac{\aleph}{N} - 3\i\tfrac{\zeta}{2}    \Big\}$ for any $a$.
\end{itemize}
When all of the above conditions are fulfilled, the vector $\bs{\Psi}\big( \la_1,\dots, \la_{N^{\prime}} \big)$, provided that it is non-vanishing, is associated with the Eigenvalue
\bem
 \wh{\La}\,\big( \xi \mid \{\la_k\}_1^{N^{\prime}} \big) \, = \,  (-1)^N \ex{\f{h}{2T} }\pl{k=1}{N^{\prime}}
 \Bigg\{ \f{ \sinh(\xi-\la_k + \i \tf{\zeta}{2} ) }{ \sinh(\xi-\la_k   -    \i \tf{\zeta}{2})  }  \Bigg\} \cdot
\Bigg( \f{ \sinh\big(\xi + \tfrac{\aleph}{N}   \big) \sinh\big(\xi - \tfrac{\aleph}{N} -\i  \zeta  \big) }{ \sinh^2(-\i\zeta) } \Bigg)^N   \\
\, + \,  (-1)^N  \ex{-\f{h}{2T} } \pl{k=1}{N^{\prime}}  \Bigg\{ \f{ \sinh(\xi-\la_k - 3 \i \tf{\zeta}{2}) }{ \sinh(\xi-\la_k -    \i \tf{\zeta}{2} ) }  \Bigg\} \cdot
 \Bigg(  \f{ \sinh\big(\xi + \tfrac{\aleph}{N} + \i   \zeta  \big) \sinh\big(\xi - \tfrac{\aleph}{N} \big)  }{ \sinh^2(-\i\zeta) }  \Bigg)^N
\label{ecriture valeur propre qtm}
\end{multline}
of $ \op{t}_{ \mf{q} }(\xi)$.

 Note that, if $\la_1,\dots, \la_{N^{\prime}}$ are all pairwise distinct and admissible, the system of Bethe Ansatz equations reduces to the usually encountered form
\beq
\ex{-\tfrac{h}{T}} (-1)^{ \mf{s}  } \pl{k=1}{N^{\prime}} \bigg\{ \f{\sinh( \i\zeta - \la_a + \la_k) }{  \sinh( \i \zeta  + \la_a - \la_k) }  \bigg\}
\cdot \Bigg\{ \f{ \sinh\big(  \la_a - \tfrac{\aleph}{N} + \i \tfrac{\zeta}{2} \big) \sinh\big(   \i \tfrac{3 \zeta}{2} +  \la_a + \tfrac{\aleph}{N}  \big) }
                    {  \sinh\big(  \la_a + \tfrac{\aleph}{N} +  \i \tfrac{\zeta}{2} \big) \sinh\big(   \i \tfrac{\zeta}{2} - \la_a + \tfrac{\aleph}{N} \big)  }    \Bigg\}^{N}
 \; = \; -1 \; , \quad a=1, \dots, N^{\prime} \,.
\label{equations de Bethe cas usuel}
\enq

The work \cite{KozGohmannGoomaneeSuzukiRigorousApproachQTMForFreeEnergy} rigorously
established that, for $N$ and $T$ large enough, the dominant Eigenvalue $\wh{\La}_{\e{max}}$
is given by $\wh{\La}\,\big( \xi \mid \{\la_k\}_1^{N^{\prime}} \big)$ for a clearly
identified admissible, pairwise distinct, solution $\la_1, \dots, \la_{N^{\prime}}$
to the system of Bethe Ansatz equations given above with $N^{\prime}=N$.

Before discussing the calculation of the free energy, \textit{viz}.\
the $N\tend + \infty$ limit of $\ln \wh{\La}_{\e{max}}$, we comment on the
physical relevance of the other Eigenvalues of the quantum transfer matrix.
One may argue, although so far not in full rigour, that static thermal correlation
functions of the XXZ chain which, in the simplest case of two-point functions, are defined as
\beq
\big<  \sg_{1}^{\a} \sg_{m+1}^{\a^{\prime}}  \big>_{T} \; = \; \lim_{L\tend +\infty}
\Bigg\{ \f{ \e{tr}_{ \mf{h}_{XXZ} }\big[ \sg_{1}^{\a} \sg_{m+1}^{\a^{\prime}} \ex{- \f{1}{T} \op{H} } \big] }{  \e{tr}_{\mf{h}_{XXZ}}\big[  \ex{- \f{1}{T} \op{H} }\big]  } \Bigg\}
\quad \e{with} \quad \a, \a^{\prime}\in \{x, y, z\} \;,
\enq
may be represented by the infinite Trotter number limit $N\tend +\infty$ of the
so-called thermal form factor expansion
\cite{KozDugaveGohmannThermaxFormFactorsXXZ,%
KozDugaveGohmannThermaxFormFactorsXXZOffTransverseFunctions}
 \beq
 \big<  \sg_{1}^{\a} \sg_{m+1}^{\a^{\prime}}  \big>_{T} \; = \; \lim_{N\tend + \infty} \Bigg\{
\wh{\mc{F}}^{\, (\a \a^{\prime})}_{0} \; + \; \sul{k=1}{2^{2N}-1}  \Bigg(  \f{ \wh{\La}_k }{ \wh{\La}_{\e{max}} }\Bigg)^m \cdot  \wh{\mc{F}}^{\,(\a \a^{\prime})}_{k}
 \Bigg\} \;.
 \label{ecriture dvpt serie FF thermaux}
 \enq
 In this picture, Trotter limits of logarithms of ratios of Eigenvalues of
 the quantum transfer matrix,
 \beq
 \xi_k \; = \;   \lim_{N\tend +\infty} \bigg\{  \f{-1 }{ \ln \big| \tf{ \wh{\La}_k }{ \wh{\La}_{\e{max}} }\big|  } \bigg\} \quad \e{and} \quad
 \phi_k \; = \;    \lim_{N\tend +\infty} \Im \ln \Big[ \tf{ \wh{\La}_k }{ \wh{\La}_{\e{max}} }\Big] \;,
 \enq
appear as the tower of correlation lengths $\xi_k$ and oscillatory phases
$\phi_k$ which drive the exponential decay in the distance of the two-point
and, more generally, multi-point correlation functions. In their turn, the
$\wh{\mc{F}}^{\,(\a \a^{\prime})}_{k}$ correspond to the so-called thermal
form factors. In the case when the Eigenvector $\wh{\Psi}_k$ of the quantum
transfer matrix is constructible in terms of some admissible solution to the
Bethe Ansatz equations, the thermal form factor $\wh{\mc{F}}^{\,(\a \a^{\prime})}_{k}$
is given by an explicit function of the associated Bethe roots, see
\cite{KozDugaveGohmannThermaxFormFactorsXXZ,%
KozDugaveGohmannThermaxFormFactorsXXZOffTransverseFunctions}.

It is also possible to argue, yet again not in full rigour, that the dynamical
thermal two-point correlation functions of the XXZ chain, defined as
\beq
\big<  \sg_{1}^{\a} \sg_{m+1}^{\a^{\prime}}(t)  \big>_{T} \; = \; \lim_{L\tend +\infty}
\Bigg\{ \f{ \e{tr}_{ \mf{h}_{XXZ} }\big[ \sg_{1}^{\a} \cdot  \ex{\i t \op{H} } \cdot \sg_{m+1}^{\a^{\prime}}  \cdot \ex{ - \i t \op{H} } \ex{- \f{1}{T} \op{H} } \big] }{  \e{tr}_{\mf{h}_{XXZ}}\big[  \ex{- \f{1}{T} \op{H} }\big]  } \Bigg\}
\quad \e{with} \quad \a, \a^{\prime}\in \{x, y, z\} \;,
\enq
may be represented by the infinite Trotter number limit $N\tend +\infty$
of the below thermal form factor expansion
\cite{KozGohmannKarbachSuzukiFiniteTDynamicalCorrFcts}
\beq
\big<  \sg_{1}^{\a} \sg_{m+1}^{\a^{\prime}}(t)  \big>_{T} \; = \; \lim_{N\tend + \infty} \Bigg\{
\wh{\mc{F}}^{\, (\a \a^{\prime})}_{0} \; + \; \sul{k=1}{2^{2N}-1}  \Bigg(  \f{ \wh{\La}_k }{ \wh{\La}_{\e{max}} }\Bigg)^m
 \cdot  \Bigg(  \f{ \wh{\La}_k\big( \tfrac{J \sin \zeta t}{N}  \big) \cdot \wh{\La}_{\e{max}}\big( - \tfrac{J \sin \zeta t}{N}  \big)  }
 { \wh{\La}_{\e{max}}\big( \tfrac{J \sin \zeta t}{N}  \big) \cdot \wh{\La}_k\big( -\tfrac{J \sin \zeta t}{N}  \big) }\Bigg)^N
 \cdot  \wh{\mc{F}}^{\,(\a \a^{\prime})}_{k}
 \Bigg\} \;.
 \label{ecriture dvpt serie FF thermaux dynamique}
 \enq

The static \eqref{ecriture dvpt serie FF thermaux} and dynamic
\eqref{ecriture dvpt serie FF thermaux dynamique} thermal form factor expansions
explain the need to access, in the most explicit way possible,  to the
sub-dominant Eigenvalues $\wh{\La}_{k}(\xi)$, at least when $\xi$ is
sufficiently close to $0$.

\subsection{The conformal field theory predictions}
In the perception of theoretical physicists the XXZ chain is
thermodynamically at a second order critical point at $T = 0$,
a so-called quantum critical point, for all parameters in the
regime $0 < h < 4J (1 + \De)$ and $-1 < \De < 1$ considered in
this work. This has strong implications. Physical systems at
second order phase transition points exhibit universal behaviour.
All static two-point correlation functions are expected to be
conformally invariant. They should decay slowly, as powers
of the distance between the two points. This has been shown to
follow for the XXZ chain, on a theoretical physics level of rigor,
first by using sophisticated expansions for the correlators in
\cite{KozKitMailSlaTerXXZsgZsgZAsymptotics} and later from
from the static form factor series \eqref{ecriture dvpt serie FF thermaux}
in \cite{KozDugaveGohmannThermaxFormFactorsXXZ,%
KozDugaveGohmannThermaxFormFactorsXXZOffTransverseFunctions,KozKitMailSlaTerRestrictedSums}.
Systems close to critical points, \textit{i.e.}\ at low temperatures
in our case, are expected to be effectivly described by conformal
quantum field theories. The spectrum of the effective conformal 
field theory (CFT) should therefore be included in the spectrum of the
the quantum transfer matrix of the considered system at low temperatures.
The low-temperature thermodynamics of the full systems, determined
by its free energy should conincide with that of the effective CFT.
In case of the XXZ chain in the antiferromagnetic massless regime
the relevant CFT is expected to be a theory of relativistic free Bosons,
which has central charge $c=1$. Based on this expectation we can provide
two precise conjectures about part of the infinite Trotter limit
of the spectrum $\Big\{\,  \wh{\La}_k \Big\}$, this in the regime
of temperatures low enough.

The first conjecture
\cite{AffleckCFTPreForLargeSizeCorrPartitionFctonAndLowTBehavior,%
BloteCardyNightingalePredictionL-1correctionsEnergyAscentralcharge}
deals with the form of the $\e{O}(T)$ corrections present in the Trotter
limit (presumed to exist) of $\wh{\La}_{\e{max}}$:
\beq
\lim_{N \tend + \infty} \ln \wh{\La}_{\e{max}} \; = \;  -\f{1}{T} e_0  \, + \,  \f{ \pi }{ 6 \op{v}_F } T \, + \, \e{O}(T^2) \;,
\label{ecriture comportement low T dominant Eigenvalue}
\enq
where $e_0$ is the per-site energy of the XXZ chain's ground state, while
$\op{v}_{F}$ is the Fermi velocity which describes the propagation of
the elementary excitations living at the edge of the model's Fermi-zone, see
\eqref{definition vitesse de Fermi du modele} for an explicit formula for $\op{v}_{F}$.
A heuristic derivation of \eqref{ecriture comportement low T dominant Eigenvalue}
was given in
\cite{DestriDeVegaAsymptoticAnalysisCountingFunctionAndFiniteSizeCorrectionsinTBAFirstpaper,%
KlumperNLIEfromQTMDescrThermoXYZOneUnknownFcton}.

The second conjecture \cite{CardyConformalExponents} describes, more generally, the
behaviour of the infinite Trotter number limit of part of the subdominant Eigenvalues.
More precisely, one may rephrase the latter as follows.
If $k$ is such that
\beq
\lim_{T\tend 0^+} \f{1}{T} \lim_{N\tend +\infty} \ln \Big[ \tf{ \wh{\La}_k }{ \wh{\La}_{\e{max}} } \Big]
\label{ecriture limite rapport vp}
\enq
exists and $ \wh{\La}_k $ belongs to the zero pseudospin $\mf{s}$ sector, \textit{c.f.} \eqref{defintion pseudospin}, then, agreeing upon
\beq
 \Ups\Big( \{p_a^{(\sg)}\}_1^{n_p^{(\sg)}} \, ; \, \{h_a^{(\sg)}\}_1^{n_h^{(\sg)}} \Big) \; = \;
 \sul{a=1}{n_p^{(\sg)}} \Big[ p_a^{(\sg)} \,+\, \f{1}{2} \Big] \, + \, \sul{a=1}{n_h^{(\sg)}} \Big[ h_a^{(\sg)} \,+\, \f{1}{2} \Big] \;,
\label{definition facteur Ups conforme}
\enq
it holds that
\beq
\lim_{N\tend +\infty}  \ln \bigg[ \f{ \wh{\La}_k }{ \wh{\La}_{\e{max}} } \bigg]  \; = \; - \f{ 2\pi T }{ \op{v}_F}
\sul{\sg=\pm}{}  \Ups\Big( \{p_a^{(\sg)}\}_1^{n_p^{(\sg)}} \, ; \, \{h_a^{(\sg)}\}_1^{n_h^{(\sg)}} \Big) \, + \, \e{O}(T^2)
\label{ecriture low T pour spectre impulsion}
\enq
for certain integers $0 \leq p_1^{(\sg)}<\cdots < p_{n_p^{(\sg)} }^{(\sg)}$ and $0 \leq h_1^{(\sg)}<\cdots < h_{n_h^{(\sg)} }^{(\sg)}$
with $n_p^{(\sg)}=n_h^{(\sg)}$. For that same $k$,
it also holds that
\beq
\lim_{N\tend +\infty}  \f{N}{t} \ln \Bigg[  \f{ \wh{\La}_k\big( \tfrac{J \sin \zeta t}{N}  \big) \cdot \wh{\La}_{\e{max}}\big( - \tfrac{J \sin \zeta t}{N}  \big)  }
 { \wh{\La}_{\e{max}}\big( \tfrac{J \sin \zeta t}{N}  \big) \cdot \wh{\La}_k\big( -\tfrac{J \sin \zeta t}{N}  \big) }\Bigg]   \; = \; -   2\pi T
\sul{\sg=\pm}{} \sg  \Ups\Big( \{p_a^{(\sg)}\}_1^{n_p^{(\sg)}} \, ; \, \{h_a^{(\sg)}\}_1^{n_h^{(\sg)}} \Big) \, + \, \e{O}(T^2)  \;.
\label{ecriture low T pour spectre energie}
\enq

Furthermore, the conjecture states that, running over all possible values of $k$
compatible with $\mf{s}=0$ such that the limit \eqref{ecriture limite rapport vp}
exists, will give rise to all possible choices of the integers $n_p^{(\sg)} = n_h^{(\sg)}$,
$ p_a^{(\sg)}$ and $h_a^{(\sg)}$ in \eqref{ecriture low T pour spectre impulsion}
and \eqref{ecriture low T pour spectre energie} that are subject to the constraint
$n_p^{(+)} + n_p^{(-)} = n_h^{(+)} + n_h^{(-)}$. The conjecture thus establishes
the precise correspondence between the low-$T$ and infinite Trotter number limit
of part of the spectrum of the quantum transfer matrix on the one hand and the
spectra of the $\op{L}_0+\ov{\op{L}}_0$ and $\op{L}_0-\ov{\op{L}}_0$ operators
in the free Boson $c=1$ conformal field theory on the other hand.

Indeed, recall
\cite{AlexandrovZaborodinTechniquesOfFreeFermions,DiFrancescoMathieuSenechalCFTKniga} that
in the free Boson $c=1$ conformal field theory, the Virasoro energetic mode operators
$\op{L}_0$, resp.\ $\ov{\op{L}}_0$, admit a complete set of Eigenvectors labelled
by the above introduced sets of integers,
\beq
\Phi \Bigl( \{p_a^{(+)}\}_1^{n_p^{(+)}} \, ; \, \{h_a^{(+)}\}_1^{n_h^{(+)}} \Bigr) \;,
   \quad \e{resp}.\quad
\ov{\Phi} \Bigl( \{p_a^{(-)}\}_1^{n_p^{(-)}} \, ; \, \{h_a^{(-)}\}_1^{n_h^{(-)}} \Bigr)    \;.
\enq
The complete family of Eigenvectors, in the sector of the theory with
$n_p^{(\sg)}-n_h^{(\sg)}=\sg \ell^{(\sg)} \in \mathbb{Z}$ fixed,
is obtained by scanning through all possible choices of $n_p^{(\sg)}$, $n_h^{(\sg)}$
compatible with the above constraint and $0 \leq p_1^{(\sg)} < \cdots
< p_{n_p^{(\sg)} }^{(\sg)}$ and $0 \leq h_1^{(\sg)}<\cdots < h_{n_h^{(\sg)} }^{(\sg)}$.
Moreover, it holds that
\beq
\op{L}_0 \cdot \Phi \Big( \{p_a^{(+)}\}_1^{n_p^{(+)}} \, ; \, \{h_a^{(+)}\}_1^{n_h^{(+)}} \Big)   \; = \;  \Ups\Big( \{p_a^{(+)}\}_1^{n_p^{(+)}} \, ; \, \{h_a^{(+)}\}_1^{n_h^{(+)}} \Big) \cdot
\Phi \Big( \{p_a^{(+)}\}_1^{n_p^{(+)}} \, ; \, \{h_a^{(+)}\}_1^{n_h^{(+)}} \Big)
\enq
and analogously for $\ov{\op{L}}_0 $ under the substitution
$\Phi  \hookrightarrow \ov{\Phi}$ and $+ \hookrightarrow -$.

\vspace{3mm}

We would also like to point out that one may reinterpret the expansion of the static correlation functions
in measure theoretic terms and then translate the statement about the universality
of the Eigenvalue ratios into one about the universality of the rate function
describing  a large deviation like phenomenon in that case. Indeed, assume that
the Trotter limit in \eqref{ecriture dvpt serie FF thermaux} may be taken pointwise
so that
\beq
\big<  \sg_{1}^{\a} \sg_{m+1}^{\a^{\prime}}  \big>_{T} \; = \;
\mc{F}^{\, (\a \a^{\prime})}_{0} \; + \; \sul{k \in \mathbb{N} }{}  \ex{- m \ga_k } \cdot  \mc{F}^{\,(\a \a^{\prime})}_{k}  \qquad \e{with} \qquad
\left\{ \ba{ccc}  \ga_k & = & \underset{N \rightarrow \infty}{\lim} \ln \Biggl[\f{\wh{\La}_k}{\wh{\La}_{\e{max}}}\Biggr]  \vspace{4mm} \\
 \mc{F}^{\,(\a \a^{\prime})}_{k} & = & \underset{N \rightarrow \infty}{\lim} \wh{\mc{F}}^{\,(\a \a^{\prime})}_{k} \ea \right. \;.
\enq
Then, one may introduce a mass 1 complex valued sequence of measures $\mathbb{M}_m^{(\a\a^{\prime})}$ on $\mathbb{N}$
with density
\beq
\mf{m}_m^{(\a\a^{\prime})}(k) \; = \; \f{  \ex{- m \ga_k } \cdot  \mc{F}^{\,(\a \a^{\prime})}_{k} }{ \big<  \sg_{1}^{\a} \sg_{m+1}^{\a^{\prime}}  \big>_{T}  } \;.
\enq
Assuming moreover that one  has $\Re(\ga_k)\not= \Re(\ga_{k^{\prime}})$ if $k\not=k^{\prime}$,
for any $\mc{E} \subset \mathbb{N}$, it holds that
\beq
\lim_{m\tend +\infty}\bigg\{\f{1}{m} \ln \bigl(\mathbb{M}_m^{(\a\a^{\prime})}( \mc{E} )\bigr)
      \bigg\} \; = \; -\ga_{k} \quad \e{with} \; k \;\; \e{such}\; \e{that} \quad
\Re(\ga_k)= \e{inf}\Big\{  \Re( \ga_{s} ) \; : \; s \in \mc{E} \Big\} \;.
\label{label ecriture mesure Mm comme un LDP}
\enq
This allows to reinterpret the behaviour of the measure in a setting analogous
to the large deviation principle which holds for probability measures.
Then $ k \mapsto  \ga_{k} $ appears as the generalisation of the concept of
a rate function to that setting. The measures $\mathbb{M}_m^{(\a\a^{\prime})}$
depend on the external parameter $T$ and the statement about universality
may be reinterpreted as a statement about the universality of the generalisation
of the rate function to that setting when $T \tend 0^+$. Of course, nothing ensures that the moduli
of the Eigenvalues of the quantum transfer matrix are all distinct, which would make the
limit \eqref{label ecriture mesure Mm comme un LDP} ill-defined. However, we
find the above picture sufficiently neat to be mentioned.

\subsection{CFT predictions from the perspective of our results}

The completeness of the Bethe Ansatz for a finite Trotter number quantum
transfer matrix considered in this work -\textit{c.f.} Subsubsection \ref{SousousSection preuve energie libre et identification vp dominante} for more details-
is, strictly speaking,
still an open issue. However, for the generic, inhomogeneous and
twisted six-vertex model completeness has been established in
\cite{TarasovVarchenkoCompletenessBAXXZ}. By choosing a slightly different
Trotter discretisation of the Boltzman statistical operator which induces 
a generic, inhomogeneous form of the quantum transfer matrix, the results of \cite{TarasovVarchenkoCompletenessBAXXZ}
apply and the Bethe Ansatz is complete. However, this information does not
help in any way to compute explicitly the ratios of Eigenvalues
$\wh{\La}_k/\wh{\La}_{\e{max}}$ in the $N \tend + \infty$ infinite Trotter
number limit. For an explicit construction, one needs sufficient insight
into the large-$N$ behaviour of the solutions to the Bethe Ansatz equations.
As already stressed, the main purpose of this work is to develop a substantial
part of the technical aspects of the analysis of the Bethe Ansatz equations
governing the spectrum of the quantum transfer matrix. This is achieved by
pushing forward the analysis of the non-linear integral equations, whose
solution allows one to access the Bethe roots, \textit{i.e.}\ the solutions
to the Bethe Ansatz equations. As opposed to the approach developed in
\cite{TarasovVarchenkoCompletenessBAXXZ} our approach does \textit{not}
allow us to discuss the issue of the completeness of the Bethe Ansatz as
this is a topic requiring a completely different set of tools.

Still, what it does allow us to construct explicitly, is a certain
\textit{subset} $\mc{B}_{\e{NLIE}}(B,\wt{B})$ of solutions to the Bethe
Ansatz equations, \textit{c.f.}\ \eqref{ecriture eqns Bethe Trotter fini}
and \eqref{equations de Bethe cas usuel}, which we believe to be
"physically complete": for any $\eps>0$, there exist $B, \wt{B}>0$
such that any Eigenvalue $\wh{\La}_s$ of the quantum transfer matrix  not described by
$\mc{B}_{\e{NLIE}}(B,\wt{B})$ is such that $\big| \wh{\La}_s\big| \, <\, \eps$.

The subset $\mc{B}_{\e{NLIE}}(B,\wt{B})$ of solutions to the Bethe Ansatz
equations is parameterised by four, possibly empty, sequences of
strictly increasing integers
\beq
0< p_1^{(\a)} <\dots < p_{ |\mc{Y}^{(\a)}| }^{(\a)}  \, \leq \, \f{ B }{ T }  \quad \e{and} \qquad 0< h_1^{(\a)} <\dots < h_{ |\mf{X}^{(\a)}| }^{(\a)} \, \leq \, \f{ B }{ T }
\label{ecriture suite croissante entiers p et h bornee en T}
\enq
with $\a \in \{ L, R \}$, and a relative integer $\mf{s} \in \mathbb{Z}$. The cardinalities
$|\mc{Y}^{(\a)}|, |\mf{X}^{(\a)}| \in \mathbb{N}$ are such that
\beq
|\mf{s}| \, + \,  \sul{\a \in L, R}{} \Big\{ |\mc{Y}^{(\a)}| \, + \,   |\mf{X}^{(\a)}| \Big\} \, \leq \, \wt{B} \; .
\label{ecriture limitation sur nombre entiers particle et trou}
\enq
Here, $B, \wt{B}$ can be taken arbitrary from the start, but should remain fixed, \textit{i.e.}\
$N, T$ independent. We refer to Theorems
\ref{Theorem classification forme racines particules et trous}
-\ref{Theorem existence solutions racines particules et trous} for more details on $\mc{B}_{\e{NLIE}}(B,\wt{B})$.

\vspace{2mm}

Moreover, we show that the solutions belonging to $\mc{B}_{\e{NLIE}}(B,\wt{B})$  always correspond to pairwise distinct Bethe roots,
\textit{i.e.}\ one is always in the framework of solutions to  \eqref{equations de Bethe cas usuel}.
We stress that this is an \textit{a posteriori} result in the sense that the analysis of the large class of admissible solutions to \eqref{ecriture eqns Bethe Trotter fini}
we focus on, leads us to conclude that, within the considered class, only solutions satisfying \eqref{equations de Bethe cas usuel} do exist.
We refer to Proposition \ref{Proposition non existence solutions NLIE avec entier coincidant qui donne une solution pblm non lineaire} for more details.
Further, we show in Proposition \ref{Proposition non nullite vecteur de Bethe via formule des normes} that the Bethe vectors built out from such solutions
are non-zero, hence ensuring that the solutions we study to the Bethe Ansatz equations do give rise to geniuine Eigenvectors and provide closed form expressions for the
associated Eigenvalues.

\vspace{2mm}

In the parameter space that we consider, that is for large Trotter number $N$
and low-temperature $T$, we are able to provide a very precise control on the
Eigenvalues of the quantum transfer matrix built from the Bethe roots belonging
to $\mc{B}_{\e{NLIE}}(B,\wt{B})$, see Propositions~\ref{Propostion DA fin pour la vp de la QTM} and
\ref{Propostion DA fin pour la fnelle energetique}. In particular, this allows
us to explicitly identify the Eigenvalue $\wh{\La}_{\e{max};\e{BA}}$ of largest
modulus among the Eigenvalues built from roots in $\mc{B}_{\e{NLIE}}(B,\wt{B})$, see
Proposition~\ref{Propostion infimum pour les vp de la QTM}. We stress that
``the'' in front of ``Eigenvalue'' emphasises that, within the considered class,
$\wh{\La}_{\e{max};\e{BA}}$ is non-degenerate. As a sample result, we provide
the characterisation of the Eigenvalue $\wh{\La}_{\e{max};\e{BA}}$ below.
For the much larger scope of results we obtain in addition, we refer to a thorough discussion
thereof that will be carried out in Section~\ref{Section Main Results}.

\begin{theorem}
\label{Theorem form close La max BA}

There exists $\eta>0$, $T_0>0$ and $C_{\mc{M}}^{(0)}>0$ such that, for any
\beq
T_0 > T > 0 \;  , \quad  \eta > \f{1}{N T^4} \;,
\enq
the quantum transfer matrix admits the Eigenvalue
\beq
\wh{\La}_{\e{max};\e{BA}} \, = \,
\bigg( \f{ \sinh(\tfrac{\aleph}{N} +\i\zeta  )  \cdot \sinh(\tfrac{\aleph}{N} +\i\zeta  ) }{ \sinh^2(  \i\zeta) }  \bigg)^{N} \cdot
\exp \Bigg\{  \f{h}{2T}      \ - \, \Oint{ \msc{C}_{\e{ref}}  }{ } \f{ \dd \mu }{ 2\pi } \: p_0^{\prime}(\mu-\xi)
\msc{L}\mathrm{n}_{\msc{C}_{\e{ref}}}\big[ 1+  \ex{ - \f{1}{T}\wh{u} } \, \big](\mu) \Bigg\} \;.
\label{ecriture Trotter fini VP dominantes BA}
\enq
In this expression
\beq
p_0(\la) \, = \, \i \ln \bigg(  \f{ \sinh(\i\tf{\zeta}{2}+ \la) }{  \sinh(\i\tf{\zeta}{2} - \la) } \bigg) \;,
\label{definition de p0}
\enq
the integration curve $\msc{C}_{\e{ref}}$ is as defined in \eqref{definition contour Cref}, \textit{c.f.}\ Fig.~\ref{contour integration Cref}, and the logarithm $\msc{L}\mathrm{n}_{ \msc{C}_{\e{ref}} }$ is
introduced in \eqref{definition logarithm de 1+a}. Finally, the function $\wh{u}$ corresponds to the unique solution to the non-linear integral equation
\beq
\wh{u}\,(\xi) \, = \,  h \, -\, T\mf{w}_N(\xi)  \; - \; T\Oint{ \msc{C}_{\e{ref}}  }{} \dd \la \: K(\xi-\la) \cdot  \msc{L}\mathrm{n}_{ \msc{C}_{\e{ref}} }\Big[ 1+  \ex{ - \f{1}{T}\wh{u} } \, \Big](\la)
\label{ecriture NLIE Trotter fini VP dominante}
\enq
on the space $\wh{\mc{E}}_{\mc{M}}$ introduced in \eqref{definition espace EM hat} with a constant $ C_{\mc{M}} \, >\,  C_{\mc{M}}^{(0)}$ and
subject to the null index condition
\beq
0 \, = \, - \Oint{ \msc{C}_{\e{ref}}  }{ }  \f{\dd \la}{2 \i \pi T} \: \f{  \wh{u}^{\, \prime}\!(\la) }{ 1+  \ex{ \f{1}{T} \wh{u}(\la)} } \;.
\label{ecriture null index condition}
\enq
The driving term $\mf{w}_N(\xi)$ appearing above is introduced in \eqref{definition fct mathfrak w N}.

\vspace{2mm}

\noindent The unique solution $\wh{u}$ satisfies
\beq
\big(\, \wh{u}(-\xi^*) \, \big)^*   = \, \wh{u}(\xi) \, ,
\label{ecriture propriete conjugaison complexe de u vp dominante}
\enq
$z^*$ being the complex conjugate of $z$,
while $\wh{\La}_{\e{max};\e{BA}} $ given above is real valued. Moreover,  $\zeta \mapsto \wh{\La}_{\e{max};\e{BA}}$ is real analytic on $\intof{0}{\pi/2}$.

Finally, $\wh{\La}_{\e{max};\e{BA}}$ is maximal with respect to the other Eigenvalues of the quantum transfer matrix computed by the Bethe Ansatz solutions belonging to $\mc{B}_{\e{NLIE}}(B,\wt{B})$.

\end{theorem}

Here we stress that the existence and uniqueness of solutions to
the mentioned non-linear integral equation and several other related
non-linear problems are addressed in
Theorem~\ref{Theorem ppl existance solution NLIE}.

In order to apply our results on the rigorous analysis of solutions
to the Bethe Ansatz equations to the \textit{per se} thermodynamics
of the XXZ chain, one still needs to resort to two conjectures.
Most probably, their proofs will rely on techniques of analysis very
different from those developed in the present work. We expect,
in particular, that these conjectures hold in a much more general
context and that their validity is not restricted to integrable models.

\begin{conj}
\label{Conjecture VP dominante matrice transfer}
For $N$ large enough and any $T \geq  a_N>0$, for some $a_N \tend 0$ as
$N\tend +\infty$, the quantum transfer matrix admits a dominant Eigenvalue
$\wh{\La}_{\e{max}}$ that is non-degenerate, \textit{i.e.}
\beq
\big| \wh{\La}_{\e{max}} \big| \, > \,  \big| \wh{\La}_{k} \big| , \qquad for \; any \quad  k \in  \intn{1}{2^{2N}-1}
\enq
and
\beq
\e{dim}\Big\{ \e{ker}\big[ \op{t}_{\mf{q}}(0) - \wh{\La}_{\e{max}} \big] \Big\} \; = \;1 \;.
\enq
\end{conj}

The second conjecture pertains to the commutativity of the thermodynamic
and Trotter limits for any $T>0$.

\begin{conj}
\label{Conjecture commutativite vp QTM}

For any $T>0$, the thermodynamic $L\tend +\infty$ and Trotter $N \tend + \infty$ limits commute,
\beq
 \lim_{L\tend +\infty}  \lim_{N\tend + \infty} \Big\{ \f{1}{L} \ln \e{tr}_{ \mf{h}_{\mf{q}} } \Big[ \big( \op{t}_{ \mf{q} }(0)\big)^{L} \Big]    \Big\}
\, = \, \lim_{N\tend +\infty}  \lim_{L\tend + \infty} \Big\{ \f{1}{L} \ln \e{tr}_{ \mf{h}_{\mf{q}} } \Big[ \big( \op{t}_{ \mf{q} }(0)\big)^{L} \Big]    \Big\}  \;.
\enq
\end{conj}

We stress that both conjectures were proven, for $T$ large enough, in \cite{KozGohmannGoomaneeSuzukiRigorousApproachQTMForFreeEnergy}.

There are several interesting implications one can draw from these conjectures.
First of all,

\begin{theorem}
\label{Theorem identification vp dominante et energie libre XXZ}
If Conjecture \ref{Conjecture VP dominante matrice transfer} holds, then
$\wh{\La}_{\e{max}} \in \R$. Moreover, there exists $T_0, \eta,  C>0$ such
that, for any $T_0 > T > CN^{-3}$, $\eta > \tf{1}{(TN^4)}$,
\beq
\wh{\La}_{\e{max};\e{BA}} \, = \, \wh{\La}_{\e{max}}\, \;.
\enq

Provided that also Conjecture \ref{Conjecture commutativite vp QTM} holds,
there exists $T_0 > 0$ such that, for any $T_0 > T > 0$,
the free energy of the $XXZ$ spin-$\tf{1}{2}$ chain introduced in
\eqref{definition energie libre}, may be expressed as
%
%
%
\beq
     f \, = \,  2 J \cos(\zeta) - \frac{h}{2} + T
                \Oint{ \msc{C}_{ \e{ref}}  }{ }
		\f{ \dd \la }{ 2   \pi } 	p_0^{\prime}(\la ) \,
		\msc{L}\mathrm{n}_{\msc{C}_{\e{ref}}}
		\big[ 1+  \ex{ -\f{1}{T}u   } \, \big](\la ) \;,
\label{ecriture forme close energie libre XXZ}
\enq
where $u$ is the unique solution on the space $\mc{E}_{\mc{M}}$
\eqref{definition espace fonctionnel principal} to the non-linear integral equation
\beq
u\,(\xi) \, = \,   \veps_0(\xi)    \; - \; T\Oint{ \msc{C}_{\e{ref}}  }{} \dd \la \: K(\xi-\la) \cdot  \msc{L}\mathrm{n}_{ \msc{C}_{\e{ref}} }\Big[ 1+  \ex{ - \f{1}{T}u } \, \Big](\la) \;.
\label{ecriture NLIE pour u VP Dominante}
\enq
The contour $\msc{C}_{\e{ref}}$ is defined in \eqref{definition contour Cref}, \textit{c.f.}\
Fig.~\ref{contour integration Cref}, and the logarithm $\msc{L}\mathrm{n}_{ \msc{C}_{\e{ref}} }$ is
introduced in \eqref{definition logarithm de 1+a}. The XXZ bare momentum $p_0$ and
energy $\veps_0$ arising in these equations are, respectively, introduced in
\eqref{definition de p0} and \eqref{definition eps 0}.

Finally, one has the $T \tend 0^+$ asymptotic behaviour
\beq
     f \, = \,  2 J \cos(\zeta) - \frac{h}{2} +
            \, \Int{-q}{q} \f{ \dd \mu }{ 2 \pi } \veps(\mu) p_{0}^{\prime}(\mu)
	    \, - \, \f{\pi}{6 \op{v}_F } T^2 \, + \e{O}\bigl(T^3\bigr) \;.
\label{ecriture DA basse T de energie libre}
\enq
The dressed energy $\veps$ appearing in the leading term is defined by
means of the solution to the linear integral equation
\eqref{definition energie habille et energie nue} and the auxiliary
conditions given below it. In its turn, the Femi velocity  $\op{v}_{F}$
is defined through \eqref{definition vitesse de Fermi du modele}.
\end{theorem}

The existence and uniqueness of solutions to the mentioned non-linear
integral equation and several other related non-linear problems are
addressed in Theorem \ref{Theorem ppl existance solution NLIE}.

Our analysis of solutions to the Bethe Ansatz equations also allows us
to show that the spectrum of the quantum transfer matrix does contain
a subset of Eigenvalues whose appropriate scaling limit reproduces the
spectrum of the free Boson model. Namely, we have the

\begin{theorem}
\label{Theorem structure conforme du spectre}
There exists a sequence of Eigenvalues $\wh{\La}_k$ of the quantum transfer matrix
at zero spectral parameter such that
\beq
\lim_{N\tend +\infty} \bigg\{  \ln \bigg[ \f{ \wh{\La}_k }{ \wh{\La}_{\e{max};\e{BA}} } \bigg]  \bigg\} \; = \; - \f{ 2\pi T }{ \op{v}_F}
\sul{\sg=\pm}{}  \Ups\Big( \{p_a^{(\sg)}\}_1^{n_p^{(\sg)}} \, ; \, \{h_a^{(\sg)}\}_1^{n_h^{(\sg)}} \Big) \, + \, \e{O}(T^2)
\enq
with $\Ups$ as introduced in \eqref{definition facteur Ups conforme}.
For that very same $k$, it holds that
\beq
\lim_{N\tend +\infty} \Bigg\{  \f{N}{t} \ln \Bigg[  \f{ \wh{\La}_k\big( \tfrac{J \sin \zeta t}{N}  \big) \cdot \wh{\La}_{\e{max;\e{BA}}}\big( - \tfrac{J \sin \zeta t}{N}  \big)  }
 { \wh{\La}_{\e{max}}\big( \tfrac{J \sin \zeta t}{N}  \big) \cdot \wh{\La}_k\big( -\tfrac{J \sin \zeta t}{N}  \big) }\Bigg]  \Bigg\} \; = \; -  2\pi T
\sul{\sg=\pm}{} \sg  \Ups\Big( \{p_a^{(\sg)}\}_1^{n_p^{(\sg)}} \, ; \, \{h_a^{(\sg)}\}_1^{n_h^{(\sg)}} \Big) \, + \, \e{O}(T^2)  \;.
\enq

Moreover, among the Eigenstates obtained from the class of Bethe Ansatz based
Eigenvalues with particle-hole parameters subject to Hypotheses
\ref{Hypotheses solubilite NLIE}-\ref{Hypotheses eqns de quantification},
these are the only limits of Eigenvalues which can be obtained when
$N\tend +\infty$ and whose low-$T$ asymptotic expansion starts, in both cases,
by a non-vanishing linear in $T$ contribution.

\end{theorem}

The details of our analysis provide a crystal clear identification of which solution to
the Bethe Ansatz equations does provide, to the leading order in $T$, the given Eigenvalue of the free Boson model
subordinate to the choice of integers
\beq
0\leq p_1^{(\pm)} \, < \, \cdots \, < \, p_{n_p^{(\pm)}}^{(\pm)} \qquad and \qquad
0\leq h_1^{(\pm)} \, < \, \cdots \, < \, h_{n_h^{(\pm)}}^{(\pm)} \;,
\enq
with $n_{p}^{(\pm)}=n_h^{(\pm)}$ and $\mf{s}=0$.

\subsection{The non-linear integral equation approach to the Bethe Ansatz equations}

While already hardly solvable in any explicit or manageable manner at finite
Trotter numbers, the system of Bethe Ansatz equations, despite several
heuristic attempts in the literature, see \textit{e.g.} \cite{TakahashiThermoXXZInfiniteNbrRootsFromQTM},
does not appear as a reasonable means for computing the infinite Trotter number
limit of the physical observables such as the dominant Eigenvalue $\wh{\La}_{\e{max}}$
and the Eigenvalue ratios $\tf{ \wh{\La}_k }{ \wh{\La}_{\e{max}} }$. It turns out
that an equivalent formulation of the Bethe Ansatz equations which appeared in
the physics literature \cite{%
BatchelorKlumperFirstIntoNLIEForFiniteSizeCorrectionSpin1XXZAlternativeToRootDensityMethod,%
DestriDeVegaAsymptoticAnalysisCountingFunctionAndFiniteSizeCorrectionsinTBAFirstpaper,%
DestriDeVegaAsymptoticAnalysisCountingFunctionAndFiniteSizeCorrectionsinTBAFiniteMagField,%
KlumperNLIEfromQTMDescrThermoRSOSOneUnknownFcton,%
KlumperNLIEfromQTMDescrThermoXYZOneUnknownFcton}
of the early 90s is the most appropriate one for studying the infinite Trotter
number limit. The matter is that there is a bijection between solutions to the
Bethe Ansatz equations and solutions to certain non-linear integral equations.
On rigorous grounds, one may formulate this correspondence as follows.

First of all, introduce the function
\beq
\mf{w}_N(\xi) \; = \;  N \ln \bigg(  \f{ \sinh(  \xi - \tf{ \aleph }{N}  + \i\tf{\zeta}{2} )  }{  \sinh(  \xi + \tf{ \aleph }{N}  + \i\tf{\zeta}{2} )   }   \bigg)  \; + \;
N \ln \bigg(  \f{ \sinh(  \xi + \tf{ \aleph }{N}  - \i\tf{\zeta}{2} )  }{  \sinh(  \xi - \tf{ \aleph }{N}  - \i\tf{\zeta}{2} )   }   \bigg) \;,
\label{definition fct mathfrak w N}
\enq
where the cuts of the logarithms are given, up to $\i\pi$-periodicity, by the segment
$\intff{\tfrac{ \aleph }{N} - \i\tfrac{\zeta}{2}}{-\tfrac{ \aleph }{N} - \i\tfrac{\zeta}{2}}$
in what concerns the first term and
$\intff{\tfrac{ \aleph }{N} + \i\tfrac{\zeta}{2}}{-\tfrac{ \aleph }{N} + \i\tfrac{\zeta}{2}}$
in what concerns the second one.

Further, introduce
\beq
\label{definition_theta}
     \th(\la) \; = \;  \left\{
        \ba{ccc}
        \i \ln \bigg( \f{ \sinh(\i\zeta+\la) }{ \sinh(\i\zeta-\la) }\bigg)
	& \text{for} &  |\Im(\la)| \, < \,  \e{min}(\zeta, \pi-\zeta) \\[3ex]
        -\pi \e{sgn}(\pi-2\zeta)   +\i \ln \bigg( \f{ \sinh(\i\zeta+\la) }{ \sinh(\la - \i\zeta) }\bigg)
        & \text{for}  &   \e{min}(\zeta, \pi-\zeta) \, < \, |\Im(\la)| \, < \, \tf{\pi}{2} \ea  \right.  \;,
\enq
where ``$\ln$'' corresponds to the principal branch of the logarithm.
The above definition makes $\th$ an $\i\pi$-periodic holomorphic function on
$\Cx \setminus \bigcup_{\ups=\pm} \Big\{ \R^+ \, + \ups \, \i \zeta \,
+\,  \i\pi \mathbb{Z} \Big\}$ with cuts along the curves $ \R^+ \, \pm \,
\i \zeta \, +\,  \i\pi \mathbb{Z}$. In the following, $\th_+(z)$
will stand for the $+$ boundary value of $\th$, \textit{viz}.\ the limit
$\th_+(z) \, = \, \lim_{\eps\tend 0^+} \th(z+\i\eps)$. This regularisation
is only needed if $z \in \Big\{ \R^+ \, \pm \, \i \zeta \,
+\,  \i\pi \mathbb{Z} \Big\}$.

Then, given an admissible solution to the Bethe Ansatz equations
$\la_1,\dots,\la_{N^{\prime}}$, if one introduces
\beq
     \wh{u}\,(\xi) \; = \; h -T \mf{w}_{N}(\xi) \, -\i\pi T \mf{s}
        + \i T N \theta_+ \big( \xi + \i \tfrac{\zeta}{2} + \tfrac{\aleph}{N} \big)
	- \i T \sul{k=1}{N^{\prime}}\th_{+}\big( \xi - \la_k \big) \;,
\enq
it is direct to see that
\beq
\ex{ - \f{ 1 }{T} \wh{u}\,(\xi)  } \; = \; \ex{-\tfrac{h}{T}} (-1)^{ \mf{s}  }   \pl{k=1}{N^{\prime}} \bigg\{ \f{\sinh( \i\zeta - \xi + \la_k) }{  \sinh( \i \zeta  +  \xi - \la_k) }  \bigg\}
\cdot  \Biggl\{ \f{ \sinh\big(   \xi - \tfrac{\aleph}{N} + \i \tfrac{\zeta}{2} \big) \sinh\big(   \i \tfrac{3 \zeta}{2} +  \xi + \tfrac{\aleph}{N}  \big) }
                    {  \sinh\big(   \xi + \tfrac{\aleph}{N} +  \i \tfrac{\zeta}{2} \big) \sinh\big(   \i \tfrac{\zeta}{2} -  \xi + \tfrac{\aleph}{N} \big)  }    \Biggr\}^{N}
\label{definiton hat u en terme racines de Bethe}
\enq
is such that the Bethe roots $\la_1,\dots, \la_{N^{\prime}}$ arise as a subset of the
zeroes, repeated according to their multiplicity, of the meromorphic function
$1+ \ex{ - \f{ 1 }{T} \wh{u}  } $. More precisely, it holds that
\beq
\Dp{\xi}^p\Big( 1+ \ex{ - \f{ 1 }{T} \wh{u}(\xi)} \,  \Big) \Bigr|_{\xi = \la_a}
= 0 \quad \e{for} \quad p=0,\dots, k_{\la_a}-1 \quad \e{and} \quad
\Dp{\xi}^{k_{\la_a}} \Big( 1+ \ex{ - \f{ 1 }{T} \wh{u}(\xi)} \,
\Big) \Bigr|_{\xi =\la_a} \not= 0  \;.
\enq
This fact, along with some information about the \textit{locii} of the given Bethe roots
allows one to reconstruct these from the knowledge of $\,\wh{u}\,(\xi)$.
The matter is that one may derive a non-linear integral equation which,
in principle, characterises $\wh{u}\,(\xi)$ in a Bethe Ansatz data independent way.
In order to state this result, we need to introduce an additional auxiliary
function, called the kernel function, which will play a prominent role in this
work:
\beq
K(\xi) \, = \, \f{1}{2\pi} \th^{\prime}(\xi) 
\, = \, \f{ \sin(2\zeta)  }{ 2 \pi \sinh(\xi-\i\zeta) \sinh(\xi+\i\zeta)  }\;.
\label{definition noyau integral K}
\enq

\begin{prop}
\label{Proposition description pblm non lineaire}
Let $\mc{D} \subset \mathbb{C}$ be any bounded, simply connected domain
with $\pm \tf{\aleph}{N}-\i\tf{\zeta}{2} \in \mc{D}$, such that $z_1, z_2 \in
\mc{D}$ and $\Re (z_1 - z_2) = 0$ implies $|\Im(z_1 - z_2)| <  \e{min}(\zeta, \pi-\zeta)$.
Then, for every admissible solution $\la_1,\dots, \la_{N^{\prime}}$ to
the Bethe Ansatz equations \eqref{ecriture eqns Bethe Trotter fini},
there exists a solution to the following non-linear problem.

\noindent Find
\begin{itemize}
\item
$\wh{u}$ piecewise continuous on $\Dp{}\mc{D}$;
\item
a collection of roots  $\wh{x}_1,\dots, \wh{x}_{ |\wh{\mf{X}}| }$ with
$\wh{x}_a \in  \ov{\mc{D}} \setminus  \Big\{  \tfrac{\aleph}{N}-\i\tfrac{\zeta}{2},
-\tfrac{\aleph}{N} -\i\tfrac{\zeta}{2} \Big\}$ for $a=1,\dots, | \wh{\mf{X}}|$,
denoted by $\wh{\mf{X}}$;
\item
a collection of admissible roots $ \wh{y}_1,  \dots, \wh{y}_{|\wh{\mc{Y}}|}$ with
$\wh{y}_a \in  \Big\{ z \in \Cx \, : \, -\tfrac{\pi}{2} < \Im(z)
\leq \tfrac{\pi}{2} \Big\} \setminus \ov{\mc{D}}$ for $a=1,\dots, | \wh{\mc{Y}}|$,
denoted by $\wh{\mc{Y}}$;
\end{itemize}
\noindent such that
\begin{itemize}
\item
$\ex{ - \f{1}{T} \wh{u} }$ extends to a meromorphic, $\i\pi$-periodic function
on $\Cx$, whose only pole in $\ov{\mc{D}}$ is $N$-fold and located at $- \aleph/N - \i \zeta/2$, of order $N$,
and at the `singular roots' defined by
\beq
    \wh{y}_{a;\e{sg}} \, = \, \wh{y}_a - \i \, \e{sgn}(\pi - 2 \zeta) \cdot   \e{min}(\zeta, \pi-\zeta) \in \ov{\mc{D}}
\enq
for some $\wh{y}_a$ in $\wh{\mc{Y}}$, the order of the pole at $\wh{y}_{a;\e{sg}}$
being given by the multiplicity of $\,\wh{y}_a$, while the collection of singular
roots is denoted $\wh{\mc{Y}}_{\e{sg}}$;
%
%
%
%
%
\item
for any $a=1,\dots, | \wh{\mf{X}}|$, resp.\ $a=1,\dots, | \wh{\mc{Y}}|$,
\beq
\ba{ccccc}
\Dp{\xi}^{r} \Big\{ \ex{ -\f{1}{T} \wh{u}( \xi ) } \Big\} \Bigr|_{\xi=\wh{x}_a} = - \de_{r,0} & \; for \; any &    r=0,\dots, k_{\, \wh{x}_a}-1 & \; and
& \Dp{\xi}^{ k_{\, \wh{x}_a} } \Big\{ \ex{ -\f{1}{T} \wh{u}( \xi ) } \Big\} \Bigr|_{\xi=\wh{x}_a} \not= 0 \;, \vspace{2mm} \\
\Dp{\xi}^{r} \Big\{ \ex{ -\f{1}{T} \wh{u}( \xi ) } \Big\} \Bigr|_{\xi=\wh{y}_a} = - \de_{r,0} & \; for \; any &     r=0,\dots, k_{\, \wh{y}_a}-1 & \quad and
& \Dp{\xi}^{ k_{\, \wh{y}_a} } \Big\{ \ex{-\f{1}{T} \wh{u}( \xi ) } \Big\} \Bigr|_{\xi=\wh{y}_a} \not= 0  \ea \;
\label{equation quantification particles et tous cas Trotter fini}
\enq
with $k_{\, \wh{x}_a}$, resp.\ $k_{\, \wh{y}_a}$, being the multiplicity of
the root $\wh{x}_a$, resp.\ $\wh{y}_a$;
\item
$\wh{u}$ is subject to the monodromy constraint
\beq
     \mf{m} \, = \, - \Oint{ \Dp{}\mc{D} }{ }
	\f{ \dd \la }{2\i\pi T } \:
        \f{  \wh{u}^{\, \prime}\!(\la) }{ 1+  \ex{ \f{1}{T} \wh{u}(\la)} }
	\; = \; - \mf{s}  - |\wh{\mc{Y}}|-|\wh{\mc{Y}}_{\e{sg}}| + |\wh{\mf{X}}| \; ,
\label{ecriture contrainte de monodromie}
\enq
with $\mf{s}  \in \mathbb{Z}$ being the spin, $|\wh{\mc{Y}}_{\e{sg}}|$ standing for
the number of singular roots counted with multiplicities and the integral being understood
in the sense of a $+$ boundary value in case that $1 + \ex{ - \f{1}{T} \wh{u} }$ vanishes
at some of the points in $\Dp{}\mc{D}$, \textit{viz}.\ the contour infinitesimally
avoids the pole while keeping it inside of $\mc{D}$;
\item $\wh{u}$ solves the non-linear integral equation
\beq
     \wh{u}\,(\xi) \, = \,  h \, -\, T\mf{w}_N(\xi)  \; - \; \i \pi \mf{s}  T
        \; - \; \i T \Theta_{\varkappa} (\xi \, | \, \wh{\mf{X}}, \wh{\mc{Y}})
     \; - \; T\Oint{ \Dp{} \mc{D}  }{} \dd \la \: K(\xi-\la)
        \msc{L}\mathrm{n}_{ \Dp{}\mc{D} }
        \Big[ 1+  \ex{ - \f{1}{T}\wh{u} } \, \Big](\la)
\label{ecriture eqn NLI forme primordiale}
\enq
valid for $\xi$ belonging to an open neighbourhood of  $\Dp{} \mc{D}$,
\beq
     \Theta_{\varkappa} (\xi \, |\, \wh{\mf{X}} , \wh{\mc{Y}}) \, = \, \sul{ a=1 }{ |\wh{\mc{Y}}| } \th_+(\xi- \wh{y}_a) + \sul{ a=1 }{ |\wh{\mc{Y}}_{\e{sg}}| } \th_+(\xi- \wh{y}_{a;\e{sg}})
\, - \, \sul{ a=1 }{ |\wh{\mf{X}}| } \th_+(\xi- \wh{x}_a) \; + \;  \mf{m} \cdot \th_+(\xi- \varkappa) \;,
\enq
and where, for $\nu \in \Dp{}\mc{D}$, one has
\beq
     \msc{L}\mathrm{n}_{ \Dp{}\mc{D} }
        \Big[ 1+  \ex{ - \f{1}{T}\wh{u} } \, \Big](\nu) \, = \,
        \Int{\varkappa}{v} \f{ \dd \mu }{ T } \:
        \f{ -  \wh{u}^{\, \prime}\!(\mu) }{ 1+  \ex{ \f{1}{T} \wh{ u }\, ( \mu )} }
     \; + \; \ln \Big[ 1+  \ex{- \f{1}{T} \wh{ u }( \varkappa )} \, \Big] \;.
\label{definition logarithm de 1+a}
\enq
The integral appearing in the \textit{rhs} of \eqref{definition logarithm de 1+a} should
be understood in the sense of a $+$ boundary value\symbolfootnote[2]{The $+$ boundary-%
value regularisation still produces logarithmic singularities -- as it should be -- for
$\msc{L}\mathrm{n}_{ \Dp{}\mc{D} } \Big[ 1+  \ex{ - \f{1}{T}\wh{u} } \, \Big]$ at the
points of $\Dp{}\mc{D}$ where $1+  \ex{ - \f{1}{T}\wh{u} }$ vanishes. These are, however,
definitely integrable, \textit{e.g.}\ as appearing in \eqref{ecriture eqn NLI forme primordiale}.}
in case that some pole of $\;\tf{\wh{u}^{\, \prime} }{ \big[ 1+  \ex{ \f{1}{T} \wh{ u }} \big]}$
lies on $\Dp{}\mc{D}$, \textit{viz}.\ the contour infinitesimally avoids the pole while
keeping it inside $\mc{D}$. Finally, the parameter $\varkappa$ appearing in
\eqref{definition logarithm de 1+a} is an arbitrary point on $\Dp{}\mc{D}$ and the
integral is taken in positive direction along $\Dp{}\mc{D}$, from $\varkappa$ to $v$. The
function ``$\ln$'' appearing above corresponds to the principal branch of the logarithm
extended to $\R^{-}$ with the convention $\e{arg}(z) \in \intfo{-\pi}{\pi}$.

\end{itemize}

Conversely, any solution to the above non-linear problem gives rise to
an admissible solution of the Bethe Ansatz equations \eqref{ecriture eqns Bethe Trotter fini}.
\end{prop}

We refer to Proposition 4.1 of \cite{KozGohmannGoomaneeSuzukiRigorousApproachQTMForFreeEnergy}
for a proof of the above proposition. Still, we stress that similar statements were
first formulated and argued in
\cite{DestriDeVegaAsymptoticAnalysisCountingFunctionAndFiniteSizeCorrectionsinTBAFirstpaper,
BatchelorKlumperFirstIntoNLIEForFiniteSizeCorrectionSpin1XXZAlternativeToRootDensityMethod,
DestriDeVegaAsymptoticAnalysisCountingFunctionAndFiniteSizeCorrectionsinTBAFiniteMagField,
KlumperNLIEfromQTMDescrThermoRSOSOneUnknownFcton,KlumperNLIEfromQTMDescrThermoXYZOneUnknownFcton},
this on a theoretical physics level of rigour.

We would like to stress an important aspect of the non-linear integral formulation. The choice of the domain $\mc{D}$ is rather arbitrary, \textit{i.e.}\
one may choose in such a way that it best suits the needs and purpose of the given analysis that one wants to carry out on the level of the
non-linear integral equation. We emphasize that different choices of $\mc{D}$, say $\mc{D}$ and $\mc{D}^{\prime}$, for a given admissible solution $\la_1,\dots, \la_{N^{\prime}}$
may lead to \textit{very} different driving terms, say $\Theta_{\varkappa} (\xi \, | \, \wh{\mf{X}}, \wh{\mc{Y}})$ and $\Theta_{\varkappa} (\xi \, | \, \wh{\mf{X}}^{\prime}, \wh{\mc{Y}}^{\prime})$,
with $\wh{\mc{Y}}$ differing from $\wh{\mc{Y}}^{\prime}$, $\wh{\mf{X}}$ differing from $\wh{\mf{X}}^{\prime}$ and $\mf{m}$ from  $\mf{m}^{\prime}$.
Hence, some choices may have better properties.
For instance, the equation where $|\wh{\mc{Y}}|$,  $|\wh{\mf{X}}|$, $|\mf{m}|$ are all uniformly bounded in $N$, where $\wh{x}_a, \wh{y}_a$ do not approach any singularity of $\varpi_{N}$
already appear simpler, in principle, to analyse as $N\tend + \infty$. So, the proper choice of the domain seems crucial in being able to carry out the analysis
of the non-linear integral equation's solutions in the given regimes of interest. One of the technical achievements of this paper is to come up with a domain $\mc{D}$
that is fit for the large-$N$, low-$T$ analysis of the solutions to the non-linear integral equation stated in
Proposition \ref{Proposition description pblm non lineaire}.

Another comment might be in order here. When analysing solutions $\la_1,\dots, \la_{N^{\prime}}$ to the Bethe Ansatz equations \eqref{ecriture eqns Bethe Trotter fini}
through the non-linear integral problem, one is not forced to to keep the same contour $\mc{D}$ when focusing on two different solutions.
However, starting from two non-linear problems as described by Proposition \ref{Proposition description pblm non lineaire}, each associated with different domains, say $\mc{D}$
and $\mc{D}^{\prime}$, and different sets $\wh{\mf{X}}, \wh{\mc{Y}}$ and $\wh{\mf{X}}^{\prime}, \wh{\mc{Y}}^{\prime}$,
subject to \eqref{equation quantification particles et tous cas Trotter fini}, it seems impossible to say
without having a thorough control on the associated solutions $\wh{u}$ and $\wh{u}^{\, \prime}$, whether $\wh{u}=\wh{u}^{\,\prime}$
or whether these solutions are different! Thus, when having in mind classifying different solutions to the non-linear problem, this in view of
constructing \textit{different} solutions to the Bethe Ansatz equations, it is best to work with a \textit{fixed} contour $\mc{D}$ and study
\textit{all} of its different solutions $\big( \wh{u},\wh{\mf{X}}, \wh{\mc{Y}}\big)$, which then do necessarily lead to different solutions
to the Bethe Ansatz equations. This path is the one chosen in this paper.

It seems finally useful to point out that, starting from $\Dp{}\mc{D}$, the non-linear
integral equation  \eqref{ecriture eqn NLI forme primordiale} allows one to
construct the analytic continuation of $\,\wh{u}$ to a dense subset of $\Cx$ as
a holomorphic function with cuts and logarithmic singularities.

The above proposition thus allows one to reduce the analysis of the Bethe
Ansatz equations to the analysis of the solvability of a non-linear problem:
the non-linear integral equation for $\wh{u}$ \eqref{ecriture eqn NLI forme primordiale}
along with the monodromy condition \eqref{ecriture contrainte de monodromie} and the quantisation conditions
\eqref{equation quantification particles et tous cas Trotter fini}
for the collections of roots $\wh{\mf{X}}$, $\wh{\mc{Y}}$, for which we
also use the notation
\beq
\wh{\mf{X}} \, = \, \big\{ \, \wh{x}_{a} \, \big\}_1^{ | \wh{\mf{X}}  | } \qquad \e{and} \qquad  \wh{\mc{Y}} \, = \, \big\{\, \wh{y}_{a} \, \big\}_1^{ | \wh{\mc{Y}}  | }  \;,
\enq
implicitly assuming that roots are repeated according to their multiplicities,
or rather that the $\wh{x}_{a}$ (the $\wh{y}_{b}$) are all distinct ``entities''
that still can possibly take the same values in $\Cx$, \textit{viz}.\ that potentially
$\wh{x}_a=\wh{x}_b$ for some distinct label $a\not=b$. We refer to
Section~\ref{Appendix Section algebraic sum of sets} of the Appendix for more
details on the construction.

We also stress that the above correspondence can be exploited such as to express
\textit{all} physically pertinent observables solely in terms of the data of
the non-linear problem. This reduces the study of finite temperature observables
of the XXZ chain to solving the non-linear problem described above. For instance, in this approach, the Eigenvalue of the quantum transfer matrix associated with a given solution
$\big( \, \wh{u}, \wh{\mc{Y}}, \wh{\mf{X}} \, \big)$ to the non-linear problem takes the form
\bem
 \wh{\La}\big( \xi \mid  \wh{u}, \wh{\mc{Y}}, \wh{\mf{X}} \big) \, = \,
      \exp\Bigg\{ \i
         \sul{ a=1 }{ |\wh{\mc{Y}} | } p_0\bigl(\xi - \wh{y}_a\bigr)
	 \, + \, \i  \sul{ a=1 }{ |\wh{\mc{Y}}_{\e{sg}} | } p_0\bigl(\xi - \wh{y}_{a;\e{sg}}\bigr)
         \, - \, \i  \sul{ a=1 }{ |\wh{\mf{X}} | } p_0\bigl(\xi - \wh{x}_{a}\bigr)
	 \; + \; \i \mf{m} p_0\bigl(\xi - \varkappa\bigr)  \Bigg\} \\ \mspace{-18.mu}
\times
\bigg( \f{ \sinh(\tf{\aleph}{N} +\i\zeta + \xi )  \cdot \sinh(\tf{\aleph}{N} +\i\zeta -  \xi ) }{ \sinh^2(  \i\zeta) }  \bigg)^{N} \cdot
\exp \Bigg\{  \f{h}{2T}      \ - \,
\Oint{ \Dp{}\mc{D}  }{ }
\f{ \dd \mu }{ 2\pi } \:
p_0^{\prime}(\mu-\xi)
\msc{L}\mathrm{n}_{\Dp{}\mc{D}}\big[ 1+  \ex{ - \f{1}{T}\wh{u} } \, \big](\mu)
\Bigg\} \;,
\label{ecriture forme vp QTM}
\end{multline}
where $p_0(\la)$ is as it has been defined in \eqref{definition de p0}.
We refer to
\cite{DestriDeVegaAsymptoticAnalysisCountingFunctionAndFiniteSizeCorrectionsinTBAFirstpaper}
for the details of the algebraic manipulations leading to the formula.
Thus, the ratios of Eigenvalues appearing in the thermal form factor expansion
-- at least in what concerns those which can be framed by the Bethe Ansatz --
are fully expressible in terms of the data of the associated solution to the
non-linear problem. While we shall not provide the explicit expression here, a
similar in spirit, although technically more involved representation
\cite{KozDugaveGohmannThermaxFormFactorsXXZ}, holds for the thermal form factors
$\wh{\mc{F}}^{\,(\a \a^{\prime})}_{k}$.

While the non-linear problem does appear more bulky to write down than the
original system of Bethe Ansatz equations, it turns out that it provides an
effective approach for studying the infinite Trotter number limit of the finite Trotter
number observables such as the dominant Eigenvalue, the Eigenvalue ratios or
the thermal form factors. On formal grounds, this simply means to send
$N\tend +\infty$ pointwise on the level of \eqref{ecriture eqn NLI forme primordiale},
leading to the replacement in the non-linear integral equation
\beq
h\, - \, T \mf{w}_N(\xi) \;\;  \hookrightarrow \;\; \veps_{0}(\xi) \; = \; \lim_{N\tend + \infty} \hspace{-2mm} \big[ h\, - \, T \mf{w}_N(\xi) \big] \;,
\enq
where $\veps_0$ is called the bare energy and takes the explicit form
\beq
\veps_{0}(\la)  \, = \,  h -  \f{ 2 J \sin^2(\zeta) }{ \sinh\big(\la + \frac{\i}{2}\zeta \big)  \sinh\big(\la - \frac{\i}{2}\zeta \big)  } \; .
\label{definition eps 0}
\enq
This produces a new non-linear integral equation for an unknown function $u$.
Likewise, one formally replaces $\wh{x}_a \hookrightarrow x_a$ and
$\wh{y}_a \hookrightarrow y_a$ on the level of the non-linear problem
\eqref{equation quantification particles et tous cas Trotter fini}-%
\eqref{ecriture eqn NLI forme primordiale} which pertains to the
characterisation of the roots. The resulting non-linear problem then
appears as a good candidate for grasping the infinite Trotter number limit
of the solutions to \eqref{ecriture eqn NLI forme primordiale} and
\eqref{equation quantification particles et tous cas Trotter fini}.
If the solutions do converge, \textit{viz}.\
$\big(\wh{u}, \wh{\mf{X}}, \wh{\mc{Y}} \big) \tend \big( u, \mf{X}, \mc{Y} \big)$,
then taking the Trotter limit on the level of the Eigenvalues becomes a triviality!
However, establishing this fact rigorously is far from obvious.

Another question which one may adress is, given a choice of the domain $\mc{D}$,
which solution $\big( \, \wh{u}, \wh{\mf{X}}, \wh{\mc{Y}} \, \big)$ to the
non-linear problem gives rise to the dominant Eigenvalue of the quantum transfer
matrix. Along the same lines, one may ask, whether there exists a clear ``ordered'' identification
of the solutions which gives rise to the ``ordered'' Eigenvalues of the quantum
transfer matrix. Again, there does not seem to have appeared, to the best of our knowledge,
a clear-cut rigorous answer so far.

The paper \cite{KozGohmannGoomaneeSuzukiRigorousApproachQTMForFreeEnergy}
provides a rigorous analysis of the above questions in the high-temperature
regime. In that case one may take the domain $\mc{D}$ to be a small disk
of radius shrinking with $1/T$, centred at $-\i{\zeta}/{2}$:
$\op{D}_{-\i {\zeta}/{2}, {c}/{T}}$.
The paper \cite{KozGohmannGoomaneeSuzukiRigorousApproachQTMForFreeEnergy}
established that the dominant Eigenvalue of the quantum transfer matrix $\wh{\La}_{\e{max}}$ is non-degenerate in the high-temperature
regime for $N\geq N_0$ and identified which solution to the non-linear problem
gives rise to the dominant Eigenvalue. In this regime, the solution giving
rise to the dominant Eigenvalue corresponds to taking both collections of
roots empty, \textit{viz}.\ $\big( \, \wh{u}, \emptyset, \emptyset  \big)$.
Further, it was shown that in the large $T$ and $N$ regime, for $|\wh{\mf{X}}|$,
$|\wh{\mc{Y}}|$ bounded in $N,T \tend +\infty$ and $\wh{x}_a$, resp.\ $\wh{y}_a$, being not too close
(in some appropriate scale in $\tf{1}{T}$ and $\tf{1}{N}$) to $-\i\tf{\zeta}{2}$,
resp.\ $\i\tf{\zeta}{2}$, the non-linear problem admits a unique solution.
Moreover, the continuity in $N\tend +\infty$, \textit{viz}.\ the convergence of
$\,\wh{u}$ toward the limiting non-linear problem's unique solution, was established.
Finally, a leading large-$T$ expansion for the limiting roots $x_a, y_a$
and the solution $u$ was given. That resulted in the proof of a closed formula
for the free energy of the XXZ chain in the large-$T$ regime.

While a considerable progress, one would still like to have a deeper understanding
of the genuine finite-temperature situation. So far, that general setting still
appears unreachable owing to the lack of an appropriate technique for showing
the existence of solutions to the non-linear integral equation. However,
the low-$T$ regime appears to be easier to deal with, and the study thereof
is precisely the subject of the present paper. More precisely, this paper
provides a setting which allows one to formulate appropriately the non-linear
problem described in Proposition~\ref{Proposition description pblm non lineaire}
in the large-$N$ low-$T$ regime and to show its unique solvability, at least for certain
classes of auxiliary roots $\wh{\mc{Y}}, \wh{\mf{X}}$ described below. As explained
above, this allows one to solve at least a subset of the Bethe equations associated
with the quantum transfer matrix and show that these do give rise to non-zero
Eigenvectors $\Psi\big(\la_1,\dots, \la_{N^{\prime}} \big)$, while providing closed
explicit expressions for the associated Eigenvalues. The latter can then be analysed
in the large-$N$ and low-$T$ regime and we are able to identify the piece of the
spectrum we compute which exhibits, in this limit, an explicit conformal structure,
as discussed in \eqref{ecriture comportement low T dominant Eigenvalue}-\eqref{ecriture low T pour spectre energie}.

This is achieved in several steps.

\begin{itemize}

 \item[i)] For an appropriate oriented contour
$\msc{C}_{\e{ref}}$, which may be interpreted as the boundary of a simply
connected domain $\mc{D}=\e{Int}\big( \msc{C}_{\e{ref}} \big) \subset \Cx$
containing an open, $N$-independent neighbourhood of $-\i\tf{\zeta}{2}$,
we show the unique solvability of the non-linear integral equation
\eqref{ecriture eqn NLI forme primordiale} for an $\textit{arbitrary}$
monodromy $\mf{m}$ (fixing $\mathfrak{s}$) and $\textit{arbitrary}$ auxiliary
parameters that will be denoted  $\op{x}_1, \dots, \op{x}_{|\op{X}|}$ and
$\op{y}_1,\dots, \op{y}_{|\op{Y}|}$ and that will be subject to certain
technical assumptions. In particular, the parameters $\op{y}_a$ may or may not
be singular in the sense that $\op{y}_a-\i\e{sgn}(\pi-2\zeta) \e{min}(\zeta, \pi-\zeta) \in
\e{Int}\big( \msc{C}_{\e{ref}} \big)$.

\item[ii)] We show that the solution converges,
when $N\tend +\infty$, to the one of the limiting non-linear integral equation for which we also establish the unique solvability.

\item[iii)] We classify, under additional constraints, all possible solutions to the
quantisation conditions \eqref{equation quantification particles et tous cas Trotter fini}
in the finite Trotter number case. We show that these do converge to the analogously
classified solutions to the associated infinite Trotter number quantisation
conditions.

\item[iv)] We provide the low-$T$ limit of these solutions.

\item[v)] We rigorously identify, within the class of the Bethe Eigenstates we are able to construct,
the dominant, \textit{viz}.\ largest in modulus, Eigenvalue. Further, we single
out those Eigenvalues which reproduce the full spectrum of the $c=1$ free Boson
conformal field theory.

 \end{itemize}

To close this introduction we would like to lay emphasis on the rather unique nature
of the analysis we develop. Indeed, despite the presence of Bethe Ansatz equations
in the literature for nearly 100 years, almost no rigorous result relative to the study
of the behaviour of their solutions in the large number of roots $N\tend +\infty$ limit appeared. In fact, the sole existing rigorous
results of this type pertain to the system of Bethe Ansatz equations governing the
spectrum of the  XXZ chain on $L$ sites -- we recall that these Bethe Ansatz equations
are sensibly different from those we consider here. Building on convex analysis
arguments, the works \cite{DorlasSamsonovThermoLim6VertexAndConvergceToDensityInSomeCases6VrtX,%
GusevCondensationGSNegativeDelta} showed the condensation properties of the real valued
solution $\la_1,\dots, \la_N$ giving rise to the model's ground state, \textit{viz}.\ that
\beq
\f{1}{L}\sul{a=1}{N} f(\la_a)   \underset{ \substack{ N,L \tend + \infty\\ N/L \tend D } }{\longrightarrow}
\Int{-q_{\mf{m}} }{ q_{\mf{m}} } \hspace{-1mm} \dd \la \: f(\la) \, \rho\big(\la\,|\, q_{\mf{m}} \big)
\qquad \e{with} \qquad D \; = \; \Int{ -q_{\mf{m}} }{ q_{\mf{m}} }\hspace{-1mm} \dd \la \: \rho\big( \la \,|\, q_{\mf{m}} \big)  \;.
\enq
In these equations $\rho\big(\la\,|\, q_{\mf{m}} \big) $ is the
unique solution to a certain Fredholm type linear integral equation,
while $q_{\mf{m}}$ is the magnetic Fermi boundary, implicitly defined
as a unique function of the fixed parameter $D \in \intff{0}{\tf{1}{2}}$
by the above integral expression for $D$. This was later generalised in
\cite{GusevCondensationExcitedStatesNegativeDelta} to certain classes
of solutions giving rise to excited states.
We stress that due to the loss of certain convexity properties when $\De \geq 0$, the
works \cite{DorlasSamsonovThermoLim6VertexAndConvergceToDensityInSomeCases6VrtX,%
GusevCondensationGSNegativeDelta,GusevCondensationExcitedStatesNegativeDelta}
could only consider the case $-1 < \De  \leq 0$. Later on, the work
\cite{KozProofOfDensityOfBetheRoots} established condensation for all
values of $\Delta$ and a large class of real valued solutions to the Bethe Ansatz
equations giving rise to so-called particle-hole excited states. Moreover, when
$D<\tf{1}{2}$, the work  \cite{KozProofOfDensityOfBetheRoots} also established
the existence of a full large $L$ asymptotic expansion for the sum appearing above
and a large class of test functions $f$. Furthermore, the works
\cite{DuminilCopinGagnebinHarelManolescuTassionDiscOfPhaseTransFKPercoAtq=4,
KozDuminilCopinKrachunManolescuTikhonovskaya6VertexFreeEnergyExpansionsRigorous}
established  the condensation, by different techniques, when $N/L=1/2$ and
provided an estimation of the first subdominant correction  to the sum in that case.
Still, the Bethe Ansatz equations for the XXZ chain are also expected
\cite{CauxHagemansDeformedStringsInXXX,%
EsslerKorepinSchoutensBAE2ParticleSectorXXXAndSomeNonStandardSols,KarbachMullerIntroToBAI},
\textit{e.g.}\ on the basis of numerical or heuristic calculations, to admit complex
valued solutions. No rigorous result ever appeared relatively to the treatment of such
types of solutions away from the trivial regime where $D=0$, rigorously treated in \cite{BabbittThomasPlancherelFormulaInfiniteXXX,ThomasGSRepForFerromagneticXXX},
for which condensation is not present due to the too few Bethe roots at play in that case.

In the case of the quantum transfer matrix at finite magnetic field $h$ 
numerical calculations indicate \cite{KozDugaveGohmannSuzukiLowTSpectrumMassiveXXZQTM,KozGohmannGoomaneeSuzukiRigorousApproachQTMForFreeEnergy,
KlumperMartinezScheerenShiroishiXXZforPositiveDeltaCrossoverAtTNicePictureCondRootsAtFermiPts} that \textit{all}
solutions are genuinely complex valued. We treat the case of such solutions
in this paper by showing their existence and proving a large amount of their
fine properties, this in the limit of large $N$ and low $T$. We are deeply
convinced that the very fact that we were able to set up a rigorous method
allowing one to obtain all that is an appreciable achievement.

The paper is organised as follows. Section~\ref{Section Main Results} provides
a detailed description of the technical results obtained in the paper.
Section~\ref{Section NLIE equivalente et contour Cu} develops a scheme allowing
one to transform one non-linear integral equation into an equivalent one which
appears simpler with respect to the analysis of the large-$N$ small-$T$ behaviour
of its solutions. Section~\ref{Section espace fnel pour la NLIE et ses ptes}
focuses on the derivation of various properties of the functional space on which
we develop the solvability theory of the non-linear equation obtained from the
Bethe Ansatz for the quantum transfer matrix.
Section~\ref{Section NLIE modifiee et solvabilite par pt fixe}
is devoted to the study of the modified non-linear integral equation discussed
in Section~\ref{Section NLIE equivalente et contour Cu}: it demonstrates
its unique solvability. Section~\ref{Section conditions de quantification} provides
a thorough discussion of the solvability of the quantisation conditions.
Section~\ref{Section Low T expansion of spectral observables} demonstrates the practical
applicability of the present analysis by establishing a detailed low-$T$ asymptotic
expansion of various physical observables related to the Bethe Ansatz of the quantum
transfer matrix. In this section it is also shown that the unique solutions to the non-linear
problem correspond to non-zero Eigenvectors each. The paper includes three appendices.
Appendix~\ref{Appendix Section algebraic sum of sets} collects our notational conventions
for collections of parameters. Appendix~\ref{Appendix Section ptes integrales et espaces fnels}
reviews certain properties of integral operators and recalls the multidimensional
Rouch\'{e} theorem. Appendix~\ref{Properties_of_the_dressed_energies}
establishes several important properties of the dressed energy function.

\section{Main results}
\label{Section Main Results}

In order to state more precisely the main achievements of this work, we first
introduce several auxiliary functions -- defined as solutions to linear
integral equations of Fredholm type -- which occur in the results we obtain.
Next, we introduce the reference contour $\msc{C}_{\e{ref}}$ which plays an
essential role for setting up the whole theory. Indeed, $\e{Int}(\msc{C}_{\e{ref}})$
will correspond to the domain $\mc{D}$ mentioned in
Proposition~\ref{Proposition description pblm non lineaire}. We will then formulate
the main hypotheses on the collection of roots $\wh{\mf{X}}$ and $\wh{\mc{Y}}$
on which our analysis builds. These hypotheses should be understood as providing
a set of conditions on the particle-hole roots for which our analysis works. In
other words, this work discusses a large albeit \textit{a priori} incomplete set
of solutions to the non-linear problem presented in Proposition
\ref{Proposition description pblm non lineaire}. Once all of the above is in place,
we introduce the functional space on which we solve the non-linear integral equation
pertaining to the original problem. The remainder of this section is devoted to a thorough
discussion of our results.

\subsection{Auxiliary linear integral equations}

The analysis developed in this work heavily builds on an auxiliary class of
special functions defined as solutions to a set of linear integral equations
driven by the operator $\e{id} + \op{K}$,
\beq
     \Big(\e{id} + \op{K} \Big)[f](\la) \; = 
        \;f(\la) \; + \; \Int{-q}{q}  \dd \mu \: K(\la-\mu) f(\mu)   \;,  
\label{definition operateur integral K}
\enq
on $L^2\big( \intff{-q}{q} \big)$. We will specify the value of $q$ below. However,
we stress that this operator is invertible, see \textit{e.g.}
\cite{KozDugaveGohmannThermoFunctionsZeroTXXZMassless,KozProofOfDensityOfBetheRoots,%
YangYangXXZStructureofGS} for any $q\geq 0$. Let $\e{id} - \op{R}$ be the inverse
operator to $\e{id} + \op{K}$. The integral kernel of $\op{R}$ will be denoted $R(\la,\mu)$ and its
dependence on $q$ will be kept implicit.

The invertibility of $\e{id} + \op{K} $ allows one to define a one parameter $Q$
family of special functions $\veps(\la\,|\,Q)$ as solutions to the below linear
integral equation
\beq
     \veps(\la\,|\, Q) \,
        + \, \Int{-Q}{Q} \dd \mu \: K\big(\la-\mu  \big) \, \veps(\mu\,|\, Q)
	  \; = \;  \veps_0(\la)  \;. 
\label{definition energie habille et energie nue}
\enq
Here, $ \veps_0$ is the bare energy introduced in \eqref{definition eps 0}. 
For $0 < h < 4J (1 + \De)$, the endpoint of the Fermi zone $q$ is defined as
the unique \cite{KozDugaveGohmannThermoFunctionsZeroTXXZMassless,%
KozFaulmannGohmannDressedEnergyInComplexPlane} positive solution $q$
to the equation $\veps(Q\,|\, Q)=0$. The associated solution of
\eqref{definition energie habille et energie nue} is called the dressed
energy and is denoted $\veps(\la) = \veps(\la\,|\, q )$. It is shown in Proposition
\ref{Proposition double recouvrement de veps} that, for $0 < \zeta < \pi/2$, the
map $\veps: \msc{U}_{\veps} \setminus \big\{ 0, \i\tfrac{\pi}{2} \big\}
\tend \veps\big( \msc{U}_{\veps} \setminus \big\{ 0, \i\tfrac{\pi}{2} \big\} \big)$ with 
\beq
     \msc{U}_{\veps} \; = \;
        \Big\{ z \in \Cx \; : \; - \tfrac{\pi}{2} < \Im(z) \leq \tfrac{\pi}{2} \Big\}
	\setminus \bigcup_{\ups= \pm} \Big\{ \intff{-q}{q} + \i \ups \zeta_{\e{m}} \Big\} \; ,
\label{definition ensemble de support pour dble rec}
\enq
is a double cover. 

Next, one introduces the dressed phase 
\beq
     \phi(\la,\mu) \, + \,
        \Int{-q}{q} \dd\nu \: K(\la-\nu)\,  \phi(\nu, \mu )
	\, = \, \f{ 1  }{ 2 \pi }  \theta\big( \la -\mu  \big) \;, 
\label{definition dressed phase}
\enq
the dressed charge 
\beq
     Z(\la)\, + \,  \Int{-q}{q}  \dd \mu \: K(\la-\mu)\,  Z(\mu ) \, = \,  1 \;,  
\label{definition dressed charge}
\enq
and the dressed momentum 
\beq
     p(\la)\, + \, \Int{-q}{q} \frac{\dd \mu}{2 \pi} \: \theta (\la-\mu)\,  p^{\prime}(\mu )
        \, = \, p_0(\la) \;,  
\label{definition dressed momentum}
\enq
where $\theta$ has been defined in \eqref{definition_theta} and
$p_0$ in \eqref{definition de p0}. Note that the existence and uniqueness of the dressed
momentum follows upon taking the derivative of \eqref{definition dressed momentum}, uniquely solving for $p^{\prime}$,
and then taking the anti-derivative under the condition that $p$ is odd. 

It will turn out to be useful to introduce another set of solutions to linear
integral equations, closely related to the above ones, which are associated with
a curved contour. In order to do so, first define the curve 
\beq
\msc{C}_{\veps} \; = \; \Big\{ \la \in \Cx \; : \;  -\tfrac{\pi}{2} \,  <  \, \Im(\la) \, \leq   \, 0  \;\;  \e{and} \; \; \Re\big[ \veps(\la) \big] = 0 \Big\} \;. 
\label{Definition de la courbe C veps}
\enq
We refer to \cite{KozFaulmannGohmannDressedEnergyInComplexPlane}, where this curve
was  characterised in full rigour for $0 \leq \De < 1$, \textit{viz}. $0< \zeta \leq \tf{\pi}{2}$.

Further, define $\veps_{\e{c}}$ as the solution to the linear integral equation
\beq
\veps_{\e{c}}(\la) \, + \, \op{K}_{\msc{C}_{\veps}}[  \veps_{\e{c};-}](\la) \, = \, \veps_{0}(\la) \qquad \e{with} \qquad
 \op{K}_{\msc{C}_{\veps}}[  f_{-} ](\la)  \, = \, \lim_{\a\tend 0^-} \Int{ \msc{C}_{\veps} }{}\hspace{-1mm} \dd \mu \,  K(\la-\mu) f(\mu+\i\a)  \;.
\label{ecriture LIE pour veps c}
\enq
Here $\veps_{\e{c};-}$ stands for the $-$ boundary value, \textit{viz}.\ from below,
of $\veps_{\e{c}}$ on the curve $\msc{C}_{\veps}$, oriented from $-q$ to $q$,
understood in the weak sense as described above. The invertibility of
$\e{id}+ \op{K}_{\msc{C}_{\veps}}$ follows from
Lemma~\ref{Lemme inversibilite de Id + K sur courbe C eps}. 
Thus, $\veps_{\e{c}}$ is well defined. It is easy to see that $\veps_{\e{c}}$
is a meromorphic function on $\Cx \setminus \bigcup_{\ups = \pm} \Big\{  \msc{C}_{\veps}
+ \i \ups \zeta + \i\pi \mathbb{Z} \Big\}$, whose only poles are simple and located 
at $\i\pi \mathbb{Z}\pm \i \tf{\zeta}{2}$. As a consequence, when
$|\Im(\la)| \, < \, \tf{\zeta}{2}$ and $0<\zeta<\tf{\pi}{2}$, one may deform
the contour $\msc{C}_{\veps}$ to $\intff{-q}{q}$ in \eqref{ecriture LIE pour veps c}
without picking up the poles of $\veps_{\e{c}}$. The latter entails that
\beq
\veps_{\e{c}}(\la)  \, = \, \veps(\la)  \qquad \e{provided}\, \e{that} \qquad  |\Im(\la)| \, < \, \f{\zeta}{2} \mod \i\pi \; . 
\label{ecriture relation entre veps c et veps}
\enq

One similarly defines analogues of the dressed charge and phase subordinate to
$\msc{C}_{\veps}$ as the unique solutions to the linear integral equations
\beq
Z_{\e{c}}(\la) \, + \,  \op{K}_{\msc{C}_{\veps}}[  Z_{\e{c}} ](\la)  \, = \, 1 \qquad  \e{and} \qquad 
\phi_{\e{c}}(\la,\mu) \, + \,  \op{K}_{\msc{C}_{\veps}}[ \phi_{\e{c}}(*, \mu ) ](\la) \, = \, \f{ 1  }{ 2 \pi }  \theta\big( \la -\mu  \big)   \;. 
\label{definiton des charges et phases c deformee} 
\enq
Further, we introduce the analogue of the dressed momentum\symbolfootnote[3]{Again,
the integral appearing below should be understood in the weak sense as in
\eqref{ecriture LIE pour veps c}.} subordinate to $\msc{C}_{\veps}$,
\beq
     p_{\e{c}}(\la)\, + \,
        \Int{ \msc{C}_{\veps} }{ }  \frac{\dd \mu}{2 \pi} \:
	   \theta(\la-\mu)\,  p^{\prime}_{\e{c};-}(\mu ) \, = \,  p_0(\la) \;,
\label{definition dressed momentum c deforme}
\enq
and the resolvent kernel associated to an integration on $\msc{C}_{\veps}$:
\beq
R_{\e{c}}(\la,\mu) \, + \,  \op{K}_{\msc{C}_{\veps}}[  R_{\e{c}}(*,\mu) ](\la)  \, = \, K(\la-\mu) \;. 
\label{definition du resolvent c deforme}
\enq
By construction, $R_{\e{c}}(\la,\mu)$ coincides with the resolvent kernel of
the inverse operator $\e{id}-\op{R}_{\e{c}}$ to  $\e{id}+ \op{K}_{\msc{C}_{\veps}}$.

\begin{figure}
\begin{center}
\includegraphics[width=0.55\textwidth]{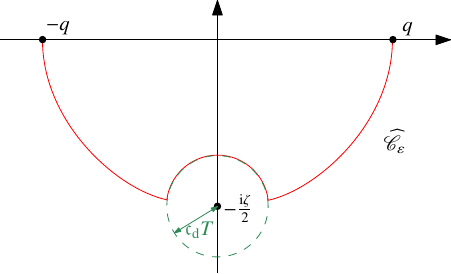}
\caption{The integration contour $\wh{\msc{C}}_{\veps}$.}
\label{contour integration hat Cveps}
\end{center}
\end{figure}

Finally, we need to introduce one last special function which arises in the
characterisation of the solutions to the non-linear integral equation
\eqref{ecriture eqn NLI forme primordiale} at large Trotter numbers and low
temperatures. Let $\wh{\msc{C}}_{\veps}$ coincide with the curve $\msc{C}_{\veps}$
with the exception of a vicinity of the point $-\i\tf{\zeta}{2}$ around which
the curve corresponds to an arc of the boundary of the disk
$\Dp{}\op{D}_{ -\i \f{\zeta}{2}, \mf{c}_{\e{d}}T}$, \textit{c.f.}\
Fig.~\ref{contour integration hat Cveps}. Here and in the following
\beq
\op{D}_{z_0,r} \; = \; \Big\{ z\in \Cx \; : \; |z-z_0|\, <\, r \Big\} \;. 
\label{definition disque ouvert generique}
\enq
The operator $\e{id}+ \op{K}_{\wh{\msc{C}}_{\veps}}$ is invertible, \textit{c.f.}\
Lemma~\ref{Lemme inversibilite de Id + K sur courbe C eps}, with resolvent kernel
$\wh{R}_{\e{c}}(\la,\mu)$, and one defines the function $\mc{W}_{N}$ -- the finite
Trotter number analogue of the dressed energy on the curved contour $\wh{\msc{C}}_{\veps}$ --
as the solution to the linear integral equation
\beq
\mc{W}_{N}(\la) \, + \, \op{K}_{ \wh{\msc{C}}_{\veps} }[  \mc{W}_{N} ](\la) \, = \, h-T\mf{w}_N(\la) \;,
\label{definition solution LIE deformee WN}
\enq
with $\mf{w}_N$ as defined through \eqref{definition fct mathfrak w N}.
It is direct to see that $\mc{W}_N\tend \veps_{\e{c}}$ as $N\tend +\infty$
uniformly on
\beq
\Big\{\la \in \Cx \; : \;  |\Im(\la)|<\tf{\zeta}{2} \Big\}\setminus
{\textstyle \bigcup}_{\ups=\pm}\op{D}_{\ups \i \tf{\zeta}{2}, \eta} \; ,  \qquad \e{for}\; \e{any} \;\;  \eta>0 \; .
\enq

\subsection{Working hypotheses}
\label{SousSection hypothese de travail}
From now on we restrict the discussion to the regime $0 \leq \Delta < 1$, namely to 
\beq
\zeta \in \intof{ 0 }{ \f{\pi}{2} } \;. 
\label{ecriture range poru zeta}
\enq
In fact, numerous findings of the present paper may be extended to the full critical
parameter regime $0<\zeta<\pi$. However, we are missing several technical results to
obtain analogues of the main theorems for $-1 < \De < 0$. This mainly pertains to our inability, so far,
to have a sufficiently good control and understanding, in that regime, on the auxiliary special function $\veps_{\e{c}}$ defined in \eqref{ecriture LIE pour veps c}.
Hence, we preferred not
to partition the discussions according to what can be done in which regime.
Therefore, from now on, unless stated otherwise, \eqref{ecriture range poru zeta}
is assumed.  

\begin{figure}
\begin{center}
\includegraphics[width=0.75\textwidth]{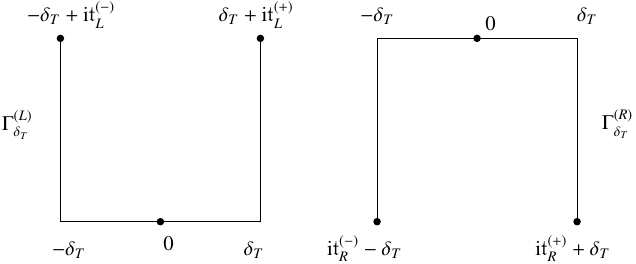}
\caption{Definition of the contours $\Ga_{\de_T}^{(L/R)}$ used in the definition
of the reference contour $\msc{C}_{\e{ref}}$. Note that one has $\mf{t}_{L}^{(\pm)}>0$
and $\mf{t}_{R}^{(\pm)}<0$. }
\label{contour integration entourant Real veps=0}
\end{center}
\end{figure}
\begin{figure}[h]
\begin{center}
\includegraphics[width=0.65\textwidth]{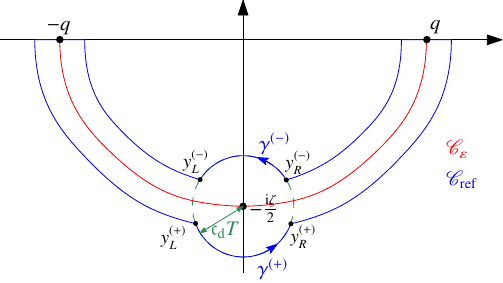}
\caption{Definition of the reference contour $\msc{C}_{\e{ref}}$.}
\label{contour integration Cref}
\end{center}
\end{figure}

To start with, we give a precise definition of the reference contour, see
Fig.~\ref{contour integration Cref}, 
\beq
\msc{C}_{\e{ref}} \, = \, \veps_R^{-1}\Big( \Ga_{\de_T}^{(R)} \Big) \cup \veps_L^{-1}\Big( \Ga_{\de_T}^{(L)} \Big) \cup\ga^{(-)}\cup \ga^{(+)} \, ,
\label{definition contour Cref}
\enq
in which $\Ga_{\de_T}^{(L/R)}$ have been depicted in
Fig.~\ref{contour integration entourant Real veps=0}, while $\veps_{L/R}$ are
the biholomorphisms subordinate to the double cover of $\veps\big( \msc{U}_{\veps}
\setminus \big\{ 0, \i\tfrac{\pi}{2} \big\} \big)$ by $\veps$, with $\msc{U}_{\veps}$
as in \eqref{definition ensemble de support pour dble rec}. We refer to
Proposition~\ref{Proposition double recouvrement de veps} for more details. 
The parameter $\de_{T}$ is chosen as 
\beq
\de_T \, =\,  - M T \ln T \;,
\label{definition du parametre delta T}
\enq
with $M > 0$. Further, we pick $\mf{c}_{\e{d}}>0$ and small enough to set
the scale of the radius $\mf{c}_{\e{d}}T$ of the disk
$\op{D}_{ -\i \f{\zeta}{2}, \mf{c}_{\e{d}}T}$ centred at $-\i\tf{\zeta}{2}$
which will play a role in the analysis. In particular, 
\begin{itemize}
 
\item the curve $\ga^{(-)}$ is given by the counterclockwise oriented arc
of $\Dp{}\op{D}_{ -\i \f{\zeta}{2}, \mf{c}_{\e{d}}T}$, joining $y_R^{(-)}$ to $y_{L}^{(-)}$;

\item the curve $\ga^{(+)}$ is given by the counterclockwise oriented arc of
$\Dp{}\op{D}_{ -\i \f{\zeta}{2}, \mf{c}_{\e{d}} T}$, joining $y_L^{(+)}$ to $y_{R}^{(+)}$. 
 
\end{itemize}
Here, the points $y_{L/R}^{(\pm)}$ are defined by the intersections 
\beq
y^{(\pm)}_{L} \; = \; \veps_{L}^{-1}\Big(\pm \de_{T}+\i \R^{+} \Big) \cap \Dp{}\op{D}_{ -\i \f{\zeta}{2}, \mf{c}_{\e{d}} T} \quad \e{and} \quad 
y^{(\pm)}_{R} \; = \; \veps_{R}^{-1}\Big(\pm \de_{T}+\i \R^{-} \Big) \cap \Dp{}\op{D}_{ -\i \f{\zeta}{2}, \mf{c}_{\e{d}} T} \;. 
\label{definition points finaux sur petit disque}
\enq
Note that, since $\op{D}_{ -\i \f{\zeta}{2}, \mf{c}_{\e{d}}T}
\subset \msc{U}_{\veps} \setminus \{0,\i\tf{\pi}{2}\}$, the points $y^{(\sg)}_{\a}$
are well defined. The points $y^{(\sg)}_{\a}$ then allow one to determine the points
\beq
\sg \de_T +\i \mf{t}^{(\sg)}_{\a} = \veps \bigl(y^{(\sg)}_{\a}\bigr) \,
\label{definition des points limites des courbes rectifiees du contour}
\enq
shown in Fig.~\ref{contour integration entourant Real veps=0}.

Let $\e{Int}\big( \msc{C}_{\e{ref}} \big)$ stand for the bounded connected component
enclosed by $\msc{C}_{\e{ref}}$. We stress that the curve $\msc{C}_{\e{ref}}$
surrounds the curve $\msc{C}_{\veps}$ introduced in \eqref{Definition de la courbe C veps}
(see Fig.~\ref{contour integration Cref}).

Further, introduce two kinds of parameters 
\beq
 \op{x}_a \in \e{Int}\big( \msc{C}_{\e{ref}} \big) \quad \e{for} \quad a=1,\dots, |\op{X}| \quad \e{and} \quad
 \op{y}_a \in \Big\{ z \in \Cx \; : \; |\Im(z)| \leq \tfrac{\pi}{2} \Big\} \setminus \ov{  \e{Int}\big( \msc{C}_{\e{ref}} \big) }
 \quad \e{for} \quad a=1,\dots, |\op{Y}|
\enq
and denote $\op{X}=\{\op{x}_a\}_{a=1}^{|\op{X}|}$ and $\op{Y}=\{\op{y}_a\}_{a=1}^{|\op{Y}|}$ the
associated collections of parameters. Here, it is understood that parameters are repeated 
according to their multiplicities. We refer to
Section~\ref{Appendix Section algebraic sum of sets} of the appendix
for more details about these definitions. It will also appear convenient to partition the
parameters $\op{y}_a$ into two disjoint families: $\op{Y}_{\e{r}}
\, = \, \{ \op{y}_{a;\e{r}} \}_{a=1}^{|\op{Y}_{\e{r}}|}$ the
collection of regular parameters, and  $\wt{\op{Y}}_{\e{sg}} \, =
\, \{ \wt{\op{y}}_{a;\e{sg}} \}_{a=1}^{|\wt{\op{Y}}_{\e{sg}}|}$,
the collection of shifted singular parameters, with
\beq
     \op{y}_{a;\e{r}} \in \Big\{ z \in \Cx \; : \; |\Im(z)| \leq \tfrac{\pi}{2} \Big\}
        \setminus \Big\{ \overline{\e{Int}\big( \msc{C}_{\e{ref}} \big)}
	+ \i \zeta \Big\}
\; , \qquad 
\wt{\op{y}}_{a;\e{sg}} \in  \e{Int}\big( \msc{C}_{\e{ref}} \big) + \i \zeta \; . 
\label{definition racines singulieres et reguliere}
\enq
Analogously to the previous discussion, it is useful to introduce the collection
of singular parameters 
\beq
\op{Y}_{\e{sg}} \, = \, \big\{ \op{y}_{a;\e{sg}} \big\}_{1}^{ |\op{Y}_{\e{sg}}| } \, ,  \quad \e{with} \quad   \op{y}_{a;\e{sg}} \, = \, \wt{\op{y}}_{a;\e{sg}} - \i \zeta \,.
\enq

The proof of the existence of solutions to the non-linear integral equation
\eqref{ecriture eqn NLI forme primordiale} subordinate to the reference contour
$\msc{C}_{\e{ref}}$ given in \eqref{definition contour Cref} relies on the
four hypotheses below about the collections of auxiliary parameters $\op{Y}$ and~$\op{X}$.
Prior to stating these, we introduce a metric on $\Cx /\big( \i\pi \mathbb{Z} \big)$:
\beq
 \e{d}_{\i\pi}(z, z^{\prime}) = \underset{r\in \mathbb{Z}}{\e{inf}}\, |z-z^{\prime}- \i r \pi| \;.
\label{definition i pi periodic distance}
\enq

\begin{hypothesis}
\label{Hypotheses solubilite NLIE} 
Assume that $\op{Y}$ and $\op{X}$ are such that 
\begin{itemize}
\item[a)] there exists $\mf{c}>0$ such that
$\e{d}_{\i\pi}\Big(\op{Y}- \ups \i  \zeta, \pm q \Big) \, > \, \mf{c}$,
with $\ups \in \{ \pm\}$;
\item[b)] there exists $\mf{c}_{\e{ref}}>0$ and large enough such that
$\e{d}_{\i\pi}\Big(\op{Y}- \ups \i   \zeta, \msc{C}_{\e{ref}} \Big) \,
> \, \mf{c}_{\e{ref}} T$, with $\ups \in \{ \pm\}$;
\item[c)] there exists $\mf{c}_{\e{sep}} > 0$ such that 
$\op{X} \cap \ov{\op{D}}_{-\i\f{\zeta}{2}, \mf{c}_{\e{sep}}T} \, = \,
\emptyset $, $\op{Y}\cap \ov{\op{D}}_{ \pm \i\f{\zeta}{2}, \mf{c}_{\e{sep}} T}
\, = \, \emptyset$, 
$\op{Y}\cap \ov{\op{D}}_{-3\i\f{\zeta}{2}, \mf{c}_{\e{sep}} T} \, = \, \emptyset$;
\item[d)] there exists $\mf{c}_{\e{loc}}>0$ such that $\e{d}_{\i\pi}\Big(\op{Y}, \pm q \Big) \, > \, \mf{c}_{\e{loc}}$
and $\e{d}_{\i\pi}\Big(\op{X}, \pm q \Big) \, > \, \mf{c}_{\e{loc}}$.
\end{itemize}

\end{hypothesis}
 
Note that hypothesis c) is to be understood modulo the natural $\i\pi$ periodicity
of the problem. Here, we also stress that $\msc{C}_{\e{ref}}$ is always taken such that
$\mf{c}_{\e{d}}<\mf{c}_{\e{sep}}$, \textit{c.f.} Fig.~\ref{contour integration hat Cveps} for the definition of $\mf{c}_{\e{d}}$.

The description of the solution set of the quantisation conditions will be made under
three additional hypotheses on the solution set. We shall denote the collection of
parameters solving the quantisation conditions associated with a given solution to
the non-linear integral equation as
\beq
\wh{\mc{Y}} \; = \; \{ \, \wh{y}_{a} \}_{1}^{ |\wh{\mc{Y}} | }  
\quad \e{and} \quad \wh{\mf{X}} \; = \; \{ \wh{x}_{a} \}_{1}^{ |\wh{\mf{X}} | } \;. 
\enq

\begin{hypothesis}
\label{Hypotheses eqns de quantification} 
The parameters composing $\wh{\mc{Y}}$ and $ \wh{\mf{X}}$ satisfy the constraints

\begin{itemize}

\item[e)] there exists $\mf{c}_{\e{rep}}>0$ such that
\[
     \e{d}_{\i\pi}\big( z , z^{\prime} \big) \, > \,  \mf{c}_{\e{rep}} T \quad
     for\, all \quad z\not= z^{\prime} \quad
     with \quad z, z^{\prime} \in \wh{\mc{Y}} \quad \e{or} \quad
     z, z^{\prime} \in \wh{\mf{X}} \;;
\]
\item[f)]
the roots are simple in the sense that the parameters building
up $\wh{\mc{Y}}$ and $ \wh{\mf{X}}$ all have multiplicity one, \textit{viz}.\ 
$k_x=1$, resp.\ $k_y=1$, for all $x \in \wh{\mf{X}}$, resp.\ $y\in \wh{\mc{Y}}$; 
\item[g)] there exists $\eps>0$ small enough such that $\wh{\mf{X}} \cap \ov{\op{D}}_{-\i\f{\zeta}{2}, \eps} \, = \, \emptyset $, $\wh{\mc{Y}} \cap \ov{\op{D}}_{\pm \i\f{\zeta}{2}, \eps}
\, = \, \emptyset$, 
$\wh{\mc{Y}} \cap \ov{\op{D}}_{-3\i\f{\zeta}{2}, \eps} \, = \, \emptyset$.
\end{itemize}

\end{hypothesis}
We shall sometimes refer to condition $\mathrm{e)}$ as the  repulsion principle. 
 
 \vspace{2mm}
 
Several comments are in order relative to the hypotheses. Some of these are purely
of technical origin and could be relaxed for the price of overburdening the analysis,
while relaxing others would most probably demand the use of other techniques and would
thus be more difficult.
\begin{itemize}

 \item[i)] Condition $\mathrm{a)}$ is of fundamental importance
to the present analysis. Should it be not valid, then one would need to alter the
scheme of analysis by carrying out a different kind of local analysis around the points
$\pm q$.

\item[ii)] Likewise, condition $\mathrm{c)}$ plays an important role for the finite
Trotter number setting. Indeed, the fact that the logarithmic singularities of $\mf{w}_N$
at $-\i\tf{\zeta}{2} \pm \i\tf{\aleph}{N}$ converge to a pole as $N\tend +\infty$
influences drastically the behaviour of the solutions to the quantisation conditions
close to $-\i\tf{\zeta}{2}$ when $N\tend + \infty$. One, however, expects that solutions
to the non-linear problem associated with such solutions will give rise to Eigenvalues
of the quantum transfer matrix which collapse to zero as $N\tend +\infty$. This and
the desire to keep the technicalities of the analysis to a reasonable minimum led
us to introduce that hypothesis.

\item[iii)]One should be able to wave-off the condition
$\mathrm{b)}$  by accommodating the parameter $M>0$ introduced in the definition of
$\de_T$, \textit{c.f.}\ \eqref{definition du parametre delta T}, provided that there is no
condensation, when $T\tend 0$, of roots to the equation $1+\e{e}^{- u(\la\,|\, \mathbb{Y})/T}=0$
close to $\msc{C}_{\veps}$, other than on $\msc{C}_{\veps}$ itself. While this seems
reasonable, it is not completely evident to prove, as seen from the discussion
developed in Section \ref{Section conditions de quantification}.

\item[iv)] Hypothesis $\mathrm{d)}$
allows one to separate the roots located at finite distance from the Fermi points $\pm q$
and those collapsing on $\pm q$ when $N\tend + \infty$ and $T \tend 0^+$. The roots collapsing
to $\pm q$ are treated in a separate way, so that this hypothesis is, in fact, no limitation
at all.

\item[v)] Further, dealing with coinciding roots, \textit{i.e.}\ when $\mathrm{f)}$ is not
valid, just as with roots not satisfying the repulsion principle, \textit{i.e.}\ when
$\mathrm{e)}$ is not valid, would make the analysis of the quantisation conditions
\eqref{equation quantification particles et tous cas Trotter fini}
carried out in Section~\ref{Section conditions de quantification} much bulkier,
but is definitely within grasp with enough effort.

\end{itemize}

\begin{figure}
\begin{center}

\includegraphics[width=0.65\textwidth]{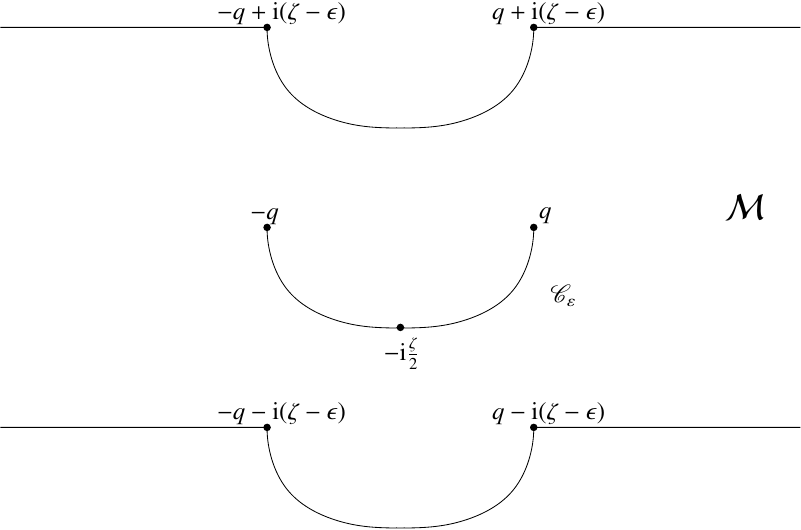}

\caption{The curve $\msc{C}_{\veps}$ and the deformed strip $\mc{M}$.}
\label{Domaine pour point fixe NLIE}
\end{center}
\end{figure}

To close this subsection, we introduce a functional space which plays a central role
in our analysis. Given the set $\mc{M}$ depicted in
Fig.~\ref{Domaine pour point fixe NLIE}, let
\beq
\mc{E}_{\mc{M}} \, = \, \Big\{ f  \; : \; f  \in \mc{O}(\mc{M})\; , \;  f(\la) \limit{ \la \in \mc{M} }{ \infty } 0  \; , \;  \norm{f}_{L^{\infty}(\mc{M})} \, \leq \, C_{\mc{M}} \cdot T^2   \Big\}
\label{definition espace fonctionnel principal}
\enq
in which $ C_{\mc{M}}$ is some not too small $T$-independent constant. We shall be
more explicit about the requirements on the magnitude of $C_{\mc{M}}$ later on.  
Moreover, the temperature $T$ will be treated as a small parameter. In particular,
for a given $C_{\mc{M}}$, we shall always focus on the regime where $T C_{\mc{M}}<1$.
The space $\mc{E}_{\mc{M}}$ endowed with the distance 
\beq
\op{d}_{\mc{E}_{\mc{M}}}\big( f, g\big) \; = \; \norm{ f - g }_{L^{\infty}(\mc{M})} 
\label{ecriture metrique pour espace EM}
\enq
is a complete metric space, see
Proposition~\ref{Proposition caractere espace metrique complet pour EM}.

\subsection{The main results}

To state our first main result relative to the unique existence of solutions and
for the purpose of further handling, it is convenient to introduce several
shorthand notations. First of all, taken that $\op{Y}, \op{Y}_{\e{sg}}, \op{X}$
are understood as collections of parameters and not as sets, it is understood throughout
the paper that $\sum_{y \in \op{Y} }{} f(y) \, = \, \sum_{a=1}^{|\op{Y}|} f(\op{y}_a)$,
\textit{etc}. We now introduce sets $\op{X}^{\prime}_{\e{var}}$, resp.\
$\op{Y}^{\prime}_{\e{var}}$, which will be complementary to $\op{X}$, resp.\ $\op{Y}$,
in the sense that they will be built out of points in $\op{D}_{\pm q, \mf{c}_{\e{loc}}}$,
with $\mf{c}_{\e{loc}}>0$ as in point $\mathrm{d)}$ of
Hypothesis~\ref{Hypotheses solubilite NLIE}. More precisely, if
\beq
x_1^{(\a)},\dots, x_{\varkappa_0^{(\a)}}^{(\a)} \in \op{D}_{\ups_{\a} q , \mf{c}_{\e{loc}} } \quad \e{and} \quad
y_1^{(\a)},\dots, y_{\mf{y}_0^{(\a)}}^{(\a)} \in \op{D}_{\ups_{\a} q , \mf{c}_{\e{loc}} }\, , \quad  \a \in \{L, R\} \; ,
\enq
where we agree upon
\beq
\ups_{L}=-1 \qquad \e{and} \qquad  \ups_{R}=1 \, ,
\label{definition de ups alpha}
\enq
then
\beq
\op{X}^{\prime}_{\e{var}} \; = \; \bigcup\limits_{\a \in \{L, R\} } \Big\{  x_{a}^{(\a)} \Big\}_{a=1}^{ \varkappa_0^{(\a)} } \quad \e{and} \quad
\op{Y}^{\prime}_{\e{var}} \; = \; \bigcup\limits_{\a \in \{L, R\} } \Big\{  y_{a}^{(\a)} \Big\}_{a=1}^{ \mf{y}_0^{(\a)} } \;.
\enq

Further, we introduce the operation of algebraic concatenation $\oplus$ and
difference $\ominus$ of collections of parameters (see
Section \ref{Appendix Section algebraic sum of sets} of the Appendix for more details)
and denote
\beq
\op{Y}_{\e{gen}} \; = \; \op{Y} \oplus \op{Y}^{\prime}_{\e{var}}   \qquad \e{and} \qquad
\op{X}_{\e{gen}} \; = \; \op{X} \oplus \op{X}^{\prime}_{\e{var}}
\enq
as well as
\beq
\mathbb{X}_{\e{gen}} = \op{Y}_{\e{gen}}\oplus \op{Y}_{\e{sg}} \ominus \op{X}_{\e{gen}} \quad \e{and} \quad
\mathbb{X}_{\e{gen};\varkappa} = \mathbb{X}_{\e{gen}} \oplus \{ \varkappa \}^{\oplus \mf{m} } \;.
\label{defintion mathbb X et son kappa shifte}
\enq
For the time being $\mf{m}\in \mathbb{Z}$, but its precise value will be specified
later on. These conventions induce the following shorthand notations relative
to sums and products:
\begin{align}
\sul{y \in \mathbb{X}_{\e{gen}} }{} f(y) \; & = \; \sul{y \in \op{Y}_{\e{gen}} }{} f(y) \, + \,  \sul{y \in \op{Y}_{\e{sg}} }{} f(y)
                                                                      \, - \,  \sul{y \in \op{X}_{\e{gen}} }{} f(y) \quad \e{and} \quad
\sul{y \in \mathbb{X}_{\e{gen};\varkappa} }{} f(y) \; = \;\sul{y \in \mathbb{X}_{\e{gen}}}{} f(y) \; + \; \mf{m} f(\varkappa) \;, \\[1ex]
%
%
%
%
%
%
%
%
\pl{y \in \mathbb{X}_{\e{gen}} }{} f(y) \; & = \; \f{ \pl{y \in \op{Y}_{\e{gen}} }{} f(y)  \cdot   \pl{y \in \op{Y}_{\e{sg}} }{} f(y) }{ \pl{y \in \op{X}_{\e{gen}} }{} f(y) }
\qquad \e{and} \qquad
\pl{y \in \mathbb{X}_{\e{gen};\varkappa} }{} f(y) \; = \;\pl{y \in \mathbb{X}_{\e{gen}} }{} f(y)  \cdot  f^{\mf{m}}(\varkappa) \;.
\end{align}
Moreover, one defines the function
\beq
u_1(\la\,|\, \mathbb{X}_{\e{gen}} ) \, = \,
-\i \pi  \mf{s} Z_{\e{c}}(\la)  \, - \,
2 \i \pi \sul{ y \in \mathbb{X}_{\e{gen}} }{} \phi_{\e{c}}(\la,y) \;.
\label{definition fonction u 1}
\enq

We are now in position to introduce the spaces of functions of interest for the
analysis of the finite Trotter number problem along with its infinite Trotter
number counterpart. These spaces depend on a selection of integers
$p_{0,s}^{(\a)} \in \mathbb{N}$, $s=1,\dots, \mf{y}_{0}^{(\a)}$
and $h_{0,\ell}^{(\a)} \in \mathbb{N}$, $\ell=1,\dots, \varkappa_{0}^{(\a)}$ in
a way that will be explained below. They take the form
\beq
\wh{\mc{E}}_{\mc{M}} \, =\, \Big\{ \, \wh{u}(\la) \, = \, \mc{W}_N(\la)\,  + \, T u_1\big(\la\,|\, \wh{\mathbb{X}}\, \big) + f(\la) \; : \; f  \in \mc{E}_{\mc{M}} \Big\} \;,
\label{definition espace EM hat}
\enq
in what concerns the finite Trotter number space, and
\beq
\wt{\mc{E}}_{\mc{M}} \, =\, \Big\{ u(\la) \, = \, \veps_{\e{c}}(\la)\,  + \, T u_1\big(\la\,|\, \mathbb{X}\big) + f(\la) \; : \; f  \in \mc{E}_{\mc{M}} \Big\} \;,
\label{definition espace EM tilde}
\enq
in what concerns the infinite Trotter number space. Above $\mc{W}_N$ is as
in \eqref{definition solution LIE deformee WN}, $u_1$ as in \eqref{definition fonction u 1},
while $\veps_{\e{c}}$ is as in~\eqref{ecriture LIE pour veps c}.

The algebraic collections of parameters $\wh{\mathbb{X}}$, resp.\ $\mathbb{X}$, appearing
in the definition of $\wh{\mc{E}}_{\mc{M}}$, resp.\ $\wt{\mc{E}}_{\mc{M}}$, are defined
as follows
\beq
\wh{\mathbb{X}} \, = \, \wh{\op{Y}}_{\e{tot}}\oplus \op{Y}_{\e{sg}} \ominus \wh{\op{X}}_{\e{tot}}  \quad \e{with} \quad
\left\{ \ba{ccc}  \wh{\op{Y}}_{\e{tot}} & = & \op{Y} \oplus \wh{\op{Y}}\, \! ^{\prime} \vspace{1mm} \\
               \wh{\op{X}}_{\e{tot}} & = & \op{X} \oplus \wh{\op{X}}\, \! ^{\prime} \ea \right.\; ,
\label{definition ensemble hat X pour somme sols part trous}
\enq
\e{resp}.
\beq
\mathbb{X} \, = \, \op{Y}_{\e{tot}}\oplus \op{Y}_{\e{sg}} \ominus \op{X}_{\e{tot}}  \quad \e{with} \quad
\left\{ \ba{ccc}  \op{Y}_{\e{tot}} & = & \op{Y} \oplus \op{Y}^{\prime} \vspace{1mm} \\
               \op{X}_{\e{tot}} & = & \op{X} \oplus \op{X}^{\prime} \ea \right. \;.
\label{definition ensemble X pour somme sols part trous}
\enq
The points building up $\wh{\op{Y}}\, \! ^{\prime}, \wh{\op{X}}\, \! ^{\prime},
\op{Y}^{\prime}, \op{X}^{\prime} \subset  \op{D}_{-q, \mf{c}_\e{loc}} \cup
\op{D}_{q, \mf{c}_\e{loc}} $ are defined as solutions
to the system of equations
\beq
\left\{ \ba{ccc}  \wh{u}\big(\, \wh{\op{x}}_{0;\ell}^{\,(\a)} \big) & = & -2\i\pi T \ups_{\a} \big(h_{0,\ell}^{(\a)}  \, +\, \tfrac{1}{2} \big)  \vspace{2mm} \\
\wh{u}\big(\, \wh{\op{y}}_{0;s}^{\,(\a)} \big) & = & 2\i\pi T \ups_{\a} \big(p_{0,s}^{(\a)}  \, +\, \tfrac{1}{2} \big)  \ea \right. \; , \quad \e{resp}. \quad
\left\{ \ba{ccc}  u\big( \op{x}_{0;\ell}^{(\a)} \big) & = & -2\i\pi T \ups_{\a} \big(h_{0,\ell}^{(\a)}  \, +\, \tfrac{1}{2} \big)  \vspace{2mm} \\
u\big( \op{y}_{0;s}^{(\a)} \big) & = & 2\i\pi T \ups_{\a} \big(p_{0,s}^{(\a)}  \, +\, \tfrac{1}{2} \big)  \ea \right. \;,
 \label{ecriture systeme eqns couplees pour fct hat u et u des espaces fnels}
\enq
with $\wh{u} \in \wh{\mc{E}}_{\mc{M}}$ and  $u \in \mc{E}_{\mc{M}}$. Therefore, it holds
\beq
\left\{ \ba{ccc} \wh{\op{X}}\, \! ^{\prime} & = & \bigcup\limits_{\a \in \{L, R\} } \Big\{ \,  \wh{\op{x}}_{0;\ell}^{\, (\a)} \Big\}_{\ell=1}^{ \varkappa_0^{(\a)} } \\
     \wh{\op{Y}}\, \! ^{\prime} & = & \bigcup\limits_{\a \in \{L, R\} } \Big\{ \,  \wh{\op{y}}_{0;s}^{\, (\a)} \Big\}_{s=1}^{ \mf{y}_0^{(\a)} }  \ea \right.  \; ,
\quad \e{resp}. \quad
\left\{ \ba{ccc} \op{X}^{\prime} & = & \bigcup\limits_{\a \in \{L, R\} } \Big\{  \op{x}_{0;\ell}^{(\a)} \Big\}_{\ell=1}^{ \varkappa_0^{(\a)} } \\
          \op{Y}^{\prime} & = & \bigcup\limits_{\a \in \{L, R\} } \Big\{  \op{y}_{0;s}^{(\a)} \Big\}_{s=1}^{ \mf{y}_0^{(\a)} }  \ea \right. \;.
\enq

Note that \eqref{ecriture systeme eqns couplees pour fct hat u et u des espaces fnels}
is a highly coupled system of equations, and thus its solvability is absolutely not evident.
However, that issue will be dealt with in the theorem given below. Still, in order
to be able to formulate the latter, we need to introduce one more construction.
Given any $\wh{u} \in \wh{\mc{E}}_{\mc{M}}$, we denote by $\wh{\mf{Z}}^{\, (\a)}$ the
set of zeroes of $\la \mapsto 1+\ex{-\f{1}{T}\wh{u}(\la)}$ that are located in
$\op{D}_{\ups_{\a} q, \mf{c}_{\e{loc}}}$ between $\msc{C}_{\e{ref}}$ and the line
$\Im\big[\, \wh{u}\, \big]=0$. As will be apparent from the discussion carried out in
Sections~\ref{Section espace fnel pour la NLIE et ses ptes} and
\ref{Section NLIE modifiee et solvabilite par pt fixe}, for each $\a\in \{L,R\}$, either
$\wh{\mf{Z}}^{\,(\a)}\cap \wh{\op{X}}\, \! ^{\prime} \, = \, \emptyset $ or
$\wh{\mf{Z}}^{\,(\a)}\cap \wh{\op{Y}}\, \! ^{\prime} \, = \, \emptyset $.
Hence, we introduce  the set of zeroes which have non-trivial intersections with
$ \wh{\op{X}}\, \! ^{\prime}$ and $ \wh{\op{Y}}\, \! ^{\prime}$
\beq
\wh{\mf{Z}}_{X} \; = \;
\bigcup_{\a \in \{L, R\}}
\Big\{ x \in \wh{\mf{Z}}^{\, (\a)} \; : \;
\wh{\mf{Z}}^{\, (\a)} \cap \wh{\op{X}}\, \! ^{\prime} \not= \emptyset \Big\} \qquad
\e{and} \qquad
\wh{\mf{Z}}_{Y} \; = \;
\bigcup_{\a \in \{L, R\}}
\Big\{ x \in \wh{\mf{Z}}^{\, (\a)} \; : \;
\wh{\mf{Z}}^{\, (\a)} \cap \wh{\op{Y}}\, \! ^{\prime} \not= \emptyset \Big\} \;.
\label{definition ensembles zeros Zx et Zy}
\enq
We then define modified local zero sets around $\pm q$
\beq
\wh{\op{X}}\, \! ^{\prime}_{\e{ref}}  \; = \;
\Big\{ \wh{\op{X}}\, \! ^{\prime} \setminus  \wh{\mf{Z}}_{X}  \Big\}
\cup \Big\{  \wh{\mf{Z}}_{Y}  \setminus \wh{\op{Y}}\, \! ^{\prime} \Big\}
\quad \e{and} \quad
\wh{\op{Y}}\, \! ^{\prime}_{\e{ref}} \; = \;
\Big\{ \wh{\op{Y}}\, \! ^{\prime} \setminus  \wh{\mf{Z}}_{Y}  \Big\}
\cup \Big\{  \wh{\mf{Z}}_{X}  \setminus \wh{\op{X}}\, \! ^{\prime} \Big\} \;.
\label{definition des ensembles hat Y et X prime ref}
\enq
\begin{figure}
\begin{center}
\includegraphics[width=0.98\textwidth]{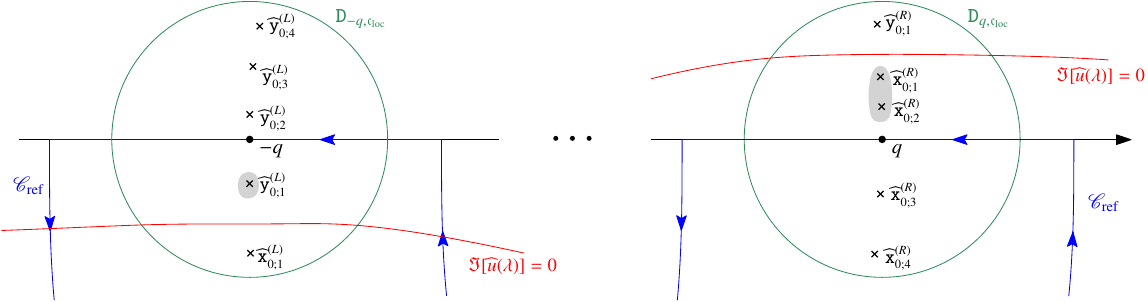}
\caption{A sketch of the particle and hole parameters (shaded) that have
to be removed from the sets $\wh{\op{Y}}'$ and $\wh{\op{X}}'$ in order to
switch to $\wh{\op{Y}}_{\e{ref}}'$ and $\wh{\op{X}}_{\e{ref}}'$.}
\label{Swapping particles and holes}
\end{center}
\end{figure}%
The situation is sketched in Fig.~\ref{Swapping particles and holes}.
Quite analogously, we define quantities for the case of the infinite Trotter
number regime by omitting the $\, \wh{} \; $. Finally, we introduce the
associated collections of parameters
\beq
\wh{\mathbb{X}}_{\e{ref}} \,  =  \, \Big\{ \op{Y}\oplus \wh{\op{Y}}\, \! ^{\prime}_{\e{ref}}  \Big\} \oplus
\op{Y}_{\e{sg}} \ominus  \Big\{ \op{X}\oplus \wh{\op{X}}\, \! ^{\prime}_{\e{ref}}  \Big\} \;, \qquad \e{resp}. \quad
\mathbb{X}_{\e{ref}} \,  =  \, \Big\{ \op{Y}\oplus \op{Y}^{\prime}_{\e{ref}}  \Big\} \oplus
\op{Y}_{\e{sg}} \ominus  \Big\{ \op{X}\oplus \op{X}^{\prime}_{\e{ref}}  \Big\}
\label{defintion mathbb X ref et son hat}
\enq
and denote
\beq
\wh{\mathbb{X}}_{ \e{ref} ; \varkappa}  \, = \,  \wh{\mathbb{X}}_{\e{ref}}\oplus \{ \varkappa \}^{\oplus \mf{m} }\;, \qquad \e{resp}. \quad
\mathbb{X}_{ \e{ref} ; \varkappa}  \, = \,  \mathbb{X}_{\e{ref}}\oplus \{ \varkappa \}^{\oplus \mf{m} } \;.
\label{defintion mathbb X ref et son hat avec un monodromie varkappa}
\enq

We are now in position to state our first result pertaining to the solvability of the non-linear integral equation with free parameters.

\begin{theorem}
\label{Theorem ppl existance solution NLIE}

Let the collections of parameters $\op{X}=\{\op{x}_a\}_{1}^{|\op{X}|}$
and $\op{Y}=\{\op{y}_a\}_{1}^{|\op{Y}|}$, with $|\op{X}|$ and $|\op{Y}|$ fixed,
satisfy Hypothesis~\ref{Hypotheses solubilite NLIE}. Let $p_{0,a}^{(\a)}, h_{0,a}^{(\a)} \in \mathbb{N}$
be such that
\beq
T p_{0,a}^{(\a)} \, = \, \e{o}(1) \;\;, \; \; a=1,\dots, \mf{y}_{0}^{(\a)} \qquad and \qquad T h_{0,a}^{(\a)} \, = \, \e{o}(1) \;\;, \; \; a=1,\dots, \varkappa_{0}^{(\a)} \; ,
\enq
with $\mf{y}_{0}^{(\a)}$, $\varkappa_{0}^{(\a)}$ fixed and such that
\beq
0 \; = \; -\mf{s} - |\op{Y}\, | - \mf{y}_{0}^{(L)} - \mf{y}_{0}^{(R)}
- |\op{Y}_{\e{sg}}|  +  |\op{X}| + \varkappa_{0}^{(L)} + \varkappa_{0}^{(R)} \;,
\enq
where $\mf{s}\in \mathbb{Z}$ is the spin.

Then, there exist
\begin{itemize}
 
\item $T_0>0$ small enough 
 
\item $\eta>0$ small enough
 
\item $C_{\mc{M}}^{(0)}>0$ large enough
  
\end{itemize}
such that the space $\wh{\mc{E}}_{\mc{M}}$ introduced in \eqref{definition espace EM hat}
is well-defined, provided that
\beq
T_0 > T > 0 \;  , \quad  \eta > \f{1}{N T^4} \quad and \quad  C_{\mc{M}} \, >\,  C_{\mc{M}}^{(0)}\; .
\enq
In particular, the system
\eqref{ecriture systeme eqns couplees pour fct hat u et u des espaces fnels}
is uniquely solvable for any given $\wh{u}\in \wh{\mc{E}}_{\mc{M}}$.

For this range of parameters, the non-linear integral equation
\beq
\wh{u}\,(\xi) \, = \,  h \, -\, T\mf{w}_N(\xi)  \; - \; \i \pi \mf{s} T  \; - \;
\i T \sul{y \in \wh{\mathbb{X}}_{ \e{ref} ; \varkappa}  }{}\th_{+}( \xi-y)
\; - \; T\Oint{ \msc{C}_{\e{ref}}  }{} \dd \la \: K(\xi-\la) \cdot  \msc{L}\mathrm{n}_{ \msc{C}_{\e{ref}} }\Big[ 1+  \ex{ - \f{1}{T}\wh{u} } \, \Big](\la) \;,
\label{ecriture NLIE a Trotter fini}
\enq
subject to the index condition
\beq
\mf{m} \, = \, - \Oint{ \msc{C}_{\e{ref}}  }{ }  \f{\dd \la}{2 \i \pi T} \:
\f{  \wh{u}^{\, \prime}\!(\la) }{ 1+  \ex{ \f{1}{T} \wh{u}(\la)} } \; = \;
-\mf{s} - |\op{Y}\, | - |\, \wh{\op{Y}}\, ^{ \prime}\!_{\e{ref}} | - |\op{Y}_{\e{sg}}|  +  |\op{X}| + |\,\wh{\op{X}}\, ^{\prime}\!_{\e{ref}} \, | \;,
\label{ecriture monodromie Trotter fini}
\enq
is well-defined on $\wh{\mc{E}}_{\mc{M}}$ and admits a unique solution
$\la \mapsto \wh{u}\, \big( \la \,|\, \wh{\mathbb{X}}\, )$ belonging to $\wh{\mc{E}}_{\mc{M}}$.
The collection of parameters $\wh{\mathbb{X}}_{ \e{ref} ; \varkappa}$ appearing in
\eqref{ecriture NLIE a Trotter fini} is as introduced in
\eqref{defintion mathbb X ref et son hat}--\eqref{defintion mathbb X ref et son hat avec un monodromie varkappa},
while $\wh{\mathbb{X}}$ has been defined in
\eqref{definition ensemble hat X pour somme sols part trous}.

For $\la$ uniformly away from $\pm \i\tf{\zeta}{2}$, the unique solution
$\wh{u}\, \big( \la \,|\,  \wh{\mathbb{X}} \, \big)$ converges, as $N\tend +\infty$,
to the unique solution $u(\la\,|\, \mathbb{X})$ to the non-linear integral equation
%
%
\beq
u\,(\xi) \, = \,   \veps_0(\xi)  \; - \; \i \pi \mf{s} T  \; - \;
\i T \sul{y \in \mathbb{X}_{ \e{ref} ; \varkappa} }{}\th_{+}( \xi-y)
\; - \; T\Oint{ \msc{C}_{\e{ref}}  }{} \dd \la \: K(\xi-\la) \cdot  \msc{L}\mathrm{n}_{ \msc{C}_{\e{ref}} }\Big[ 1+  \ex{ - \f{1}{T}u } \, \Big](\la) \;,
\label{ecriture NLIE a Trotter infini}
\enq
subject to the index condition
\beq
\mf{m} \, = \, - \Oint{ \msc{C}_{\e{ref}}  }{ } \f{ \dd \la }{2\i\pi T }
\f{  u^{ \prime}\!(\la) }{ 1+  \ex{ \f{1}{T} u(\la)} } \; = \;
- \mf{s} - |\op{Y}\,| - |\, \op{Y}^{\, \prime}_{\e{ref}} | -|\op{Y}_{\e{sg}}| + |\op{X}| + |\, \op{X}^{\, \prime}_{\e{ref}} |\;,
\label{ecriture monodromie Trotter infini}
\enq
on the space of functions $\wt{\mc{E}}_{\mc{M}}$, \textit{c.f.}\
\eqref{definition espace EM tilde}, on which it is well defined for the range of
parameters considered. Finally, $\mathbb{X}_{ \e{ref} ; \varkappa}$ appearing
in \eqref{ecriture NLIE a Trotter fini} has been introduced in
\eqref{defintion mathbb X ref et son hat}--\eqref{defintion mathbb X ref et son hat avec un monodromie varkappa},
while $\mathbb{X}$ is given by \eqref{definition ensemble X pour somme sols part trous}.
\end{theorem}

The above theorem produces the unique solution to the non-linear integral
equation provided that the sets $\op{X}, \op{Y}$ of auxiliary parameters satisfy
Hypothesis~\ref{Hypotheses solubilite NLIE}. One may therefore consider the
solvability of the quantisation conditions
\eqref{equation quantification particles et tous cas Trotter fini}
for these collections of solutions $\wh{u}\, \big(\la\,|\, \wh{\mathbb{X}}\big)$.
The quantisation conditions \eqref{equation quantification particles et tous cas Trotter fini} will thus impose specific values
$\wh{\op{X}}, \wh{\op{Y}}$ to the $\op{X}, \op{Y}$ parameters arising as the building blocks of $ \wh{\mathbb{X}}$
through \eqref{definition ensemble hat X pour somme sols part trous}.
Clearly, the solution sets
\beq
\wh{\op{Y}} \oplus \wh{\op{Y}}\, ^{ \prime}\!_{\e{ref}} \, =\, \wh{\mc{Y}} \qquad \e{and} \qquad
\wh{\op{X}} \oplus \wh{\op{X}}\, ^{ \prime}\!_{\e{ref}} \, =\, \wh{\mf{X}}
\label{ecriture de hat Y et hat X comme somme directe X et X ref}
\enq
have to fulfil the requirements $\mathrm{a)-d)}$
of Hypothesis~\ref{Hypotheses solubilite NLIE} for this consideration to make sense
in the setting provided by Theorem~\ref{Theorem ppl existance solution NLIE}.
It is direct to see from the very procedure that constructs $\wh{\op{Y}}^{\, \prime}_{\e{ref}}$, $\wh{\op{X}}^{\, \prime}_{\e{ref}}$
that these sets of parameters do satisfy points a)-c) of Hypothesis~\ref{Hypotheses solubilite NLIE}.
Point d) is not satisfied by construction as  $\wh{\op{Y}}^{\, \prime}_{\e{ref}}$, $\wh{\op{X}}^{\, \prime}_{\e{ref}}$
do precisely describe that part of the collection of points building up the solution sets $\wh{\mc{Y}}, \wh{\mf{X}}$
which are located in some small neighbourhood of $\pm q$.

In the
low-$T$ limit, we are able to provide a full classification of the solution set, under
the additional conditions of Hypothesis~\ref{Hypotheses eqns de quantification}.
The derivation of this classification result proceeds in several steps. First, in
Proposition~\ref{Proposition convergence trotter infini systeme de racines}, we
establish that all solutions $\big(\,  \wh{\mc{Y}}, \wh{\mf{X}} \, \big)$ to the finite
Trotter number quantisation conditions which satisfy
Hypothesis~\ref{Hypotheses eqns de quantification} converge, as $N\tend +\infty$,
towards a solution $\big( \mc{Y}$, $\mf{X} \big)$ to the infinite Trotter number quantisation
conditions. Then, in
Proposition~\ref{Proposition caracterisation dvpt basse T pour part et trous},
we provide a thorough characterisation of the latter.
 

Analogously to \eqref{definition racines singulieres et reguliere}, we partition the
collection $\wh{\mc{Y}}$ into a collection of regular roots $\wh{\mc{Y}}_{\e{r}}$
and of the shifted singular ones $\wh{\wt{\mc{Y}}}_{\e{sg}}$. In particular, we introduce
the set $\wh{\mc{Y}}_{\e{sg}}$. We further set
\beq
\wh{\mathbb{Y}} \, = \, \wh{\mc{Y}} \oplus \wh{\mc{Y}}_{\e{sg}} \ominus \wh{\mf{X}}  \qquad \e{and} \qquad
\wh{\mathbb{Y}}_{\varkappa} \, = \,\wh{\mathbb{Y}} \, \oplus \{ \varkappa \}^{\oplus \mf{m} } \;.
\label{definition ensemble Y hat et Y hat q}
\enq

\begin{prop}
\label{Proposition convergence trotter infini systeme de racines}
Let the collection of parameters $\wh{\mc{Y}}$ and $\wh{\mf{X}}$ solving the
quantisation conditions,
\beq
\ex{- \f{1}{T} \wh{u}(s\,|\, \wh{\mathbb{Y}}) } \; = \; -1 \quad  and \quad  \wh{u}^{\, \prime}(s \,|\, \wh{\mathbb{Y}})  \; \not= \; 0 \qquad \text{for all} \qquad
s \in \wh{\mc{Y}} \cup \wh{\mf{X}} \;, 
\label{ecriture conditions de quantifications Trotter fini}
\enq
fulfil Hypothesis~\ref{Hypotheses eqns de quantification}. Moreover, let $\wh{\op{Y}}, \wh{\op{X}}$ defined through \eqref{ecriture de hat Y et hat X comme somme directe X et X ref}
fulfil Hypothesis~~\ref{Hypotheses solubilite NLIE}.

Then, there exists $T_0>0$ small enough and $\eta>0$ small enough such that, for any
$T_0>T>0$ and $\eta \, > \, \tf{1}{(NT^4)}$,
\beq
\wh{x}_a \; = \; x_a \, + \, \e{O}\biggl( \f{1}{NT^3} \biggr) \qquad
\text{and} \qquad \wh{y}_a \; = \; y_a \, + \, \e{O}\biggl( \f{1}{N T^3} \biggr) 
\label{ecriture convergence position N fini vers N infini particle hole Bethe roots}
\enq
with a remainder that is uniform in $T$. The collections of parameters
$\mf{X}=\{x_a\}_{1}^{|\mf{X}|}$ and $\mc{Y}=\{y_a\}_{1}^{|\mf{X}|}$ 
are such that the ``infinite Trotter number quantisation conditions'' are fulfilled,
\beq
\ex{- \f{1}{T} u(s \,|\, \mathbb{Y}) } \; = \; -1 \quad 
\text{and} \quad  u^{\prime}(s \,|\, \mathbb{Y})  \; \not= \; 0 \qquad \text{for all} \qquad
s \in \mc{Y} \cup \mf{X} \;.  
\label{ecriture condition de quantification Trotter infini}
\enq
Above $ u(s \,|\, \mathbb{Y})$ is the unique solution of
\eqref{ecriture NLIE a Trotter infini}-\eqref{ecriture monodromie Trotter infini}
on the space $\wt{\mc{E}}_{\mc{M}}$, introduced in \eqref{definition espace EM tilde}.
\end{prop}

Theorem~\ref{Theorem ppl existance solution NLIE} and
Proposition~\ref{Proposition convergence trotter infini systeme de racines} already
allow one to take the infinite Trotter number limit of the Eigenvalues of the
quantum transfer matrix. In order to state this result, we introduce analogues of
the sets $\wh{\mathbb{X}}_{\e{ref}}$ and $\mathbb{X}_{\e{ref}}$, \textit{c.f.}\
\eqref{defintion mathbb X ref et son hat},
\eqref{defintion mathbb X ref et son hat avec un monodromie varkappa},
on the level of parameters solving the quantisation conditions, namely
\beqa
\wh{\mathbb{Y}}_{\e{ref}} \; = \; \wh{\mc{Y}}_{\e{ref}} \oplus \wh{\mc{Y}}_{\e{sg}} \ominus \wh{\mf{X}}_{\e{ref}}
\quad \e{with} \quad
\left\{ \ba{cc c }   \wh{\mf{X}}_{\e{ref}}  & = & \Big\{ \wh{\mf{X}} \setminus \wh{\mf{Z}}_{X} \Big\} \bigcup \Big\{ \wh{\mf{Z}}_{Y}  \setminus  \wh{\mc{Y}} \Big\}  \vspace{2mm} \\
                     \wh{\mc{Y}}_{\e{ref}}  & = & \Big\{ \wh{\mc{Y}} \setminus \wh{\mf{Z}}_{Y} \Big\} \bigcup \Big\{ \wh{\mf{Z}}_{X}  \setminus  \wh{\mf{X}} \Big\} \ea \right.
\label{definition Y ref}
\eeqa
in which the sets $\wh{\mf{Z}}_{X,Y}$ are defined analogously to
\eqref{definition ensembles zeros Zx et Zy} with now $\wh{\mf{Z}}^{\,(\a)}$
corresponding to the zeroes of $\la \mapsto 1 \, + \, \ex{-\f{1}{T} \wh{u}(\la\,|\, \mathbb{Y})}$
located between $\msc{C}_{\e{ref}}$ and the line
$\Im\big[ \, \wh{u}(*\,|\, \wh{\mathbb{Y}} ) \big]=0$
inside of the disc $\op{D}_{\ups_{\a} q , \mf{c}_{\e{loc}}}$. The corresponding
quantities associated with the infinite Trotter number regime are defined
analogously but without the $\; \wh{} \,$.

\begin{cor}
\label{Corolaire limit Trotter vp QTM} 
Let a Bethe Eigenstate of the quantum transfer matrix be associated with the
triple $\Big(\, \wh{u}\,(*\,|\, \wh{\mathbb{Y}}), \wh{\mc{Y}}, \wh{\mf{X}} \, \Big)$
defining a solution to the finite Trotter number non-linear problem associated
with the reference contour $\msc{C}_{\e{ref}}$ and such that $\wh{\mc{Y}}, \wh{\mf{X}}$
fulfil Hypothesis~\ref{Hypotheses eqns de quantification} with $\wh{\op{Y}}, \wh{\op{X}}$ defined through \eqref{ecriture de hat Y et hat X comme somme directe X et X ref}
fulfilling Hypothesis~~\ref{Hypotheses solubilite NLIE}. The corresponding Eigenvalue of the
quantum transfer matrix admits the infinite Trotter number limit and the latter is
expressed in terms of the triple $\Big( \, u\,(*\,|\, \mathbb{Y}), \mc{Y}, \mf{X}\, \Big)$
defining a solution to the infinite Trotter number limit of the non-linear problem
associated with $\msc{C}_{\e{ref}}$:
\beq
     \lim_{N\tend +\infty}
        \bigg\{ \wh{\La}\, \Big( 0 \big|\, \wh{u}(*\,|\, \wh{\mathbb{Y}}_{\e{ref}; \varkappa}),
	        \wh{\mc{Y}}_{\e{ref}}, \wh{\mf{X}}_{\e{ref}} \Big) \bigg\}\, = \,
     \La\Big( 0 \big| u(*\,|\, \mathbb{Y}), \mc{Y}_{\e{ref}}, \mf{X}_{\e{ref}} \Big)
        \, = \, (-1)^{\mf{s}}  \cdot \ex{  \f{h}{2T}   -\f{2J}{T}\cos(\zeta) }
	\cdot \ex{\i \mc{P}(\mathbb{Y}) } \;. 
\label{ecriture forme explicite basse T pour  vp QTM}
\enq
Here we agree upon 
\beq
     \mc{P}( \mathbb{Y} ) \, = \,
        \sul{ y \in \mathbb{Y}_{\e{ref}; \varkappa} }{} p_0( y  )  \ - \,
	\Oint{ \msc{C}_{ \e{ref}}  }{ } \f{ \dd \la }{ 2 \i \pi } \:
	p_0^{\prime}(\la ) \,
     \msc{L}\mathrm{n}_{\msc{C}_{\e{ref}}}
        \big[ 1+  \ex{ -\f{1}{T}u   } \, \big](\la \,|\, \mathbb{Y})
	\;,
\label{definition impulsion thermale}
\enq
where $p_0$ was introduced in \eqref{definition de p0}, while $\mathbb{Y}_{\e{ref}; \varkappa}$
is as in \eqref{defintion mathbb X ref et son hat avec un monodromie varkappa} and corresponds
to the index of $u$ with respect to $\msc{C}_{\e{ref}}$, \textit{c.f.}\
\eqref{ecriture monodromie Trotter infini}. It further holds that 
\beq
\lim_{N\tend +\infty} \Biggl\{  \f{ \wh{\La}\, \Big(  \tfrac{J t \sin\zeta }{N} \big|\, \wh{u}(*\,|\, \wh{\mathbb{Y}}) , \wh{\mc{Y}}_{\e{ref}}, \wh{\mf{X}}_{\e{ref}} \,  \Big)  }
{ \wh{\La}\, \Big(  -\tfrac{J t \sin\zeta }{N} \big|\, \wh{u}(*\,|\, \wh{\mathbb{Y}}) , \wh{\mc{Y}}_{\e{ref}}, \wh{\mf{X}}_{\e{ref}} \, \Big)  }\Biggr\}^{N} \; = \; \ex{\i  t \mc{E}(\mathbb{Y})  }\;,
\label{definition eto E de Y}
\enq
where we agree upon 
\beq
     \mc{E}(\mathbb{Y}) \, = \, \sul{ y \in \mathbb{Y}_{\e{ref}; \varkappa} }{} \veps_0( y  )  \ - \,
        \Oint{ \msc{C}_{ \e{ref}}  }{} \f{ \dd \la }{ 2 \i \pi } \:
	\veps_0^{\prime}(\la ) \,
     \msc{L}\mathrm{n}_{\msc{C}_{ \e{ref}}} 
        \big[ 1+  \ex{ -\f{1}{T}u   } \, \big](\la \,|\, \mathbb{Y}) \;,
\label{definition mc E de Y}
\enq
and $\veps_{0}$ was defined in \eqref{definition eps 0}. 
\end{cor}

\Proof 
The results are an immediate consequence of the integral representation
\eqref{ecriture forme vp QTM} and the results relative to the infinite Trotter limit 
provided by Theorem~\ref{Theorem ppl existance solution NLIE} and
Proposition~\ref{Proposition convergence trotter infini systeme de racines}. \qed

\vspace{2mm}

The next result provides a full characterisation of the solution set to the finite and infinite
Trotter number quantisation conditions
\eqref{ecriture condition de quantification Trotter infini}. 
For classification purposes, it is convenient to partition the solution set according
to the sign of their real parts  
\beq
     \left\{ \ba{ccc} \wh{\mf{X}} &= &
        \Big\{ \wh{x}_a^{\; (R)} \Big\}_{1}^{|\mf{X}^{(R)}|} \cup
	\Big\{ \wh{x}_a^{\; (L)} \Big\}_{1}^{|\mf{X}^{(L)}|} \vspace{1mm} \\
     \wh{\mc{Y}} &= & \Big\{ \wh{y}_a^{\; (R)} \Big\}_{1}^{|\mc{Y}^{(R)}|} \cup
                 \Big\{ \wh{y}_a^{\; (L)} \Big\}_{1}^{|\mc{Y}^{(L)}|}  \ea \right.
\qquad \e{where}\, , \; \e{for} \;\; \a\in \{L,R\},   \qquad 
\left\{ \ba{c} \ups_{\a} \Re\big( \wh{x}_a^{\; (\a)} \big) >0  \vspace{1mm}  \\
\ups_{\a}  \Re\big( \wh{y}_a^{\; (\a)} \big) >0  \ea \right.  \;.
\label{decomposition particules et trous selon partie reelle}
\enq
We recall that $\ups_{\a}$ was introduced in \eqref{definition de ups alpha}.
The infinite Trotter number counterparts of the above sets are denoted without the
$\;\wh{}\,$. Note that we keep the same notation for the cardinalities $|\mf{X}^{(\a)}|$
and $|\mc{Y}^{(\a)}|$ occurring in the finite and infinite Trotter number cases.

Given a solution $\la\mapsto \wh{u}\, \big(\la\,|\, \wh{\mathbb{X}}\, \big)$ to
\eqref{ecriture NLIE a Trotter fini}, resp.\ $\la\mapsto u\big(\la\,|\, \mathbb{X}\big)$
to \eqref{ecriture NLIE a Trotter infini}, the quantisation conditions
\eqref{ecriture conditions de quantifications Trotter fini}, resp.\
\eqref{ecriture condition de quantification Trotter infini}, fixing $\wh{\mathbb{X}}$ to
$\wh{\mathbb{Y}}\, =\, \wh{\mc{Y}} \ominus \wh{\mf{X}}$ are equivalent to
the coupled system of equations,
\beq
\wh{u}\, \big(\,  \wh{y}_a^{\; (R)} \,|\, \wh{\mathbb{Y}} \big)
\, = \, 2\i\pi T \, \big(  p_a^{(R)} \, + \,  \tfrac{1}{2}  \big)  \qquad \e{and} \qquad
\wh{u}\, \big(\, \wh{x}_a^{\; (R)} \,|\, \wh{\mathbb{Y}} \big)
\, = \, - 2\i\pi T \, \big( h_a^{(R)} \,+ \,  \tfrac{1}{2} \big) \;,
\label{equation pour part trous type + v2}
\enq
in what concerns the roots with positive real part and 
\beq
\wh{u}\, \big(\,  \wh{y}_a^{\; (L)} \,|\, \wh{\mathbb{Y}} \big)  \, = \,  - 2\i\pi T\,  \big(   p_a^{(L)} \, + \,  \tfrac{1}{2}  \big)  \qquad \e{and} \qquad
\wh{u}\, \big(\,  \wh{x}_a^{\; (L)} \,|\, \wh{\mathbb{Y}} \big) \, = \,   2\i\pi T \, \big(   h_a^{(L)} \, + \,  \tfrac{1}{2} \big)  \;,
\label{equation pour part trous type - v2}
\enq
as far as the roots with negative real part are concerned. These equations should be
complemented with the non-vanishing condition for the derivatives,
\beq
\wh{u}^{\; \prime}\big( \, \wh{y}_a^{\, (\a)} \,|\, \wh{\mathbb{Y}} \big) \; \not= \; 0  \quad \e{and} \quad \wh{u}^{\, \prime}\big( \wh{x}_a^{\; (\a)} \,|\, \wh{\mathbb{Y}} \big) \; \not= \; 0
\enq
for $\a \in \{L, R\}$.

The equations corresponding to the infinite Trotter number setting are obtained by removing
the \; $\wh{}$\, everywhere.

The integers $p_a^{(\a)}, h_a^{(\a)} \in \mathbb{N}$ occurring in
\eqref{equation pour part trous type + v2}--\eqref{equation pour part trous type - v2}
produce a lifting of the ``exponentiated quantisation conditions''
\eqref{ecriture condition de quantification Trotter infini} in that, as established below, these equations
already admit a unique solution as soon as the integers $p_a^{(\a)}, h_a^{(\a)}
\in \mathbb{N}$ are specified. We stress that the change of signs between
\eqref{equation pour part trous type + v2} and \eqref{equation pour part trous type - v2}
is in accordance with the strictly increasing behaviour of $\Im\big[ \veps_{\e{c}} \big]$
along the counterclockwise oriented curve
$\big\{\la \; : \;  \Re\big[ \veps(\la)\big]\,=\,0 \big\}$.

\vspace{2mm}

From now on, we introduce a convenient convention. We shall say that solution
sets $\big(  \wh{\mc{Y}}, \wh{\mf{X}} \, \big)$ -- resp.\
$\big( \mc{Y}, \mf{X} \big)$ in the infinite Trotter number setting --
satisfy Hypotheses~\ref{Hypotheses solubilite NLIE}-\ref{Hypotheses eqns de quantification}
if $\wh{\mf{X}} \, =\,  \wh{\op{X}} \oplus  \wh{\op{X}}^{\, \prime}_{\e{ref}}$,
$\wh{\mc{Y}} \, = \, \wh{\op{Y}} \oplus  \wh{\op{Y}}^{\, \prime}_{\e{ref}}$
and $\wh{\op{X}}, \wh{\op{Y}}$ satisfy Hypothesis~\ref{Hypotheses solubilite NLIE},
while, taken as a whole, $\wh{\mc{Y}}, \wh{\mf{X}}$ satisfy
Hypotheses~\ref{Hypotheses eqns de quantification} -- resp.\ with the \,
$\wh{}$\, dropped in the infinite Trotter number setting.

\vspace{2mm}

\begin{prop}
\label{Proposition caracterisation dvpt basse T pour part et trous}
\begin{enumerate}
\item
Given four fixed integers $|\mc{Y}^{(R)}|, |\mc{Y}^{(L)}|, |\mf{X}^{(R)}|,
|\mf{X}^{(L)}|$ and $\mf{s}\in \mathbb{Z}$ so that
\beq
\mf{s} \, + \, |\mc{Y}^{(R)}| \, + \, |\mc{Y}^{(L)}| \, = \,  |\mf{X}^{(R)}|
\, + \,  |\mf{X}^{(L)}| \;,
\enq
there exists $T_0>0$ such that
for any choice of a collection of integers, possibly depending on $T$,  
\beq
0\leq p_1^{(\a)}< \cdots <  p_{|\mc{Y}^{(\a)}|} ^{(\a)}      \qquad and \qquad 0\leq h_1^{(\a)}< \cdots <  h_{|\mf{X}^{(\a)}|} ^{(\a)}  
\label{ecriture entiers pa et ha}
\enq
with $\a \in \{L, R\}$, the logarithmic quantisation conditions 
\eqref{equation pour part trous type + v2}--\eqref{equation pour part trous type - v2}
admit unique solution sets $\wh{\mf{X}}$ and $\wh{\mc{Y}}$ subject to Hypotheses
\ref{Hypotheses solubilite NLIE} and \ref{Hypotheses eqns de quantification} for
$T_0>T>0$. The latter are such that there are no singular roots,
\beq
\wh{\mc{Y}}_{\e{sg}}=\emptyset \; , \quad while  \qquad \wh{\mf{X}} \, = \, \wh{\mf{X}}^{\, (R)} \cup \wh{\mf{X}}^{\, (L)} \;, \qquad
\wh{\mc{Y}} \, = \, \wh{\mc{Y}}^{\, (R)} \cup \wh{\mc{Y}}^{\, (L)}  \;,
\enq
with $\wh{\mf{X}}^{\, (\a)} \, =\, \big\{ \, \wh{x}_a^{\; (\a)} \big\}_1^{|\mf{X}^{(\a)}|}$
and $\wh{\mc{Y}}^{\, (\a)} \, =\,  \big\{ \, \wh{y}_a^{\; (\a)} \big\}_1^{|\mc{Y}^{(\a)}|}$.
\item
Any such solution gives rise to a non-zero Bethe Eigenvector of the quantum transfer matrix.
\item
The solutions to the non-linear integral equation \eqref{ecriture NLIE a Trotter fini}
subject to the logarithmic quantisation conditions
\eqref{equation pour part trous type + v2}--\eqref{equation pour part trous type - v2}
associated with two distinct choices of integers $\Big\{ \{p_a^{(\a)} \} ;
\{ h_a^{(\a)}\} \Big\}$ and $\Big\{ \{\wh{p}_a^{(\a)} \} ; \{ \wh{h}_a^{(\a)}\} \Big\}$
are distinct. Moreover, these solutions give rise to two distinct Bethe Eigenvectors of
the quantum transfer matrix.
\item
There exist solutions to the logarithmic quantisation conditions
\eqref{equation pour part trous type + v2}--\eqref{equation pour part trous type - v2}
corresponding to not necessarily pairwise distinct choices of integers
$\{p_a^{(\a)} \}$ and  $\{ h_a^{(\a)}\} $. However, these do not give rise to
Bethe Eigenvectors of the quantum transfer matrix.
\item
The same statements hold relatively to the infinite Trotter number case, upon
omitting the \, $\wh{}$\,  notation.
\item
Finally, the infinite Trotter number roots  $y_a^{(\a)}$ and $x_a^{(\a)}$,
$\a \in \{L, R\}$, admit all-order low-$T$ asymptotic expansions
\beq
y_a^{(\a)} \, \simeq  \, \sul{k \geq 0 }{} y_{a;k}^{(\a)} \cdot T^{k} \qquad and \qquad 
x_a^{(\a)} \, \simeq  \, \sul{k \geq 0 }{} x_{a;k}^{(\a)} \cdot T^{k} \;. 
\label{ecriture DA low T des racines part et trous}
\enq
The leading terms of these expansions are given by 
\beq
y_{a;0}^{(\a)} \, = \, \veps^{-1}_{\a}\Big( \ups_\a  2\i\pi T \big( p_a^{(\a)} \, + \,  \tfrac{1}{2}  \big)   \Big) \qquad and \qquad 
x_{a;0}^{(\a)} \, = \, \veps^{-1}_{\a}\Big( - \ups_\a  2\i\pi T \big( h_a^{(\a)} \, + \,  \tfrac{1}{2}  \big)   \Big) \;. 
\label{ecriture terms dominants dans DA basse T part trous}
\enq
The first two subleading corrections involve the function
$u_1\big( \la  \,|\, \mathbb{Y} \big)$ introduced in \eqref{definition fonction u 1}
and take the explicit form 
\beq
     y_{a;1}^{(\a)} \, = \, - \f{1}{\veps^{\prime}\big(y_{a;0}^{(\a)} \big)}
        \cdot   u_1\big( y_{a;0}^{(\a)}  \,|\, \mathbb{Y}_0 \big)   \;, \quad
     x_{a;1}^{(\a)} \, = \, - \f{1}{\veps^{\prime}\big(x_{a;0}^{(\a)} \big)}
        \cdot  u_1\big( x_{a;0}^{(\a)}  \,|\, \mathbb{Y}_0 \big)
\enq
and
\begin{multline} \label{y_second_correction}
     y_{a;2}^{(\a)} =
        - \frac{\veps''(y_{a;0}^{(\a)})}{2 \veps'(y_{a;0}^{(\a)})}\bigl(y_{a;1}^{(\a)}\bigr)^2
        - \frac{u_1'(y_{a;0}^{(\a)}\,|\,\mathbb{Y}_0)}{\veps'(y_{a;0}^{(\a)})} y_{a;1}^{(\a)}
        - \frac{1}{\veps'(y_{a;0}^{(\a)})}
	  \sul{\sg= \pm }{}  \sg \f{ R_{\e{c}}\bigl(y_{a;0}^{(\a)}, \sg q\bigr)}
	                           { 2 \veps^{\prime}(\sg q)}
          \cdot \Biggl\{\Bigl(u_1\bigl(\sg q \,|\, \mathbb{Y}_0 \bigr)\Bigr)^2
	                + \f{\pi^2}{3}  \Biggr\} \\
        - \frac{1}{\veps'(y_{a;0}^{(\a)})}
        \sum_{\beta\in\{L,R\}}\Biggl\{
	  \sum_{b=1}^{|\mf{X}^{(\beta)}|}
	  \bigl(\partial_{x_b^{(\beta)}} u_1(z\,|\,\mathbb{Y})\bigr) \, x_{b;1}^{(\beta)} +
	  \sum_{b=1}^{|\mc{Y}^{(\beta)}|}
	  \bigl(\partial_{y_b^{(\beta)}} u_1(z\,|\,\mathbb{Y})\bigr) \, y_{b;1}^{(\beta)}
	  \Biggr\}_{\mathbb{Y} = \mathbb{Y}_0,\ z = y_{a;0}^{(\a)}} \;,
\end{multline}
where $\mathbb{Y}_0$ is obtained from $\mathbb{Y}$ by replacing
$x_a^{(\a)}$ and $y_a^{(\a)}$ with $x_{a;0}^{(\a)}$ and $y_{a;0}^{(\a)}$, respectively.
The expression for $x_{a;2}^{(\a)}$ is obtained from \eqref{y_second_correction} 
upon the substitution $y_{a;b}^{(\a)} \hookrightarrow x_{a;b}^{(\a)}$, where $b \in \{ 0, 1\}$.
\end{enumerate}
\end{prop}

One of the most striking conclusions of the low-$T$ characterisation provided by
Proposition~\ref{Proposition caracterisation dvpt basse T pour part et trous}
is that the hole roots $\mf{X}$ and particle roots $\mc{Y}$ \textit{only} lie in
an $\e{O}(T)$-sized tubular neighbourhood of the curve $\big\{\la \, : \,
\Re\big[ \veps(\la)\big]\,=\,0 \big\}$. In particular, quite unexpectedly,
this means that there are \textit{no} string patterns for the particle roots
nor there are shifted singular roots. This is in stark contrast to the many heuristic,
non-rigorous results pertaining to the string patterns of Bethe roots for the
excited states of numerous integrable models, argued earlier in the physics literature
\cite{BetheSolutionToXXX,GaudinFonctionOndeBethe,GaudinFonctionOndeBethe,%
FaddeevABALesHouchesLectures,KorepinAnalysisofBoundStateConditionMassiveThirring,%
TakahashiThermodynamics1DSolvModels}. Still, such a lack of strings has already been
heuristically argued for the zero-magnetisation sector of the zero-temperature XXZ chain
\cite{BabelondeVegaVialletStringHypothesisWrongXXZ,%
DestriLowensteinFirstIntroHKBAEAndArgumentForStringIsWrong,%
VirosztekWoynarovichStudyofExcitedStatesinXXZHigherLevelBAECalculations,%
WoynaorwiczHLBAEMAsslessXXZ0Delta1} as well as for the spectrum of the quantum
transfer matrix attached to that model in the regime $\De >1$
\cite{KozDugaveGohmannSuzukiLowTSpectrumMassiveXXZQTM}.

In fact, Proposition~\ref{Proposition caracterisation dvpt basse T pour part et trous}
ensures the existence of an all order low-$T$ asymptotic expansion of the hole
and particle roots. Here, we stress that the integers $p_a^{(\a)}, h_a^{(\a)}
\in \mathbb{N}$ occurring in \eqref{ecriture entiers pa et ha} may evolve with $T$, which
allows one to consider the situation when $Tp_a^{(\a)}$ and  $Th_a^{(\a)}$ approach
some fixed real number for $T\tend 0^+$. Then one sees that in the low-$T$ limit,
by considering different solutions to the non-linear problem, one gets that
the possible values for roots $x_a$ and $y_a$ described in this way condense
densely on the curve $\big\{\la \, : \,  \Re\big[ \veps(\la)\big]\,=\,0 \big\}$.

It is also important to note that for \textit{fixed} integers $p_a^{(\a)}$ or
$h_a^{(\a)}$ -- \textit{viz}.\ such that are \textit{not} scaling with $T$ -- the 
leading term given in \eqref{ecriture terms dominants dans DA basse T part trous}
reduces to $\ups_{\a} q + \e{O}(T)$. Thus, such roots simply collapse on $\pm q$
in the low-$T$ limit. This class of roots actually grasps the spectral structure
of the $c=1$ free Boson model and thus highlights the portion of the infinite
Trotter number limit of the spectrum of the quantum transfer matrix which has a
direct interpretation as the spectrum of a conformal field theory. Indeed, for
fixed integers $p_a^{(\a)}$ or $h_a^{(\a)}$ -- \textit{viz}.\ such that are \textit{not}
scaling with $T$ -- one may write the particle/hole roots expansion more explicitly as
\begin{align}
  x_a^{(\a)} & = \ups_{\a} q - \f{ 2\i\pi  T }{ \veps^{\prime}( \ups_{\a} q) } 
  \bigg\{ \ups_{\a} \Big[ h_a^{(\ups_{\a})}+\f{1}{2} \Big]  \, + \, \f{1}{2\i\pi}    u_1(\ups_{\a} q \,|\, \mathbb{Y})   \bigg\}  \,  + \, \e{O}\big( T^2 \big)  \;,
 \label{DA basse T des trous CFT} \\[1ex]
 y_a^{(\a)}   & = \ups_{\a} q + \f{ 2\i\pi  T }{ \veps^{\prime}( \ups_{\a} q) } 
 \bigg\{ \ups_{\a} \Big[ p_a^{(\ups_{\a})}+\f{1}{2} \Big]  \, - \,  \f{1}{2\i\pi}   u_1(\ups_{\a} q \,|\, \mathbb{Y})  \bigg\}    \,  + \, \e{O}\big( T^2 \big)  \;,
\label{DA basse T des particules CFT} 
\end{align}
with $\ups_{\a}$ as introduced in \eqref{definition de ups alpha}.

In order to discuss the last result of the paper, which pertains to the identification
of a CFT structure in the low-$T$ spectrum of the quantum transfer matrix, we first
need to introduce some more notation. Indeed, due to their very different interpretation
and structure, one should partition a particle-hole root solution set to
\eqref{equation pour part trous type + v2}--\eqref{equation pour part trous type - v2}
into sets containing particle, resp.\ hole, roots that are uniformly away from the
endpoints of the Fermi zone $\mc{Y}^{(\e{fR})}, \mc{Y}^{(\e{fL})}$, resp.\
$\mf{X}^{(\e{fR})}, \mf{X}^{(\e{fL})}$, and sets $\mc{Y}^{(\pm)}$, resp.\
$\mf{X}^{(\pm)}$, containing roots which, on the contrary, collapse, as $T\tend 0^+$,
on the endpoints $\pm q$ of the Fermi zone:
\beq
\mc{Y} \, = \, \mc{Y}^{(\e{fR})} \oplus \mc{Y}^{(\e{fL})} \oplus  \mc{Y}^{(+)}\oplus \mc{Y}^{(-)} \qquad \e{with} \qquad 
\mc{Y}^{(\a)} \; = \; \Big\{  y_a^{(\a)} \Big\}_{a=1}^{n_p^{(\a)} } \;, \quad \a\in \Big\{ +, -, \e{fR}, \e{fL} \Big\}
\label{definition ensemble Y partitiones divers types particules}
\enq
and 
\beq
\mf{X} \, = \, \mf{X}^{(\e{fR})} \oplus \mf{X}^{(\e{fL})} \oplus \mf{X}^{(+)}\oplus \mf{X}^{(-)}
\qquad \e{with} \qquad \mf{X}^{(\a)} \; = \; \Big\{  x_a^{(\a)} \Big\}_{a=1}^{n_h^{(\a)} } \;,
\quad \a\in \Big\{ +, -, \e{fR}, \e{fL} \Big\} \;. 
\label{definition ensemble X partitiones divers types trous}
\enq
Note that these equations also define the integers $n^{(\alpha)}_{p}$, resp.\
$n^{(\alpha)}_{h}$, as the cardinalities of the sets $\mc{Y}^{(\a)}$, resp.\ $\mf{X}^{(\a)}$.

Analogously, we partition the sets of integers \eqref{ecriture entiers pa et ha}
describing these roots in four families
\beq
\Big\{  p_a^{(\a)} \Big\}_{a=1}^{n_p^{(\a)} } \; , \quad \e{resp}. \quad  \Big\{  h_a^{(\a)} \Big\}_{a=1}^{n_h^{(\a)} } \;, \quad  \a\in \Big\{ +, -, \e{fR}, \e{fL} \Big\} \;.  
\label{ecriture partition des ensembles d entiers selon far et close left and right}
\enq
The integers may vary with $T$ but in such a way that
\beq
\limsup_{T\tend 0^+} T p_a^{(\pm)} \, = \, \limsup_{T\tend 0^+} T h_a^{(\pm)} \; = \; 0 \qquad \e{and} \qquad 
\liminf_{T\tend 0^+} T p_a^{(\a)} \, = \, \liminf_{T\tend 0^+} T h_a^{(\a)} \; > \; 0 \; , \quad \a \in \big\{ \e{fR}, \e{fL} \big\} \;. 
\label{ecriture limite sur les entiers}
\enq
Therefore, as $T\tend 0^+$,
\begin{enumerate}
\item
the roots $y_a^{(\sg)}$, $x_a^{(\sg)}$, $\sg = \pm$, will collapse on $\sg q$
\item
the roots $y_a^{(\a)}$, $x_a^{(\a)}$, $\a \in \{fL,fR\}$, will collapse on the curve
$\big\{\la \; : \;  \Re\big[ \veps(\la)\big]\,=\,0 \big\}$ while being located
uniformly away from the points $\pm q$.
\end{enumerate}
To leading order in $T$ in the asymptotics of the Eigenvalues, one may only distinguish
the overall differences between the numbers of particles and holes collapsing on $\pm q$,
\beq
\ell^{(\sg)} \; = \; \sg \Big(n_p^{(\sg)} \, - \, n_h^{(\sg)} \Big)  \;. 
\label{definition Umklapp integer}
\enq
These relative integers are called Umklapp integers in the physics literature. 

It will turn out to be useful to gather all the ``macroscopic'' data in the set
\beq
\mathbb{Y}^{(\e{far})}\; = \;\mc{Y}^{(\e{fR})} \oplus \mc{Y}^{(\e{fL})} \ominus  \mf{X}^{(\e{fR})} \oplus \mf{X}^{(\e{fL})} \oplus \{ q \}^{\oplus \ell^{(+)} }
\oplus \{ - q \}^{\oplus - \ell^{(-)}} \;.
\label{definition ensemble reduit YM mathbb}
\enq
Moreover, it is easy to see owing to \eqref{ecriture DA low T des racines part et trous} that
\beq
u_1(\la\,|\, \mathbb{Y}) \, = \, u_1(\la\,|\, \mathbb{Y}^{(\e{far})})  \;  + \, \e{O}\big( T  \big) \;, 
\label{ecriture partition de u1 via racines far et close}
\enq
where the control holds for fixed integers $p_a^{(\pm)}, h_a^{(\pm)}$. In case
when only \eqref{ecriture limite sur les entiers} holds, the control goes rather as 
$ \e{O}\Big( T \e{max}\big\{p_a^{(\pm)}, h_a^{(\pm)}\big\} \Big)$.

\begin{prop}
\label{Propostion DA fin pour la vp de la QTM}

Let $\mathbb{Y}$, the solution set of the infinite Trotter number quantisation conditions
\eqref{ecriture condition de quantification Trotter infini}, be partitioned into the
sets introduced in \eqref{definition ensemble Y partitiones divers types particules}--\eqref{definition ensemble X partitiones divers types trous}.
Then the non-trivial part
\eqref{definition impulsion thermale} of the infinite Trotter number limit of the
Eigenvalues of the quantum transfer matrix associated with $\mathbb{Y}$ according to
Proposition \ref{Proposition convergence trotter infini systeme de racines} admits
the low-$T$ expansion
\beq
\mc{P}( \mathbb{Y} ) \, = \,  \f{1}{T} \mc{P}_{-1}   \; + \;   \msc{P}_{0}( \mathbb{Y}^{(\e{far})} ) \; + \; T \Big\{ \mc{P}_{1}( \mathbb{Y}^{(\e{far})} ) \; + \; \varpi_{1}(\mathbb{H}) \Big\}  \; + \; \e{O}(T^2) \;, 
\label{ecriture dvpt basse tempe pour P de Y}
\enq
with $\mathbb{Y}^{(\e{far})}$ as defined in \eqref{definition ensemble reduit YM mathbb} and 
\beq
\mathbb{H} \; = \; \bigcup_{\sg = \pm} \mathbb{H}^{(\sg)} \qquad where \qquad \mathbb{H}^{(\sg)} \; = \; \Big\{ \{p_a^{(\sg)}\}_{1}^{n_p^{(\sg)} } \; ; \;  \{ h_a^{(\sg)}\}_{1}^{n_h^{(\sg)} }   \Big\} \; .
\label{definition ensembles CFT particules trous}
\enq
Here we agree upon 
\beq
\mc{P}_{-1}  \; = \;  -  \Int{-q}{q}  \f{\dd \mu }{ 2\i\pi} \veps_{\e{c}}(\mu) p_0^{\prime}(\mu) 
\label{definition P moins 1}
\enq
as well as 
\beq
\msc{P}_{0}( \mathbb{Y}^{(\e{far})} )\; = \hspace{-5mm} \sul{ y \in \mc{Y}^{(\e{f}R)} \oplus \mc{Y}^{(\e{f}L)} \ominus \mf{X}^{(\e{f}R)} \ominus \mf{X}^{(\e{f}L)} }{} \hspace{-5mm}  p_{\e{c}}( y  )
\; + \; \big( \ell^{(+)}+\ell^{(-)} + \mf{s} \big) p_{\e{c}}(q) \;,
\label{definition mas P0}
\enq
in which $p_{\e{c}}$ was introduced in \eqref{definition dressed momentum c deforme}. 
Finally, agreeing upon 
\beq
\op{v}_F  \, =  \, \f{ \veps^{\prime}_{\e{c}}(q) }{ p^{\prime}_{\e{c}}(q) } \;, 
\label{definition vitesse de Fermi du modele}
\enq
one has
\beq
\mc{P}_{1}( \mathbb{Y}^{(\e{far})} )  \; = \; \f{1}{4\i\pi \op{v}_F}\sul{ \sg = \pm }{}  
\bigg\{    u_1(\sg q \mid \mathbb{Y}^{(\e{far})})  \Big( u_1(\sg q \mid\mathbb{Y}^{(\e{far})})  -4\i\pi \ell^{(\sg)} \Big)  \, + \, \f{\pi^2}{3}  \bigg\} \;,
\enq
while 
\beq
\varpi_{1}(\mathbb{H}) \; = \; \f{ 2 \i \pi }{ \op{v}_F } \sul{ \sg = \pm }{}  \sul{ \a \in \mathbb{H}^{(\sg)} }{} \Big( \a \, + \, \f{1}{2} \Big) \;. 
\label{definition fct varpi1 de H}
\enq

\end{prop}

The last result provides an explicit control over the relative magnitude of the
Eigenvalues in the infinite Trotter number limit. In particular, it allows one
to clearly identify the configuration $\mathbb{Y}$, solving the infinite Trotter
number quantisation constraints \eqref{ecriture condition de quantification Trotter infini}
in the class of solutions satisfying the Hypotheses
\ref{Hypotheses solubilite NLIE} and \ref{Hypotheses eqns de quantification}, which gives
rise to the Eigenvalue of largest modulus. For that purpose, one should solve for the
minimising configuration $\mathbb{Y}$ of the imaginary parts of
$ \msc{P}_{0}( \mathbb{Y}^{(\e{far})} )$ and $\mc{P}_{1}( \mathbb{Y}^{(\e{far})} ) \; + \;
\varpi_{1}(\mathbb{H})$. The first minimisation problem will fix the set of ``far'' roots,
while the second one will fix the set of ``close'' roots collapsing to $\pm q$. One obtains
the following result:
\begin{prop}
\label{Propostion infimum pour les vp de la QTM}
$\Im\Big[\msc{P}_{0}( \mathbb{Y}^{(\e{far})} ) \Big]$ attains its minimum for 
\beq
\mc{Y}^{(\e{f}R)} \, = \,   \mc{Y}^{(\e{f}L)}  \, = \,   \mf{X}^{(\e{f}R)}
\, = \,    \mf{X}^{(\e{f}L)}  \, = \, \emptyset  \;, 
\label{ecriture tous les ensembles far sont vides}
\enq
that is when there are no ``far'' roots.

When \eqref{ecriture tous les ensembles far sont vides} holds,
$\Im\Big[\mc{P}_{1}( \mathbb{Y}^{(\e{far})} ) \; + \; \varpi_{1}(\mathbb{H}) \Big]$
attains its minimum for 
\beq
n_{p}^{(+)} \, = \,  n_{p}^{(-)} \, = \, n_{h}^{(+)} \, = \,
n_{h}^{(-)} =0 \quad and \quad \mf{s}=0 
\enq
that is when there are no ``close'' roots and the spin is zero.
For the minimising configuration with $\mf{X}=\mc{Y}=\emptyset$, \textit{viz}. $\mathbb{Y}=\emptyset$, it holds that
\beq
\mc{P}( \emptyset )  \, = \, \f{1}{T} \mc{P}_{-1}   \; - \;  \f{\i\pi }{6 \op{v}_F } T    \; + \; \e{O}(T^2) \;. 
\label{ecriture DA P emptyset}
\enq
\end{prop}
Here one should note that the prefactor $- \tf{ \i\pi }{ 6 \op{v}_F }$ in front of
the linear-in-$T$ correction is exactly the one conjectured for the low-$T$
expansion of the free energy of one dimensional quantum models belonging to the
universality class of a conformal field theory with central charge $1$
\cite{AffleckCFTPreForLargeSizeCorrPartitionFctonAndLowTBehavior,%
CardyConformalDimensionsFromLowLSpectrum}. While we cannot fully prove that the
minimising configuration does correspond to the actual dominant Eigenvalue,
due to the restrictions imposed by our use of Hypotheses \ref{Hypotheses solubilite NLIE},
\ref{Hypotheses eqns de quantification}, we find this last coincidence rather striking.

We are also able to provide a similar to
Proposition~\ref{Propostion DA fin pour la vp de la QTM} characterisation of the
low-$T$ behaviour of $\mc{E}(\mathbb{Y})$, as introduced in \eqref{definition mc E de Y}.

\begin{prop}
\label{Propostion DA fin pour la fnelle energetique}
With the notation of Proposition \ref{Propostion DA fin pour la vp de la QTM} one has the
low-$T$ expansion
\beq
     \mc{E}(\mathbb{Y}) \, = \, \f{1}{T} \mc{E}_{-1} \, + \,
        \msc{E}_{0}( \mathbb{Y}^{(\e{far})} ) \; + \;
	T \Big\{ \mc{E}_{1}( \mathbb{Y}^{(\e{far})} ) \; + \;
	\vsg_{1}(\mathbb{H}) \Big\}  \; + \; \e{O}(T^2) \;,
\label{ecriture dvpt basse tempe pour E de Y}
\enq
where $\mathbb{Y}^{(\e{far})}$ was defined in \eqref{definition ensemble reduit YM mathbb}
and $\mathbb{H}$ by \eqref{definition ensembles CFT particules trous}.
The two lowest-order terms in \eqref{ecriture dvpt basse tempe pour E de Y} have
the explicit forms
\beq
     \mc{E}_{-1}  \; = \;
        -  \Int{-q}{q}  \f{\dd \mu }{ 2\i\pi} \veps(\mu) \veps_0^{\prime}(\mu)
\quad \text{and} \quad
     \msc{E}_{0}( \mathbb{Y}^{(\e{far})} )\; = \;
        \hspace{-1.2cm}
	   \sul{y \in \mc{Y}^{(\e{f}R)} \oplus \mc{Y}^{(\e{f}L)} \ominus
	        \mf{X}^{(\e{f}R)} \ominus \mf{X}^{(\e{f}L)} }{}
        \hspace{-1.2cm}
	\veps_{\e{c}}( y  ) \;.
\enq
The remaining terms can be written as
\begin{align}
     & \mc{E}_{1}( \mathbb{Y}^{(\e{far})} ) \; = \; \f{1}{4\i\pi}\sul{\sg = \pm}{} \sg \,
       u_1(\sg q \mid \mathbb{Y}^{(\e{far})})  \Big( u_1(\sg q \mid\mathbb{Y}^{(\e{far})}) - 4\i\pi \ell^{(\sg)} \Big) \;, \\[1ex]
     & \vsg_{1}(\mathbb{H}) \; = \;
        2 \i \pi   \sul{ \sg = \pm }{} \sg \sul{ \a \in \mathbb{H}^{(\sg)} }{}
	              \Big( \a \, + \, \f{1}{2} \Big) \;.
\end{align}

\end{prop}

Propositions \ref{Propostion DA fin pour la vp de la QTM} and \ref{Propostion DA fin pour la fnelle energetique} can be put together so as to
demonstrate the presence of the spectrum of the $c=1$ free Boson conformal field theory within the low-$T$ limit of the spectrum
of the quantum transfer matrix in the infinite Trotter number limit. Let $\wh{\La}_{\e{max};\e{BA}}$ be the dominant Eigenvalue of the quantum transfer matrix among the
class of Bethe Ansatz based Eigenvalues with particle-hole parameters $\big(  \wh{\mc{Y}}, \wh{\mf{X}} \, \big)$ subject to Hypotheses
\ref{Hypotheses solubilite NLIE}-\ref{Hypotheses eqns de quantification} as identified in Proposition
\ref{Propostion infimum pour les vp de la QTM}. Then, Theorem \ref{Theorem structure conforme du spectre} stated earlier on does hold.

\section{An equivalent non-linear integral equation}
\label{Section NLIE equivalente et contour Cu}

\subsection{An auxiliary contour with better properties}
\label{Sous Section amelioration du contour}

While suggested by the existing physical literature, the way of writing
\eqref{ecriture NLIE a Trotter fini} or its infinite Trotter number
counterpart \eqref{ecriture NLIE a Trotter infini}, as it originally appeared
in the physics literature, is not appropriate for the study of the non-linear
integral equation's solvability in the low-$T$ regime.

First of all, one expects -- and this fact will be backed up by the analysis
to come -- that the non-linear logarithmic term in \eqref{ecriture NLIE a Trotter fini},
or in \eqref{ecriture NLIE a Trotter infini}, will exhibit 
\begin{itemize}
 
\item a finite in $N,T$ behaviour on a non-empty portion $\msc{C}_{\e{ref};\e{lead}}$
of $\msc{C}_{\e{ref}}$, where $\Re\big[\, \wh{u}(\la) \, \big]<0$; 
 
\item a subdominant $\e{o}(1)$ behaviour on its complement $\msc{C}_{\e{ref}}\setminus
\msc{C}_{\e{ref};\e{lead}}$, where $\Re\big[\, \wh{u}(\la) \, \big]>0$. 

\end{itemize}
This finite behaviour should be extracted at the very start of the analysis in order to
be able to reduce the question of the solvability of the problem to the construction of
the fixed point of a contractive mapping. The possibility to do so in an efficient manner
strongly depends on whether the given solution does exhibit an appropriate behaviour on the
reference contour $\msc{C}_{\e{ref}}$. It seems rather clear that one cannot choose
the reference contour $\msc{C}_{\e{ref}}$ such that it has, uniformly for all the
auxiliary sets of parameters $\mathbb{Y}$, the optimal properties for the analysis
of the low-$T$ behaviour of the non-linear terms involving $\wh{u}$ in
\eqref{ecriture NLIE a Trotter fini} or \eqref{ecriture NLIE a Trotter infini}. 
The strategy, whose validity will be backed up by the analysis to come, consists in
setting into one-to-one correspondence the original non-linear problem with one living
on a contour $\msc{C}_{\, \wh{u}}$ that depends on the function $\wh{u}$ and the very choice
of the parameters $\mathbb{Y}$. The possibility to accommodate the contour stems from
the fact that, upon adopting Hypothesis~\ref{Hypotheses solubilite NLIE},
any solution $\wh{u}$ or $u$ to the non-linear integral equation at finite
\eqref{ecriture NLIE a Trotter fini} or infinite  \eqref{ecriture NLIE a Trotter infini}
Trotter numbers, will be holomorphic in some small tubular neighbourhood of the 
reference contour $\msc{C}_{\e{ref}}$. This section is devoted to a thorough discussion
of the deformation procedure $\msc{C}_{\e{ref}} \hookrightarrow \msc{C}_{\, \wh{u}}$ 
of the original contour. This will be done  by  assuming certain properties of the
function $\wh{u}$ solving this non-linear equation. We will later establish in
Section~\ref{Section espace fnel pour la NLIE et ses ptes} that these properties
are indeed satisfied for any function belonging to the functional space on which we
formulate the problem, $\wh{\mc{E}}_{\mc{M}}$ or $\wt{\mc{E}}_{\mc{M}}$, depending on
whether one focuses on the finite or infinite Trotter number setting.

As already mentioned on the level of the discussed examples, the description of the
thermodynamics of the XXZ chain -- the non-linear integral equations in particular
-- involves integrals of the type 
\beq
     \Oint{ \msc{C}_{ \e{ref} }  }{} \f{\dd \mu }{ 2\i\pi } \:
        f(\mu) \f{u^{ \prime}(\mu)  }{ 1 + \ex{\f{1}{T} u(\mu)  }  } \;,
\label{ecriture rep int generique f contre ln prime 1+a}
\enq
where $f$ is holomorphic in some fixed, $T$-independent neighbourhood of
$\msc{C}_{\e{ref}}$. Exactly as in the discussion following \eqref{definition logarithm de 1+a},
if some pole of $\tf{u^{ \prime}  }{ \big[ 1 + \ex{\f{1}{T} u  }  \big] } $ is located on
$\msc{C}_{ \e{ref} }$, then the integral should be understood in the sense of a~$+$,
\textit{viz}.\ from the interior of $\msc{C}_{ \e{ref} }$, boundary value.

In particular, given a function $u$ satisfying the index
condition \eqref{ecriture monodromie Trotter fini} or
\eqref{ecriture monodromie Trotter infini}, when $\la \pm \i \zeta
\not\in \ov{\e{Int}\big( \msc{C}_{\e{ref}} \big) }$, the non-linear logarithmic
term in \eqref{ecriture NLIE a Trotter fini} or \eqref{ecriture NLIE a Trotter infini}
is of this type as can be readily seen by carrying out an integration by parts:
\beq
     \, - \, \i  \mf{m} \th(\la- \varkappa) \, - \,
        \Int{ \msc{C}_{\e{ref}} }{} \dd \mu \: K(\la-\mu)\,
	   \msc{L}n_{\msc{C}_{\e{ref}}}\big[1+\ex{-\frac{1}{T}u } \big](\mu  ) \, = \, 
     \,  \Int{ \msc{C}_{\e{ref}} }{} \f{ \dd \mu }{ 2\pi T} \:
        \th(\la-\mu)  \f{ u^{\prime}(\mu )  }{ 1+\ex{\frac{1}{T}u(\mu)}  } \;. 
\enq

\begin{figure}[h]
\begin{center}

 \includegraphics{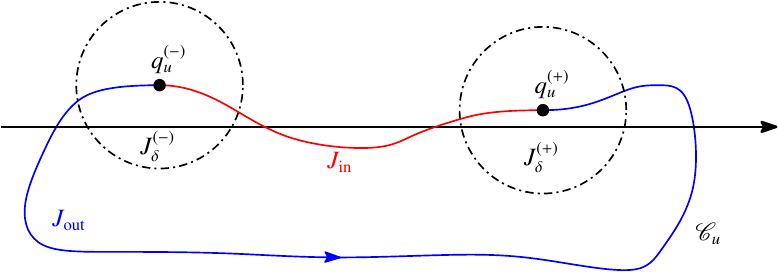}
 
\caption{Definition of the deformed contour $\msc{C}_{ u }$.
}
\label{contour integration Cu}
\end{center}
\end{figure}

The structure \eqref{ecriture rep int generique f contre ln prime 1+a} opens up
the possibility of trading the original contour $\msc{C}_{\e{ref}}$ for a more
convenient one $\msc{C}_{ u }$ by means of some contour deformation and residue
calculation. More precisely, provided the function $u$ fulfills reasonable
properties, in any integral representation of the above type, one may slightly
accommodate the contour by deforming it $\msc{C}_{ \e{ref} } \hookrightarrow \msc{C}_{ u }$,
with $\msc{C}_{ u }$ as depicted in Fig.~\ref{contour integration Cu}. Ideally,
one demands that the contour $ \msc{C}_{ u}$ adapted to $u$ will satisfy the 

\begin{property}
\label{Propriete pour deformer les contours}
\begin{enumerate}
\item
$\msc{C}_{ u }$ passes through two zeroes $q^{(\pm)}_{u }$ of
$1-\ex{-\frac{1}{T}u}$  satisfying $u\big(  q^{(\pm)}_{ u }   \big)
\, = \, 0 $ and these are the only zeroes of $1-\ex{-\frac{1}{T}u}$ on 
$\msc{C}_{ u }$. Moreover, these zeroes are such that
$\pm \Re\big[ u^{\prime}\big(  q^{(\pm)}_{ u }  \big) \big]
> c> 0$ for some $c>0$, uniformly in $T,N^{-1}$ small enough ;
\item
$u \in \mc{O}(\mc{V}_{\pm q} )$ with $\mc{V}_{\pm q}$ an open neighbourhood
of $\pm q$ containing $\op{D}_{\pm q , \eps}$ for some $T, N$ independent
$\eps>0$ and such that $u: \mc{V}_{\pm q} \tend \op{D}_{0,\varrho}$, $\varrho>0$
and $T, N$ independent, is a biholomorphism;
\item
there exists $J_{\de}^{(\pm)}\subset \msc{C}_{ u }$ such that
$u\big( J_{\de}^{(-)}   \big)= \intff{-\de}{\de} $ and
$u\big(J_{\de}^{(+)} \big)= \intff{\de}{-\de} $,
for some $\de>0$, possibly depending on $T$, but such that $\tf{\de}{T} > -M \ln T$
as $T \tend 0^+$;
\item
the complementary set $J_{\de} \, = \, \msc{C}_{ u }
\setminus \Big\{ J_{\de}^{(-)} \cup J_{\de}^{(+)}  \Big\}$ is such that 
$\big| \Re\big[ u(\la ) \big] \big|  > \tf{\de}{2}$  for all $\la \in J_{\de}$;
\item
$1+\ex{-\frac{1}{T}u}$ has no poles in the bounded
domain $\mc{U}_{u}$ such that $\Dp{}\mc{U}_{u} =\msc{C}_{ \e{ref} } - \msc{C}_{ u }$;
\item
the zeroes of $1+\ex{-\frac{1}{T}u}$ in the bounded domain $\mc{U}_{u}$
correspond to the unique solutions $\mf{x}_{k}^{(\a)}$
in $\mc{V}_{\ups_\a q}$ to the equation $u\big( \mf{x}_{k}^{(\a)} \big) \, = \,
2\i\pi T \bigl( k - \tfrac{1}{2} \bigr)$ with $k \in \mathbb{I}_{u}^{(\a)} \subset{\mathbb{Z}}$,
$\a \in \{ L, R \}$ and $|\mathbb{I}_{u}^{(\a)}|$ is bounded uniformly in $T,N^{-1}$
small enough;
\item
there exist unique $\tau_{\a} \in \msc{C}_{\e{ref}}$, $\tau_{\a} \in \mc{V}_{\ups_{\a} q}$
with $\a\in \{L,R\}$, such that
\beq
u\big(\tau_{\a} \big) \, = \, 2\i\pi T \bigl( \mf{p}_{ \a } - \tfrac{1}{2} + \eps_{ \a } \bigr)
\qquad with \qquad
\mf{p}_{\a} \in \mathbb{Z} \;,  \quad \eps_{\a} \in \intfo{0}{1}
\label{definition du parametre t alpha dans le cas de propriete bonnes pour fct u}
\enq
and $\mf{p}_{\a} $ bounded uniformly in $T,N^{-1}$ small enough.
\end{enumerate}
Above, we have denoted by $\intff{a}{b}$ the oriented segment run through from $a$ to $b$.  
\end{property}

Basically what these properties translate into is that, on $\msc{C}_{u}$,
the function $1-\ex{ -\f{1}{T}u }$ has two zeroes and that the real part of
$u$ changes sign between the two portions of the curve delimited by the two
zeroes. Moreover, the real part of $u$ is away from zero in a controlled way as
soon as one is uniformly away from the roots. The reason why one only
focuses on the case of two zeroes stems from the fact that $\veps_{0}$,
but also $\veps_{\e{c}}$ and $\mc{W}_N$, have only two roots of the mentioned
type and that the non-linear integral equation -- as indicated by a heuristic
analysis -- seems to preserve these properties on the level of its solutions.
 
In particular, we believe that all physically pertinent solutions to
\eqref{ecriture NLIE a Trotter fini}, \eqref{ecriture monodromie Trotter fini}
and  \eqref{ecriture conditions de quantifications Trotter fini} in the finite
Trotter number case, or \eqref{ecriture NLIE a Trotter infini},
\eqref{ecriture monodromie Trotter infini} and
\eqref{ecriture condition de quantification Trotter infini}
in the infinite Trotter number case, do belong to the class of functions 
which enjoys $\mathrm{i)-vii)}$ above and thus allows for such a contour
deformation. Irrespective of our beliefs relative to the general setting,
the solutions we shall construct later on do precisely belong to this class.

\begin{figure}
\begin{center}
\includegraphics[width=0.72\textwidth]{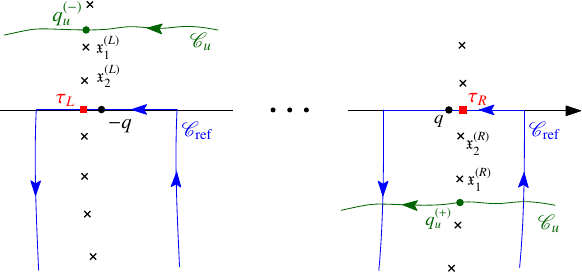}
\caption{Sketch of zeroes crossed when the contour is deformed from
$\msc{C}_{\e{ref}}$ to $\msc{C}_{u}$.}
\label{zeroes and contour deformations}
\end{center}
\end{figure}

We stress that the subset $ \mathbb{I}_{u}^{(\a)}$ defining the zeroes
contained in $\mc{U}_{u}$ may be empty. In fact, one may determine
$\mathbb{I}_{u}^{(\a)}$ explicitly in terms of certain data related
to the behaviour of $u$ on $\msc{C}_{\e{ref}}$. Recalling point $\mathrm{vii)}$
of Properties \ref{Propriete pour deformer les contours} and the $+$ boundary
value interpretation of the integral \eqref{ecriture rep int generique f contre ln prime 1+a},
one infers that
\beq
\mathbb{I}_{u}^{(\a)} \; = \; 
     \begin{cases}
        \intn{1}{\mf{p}_{\a} } & \mf{p}_{\a} \, > \,  0 \\
        \emptyset & \mf{p}_{\a} \, = \,  0 \\
        \intn{1+\mf{p}_{\a} }{0} & \mf{p}_{\a} \, < \,  0 \,.
     \end{cases}
\enq
Indeed, if $\mf{p}_{\a} \, = \,  0 $, then no zero is crossed upon deforming
$\msc{C}_{\e{ref}}$ to $\msc{C}_{u}$ in the neighbourhood of $\mc{V}_{\ups_{\a} q}$.
If  $\mf{p}_{\a} \, > \,  0 $, then one crosses necessarily the zeroes
$\mf{x}_k^{(\a)}$ with $k=1,\dots,  \mf{p}_{\a}$ upon deforming $\msc{C}_{\e{ref}}$
to $\msc{C}_{u}$ in the neighbourhood of $\mc{V}_{\ups_{\a} q}$. Finally,
if $\mf{p}_{\a} \, < \,  0 $ then one crosses necessarily the zeroes $\mf{x}_k^{(\a)}$
with $k=1 + \mf{p}_{\a},\dots, 0$ upon deforming $\msc{C}_{\e{ref}}$ to $\msc{C}_{u}$
in the neighbourhood of $\mc{V}_{\ups_{\a} q}$, see Fig.~\ref{zeroes and contour deformations}.

One should also note that the pole contribution associated with the crossing
of the zeroes $\mf{x}_k^{(\a)}$ when deforming $\msc{C}_{\e{ref}}$ to $\msc{C}_{u}$
will come with an index $\ups_{\a} \, \e{sgn}(\mf{p}_{\a})$. The latter stems from the direction along which
on has to deform $\msc{C}_{\e{ref}}$ so as to reach $\msc{C}_{u}$ and the fact that $\la \mapsto \Im[ u(\la) ]$
increases with $\Im \la$ in a neighbourhood of $q$ but decreases in a neighbourhood of $-q$.  Hence, it holds that
\bem
     \Oint{ \msc{C}_{ \e{ref} }  }{} \f{\dd \mu }{ 2\i\pi T} \:
        f(\mu) \f{u^{ \prime}(\mu)  }{ 1 + \ex{\f{1}{T} u(\mu)  }  } \; = \;
     \Oint{ \msc{C}_{ u }  }{} \f{\dd \mu }{ 2\i\pi T} \:
        f(\mu) \f{u^{ \prime}(\mu)  }{ 1 + \ex{\f{1}{T} u(\mu)  }  }   \\
\; - \; \sul{\a \in \{L,R\} }{} \ups_{\a}
\bigg\{ \bs{1}_{\mathbb{N}^*}\big(\mf{p}_{\a}\big) \sul{ a=1 }{ \mf{p}_{\a} } f\big( \mf{x}_a^{(\a)} \big)
\; - \; \bs{1}_{-\mathbb{N}^*}\big(\mf{p}_{\a}\big) \sul{ a=1 +\mf{p}_{\a} }{ 0 } f\big( \mf{x}_a^{(\a)} \big) \bigg\} \;.
\label{ecriture deformation ctr C ref vers C u ac residus}
\end{multline}

The possibility to carry out such a contour deformation already allows one to
connect some of the constants arising in the problem.

\begin{lemme}
\label{Lemme lien entre p pm et monodromie}
Let $u(\la)$ satisfy the requirements $\mathrm{i)-vii)}$ of
Properties~\ref{Propriete pour deformer les contours}. Then
$1+\ex{-\frac{1}{T}u(\lambda)}$ has $0$ index, \textit{c.f.}\
\eqref{ecriture monodromie Trotter infini}, relatively to $\msc{C}_{u}$.
\end{lemme}

\Proof 
Denote by $J_{\e{in}}$ the part of $\msc{C}_{u}$ joining $q^{(+)}_{ u }$ to
$q^{(-)}_{ u }$ on which $\Re (u)$ is negative and by $J_{\e{out}} = \msc{C}_{u} 
\setminus J_{\e{in}}$ its complement. Then it holds that
\beq
     \Oint{  \msc{C}_{ u } }{}  \f{\dd \mu }{ 2\i\pi T } \:
        \f{ - u^{\prime}(\mu)  }{1+ \ex{\f{1}{T} u(\mu) }  }
     \, = \, - \Int{  J_{\e{in}} }{}  \f{\dd \mu }{ 2\i\pi T } \: u^{\prime}(\mu)
        \,  +  \, \Int{  J_{\e{in}}  }{}  \f{\dd \mu }{ 2\i\pi T } \:
	          \f{   u^{\prime}(\mu)  }{1+ \ex{- \f{  1}{T} u(\mu) }  } 
      \,  -  \, \Int{  J_{\e{out}}  }{}  \f{\dd \mu }{ 2\i\pi T } \:
         \f{   u^{\prime}(\mu)  }{1+ \ex{\f{  1}{T} u(\mu) }  } \;.
\enq
The three integrals may be taken explicitly given the assumptions on $\msc{C}_{u}$
and $u$. Indeed,
\beq
     \Int{  J_{\e{in}}  }{}  \f{\dd \mu }{ 2\i\pi T } \:
        \f{   u^{\prime}(\mu)  }{1+ \ex{- \f{1}{T} u(\mu) }  } \; = \; \f{1}{2\i\pi}
     \ln \Big[ 1 + \ex{ \f{1}{T} u(\mu) } \Big]
         \biggr|_{ q^{(+)}_{ u } }^{ q^{(-)}_{ u }  } \; = \; 0 \;,  
\enq
where the antiderivative is taken in terms of the principal branch of the
logarithm which is possible owing to $\Re\big[ u(\mu) \big] \, < \, 0$
for $\mu \in J_{\e{in}}$, and where we used point $\mathrm{i)}$ of
Properties~\ref{Propriete pour deformer les contours} which ensures that
\beq
\ln \Big[  1 + \ex{ \f{1}{T} u( q^{(\pm)}_{ u } ) }  \Big] \;= \; \ln 2 \;.
\enq
Likewise it holds that
\beq
     - \Int{  J_{\e{out}}  }{}  \f{\dd \mu }{ 2\i\pi T } \:
        \f{   u^{\prime}(\mu)  }{1+ \ex{\f{ 1}{T} u(\mu) }  }  \; = \;\f{1}{2\i\pi}
     \ln \Big[ 1 + \ex{ -\f{1}{T} u(\mu) }  \Big]
         \biggr|_{ q^{(-)}_{ u } }^{ q^{(+)}_{ u }  } \; = \; 0\;. 
\enq
Thus, all-in-all,
\beq
    - \Oint{  \msc{C}_{ u } }{}  \f{\dd \mu }{ 2\i\pi T } \:
       \f{ u^{\prime}(\mu)  }{1+\ex{\f{ 1}{T} u(\mu) } } \, = \,
      \f{ 1 }{ 2\i\pi T }
      \Big\{   u\big(q^{(+)}_{ u }  \big) \, - \, u\big(q^{(-)}_{ u }  \big)  \Big\}
      \; = \; 0 \;.
\enq
\qed


\subsection{An equivalent non-linear integral equation}
In this section, we are going to recast the original non-linear integral equation,
after having  accommodated the contour to $\msc{C}_{u}$, into an equivalent one
whose analysis is simpler. For that purpose, we introduce the notation
\beq
|u|(\la)  \; = \;  u (\la) \cdot \e{sgn}\Big\{ \Re \big[ u(\la)  \big] \Big\}  \;.
\label{definition u absolut et bar u}
\enq
Also, we denote by $\msc{L}n_{\msc{C}_{u}}\Big[ 1 + \ex{-\f{u}{T} } \Big] \big( \la  \big)$
the logarithm of $1+\ex{-\f{1}{T}u}$ relatively to $\msc{C}_{u}$, analogously to the
definition given in \eqref{definition logarithm de 1+a}.

We first establish a technical lemma that will allow us to transform some of the integrals
of interest.

\begin{lemme}
\label{Lemme DA fonction holomorphe versis log}
Let $f$ be holomorphic in an open neighbourhood of $\msc{C}_{u}$ and have zero
index with respect to $\msc{C}_{u}$, \textit{viz}.\ $\Int{ \msc{C}_{u} }{} \dd s \: f(s) =0$.
Let $F$ be an antiderivative of $f$ on $\msc{C}_{u}$. Then, for any
$\Big\{ u\big( \la \big), \msc{C}_{u} \Big\}$ enjoying the
Properties~\ref{Propriete pour deformer les contours} and having index $\mf{m}$,
\textit{c.f.} \eqref{ecriture monodromie Trotter infini},  with respect to
$\msc{C}_{\e{ref}}$, it holds that
\bem
     -2\i\pi \mf{m} F(\varkappa) \, + \,
        \Int{ \msc{C}_{ \e{ref} } }{}  \dd \mu \: f\big( \mu \big)\,
	\msc{L}n_{ \msc{C}_{ \e{ref} } }\Big[ 1 + \ex{-\f{u}{T} } \Big] \big( \mu  \big)
     \;  =  \;
        \Int{ \msc{C}_{ u } }{}  \dd \mu \: f\big( \mu \big)\,
	\msc{L}n_{ \msc{C}_{u} }\Big[ 1 + \ex{-\f{u}{T} } \Big] \big( \mu  \big) \\
\; - \; 2\i\pi \sul{\a \in \{L,R\} }{} \ups_{\a}
\bigg\{ \bs{1}_{\mathbb{N}^*}\big(\mf{p}_{\a}\big) \sul{ a=1 }{ \mf{p}_{\a} } F\big( \mf{x}_a^{(\a)} \big)
\; - \; \bs{1}_{-\mathbb{N}^*}\big(\mf{p}_{\a}\big) \sul{ a=1 +\mf{p}_{\a} }{ 0 } F\big( \mf{x}_a^{(\a)} \big) \bigg\} \;,
\label{deformation integrale sur C ref vers integrale sur Cu}
\end{multline}
where
\bem
 \Int{ \msc{C}_{ u } }{}  \dd \mu \: f\big( \mu \big)\,
	\msc{L}n_{ \msc{C}_{u} }\Big[ 1 + \ex{-\f{u}{T} } \Big] \big( \mu  \big) \; = \;
        \Int{ \msc{C}_{-q;q}   }{  }  \f{\dd \mu }{  T } \: f(\mu)  u( \mu )   \\
     \, + \, \Int{ q }{ q^{(+)}_{u} }  \f{\dd \mu }{  T } \: f(\mu)  u( \mu )
     \, + \, \Int{ q^{(-)}_{u} }{ - q }  \f{\dd \mu }{  T } \: f(\mu)  u( \mu )
     \, + \,  \Int{  \msc{C}_{u}  }{}  \dd \mu \:
                 f(\mu) \ln \bigg[ 1 + \ex{  - \f{1}{T} |u|(\mu)   }  \bigg]  \;. 
\label{decomposition fondamentale integrale contre log}
\end{multline}
Here $\msc{C}_{-q;q}$ is any curve joining $-q$ to $q$ such that
$\mu \mapsto f(\mu) u(\mu)$ is holomorphic in the domain delimited by the
curve $J_{\e{in}}$ and $\intff{ q^{(-)}_{u} }{ -q } \cup \msc{C}_{-q;q} \cup
\intff{q}{ q^{(+)}_{u} }$.  

Furthermore, one has 
\beq
     \Int{  \msc{C}_{ u }  }{}  \dd \mu \:
        f(\mu) \ln \bigg[ 1 + \ex{  - \f{1}{T} |u|(\mu)   }  \bigg]  \;  =  \; 
     -  \f{\pi^2 T}{6}
     \Bigg\{ \f{  f\big( q^{(+)}_{u} \big)  }{u^{\prime}\big(  q^{(+)}_{u}    \big) }  
     \; - \; \f{ f\big( q^{(-)}_{u} \big) }{u^{\prime}\big(  q^{(-)}_{u}   \big) }   \Bigg\} 
     \; + \;  \e{O}\big( T^3 \big) \;. 
\label{ecriture DA integrale avec u lisse}
\enq
\end{lemme}

\Proof 
By definition of $\msc{L}n_{ \msc{C}_{u} }$ and upon using the index condition
\eqref{ecriture monodromie Trotter infini},  it holds that
\bem
     -2\i\pi \mf{m} F(\varkappa) \, + \,  \Int{ \msc{C}_{\e{ref}} }{}  \dd \mu \:
        f\big( \mu \big)\,
        \msc{L}n_{ \msc{C}_{\e{ref}} }\Big[ 1 + \ex{-\f{u}{T} } \Big] \big( \mu  \big)   \;  =  \;
 -\Int{  \msc{C}_{ \e{ref} } }{}  \f{\dd \mu }{  T } \: F(\mu)
        \f{ - u^{\prime}(\mu)  }{1+ \ex{\f{1}{T} u(\mu) }  }    \\
     \;  =  \;  \Int{  \msc{C}_{ u } }{}  \f{\dd \mu }{  T } \: F(\mu)
        \f{u^{\prime}(\mu)  }{1+ \ex{\f{1}{T} u(\mu) }  }
\; - \; 2\i\pi \sul{\a \in \{L,R\} }{} \ups_{\a}
\bigg\{ \bs{1}_{\mathbb{N}^*}\big(\mf{p}_{\a}\big) \sul{ a=1 }{ \mf{p}_{\a} } F\big( \mf{x}_a^{(\a)} \big)
\; - \; \bs{1}_{-\mathbb{N}^*}\big(\mf{p}_{\a}\big) \sul{ a=1 +\mf{p}_{\a} }{ 0 } F\big( \mf{x}_a^{(\a)} \big) \bigg\}  \;.
\end{multline}
Here we used \eqref{ecriture deformation ctr C ref vers C u ac residus} in the
intermediate equality. Finally, integrating by parts backwards and using the zero
index with respect to $\msc{C}_{u}$ ensured by Lemma~\ref{Lemme lien entre p pm et monodromie}
leads to
\beq
\Int{  \msc{C}_{ u } }{}  \f{\dd \mu }{  T } \:
   F(\mu)  \f{u^{\prime}(\mu)}{1 + \ex{\f{1}{T} u(\mu) }  } \; = \;
        \Int{ \msc{C}_{u} }{}  \dd \mu \: f\big( \mu \big)\,
        \msc{L}n_{ \msc{C}_{u} }\Big[ 1 + \ex{-\f{u}{T} } \Big] \big( \mu  \big)   \;.
\enq

Next, it holds that
\bem
\Int{  \msc{C}_{ u } }{}  \f{\dd \mu }{  T } \: F(\mu)  \f{u^{\prime}(\mu)  }{1+ \ex{\f{1}{T} u(\mu) }  }
      \,  = \, \Int{  J_{\e{in}} }{}  \f{\dd \mu }{  T } \: F(\mu)
         u^{\prime}( \mu )   \,  -  \, \Int{  J_{\e{in}}  }{}  \f{\dd \mu }{   T } \:
	    F(\mu)  \f{   u^{\prime}(\mu)  }{1+ \ex{ - \f{  1}{T} u(\mu) }  } 
      \,  +  \, \Int{  J_{\e{out}}  }{}  \f{\dd \mu }{  T } \: F(\mu)
         \f{   u^{\prime}(\mu)  }{1+ \ex{\f{  1}{T} u(\mu) }  } \\
     \,  = \, -  \Int{  J_{\e{in}} }{}  \f{\dd \mu }{  T } \: f(\mu)  u( \mu )
        \, + \, \f{1}{T} \bigg\{  F( q^{(-)}_{u} ) u(q^{(-)}_{u}) 
     \, - \,  F(q^{(+)}_{u}) u(q^{(+)}_{u})   \bigg\} 
     -  \bigg\{ F(\mu) \ln \Big[ 1 + \ex{  \f{1}{T} u(\mu) }  \Big] \bigg\}
        \biggr|_{ q^{(+)}_{ u } }^{ q^{(-)}_{ u }  }\\
     -  \bigg\{ F(\mu) \ln \Big[ 1 + \ex{ -\f{1}{T} u(\mu) }  \Big] \bigg\}
        \biggr|_{ q^{(-)}_{ u } }^{ q^{(+)}_{ u }  }   
     \,  +  \, \Int{  J_{\e{in}}  }{}   \dd \mu \:
        f (\mu)  \ln \Big[ 1 + \ex{  \f{1}{T} u(\mu) }  \Big] 
     \,  +  \, \Int{  J_{\e{out}}  }{}  \dd \mu \: f(\mu)
        \ln \Big[ 1 + \ex{  - \f{1}{T} u(\mu) }  \Big]     \;. 
\end{multline}
All-in-all, using point {\rm i)} of Properties~\ref{Propriete pour deformer les contours},
this yields the representation \eqref{decomposition fondamentale integrale contre log}. 

One may then decompose the integral involving the logarithm as
\beq
     \Int{  \msc{C}_{ u }  }{}  \dd \mu \: f(\mu)
        \ln \bigg[ 1 + \ex{  - \f{1}{T} |u|(\mu)    }  \bigg]  \; = \; 
     I_{\e{ext}}  \; +  \;  I_{\e{in}} \;.
\enq
The integrands in $I_{\e{ext}}$ and $I_{\e{in}}$ solely involve the principal
branch of the logarithm. Furthermore, due to the estimates on $f$ and $u$, 
\beq
     I_{\e{ext}} \; = \; \Int{J_{\de} }{} \dd \mu \: f(\mu)
        \ln \bigg[ 1 + \ex{ - \f{1}{T} |u|(\mu)    }  \bigg]      \; = \; 
     \e{O}\big( T^{\tf{M}{2}} \big) \; ,
\enq
with $|u|$ as defined in \eqref{definition u absolut et bar u} and $J_{\de}= \msc{C}_{ u }\setminus \{J^{(-)}_{\de} \cup J^{(+)}_{\de}\} $
\textit{c.f.} point $\mathrm{iv)}$ of Properties \ref{Propriete pour deformer les contours}.

To estimate  $I_{\e{in}}$, upon recalling the definition of $J^{(\pm)}_{\de}$ in point $\mathrm{iii)}$ of Properties \ref{Propriete pour deformer les contours},
we perform a local change of variables $z=u(\la )$ around $q^{(\pm)}_{u}$:
\beq
     I_{\e{in}} \; = \hspace{-3mm}  \Int{   J_{\de}^{(+)}\cup J_{\de}^{(-)}  }{} \hspace{-3mm} \dd \mu \: f(\mu)
        \ln \bigg[ 1 + \ex{ - \f{1}{T} |u|(\mu)    }  \bigg]   
     \; = \; - \sul{\eps= \pm}{} \Int{ - \eps \de}{ \eps \de} \dd v \:
        \f{ f }{u^{\prime}} \circ u^{-1}_{\eps}(v   )
	\ln\Big[ 1 + \ex{ - \f{ |v| }{T} } \Big]     \;.
\enq
Above, we have taken into account the change of orientation induced by $u$
around $q^{(-)}_{ u }$. Also $u_{\pm  }^{-1}$ stands for the local inverse of
$u$ around $q^{(\pm)}_{ u }$.  Note that, by point $\mathrm{ii)}$ of
Properties~\ref{Propriete pour deformer les contours} $u_{\pm}^{-1}$ does
exist in some sufficiently small neighbourhood of $0$. After applying the results of
Lemma~\ref{lemme Natte integrale elementaire} to $I_{\e{in}}$, we get that 
\beq
I_{\e{in}} \; = \;  - \f{\pi^2 T}{6}
 \Bigg\{ \f{  f\big( q^{(+)}_{u} \big)  }{u^{\prime}\big(  q^{(+)}_{u}   \big) }  
 \; - \; \f{ f\big( q^{(-)}_{u} \big) }{u^{\prime}\big(  q^{(-)}_{u}   \big) }   \Bigg\} 
 \; + \;  \e{O}\big( T^3 \big)   \;.
\enq
By putting all the bits together, the claim follows. \qed

\subsection{The low-$T$ expansion of the solution to the non-linear integral equation}
One of the consequences of the previous considerations is that \textit{given} a solution
to the non-linear problem which satisfies
Properties~\ref{Propriete pour deformer les contours}, one immediately obtains
the low-$T$ and large-$N$, resp.\ low-$T$, expansion of the solution to
\eqref{ecriture NLIE a Trotter fini}, \eqref{ecriture monodromie Trotter fini}, resp.\
\eqref{ecriture NLIE a Trotter infini}, \eqref{ecriture monodromie Trotter infini}.
 
To start with, observe that, owing to the previous discussions, given any solution
to \eqref{ecriture NLIE a Trotter fini} and \eqref{ecriture monodromie Trotter fini},
satisfying  the requirements $\mathrm{i)-vii)}$ of
Properties \ref{Propriete pour deformer les contours},
Lemma~\ref{Lemme DA fonction holomorphe versis log} ensures that
one may recast the non-linear
integral equation \eqref{ecriture NLIE a Trotter fini} in the form 
\beq
     \wh{u}\,\big(\la \,|\, \wh{\mathbb{X}} \, \big) \; = \; h-T\mf{w}_{N}(\la) \, - \, \i \pi  T  \mf{s}
     \, - \, \i T \sul{ y \in \wh{\mathbb{X}}  }{} \th_+(\la-y) \, - \,
     T \Int{ \msc{C}_{ \wh{u} } }{} \dd \mu \: K(\la-\mu)\,
     \msc{L}n_{ \msc{C}_{\wh{u}}}\big[1+\ex{-\frac{1}{T} \wh{u}} \big]\big(\mu \,|\,  \wh{\mathbb{X}}\, \big)
     \;. 
\label{ecriture NLIE associee a u}
\enq
Here $\wh{\mathbb{X}}\, =\, \wh{\op{Y}}_{\e{tot}}\oplus \op{Y}_{\e{sg}} \ominus \wh{\op{X}}_{\e{tot}}$ is as defined in \eqref{definition ensemble hat X pour somme sols part trous}, while
$\wh{\op{Y}}_{\e{tot}} \, = \, \wh{\op{Y}} \oplus \wh{\op{Y}}\; \!^{\prime}$ and $\wh{\op{X}}_{\e{tot}} \, = \, \wh{\op{X}} \oplus \wh{\op{X}}\;\!^{\prime}$, where
\begin{align}
& \wh{\op{X}}\;\!^{\prime}\; = \;
   \bigg\{\bigcup\limits_{ \substack{\a\in \{L,R\} \\ : \ups_{\a} \mf{p}_{\a}<0} }{}
          \big\{ \, \wh{\mf{x}}_{k}^{\,(\a)}  \big\}_{k=1+\mf{p}_{\a}}^{0}
          \setminus \wh{\op{Y}}\, \! ^{\prime}_{\e{ref}} \bigg\} \cup
   \bigg\{\wh{\op{X}}\, \! ^{\prime}_{\e{ref}}  \setminus
          \bigcup\limits_{ \substack{  \a\in \{L,R\}  \\ : \ups_{\a} \mf{p}_{\a}>0} }{}
	   \big\{ \, \wh{\mf{x}}_{k}^{\,(\a)} \big\}_{k=1}^{\mf{p}_{\a}}  \bigg\} \;,  \\
& \wh{\op{Y}}\; \!^{\prime}\; = \;
     \bigg\{\bigcup\limits_{\substack{\a\in \{L,R\} \\ : \ups_{\a} \mf{p}_{\a}>0} }{}
     \big\{ \, \wh{\mf{x}}_{k}^{\,(\a)}  \big\}_{k=1}^{\mf{p}_{\a}}
     \setminus \wh{\op{X}}\, \! ^{\prime}_{\e{ref}}  \bigg\} \cup
\bigg\{ \wh{\op{Y}}\, \! ^{\prime}_{\e{ref}}   \setminus
          \bigcup\limits_{ \substack{  \a\in \{L,R\}  \\ : \ups_{\a} \mf{p}_{\a}< 0} }{} \big\{ \, \wh{\mf{x}}_{k}^{\,(\a)} \big\}_{k=1+\mf{p}_{\a}}^{0}  \bigg\} \;,
\end{align}
\textit{c.f.}~\eqref{definition des ensembles hat Y et X prime ref}.
In this context, the roots $\wh{\mf{x}}_{k}^{\,(\a)}$ solve
$\wh{u}\,\big(\,\wh{\mf{x}}_{k}^{\,(\a)} \,|\, \wh{\mathbb{X}} \, \big)
\; = \; 2\i\pi T \big(k - \tfrac{1}{2}\big)$
and $\wh{\mf{x}}_{k}^{\,(\a)}$ belongs to a small neighbourhood of $\ups_{\a} q$.

We also insist that the contour is such that, given $\la \in \msc{C}_{\wh{u}}$, the poles
of $\mu \mapsto K(\la-\mu)$ at $\mu=\la \pm \i \zeta$  are located outside of $\msc{C}_{\wh{u}}$.
Finally, the equation for a solution $u(\la \,|\, \mathbb{X} )$ to
\eqref{ecriture NLIE a Trotter infini}, \eqref{ecriture monodromie Trotter infini}
follows upon the replacements $\wh{u} \hookrightarrow u$,
$h-T\mf{w}_{N}(\la) \hookrightarrow \veps_{0}(\la)$ and appropriate substitutions of the zeroes
$\wh{\mf{x}}_{k}^{\,(\a)} \hookrightarrow \mf{x}_{k}^{\,(\a)}$ and algebraic sums
$\wh{\mathbb{X}} \, \hookrightarrow \, \mathbb{X}$.

The above equation is the starting point for obtaining the large-$N$, low-$T$ asymptotic
expansions of the solutions.

\begin{prop}
\label{Proposition correspondance entre NLIE originelle et NLIE point fixe}
\begin{enumerate}
\item
Every solution $\wh{u}\,\big(\la \,|\, \wh{\mathbb{X}} \, \big)$ to the non-linear
integral equation \eqref{ecriture NLIE a Trotter fini} and
\eqref{ecriture monodromie Trotter fini} satisfying the
Properties~\ref{Propriete pour deformer les contours} solves the
non-linear integral equation 
\beq
     \wh{u}\,\big(\la \,|\, \wh{\mathbb{X}} \, \big) \; = \;
        \mc{W}_N(\la) \, + \,
	T  u_1\big(\la \,|\, \wh{\mathbb{X}} \, \big)
     \, + \,  \wh{\mc{R}}_{T}\big[ \, \wh{u}\,\big(* \,|\, \wh{\mathbb{X}} \, \big) \big](\la) \;.
\label{ecriture eqn NL pour u forme asymptotique}
\enq
Here $\mc{W}_N(\la)$ is defined through \eqref{definition solution LIE deformee WN},
$u_1$ has been introduced in \eqref{definition fonction u 1}, while the operator
$\wh{\mc{R}}_{T}$ decomposes as $\wh{\mc{R}}_{T} = \wh{\mc{R}}_{T}^{(1)}
+ \wh{\mc{R}}_{T}^{(2)}$ with 
\beq
     \wh{\mc{R}}_{T}^{\, (1)} \big[\, \wh{u}\, \big(* \,|\, \wh{\mathbb{X}} \, \big) \big](\la)   \; = \;
     - \,  \bigg\{  \Int{ q }{ q^{(+)}_{ \wh{u} } }\! \!  \dd \mu \:
           \wh{u}\,\big(\mu \,|\, \wh{\mathbb{X}} \, \big)
     \, + \, \Int{ q^{(-)}_{ \wh{u} } }{ - q }  \! \! \dd \mu \:
             \wh{u}\,\big(\mu \,|\, \wh{\mathbb{X}} \, \big)  \bigg\} \cdot  R_{\e{c}} (\la,\mu)
\label{definition operateur reste de type 1}
\enq
and 
\beq
     \wh{\mc{R}}_{T}^{\, (2)} \big[\, \wh{u}\,\big(* \,|\, \wh{\mathbb{X}} \, \big) \big](\la)
        \; = \;  - \, T \Int{ \msc{C}_{ \wh{u} } }{} \dd \mu \: R_{\e{c}}(\la,\mu)
	\ln\Big[ 1 + \ex{-\frac{1}{T}| \, \wh{u} \, |(\mu \,|\, \wh{\mathbb{X}} \, ) } \Big] \;.
\label{definition operateur reste de type 2}
\enq
$R_{\e{c}}(\la,\mu)$ corresponds to the integral kernel of the operator
$\e{id} - \op{R}_{\e{c}}$ inverse to $\e{id}+\op{K}_{\msc{C}_{\veps}}$,
\textit{c.f.}~\eqref{definition du resolvent c deforme}. We refer to
\eqref{definition u absolut et bar u} for the definitions of $|\, \wh{u} \, |$. Finally, we stress that the integration in
\eqref{definition operateur reste de type 1} runs through segments. 
\item
Conversely, any solution to \eqref{ecriture eqn NL pour u forme asymptotique} that
is subject to the monodromy condition \eqref{ecriture monodromie Trotter fini} and
satisfies the Properties~\ref{Propriete pour deformer les contours} solves
\eqref{ecriture NLIE a Trotter fini} with the range of parameters $\wh{\mathbb{X}}_{\e{ref}}$
defined according to \eqref{definition ensembles zeros Zx et Zy}-\eqref{defintion mathbb X ref et son hat}.
\item
The same statements hold for any solution $u(\la\,|\, \mathbb{X})$ to the non-linear
integral equation \eqref{ecriture NLIE a Trotter infini} and \eqref{ecriture monodromie Trotter infini}
satisfying the Properties~\ref{Propriete pour deformer les contours} provided one makes the replacement
\beq
     \mc{W}_N \; \; \hookrightarrow \;\;\veps_{\e{c}}  
\enq
in which $\veps_{\e{c}}$ is defined through \eqref{ecriture LIE pour veps c}.
\end{enumerate}
\end{prop}

\Proof 
By virtue of equation \eqref{decomposition fondamentale integrale contre log} given
in  Lemma~\ref{Lemme DA fonction holomorphe versis log}, upon taking the contour
$\msc{C}_{-q;q}$ appearing there to coincide with $\wh{\msc{C}}_{\veps}$ introduced
above of \eqref{definition solution LIE deformee WN}, one may recast the non-linear
integral equation \eqref{ecriture NLIE associee a u} in the form 
\beq
     \Big( \e{id} + \op{K}_{ \wh{\msc{C}}_{\veps} } \Big)
        \big[ \, \wh{u}\,\big(* \,|\, \wh{\mathbb{X}} \, \big) \big](\la)
     \; = \; h - T \mf{w}_N(\la) \, - \, \i \pi  T  \mf{s}
     \, - \, \i T \sul{ y \in \wh{\mathbb{X}}}{} \th_+(\la-y)
     + \mc{K}_{T}\big[ \, \wh{u}\,\big(* \,|\, \wh{\mathbb{X}} \, \big) \big](\la) \;.
\label{ecriture eqn NL pour u legerement transformee}
\enq
Here $ \op{K}_{ \wh{\msc{C}}_{\veps} }$ is as introduced
in \eqref{ecriture LIE pour veps c} provided one makes the substitution $\msc{C}_{\veps} \hookrightarrow \wh{\msc{C}}_{\veps}$. Further,
the operator $\mc{K}_{T}$ decomposes as $\mc{K}_{T}=\mc{K}_{T}^{(1)}+\mc{K}_{T}^{(2)}$, where 
\beq
     \mc{K}_{T}^{(1)} \big[ \,  \wh{u}\,\big(* \,|\, \wh{\mathbb{X}} \, \big) \big](\la)   \; = \;
     - \,   \bigg\{  \Int{ q }{ q^{(+)}_{  \wh{u} } }\! \!  \dd \mu \:
            \wh{u}\,\big(\mu \,|\, \wh{\mathbb{X}} \, \big)
     \, + \, \Int{ q^{(-)}_{  \wh{u} } }{ - q }  \! \! \dd \mu \:
             \wh{u} \, \big( \mu \,|\, \wh{\mathbb{X}} \, \big)  \bigg\}
\cdot  K(\la-\mu)  
\enq
and 
\beq
     \mc{K}_{T}^{(2)} \big[ \,  \wh{u}\,\big(* \,|\, \wh{\mathbb{X}} \, \big) \big](\la) \; = \;
        - \, T \Int{ \msc{C}_{  \wh{u} } }{} \dd \mu \: K(\la-\mu)
	       \ln\Big[ 1 + \ex{-\frac{1}{T}|\,  \wh{u} \, |(\mu \,|\, \wh{\mathbb{X}}\, ) } \Big] \;.
\enq
By inverting $\e{id} + \op{K}_{ \wh{\msc{C}}_{\veps} }$ as ensured by Lemma \ref{Lemme inversibilite de Id + K sur courbe C eps}
and using \eqref{definiton des charges et phases c deformee}-\eqref{definition du resolvent c deforme} and \eqref{definition solution LIE deformee WN},
one recasts
\eqref{ecriture eqn NL pour u legerement transformee} in the form 
\bem
     \wh{u}\,\big(\la \,|\, \wh{\mathbb{X}} \, \big) \; = \;   \mc{W}_N(\la) \, + \, T \, \wh{u}_1 \big(\la \,|\, \wh{\mathbb{X}} \, \big)
- \,  \bigg\{  \Int{ q }{ q^{(+)}_{ \wh{u} } }\! \!  \dd \mu \: \wh{u}\,\big(\mu \,|\, \wh{\mathbb{X}} \, \big)
\, + \, \Int{ q^{(-)}_{ \wh{u} } }{ - q }  \! \! \dd \mu \:  \wh{u}\,\big(\mu \,|\, \wh{\mathbb{X}} \, \big) \bigg\} \cdot  \wh{R}_{\e{c}} (\la,\mu)   \\
 - \, T \Int{ \msc{C}_{ \wh{u} } }{} \dd \mu \: \wh{R}_{\e{c}}(\la,\mu) \ln\Big[ 1 + \ex{-\frac{1}{T}| \, \wh{u} \, |(\mu \,|\, \wh{\mathbb{X}}\, ) } \Big]  \;,
\label{ecriture eqn NL pour u Tortter fini a transformer}
\end{multline}
where $\wh{R}_{\e{c}}(\la,\mu)$ is the resolvent kernel of the operator
$\wh{\op{R}}_{\e{c}}$ giving the inverse $\e{id}-\wh{\op{R}}_{\e{c}}$ of
$\e{id} + \op{K}_{ \wh{\msc{C}}_{\veps} } $, \textit{c.f.}\ above of
\eqref{definition solution LIE deformee WN}. Also, 
\beq
\wh{u}_1 \big(\la \,|\, \wh{\mathbb{X}} \, \big) \, = \,  -\i \pi  \mf{s} \, \wh{Z}_{\e{c}}(\la)  \, - \, 2 \i \pi \sul{ y \in  \wh{\mathbb{X}} }{} \wh{\phi}_{\e{c}}(\la,y)
\enq
in which $\wh{Z}_{\e{c}}$ and $\wh{\phi}_{\e{c}}$ solve
\eqref{definiton des charges et phases c deformee} with $\msc{C}_{\veps}$
replaced by $\wh{\msc{C}}_{\veps}$. Then one deforms the contours
$\wh{\msc{C}}_{\veps} \hookrightarrow \msc{C}_{\veps}$ in the action of the
inverse $\e{id}-\wh{\op{R}}_{\e{c}}$ in all functions except $\mc{W}_N(\la)$
which leads to the replacement 
\beq
\wh{u}_1\big(\la \,|\, \wh{\mathbb{X}} \, \big) \;\; \hookrightarrow \;\; u_1\big(\la \,|\, \wh{\mathbb{X}} \, \big)  \qquad \e{and} \qquad  \wh{R}_{\e{c}} (\la,\mu) \;\; \hookrightarrow \;\; R_{\e{c}} (\la,\mu)
\enq
in \eqref{ecriture eqn NL pour u Tortter fini a transformer}. This completes the proof
relative to $\wh{u}\, \big( \la \,|\,  \wh{\mathbb{X}}\,\big)$.


The proof of the statement for $u\big( \la \,|\, \mathbb{X}\big)$ goes along identical lines,
with the exception that one may directly deform $\msc{C}_{-q;q}$ to $\msc{C}_{\veps}$ in the
early steps of the analysis owing to the lack of cuts along
$\intff{-\i\tf{\zeta}{2} + \tf{\aleph}{N} }{ -\i\tf{\zeta}{2} - \tf{\aleph}{N} }$
of $\veps_0$ as opposed to $h-T\mf{w}_N$. \qed

\begin{cor}
\label{Corollaire DA pour une solution de la NLIE}
Any solution $\wh{u}\,\big( \la \,|\,  \wh{\mathbb{X}}\,\big)$ to the non-linear integral equation
\eqref{ecriture NLIE a Trotter fini} and \eqref{ecriture monodromie Trotter fini}
exhibiting the Properties~\ref{Propriete pour deformer les contours} admits the
low-$T$, low-$\tf{1}{(NT^3)}$ asymptotic expansion
\beq
\wh{u}\, \big( \la \,|\,  \wh{\mathbb{X}}\,\big) \;  = \; \veps_{\e{c}}(\la) \, + \, \sul{k = 1}{ \lfloor \tf{M}{2} \rfloor -1} T^k \,
u_{k}\big( \la \,|\,  \wh{\mathbb{X}}\,\big)
\; + \; \e{O}\biggl( \f{1}{NT^3} + T^{ \lfloor \tf{M}{2} \rfloor } \biggr)
\label{ecriture DA de u Trotter fini low T ordre 2}
\enq
in which $M$ has been introduced in point $\mathrm{iii)}$ of
Properties~\ref{Propriete pour deformer les contours},
$u_{1}(\la\,|\, \wh{\mathbb{X}}\, )$ is given by \eqref{definition fonction u 1} and
\beq
u_2\big( \la \,|\,  \wh{\mathbb{X}}\,\big)\, = \,
     \sul{\sg= \pm }{}  \sg \f{ R_{\e{c}}(\la, \sg q  ) }{ 2 \veps_{\e{c}}^{\prime}(\sg q)} 
        \cdot \Biggl\{\Bigl(u_1\bigl(\sg q \,|\, \wh{\mathbb{X}}\, \bigr)\Bigr)^2
	              + \f{\pi^2}{3}  \Biggr\} \;.
\label{function_u2_explicitly}
\enq
Moreover, the zeroes $ q^{(\pm)}_{ \wh{u} } $ admit the expansion 
\beq
     q^{( \sg  )}_{ \wh{u} } \, =  \,  \sg   q \, + \,
        \sul{k = 1 }{ \lfloor \tf{M}{2} \rfloor -1 } q_{k}^{\,( \sg  )} \, T^k
     \; + \; \e{O}\biggl( \f{1}{NT^3} \, + \, T^{ \lfloor \tf{M}{2} \rfloor } \biggr)
     \qquad  with  \qquad
     q_{1}^{\,( \sg  )} \; = \;
        - \f{  u_1\big(\sg q \,|\, \wh{\mathbb{X}}\, \big)}{\veps_{\e{c}}^{\prime}( \sg q)} \;.
\label{expression racines q pm mathbbY}
\enq

Similarly, any solution $u(\la\,|\, \mathbb{Y})$ to the non-linear integral equation
\eqref{ecriture NLIE a Trotter infini} and \eqref{ecriture monodromie Trotter infini}
satisfying the Properties \ref{Propriete pour deformer les contours} admits the same
kind of low-$T$ asymptotic expansion in which the $\e{O}(\tf{1}{(NT^3)})$ remainder is dropped.
Moreover, uniformly away from the cuts and singularities of the functions, and in particular
on  $\msc{C}_{\e{ref}}$, it holds -- in the sense of asymptotic low-$T$ large-$NT$
expansions -- that 
\beq
\wh{u}\, \big( \la \,|\,  \wh{\mathbb{X}}\,\big) \; = \; u\big( \la \,|\,  \wh{\mathbb{X}}\,\big) \; + \; \e{O}\biggl( \f{1}{NT^3} \biggr)
\label{ecriture DA de u hat en term de u avec corrections en N}
\enq
with a remainder that is uniform in $T\tend 0^+$, $NT^3\tend +\infty$.  
\end{cor}
\Proof 
\noindent  From \eqref{ecriture eqn NL pour u forme asymptotique}, one may access
the first few terms of the low-$T$ expansion of $\wh{u}\,( *  \,|\, \wh{\mathbb{X}})$. Indeed,
it holds that
\bem
\mc{R}_{T}^{(1)} \big[ \, \wh{u}\, \big( * \,|\,  \wh{\mathbb{X}}\,\big) \big](\la)   \; = \;
-  R_{\e{c}}(\la,  q^{(+)}_{ \wh{u} } ) \, \wh{u}^{\,\prime}\big(  q^{(+)}_{ \wh{u} }   \,|\,  \wh{\mathbb{X}}\,\big)
\Int{ q }{ q^{(+)}_{ \wh{u} } }  \dd \mu \: \big( \mu  -  q^{(+)}_{ \wh{u} } \big)\\
\hspace{2cm} - R_{\e{c}}(\la,q^{(-)}_{ \wh{u} } ) \, \wh{u}^{\,\prime}\big( q^{(-)}_{ \wh{u} } \,|\, \wh{\mathbb{X}}\,\big)
                                                  \Int{ q^{(-)}_{ \wh{u} } }{ - q } \dd \mu \: \big( \mu  -  q^{(-)}_{ \wh{u} } \big)
\;+ \; \e{O}\Big( \big(  q - q^{(+)}_{ \wh{u} }  \big)^3 \Big) \, + \, \e{O}\Big( \big(  q + q^{(-)}_{ \wh{u} }  \big)^3 \Big) \\
\, = \, \f{1}{2}  \Big(  q - q^{(+)}_{ \wh{u} }   \Big)^2  R_{\e{c}}(\la,q^{(+)}_{ \wh{u} } ) \, \wh{u}^{\,\prime}\,\big( q^{(+)}_{ \wh{u} }   \,|\, \wh{\mathbb{X}}\,\big)
 \, - \, \f{1}{2} \Big(  q + q^{(-)}_{ \wh{u} }   \Big)^2   R_{\e{c}}(\la,q^{(-)}_{ \wh{u} } ) \, \wh{u}^{\,\prime}\big( q^{(-)}_{ \wh{u} }   \,|\,  \wh{\mathbb{X}}\,\big)
\;+ \;  \e{O}\Big( \sul{ \sg = \pm }{}  \big|  q -\sg q^{(\sg)}_{ \wh{u} }  \big|^3 \Big)   \;. 
\end{multline}
In its turn, by Lemma \ref{Lemme DA fonction holomorphe versis log},
\beq
\mc{R}_{T}^{(2)} \big[ \, \wh{u}\, \big( *  \,|\,  \wh{\mathbb{X}}\,\big)  \big](\la)   \; = \;
\, \f{\pi^2 T^2 }{ 6 } \sul{\sg= \pm }{} \sg  \f{ R_{\e{c}}\big( \la,q^{(\sg)}_{ \wh{u} } \big) }{ \wh{u}^{\,\prime}\big( q^{(\sg)}_{ \wh{u} }   \,|\,  \wh{\mathbb{X}}\,\big)   }
\; + \; \e{O}\big( T^4 \big)  \;. 
\enq
All-in all, this yields
\bem
\mc{R}_{T} \big[ \, \wh{u}\, \big( *  \,|\,  \wh{\mathbb{X}}\,\big)  \big](\la)   \; = \;     \, \f{\pi^2 T^2 }{ 6 } \sul{ \sg = \pm }{} \sg
\f{ R_{\e{c}}\big( \la,q^{(\sg)}_{ \wh{u} } \big) }{ \wh{u}^{\,\prime}\big( q^{(\sg)}_{\wh{u}} \,|\,  \wh{\mathbb{X}}\,\big)   }
\, + \, \frac12 \sul{\sg= \pm }{}\sg    R_{\e{c}}\big( \la,q^{(\sg)}_{\wh{u}} \big) \, \wh{u}^{\,\prime}\big( q^{(\sg)}_{\wh{u}}   \,|\,  \wh{\mathbb{X}}\,\big)
\cdot    \Big(  q - \sg q^{(\sg)}_{\wh{u}}   \Big)^2 \\
   \; + \; \e{O}\bigg( T^4 + \sul{ \sg= \pm }{}  \big|  q -\sg q^{(\sg)}_{ \wh{u} }  \big|^3  \bigg)  \;.
\label{ecriture DA RT} 
\end{multline}

One may improve the expansion by gaining a better control on the low-$T$ expansion of the zeroes $q^{(\sg)}_{ \wh{u} }$. Indeed, so far we have established that  
\beq
\wh{u}\,\big(\la \,|\,  \wh{\mathbb{X}}\,\big)  \; = \;
\mc{W}_N(\la) \, + \,  T u_1\big(\la\,|\,  \wh{\mathbb{X}}\,\big)  \, + \,
\e{O}\Big( T^2 \, + \, \sul{\sg=\pm}{} \big|  q - \sg q^{(\sg)}_{\wh{u}}   \big|^2  \Big) \;.
\enq
Thus, upon using that  $\mc{W}_N(\la)=\veps_{\e{c}}(\la) \; + \; \e{O}\big( \tf{1}{(NT)}\big) $ uniformly away from the limiting singularities at $\pm \i\tf{\zeta}{2}$, 
\beq
0=\wh{u}\,\big( q^{(\sg)}_{ \wh{u} } \,|\, \wh{\mathbb{X}}\,\big)  \; = \;
\veps_{\e{c}}^{\prime}(\sg q )\Big(  q^{(\sg)}_{\wh{u}} - \sg q \Big)
\, + \,  T   u_1\big(\sg q \,|\,  \wh{\mathbb{X}}\,\big)
\, + \,  \e{O}\biggl( \f{1}{NT}  \, + \, T^2 \, + \,
T \sul{\sg=\pm}{} \big|  q - \sg q^{(\sg)}_{\wh{u}} \big| \, + \,
\sul{\sg=\pm}{} \big|  q - \sg q^{(\sg)}_{\wh{u}} \big|^2
\biggr)
\enq
which yields \eqref{expression racines q pm mathbbY} upon direct reasoning. Plugging
this expansion back into \eqref{ecriture DA RT} allows one to conclude with respect
to the form of the expansion up to order $T^2$. The existence of the all-order low-$T$
expansion then follows upon implementing an order-by-order bootstrap analysis, see
\textit{e.g.}~\cite{KozProofexistenceAEYangYangEquation,KozProofOfDensityOfBetheRoots}.
These details are left to the reader.

Note that the $\e{O}\bigl( \tf{1}{NT^3}\bigr)$ loss of precision on $\msc{C}_{\e{ref}}$
in \eqref{ecriture DA de u Trotter fini low T ordre 2} or
\eqref{ecriture DA de u hat en term de u avec corrections en N} stems from $\msc{C}_{\e{ref}}$
being at distance $\mf{c}_{\e{d}} T$ from the singularity at $-\i\tf{\zeta}{2}$, so that,
uniformly on $\msc{C}_{ \wh{u} }$, one has rather $\mc{W}_N=\veps_{\e{c}} \, + \,
\e{O}\bigl( \tf{1}{NT^3}\bigr)$.\qed

It will be useful for further purposes to observe that the analytic continuation of
the solution to the non-linear integral equation ensures that, for almost all 
$\la\in \Cx\setminus\bigcup_{n\in \mathbb{Z}, \ups= \pm}^{}
\op{D}_{\ups\i\tf{\zeta}{2}+\i\pi n, \eta}$, $\eta>0$,   
\bem
\wh{u}\,\big(\la \,|\, \wh{\mathbb{X}}\,\big) \; = \; \veps_{\e{c}}(\la) \, + \,  T u_1\big(\la\,|\, \wh{\mathbb{X}}\,\big) \, + \,
T^2 u_2\big(\la\,|\, \wh{\mathbb{X}}\,\big)
\, - \,  T \sul{\sg=\pm}{} \sg \ln \Big[ 1+\ex{ -\f{1}{T}\wh{u}\,(\la-\i \sg \zeta\,|\, \wh{\mathbb{X}}\,) } \Big] \cdot \bs{1}_{\e{Int}( \msc{C}_{ \wh{u} } )} (\la-\i \sg \zeta) \\
\, + \, 2\i\pi T \sul{\sg=\pm }{} n_{\la;\sg}\cdot \bs{1}_{ \e{Int}( \msc{C}_{ \wh{u} } ) }(\la-\i \sg \zeta) \; + \; 
\e{O}\biggl( T^3 \, + \, \f{1}{NT} \biggr) \;.
\label{ecriture DA uniforme u- sur tout le plan cplx}
\end{multline}
Above, $n_{\la;\sg} \in \mathbb{Z}$ and may depend on $\la$ and $\sg$.

\section{Fine properties of the functional space for the non-linear integral equation}
\label{Section espace fnel pour la NLIE et ses ptes}

Throughout this section we shall focus on collections of parameters $\op{X}, \op{Y}$
and $\op{Y}_{\e{sg}}$ such that Hypothesis~\ref{Hypotheses solubilite NLIE}
holds. For the time being, we consider two collections of arbitrary points in
$\bigcup_{\sg = \pm} \op{D}_{\sg q, \mf{c}_{\e{loc}}}$ $\op{X}^{\prime}_{\e{var}}$
and $\op{Y}^{\prime}_{\e{var}}$, with $\mf{c}_{\e{loc}}$ as in point $\mathrm{d)}$
of Hypothesis~\ref{Hypotheses solubilite NLIE}. Let
\beq
\mathbb{X}_{\e{gen}} \; = \; \op{Y}_{\e{gen}} \oplus \op{Y}_{\e{sg}} \ominus \op{X}_{\e{gen}}
\quad \e{and} \quad \left\{ \ba{ccc}  \op{Y}_{\e{gen}} &  = &  \op{Y} \oplus \op{Y}^{\prime}_{\e{var}} \vspace{2mm}  \\
                                   \op{X}_{\e{gen}} &  = &  \op{X} \oplus \op{X}^{\prime}_{\e{var}} \ea \right. \;.
\label{definition Y gen}
\enq

Point $a)$ of Hypothesis~\ref{Hypotheses solubilite NLIE}, along with
$\op{X}^{\prime}_{\e{var}}$ and $\op{Y}^{\prime}_{\e{var}}$ being built out
of points in $\bigcup_{\sg = \pm} \op{D}_{\sg q, \mf{c}_{\e{loc}}}$, ensures
that $\la \mapsto u_1\big(\la\mid \mathbb{X}_{\e{gen}} \big)$ is holomorphic in an open $T$-independent neighbourhood $\mc{V}_{\a}$,
$\a \in \{L, R\}$, of $\ups_{\a} q$, where $\ups_{\a}$ was introduced
in~\eqref{definition de ups alpha}. In particular, this means that there exists a
$T$-independent constant $C_u>0$ such that  
\beq
\norm{ u_1\big(\ast \,|\, \mathbb{X}_{\e{gen}} \big) }_{ L^{\infty}(\mc{V}_{\a} ) } \, \leq \, C_u \;.
\label{definition de la borne sur u1 voisinage de pm q}
\enq
Since $\veps^{\prime}_{\e{c}}(\pm q) \not = 0$, the neighbourhoods $\mc{V}_{\a}$
can be chosen such that
\beq
\e{inf}\Big\{ |\veps^{\prime}_{\e{c}}(\la)| \; : \; \la \in \mc{V}_L\cup \mc{V}_{R}  \Big\} \, > \, m_{\veps}
\label{ecriture borne inf sur veps prime voisinage pm q}
\enq
for some $m_\veps > 0$.

\subsection{The main building ingredients of the functional space}

Given $f \in \mc{E}_{\mc{M}}$
(\textit{c.f.} \eqref{definition espace fonctionnel principal}), we define two functions
\begin{align}
     \wh{f}\big(\la \,|\, \mathbb{X}_{\e{gen}}\big) & =
        \mc{W}_N(\la) \, + \, T u_1\big( \la \,|\, \mathbb{X}_{\e{gen}} \big) \, + \, f(\la) \;, \\[1ex]
     f\big(\la \,|\, \mathbb{X}_{\e{gen}}\big) & =
        \veps_{\e{c}}(\la) \, + \, T u_1\big( \la \,|\, \mathbb{X}_{\e{gen}} \big) \, + \, f(\la) \; .
\end{align}
The first one will be relevant for the analysis of the finite Trotter
number non-linear integral equation  \eqref{ecriture NLIE a Trotter fini}, while the
second one will be relevant for the infinite Trotter number counterpart
\eqref{ecriture NLIE a Trotter infini}. We first establish the existence of the points
$\tau_{\a}$, in the sense of point $\mathrm{vii)}$ of
Properties~\ref{Propriete pour deformer les contours}, for these auxiliary functions.
Before stating the result, we however recall that $\veps_{\a}^{-1}$, $\a \in \{L,R\}$,
are the two inverses of $\veps$ as discussed in
Proposition~\ref{Proposition double recouvrement de veps}.

\begin{lemme}
\label{Lemme identification des entiers definissant p pm}
Let $\op{X}, \op{Y}$ fulfil  Hypotheses~\ref{Hypotheses solubilite NLIE}
and let $\op{X}^{\prime}_{\e{var}}\, , \op{Y}^{\prime}_{\e{var}}$ be built out
of points in $\bigcup_{\sg = \pm} \op{D}_{\sg q, \mf{c}_{\e{loc}} }$.
Then there exists $T_0$ small enough such that, for any $T_0>T>0$,
$f \in \mc{E}_{\mc{M}}$ and $\a\in \{L, R\}$, there exist unique
$\tau_{\a} \in \intoo{ -\de_T }{ \de_T }$, $\mf{p}_{\a} \in \mathbb{Z}$ and $\eps_{\a} \in \intfo{0}{1}$ such that
\beq
 f\big( \veps_{\a}^{-1}\big( \tau_{\a} \big) \,|\, \mathbb{X}_{\e{gen}} \big)   \, = \, 2\i\pi T\Big( \mf{p}_{ \a }\, - \, \tfrac{1}{2} \, + \, \eps_{\a} \Big) \;.
\label{ecriture identification zero cas trotter infini}
\enq
Moreover, in this regime, $|\mf{p}_{\a}|$ is bounded by a constant only depending on
$|\op{Y}_{\e{gen}}|$, $|\op{X}_{\e{gen}}|$ and $|\op{Y}_{\e{sg}}|$ while $\tau_{\a}=\e{O}(T)$.

Similarly, there exists $\eta > 0$ small enough such that, for any $T_0>T>0$ and
$\eta > \tf{1}{(NT^2)} > 0$, there exists a unique $\wh{\tau}_{\a} \in \intoo{ -\de_T }{ \de_T }$,
with $\wh{\tau}_{\a}\, = \, \tau_{\a}\, + \, \e{O}\big( \tf{1}{(NT)}\big)$, $\wh{\mf{p}}_{\a} \in \mathbb{Z}$ and $\wh{\eps}_{\a} \in \intfo{0}{1}$
with $ \wh{ \mf{p} }_{ \a }\, + \, \wh{\eps}_{\a}  \, = \,   \mf{p}_{ \a }  \, + \, \eps_{\a}\, + \, \e{O}\big( \tf{1}{(NT)}\big)$
such that
\beq
 \wh{f}\big( \veps_{\a}^{-1}\big( \, \wh{\tau}_{\a} \big) \,|\, \mathbb{X}_{\e{gen}} \big)   \, = \, 2\i\pi T\Big( \wh{ \mf{p} }_{\a }\, - \, \tfrac{1}{2} \, + \, \wh{\eps}_{\a} \Big) \;.
\label{ecriture identification zero cas trotter fini}
\enq
Moreover, in this regime, $|\wh{\mf{p}}_{\a}|$ is bounded by a constant only depending on $|\op{Y}_{\e{gen}}|$, $|\op{X}_{\e{gen}}|$ and $|\op{Y}_{\e{sg}}|$.
\end{lemme}

\Proof 
We first prove \eqref{ecriture identification zero cas trotter infini} and then
discuss the differences occurring in the proof of
\eqref{ecriture identification zero cas trotter fini}. 


For $t \in \intff{-\de_T}{\de_T} $, with $\de_{T}$ as defined through
\eqref{definition du parametre delta T}, one has 
\beq
\label{fepsinverseoft}
f\big( \veps_{\a}^{-1}(t)\,|\, \mathbb{X}_{\e{gen}} \big) \, = \, t \, + \,   T  u_1\big( \veps_{\a}^{-1}(t) \,|\, \mathbb{X}_{\e{gen}} \big) \, + \, f\big( \veps_{\a}^{-1}(t) \big) \;, \qquad  \e{where} \quad \a \in \{L,R\} \; .
\enq
Owing to the boundedness of $u_1$ and $f$ the map $t \, \mapsto 
\, \Re\Big[ f\big( \veps_{\a}^{-1}(t)\,|\, \mathbb{X}_{\e{gen}} \big) \Big]$
is strictly increasing on $\intff{-\de_T}{\de_T}$ for $T$ small enough. Indeed, 
\beq
\Dp{t}\Re\Big[ f\big( \veps_{\a}^{-1}(t)\,|\, \mathbb{X}_{\e{gen}} \big) \Big]   \; = \; 1 \, + \,
               \Re \bigg\{ T \f{ u_1^{\prime}\big( \veps_{\a}^{-1}(t) \,|\, \mathbb{X}_{\e{gen}} \big) }{ \veps^{\prime}_{\e{c}}\big( \veps_{\a}^{-1}(t) \big) }
\, + \, \f{ f^{\prime}\big( \veps_{\a}^{-1}(t) \big) }{ \veps^{\prime}_{\e{c}}\big( \veps_{\a}^{-1}(t) \big) } \bigg\} \;, 
\enq
so that, owing to \eqref{ecriture borne inf sur veps prime voisinage pm q} and since uniformly in $T$ small enough it holds
$\e{d}\big(\veps_{\a}^{-1}\big( \intff{-\de_T}{\de_T} \big) , \Dp{}\mc{V}_{\a} \big) >0$,
because $\veps_{\a}^{-1}\big( \intff{-\de_T}{\de_T} \big) \subset \mc{V}_{\a}$ for such $T$s, one has the lower bound
\beq
\Dp{t}\Re\Big[ f\big( \veps_{\a}^{-1}(t)\,|\, \mathbb{X}_{\e{gen}} \big) \Big] \, > \, 1 \, - \, T \f{ C \cdot C_u }{ m_{\veps} } \, - T^2 \f{ C_{\mc{M}} }{ m_{\veps} }
\enq
for some $C>0$. This lower bound is manifestly strictly positive provided that $T$
is small enough. 

Moreover, using that $\veps_{\a}^{-1} (\pm \de_T) \; = \; \ups_{\a} q + \e{O}(\de_T)$ one has
\beq
f\big( \veps_{\a}^{-1}( \pm \de_T )\,|\, \mathbb{X}_{\e{gen}} \big) \, = \, \pm \de_T \, + \, T u_1\big( \ups_{\a} q \,|\, \mathbb{X}_{\e{gen}} \big) \, + \, f\big( \ups_{\a}  q \big)
\;+\;\e{O}\Big( T \de_T \, + \, T^2 \de_T    \Big)
\enq
with $\ups_{\a}$ as in \eqref{definition de ups alpha}. This entails, for $T$ small enough,
that $\Re\Big[ f\big( \veps_{\a}^{-1}(t)\,|\, \mathbb{X}_{\e{gen}} \big) \Big]$ changes sign
on $\intoo{-\de_T}{\de_T}$ at a unique $\tau_{\a}$.

Setting $t = \tau_\a$ in \eqref{fepsinverseoft} and taking the real part, one infers that
\beq
\tau_{\a} \, = \,   - \,   T  \Re\big[ u_1\big( \veps_{\a}^{-1}(\tau_{\a}) \,|\, \mathbb{X}_{\e{gen}} \big) \big] \, - \, \Re\big[ f\big( \veps_{\a}^{-1}(\tau_{\a}) \big)  \big]
\, = \, \e{O}(T) \;.
\enq
Using once more \eqref{fepsinverseoft} one obtains
\begin{multline}
\frac1T \Im[ f\big( \veps_{\a}^{-1}(\tau_{\a})\,|\, \mathbb{X}_{\e{gen}} \big)] =
2\pi \Big( \mf{p}_{ \a }\, - \, \tfrac{1}{2} \, + \, \eps_{\a} \Big) \\ \; = \;
    \Im\big[ u_1\big( \veps_{\a}^{-1}(\tau_{\a}) \,|\, \mathbb{X}_{\e{gen}} \big) \big] \, + \, \f{1}{T} \Im\big[ f\big( \veps_{\a}^{-1}(\tau_{\a}) \big)  \big]
\; = \;   \Im\big[ u_1\big( \ups_{\a} q \,|\, \mathbb{X}_{\e{gen}} \big) \big] \; + \; \e{O}(T)
\end{multline}
for unique $\mf{p}_{ \a } \in {\mathbb Z}$, $\eps_\a \in [0;1[$.
In the last equality, we have used the estimate $\veps_{\a}^{-1}(\tau_{\a})=\ups_{\a} q \, + \, \e{O}(T)$ as inferred from the behaviour of $\tau_{\a}$
obtained above. This ensures the boundedness property of $|\mf{p}_{\a}|$.

Now, observe that $\mc{W}_N=\veps_{\e{c}}+\e{O}\big( \tf{1}{ NT }\big)$ uniformly
away from $\pm \i\tf{\zeta}{2}$. Thus, $\mc{W}_N$ is invertible  on some open,
$N,T$-independent neighbourhood of $\ups_{\a} q$ provided that $NT$ is large enough.
Moreover, it holds  that
\beq
\mc{W}_{N;\a}^{-1} \, =\, \veps_{\a}^{-1} + \e{O}\big( \tf{1}{ N T }\big)
\enq
on the image under $\mc{W}_N$ of that neighbourhood.
The rest follows from handlings analogous to the infinite Trotter number case. \qed  

\vspace{3mm}







We now need to introduce a specific class of functions on which we will focus. Indeed, in
the following, instead of keeping the parameters building up $\op{X}^{\prime}_{\e{var}}$
or $\op{Y}^{\prime}_{\e{var}}$ arbitrary in $\bigcup_{\sg=\pm}\op{D}_{\sg q , \mf{c}_{\e{loc}} }$,
we will focus on parameters
\beq
\wh{\op{X}}^{\,\prime} \; = \; \bigcup\limits_{\a \in \{L,R\} } \! \Big\{\, \wh{\op{x}}_{0; \ell}^{\, (\a)}  \Big\}_{\ell=1}^{ \varkappa^{(\a)}_0} \quad \e{and} \quad
\wh{\op{Y}}^{\,\prime} \; = \; \bigcup\limits_{\a \in \{L,R\} }\!  \Big\{ \, \wh{\op{y}}_{0;\ell}^{\, (\a)} \Big\}_{\ell=1}^{ \mf{y}^{(\a)}_0 }  \;,
\label{definition ensemble trous et particules locales en pm q}
\enq
with $\wh{\op{x}}_{0; \ell}^{\, (\a)}, \wh{\op{y}}_{0; \ell}^{\, (\a)} \in
\op{D}_{\ups_{\a} q , \mf{c}_{\e{loc}} }$ solving the below set of coupled equations
in many variables
\begin{align}
 \wh{f}\big(\,  \wh{\op{x}}_{0; \ell}^{\, (\a)} \,|\, \wh{\mathbb{X}} \big) & \; = \; - 2\i\pi T \ups_{\a} \big(  h_{0;\ell}^{(\a)} \, + \, \tfrac{1}{2} \big) \;,
                                                                                     \label{ecriture eqns definisant trou locaux en pm q} \\[1ex]
 \wh{f}\big( \, \wh{\op{y}}_{0;\ell}^{\, (\a)} \,|\, \wh{\mathbb{X}} \big) & \; =  \; 2\i\pi T \ups_{\a} \big(  p_{0;\ell}^{(\a)} \, + \, \tfrac{1}{2} \big)
  \label{ecriture eqns definisant particules locales en pm q}
\end{align}
for $h_{0;\ell}^{(\a)}, p_{0;\ell}^{(\a)} \in {\mathbb N}$, $\a \in \{L, R\}$.
Analogously, we define quantities for the infinite Trotter number setting with the
$\, \wh{}\, $ omitted everywhere. Above and in the rest of this section we shall
agree upon the notation
\beq
\wh{\mathbb{X}} \;  = \;  \wh{\op{Y}}_{\e{tot}} \oplus \op{Y}_{\e{sg}} \ominus \wh{\op{X}}_{\e{tot}}
\quad \e{and} \quad
     \begin{cases}
        \wh{\op{Y}}_{\e{tot}} = \op{Y} \oplus \wh{\op{Y}}\;\!^{\prime} \\[1ex]
        \wh{\op{X}}_{\e{tot}} = \op{X} \oplus \wh{\op{X}}\;\!^{\prime}
     \end{cases}\,.
\label{definition hat Y ensemble}
\enq
It is clear that these parameters definitely depend on the choice of $f \in \mc{E}_{\mc{M}}$,
as well as on the choice of the remaining parameters $\op{X}, \op{Y}$. However, unless
specified otherwise, \textit{viz}.\ $\wh{\op{x}}_{0; \ell}^{\, (\a)}[f]$ \textit{etc}.,
we shall keep this dependence implicit.

The next result shows the unique solvability for the parameters
$\wh{\op{x}}_{0; \ell}^{\, (\a)} , \wh{\op{y}}_{0; \ell}^{\, (\a)}$ and discusses
their dependence on $f$ and $\op{X}, \op{Y}$.

\begin{prop}
\label{Proposition existence et continuite parameters particule trou partiels}
Let $\op{X}=\big\{ \op{x}_a \big\}_{a=1}^{|\op{X}|}$,
$\op{Y}=\big\{ \op{y}_a \big\}_{a=1}^{|\op{Y}|}$
fulfil Hypothesis~\ref{Hypotheses solubilite NLIE}. Let $p_{0,a}^{(\a)},
h_{0,a}^{(\a)} \in \mathbb{N}$ be such that
\beq
T \, p_{0,a}^{(\a)} \; = \; \e{o}(1) \;,\ a=1,\dots, \mf{y}^{(\a)}_0 \quad and \quad
T \, h_{0,a}^{(\a)} \; = \; \e{o}(1) \;,\ a=1,\dots, \varkappa^{(\a)}_0 \;, \quad \a\in \{L,R\}
\enq
as $T\tend 0^+$ and with $\mf{y}^{(\a)}_0, \varkappa^{(\a)}_0$ bounded uniformly
in $T, N$.
\begin{enumerate}
\item
Then there exist $T_0, \eta>0$ and small enough such that
for any $0<T<T_0$, $ \tf{1}{(NT^2)} < \eta $ and $f \in \mc{E}_{\mc{M}}$ the system
\eqref{ecriture eqns definisant trou locaux en pm q}--\eqref{ecriture eqns definisant particules locales en pm q} admits a unique solution with
$\wh{\op{x}}_{0; \ell}^{\, (\a)},  \wh{\op{y}}_{0; \ell}^{\, (\a)} \in
\op{D}_{\ups_{\a}q, \mf{c}_{\e{loc}} }$.
\item
There exist $\varrho>0$ and $C>0$ such that
\begin{align}
\big|\, \wh{\op{x}}_{0; \ell}^{\, (\a)} [f] \, - \, \wh{\op{x}}_{0; \ell}^{\, (\a)}[g] \big| & \leq
C  \norm{f-g }_{ L^{\infty}\big(   \cup_{\a \in \{L,R\}}^{} \veps_{\a}^{-1}( \ov{\op{D}}_{0,3\varrho} )   \big) } \;, \notag \\[1ex]
\big|\, \wh{\op{y}}_{0; \ell}^{\, (\a)}[f] \, - \, \wh{\op{y}}_{0; \ell}^{\, (\a)}[g] \big| & \leq
                                   C  \norm{f-g }_{ L^{\infty}\big(   \cup_{\a \in \{L,R\}}^{} \veps_{\a}^{-1}( \ov{\op{D}}_{0,3\varrho} )   \big) } \;.
\label{ecriture estimees sur la variation des parametres}
\end{align}
\item
Let
$\bs{x} \, = \, \bigl(\op{x}_1, \dots, \op{x}_{|\op{X}|} \bigr)$,
$\bs{y} \, = \, \bigl(\op{y}_1, \dots, \op{y}_{|\op{Y}|} \bigr)$ and let $\bs{x}^{(0)} \, = \, \bigl(\op{x}_1^{(0)}, \dots, \op{x}_{|\op{X}|}^{(0)} \bigr)$,
$\bs{y}^{(0)} \, = \, \bigl(\op{y}_1^{(0)}, \dots, \op{y}_{|\op{Y}|}^{(0)} \bigr)$. Assume that $\{\op{x}_a^{(0)}\}_{1}^{|\op{X}| }$ and
$\{\op{y}_a^{(0)}\}_{1}^{|\op{Y}| }$ fulfil Hypothesis~\ref{Hypotheses solubilite NLIE}.
Then, the exists a open neighbourhood of
$(\bs{x}^{(0)}; \bs{y}^{(0)})^{\op{t}}$ that does not depend on $T, \tf{1}{NT^2}$ small enough such that for any $\bs{\nu}=(\bs{x}; \bs{y})^{\op{t}}$
the associated collections of parameters fulfil Hypothesis~\ref{Hypotheses solubilite NLIE} and the maps
\beq
     \bs{v} \mapsto \wh{\op{x}}_{0; \ell}^{\, (\a)} \;, \quad resp.\quad
     \bs{v} \mapsto  \wh{\op{y}}_{0; \ell}^{\, (\a)} \;,
\enq
induced by \eqref{ecriture eqns definisant trou locaux en pm q}--\eqref{ecriture eqns definisant particules locales en pm q},
are holomorphic on it.

Moreover, their partial derivatives enjoy the bounds
\beq
     \big|\Dp{\op{v}_a}^{s} \wh{\op{x}}_{0; \ell}^{\, (\a)} \big| \, \leq  \,  C T
\quad and \quad
     \big|\Dp{\op{v}_a}^{s} \wh{\op{y}}_{0; \ell}^{\, (\a)} \big| \, \leq  \,  C T
\label{ecriture estimees en T sur les trous et particules proches de pm q}
\enq
for some $C>0$ depending on $s \in \mathbb{N}^{*}$, this uniformly in $T, N$ and
$f \in \mc{E}_{\mc{M}}$, provided that $C_{\mc{M}} T <1$.
\item
Exactly the same conclusions hold for the parameters associated with the infinite
Trotter number setting.
\end{enumerate}
\end{prop}

\Proof
Let $\mf{h}_{\a}=(\mf{y}^{(\a)}_0 +\varkappa^{(\a)}_0 )$, $\mf{h}=\mf{h}_{L} + \mf{h}_{R}$ and
$\mc{W}= \op{D}_{-q, \eps}^{\mf{h}_{L}}\times \op{D}_{q, \eps}^{\mf{h}_{R}} \subset \Cx^{\mf{h}}$ with $\eps>0$ and small enough, in particular $\eps < \mf{c}_{\e{loc}}$.
We shall make use of the vectors
\beq
\bs{z} \, = \, \big( \bs{z}^{(L)} , \bs{z}^{(R)} \big)^{\op{t}} \qquad \e{and} \qquad
\bs{z}^{(\a)} \, = \, \big( \bs{x}^{(\a)} , \bs{y}^{(\a)} \big)^{\op{t}} \;,
\enq
%
%
%
while
\beq
\bs{x}^{(\a)}\; = \; \Big( x_1^{(\a)},\dots, x_{ \varkappa^{(\a)}_0 }^{(\a)}  \Big)^{\op{t}} \qquad \e{and} \qquad
\bs{y}^{(\a)}\; = \; \Big( y_1^{(\a)},\dots, y_{ \mf{y}^{(\a)}_0 }^{(\a)}  \Big)^{\op{t}} \;.
\enq
Further, given $\bs{z} \in \mc{W}$, we denote
\beq
\op{X}^{\,\prime}_{\e{var}} \; = \; \bigcup\limits_{\a \in \{L,R\} } \big\{ x_{k}^{\, (\a)} \big\}_{k=1}^{ \varkappa^{(\a)}_0 } \qquad \e{and} \qquad
\op{Y}^{\,\prime}_{\e{var}} \; = \; \bigcup\limits_{\a \in \{L,R\} } \big\{ y_{k}^{\, (\a)} \big\}_{k=1}^{ \mf{y}^{(\a)}_0 }
\enq
and define $\mathbb{X}_{\e{gen}}$ as in \eqref{definition Y gen}. We have at our
disposal enough notations so as to introduce the holomorphic map
\beq
\wh{\Psi}_T : \mc{W} \;   \tend  \;  \Cx^{\mf{h}} \;,
\quad \e{where} \quad
\wh{\Psi}_T \big( \bs{z} \big) \; = \;
\begin{pmatrix}
\wh{\Psi}_T^{(L)} \big( \bs{z} \big) \\[1ex]
\wh{\Psi}_T^{(R)} \big( \bs{z} \big)
\end{pmatrix}
\label{definition composantes left and right de la map Psi T}
\enq
and
\beq
\wh{\Psi}_T^{(\a)} \big( \bs{z} \big) \; = \;
\Big( \wh{f}\, \big( \, x_1^{(\a)} \,|\, \mathbb{X}_{\e{gen}} \big)   \; , \dots, \; \wh{f}\, \big( \, x^{(\a)}_{ \varkappa^{(\a)}_0 } \,|\, \mathbb{X}_{\e{gen}} \big)  , \;
\wh{f}\, \big( \, y_1^{(\a)} \,|\, \mathbb{X}_{\e{gen}} \big)   \; , \dots,
                         \;\wh{f}\, \big( \, y^{(\a)}_{ \mf{y}^{(\a)}_0 } \,|\, \mathbb{X}_{\e{gen}} \big) \Big)^{\op{t}} \;.
\label{definition fct Psi}
\enq

The existence and uniqueness of the roots $\op{x}_{0,k}^{(\a)}$ and
$\op{y}_{0,k}^{(\a)}$ would follow from the unique solvability of the equation
\beq
\wh{\Psi}_T\big( \bs{z}  \big)\, = \,
 \Big( \bs{h}^{(L)}_{  \varkappa^{(L)}_0 } \, , \,   \bs{p}^{(L)}_{  \mf{y}^{(L)}_0 } \, ,
                              \,  \bs{h}^{(R)}_{ \varkappa^{(R)}_0 } \, , \,   \bs{p}^{(R)}_{ \mf{y}^{(R)}_0 }  \Big)^{\op{t}}
\label{ecriture equation sur solvabilite conditions auxiliaires sur racines}
\enq
in which
\begin{equation}
\bs{h}^{(\a)}_{ \varkappa^{(\a)}_0 } \, = \, -  \ups_{\a} 2\i\pi T \Big(  h_{1}^{(\a)} \, + \, \tfrac{1}{2} \, , \, \cdots \, , \,
 h_{ \varkappa^{(\a)}_0 }^{(\a)} \, + \, \tfrac{1}{2}  \Big)  \; ,  \quad
\bs{p}^{(\a)}_{ \mf{y}^{(\a)}_0 } \, =  \, \ups_{\a} 2\i\pi T \Big(  p_{1}^{(\a)} \, + \, \tfrac{1}{2} \, , \, \cdots \, , \,
 p_{ \mf{y}^{(\a)}_0 }^{(\a)} \, + \, \tfrac{1}{2}  \Big) \;.
\end{equation}

In order to study the properties of the map $\wh{\Psi}_T$, it is convenient to
represent $\wh{f}$ in the form
\beq
\label{decompositionfhat}
 \wh{f}\, \big(  \la   \,|\, \mathbb{X}_{\e{gen}} \big)  \; = \; \veps_{\e{c}} \big(  \la   \big) \, + \,
 T \wh{ \de f}\big(    \la  \,|\, \bs{z}  \big) \;,
\enq
where we have introduced
\beq
   \wh{ \de f}\big(    \la  \,|\, \bs{z}  \big) \; = \;
  u_{1}\big( \la\,|\,  \mathbb{X}_{\e{gen}}  \, \big) \, + \, \f{1}{T}f(\la)
\, +\, \f{1}{T}\Big( \mc{W}_N(\la)-\veps_{\e{c}}(\la) \Big) \;.
\enq
Then, one readily gets to the representation for the differential of $\wh{\Psi}_T$
\beq
\op{D}_{ \bs{z}  }\wh{\Psi}_T \; = \;
\begin{pmatrix}
\op{D}_{ \bs{z}^{(L)} }\wh{\Psi}_T^{(L)}  & \op{D}_{ \bs{z}^{(R)}  }\wh{\Psi}_T^{(L)} \\[1ex]
\op{D}_{ \bs{z}^{(L)}  }\wh{\Psi}_T^{(R)}  & \op{D}_{ \bs{z}^{(R)} }\wh{\Psi}_T^{(R)}
\end{pmatrix} \;.
\enq
The partial differentials appearing above take the form
\beq
\op{D}_{ \bs{z}^{(\a)}  }\wh{\Psi}_T^{(\be)}    \; = \;
\begin{pmatrix}
 \veps_{\e{c}}^{\, \prime} \big(   x_a^{(\be)}  \big) \,  \de_{ab} \de_{\a \be}
                                   \, + \, T  \Dp{x_b^{(\a)}} \big[ \wh{ \de f}\big( x_a^{(\be)}  \,|\, \bs{z}   \big) \big]
   &   T  \Dp{ y_b^{(\a)} } \big[ \wh{ \de f}\big( x_a^{(\be)}  \,|\, \bs{z}   \big) \big]   \vspace{3mm}  \\
 T  \Dp{ x_b^{(\a)} } \big[ \wh{ \de f}\big( y_a^{(\be)}  \,|\, \bs{z}   \big) \big]
    &   \veps_{\e{c}}^{\, \prime} \big(  y_a^{(\be)}  \big) \, \de_{ab}  \de_{\a \be}
                         \, + \, T  \Dp{y_b^{(\a)}} \big[ \wh{ \de f}\big( y_a^{(\be)}  \,|\, \bs{z}   \big) \big]
\end{pmatrix} \;.
\label{ecriture differentielle de Psi}
\enq
Since, $\bs{z} \mapsto \wh{ \de f}\big( z_a  \,|\, \bs{z} \big) $ is holomorphic
on $\mc{W}$ and bounded there uniformly in $T, \tf{1}{NT^2}$ small enough,
it follows that, uniformly throughout any open relatively compact subset
$\mc{U} \subset \mc{W}$, one has
\beq
\big| \Dp{x_b^{(\a)}} \big[ \wh{ \de f}\big( z_a  \,|\, \bs{z}   \big) \big] \big| \; + \;
\big| \Dp{y_b^{(\a)}} \big[ \wh{ \de f}\big( z_a  \,|\, \bs{z}   \big) \big] \big| \, =  \,
\e{O}\big( 1 \big)
\enq
as soon as $T<T_0$ and $\tf{1}{(NT^2)}<\eta$ with $T_0, \eta>0$ and small enough.
Using that $\veps_\e{c}^{\, \prime} (\pm q) \ne 0$ and employing a similar argument as in 
\eqref{ecriture borne inf sur veps prime voisinage pm q} then entails that, for
some $c>0$,
\beq
c^{-1} \; \geq \; \Big| \det\big[ \op{D}_{ \bs{z}  }\wh{\Psi}_T  \big] \Big| \; \geq  \; c >0
\label{ecriture bornes uniformes sup et inf sur det phi hat T}
\enq
uniformly in $T, \tf{1}{NT^2}$ small enough. The latter ensures that
$\wh{\Psi}_{T}$ is a local biholomorphism around any $\bs{z} \in \mc{U}$,
see \textit{e.g.}\ \cite{RangeHolomorphicFctsInManyVariables}.

One may obtain a characterisation of the range of this biholomorphism by
employing the decomposition \eqref{decompositionfhat} on $\wh{\Psi}_{T}$. Then
\beq
\wh{\Psi}_T \, = \, \Psi_{0} \, + \, T \de \wh{\Psi}_T
\qquad \e{with} \qquad
\Psi_0 \, = \, \begin{pmatrix} \Psi_0^{(L)} \\[1ex]  \Psi_0^{(R)}  \end{pmatrix}
\enq
and
\beq
\Psi_0^{(\a)}\big( \bs{z} \big) \, = \, \Big( \veps_{\e{c}}\big( x_1^{(\a)} \big) \, , \,  \dots \, , \, \veps_{\e{c}}\big( x_{ \varkappa^{(\a)}_0 }^{(\a)} \big) \, , \,
 \veps_{\e{c}}\big( y_1^{(\a)} \big) \, , \,  \dots \, , \, \veps_{\e{c}}\big( y_{ \mf{y}^{(\a)}_0 }^{(\a)} \big) \Big)^{\op{t}} \;.
\enq
It follows from Proposition \ref{Proposition double recouvrement de veps} that $\Psi_0: \mc{W} \tend \Psi_0\big( \mc{W} \big) $ is a biholomorphism.
Next, reduce $\eps$ if need be so that $\ov{\op{D}}_{\ups_{\a} q ,\eps } \subset \mc{V}_{\a}$ and recall
the bound \eqref{definition de la borne sur u1 voisinage de pm q} with $\mc{V}_{\a}$ as introduced around that equation.
Then, for $\varrho>0$ and small enough, so that the special analytic polyhedron
\beq
\mc{D}_{2\varrho} \; = \; \Big\{ \bs{z} \in \mc{W} \; : \; \big| \veps_{\e{c}}(z_a)\big|\; < \; 2 \varrho \; , \; a=1,\dots, \mf{h} \Big\}
\enq
is a relatively compact subset of  $\mc{W}$, there exists $T_0, \eta>0$ such that,
uniformly in $0<T<T_0$ and $0<\tf{1}{(NT^2)}<\eta$ it holds for every
$\bs{w}\in \op{D}_{0,\varrho}^{\mf{h}}$ and
\beq
\label{defgammarho}
\bs{z} \in
\Ga_{2\varrho} \; = \;  \Big\{ \bs{z} \in \ov{\mc{D}}_{2\varrho} \; : \;
   \big| \veps_{\e{c}}(z_a)\big|\; = \; 2 \varrho \; , \; a=1,\dots, \mf{h} \Big\}
\enq
that
\beq
T \big| \big[\de \wh{\Psi}_T(\bs{z}) \big]_{a} \big| \; \leq  \; \f{\varrho}{2} \, < \, \big| \veps_{\e{c}}(z_a)-w_a \big| \; = \; \big| \big[ \Psi_0(\bs{z})-\bs{w} \big]_{a} \big| \;.
\label{borne sur coordonnee delta Psi T hat}
\enq
As a consequence, it follows from the multi-dimensional Rouch\'{e}
Theorem~\ref{Theoreme Rouche generalise poru plusieurs vars complexes} that
$\wh{\Psi}_T-\bs{w}$ and $\Psi_0-\bs{w}$ have the same number of zeroes in
$\mc{D}_{2\varrho}$, counted with multiplicities. Since
$\Psi_0:\mc{D}_{2\varrho} \mapsto \op{D}_{0,2\varrho}^{\mf{h}}$ is a
biholomorphism, there is only one zero. Thus, since $\wh{\Psi}_T$ is
one-to-one on $\mc{D}_{2\varrho}$ and a local bi-holomorphism, it follows
that there exists $\mc{S}\subset \mc{D}_{2\varrho}$ such that
\beq
\wh{\Psi}_{T} \; : \; \mc{S} \tend    \op{D}_{0,\varrho}^{\mf{h}}
\label{ecriture biholomorphisme hat psi T avec domaines}
\enq
is a biholomorphism. Note that by construction, $\varrho$ is $T$ and $N$ independent
as long as we are in the range of interest for these parameters. We now show that
$\mc{S} \supset \mc{D}_{\varrho/2}$, and thus contains a fixed, $T, N$ independent
open neighbourhood of $\bs{z}_0 = \big( \bs{z}_{0}^{(L)}, \bs{z}_{0}^{(R)} \big)^{\op{t}}$
with $\bs{z}_{0}^{(\a)}=\ups_{\a} q \big( 1,\dots, 1)^{\op{t}} \in \Cx^{\mf{h}_{\a}}$.
Indeed, for any $\bs{z}\in  \mc{D}_{\varrho/2} $, it follows from the first bound in
\eqref{borne sur coordonnee delta Psi T hat} that
\beq
     \big| \big[ \wh{\Psi}_{T}(\bs{z}) \big]_a \big| \; \leq \;
        \big| \big[ \Psi_{0}(\bs{z}) \big]_a \big| \, + \,
	T \big| \big[\de \wh{\Psi}_T(\bs{z}) \big]_{a} \big| \, < \, \varrho \; ,
     \quad \e{so} \; \e{that} \quad
     \wh{\Psi}_{T}(\bs{z}) \in \op{D}_{0,\varrho}^{\mf{h}}\;.
\enq
We have thus established the unique solvability of the system
\eqref{ecriture equation sur solvabilite conditions auxiliaires sur racines}.

To get the remaining bounds, one observes that the multi-dimensional Rouch\'{e}
Theorem \ref{Theoreme Rouche generalise poru plusieurs vars complexes} provides
one with an integral representation of the inverse of $\wh{\Psi}_{T}$. For any
$\bs{w}\in \op{D}_{0,\varrho}^{\mf{h}}$, it holds that
\beq
     \Big[ \wh{\Psi}_{T}^{\,-1}(\bs{w})\Big]_{j} \; = \; \Int{ \Ga_{2\varrho} }{}
       \f{ \dd^{\mf{h}} \nu }{ (2\i\pi)^{\mf{h}} } \:
     \f{\nu_j \det\bigl[\op{D}_{\bs{\nu}} \wh{\Psi}_{T}\bigr]}
       {\pl{a=1}{\mf{h} } \Bigl[ \wh{\Psi}_{T}(\bs{\nu}) \, - \, \bs{w} \Bigr]_a}
\label{ecriture representation integrale pour hat psi moins 1}
\enq
in which $\Ga_{2\varrho}$ was defined in \eqref{defgammarho}.
Now, for $f,g \in \mc{E}_{\mc{M}}$, denoting by $\wh{\Psi}_{T}[f], \wh{\Psi}_{T}[g]$
the associated biholomorphisms, one obtains
\beq
     \Big[ \wh{\Psi}_{T}^{\,-1}[f](\bs{w}) \, - \,  \wh{\Psi}_{T}^{\,-1}[g](\bs{w}) \Big]_{j}
        \; = \; \Int{ \Ga_{2\varrho} }{} \f{\dd^{\mf{h}} \nu}{(2\i\pi)^{\mf{h}}} \:
	\nu_j \, \De[f,g](\bs{\nu}) \;.
\label{ecriture representation integrale pour difference hat psi moins 1}
\enq
Here the non-trivial part of the integrand decomposes as
$\De[f,g]=\De^{(1)}[f,g]+\De^{(2)}[f,g]$ with
\begin{align}
     \De^{(1)}[f,g] & = \f{ \det\big[ \op{D}_{\bs{\nu}} \wh{\Psi}_{T}[f]\big] \, - \,
        \det\big[ \op{D}_{\bs{\nu}} \wh{\Psi}_{T}[g]\big]}
        {\prod_{a=1}^{\mf{h}} \Big[ \wh{\Psi}_{T}[f](\bs{\nu}) \, - \, \bs{w} \Big]_a} \;, \\[1ex]
     \De^{(2)}[f,g] & = \det\big[ \op{D}_{\bs{\nu}} \wh{\Psi}_{T}[g]\big]
     \Bigg\{\pl{a=1}{\mf{h}} \f{1}{\Big[ \wh{\Psi}_{T}[f](\bs{\nu}) \, - \, \bs{w} \Big]_a}
        \; - \;  \pl{a=1}{\mf{h} } \f{1}{ \Big[ \wh{\Psi}_{T}[g](\bs{\nu})
	\, - \, \bs{w} \Big]_a   }   \Bigg\} \;.
\end{align}

We first bound $\De^{(2)}[f,g]$ and, for that purpose, decompose it as
$\De^{(2)}[f,g] \; = \; \sul{s=1}{\mf{h} } \De^{(2)}_{s}[f,g]$, where
\beq
 \De^{(2)}_{s}[f,g]  \; = \;
\pl{a=1}{s-1 }   \f{ 1  }{ \Big[ \wh{\Psi}_{T}[f](\bs{\nu}) \, - \, \bs{w} \Big]_a   }
\cdot \Bigg\{   \f{  \det\big[ \op{D}_{\bs{\nu}} \wh{\Psi}_{T}[g]\big]  }{ \Big[ \wh{\Psi}_{T}[f](\bs{\nu}) \, - \, \bs{w} \Big]_s   }
\, - \,   \f{  \det\big[ \op{D}_{\bs{\nu}} \wh{\Psi}_{T}[g]\big]  }{ \Big[ \wh{\Psi}_{T}[g](\bs{\nu}) \, - \, \bs{w} \Big]_s   }   \Bigg\}
\cdot   \pl{a=s+1}{\mf{h} }   \f{ 1  }{ \Big[ \wh{\Psi}_{T}[g](\bs{\nu}) \, - \, \bs{w} \Big]_a   } \;.
\enq
Next, one observes that for any $h \in \mc{E}_{\mc{M}}$, $\nu \in \Ga_{2\varrho}$
and $\bs{w}\in \op{D}_{0,\varrho}^{\mf{h}}$ one has the lower bound
\beq
     \Big| \Big[ \wh{\Psi}_{T}[h](\bs{\nu}) \, - \, \bs{w} \Big]_s \Big| \, = \,
     \Big|\veps_{\e{c}}(\nu_s) \, +  \, T  \Big[ \de \wh{\Psi}_T[h](\bs{\nu})\Big]_s
        \, - \, w_s \Big|
     \, \geq \, 2 \varrho - \f{\varrho}{2} \, - \, \varrho \, \geq   \,  \f{\varrho}{2} \;.
\label{ecriture lower bound pour integrande Rouche multidimensionnel}
\enq
Thus, by invoking \eqref{ecriture bornes uniformes sup et inf sur det phi hat T}, one concludes
that
\beq
     \big| \De^{(2)}[f,g]  \big| \; \leq  \;
        \mf{h}\cdot c^{-1} \cdot \Big( \f{2}{\varrho} \Big)^{\mf{h}+1} 
	\cdot \underset{s=1,\dots, \mf{h}}{\e{sup}} \big| f(\nu_s)-g(\nu_s) \big| \; \leq \;
     C \cdot \norm{ f-g }_{ L^{\infty}\big(\cup_{\a \in \{L,R\}}^{} \veps_{\a}^{-1}( \ov{\op{D}}_{0,2\varrho} )   \big) } \;.
\enq
In the last bound we used the maximum modulus principle and that
$| \veps_{\e{c}}(\nu_s) | \, = \, 2\varrho $ implies that
$\nu_s \in \bigcup\limits_{\a \in \{L,R\}}^{} \veps_{\a}^{-1}\big(\ov{\op{D}}_{0,2\varrho} \big)$.

Finally, by invoking the multilinearity of the determinant, one decomposes
$\De^{(1)}[f,g] \; = \; \sul{s=1}{\mf{h} } \De^{(1)}_{s}[f,g] $ with
\beq
\De^{(1)}_{s}[f,g] \; = \; \f{ 1   }
          { \prod_{a=1}^{\mf{h} }    \Big[ \wh{\Psi}_{T}[f](\bs{\nu}) \, - \, \bs{w} \Big]_a   }
\cdot \det\bigg[  \underbrace{ \Dp{\nu_a} \wh{\Psi}_{T}[f](\bs{\nu}) }_{a=1,\dots, s-1} \; , \;  \Dp{\nu_s} \wh{\Psi}_{T}[f](\bs{\nu}) \, -\, \Dp{\nu_s} \wh{\Psi}_{T}[g](\bs{\nu})
\; , \;  \underbrace{ \Dp{\nu_a} \wh{\Psi}_{T}[g](\bs{\nu}) }_{a=s+1,\dots, \mf{h} } \bigg] \;.
\enq
Since
\beq
     \Bigl[ \Dp{\nu_s}
     \wh{\Psi}_{T}[f](\bs{\nu}) \, -\, \Dp{\nu_s} \wh{\Psi}_{T}[g](\bs{\nu}) \Bigr]_{a}
     \, = \,  \bigl[f^{\prime}(\nu_s) - g^{\prime}(\nu_s)\bigr]\de_{sa} \;,
\enq
Hadamard bounds for determinant along with
\eqref{ecriture lower bound pour integrande Rouche multidimensionnel} lead to
\begin{equation}
\big| \De^{(1)}[f,g]  \big| \; \leq  \; \mf{h} \cdot \Big( \f{2}{\varrho} \Big)^{\mf{h}} \cdot \underset{s=1,\dots, \mf{h} }{\e{max}}
\bigg\{ \pl{a=1}{s-1} \norm{ \Dp{\nu_a} \wh{\Psi}_{T}[f](\bs{\nu})  } \cdot \big| f^{\prime}(\nu_s)-g^{\prime}(\nu_s) \big|
\cdot  \pl{a=s+1}{\mf{h} } \norm{ \Dp{\nu_a} \wh{\Psi}_{T}[g](\bs{\nu})  }  \bigg\}  \;.
\end{equation}
Then, since for any $h \in \mc{E}_{\mc{M}}$ one has the upper bound
$\big| \big[ \wh{\Psi}_{T}[f](\bs{\nu}) \big]_{a} \big| \leq C$ uniformly in $T,\tf{1}{(NT)}$
small enough on $\mc{D}_{3\varrho}$, which contains $\ov{\mc{D}}_{2\varrho}$ properly,
one invokes that it holds that
\beq
\norm{ \Dp{\nu_a} \wh{\Psi}_{T}[h](\bs{\nu})  }  \; \leq \; C^{\prime} \quad \e{and}  \quad
\norm{h^{\prime}}_{ L^{\infty}\big(   \cup_{\a \in \{L,R\}}^{} \veps_{\a}^{-1}( \ov{\op{D}}_{0,2\varrho} )   \big) }  \; \leq \; C^{\prime\prime}
\norm{ h }_{ L^{\infty}\big(   \cup_{\a \in \{L,R\}}^{} \veps_{\a}^{-1}( \ov{\op{D}}_{0,3\varrho} )   \big) }  \;.
\enq
All-in-all this entails that for some $C>0$
\beq
\big| \De^{(1)}[f,g]  \big| \; \leq  \; C \cdot \norm{f-g}_{ L^{\infty}\big(   \cup_{\a \in \{L,R\}}^{} \veps_{\a}^{-1}( \ov{\op{D}}_{0,3\varrho} )   \big) }  \;.
\enq
After inserting $\bs{w}=\Big( \bs{h}^{(L)}_{  \varkappa^{(L)}_0 } \, , \,  
\bs{p}^{(L)}_{  \mf{y}^{(L)}_0 } \, , \,
\bs{h}^{(R)}_{ \varkappa^{(R)}_0 } \, , \,   \bs{p}^{(R)}_{ \mf{y}^{(R)}_0 }  \Big)^{\op{t}}$,
\textit{c.f.}~\eqref{ecriture equation sur solvabilite conditions auxiliaires sur racines},
in the integral representation
\eqref{ecriture representation integrale pour difference hat psi moins 1}
for $\wh{\Psi}_{T}^{\,-1}[f]-\wh{\Psi}_{T}^{\,-1}[g]$, the bound
\eqref{ecriture estimees sur la variation des parametres} follows owing to the compactness of
$\Ga_{2\varrho}$.

It thus remains to establish the property relative to the holomorphic dependence on the
parameters building up $\op{Y}$ and $\op{X}$. The holomorphic dependence is immediate on the
level of the integral representation
\eqref{ecriture representation integrale pour hat psi moins 1}
for $\wh{\Psi}_{T}^{\,-1}[f]$. To close, the estimates
\eqref{ecriture estimees en T sur les trous et particules proches de pm q}
follow from straightforward bounds after differentating under the integral
\eqref{ecriture representation integrale pour hat psi moins 1}
with  $\bs{w}=\Big( \bs{h}^{(L)}_{  \varkappa^{(L)}_0 } \, , \,
\bs{p}^{(L)}_{  \mf{y}^{(L)}_0 } \, , \,
\bs{h}^{(R)}_{ \varkappa^{(R)}_0 } \, , \,   \bs{p}^{(R)}_{ \mf{y}^{(R)}_0 }  \Big)^{\op{t}}$
as introduced in \eqref{ecriture equation sur solvabilite conditions auxiliaires sur racines}.
\qed

As already stated, we focus below on sets $\wh{\mathbb{X}}$, resp. $\mathbb{X}$,
\textit{c.f.}~\eqref{definition hat Y ensemble} defined in terms of the parameters
$\wh{\op{x}}_{0;\ell}^{\,(\a)}$ and $\wh{\op{y}}_{0;\ell}^{\,(\a)}$, resp.\
$\op{x}_{0;\ell}^{\,(\a)}$ and $\op{y}_{0;\ell}^{\,(\a)}$, solving the local
quantisation conditions at $\pm q$, \textit{c.f.}\
\eqref{definition ensemble trous et particules locales en pm q}. In order to provide
enough tools for the analysis to come, we need to obtain more precise information about
the local inverses of $f(\la\,|\, \mathbb{X} )$ and $\wh{f} (\la\,|\, \wh{\mathbb{X}} )$ close
to $\pm q$. Prior to stating the result, we recall that $\mc{V}_{\a}$, as appearing below,
was introduced at the beginning of Section~\ref{Section espace fnel pour la NLIE et ses ptes}.

\begin{prop}
\label{Proposition continuite inverse f et sa comparaison a veps inverse}
For a given $C_{\mc{M}}>0$ there exists $T_0>0$, $\eta > 0$ and $\rho > 0$ small enough,
such that $C_{\mc{M}}T_0<1$, and such that
for all $T<T_0$, $\a\in \{L,R\}$ and $\eta > \tf{1}{(NT)}$ there exist
\begin{itemize}
%
\item an open neighbourhood $\mc{U}_{\a}$  of $\ups_{\a} q $, satisfying
$\mc{U}_{\a}  \subset \mc{V}_{\a}$ and $\e{d}(\mc{U}_{\a} ,  \Dp{}\mc{V}_{\a})>c$
for some $T$-independent constant $c>0$;

\item an open neighbourhood $\wh{\mc{U}}_{\a}$  of $\ups_{\a} q $, satisfying
$\wh{\mc{U}}_{\a}  \subset \mc{V}_{\a}$ and $\e{d}(\wh{\mc{U}}_{\a} ,  \Dp{}\mc{V}_{\a})>c$
for some $T$-independent constant $c>0$;
\end{itemize}
such that
\begin{itemize}
\item the map $\la \mapsto f\bigl(\la \,|\, \mathbb{X}\bigr)$ is a biholomorphism
$\mc{U}_{\a} \tend  \op{D}_{0,\rho}$;

\item the map $\la \mapsto \wh{f}\bigl(\la \,|\, \wh{\mathbb{X}}\bigr)$ is a biholomorphism
$\wh{\mc{U}}_{\a} \tend  \op{D}_{0,\rho}$.
\end{itemize}
Furthermore, the neighbourhood $\mc{U}_{\a}$, resp.\ $\wh{\mc{U}}_{\a}$,
contains a $T$-independent, resp.\ a $T$- and $NT$-independent, neighbourhood
of $\ups_{\a} q$.

For all $f, g \in \mc{E}_{\mc{M}}$ one has the norm estimates
\begin{align}
     \norm{ f^{-1}_{\a}(*\,|\,\mathbb{X})
        \, - \, \veps_{\a}^{-1} }_{ L^{\infty}(  \op{D}_{0,\, \rho})  }
	& \leq \f{ C T }{ \rho^2 } \;,
\label{ecriture bornage f inverse contre veps inverse} \\
     \norm{ f^{-1}_{\a}(*\,|\,\mathbb{X}[f])
        \, - \, g^{-1}_{\a}(*\,|\,\mathbb{X}[g]) }_{ L^{\infty}(  \op{D}_{0,\, \rho})  }
	& \leq
     \f{ C^{\prime} }{ \rho^2 }\cdot  \norm{f-g}_{L^{\infty}\big( \cup_{\a\in \{L,R\}} \veps_{\a}^{-1}(\op{D}_{0,2\rho} ) \big) }
\label{ecriture continuite inverse f}
\end{align}
for some constants $C, C^{\prime} >0$. Likewise, it holds that
\begin{align}
     \norm{ \wh{f}^{-1}_{\a}(*\,|\, \wh{\mathbb{X}})
        \, - \, \veps_{\a}^{-1} }_{ L^{\infty}(  \op{D}_{0,\, \rho})  }
	& \leq \f{ C  }{ \rho^2 } \Big( T \, + \, \f{1}{NT}\Big) \;,
\label{ecriture bornage f inverse contre veps inverse hat} \\
     \norm{ \wh{f}^{-1}_{\a}(*\,|\,\wh{\mathbb{X}}[f])
        \, - \, \wh{g}^{\, -1}_{\a}(*\,|\, \wh{\mathbb{X}}[f] ) }_{ L^{\infty}(  \op{D}_{0, \, \rho}) }
	& \leq
     \f{ C^{\prime} }{ \rho^2 }\cdot  \norm{f-g}_{L^{\infty}\big( \cup_{\a\in \{L,R\}} \veps_{\a}^{-1}(\op{D}_{0,2\rho} ) \big) }  \; .
\label{ecriture continuite inverse f hat}
\end{align}
All constants appearing in the bounds do not depend on $C_{\mc{M}}$ as long as $C_{\mc{M}} T<1$.
\end{prop}

We stress that in the statement of the bounds \eqref{ecriture continuite inverse f} and \eqref{ecriture continuite inverse f hat} we have explicitly
insisted on the dependence of $\mathbb{X}$ and $\wh{\mathbb{X}}$ on the considered function.

Moreover, we remind that in each of the above norm bounds, one may trivially upper bound as
\beq
 \norm{f-g}_{L^{\infty}\big( \cup_{\a\in \{L,R\}} \veps_{\a}^{-1}(\op{D}_{0,2\rho} ) \big) } \, \leq \, \  \norm{f-g}_{L^{\infty}(\mc{M})}
\enq
if need be.

\Proof
We provide the proof for $f(*\,|\, \mathbb{X})$ and leave the details
of the case $\wh{f}(*\,|\, \wh{\mathbb{X}})$ to the reader.

%
%
%
%
%
%
%
%
%
%
%

Let $\mc{W}_{\a}=\veps^{-1}_{\a}\big( \op{D}_{0,2\rho} \big)$. Since $\veps^{-1}_{\a}$
is a biholomorphism, mapping $\op{D}_{0,2\rho}$ to an open neighbourhood of $\ups_\alpha q$,
we may choose $\rho$ small enough such that $\mc{W}_{\a} \subset \mc{V}_\a$ and
$\e{d} (\mc{W}_{\a}, \Dp{} \mc{V}_\a) > c$ for some $c > 0$.

Then $\la \mapsto f(\la \,|\, \mathbb{X})$ is holomorphic in $\mc{W}_{\a}$ and,
since $\veps_{\e{c}}=\veps$ in this domain, we have, for all
$\la \in \Dp{} \mc{W}_{\a}$ and any $z \in \op{D}_{0,\rho}$, that for $T$ small enough
\beq
     \big| f(\la\,|\, \mathbb{X}) \, - \, \veps (\la) \big|
        \; \leq \; T \norm{u_1(*\,|\, \mathbb{X}) }_{ L^{\infty}(\mc{W}_\a)}
     \, + \, \norm{f }_{ L^{\infty}(\mc{W}_\a)}  \, \leq \, C_u T \, + \, C_{\mc{M}}T^2
     \, < \,  \rho  \, < \, |\veps (\la)- z | \;.
\enq
Hence, by Rouch\'e's theorem, the holomorphic maps $\la \mapsto f(\la\,|\, \mathbb{X})-z$
and $\la \mapsto \veps(\la)-z$ have the same number of zeroes in $\mc{W}_{\a}$.
Proposition~\ref{Proposition double recouvrement de veps} ensures that $\la \mapsto \veps(\la)-z$ has a single zero in $\mc{W}_{\a}$,
hence so does $\la \mapsto f(\la\,|\, \mathbb{X})-z$. Thus, the function
$f\big(* \,|\, \mathbb{X}\big)$ has a local holomorphic inverse
$f^{-1}_\a \big(* \,|\, \mathbb{X}\big): \op{D}_{0,\rho} \rightarrow
f^{-1}_\a \big(\op{D}_{0,\rho} \,|\, \mathbb{X}\big) = \mc{U}_\a \subset \mc{W}_\a$,
and, by construction, $\mc{U}_\a \subset \mc{V}_\a$,
$\e{d} (\mc{U}_{\a}, \Dp{} \mc{V}_\a) > c$. Moreover, $| f(\ups_\a q \,|\, \mathbb{X})| \leq
C_u T + C_{\mc{M}}T^2 < \rho$ for $T$ small enough implies that $\ups_\a q \in \mc{U}_\a$.

Using the fact that for any $z \in \op{D}_{0,\rho}$,  $f(\la\,|\, \mathbb{X}) - z$
has a single zero inside $\mc{W}_\a$  we conclude that, for any $z \in \op{D}_{0,\rho}$,
\beq
     f^{-1}_{\a} (z \,|\, \mathbb{X}) \, = \Int{ \Dp{}\mc{W}_\a}{} \f{ \dd s }{ 2\i\pi } \:
        \f{ s f^{\prime}(s \,|\, \mathbb{X})  }{ f(s \,|\, \mathbb{X}) \, - \, z }
\enq
and similarly for $\veps_\a^{-1}$. Thus,
\beq
     f^{-1}_{\a} (z \,|\, \mathbb{X})\, - \, \veps_{\a}^{-1}(z) \,
        =  \Int{ \Dp{}\mc{W}_\a}{}  \f{ \dd s }{ 2\i\pi } \:
   s \f{ f^{\prime}(s \,|\, \mathbb{X})\, (\veps(s)-z) \, - \, \veps^{\prime}(s) \,  \big( f(s \,|\, \mathbb{X}) \, - \, z \big)  }{ \big[\veps(s)-z \big] \cdot \big[ f(s \,|\, \mathbb{X}) \, - \, z \big] } \;.
\enq
Now, for the range of parameters involved, one has that
\bem
\big| f^{\prime}(s \,|\, \mathbb{X})\, (\veps(s)-z) \, - \, \veps^{\prime}(s) \,  \big( f(s \,|\, \mathbb{X}) \, - \, z \big) \big| \, = \,
\big|  \Big( T u_1^{\prime}(s \,|\, \mathbb{X}) + f^{\prime}(s) \Big) \, (\veps(s)-z) \, - \, \veps^{\prime}(s) \Big( Tu_1(s \,|\, \mathbb{X}) + f(s) \Big)  \big|  \\
\, \leq \,  \big| \veps(s)-z \big| \cdot \Big(   T \norm{ u_1^{\prime}(* \,|\, \mathbb{X}) }_{L^{\infty}(\mc{W}_\a)}
+ \norm{f^{\prime}}_{L^{\infty}(\mc{W}_\a)}  \Big)  \\
\; + \; |\veps^{\prime}(s)| \cdot \Big(   T \norm{ u_1(* \,|\, \mathbb{X}) }_{L^{\infty}(\mc{W}_\a) }
+ \norm{f}_{L^{\infty}(\mc{W}_\a)}  \Big)
\, \leq \, T C \, C_1 \, (C_u+C_{\mc{M}}T) \;.
\end{multline}
Above, we used that owing to $\e{d}\big(\mc{W}_{\a}, \Dp{}\mc{V}_{\a} \big)>0$
uniformly in $T$, there exist constants $C_k$ such that for any holomorphic
function $g$ on an open neighbourhood of $\mc{V}_{\a}$
\beq
\norm{ g^{(k)} }_{L^{\infty}(\mc{W}_\a)} \;  \leq \; C_k \cdot \norm{ g }_{L^{\infty}(\mc{V}_{\a})} \;.
\enq
Finally, one also has $|\veps(s)-z|>\rho$, leading to
\beq
\big|  f(s \,|\, \mathbb{X}) \, - \, z  \big|\; \geq \;  \big| |\veps(s)-z|  \, - \, T  \norm{ u_1(* \,|\, \mathbb{X}) }_{L^{\infty}(\mc{V}_{\a}) }   - \norm{ f }_{L^{\infty}(\mc{V}_{\a}) } \big|
\; \geq \; \rho - C_u T - \mc{C}_{\mc{M}}T^2 \;.
\enq
Putting these bounds together leads to \eqref{ecriture bornage f inverse contre veps inverse},
since $\Dp{}\mc{W}_\a $ is a compact subset of $\Cx$.

Finally, one has
\beq
f^{-1}_{\a} \big(z \,|\, \mathbb{X}[f] \big)\, - \, g^{-1}_{\a} \big(z \,|\, \mathbb{X}[g] \big)
   \, =  \Int{ \Dp{}\mc{W}_\a}{} \f{ \dd s }{ 2\i\pi } \:
s \f{ f^{\prime}(s \,|\, \mathbb{X}[f] )\, \big( g(s \,|\, \mathbb{X}[g] )-z\big) \, - \, g^{\prime}(s \,|\, \mathbb{X}[g] ) \,  \big( f(s \,|\, \mathbb{X}[f] ) \, - \, z \big)  }
{  \big[ g(s \,|\, \mathbb{X}[g] ) \, - \, z \big]  \cdot \big[ f(s \,|\, \mathbb{X}[f] ) \, - \, z \big] } \;.
\enq
Here, we have made explicit the dependence of the parameters building up $\mathbb{X}$
on the function of interest.

Similarly to the above, the denominator is bounded from below in modulus by
$\big( \rho - C_u T - \mc{C}_{\mc{M}}T^2\big)^2 $, while, for $s \in \Dp{}\mc{W}_\a$,
$z \in \op{D}_{0, \rho}$,
\bem
\big| f^{\prime}(s \,|\, \mathbb{X}[f] )\, \big( g(s \,|\, \mathbb{X}[g] )-z\big) \, - \, g^{\prime}(s \,|\, \mathbb{X}[g] ) \,  \big( f(s \,|\, \mathbb{X}[f] ) \, - \, z \big)  \big|  \\
\, = \, \Big| \Big[ f^{\prime}(s)-g^{\prime}(s)+ T u_1^{\prime}(s \,|\, \mathbb{X}[f] ) - T u_1^{\prime}(s \,|\, \mathbb{X}[g] ) \Big]
\cdot \Big[ g(s \,|\, \mathbb{X}[g])-z\Big]  \\
\, - \, g^{\prime}(s \,|\, \mathbb{X}[g]) \cdot \Big[ f(s) \, - \, g(s)   \, + \,  T u_1 (s \,|\, \mathbb{X}[f] ) - T u_1 (s \,|\, \mathbb{X}[g] ) \Big] \Big| \\
\; \leq \; \Big\{ \norm{ f^{\prime}-g^{\prime} }_{ L^{\infty}(\mc{W}_{\a})  } \, + \, C T   \norm{f-g}_{L^{\infty}\big( \cup_{\a\in \{L,R\}} \veps_{\a}^{-1}(\op{D}_{0,2\rho} ) \big) } \Big\}
\cdot \Big( \norm{\veps}_{L^{\infty} (\mc{W}_{\a}) } + \rho \, + \, T \norm{ u_1(* \,|\, \mathbb{X}[g]) }_{L^{\infty}(\mc{W}_{\a}) }
\, + \,   \norm{ g }_{L^{\infty}(\mc{W}_{\a}) }  \Big)  \\
\; + \; \Big\{ \norm{ f -g }_{ L^{\infty}(\mc{W}_{\a})  } \, + \, C T \norm{f-g}_{L^{\infty}\big( \cup_{\a\in \{L,R\}} \veps_{\a}^{-1}(\op{D}_{0,2\rho} ) \big) } \Big\}
\cdot \Big( \norm{\veps^{\prime}}_{L^{\infty} (\mc{W}_{\a}) } \, + \, T \norm{ u_1^{\prime}(* \,|\, \mathbb{X}[g]) }_{L^{\infty}(\mc{W}_{\a}) }
\, + \,   \norm{ g^{\prime} }_{L^{\infty}(\mc{W}_{\a}) }  \Big) \\
\; \leq \;
2 C_1  \norm{f-g}_{L^{\infty}\big( \cup_{\a\in \{L,R\}} \veps_{\a}^{-1}(\op{D}_{0,2\rho} ) \big) } \cdot \Big( C + T C_u + T^2 C_{\mc{M}} \Big)  \;.
\end{multline}
Here we have remembered that $\veps^{-1}_\a (\op{D}_{0, 2\rho}) = \mc{W}_\a$.
In the intermediate bounds, we have used that, owing to the estimate
\eqref{ecriture estimees sur la variation des parametres} obtained in Proposition
\ref{Proposition existence et continuite parameters particule trou partiels} it holds
that, upon reducing $\rho$ if need be so as to adjust with $\varrho$ appearing there,
\beq
 \norm{ u_1^{(k)}(* \,|\, \mathbb{X}[f] ) - u_1^{(k)}(* \,|\, \mathbb{X}[g] )  }_{L^{\infty}(\mc{W}_{\a}) } \; \leq \;
C   \norm{f-g}_{L^{\infty}\big( \cup_{\a\in \{L,R\}} \veps_{\a}^{-1}(\op{D}_{0,2\rho} ) \big) }
\enq
for $k=0$ or $1$.

This entails \eqref{ecriture continuite inverse f}. \qed

\subsection{The integration contour}

\vspace{3mm}

We now associate a contour $\msc{C}_{\mathbb{X}}[f]$, resp.\
$\wh{\msc{C}}_{\mathbb{X}}[f]$ to $f(\la\,|\, \mathbb{X})$,
resp.\ $\wh{f}(\la\,|\, \wh{\mathbb{X}})$. We will establish later on, in
Proposition~\ref{Proposition domaine local holomorphie de F}, that
the pairs $\Big( f(\la\,|\, \mathbb{X}), \msc{C}_{\mathbb{X}}[f] \Big)$, resp.\
$\Big( \wh{f}(\la\,|\, \wh{\mathbb{X}}), \wh{\msc{C}}_{\mathbb{X}}[f] \Big)$,
enjoy Properties~\ref{Propriete pour deformer les contours}.

It has been established in Proposition \ref{Proposition continuite inverse f et sa comparaison a veps inverse}
 that   there exist $T$- and $NT$-independent $\rho>0$ and open neighbourhoods $ \mc{U}_{\a} , \wh{\mc{U}}_{\a}\subset \mc{V}_{\a}$ of $\ups_\alpha q$,
$\alpha \in \{R, L\}$, such that $f(*\,|\, \mathbb{X}): \mc{U}_{\a} \tend
\op{D}_{0,\, \rho}$, resp.\ $\wh{f}(*\,|\, \wh{\mathbb{X}}): \wh{\mc{U}}_{\a} \tend
\op{D}_{0, \, \rho}$,  is a biholomorphism. The inverse on $\mc{U}_{\a}$, resp.\
$\wh{\mc{U}}_{\a}$,  will be denoted by $f^{-1}_{\a}(*\,|\, \mathbb{X})$,
resp.\ $\wh{f}^{-1}_{\a}(*\,|\, \wh{\mathbb{X}})$.

In particular, this ensures that there exists a unique $q_{\mathbb{X}}^{(\ups_{\a})}[f]
\in \mc{U}_{\a}$, resp.\ $\wh{q}_{\mathbb{X}}^{\,(\ups_{\a})}[f] \in \wh{\mc{U}}_{\a}$,
such that 
\beq
f \Big( q_{\mathbb{X}}^{(\ups_\alpha)}[f] \,|\, \mathbb{X} \Big) \, = \, 0
\qquad \e{and} \qquad \wh{f} \, \Big( \, \wh{q}_{\mathbb{X}}^{\, (\ups_\alpha)}[f] \,|\, \wh{\mathbb{X}} \Big) \, = \, 0\;.
\label{definition q pm de Y et f}
\enq
The invertibility of $f(*\,|\, \mathbb{X})$, resp.\ $\wh{f}(*\,|\, \wh{\mathbb{X}})$, on
a domain whose image contains a disc of fixed radius allows one to unambiguously
define, for $T$ low enough, two auxiliary complex numbers as
\beq
\mf{z}_{\a}^{(\pm)}[f] \, = \,  \veps\Big( f^{-1}_{\a}\big( \pm \de_T    \,|\, \mathbb{X} \big) \,   \Big) \;, \quad \e{resp}. \quad
\wh{\mf{z}}_{\a}^{\, (\pm)}[f] \, = \,  \veps\Big( \wh{f}^{\,-1}_{\a}\big( \pm \de_T   \,|\, \wh{\mathbb{X}} \big) \,   \Big) \,.
\label{definition tau pm de f et alpha}
\enq
Then, we introduce the points $y_{\a}^{(\sg)}[f]$, resp.\ $\wh{y}_{\a}^{\,(\sg)}[f]$,
as the intersections of the curves
$\veps_{\a}^{-1}\Big(\mf{z}_{\a}^{(\sigma)}[f]+ \i \R^{-\ups_{\a}} \Big)$, resp.\
$\veps_{\a}^{-1}\Big(\, \wh{\mf{z}}_{\a}^{\, (\sigma)}[f]+ \i \R^{-\ups_{\a}} \Big)$,
with the circle of radius $\mf{c}_{\e{d}} T$ centred at $-\i\tf{\zeta}{2}$:
\beq
     y_{\a}^{(\sg)}[f] \, = \,
        \veps_{\a}^{-1}\Big(\mf{z}_{\a}^{(\sg)}[f]+ \i \R^{-\ups_{\a}}  \Big)
	\cap \Dp{}\op{D}_{-\i\f{\zeta}{2}, \mf{c}_{\e{d}}} \; , \quad \e{resp}. \quad 
     \wh{y}_{\a}^{\,(\sg)}[f] \, = \,
        \veps_{\a}^{-1}\Big(\, \wh{\mf{z}}_{\a}^{\, (\sg)}[f]+ \i \R^{-\ups_{\a}}  \Big)
	\cap \Dp{}\op{D}_{-\i\f{\zeta}{2}, \mf{c}_{\e{d}}} \;,  
\label{definition probleme intersection Trotter fini et infini}
\enq
where $\sg \in \{\pm \}$. 
We further define  $\mf{t}_{\a}^{(\sg)}[f]$, resp.  $\wh{\mf{t}}_{\a}^{\,(\sg)}[f]$, through 
\beq
\mf{z}_{\a}^{(\sg)}[f] \, + \, \i \cdot  \mf{t}_{\a}^{(\sg)}[f] \, = \, \veps \Big( y_{\a}^{(\sg)}[f]  \Big) \; , \quad \e{resp}. \quad 
\wh{\mf{z}}_{\a}^{\, (\sg)}[f] \, + \, \i \cdot \wh{\mf{t}}_{\a}^{\,(\sg)}[f]  \, = \, \veps \Big( \, \wh{y}_{\a}^{\,(\sg)}[f]   \Big) \; . 
\label{definition des points limites des arcs ga sg et hat ga sg}
\enq
We denote by 
\begin{itemize}

\item $\ga^{(-)}[f]$ the anti-clockwise oriented arc of $\Dp{}\op{D}_{-\i \frac{\zeta}{2}, \mf{c}_{\e{d}} T} $ joining $y_{R}^{(-)}[f]$ to $y_{L}^{(-)}[f]$;

\item $\ga^{(+)}[f]$ the anti-clockwise oriented arc of $\Dp{}\op{D}_{-\i \frac{\zeta}{2}, \mf{c}_{\e{d}} T} $ joining $y_{L}^{(+)}[f]$ to $y_{R}^{(+)}[f]$;  

\end{itemize}
resp.\
\begin{itemize}

\item $\wh{\ga}^{\, (-)}[f]$ the anti-clockwise oriented arc of $\Dp{}\op{D}_{-\i \frac{\zeta}{2}, \mf{c}_{\e{d}} T} $ joining $\wh{y}_{R}^{\, (-)}[f]$ to $\wh{y}_{L}^{\,(-)}[f]$;

\item $\wh{\ga}^{\, (+)}[f]$ the anti-clockwise oriented arc of $\Dp{}\op{D}_{-\i \frac{\zeta}{2}, \mf{c}_{\e{d}} T} $ joining $\wh{y}_{L}^{\, (+)}[f]$ to $\wh{y}_{R}^{\, (+)}[f]$. 
\end{itemize}
Finally, we introduce the oriented segments
\beq
\mc{I}_{R}^{(+)}[f] \; = \; \Big[ \mf{z}_{R}^{(+)}[f] \, + \, \i \cdot  \mf{t}_{R}^{(+)}[f] \, ; \,  \mf{z}_{R}^{(+)}[f] \Big] \; , \quad 
\mc{I}_{R}^{(-)}[f] \; = \; \Big[ \mf{z}_{R}^{(-)}[f] \, ; \,  \mf{z}_{R}^{(-)}[f] \, + \, \i \cdot  \mf{t}_{R}^{(-)}[f]   \Big] 
\label{definition contour I R pm}
\enq
and 
\beq
\mc{I}_{L}^{(+)}[f] \; = \;  \Big[ \mf{z}_{L}^{(+)}[f] \, ; \,  \mf{z}_{L}^{(+)}[f] \, + \, \i \cdot  \mf{t}_{L}^{(+)}[f]   \Big]   \; , \quad 
\mc{I}_{L}^{(-)}[f] \; = \; \Big[ \mf{z}_{L}^{(-)}[f] \, + \, \i \cdot  \mf{t}_{L}^{(-)}[f] \, ; \,  \mf{z}_{L}^{(-)}[f] \Big] \;. 
\label{definition contour I L pm}
\enq
The oriented segments $\wh{\mc{I}}_{\a}^{\,(\sg)}[f]$ are defined analogously upon
adding the hats to the endpoints. 

All of the above allows one to introduce the below curves,
\begin{align}
     \msc{C}_{\mathbb{X};L}[f] & =
        \veps_L^{-1}\Big( \mc{I}_{L}^{(+)}[f] \Big)
	\cup \veps_L^{-1}\Big( \mc{I}_{L}^{(-)}[f] \Big) 
        \cup f^{-1}_{L}\Big( I_{\de_T}   \,|\, \mathbb{X} \Big) \;, \\[1ex]
     \msc{C}_{\mathbb{X};R}[f] & =
        \veps_R^{-1}\Big( \mc{I}_{R}^{(+)}[f] \Big)
	\cup \veps_R^{-1}\Big( \mc{I}_{R}^{(-)}[f] \Big) 
        \cup f^{-1}_{R}\Big( - I_{\de_T}   \,|\, \mathbb{X} \Big) \;.
\end{align}
Here and in the following $\pm I_{\de_T} = [\mp \de_T; \pm \de_T]$. Then, one sets 
\beq
     \msc{C}_{\mathbb{X}}[f]
        \, = \, \msc{C}_{\mathbb{X};L}[f] \cup \ga^{(-)}[f]
	\cup \ga^{(+)}[f] \cup \msc{C}_{\mathbb{X};R}[f] \;.
\label{definition contour C de Y}
\enq
Similarly, one defines 
\begin{align}
     \wh{\msc{C}}_{\mathbb{X};L}[f] & =
        \veps_L^{-1}\Big( \, \wh{\mc{I}}_{L}^{\, (+)}[f] \Big)
	   \cup \veps_L^{-1}\Big( \, \wh{\mc{I}}_{L}^{\, (-)}[f] \Big) 
        \cup \wh{f}^{\, -1}_{L}\Big( I_{\de_T}
	             \,|\, \wh{\mathbb{X}} \Big) \;, \\[1ex]
     \wh{\msc{C}}_{\mathbb{X};R}[f] & =
        \veps_R^{-1}\Big( \, \wh{\mc{I}}_{R}^{\,(+)}[f] \Big)
	   \cup \veps_R^{-1}\Big( \, \wh{\mc{I}}_{R}^{\,(-)}[f] \Big) 
        \cup \wh{f}^{\, -1}_{R}\Big( - I_{\de_T}   \,|\, \wh{\mathbb{X}} \Big) \;.
\end{align}
The above allows us to define  the finite Trotter number contour
\beq
\wh{\msc{C}}_{\mathbb{X}}[f] \, = \, \wh{\msc{C}}_{\mathbb{X};L}[f] \cup \wh{\ga}^{\, (-)}[f] \cup  \wh{\ga}^{\, (+)}[f]  \cup \wh{\msc{C}}_{\mathbb{X};R}[f]  \;.
 \label{definition contour C de Y Trotter fini}
\enq
%
%
%


Note that Proposition~\ref{Proposition continuite inverse f et sa comparaison a veps inverse}
readily allows one to obtain the estimates
\begin{equation}
\begin{split}
     \mf{z}_{\a}^{(\pm)}[f] & =
        \veps\Big( \veps^{-1}_{\a}\big( \pm \de_T    \big)
	\, + \, \e{O}(T)   \Big) 
     = \pm \de_T   \, + \, \e{O}(T) \;,
       \\[1ex]
     \wh{\mf{z}}_{\a}^{\, (\pm)}[f] & =
        \veps\Big( \veps^{-1}_{\a}\big( \pm \de_T    \big)
	           \, + \, \e{O}\big(T + \tfrac{1}{NT} \big)    \Big) 
     = \pm \de_T  \, + \, \e{O}\big(T + \tfrac{1}{NT} \big)
       \;.
\end{split}
\label{ecriture DA basse tempe de tau et eta f}
\end{equation}

\subsection{Local behaviour of the maps
$f(\la\,|\, \mathbb{X} )$ and $\wh{f}(\la\,|\, \wh{\mathbb{X}} )$}

We now provide an estimate for the magnitudes of $\ex{\mp \f{1}{T} f(\la \,|\, \mathbb{X}) }$
and $\ex{\mp \f{1}{T} \wh{f}(\la \,|\, \wh{\mathbb{X}} \, ) }$ in certain domains of $\Cx$.
In particular, we will establish that
\beq
\la \mapsto 1+\ex{-\f{1}{T} f(\la\,|\, \mathbb{X} ) } \; , \quad \e{resp.} \quad   \la \mapsto 1+\ex{-\f{1}{T} \wh{f}(\la\,|\, \wh{\mathbb{X}} \, ) }  \;,
\enq
is holomorphic and non-zero in the region located in between $\msc{C}_{\e{ref}}$ and
$\msc{C}_{\mathbb{X}}[f]$, resp.\ $\msc{C}_{\e{ref}}$ and $\wh{\msc{C}}_{\mathbb{X}}[f]$.

\begin{prop}
\label{Proposition borne sup sur exponentielle en tempe de la fct f lambda mathbbY} 
Recall the definition \eqref{definition du parametre delta T}  of $\de_{T}$ and
introduce the domains $\mc{G}_{\a}= \mc{G}_{\a}^{(-)} \cup \mc{G}_{\a}^{(+)}$,
$\a \in \{L,R\}$, with 
\beq
\begin{split}
\mc{G}_{L}^{(\pm)}  &   =  \;  \Big\{  x+\i y \, : \, |x \, \mp \, \de_T| \, < \,  C_{\mf{r}} T \, , \, y \in \intoo{ - \eps }{ \tfrac{1}{T} C_{g} }  \Big\}  \; ,   \\[1ex]
\mc{G}_{R}^{(\pm)}   &  =  \;  \Big\{  x+\i y \, : \, |x \, \mp \, \de_T| \, < \,  C_{\mf{r}} T \, , \, y \in \intoo{- \tfrac{1}{T} C_{g} }{ \eps }  \Big\} \;,
\end{split}
\label{ecriture domain bornage exponentielle de proche de C ref}
\enq
where $C_\mf{r} > 0$, $\eps>0$ small enough and $C_{g}>0$ and large enough. Then there exist
$\eta>0$ and $T_0$ small enough, such that 
\beq
     \begin{cases}
        y_{\a}^{(\sg)} \, , \; y_{\a}^{(\sg)}[f]
        \in \veps_\a^{-1} \bigl(\mc{G}_{\a}^{(\sg)}\bigr) &
	for \; any \;  T_0>T>0 \;, \\[1ex]
        \wh{y}_{\a}^{\, (\sg)}[f]  \in \veps_\a^{-1} \bigl(\mc{G}_{\a}^{(\sg)}\bigr)
        & for \; any \;  T_0>T>0 \; and \; \eta \, > \,  \f{1}{NT} \, > \, 0 \;.
     \end{cases}
\label{propriete appartenance a cal G alpha pm}
\enq
Let 
\beq
     C_{\veps} \, = \, 2 \cdot
        \e{sup} \Big\{ \big| \big(\veps_{\a}^{-1} \big)^{\prime}(z) \big|
        \; : \;  z \in \mc{G}_{\a}^{(\sg)}, \a\in \{L,R\} \, , \sg \in \{ \pm \} \Big\}
	\in \R^+ \;. 
\enq
Let $T < T_0$ and $C_{\mf{r}} C_{\veps} < \mf{c}_{\e{ref}}$, $\mf{c}_{\e{ref}}$ being
the constant arising in point $\mathrm{b)}$ of Hypothesis~\ref{Hypotheses solubilite NLIE}.
Then, given $\la \in \veps_{\a}^{-1}\big( \mc{G}_{\a}^{(\pm)} \big)$, it holds that
\beq
     \f{ T^{M+|\op{Y}|} }{ C }
        \bigg( \f{ \mf{c}_{\e{ref}} \, - \,
	   C_{\mf{r}} C_{\veps} }{ C^{\prime} } \bigg)^{|\op{Y}|}
	   \, \ex{ -(C_{\mf{r}}+C_{\mc{M}} T) } \, \leq \, 
     \Big| \ex{ \mp  \f{1}{T} f(\la\,|\, \mathbb{X}) } \Big| \, \leq \, C \, T^{M-|\op{Y}|}
        \bigg( \f{ C^{\prime} }{ \mf{c}_{\e{ref}} \, - \,
	   C_{\mf{r}} C_{\veps} } \bigg)^{|\op{Y}|}  
           \, \ex{ (C_{\mf{r}}+C_{\mc{M}} T) } \;,
\label{borne sup et inf sur partie laterale contour pour exp de f de Y mathbb}
\enq
where $M$ appears in the definition of $\de_T$, \textit{c.f.} \eqref{definition du parametre delta T}, and $C, C^{\prime}>0$
are constants such that $C'$ is independent of $\mathbb{X}$, while $C$ only depends
on $|\op{X}_{\e{tot}}|$, $|\op{Y}_{\e{tot}}|$  and $|\op{Y}_{\e{sg}}|$. Moreover, the sole dependence on the
function $f$ lies in the exponential terms. 

The same bounds hold with 
\beq
f(\la\,|\, \mathbb{X}) \hookrightarrow  \wh{f}(\la\,|\, \wh{\mathbb{X}})
\enq
provided that $\eta>\tf{1}{(NT^4)}$ and $T_0>T>0$. 
\end{prop}

\Proof 
It is direct to see, with the help of \eqref{ecriture DA basse tempe de tau et eta f}
and Proposition \ref{Proposition DA des angles vth et Lipschitz pour angles et t sg alpha}, eq. \eqref{ecriture DA mf t a basse T},
that by taking $C_{g}$ large enough and $\tf{1}{(NT^4)}$ small enough,
\eqref{propriete appartenance a cal G alpha pm} does hold. 

We now focus on the proof of
\eqref{borne sup et inf sur partie laterale contour pour exp de f de Y mathbb}.
To start with, it is useful to observe that 
\beq
     \phi_{\e{c}}(\la,\mu) \; = \;
        \phi_{\e{c};\e{reg}}(\la,\mu)
	+ \frac{1}{2\pi}\th(\la-\mu) \qquad \e{with} \qquad \phi_{\e{c};\e{reg}}(\la,\mu)
	\; = \; - \Int{\msc{C}_{\veps} }{} \dd \nu \: K(\la-\nu) \phi_{\e{c}}(\nu,\mu)  \;. 
\enq
The function $\phi_{\e{c};\e{reg}}$ is smooth on 
\beq
\Cx \setminus\Big\{  \msc{C}_{\veps} \, \pm  \, \i\zeta  + \i \pi \mathbb{Z} \Big\} \times \Cx \setminus\Big\{  \msc{C}_{\veps} \, + \,  \R^{-} \, \pm \,  \i\zeta  + \i \pi \mathbb{Z} \Big\}
\enq
and has jump discontinuities on the complement of this set and logarithmic singularities
on the endpoints of the discontinuity curves component-wise. 

Then, one introduces 
\beq
     u_{1;\e{reg}}(\la\,|\, \mathbb{X}) \, = \,
        - \i \pi   \mf{s}Z_{\e{c}}(\la)
	\, - \, 2 \i \pi \sul{ y \in \mathbb{X} }{} \phi_{\e{c};\e{reg}}(\la,y)
\label{definition u1 reg}
\enq
so that one obtains the factorisation 
\bem
\label{factor_exp_fovert}
     \ex{- \f{1}{T}  f(\la\,|\, \mathbb{X}) } \; = \; \ex{ -\f{1}{T} \veps_{\e{c}}(\la) }
     \pl{y \in \op{Y}_{\e{tot}} }{}
        \bigg\{\f{\sinh(\i\zeta + y - \la)}{\sinh(\i\zeta + \la -y)}\bigg\} \cdot
     \pl{y \in \op{Y}_{\e{sg}}}{}
        \bigg\{ \f{\sinh(\i\zeta + y - \la)} {\sinh(\i\zeta + \la - y )}\bigg\}  \\
\times  
        \pl{x \in \op{X}_{\e{tot}} }{}
	\bigg\{\f{\sinh(\i\zeta + \la - x)}{\sinh(\i\zeta + x - \la)}\bigg\}
        \cdot \ex{-\f{1}{T}f(\la) -  u_{1;\e{reg}}(\la\,|\, \mathbb{X})} \;.
\end{multline}
We remind the reader that $\op{Y}_{\e{tot}}$, $\op{X}_{\e{tot}}$ are as
defined in \eqref{definition hat Y ensemble}.

Given $\la \in \veps_{\a}^{-1}\big(\mc{G}_{\a})$, one then bounds each of the factors
on the right hand side. Owing to the continuity of $\veps_\a^{-1}$ we can choose for
every $\eps' > 0$ the control parameter $\eps > 0$, arising in the definition of
$\mc{G}_{\a}$, in such a way that $- \zeta/2 < \Im (\lambda) < \eps'$. For $z \in
\op{X}_{\e{tot}} \cup \op{Y}_{\e{sg}}$ we have $- \zeta/2 - \mf{c}_{\e{d}} T
< \Im (z) < \mf{c}_{\e{loc}}$ by definition, \textit{c.f.}\ Fig.~\ref{contour integration Cref}.
Hence, $|\Im(\la - z)| < \tf{\zeta}{2} + 2 \eps^{\prime}$, if $T$ and $\mf{c}_{\e{loc}}$
are small enough, and one concludes that, for some $C>0$,  
%
%
%
\beq
     \pl{x \in \op{X}_{\e{tot}} }{}
        \bigg|\f{ \sinh(\i\zeta + \la - x)}{\sinh(\i\zeta + x - \la)}\bigg|^{\pm 1}
	   \; \leq \: C^{|\op{X}_{\e{tot}}|} \;, \quad \quad
     \pl{y \in \op{Y}_{\e{sg}}}{}
        \bigg|\f{\sinh(\i\zeta + y - \la)}{\sinh(\i\zeta + \la - y )}\bigg|^{\pm 1}
	   \; \leq \: C^{|\op{Y}_{\e{sg}}|} \;,
\enq
this uniformly in $\la \in \veps_{\a}^{-1}\big(\mc{G}_{\a})$.

It remains to bound the contribution of the particle roots. This can
be done in two steps, depending on whether $\la \in \veps_{\a}^{-1}\big(\mc{G}_{\a})$
is located close to $\ups_\a q$ or not. In step one we choose $T$, $\varrho$, $\eps$
small enough, such that for all $\la = \veps_\a^{-1} (z)$, $z \in \mc{G}_{\a}$
with $\varrho > - \ups_\a \Im(z) > - \eps$ we have $|\la - \ups_\a q| < \mf{c}/2$,
where $\mf{c}$ is the constant introduced in item $\mathrm{a)}$ of
Hypothesis~\ref{Hypotheses solubilite NLIE}. Then, for any $y \in \op{Y}_{\e{tot}}$,
$\ups \in \{\pm \}$,
\beq
     |\ups \i \zeta + \la - y | \, >
        \, \big|\, |\ups \i \zeta - y + \ups_\a q| - |\la - \ups_\a q| \, \big| \, > \,
	   \e{d}\big( \op{Y}_{\e{tot}}-\i\ups \zeta, \ups_\a q \big)
	\, - \, \f{\mf{c}}{2} \, > \, \f{\mf{c}}{2} \;.
\enq
Above, we have also used that since the elements building up $\op{Y}^{\prime}$ belong to a small open neighbourhood of
$\pm q$, one also has that $\e{d}\big( \op{Y}_{\e{tot}} \mp \i \zeta, \sg q \big)>\mf{c}$ with $\sg=\pm$.

In step two assume that $\la = \veps_\a^{-1} (z)$, $z \in \mc{G}_{\a}^{(\pm)}$,
$- \ups_{\a} \Im (z) \geq \varrho$.  First, since $-\tf{\zeta}{2} <\Im(\la)<\eps^{\prime}$
taken that elements building up $\op{Y}^{\prime}$ belong to a small neighbourhood of $\pm q$,
one gets that for $y\in \op{Y}^{\prime}$ such that $|y \mp q|<\eps$ with $\eps$ small enough
\beq
     |\ups \i \zeta + \la - y | \, > \,
        \big|\, |\ups \i \zeta + \la \mp q  | - |y \mp  q| \, \big| \, > \,
	\f{\zeta}{4} -\eps \, >\, c >0 \;.
\enq
In order to lower bound the same quantity for $y\in \op{Y}$, one first sets $\mu= \veps_{\a}^{-1}\big( \i \Im (z) \mp \de_T \big)$.
Note that if $\Im(z)$ is not too large, $- \ups_{\a} \Im(z) \leq t_{\a}^{(\pm)}$ as introduced in
Fig.~\ref{contour integration entourant Real veps=0}, then $\mu \in  \msc{C}_{\e{ref}}$ by construction.
Else, if $- \ups_{\a} \Im(z) >  t_{\a}^{(\pm)}$, then one has $\mu \in \op{D}_{-\i\tf{\zeta}{2}, \mf{c}_{\e{d}} T }$.
For any $y \in \op{Y}$,
$\ups \in \{\pm \}$, one has the lower bound
\beq
     |\ups \i \zeta + \la - y | \, > \,
        \big|\, |\ups \i \zeta + \mu - y| - |\mu - \la| \, \big| \;. 
\enq
Furthermore, one has the upper bound 
\beq
     \big|\veps_{\a}^{-1}(z) \, - \, \veps_{\a}^{-1}( \mp \de_T + \i \Im(z) ) \big| \; \leq \;
        2 \cdot \e{sup}\Big\{|(\veps_{\a}^{-1})' (z )| \, : \, z \in \mc{G}_{\a} \Big\}
     \cdot  |\Re(z) \pm \de_T| \,  \leq \, C_{\veps}  C_{\mf{r}} T \;. 
\enq
Thus, owing to assumption $\mathrm{b)}$ of
Hypothesis~\ref{Hypotheses solubilite NLIE}, one infers that
\beq
     |\ups \i \zeta + \la - y| \, > \,
        \e{d}\big(\op{Y}_{\e{tot}} - \i \ups \zeta, \mu \big) \, - \,
	C_{\veps}  C_{\mf{r}} T \, \geq  \,
	\big( \mf{c}_{\e{ref}}  \, - \, C_{\veps}  C_{\mf{r}} \big) T \;. 
\label{ecriture borne inf sur une config}
\enq
Note that when $\mu \in \msc{C}_{\e{ref}}$, \eqref{ecriture borne inf sur une config} is direct.
Else, when $\mu \in \op{D}_{-\i\tf{\zeta}{2}, \mf{c}_{\e{d}} T}$, \eqref{ecriture borne inf sur une config} follows
from the fact that elements of $\op{Y}$ do not belong to
$\op{D}_{\i\tf{\zeta}{2},\mf{c}_{\e{sep}}T } \cup \op{D}_{-3\i\tf{\zeta}{2},\mf{c}_{\e{sep}}T } $ by point $\mathrm{c)}$
of Hypotheses \ref{Hypotheses solubilite NLIE}, and hence do not belong to
$\op{D}_{\i\tf{\zeta}{2},\mf{c}_{\e{d}}T } \cup \op{D}_{-3\i\tf{\zeta}{2},\mf{c}_{\e{d}}T } $ as well since $\mf{c}_{\e{d}}< \mf{c}_{\e{sep}}$.
This ensures that $\e{d}(\op{Y}_{\e{tot}}-\i\ups \zeta, \mu) \geq \e{d}\big(\op{Y}-\i\ups \zeta,  \msc{C}_{\e{ref}}\big)$.

Thus, all-in-all, if $\mf{c}_{\e{ref}} > C_{\veps}  C_{\mf{r}}$,
\beq
     \pl{y \in \op{Y}_{\e{tot}} }{}
        \bigg|\f{ \sinh(\i\zeta + y -\la)}{\sinh(\i\zeta + \la - y)}\bigg|^{\pm 1}
	\, \leq \,
     \f{\big(C^{\prime}/T \big)^{ |\op{Y} |} }
       {\big( \mf{c}_{\e{ref}} \, - \, C_{\veps} C_{\mf{r}} \big)^{|\op{Y} |}} \;.
\enq
%
%
%
%
%
%
%
%
%
%
%
%
%
%
%
%
%
%
%
%

Since $u_{1;\e{reg}}$ is bounded on $\veps_{\a}^{-1}(\mc{G}_{\a})$, one obtains,
for all $\la \in \veps_{\a}^{-1}(\mc{G}_{\a}^{(\pm)})$, the estimate
\beq
     \f{ T^{ |\op{Y}|} }{ C } \bigg( \f{ \mf{c}_{\e{ref}} \, - \,
        C_{\mf{r}} C_{\veps} }{ C^{\prime} } \bigg)^{|\op{Y}|} \,
	\ex{ - C_{\mc{M}} T  }\Big|\ex{ \mp\f{1}{T} \veps_{\e{c}}(\la) } \Big|  \, \leq \, 
     \Big|\ex{ \mp  \f{1}{T} f(\la\,|\, \mathbb{X}) } \Big| \, \leq \,
        \Big|\ex{ \mp\f{1}{T} \veps_{\e{c}}(\la) } \Big|  C \, T^{- |\op{Y}|}
     \bigg( \f{ C^{\prime} }{ \mf{c}_{\e{ref}} \, - \,
        C_{\mf{r}} C_{\veps} } \bigg)^{|\op{Y}|}
     \, \ex{  C_{\mc{M}} T  } \;. 
\enq
It solely remains to observe that, for $\la=\veps_{\a}^{-1}(x+\i y)$ with
$|x\mp \de_T| < C_{\mf{r}} T$, one has that 
\beq
     \Big|\ex{\mp\f{1}{T} \veps_{\e{c}}\circ\veps_{\a}^{-1}(x+\i y)}\Big| \; = \;
        \ex{\mp\f{1}{T}(\pm \de_{T} + (x \mp \de_{T}))} 
     \begin{cases}
        \leq \ex{-\f{1}{T} \de_T + C_{\mf{r}}} = T^{M} \ex{C_{\mf{r}}} \\[1ex]
        \geq \ex{-\f{1}{T} \de_T - C_{\mf{r}}} = T^{M} \ex{- C_{\mf{r}}}
     \end{cases} \;. 
\enq
This allows one to conclude with respect to the bounds involving $f(\la \,|\, \mathbb{X})$.

The bounds for $\wh{f}(\la \,|\, \wh{\mathbb{X}}\,)$ then follow upon observing that
\beq
     \wh{f}(\la \,|\,  \wh{\mathbb{X}}\,) \; = \;
        f(\la \,|\,  \wh{\mathbb{X}}\, ) \; + \;
	\e{O}\biggl( \f{1}{NT^3} \biggr)  \quad \text{uniformly throughout} \quad 
     \bigcup\limits_{\a \in \{L,R\}}^{} 
     \mc{G}_{\a} \,. 
\enq
%
%
%
\qed

The proposition below provides convenient bounds for
$\ex{\mp \f{1}{T} f(\la\,|\, \mathbb{X}) }$ and
$\ex{\mp \f{1}{T} \wh{f}(\la\,|\, \wh{\mathbb{X}} \, ) }$
on portions of $\Dp{}\op{D}_{-\i\tf{\zeta}{2}, \mf{c}_{\e{d}} T}$.

\begin{prop}
\label{Proposition bornage f et f hat sur arc circulaires rayon ordre T}
Let 
\beq
     \tau \, = \, h  \,   - \,  2 J \cos(\zeta) \, - \,
        \Int{-q}{q} \dd \mu \:
	   \veps(\mu) K\big( \i\tfrac{\zeta}{2}+\mu\big)  \;. 
\label{definition tau}
\enq
Introduce two arcs $\mc{A}^{(\pm)}$ of $\Dp{}\op{D}_{-\i\f{\zeta}{2}, \mf{c}_{\e{d}} T}$ :
\beq
\mc{A}^{(\sg)} \, = \, \bigg\{-\i \f{\zeta}{2} \, + \, \mf{c}_{\e{d}} T \ex{\i\vth} \; : \; \vth \in \big[ - \sg \pi  +T \vth_{L}^{( \sg)} \; ; \; T  \vth_{R}^{( \sg)}  \big] \bigg\} \; , \quad \sg \in \{ \pm \} \; ,
\label{definition arcs A sg}
\enq
delimited by the angles 
\beq
\vth_{\a}^{(\sg)} \; = \; \f{ \mf{c}_{\e{d}} \ups_{\a}  }{ 2J \sin \zeta    } \Re\Big( \tau - \sg \de_T \Big)  \, + \, \sg C_{\mc{A}} T\ups_{\a} \;, 
\label{definition angles delimitant les arcs de A sg}
\enq
where $\a \in \{L, R\}$ and $C_{\mc{A}}$ is a fixed large enough constant. Assume that
Hypothesis~\ref{Hypotheses solubilite NLIE} holds.

Then, there exist $T_0, \eta >0$ such that, uniformly in 
$ T_0 > T > 0$ and $\eta > 1/(NT^4)$,
the following estimates hold true:
\begin{equation}
%
     \Big| \ex{\mp \f{1}{T} f(\la\,|\, \mathbb{X}) } \Big| \, + \,
        \Big| \ex{\mp  \f{1}{T} \wh{f}(\la\,|\, \wh{\mathbb{X}}) } \Big|
     \leq C \, T^{ M-|\op{Y}| } \,
        \bigg( \f{ C^{\prime} }{\mf{c}_{\e{sep}} \, - \, \mf{c}_{\e{d}}} \bigg)^{|\op{Y}|}  
     \, \ex{C_{\mc{M}} T}  \quad for \quad \la \in \mc{A}^{(\pm)} \;.
        \label{borne sur exp de f le long de arc +}
%
%
%
%
\end{equation}
$C, C^{\prime}$ are constants such that $C^{\prime}$ does not depend on $\mathbb{X}$ or $\wh{\mathbb{X}}$
while $C$ only depends on $|\op{X}_{\e{tot}}|$, $|\op{Y}'|$ and $|\op{Y}_{\e{sg}}|$. Also
$\mf{c}_{\e{sep}} > \mf{c}_{\e{d}}$ is the constant arising in point $\mathrm{c)}$
of Hypothesis \ref{Hypotheses solubilite NLIE}. 

\end{prop}

\Proof 
Let 
\beq
     \mf{e}(\vth) \, = \,
        \veps_{\e{c}}\Big(  -\i \tfrac{\zeta}{2} \, + \,
	\mf{c}_{\e{d}} T \ex{\i\vth} \Big)  \;. 
\enq
Observe that, when $T$ is small enough, owing to the
expansion~\eqref{ecriture dvpm petit t pour veps}, it holds that 
\beq
     \Re\big[  \mf{e}(\vth) \big] \; = \;
        - \f{ 2J \sin \zeta  }{ \mf{c}_{\e{d}} T} \sin \vth \, + \,
	\Re(\tau) \, + \, \e{O}(T) \;. 
\label{ecriture perturbative de dressed energy proche pole en moins i zeta sur 2}
\enq
In the following, one should discuss the cases $\Re(\tau)>0$,
$\Re(\tau)<0$ and $\Re(\tau)=0$ separately. Indeed, for $T$ small enough, 
\beq
     \e{sgn}\big(  \vth_{\a}^{( \sg)}  \big) \; = \;
        \ups_{\a} \e{sgn}\Big[ \Re(\tau) \Big] \quad \e{if} \quad \Re(\tau) \not= 0 \qquad 
     \e{and} \qquad 
     \e{sgn}\big(  \vth_{\a}^{( \sg)}  \big) \; = \;
        -\sg \ups_{\a} \ \quad \e{if} \quad \Re(\tau)  = 0 \;. 
\enq
Therefore, the sign of $\Re(\tau)$ determines whether the length of the intervals 
$ \big[ - \sg \pi  +T \vth_{L}^{( \sg)} \; ; \; T  \vth_{R}^{( \sg)}  \big]$ is
smaller or larger than $\pi$. That fact plays a role in the analysis owing to the 
potential change in the sign of $\sin \vth$
in~\eqref{ecriture perturbative de dressed energy proche pole en moins i zeta sur 2}.

We first focus on the $\Re(\tau)>0$ case. Then, for $T$ small enough $\vth_{L}^{(+)}<0$
and $ \vth_{R}^{( + )} >0$. Thus, it holds that 
\beq
     \Re\big[ \mf{e}(\vth) \big] \; \ge \;
	\begin{cases}
	   \Re\big[\mf{e}\big(- \pi + T \vth_{L}^{(+)} \big) \big]  \; + \; \e{O}(T)
	   & \text{if}\ \vth \in \big[- \pi  +T \vth_{L}^{(+)} \; ; \; -\pi \big]
	     \;, \\[1ex] 
          \Re\big(\tau \big) \; + \; \e{O}(T) &
	  \text{if}\ \vth \in \big[- \pi \; ; \; 0 \big] \;, \\[1ex] 
          \Re\big[\mf{e}\big( T \vth_{R}^{(+)} \big) \big]  \; + \; \e{O}(T) &
	  \text{if}\ \vth \in  \big[ 0 \; ; \; T  \vth_{R}^{( + )} \big] \;.
        \end{cases}
\enq
This leads to the overall bound on $\big[- \pi + T \vth_{L}^{(+)} \; ;
\; T \vth_{R}^{(+)} \big]$:
\beq
     \Re\big[ \mf{e}(\vth) \big] \; \ge \;
        \e{min} \bigg\{ \Re\big[\mf{e}\big(- \pi + T \vth_{L}^{(+)} \big) \big] \, , \,
	   \Re\big( \tau \big) \, , \, \Re\big[\mf{e}\big( T \vth_{R}^{(+)} \big) \big]
	   \bigg\} \; + \; \e{O}(T)  
     \; > \; - MT \ln T - CT
\enq
for some $C>0$ if $T$ is small enough.

Similarly, for $T$ small enough, one gets $\vth_{L}^{(-)}<0$ and $ \vth_{R}^{( - )} >0$
so that, on $\big[  \pi  +T \vth_{L}^{(-)} \; ; \; T  \vth_{R}^{(-)}  \big]$ it holds that
\beq
     \Re\big[ \mf{e}(\vth) \big] \; \le \;
        \e{max}\bigg\{ \Re\big[\mf{e}\big( \pi  +T \vth_{L}^{(-)} \big) \big] \, , \, \Re(\tau) \, , \,
	\Re\big[\mf{e}\big( T \vth_{R}^{(-)} \big) \big]  \bigg\} \; + \; \e{O}(T)  
     \; < \; MT \ln T + CT
\enq
for some $C>0$ if $T$ is small enough.

When $\Re(\tau) < 0$ and for $T$ small enough, one rather has $\vth_{L}^{(\pm)}>0$
and $ \vth_{R}^{( \pm )} < 0$ so that the lengths of the two intervals considered
above change. Still, the analysis is similar, and, for $T$ small enough, one
obtains as above that
\begin{align}
     \Re\big[ \mf{e}(\vth) \big] &  > - MT \ln T - CT
        \quad \text{for}\ \vth \in \big[ -\pi +T \vth_{L}^{(+)} \; ;
	                              \; T  \vth_{R}^{(+)}  \big]\; , \\[1ex]
     \Re\big[  \mf{e}(\vth) \big] &  < MT \ln T + CT
        \quad \text{for}\ \vth \in \big[ \pi + T \vth_{L}^{(-)} \; ;
	                                 \; T  \vth_{R}^{(-)}  \big] \;. 
\end{align}
Finally, for $\Re(\tau)=0$, one may check that the same upper and lower bounds
hold as well. 

Now all is in place to establish the desired bounds involving
$\wh{f}(\la \,|\, \wh{\mathbb{X}})$. In a similar way as in \eqref{factor_exp_fovert}
\begin{multline}
     \ex{- \f{1}{T}  \wh{f}(\la\,|\, \wh{\mathbb{X}}) } \; = \;
        \ex{ -\f{1}{T} \veps_{\e{c}}(\la) -\f{1}{T} \de \mc{W}_N(\la)} 
     \pl{y \in \wh{\op{Y}}_{\e{tot}} }{}
        \bigg\{\f{\sinh(\i\zeta + y - \la)}{\sinh(\i\zeta + \la -y)}\bigg\} \cdot
     \pl{y \in \op{Y}_{\e{sg}}}{}
        \bigg\{ \f{\sinh(\i\zeta + y - \la)} {\sinh(\i\zeta + \la - y )}\bigg\}  \\
\times  
        \pl{x \in \wh{\op{X}}_{\e{tot}} }{}
	\bigg\{\f{\sinh(\i\zeta + \la - x)}{\sinh(\i\zeta + x - \la)}\bigg\}
        \cdot \ex{-\f{1}{T}f(\la) -  u_{1;\e{reg}}(\la\,|\, \wh{\mathbb{X}})} \;,
\end{multline}
%
%
%
%
%
%
%
%
 %
%
%
%
%
where $\de \mc{W}_N(\la) = \mc{W}_N(\la)-\veps_{\e{c}}(\la)$ while $\mc{W}_N(\la)$ is as
defined in \eqref{definition solution LIE deformee WN}. 

It remains to upper and lower bound each of the terms building up the product given above. 
For $\la \in \Dp{}\op{D}_{ -\i\tf{\zeta}{2}, \mf{c}_{\e{d}} T  }$ and
$z \in \wh{\op{X}}_{\e{tot}}\cup \wh{\op{Y}}^{\prime}
\cup \op{Y}_{\e{sg}}$, it holds that $\e{max}\Big\{ |\Im(\la - z)|,  |\Im(\la-q)| \Big\}
< \tf{\zeta}{2}+\eps^{\prime}$ with $\eps^{\prime}>0$ but as small as need be. This
entails that, uniformly in $\la \in \Dp{}\op{D}_{ -\i\tf{\zeta}{2}, \mf{c}_{\e{d}} T}$,    
\beq
     \pl{x \in \wh{\op{X}}_{\e{tot}} }{}
        \bigg|\f{ \sinh(\i\zeta + \la - x)}{\sinh(\i\zeta + x - \la)}\bigg|^{\pm 1}
	   \; \leq \: C^{|\op{X}_\e{tot}|} \;, \quad 
     \pl{y \in \op{Y}_{\e{sg}}\cup \wh{\op{Y}}^{\prime} }{}
        \bigg|\f{\sinh(\i\zeta + y - \la)}{\sinh(\i\zeta + \la - y )}\bigg|^{\pm 1}
	   \; \leq \: C^{|\op{Y}_{\e{sg}}| + |\op{Y}^{\prime}|}  \;.
\enq

Further observe that, for $\la \in \Dp{}\op{D}_{ -\i\tf{\zeta}{2}, \mf{c}_{\e{d}} T}$
and $y \in \op{Y}$, owing to point $\mathrm{c)}$ of
Hypothesis~\ref{Hypotheses solubilite NLIE},
\beq
     \e{min} \Big\{|\i\zeta + \la - y |\, , \, |\i\zeta + y - \la |\Big\}
        \; \geq \; \big( \mf{c}_{\e{sep}} \, - \, \mf{c}_{\e{d}} \big)\cdot  T \;.
\label{ecriture borne inf sur distance lambda  moins y a pm i zeta}
\enq
%
%
%
Taken that  $\mf{c}_{\e{sep}} > \mf{c}_\e{d}$, this leads, for some $C>0$,
to the upper bound
\beq
     \pl{y \in \op{Y}}{}
        \bigg|\f{\sinh(\i\zeta + y - \la)}{\sinh(\i\zeta + \la - y)}   \bigg|^{\pm 1}
	\, \leq \,
     \bigg(\f{\tf{C}{T}}{\mf{c}_{\e{sep}} \, - \, \mf{c}_{\e{d}}} \bigg)^{ |\op{Y}|} \;. 
\enq
%
%
%
%
%
%
%
%
%
%
%
%
%
%

Since $\la \mapsto u_{1;\e{reg}}\big( \la \,|\, \wh{\mathbb{X}} \big)$ is bounded on
$\Dp{}\op{D}_{ -\i\tf{\zeta}{2}, \mf{c}_{\e{d}} T}$ and since direct bounds lead to 
\beq
     \Big| \de \mc{W}_N(\la) \Big| \, \leq  \, \f{C}{  N T^3 } \qquad
        \e{uniformly} \, \e{in} \qquad
	\la \in \Dp{}\op{D}_{ -\i\tf{\zeta}{2}, \mf{c}_{\e{d}} T  } \;, 
\enq
all-in-all, by virtue of the bounds on $\veps_{\e{c}}$ along $\mc{A}^{(\pm)}$, one
sees that the estimates \eqref{borne sur exp de f le long de arc +}
hold for $0 \, < \, T \, < \, T_0$ and $\tf{1}{(NT^4)} \, < \, \eta$ with $T_0$ and
$\eta$ small enough. 

With similar reasoning one establishes analogous bounds involving
$f(\la\,|\, \mathbb{X})$. \qed

\subsection{Compatibility with Properties~\ref{Propriete pour deformer les contours}}

The next result allows one to ascertain that, given $f\in \mc{E}_{\mc{M}}$
and the contour $\msc{C}_{\e{ref}}$, the contours $\wh{\msc{C}}_{\mathbb{X}}[f]$
defined in \eqref{definition contour C de Y Trotter fini} -- resp.\
$\msc{C}_{\mathbb{X}}[f]$  given in \eqref{definition contour C de Y} --
are adapted to the functions  $\wh{f}(\la\,|\, \wh{\mathbb{X}})$ -- resp.\
$f(\la\,|\, \mathbb{X})$ -- in the sense that
\beq
\Big( \wh{f}\big( \la\,|\, \wh{\mathbb{X}}\, \big), \, \wh{\msc{C}}_{\mathbb{X}}[f], \, \msc{C}_{\e{ref}} \Big) \;\; -- \;\; \e{resp.} \; \; \Big( f\big(\la\,|\, \mathbb{X}\big), \, \msc{C}_{\mathbb{X}}[f], \, \msc{C}_{\e{ref}} \Big) \;\; --
\enq
enjoy all the requirements $\mathrm{i)}$-$\mathrm{vii)}$ of
Properties~\ref{Propriete pour deformer les contours}.

\begin{prop}
\label{Proposition domaine local holomorphie de F}
Provided Hypothesis~\ref{Hypotheses solubilite NLIE} holds for $\op{X}, \op{Y}$, there exist $T_0 > 0$ and $\eta > 0$
such that, for any $f\in \mc{E}_{\mc{M}}$,
\beq
     for \; T_0>T>0 \; \; and \;\; \eta> \f{1}{NT^4} \quad the \; function
        \;\; \la \, \mapsto \, 1 + \ex{-\f{1}{T} \wh{f}(\la\,|\,\wh{\mathbb{X}} \, ) } \; ,
\enq
resp.
\beq
      for \; T_0 > T > 0 \quad the \; function
         \;\; \la \, \mapsto \, 1 + \ex{-\f{1}{T}f(\la\,|\, \mathbb{X}) } \;,
\enq
is holomorphic in the bounded domain the boundary of which is delimited
by the curves $\msc{C}_{\e{ref}}\cup \wh{\msc{C}}_{\mathbb{X}}[f]$, resp.\
$\msc{C}_{\e{ref}}\cup \msc{C}_{\mathbb{X}}[f]$.

The only two zeroes of $ 1 - \ex{-\f{1}{T} \wh{f}(\la\,|\, \wh{\mathbb{X}}\, ) } $,
resp.\ $1 - \ex{-\f{1}{T}f(\la\,|\, \mathbb{X}) }$, on
$  \wh{\msc{C}}_{\mathbb{X}}[f]$, resp.\ $  \msc{C}_{\mathbb{X}}[f]$, are
located at $\wh{q}_{\mathbb{X}}^{\,(\pm)}[f]$, resp.\ $q_{\mathbb{X}}^{\,(\pm)}[f]$, and satisfy
\eqref{definition q pm de Y et f} as well as
\beq
 \sg \wh{f}^{\prime}\big( \,  \wh{q}_{\mathbb{X}}^{\,(\sg)}[f]    \,|\, \wh{\mathbb{X}} \big) > c\; , \; \;   resp. \qquad
 \sg f^{\prime}\big(  q_{\mathbb{X}}^{\,(\sg)}[f]    \,|\, \mathbb{X} \big) > c \; ,
\label{ecriture bornes inf directe sur derivee en les pts de Fermi modifies}
\enq
for some $c>0$ and uniformly in $T_0>T>0, \,  \eta > \tf{1}{(NT^4)}$, resp.\ $T_0>T>0$\,.


The map $\wh{f}(*\,|\, \wh{\mathbb{X}} \, )$, resp.\ $f(*\,|\, \mathbb{X})$, is
holomorphic on an open neighbourhood $\wh{\mc{U}}_{\a}$, resp.\ $\mc{U}_{\a}$,
containing $\op{D}_{\pm q, \eps}$ for some $\eps>0$ small enough and such that
$\wh{f}(*\,|\, \wh{\mathbb{X}}\, ) \, : \, \wh{\mc{U}}_{\a} \tend \op{D}_{0,\varrho}$, resp.\
$f(*\,|\, \mathbb{X}) \, : \,  \mc{U}_{\a} \tend \op{D}_{0,\varrho}$,
for some $\varrho >0$ is a biholomorphism.


There exists $\wh{\tau}_{\a} \in \Interior(I_{\de_T}) \subset \msc{C}_{\e{ref}}$, $\wh{\eps}_{\a}
\in \intfo{0}{1}$ and $\wh{\mf{p}}_{\a} \in \mathbb{Z}$ that is uniformly bounded
in $0<T<T_0$ and $0< \tf{1}{(NT^4)} < \eta$, resp.\ $\tau_{\a} \in \Interior(I_{\de_T})
\subset \msc{C}_{\e{ref}}$, $\eps_{\a} \in \intfo{0}{1}$ and $\mf{p}_{\a} \in
\mathbb{Z}$ that is uniformly bounded in $0<T<T_0$, such that
\beq
\wh{f}\big( \, \wh{\tau}_{\a}  \,|\, \wh{\mathbb{X}} \,\big) \; = \; 
2\i\pi T \big( \, \wh{\mf{p}}_{\a} - \tfrac{1}{2}  + \wh{\eps}_{\a} \big) \;, \quad resp.\  \quad
f\big( \tau_{\a}  \,|\, \mathbb{X} \big) \; = \;
2\i\pi T \big( \mf{p}_{\a} - \tfrac{1}{2}  + \eps_{\a} \big) \;.
\label{ecriture existence point ou f hat est im pur sur C ref et sa valeur}
\enq

The zeroes of  $ 1 + \ex{-\f{1}{T} \wh{f}(\la\,|\, \wh{\mathbb{X}}\, ) }$, resp.\
$1 + \ex{-\f{1}{T}f(\la\,|\, \mathbb{X}) }$, in the bounded domain whose boundary
is delimited by the curves $\msc{C}_{\e{ref}}\cup \wh{\msc{C}}_{\mathbb{X}}[f]$, resp.\
$\msc{C}_{\e{ref}}\cup \msc{C}_{\mathbb{X}}[f]$, are the unique solutions in
$\wh{\mc{U}}_{\a}$, resp.\ $\mc{U}_{\a}$, to
\beq
\wh{f}\big( \, \wh{\mf{x}}_{k}^{\, (\a)}  \,|\, \wh{\mathbb{X}} \big) \; = \;
2\i\pi T \big(k - \tfrac{1}{2} \big) \quad with \quad
k \in \begin{cases}
         \intff{ 1 }{ \wh{\mf{p}}_{\a} }  & \wh{\mf{p}}_{\a} >0   \\
         \emptyset   &  \wh{\mf{p}}_{\a}=0  \\
         \intff{ 1 + \wh{\mf{p}}_{\a} }{0}  & \wh{\mf{p}}_{\a} <0 \; ,
      \end{cases}
\label{equation donnant les zeros enlaces hat}
\enq
resp.
\beq
f\big( \, \mf{x}_{k}^{(\a)}  \,|\, \mathbb{X} \big) \; = \;
2\i\pi T \big( k - \tfrac{1}{2} \big) \quad with \quad
k \in \begin{cases}
         \intff{ 1 }{ \mf{p}_{\a} }  & \mf{p}_{\a} >0   \\
         \emptyset   &  \mf{p}_{\a}=0  \\
         \intff{ 1 + \mf{p}_{\a} }{0}  & \mf{p}_{\a} <0 \; .
      \end{cases}
\label{equation donnant les zeros enlaces}
\enq

Finally, upon setting
\beq
\wh{J}^{\, (\a)}[f] \; = \;
\wh{f}_{\a}^{-1}\Big(I_{\de_T}  \,|\, \wh{\mathbb{X}} \Big) \; , \; \;  resp. \qquad
J^{(\a)}[f] \; = \; f_{\a}^{-1}\Big(I_{\de_T}   \,|\, \mathbb{X} \Big)
\enq
one gets that $\wh{f}\,\Big( \wh{J}^{\, (\a)}[f] \,|\,  \wh{\mathbb{X}} \Big) \, = \,
I_{\de_T}$, resp.\
$f\Big( J^{\, (\a)}[f] \,|\,  \mathbb{X} \Big) \, = \, I_{\de_T}$, and
\beq
\Big| \Re\Big[ \wh{f}\, \big( \la  \,|\,  \wh{\mathbb{X}} \big)   \Big] \Big| \; > \;
\f{\de_T}{2} \quad for \; \la \in  \wh{\msc{C}}_{\mathbb{X}}[f] \setminus \wh{J}^{\, (\a)}[f] \; ,
\label{ecriture bornes inf sur parties relles hors du voisiage des pts de Fermi modifies}
\enq
resp.
\beq
\Big| \Re\Big[ f\big( \la  \,|\,  \mathbb{X} \big)   \Big] \Big|
\; > \; \f{\de_T}{2} \quad for \; \la \in  \msc{C}_{\mathbb{X}}[f] \setminus J^{\, (\a)}[f] \;.
\enq
\end{prop}

\Proof 
We only discuss the proof in the case of the finite Trotter number setting. The
other case is treated similarly, upon introducing the analogous quantities with the\quad
$\wh{}$\quad being removed.

\begin{figure}[h]
\begin{center}
\includegraphics[width=0.85\textwidth]{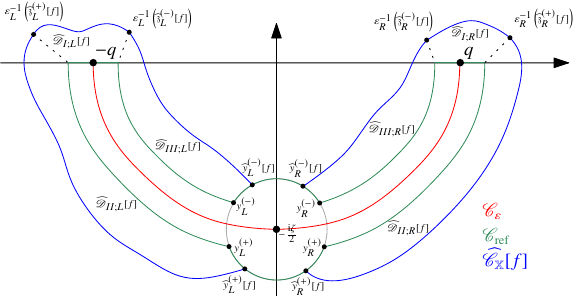}
\caption{The domain $\wh{\msc{D}}[f]$.}
\label{Domaine integration pour deformation contour}
\end{center}
\end{figure}

First observe that the statement relative to $\wh{f}\, \big( * \,|\, \wh{\mathbb{X}}\big):
\wh{\mc{U}}_{\a} \tend \op{D}_{0,\varrho}$ being a biholomorphism is ensured by
Proposition~\ref{Proposition continuite inverse f et sa comparaison a veps inverse}. Next,
the statement \eqref{ecriture existence point ou f hat est im pur sur C ref et sa valeur}
relative to $\wh{\tau}_{\a}$ follows from
Lemma~\ref{Lemme identification des entiers definissant p pm}. The biholomorphism property
and direct bounds immediately lead to
\eqref{ecriture bornes inf directe sur derivee en les pts de Fermi modifies} and the
existence of the zeroes $\wh{q}_{\mathbb{X}}^{\,(\sg)}[f]$. Also, it ensures that in the
neighbourhood $\wh{\mc{U}}_{\a}$, the $ \wh{\mf{x}}_{k}^{\, (\a)}$ given in
\eqref{equation donnant les zeros enlaces hat} are the sole zeroes
of  $1 + \ex{-\f{1}{T} \wh{f}(\la\,|\, \wh{\mathbb{X}}) }$  contained in the domain
$\wh{\msc{D}}[f]$ delimited by the curves $\msc{C}_{\e{ref}} \cup \wh{\msc{C}}_{\mathbb{X}}[f]$.

Clearly, $\la \, \mapsto \, 1 + \ex{-\f{1}{T} \wh{f}(\la\,|\, \wh{\mathbb{X}}) }$ is meromorphic
on the domain $\wh{\msc{D}}[f]$ delimited by the curves $\msc{C}_{\e{ref}} \cup
\wh{\msc{C}}_{\mathbb{X}}[f]$. We thus first focus on establishing the no-pole property
throughout the whole domain as well as the lack of zeroes away from $\wh{\mc{U}}_{\a}$.
First, observe that the domain $\wh{\msc{D}}[f]$ may be partitioned into six sub-domains,
\textit{c.f.}\ Figure~\ref{Domaine integration pour deformation contour}, 
\beq
     \wh{\msc{D}}[f] \; = \; \bigcup\limits_{A \in \{I, II, III\} }^{} 
        \bigcup\limits_{ \a \in \{L,R\} } \wh{\msc{D}}_{A;\a}[f] \;. 
\label{definition domain Df pour la deformation de contours}
\enq
Given $\a \in \{L,R\}$, the sub-domains are limited by the boundaries
\beq
     \Dp{}\wh{\msc{D}}_{I;\a}[f] \, = \,
        \wh{f}^{\, -1}_{\a}\Big( \intff{\de_T}{ - \de_{T} } \,|\, \wh{\mathbb{X}} \Big)  \cup
     \veps_{\a}^{-1}\Big(\big[\, \wh{\mf{z}}_{\a}^{\, (-)}[f] \, ; \, -\de_{T} \big] \Big)
     \cup \veps^{-1}_{\a}\Big( \intff{-\de_T}{ \de_{T} } \Big)
        \cup \veps_{\a}^{-1}\Big( \big[ \de_{T} \, ; \,
	   \wh{\mf{z}}_{\a}^{\, (+)}[f] \, \big]  \Big)  
\enq
and 
\bem
\Dp{}\wh{\msc{D}}_{II;\a}[f] \, = \,   \veps_{\a}^{-1}\Big( \big[ \de_{T}  \, ; \, \wh{\mf{z}}_{\a}^{\,(+)}[f]  \big] \Big) \cup
 \veps_{\a}^{-1}\Big( \big[ \,  \wh{\mf{z}}_{\a}^{\,(+)}[f]\, ;\,  \wh{\mf{z}}_{\a}^{\,(+)}[f] + \i \cdot \wh{\mf{t}}_{\a}^{\,(+)}[f]  \big] \Big) \\
\cup \ga\Big( \,  \wh{y}^{\,(+)}_{\a}[f] \, , \, y_{\a}^{(+)}   \Big) 
\cup  \veps_{\a}^{-1}\Big( \intff{ \de_T + \i \cdot \mf{t}_{\a}^{(+)} }{ \de_T } \Big) \;,
\end{multline}
where $ \veps_{\a}^{-1} \big( \de_T + \i \cdot \mf{t}_{\a}^{(\pm)} \big) \, =  \,
y_{\a}^{(\pm)}$, as well as by
%
%
%
\bem
\Dp{}\wh{\msc{D}}_{III;\a}[f] \, = \,   \veps_{\a}^{-1}\Big( \big[ \,  \wh{\mf{z}}_{\a}^{\,(-)}[f] \, ; \, -\de_{T}    \big] \Big) 
\cup  \veps_{\a}^{-1}\Big( \intff{ -\de_T  }{ -\de_T + \i \cdot \mf{t}_{\a}^{(-)}  } \Big) \\
\cup \ga\Big(  y_{\a}^{(-)}  \, , \,   \wh{y}^{\,(-)}_{\a}[f] \Big)  
\cup  \veps_{\a}^{-1}\Big( \big[ \,   \wh{\mf{z}}_{\a}^{\,(-)}[f] + \i \cdot \wh{\mf{t}}_{\a}^{\,(-)}[f] \, ; \,  \wh{\mf{z}}_{\a}^{\,(-)}[f]   \big] \Big) \;. 
\end{multline}

By construction, it holds that, if $C_\mf{r}, C_{g}$ occurring in
\eqref{ecriture domain bornage exponentielle de proche de C ref} are large
enough and $T$ is small enough, then
\beq
\wh{\msc{D}}_{II;\a}[f] \, \subset \, \veps_{\a}^{-1}\big( \mc{G}^{(+)}_{\a} \big) \qquad \e{and} \qquad
\wh{\msc{D}}_{III;\a}[f] \, \subset \, \veps_{\a}^{-1}\big( \mc{G}^{(-)}_{\a} \big)  \;,
\enq
where the domains $ \mc{G}^{(\pm)}_{\a} $ are as defined in
Proposition~\ref{Proposition borne sup sur exponentielle en tempe de la fct f lambda mathbbY},
see \eqref{ecriture domain bornage exponentielle de proche de C ref}. Thus, by the
estimates~\eqref{borne sup et inf sur partie laterale contour pour exp de f de Y mathbb}
obtained in
Proposition~\ref{Proposition borne sup sur exponentielle en tempe de la fct f lambda mathbbY}  
with $M> | \op{Y} | + 2$ one has that $ 1 + \ex{-\f{1}{T} \wh{f}(\la\,|\, \wh{\mathbb{X}}) }
=1+\e{O}(T^2)$ uniformly throughout $\wh{\msc{D}}_{II;\a}[f]$ so that this function cannot
have zeroes there. These bounds do ensure as well that the function cannot have poles there. 

In what concerns $\wh{\msc{D}}_{III;\a}[f]$, one first decomposes
\beq
1 + \ex{-\f{1}{T} \wh{f}(\la\,|\, \wh{\mathbb{X}}\, ) } \; = \; \ex{-\f{1}{T} \wh{f}(\la\,|\, \wh{\mathbb{X}}\, ) } \cdot
\Big( 1 + \ex{\f{1}{T} \wh{f}(\la\,|\, \wh{\mathbb{X}}\,) } \Big) \;.
\enq
Then, the lower and upper bounds in \eqref{borne sup et inf sur partie laterale contour pour exp de f de Y mathbb} ensure, respectively, that the function
has neither poles nor zeroes in $\wh{\msc{D}}_{III;\a}[f]$.

Finally, in what concerns the domain $\wh{\msc{D}}_{I;\a}[f]$,
condition $\mathrm{a)}$ of Hypothesis~\ref{Hypotheses solubilite NLIE}
ensures that $ 1 + \ex{-\f{1}{T} \wh{f}(\la\,|\, \wh{\mathbb{X}} \, ) }$ has no poles in an open
neighbourhood of $\pm q$ and thus in particular not in $\wh{\msc{D}}_{I;\a}[f]$.
Since $\wh{\msc{D}}_{I;\a}[f]\subset \wh{\mc{U}}_{\a}$, all potential zeroes in this domain
have been already identified.

The estimates on $\wh{\msc{D}}_{II;\a} [f]  \cup  \wh{\msc{D}}_{III;\a} [f]$ and the invertibility on $\wh{J}^{(\a)}[f]$ ensure
that the only zeroes of $1 - \ex{-\f{1}{T} \wh{f}(\la\,|\, \wh{\mathbb{X}}\,) }$ are located at $\wh{q}_{\mathbb{X}}^{\,(\pm)}[f]$
as introduced through \eqref{definition q pm de Y et f}.
Finally, the estimates \eqref{borne sup et inf sur partie laterale contour pour exp de f de Y mathbb}
ensure the validity of \eqref{ecriture bornes inf sur parties relles hors du voisiage des pts de Fermi modifies}. \qed

\section{The modified non-linear integral operator: existence and uniqueness of solutions}
\label{Section NLIE modifiee et solvabilite par pt fixe}
We are now in position to introduce the finite and infinite Trotter number
non-linear operators of interest on $\mc{E}_{\mc{M}}$, namely $ \wh{\mc{L}}_{T}$ and
$\mc{L}_{T}$. The operator $\wh{\mc{L}}_{T}$ decomposes as $\wh{\mc{L}}_{T} =
\wh{\mc{L}}_{T}^{\, (1)} + \wh{\mc{L}}_{T}^{\,(2)}$ with 
\beq
     \wh{\mc{L}}_{T}^{\, (1)} \big[ f \big](\la) \; = \;  
        - \,  \bigg\{  \Int{ q }{ \wh{q}^{\, (+)}_{\mathbb{X}}[f] }\! \!  \dd \mu \,
	\wh{f}( \mu  \,|\, \wh{\mathbb{X}} \,)
     \, + \, \Int{ \wh{q}^{\, (-)}_{\mathbb{X}}[f] }{ - q }  \! \!  \dd \mu \,
        \wh{f}( \mu  \,|\,  \wh{\mathbb{X}} \,)  \bigg\} \cdot  R_{\e{c}} (\la,\mu)
\enq
and 
\beq
   \wh{\mc{L}}_{T}^{\,(2)} \big[ f\big](\la) \; = \;
      - \, T \Int{ \wh{\msc{C}}_{\mathbb{X}}[f] }{} \dd \mu \: R_{\e{c}}(\la,\mu)
        \ln\Big[ 1 + \ex{-\frac{1}{T}|\wh{f}\,|(\mu \,|\, \wh{\mathbb{X}} \,) } \Big] \;.
\enq
The operator $\mc{L}_{T}$ is obtained from $\wh{\mc{L}}_{T}$ upon the substitution
\beq
 \wh{q}^{\, (\pm)}_{\mathbb{X}}[f] \; \hookrightarrow  \;  q^{\, (\pm)}_{\mathbb{X}}[f]  \;, \qquad \wh{\msc{C}}_{\mathbb{X}}[f] \; \hookrightarrow  \; \msc{C}_{\mathbb{X}}[f] \; ,
 \qquad \wh{f}( \mu  \,|\, \wh{\mathbb{X}} \,) \; \hookrightarrow \;  f( \mu  \,|\, \mathbb{X} ) \;.
\enq

Also, we recall that $ \wh{q}^{\, (\pm)}_{\mathbb{X}}[f]$ and $q^{(\pm)}_{\mathbb{X}}[f]$
are defined through \eqref{definition q pm de Y et f} while the $f$-dependent integration
contours $\wh{\msc{C}}_{\mathbb{X}}[f]$ and $\msc{C}_{\mathbb{X}}[f]$ are given by
\eqref{definition contour C de Y Trotter fini} and \eqref{definition contour C de Y},
respectively.

The operators $\mc{L}_{T}^{(a)}$, $a = 1, 2$, are closely related to the operators
$\mc{R}_{T}^{(a)}$ introduced earlier in \eqref{definition operateur reste de type 1},
\eqref{definition operateur reste de type 2} in that 
\beq
     \wh{\mc{L}}_{T}^{\, (a)}[f] \, = \, \wh{\mc{R}}_{T}^{\, (a)}\Big[\wh{f}(*\,|\, \wh{\mathbb{X}}\,)
        \Big] \biggr|_{q^{(\pm)}_{\wh{u}}
	               \hookrightarrow  \wh{q}^{\, (\pm)}_{\mathbb{X}}[f],\
                       \msc{C}_{ \wh{u} } \hookrightarrow  \wh{\msc{C}}_{\mathbb{X}}[f]} \;.
\enq

The reason for focusing on the operators $\wh{\mc{L}}_{T}$ and $\mc{L}_{T}$ is that their fixed
points allow one to construct solutions to the original non-linear integral equations,
be it at finite \eqref{ecriture NLIE a Trotter fini} or infinite
\eqref{ecriture NLIE a Trotter infini} Trotter numbers, and subject to the associated
index conditions \eqref{ecriture monodromie Trotter fini} or
\eqref{ecriture monodromie Trotter infini}. 

\begin{lemme}
There exists $T_0, \eta>0$ such that, given $0<T<T_0$ and $\eta > \tf{1}{(NT^4)}$:
\begin{itemize}

 \item any fixed point $g$ of\ \ $\wh{\mc{L}}_{T}$, resp.~$\mc{L}_{T}$, gives rise to a solution
$\wh{u}(\la\,|\, \wh{\mathbb{X}}\,)$ to  \eqref{ecriture NLIE a Trotter fini},
\eqref{ecriture monodromie Trotter fini}, resp.\ $u(\la\,|\, \mathbb{X})$
to \eqref{ecriture NLIE a Trotter infini}, \eqref{ecriture monodromie Trotter infini};

\item any solution to \eqref{ecriture NLIE a Trotter fini},
\eqref{ecriture monodromie Trotter fini}, resp.\
\eqref{ecriture NLIE a Trotter infini}, \eqref{ecriture monodromie Trotter infini},
gives rise to a fixed point of\ \ $\wh{\mc{L}}_{T}$, resp.\ $\mc{L}_{T}$.

\end{itemize}

Finally, the uniqueness of the fixed points of\ \ $\wh{\mc{L}}_{T}$ or $\mc{L}_{T}$ implies the
uniqueness of the solutions to the associated non-linear integral equations.
\end{lemme}

\Proof 
Let $N, T $ be as given in the statement of the lemma with $T_0, \eta >0$ as appearing
in Proposition~\ref{Proposition domaine local holomorphie de F}. 

Assume one is given a fixed point $g$ of\ \ $\wh{\mc{L}}_{T}$ on the space $\mc{E}_{\mc{M}}$
and set $\wh{u}\,\big(\la\,|\, \wh{\mathbb{X}}\,\big)\,=\,\wh{g}\,\big(\la\,|\, \wh{\mathbb{X}}\,
\big)$.  Then, Proposition~\ref{Proposition domaine local holomorphie de F} ensures that
$\la \mapsto \wh{u}\,\big(\la\,|\, \wh{\mathbb{X}}\,\big)$ does enjoy the requirements
$\mathrm{i)-vii)}$ that are listed in Properties~\ref{Propriete pour deformer les contours}.
By virtue of
Proposition~\ref{Proposition correspondance entre NLIE originelle et NLIE point fixe}, 
one obtains a solution to the non-linear integral equation \eqref{ecriture NLIE a Trotter fini}.
By construction, the latter satisfies the monodromy condition
\eqref{ecriture monodromie Trotter fini}.

Reciprocally, Proposition~\ref{Proposition domaine local holomorphie de F} does ensure
that any solution  $\wh{u}(\la\,|\, \wh{\mathbb{X}})$ to \eqref{ecriture NLIE a Trotter fini},
\eqref{ecriture monodromie Trotter fini} on $\wh{\mc{E}}_{\mc{M}}$
satisfies the requirements $\mathrm{i)-vii)}$ of
Properties~\ref{Propriete pour deformer les contours}. Then,
Proposition~\ref{Proposition correspondance entre NLIE originelle et NLIE point fixe}
ensures that this solutions gives rise to a fixed point $g$ of\ \ $\wh{\mc{L}}_{T}$
by the substitution
\beq
g(\la)\, = \,
\wh{u}\,\big(\la\,|\, \wh{\mathbb{X}}\,\big) - \mc{W}_N(\la) -u_1\big(\la\,|\, \wh{\mathbb{X}}\,\big) \; .
\enq

Since any solution to \eqref{ecriture NLIE a Trotter fini},
\eqref{ecriture monodromie Trotter fini} on $\wh{\mc{E}}_{\mc{M}}$ gives rise to a fixed point
of\ \ $\wh{\mc{L}}_{T}$ on $\mc{E}_{\mc{M}}$, it will be unique as soon as\ \ $\wh{\mc{L}}_{T}$
admits a unique fixed point. 

\vspace{2mm}

The reasoning in the infinite Trotter number case is basically the same. 
\qed

\subsection{Stability of the operator}
\label{SousSection Stabilite operateur}

In this subsection we show that the operators $\wh{\mc{L}}_{T}$ and $\mc{L}_{T}$ both stabilise the
space $\mc{E}_{ \mc{M} }$ provided that the constant $C_{\mc{M}}$ arising in the
very definition of the space $\mc{E}_{\mc{M}}$, \textit{c.f.}\
\eqref{definition espace fonctionnel principal}, is large enough, while $T$
and $\tf{1}{(NT^4)}$ are taken small enough. 
 
\begin{prop}
\label{Proposition stabilite operateur O}
There exists $C_{\mc{M}}^{(0)}>0$ and $T_0, \eta >0$ such that for any
$C_{\mc{M}}> C_{\mc{M}}^{(0)}$, $0<T<T_{0}$ and $\eta > \tf{1}{(NT^4)}$, 
for any choices of parameters sets $\op{X}$ and $\op{Y}$ satisfying
Hypothesis~\ref{Hypotheses solubilite NLIE} with $|\op{X}_{\e{tot}}|$ and $|\op{Y}_{\e{tot}}|$ fixed, \textit{c.f.}
\eqref{definition ensemble trous et particules locales en pm q}, \eqref{definition hat Y ensemble},
\beq
     \wh{\mc{L}}_{T}\, \big[ \mc{E}_{\mc{M}} \big] \, \subset \, \mc{E}_{\mc{M}}   \qquad and \qquad
\mc{L}_{T} \, \big[ \mc{E}_{\mc{M}} \big] \, \subset \, \mc{E}_{\mc{M}}   \; ,
\enq
provided that one takes $M \ge |\op{Y}| + 1$ with $M$ as appearing
in \eqref{definition du parametre delta T}.   
 
\end{prop}

\Proof
We will only discuss the case of finite Trotter number, the infinite Trotter number case can be
treated quite analogously.
 
 
It is clear that for any $f \in \mc{E}_{\mc{M}}$, one has that
$\wh{\mc{L}}_{T}\,\big[ f \big] \in \mc{O}( \mc{M} ) $ and that
$\wh{\mc{L}}_{T} \big[ f \big](\la) \tend 0$ as $\la \tend \infty$ in $\mc{M}$.
Hence, it remains to establish the boundedness property. This will be done for each
of the two operators $\wh{\mc{L}}_{T}^{\,(a)}$, $a\in \{1,2\}$, separately. 
 
From now on, for a fixed value of $C_{\mc{M}}$, we consider a range of temperatures
$0<T<T_0$ such that $C_{\mc{M}} T_0<1$. 
 
\subsubsection*{Stability of $\wh{\mc{L}}_{T}^{\, (1)}$}
Since $\wh{f}\big( \,\wh{q}_{\mathbb{X}}^{\, (\pm)}[f] \,|\, \wh{\mathbb{X}} \big)
\, = \, 0$, one has the Taylor integral representation 
\beq
\wh{f}\big( \mu \,|\, \wh{\mathbb{X}} \big) \, = \,
\Int{0}{1} \! \dd s  \, \wh{f}^{\,\prime}\Big( \wh{q}_{\mathbb{X}}^{\, (\pm)}[f] + s \big(\mu - \wh{q}_{\mathbb{X}}^{\, (\pm)}[f]  \big) \,|\, \wh{\mathbb{X}} \Big) \cdot
\big(\mu - \wh{q}_{\mathbb{X}}^{\, (\pm)}[f]  \big) \;.
\enq
By virtue of Proposition
\ref{Proposition continuite inverse f et sa comparaison a veps inverse}, 
$\big[ \ups_{\a} q \, ; \,  \wh{q}_{\mathbb{X}}^{\, (\ups_{\a})}[f] \, \big]
\subset \wh{\mc{U}}_{\a} \subset \mc{V}_{\a}$ with
$\e{d}\big( \wh{\mc{U}}_{\a} , \Dp{}\mc{V}_{\a})>0$ uniformly in $T<T_0$, 
$T_0$ small enough, and $NT$ large enough. Hence, there exists $C_1>0$ such that 
\bem
\norm{ \wh{f}^{\,\prime}( * \,|\, \wh{\mathbb{X}} ) }_{ L^{\infty}( \wh{\mc{U}}_{\a}) }
\, \leq \, C_1 \Big( \norm{ \mc{W}_N }_{ L^{\infty}( \mc{V}_{\a} ) } \, + \, T \norm{ u_1(*\,|\, \wh{\mathbb{X}} ) }_{ L^{\infty}( \mc{V}_{\a} ) }
\, + \, \norm{ f }_{ L^{\infty}( \mc{V}_{\a} ) }  \Big) \\
\leq C_1 \Big( C \, + \, C_u T \, + \, C_{\mc{M}} T^2 \Big) \;. 
\end{multline}
Moreover, one also has the direct estimate   
\bem
     \Int{\pm q }{ \wh{q}_{\mathbb{X}}^{\, (\pm)}[f]    }   \hspace{-3mm} | \dd \mu | \big|\mu - \wh{q}_{\mathbb{X}}^{\,(\pm)}[f]  \big|   \; = \;
\Int{0}{1} \hspace{-2mm} \dd t \, \big|\,  \wh{q}_{\mathbb{X}}^{\, (\pm)}[f]  \mp q
\big|  \cdot \big| \, \wh{q}_{\mathbb{X}}^{\,(\pm)}[f]  \mp q  \, - \, t \big( \, \wh{q}_{\mathbb{X}}^{\,(\pm)}[f]  \mp q  \big) \big|   \\
\, = \, \f{1}{2} \big| \, \wh{q}_{\mathbb{X}}^{\,(\pm)}[f]  \mp q  \big|^2 \, \leq \, \f{ C T^2 }{ \rho^4} \;,
%
\end{multline}
as is readily inferred from \eqref{ecriture bornage f inverse contre veps inverse hat}, upon restricting to the range of parameters 
$\eta > \tf{ 1 }{ (NT^4) }$ with $\eta>0$ and small enough.

Thus, for any $\la \in \mc{M}$, 
\beq
\Bigg|  \Int{ \ups_{\a} q }{ \wh{q}^{ \, ( \ups_{\a} )}_{\mathbb{X}} [f] }\! \! \!  \dd \mu \, \wh{f}\big( \mu  \,|\, \wh{\mathbb{X}} \,\big)   R_{\e{c}} (\la,\mu)   \Bigg| \, \leq \,
\norm{ R_{\e{c}} }_{ L^{\infty}(\mc{M}\times \wh{\mc{U}}_{\a}) }  
\cdot   \norm{ \wh{f}^{\, \prime}\big( *  \,|\, \wh{\mathbb{X}} \,\big)  }_{ L^{\infty}( \wh{\mc{U}}_{\a}) } \cdot
\bigg|  \Int{ \ups_{\a} q }{ \wh{q}_{\mathbb{X}}^{ \, (  \ups_{\a} )}[f]    }   \hspace{-3mm} | \dd \mu | \big|\mu - \wh{q}_{\mathbb{X}}^{ \, (  \ups_{\a} )}[f]  \big|  \bigg|
\; \leq \; \wt{C} T^2  \; . 
\enq
All-in-all, this leads to
\beq
     \norm{\wh{\mc{L}}_{T}^{\,(1)} \big[ f \big] }_{ L^{\infty}(\mc{M})}
        \, \leq \, C T^2
%
\enq
for some constant $C>0$ that \textit{does not} depend on $C_{\mc{M}}$ as long as
$C_{\mc{M}}T_0<1$.

\subsubsection*{Stability of $\wh{\mc{L}}_{T}^{\,(2)}$}
One starts by decomposing the operator as 
\beq
     \wh{\mc{L}}_{T}^{\,(2)} \, = \,
        \sul{\a \in \{L,R\} }{} \big( \wh{\mc{Q}}_{T;\a}^{\, (1)}
	   + \wh{\mc{Q}}_{T;\a}^{\, (2)}\big)
	   \; + \; \sul{ \sg \in \{\pm \} }{}  \wh{\mc{Q}}_{T;\sg}^{\,(3)} \, ,  
\label{ecriture decomposition LT 2}
\enq
where we agree upon
\begin{align}
\wh{\mc{Q}}_{T;\a}^{\, (1)} \big[ f\big](\la)   & =
     - \, T \hspace{-3mm} \Int{ \wh{\msc{C}}_{\mathbb{X};\a}^{\,(\e{ext})}[f] }{}\hspace{-3mm}
        \dd \mu \: R_{\e{c}}(\la,\mu)
	\ln\Big[ 1 + \ex{-\frac{1}{T}|\wh{f}\, |(\mu \,|\, \wh{\mathbb{X}} \,) } \Big] \;,
\label{definition Q T alpha 1} \\ 
     \wh{\mc{Q}}_{T;\a}^{\, (2)} \big[ f\big](\la)   & =
        \ups_{\a} \, T \Int{ -\de_{T} }{ \de_{T} } \hspace{-1mm} \dd s  \,
        \f{ R_{\e{c}}\Big(\la, \wh{f}_{\a}^{\,-1}\big( s
	                                               \,|\, \wh{\mathbb{X}} \,\big)   \Big)  }
        { \wh{f}^{\,\prime}
	  \Big( \wh{f}_{\a}^{\, -1}\big( s   \,|\,\wh{\mathbb{X}} \, \big)
	        \,|\, \wh{\mathbb{X}} \, \Big)  }    \ln\Bigl[ 1 \, + \, \ex{- \f{|s|}{T} } \Bigr]
\label{definition Q T alpha 2}
\end{align}
and 
\beq
     \wh{\mc{Q}}_{T;\sg}^{\, (3)} \big[ f\big](\la)   \; = \;  
     - \, T\hspace{-3mm} \Int{ \wh{\ga}^{\,(\sg)}[f] }{}\hspace{-3mm}  \dd \mu \:
        R_{\e{c}}(\la,\mu)
	\ln\Big[ 1 + \ex{-\frac{1}{T}|\wh{f}\, |(\mu \,|\, \wh{\mathbb{X}} \, ) } \Big]  \;.
\label{definition Q T sigma 3}
\enq
$\wh{\mc{Q}}_{T;\a}^{\,(2)}$ is obtained by writing out explicitly the integration along
the interval $I_{\delta_T}$ deformed by $f_{\a}^{-1}$, which is part of
$\wh{\msc{C}}_{\mathbb{X}}[f]$ as it was defined in
\eqref{definition contour C de Y Trotter fini}. The integration curve
$\wh{\msc{C}}_{\mathbb{X};\a}^{\, (\e{ext})}[f]$ appearing in $\wh{\mc{Q}}_{T;\a}^{\, (1)}$
corresponds to 
\beq
     \wh{\msc{C}}_{\mathbb{X};\a}^{\, (\e{ext})}[f]
        \, = \, \bigcup_{\sg=\pm} \veps_{\a}^{-1}\Big( \wh{\mc{I}}^{\, (\sg)}_{\a} \Big) \;,
\enq
where $ \wh{\mc{I}}^{\, (\sg)}_{\a} $ is defined by
\eqref{definition contour I R pm}, \eqref{definition contour I L pm}. 
Finally, the arcs $\wh{\ga}^{\,(\sg)}[f]$ are defined just below
\eqref{definition des points limites des arcs ga sg et hat ga sg}.


We shall first estimate the contribution of $\wh{\mc{Q}}_{T;\a}^{\, (1)} \big[ f\big]$.
The endpoints $\wh{\mf{z}}^{\, (\sg)}_{\a}[f]$ admit the low-$T$ expansion
\eqref{ecriture DA basse tempe de tau et eta f}, where the remainder can be bounded by
$C T $ provided that $\eta > \tf{1}{(NT^4)}$ is small enough. Moreover, $C$ does not
depend on $C_{\mc{M}}$ provided that $T C_{\mc{M}}<1$ as follows from
Proposition~\ref{Proposition continuite inverse f et sa comparaison a veps inverse}. 
It follows that $\veps_{\a}^{-1}\Big( \wh{\mc{I}}^{\, (\sg)}_{\a} \Big)
\subset \mc{G}^{(\sg)}_{\a}$ introduced in
Proposition~\ref{Proposition borne sup sur exponentielle en tempe de la fct f lambda mathbbY}. 

Also, it is easy to see that there exists a $T$-independent, sufficiently small, open,
relatively compact, neighbourhood $\mc{W}$ of the curve $\msc{C}_{\veps}$ such that 
$\wh{\msc{C}}_{\mathbb{X};\a}^{\,(\e{ext})}[f] \subset \mc{W}$. Then, reducing $T_0>0$
and $\eta>0 $ if need be and also using that $\veps_{\a}^{-1}\Big( \wh{\mc{I}}^{\, (\sg)}_{\a} \Big)
\subset \mc{G}^{(\sg)}_{\a}$, direct bounds based on
Proposition~\ref{Proposition borne sup sur exponentielle en tempe de la fct f lambda mathbbY}
yield, for any $\la \in \mc{M}$,
\beq
     \Big| \wh{\mc{Q}}_{T;\a}^{\, (1)} \big[ f\big](\la)   \Big|
        \, \leq \,  \big| \wh{\msc{C}}_{\mathbb{X};\a}^{\,(\e{ext})}[f] \big|
	   \cdot \norm{ R_{\e{c}} }_{L^{\infty}(\mc{M}\times \mc{W})} 
        \cdot \bigg\{ - T \ln \Big[ 1- C T^{M-|\op{Y}|} \ex{C_{\mc{M}}T } \Big] \bigg\} \;. 
\enq
Furthermore,
\beq
     \big|\wh{\msc{C}}_{\mathbb{X};\a}^{(\e{ext})}[f] \big| \;  \leq   \;
%
%
	   \sum_{\sigma = \pm}
	   \Int{ 0 }{ \wh{\mf{t}}^{\, (\sg)}_{\a}[f] }
	   \f{ \dd  s }{ \Big| \veps^{\prime}\Big(
	                       \veps_{\a}^{-1}\big( \, \wh{\mf{z}}^{\, (\sg)}_{\a}[f]
			       + \i s\big) \Big)  \Big| }  
     \; \leq  \; \Int{-\eps }{ + \infty} \f{ C \dd s  }{1+s^2} \,< \, C^{\prime} \;. 
\enq
Here $\eps>0$, and we used the fact that $\wh{\mf{z}}^{\, (\sg)}_{\a}[f]$ is uniformly
bounded in the considered range of $T$ and $N$ as well as the 
asymptotics \eqref{ecriture comportement asymptotique de veps prime le long de veps alpha}. 

Thus, one eventually obtains the upper bound
\beq
     \norm{ \wh{\mc{Q}}_{T;\a}^{\,(1)}
        \big[ f\big] }_{ L^{\infty}(\mc{M})}
	\, \leq \,  C \cdot T^{M +1 - |\op{Y}|} \;.
%
\enq

Now we move on to estimating the second contribution
$\wh{\mc{Q}}_{T;\a}^{\, (2)} \big[ f\big]$. It follows from
Proposition~\ref{Proposition continuite inverse f et sa comparaison a veps inverse},
upon using $T^3 \eta > \tf{1}{(NT)}$, that
\beq
     \wh{f}_{\a}^{\, -1}\big( \intff{-\de_T}{\de_T}   \,|\, \wh{\mathbb{X}} \big)
	\subset \wh{\mc{U}}_{\a}
	\subset \mc{V}_{\a} \;.
\enq
%
%
%
%
%
%
%
%
%
%
%
This leads to 
\beq
     \Big| \wh{\mc{Q}}_{T;\a}^{\, (2)} \big[ f\big](\la) \Big|
        \, \leq \, \f{ T^2 \int_{\R}^{} \dd s \:
	\ln\big[1+\ex{-|s|} \big]
	\cdot \norm{R_{\e{c}} }_{L^{\infty}(\mc{M} \times \wh{\mc{U}}_{\a})}}
     {\e{inf}\big\{|\veps^{\prime}(s)|: s\in \mc{V}_{\a} \big\}
        - T \norm{ u_1^{\prime}\big( *  \,|\, \wh{\mathbb{X}} \,\big)  } _{ L^{\infty}(\wh{\mc{U}}_{\a}) }
     \, - \, \norm{f^{\prime}}_{L^{\infty}(\wh{\mc{U}}_{\a})}} \;. 
\enq
Upon using that $T$ is such that $T C_{\mc{M}} < 1 $ and invoking
\eqref{ecriture borne inf sur veps prime voisinage pm q} along with the classical
bounds for holomorphic functions $\norm{ g^{\prime} }_{ L^{\infty}( U) }
\leq C \norm{ g }_{ L^{\infty}( V) } $ with $U\subset V$ and $\e{d}(U, \Dp{}V)>0$,
one obtains that 
\beq
     \Big| \wh{\mc{Q}}_{T;\a}^{\, (2)} \big[ f\big](\la)   \Big|
        \, \leq \, C^{\prime} T^2
%
\enq
with $C^{\prime}>0$. 

Finally, we focus on estimating the third contribution
$\wh{\mc{Q}}_{T;\sg}^{\, (3)} \big[ f\big]$. For that purpose recall that 
the endpoints $\wh{y}_{\a}^{\,(\sg)}[f]$ of the arcs $\wh{\ga}^{\,(\sg)}$ are given by 
\beq
     \wh{y}_{\a}^{\,(\sg)}[f]
        \, = \,  - \i \f{\zeta}{2}
	\; + \; \mf{c}_{\e{d}} T \ex{ \i \wh{\vth}_{\a}^{\,(\sg)}[f] } \;,
\label{ecriture rep param en terme angle pour pts y sg alpha}
\enq
in which the angles are as defined in
Proposition~\ref{Proposition DA des angles vth et Lipschitz pour angles et t sg alpha}, eq. \eqref{equation donnant forme solution angles intersection avec D - i zeta sur 2 cd T}.
The low-$T$ expansion for $\wh{\vth}_{\a}^{\,(\sg)}[f]$ given in
\eqref{ecriture DA angle vth a basse T} ensures that, provided $C_{\mc{A}}$ is taken
large enough in \eqref{definition angles delimitant les arcs de A sg}, for $T$ small
enough and uniformly in  $f \in \mc{E}_{\mc{M}}$ it holds that
$\wh{\ga}^{\,(\sg)} \subset \mc{A}^{(\sg)}$. Here, we recall that  
the wider arcs $\mc{A}^{(\sg)}$ are defined through \eqref{definition arcs A sg}.
Thus, by virtue of
Proposition~\ref{Proposition bornage f et f hat sur arc circulaires rayon ordre T}, 
one immediately concludes that 
\beq
     \Big| \wh{\mc{Q}}_{T;\sg}^{\, (3)} \big[ f\big](\la) \Big|
        \, \leq \,  2\pi T^{1+M-|\op{Y}|}  \wt{C}   \cdot
	\norm{ R_{\e{c}} }_{  L^{\infty}(\mc{M} \times
	\op{D}_{- \i\tf{\zeta}{2}, \mf{c}_{\e{d}} T}) } \;,
\enq
in which $\wt{C}$ only depends on $|\op{X}_{\e{tot}}|$, $|\op{Y}_{\e{tot}}|$ and $|\op{Y}_{\e{sg}}|$.

This yields that for some constant $C>0$ and $0<T<T_0$, $\eta > \tf{1}{ (NT^4)}$,
$T_0, \eta>0$ and small enough 
\beq
     \norm{\wh{\mc{L}}_{T}^{\, (2)} \big[ f \big] }_{ L^{\infty}(\mc{M})}
        \, \leq \, C T^2 \;,
\enq
provided that $M \ge |\op{Y}| + 1$.

At this stage, it is straightforward to conclude, by putting together the bounds
on $\wh{\mc{L}}_{T}^{(a)}[f]$, $a = 1, 2$, that there exists $\mc{C}_{\mc{M}}^{(0)}>0$ large
enough and $T_0, \eta >0$ small enough such that, for all $\mc{C}_{\mc{M}}>\mc{C}_{\mc{M}}^{(0)}$
and $0<T<T_0$, $\eta > \tf{1}{(NT^4)}$ -- in particular such that $\mc{C}_{\mc{M}} T_0 <1$ --
it holds that
\beq
\norm{ \wh{\mc{L}}_{T} [f] }_{ L^{\infty}(\mc{M}) } \,  \leq \, C_{\mc{M}} T^2 \;,
\enq
which is precisely the sought result. \qed





\subsection{Contractivity of the operator}
\label{SousSection Contractivite Operateur}

\begin{prop}
\label{Proposition contractivite operateur}
Let $M \ge |\op{Y}| + 1$, where $M$ is the constant defined in
\eqref{definition du parametre delta T}. Then there exist $C>0$ and $T_0, \eta>0$ small
enough such that, for any $0<T<T_0$ and $\eta > \tf{1}{(NT^4)}$,
\beq
\norm{ \, \wh{\mc{L}}_{T} [f] - \wh{\mc{L}}_{T}[g] \,  }_{ L^{\infty}(\mc{M}) } \, \leq \, C  \cdot T \cdot \norm{ f -g \, }_{ L^{\infty}(\mc{M}) } \qquad for\;  all \quad f, g \in \mc{E}_{\mc{M}} \;.
\enq
Likewise, for any  $0<T<T_0$,
\beq
\norm{ \, \mc{L}_{T}[f] - \mc{L}_{T}[g] \,  }_{ L^{\infty}(\mc{M}) } \, \leq \, C  \cdot T \cdot \norm{ f -g \, }_{ L^{\infty}(\mc{M}) } \qquad for\;  all \quad f, g \in \mc{E}_{\mc{M}} \;.
\enq
\end{prop}

\Proof 
For the sake of clarity, we shall insist below on the function-dependence of the
sets $\wh{\mathbb{X}}$, \textit{viz}.\ use the notation $\wh{\mathbb{X}}[f]$,
$\wh{\mathbb{X}}[g]$.

\subsubsection*{Contractivity of $\wh{\mc{L}}_{T}^{(1)}$ }
One starts by decomposing
\bem
     \wh{\mc{L}}_{T}^{\,(1)} \big[ f \big](\la) \, - \,
        \wh{\mc{L}}_{T}^{\,(1)} \big[ g \big](\la)  \; = \;  
     - \, \sul{\sg = \pm  }{}  \sg  \Int{  \sg q }{ \wh{q}^{\,(\sg)}_{\mathbb{X}}[f] }\! \!\dd \mu \:
     \Big[  \wh{f}\, \big( \mu  \,|\, \wh{\mathbb{X}}[f] \big) \, - \, \wh{g}\, \big( \mu  \,|\, \wh{\mathbb{X}}[g] \big)\Big] R_{\e{c}}(\la,\mu) \\[-1ex]
     \, + \, \sul{\sg = \pm  }{}  \sg
        \Int{ \wh{q}^{\,(\sg)}_{\mathbb{X}}[f] }{ \wh{q}^{\,(\sg)}_{\mathbb{X}}[g] }\! \!\dd \mu \:
        \wh{g}\,\big( \mu  \,|\, \wh{\mathbb{X}}[g] \big)  R_{\e{c}}(\la,\mu) \;.
\end{multline}
Then, one bounds each term separately. For any $\la \in \mc{M}$, it holds that
%
%
%
%
%
%
%
%
%
%
%
%
%
%
%
%
\bem
\bigg|   \Int{  \ups_{\a} q }{ \wh{q}^{\, (\ups_{\a} )}_{\mathbb{X}}[f] }\! \!  \dd \mu \,
\Big[  \wh{f}\,\big( \mu\,|\,  \wh{\mathbb{X}}[f] \big)    \, - \, \wh{g}\, \big(\mu\,|\,  \wh{\mathbb{X}}[g] \big) \Big] \, R_{\e{c}}(\la, \mu)   \bigg| \\
\; = \; \bigg|   \Int{  \ups_{\a} q }{ \wh{q}^{\,(\ups_{\a} )}_{\mathbb{X}}[f] }\! \!  \dd \mu \,
\Big[  f( \mu  )  \, + \,  T u_1\big( \mu\,|\,  \wh{\mathbb{X}}[f] \big)  \, - \, g( \mu  ) \, - \,  T u_1\big(\mu\,|\,  \wh{\mathbb{X}}[g] \big)\Big]\,  R_{\e{c}}(\la, \mu)   \bigg| \\
\, \leq \, \norm{R_{\e{c}}}_{L^{\infty}(\mc{M}\times \mc{V}_{\a} ) } \cdot
\Big\{  \norm{f-g}_{ L^{\infty}( \mc{W}_{\a} ) } \, + \, C  T  \norm{f-g}_{ L^{\infty}\big( \cup_{\a \in \{L,R\}} \mc{W}_{\a} \big) }  \Big\}
\cdot \big| \, \wh{q}^{\,(\ups_{\a})}_{\mathbb{X}}[f] - \ups_{\a} q \, \big|  \\
\, \leq \,  C T \cdot  \underset{\a\in \{L,R\}}{\e{max}}\norm{f-g}_{ L^{\infty}( \mc{W}_{\a} ) } \;.
%
%
%
%
%
\end{multline}
Above we made use of
Proposition~\ref{Proposition continuite inverse f et sa comparaison a veps inverse}, have set
\beq
\mc{W}_{\a} \; = \; \veps_{\a}^{-1}(\op{D}_{0,2\rho} ) \;, \quad \a \in \{L,R\} \;,
\label{definition voisinage W alpha}
\enq
with $\rho>0$ small enough as introduced in Proposition~\ref{Proposition continuite inverse f et sa comparaison a veps inverse},
and bounded $\tf{1}{(NT)} \leq \eta T^3$
by using $\eta > \tf{ 1 }{ (NT^4) }$ and small enough so as to infer
\beq
     \Big| \, \wh{q}^{\,(\ups_{\a})}_{\mathbb{X}}[f] - \ups_{\a} q   \Big|  \, \leq \,
        C T  \quad \text{which also implies that} \qquad 
        \big[  \ups_{\a} q \, ; \,  \wh{q}^{\,(\ups_{\a})}_{\mathbb{X}}[f]\, \big]
	\subset \wh{\mc{U}}_{\a} \subset \mc{V}_{\a} \subset \mc{W}_{\a}\;.
\enq
Furthermore, by virtue of Proposition \ref{Proposition existence et continuite parameters particule trou partiels}, one has
\beq
\norm{  u_1(*\,|\,  \wh{\mathbb{X}}[f] ) \, - \, u_1(*\,|\,  \wh{\mathbb{X}}[g] ) }_{ L^{\infty}( \mc{W}_{\a} ) } \; \leq \; C \norm{f-g}_{ L^{\infty}\big( \cup_{\a \in \{L,R\}} \mc{W}_{\a} \big) }
\label{ecriture bornes sur difference fnelle des u1}
\enq

Since the open neighbourhood $\wh{\mc{U}}_{\a}$ introduced in
Proposition~\ref{Proposition continuite inverse f et sa comparaison a veps inverse} 
contains a $T$ and $\tf{1}{(NT)}$-independent neighbourhood of $\ups_{\a} q$, provided
that these are both small enough, one obtains from the low-$T$, large-$NT^4$ estimates
given above that 
\beq
\wh{q}_{\mathbb{X}}^{\, (\ups_{\a})}[f] + s \big(\mu - \wh{q}_{\mathbb{X}}^{\, (\ups_{\a})}[f] \big) \in \wh{\mc{U}}_{\a}
                                   \qquad \e{for} \qquad s \in \intff{0}{1}\quad \e{and} \quad \mu \in
\big[\,   \ups_{\a} q \, ; \,  \wh{q}^{\, (\ups_{\a})}_{\mathbb{X}}[f]\, \big] \;.
\enq
Hence, the Taylor-integral formula applied to the second term yields
\bem
\bigg|   \Int{ \wh{q}^{\,(\ups_{\a})}_{\mathbb{X}}[f] }{ \wh{q}^{\, (\ups_{\a})}_{\mathbb{X}}[g] }\hspace{-3mm} \dd \mu \,
               \wh{g}\, \big(\mu\,|\,  \wh{\mathbb{X}}[g] \big)  R_{\e{c}}(\la, \mu) \bigg| \, = \,
\bigg|   \Int{ \wh{q}^{\, (\ups_{\a})}_{\mathbb{X}}[f] }{ \wh{q}^{\, (\ups_{\a})}_{\mathbb{X}}[g] }\hspace{-3mm} \dd \mu \, R_{\e{c}}(\la, \mu)
\Int{0}{1} \! \dd s  \,   \wh{g}\;'\,\Big( \, \wh{q}_{\mathbb{X}}^{\, (\ups_{\a})}[g] + s \big(\mu - \wh{q}_{\mathbb{X}}^{\, (\ups_{\a})}[g]  \big) \,|\, \wh{\mathbb{X}}[g] \Big)
\cdot \big(\mu - \wh{q}_{\mathbb{X}}^{\, (\ups_{\a})}[g] \,  \big)   \bigg| \\
\; \leq \; C  \norm{\, \wh{g}\,\big( *\,|\,  \wh{\mathbb{X}}[g]\big)  }_{ L^{\infty}(\mc{V}_{\a}) } \big| \,  \wh{q}^{\, (\ups_{\a})}_{\mathbb{X}}[f] - \wh{q}^{\, (\ups_{\a})}_{\mathbb{X}}[g]  \big|^2
\; \leq \; C^{\prime }   \norm{ f-g}_{ L^{\infty}(\mc{W}_{\a}) }^2  \;.
\end{multline}
Above we have used that, since $\e{d}\Big( \wh{\mc{U}}_{\a}\, , \, \Dp{}\mc{V}_{\a} \Big)>0$
uniformly in $T$, $\tf{1}{(NT^4)}$ small enough, for any holomorphic function 
on $\mc{V}_{\a}$, $\norm{f^{\prime} }_{ L^{\infty}(\wh{\mc{U}}_{\a}) } \, \leq \,
C \cdot \norm{ f }_{ L^{\infty}(\mc{V}_{\a}) }$ for a geometric constant that only depends on 
$\wh{\mc{U}}_{\a}$ and $\mc{V}_{\a}$. Further,  we invoked
\eqref{ecriture continuite inverse f} in the penultimate bound.

All-in-all, this entails that  for $0<T<T_0$ and $\eta > \tf{1}{(NT^4)}$ with
$T_0, \eta>0$ and small enough
\beq
\norm{  \wh{\mc{L}}_{T}^{\, (1)} \big[ f \big]  \, - \,  \wh{\mc{L}}_{T}^{\,(1)} \big[ g \big]  \, }_{ L^{\infty}(\mc{M}) }  \, \leq \,
 C \cdot \Big( T \, + \, \underset{\a\in \{L,R\} }{\e{max}} \big\{ \norm{ f - g \, }_{ L^{\infty}(\mc{W}_{\a}) } \big\} \Big)
\cdot \underset{\a\in \{L,R\} }{\e{max}} \big\{ \norm{ f - g \, }_{ L^{\infty}(\mc{W}_{\a}) } \big\}
\label{ecriture contractivite L1 avec un control precis sur les domaines}
\enq
for some $C>0$.

Then, one may further simplify the bound by noticing that
\beq
\norm{ f - g \, }_{ L^{\infty}(\mc{W}_{\a}) }  \; \leq \; \norm{ f - g \, }_{ L^{\infty}(\mc{M}) }  \; \leq \; 2  C_{\mc{M}} T^2 \;.
\enq
Therefore,  for $0<T<T_0$ and $\eta > \tf{1}{(NT^4)}$ with
$T_0, \eta>0$ and small enough
\beq
\norm{  \wh{\mc{L}}_{T}^{\, (1)} \big[ f \big]  \, - \,  \wh{\mc{L}}_{T}^{\,(1)} \big[ g \big]  \, }_{ L^{\infty}(\mc{M}) }  \, \leq \, C  T \,  \norm{ f-g \, }_{ L^{\infty}(\mc{M}) }
\label{ecriture contractivite L1}
\enq
for some $C>0$.

\subsubsection*{Contractivity of $\wh{\mc{L}}_{T}^{\, (2)}$ }

Recall the decomposition \eqref{ecriture decomposition LT 2} of $\wh{\mc{L}}_{T}^{\, (2)}$. 
Below, we treat each of the building blocks separately, first focusing on
$ \wh{\mc{Q}}_{T;\a}^{\,(1)}$, see \eqref{definition Q T alpha 1}. 
The latter operator takes the explicit form 
\beq
     \wh{\mc{Q}}_{T;\a}^{\,(1)} \big[ f\big](\la)
        \, = \,  - \, T \sul{\sg = \pm }{}
	           \Int{  \veps_{\a}^{-1}\big( \wh{\mc{I}}^{\,(\sg)}_{\a} [f] \big)}{}
     \hspace{-5mm}  \dd \mu \: R_{\e{c}}(\la,\mu)
        \ln\Big[ 1 + \ex{-\frac{1}{T} \sg  \wh{f}\,  (\mu \,|\, \wh{\mathbb{X}}[f] ) } \Big]  \;.
\enq
Upon using
Proposition~\ref{Proposition borne sup sur exponentielle en tempe de la fct f lambda mathbbY}
and the fact that
\bem
\veps_{\a}^{-1}\Big( \Big[ \, \wh{\mf{z}}_{\a}^{\,(\sg)}[f] \, +\, \i \, \wh{\mf{t}}_{\a}^{\, (\sg)}[f]
	\,;\, \wh{\mf{z}}_{\a}^{\, (\sg)}[f]  \Big] \Big) \; = \;
\veps_{\a}^{-1}\Big( \wh{\mc{J}}_{\a;3}^{\, (\sg)}[f,g]  \Big) \\
\cup
\veps_{\a}^{-1}\Big( \Big[ \, \wh{\mf{z}}_{\a}^{\,(\sg)}[g] \, +\, \i \, \wh{\mf{t}}_{\a}^{\, (\sg)}[g]
	\,;\, \wh{\mf{z}}_{\a}^{\, (\sg)}[g]  \Big] \Big)
\cup
\veps_{\a}^{-1}\Big( \wh{\mc{J}}_{\a;1}^{\, (\sg)}[f,g]  \Big)
\end{multline}
with
\beq
     \wh{\mc{J}}_{\a;1}^{\, (\sg)}[f,g] \, = \,
        \Big[ \, \wh{\mf{z}}_{\a}^{\, (\sg)}[g] \,;\, \wh{\mf{z}}_{\a}^{\, (\sg)}[f] \,\Big]
	\qquad \e{and} \qquad
     \wh{\mc{J}}_{\a;3}^{\, (\sg)}[f,g] \, = \,
        \Big[ \, \wh{\mf{z}}_{\a}^{\,(\sg)}[f] \, +\, \i \, \wh{\mf{t}}_{\a}^{\, (\sg)}[f]
	\,;\, \wh{\mf{z}}_{\a}^{\, (\sg)}[g] \, +\, \i \,  \wh{\mf{t}}_{\a}^{\, (\sg)}[g]
	   \Big]  \:,
\enq
one obtains
\beq
     \wh{\mc{Q}}_{T;\a}^{\,(1)} \big[ f\big](\la)
        \, - \, \wh{\mc{Q}}_{T;\a}^{\, (1)} \big[ g \big](\la)
	\, = \, \wh{\mc{T}}_{T;\a}^{\,(1)}[f,g](\la)
	\, + \, \wh{\mc{T}}_{T;\a}^{\,(2)}[f,g](\la)
	\, + \, \wh{\mc{T}}_{T;\a}^{\,(3)}[f,g](\la) \;, 
\enq
where, for $k\in \{1,3\}$, 
\beq
     \wh{\mc{T}}_{T;\a}^{\,(k)}[f,g](\la) \; = \;
        - \, T \sul{\sg = \pm }{} \sg \ups_{\a} \hspace{-6mm}
	   \Int{  \veps_{\a}^{-1}\big( \wh{\mc{J}}_{\a;k}^{\, (\sg)}[f,g] \big) }{  }
        \hspace{-5mm} \dd \mu \: R_{\e{c}}(\la,\mu)
	   \ln\Big[ 1 + \ex{-\frac{1}{T} \sg   \wh{f}\, (\mu \,|\, \wh{\mathbb{X}}[f] ) } \Big]
\enq
and where
\beq
     \wh{\mc{T}}_{T;\a}^{\,(2)}[f,g](\la) \, = \,
        \sul{\sg = \pm }{}  \hspace{-2mm}
	\Int{ \veps_{\a}^{-1}\big( \wh{\mc{I}}_{\a}^{\,(\sg)}[g] \big) }{}
     \hspace{-6mm}  \dd \mu \: R_{\e{c}}(\la,\mu)  \Int{0}{1} \dd s \cdot
     \f{\sg \, \big[ \wh{f}\, \big( \mu \,|\, \wh{\mathbb{X}}[f] \big) - \wh{g}\, \big( \mu \,|\, \wh{\mathbb{X}}[g] \big) \big]}
       {  1 + \ex{ \frac{\sg }{T} s  \wh{f}\,( \mu \,|\, \wh{\mathbb{X}}[f] )
                 + \frac{\sg }{T} (1-s)  \wh{g}\, ( \mu \,|\, \wh{\mathbb{X}}[g] )  }  } \;.
\enq

Each term of the decomposition should be analysed separately. Since one has that
$\wh{\mf{z}}_{\a}^{\, (\sg)}[f] = \sg \de_{T} \, + \, \e{O}(T)$ uniformly in $f$ and in
$\eta> \tf{1}{(NT^4)}$ and with a remainder that does only depend on
$|\op{X}_{\e{tot}}|$, $|\op{Y}_{\e{tot}}|$ and $|\op{Y}_{\e{sg}}|$, it follows that $\wh{\mc{J}}_{\a;1}^{\, (\sg)}[f,g]
\subset \mc{G}^{(\sg)}_{\a}$, see
\eqref{ecriture domain bornage exponentielle de proche de C ref}.  
Thus, since $\veps_{\a}^{-1}\big( \big[ \, \wh{\mf{z}}_{\a}^{\, (\sg)}[g] \,;\,
\wh{\mf{z}}_{\a}^{\, (\sg)}[f] \big] \big)$ is homotopic to 
$\big[ \, \veps_{\a}^{-1}\big( \, \wh{\mf{z}}_{\a}^{\,(\sg)}[g] \big) \,;\,
\veps_{\a}^{-1}\big( \, \wh{\mf{z}}_{\a}^{\,(\sg)}[f]\big) \big]$
on $\veps_{\a}^{-1}\big( \mc{G}^{(\sg)}_{\a} \big)$, taken that
Proposition~\ref{Proposition borne sup sur exponentielle en tempe de la fct f lambda mathbbY}
ensures $\mu \mapsto
\ln\Big[ 1 + \ex{-\frac{1}{T} \sg   \wh{f}\, (\mu \,|\, \wh{\mathbb{X}}[f] ) } \Big]$ to be
holomorphic on $\veps_{\a}^{-1}\big( \mc{G}^{(\sg)}_{\a} \big)$, one obtains
\beq
 \wh{\mc{T}}_{T;\a}^{\, (1)}[f,g](\la)  \; = \; - \, T \sul{\sg = \pm }{} \sg \ups_{\a} \hspace{-6mm}
               \Int{   \veps_{\a}^{-1}\big( \mf{z}_{\a}^{(\sg)}[g] \big)  }{ \veps_{\a}^{-1}\big( \mf{z}_{\a}^{(\sg)}[f]\big)   }
 \hspace{-5mm}  \dd \mu \: R_{\e{c}}(\la,\mu) \ln\Big[ 1 + \ex{-\frac{1}{T} \sg   \wh{f}\, (\mu \,|\, \wh{\mathbb{X}}[f] ) } \Big] \;.
\enq
At this stage, it remains to observe that $\big[ \, \veps_{\a}^{-1}\big(
\, \wh{\mf{z}}_{\a}^{\, (\sg)}[g] \big) \,;\,
\veps_{\a}^{-1}\big(\, \wh{\mf{z}}_{\a}^{\, (\sg)}[f]\big) \big]
\subset \mc{V}_{\a}$ as introduced
in Proposition~\ref{Proposition continuite inverse f et sa comparaison a veps inverse}. Upon
using \eqref{definition tau pm de f et alpha} and the bounds provided by 
Propositions \ref{Proposition continuite inverse f et sa comparaison a veps inverse} and
\ref{Proposition borne sup sur exponentielle en tempe de la fct f lambda mathbbY}, one
concludes that, given $\la \in \mc{M}$,
\bem
\Big| \wh{\mc{T}}_{T;\a}^{\, (1)}[f,g](\la) \Big| \, \leq  \,  C T^{M+1-|\op{Y}|} \norm{ R_{\e{c}} }_{L^{\infty}(\mc{M}\times \mc{V}_{\a}) } \cdot 
\sul{\sg=\pm}{}\big| f_{\a}^{-1}\big(\sg \de_T  \,|\, \wh{\mathbb{X}}[f] \big)
                         \, -\,  g_{\a}^{-1}\big(\sg \de_T   \,|\, \wh{\mathbb{X}}[g]\big) \big|  \\
\; \leq  \; C T^{M+1-|\op{Y}|} \cdot \norm{ f- g }_{ L^{\infty}\big( \cup_{\a\in \{L,R\} } \mc{W}_{\a} \big) } \;,
\end{multline}
where we recall that $\mc{W}_{\a}$ is as introduced in \eqref{definition voisinage W alpha}.
\vspace{2mm}

Similarly, since $\wh{\mf{t}}_{\a}^{\, (\sg)}[f]$, $\wh{\mf{t}}_{\a}^{\, (\sg)}[g]$ admit the low-$T$, low-$\tf{1}{(NT^4)}$
expansion \eqref{ecriture DA mf t a basse T}, it follows that
$\wh{\mc{J}}_{\a;3}^{\, (\sg)}[f,g] \subset \mc{G}^{(\sg)}_{\a}$ as can be seen from
\eqref{ecriture domain bornage exponentielle de proche de C ref}. 
Furthermore,
$
 \veps_{\a}^{-1}\Big( \Big[\, \wh{\mf{z}}_{\a}^{\,(\sg)}[f] \, +\, \i \, \wh{\mf{t}}_{\a}^{\, (\sg)}[f] 
\,;\, \, \wh{\mf{z}}_{\a}^{\,(\sg)}[g] \, +\, \i \, \wh{\mf{t}}_{\a}^{\, (\sg)}[g] \Big] \Big)
$
is clearly homotopic to 
$
\Big[ \, \veps_{\a}^{-1}\Big( \, \, \wh{\mf{z}}_{\a}^{\,(\sg)}[f] \, +\, \i \, \wh{\mf{t}}_{\a}^{\, (\sg)}[f]  \Big)
 \,;\, \veps_{\a}^{-1}\Big( \,   \wh{\mf{z}}_{\a}^{\,(\sg)}[g] \, +\, \i \, \wh{\mf{t}}_{\a}^{\, (\sg)}[g]\Big) \Big] 
$
on $\veps_{\a}^{-1}\big( \mc{G}^{(\sg)}_{\a} \big)$. Since
Proposition~\ref{Proposition borne sup sur exponentielle en tempe de la fct f lambda mathbbY}
ensures that the map $\mu \mapsto
\ln\Big[ 1 + \ex{-\frac{1}{T} \sg   \wh{f}\, (\mu \,|\, \wh{\mathbb{X}}[f]) } \Big]$ is holomorphic
on $\veps_{\a}^{-1}\big( \mc{G}^{(\sg)}_{\a} \big)$, by using that $(\veps_{\a}^{-1})^{\prime}$
is bounded on $ \mc{G}^{(\sg)}_{\a}$, given $\la \in \mc{M}$, one obtains the upper bound
\bem
\Big| \wh{\mc{T}}_{T;\a}^{\, (3)}[f,g](\la) \Big| \, \leq  \,  C T^{M+1-|\op{Y}|} \cdot \underset{\sg=\pm}{\e{max}} \norm{ R_{\e{c}} }_{L^{\infty}\big(\mc{M}\times \veps_{\a}^{-1}( \mc{G}_{\a}^{(\sg)} ) \big) } \\
\times\sul{\sg=\pm}{}\Big|\, \wh{\mf{z}}_{\a}^{\,(\sg)}[f] \, +\, \i \, \wh{\mf{t}}_{\a}^{\, (\sg)}[f]  \,-\, \wh{\mf{z}}_{\a}^{\, (\sg)}[g] \, - \, \i \,  \wh{\mf{t}}_{\a}^{\, (\sg)}[g]   \Big|  
\; \leq  \; C T^{M+1-|\op{Y}|} \cdot \norm{ f- g }_{ L^{\infty}\big( \cup_{\a\in \{L,R\}} \mc{W}_{\a} \big) } \;.
\end{multline}
The last bound here follows from
Proposition~\ref{Proposition DA des angles vth et Lipschitz pour angles et t sg alpha}.

\vspace{2mm}

Finally, it is easy to see that
\beq
\Big| \veps_{\a}^{-1}\big(\mf{z}_{\a}^{(\sg)}[g]+\i \R^{-\ups_{\a}} \big) \Big| \leq C
\quad \e{and} \quad
 \veps_{\a}^{-1}\big( \wh{\mc{I}}_{\a}^{\,(\sg)}[g] \big) \subset \mc{M}\setminus \op{D}_{ - \i \frac{\zeta}{2}, \mf{c}_{\e{d}} T }
\enq
so that, employing the lower and upper bounds provided by
Proposition~\ref{Proposition borne sup sur exponentielle en tempe de la fct f lambda mathbbY},
one immediately infers that 
\beq
\Big| \wh{\mc{T}}_{T;\a}^{\,(2)}[f,g](\la) \Big| \, \leq  \,  C T^{M -|\op{Y}|}\cdot
\norm{ f- g }_{ L^{\infty}\big( \mc{M}\setminus \op{D}_{ - \i \tf{\zeta}{2}, \mf{c}_{\e{d}} T } \big) } \;.
\enq
Hence, all-in-all, we have established that, provided that $0<T<T_0$ and $\eta > \tf{1}{(NT^4)}$,  
\beq
     \norm{ \, \wh{\mc{Q}}_{T;\a}^{\,(1)} \big[ f\big]
        \, - \, \wh{\mc{Q}}_{T;\a}^{\,(1)} \big[ g \big]  \, }_{ L^{\infty}(\mc{M}) }
	\, \leq  \,  C T^{M -|\op{Y}|} \cdot \norm{ f- g }_{ L^{\infty}\big( \mc{M}\setminus \op{D}_{ - \i \tf{\zeta}{2}, \mf{c}_{\e{d}} T } \big) } \;.
\label{ecriture borne contractivite Q1}
\enq
%
%
%


We now turn to $\wh{\mc{Q}}_{T;\a}^{\,(2)}$ introduced in \eqref{definition Q T alpha 2}.
One has the decomposition
\beq
     \wh{\mc{Q}}_{T;\a}^{\, (2)} \big[ f\big](\la)  \, - \,
        \wh{\mc{Q}}_{T;\a}^{\, (2)} \big[ g \big](\la) \, = \,
	\wh{\mc{M}}_{T;\a}^{\,(1)}[f,g](\la) \, + \,
	\wh{\mc{M}}_{T;\a}^{\, (2)}[f,g](\la) \;,
\enq
where we have introduced 
\beq
     \wh{\mc{M}}_{T;\a}^{\, (1)}[f,g](\la) \, = \,   \ups_{\a} T \Int{ -\de_{T} }{ \de_{T} } \hspace{-1mm} \dd s  \:
\f{ \ln\Bigl[ 1 \, + \, \ex{- \f{|s|}{T} } \Bigr] } { \wh{f}^{\, \prime}\Big( \wh{f}_{\a}^{\, -1}\big( s  \,|\, \wh{\mathbb{X}}[f] \big)\,|\, \wh{\mathbb{X}}[f] \Big)}
     \Int{ \wh{g}_{\a}^{\, -1}\big( s  \,|\, \wh{\mathbb{X}}[f] \big) }
         { \wh{f}_{\a}^{\, -1}\big( s  \,|\, \wh{\mathbb{X}}[g] \big) }
	 \hspace{-5mm} \dd t \: \Dp{t} R_{\e{c}}(\la, t )
\enq
and 
\bem
     \wh{\mc{M}}_{T;\a}^{\, (2)}[f,g](\la) \, = \,
        \ups_{\a} T \Int{ -\de_{T} }{ \de_{T} } \hspace{-1mm} \dd s \:
        \ln\big[ 1 \, + \, \ex{- \f{|s|}{T} } \big]   
        R_{\e{c}}\Big(\la,  \wh{g}_{\a}^{\, -1}\big(s    \,|\, \wh{\mathbb{X}}[g] \big) \Big) \\
        \times \f{ \wh{g}^{\, \prime}\Big( \wh{g}_{\a}^{\, -1}\big( s   \,|\, \wh{\mathbb{X}}[g] \big) \,|\, \wh{\mathbb{X}}[g] \Big)
        \, - \, \wh{f}^{\, \prime}\Big( \wh{f}_{\a}^{\, -1}\big( s  \,|\, \wh{\mathbb{X}}[f] \big)  \,|\, \wh{\mathbb{X}}[f] \Big)}
        {\wh{f}^{\, \prime}\Big( \wh{f}_{\a}^{\, -1}\big( s  \,|\, \wh{\mathbb{X}}[f] \big)\,|\, \wh{\mathbb{X}}[f] \Big)
        \cdot \wh{g}^{\prime}\Big( \wh{g}_{\a}^{\, -1}\big( s \,|\, \wh{\mathbb{X}}[g] \big)\,|\, \wh{\mathbb{X}}[g] \Big)} \;.
\end{multline}

In order to estimate $ \wh{\mc{M}}_{T;\a}^{\, (1)}[f,g]$, one observes that, owing to
Proposition~\ref{Proposition continuite inverse f et sa comparaison a veps inverse}, it
holds that
\beq
\wh{f}_{\a}^{-1}\big( \intff{-\de_T}{\de_T} \,|\, \wh{\mathbb{X}}[f] \big) \subset \wh{\mc{U}}_{\a} \subset \mc{V}_{\a} \, .
\enq
Since, moreover, $\intff{-\de_T}{\de_T}  \subset \op{D}_{0,\rho}$ provided that
$T$ is small enough, direct bounds and \eqref{ecriture continuite inverse f} yield that,
for $\la \in \mc{M}$, 
\bem
 \big| \wh{\mc{M}}_{T;\a}^{\,(1)}[f,g](\la) \big| \, \leq  \; 
 \norm{ \wh{f}_{\a}^{\,-1}\big( * \,|\, \wh{\mathbb{X}}[f] \big) \, - \, \wh{g}_{\a}^{\, -1}\big( * \,|\, \wh{\mathbb{X}}[g] \big) }_{ L^{\infty} ( \op{D}_{0,\rho} )  } \\
\times\f{  T^2  \cdot\norm{\Dp{2} R_{\e{c}} }_{ L^{\infty}(\mc{M}\times \wh{\mc{U}}_{\a}) }
\cdot \int_{\R}{} \dd s \: \ln \bigl[1 + \ex{-|s|}\bigr] }
{  \e{inf}\{|\mc{W}_N^{\prime}(\la)| \, : \, \la \in \wh{\mc{U}}_{\a}\} \, - \, T \norm{ u_1^{\prime}(*\,|\,\wh{\mathbb{X}}[f]) }_{   L^{\infty}( \wh{\mc{U}}_{\a}) }
\, -\, \norm{ f^{\prime}  }_{   L^{\infty}( \wh{\mc{U}}_{\a}) }     } 
\, \leq \, C T^2 \norm{f-g }_{ L^{\infty}\big(  \cup_{\a\in \{L,R\}} \mc{W}_{\a}\big) }  \;.
\end{multline}
In the intermediate step, we have used that $\mc{W}^{\prime}_N=\veps^{\prime}
+ \e{O}\big( \tfrac{1}{NT} \big)$ uniformly on $\mc{V}_{\a}$, and that $0<T<T_0$, 
$\eta > \tf{1}{(NT^4)}$ are small enough. In the last step we have used
\eqref{ecriture borne inf sur veps prime voisinage pm q}.

\vspace{2mm}

To bound $\wh{\mc{M}}_{T;\a}^{\, (2)}[f,g]$, one uses that
\bem
 \Big|\, \wh{g}^{\, \prime}\Big(\, \wh{g}_{\a}^{\, -1}\big( s   \,|\, \wh{\mathbb{X}}[g] \big)\,|\, \wh{\mathbb{X}}[g] \Big)
\, - \, \wh{f}^{\, \prime}\Big( \, \wh{f}_{\a}^{\, -1}\big( s  \,|\, \wh{\mathbb{X}}[f]  \big)\,|\, \wh{\mathbb{X}}[f] \Big) \Big| \\
\; \leq \;  \bigg|   \wh{g}^{\, \prime}  \Big(\, \wh{g}_{\a}^{\, -1}\big( s   \,|\, \wh{\mathbb{X}}[g] \big)\,|\, \wh{\mathbb{X}}[g] \Big)
\, - \,  \wh{f}^{\, \prime}  \Big(\, \wh{g}_{\a}^{\, -1}\big( s   \,|\, \wh{\mathbb{X}}[g] \big)\,|\, \wh{\mathbb{X}}[f] \Big) \bigg|
\; + \; \bigg|  \Int{ \wh{g}_{\a}^{\, -1}\big( s \,  \,|\, \wh{\mathbb{X}}[g]   \big) }{ \wh{f}_{\a}^{\, -1}\big( s  \,|\, \wh{\mathbb{X}}[f]  \big) } \hspace{-5mm} \dd t \:
\wh{f}^{\, \prime \prime}\big(t \,|\, \wh{\mathbb{X}}[f]  \big)\bigg| \\
\; \leq \; C \norm{f-g }_{ L^{\infty}\big( \cup_{\a\in \{L,R\}} \mc{W}_{\a} \big) } \, + \, C^{\prime}
\Big( \norm{ \mc{W}_{N}^{ \prime \prime}  }_{   L^{\infty}( \wh{\mc{U}}_{\a}) }  \, + \, T  \norm{u_1^{ \prime \prime}(*\,|\, \wh{\mathbb{X}}[f] ) }_{   L^{\infty}(\wh{\mc{U}}_{\a}) }
\, +\,  \norm{ f^{ \prime \prime}  }_{   L^{\infty}(\wh{\mc{U}}_{\a}) }   \Big)  \cdot \norm{f-g }_{ L^{\infty}\big( \cup_{\a\in \{L,R\}} \mc{W}_{\a} \big)  } \\
\, \leq \,  C'' \norm{f-g }_{ L^{\infty}\big( \cup_{\a\in \{L,R\}} \mc{W}_{\a} \big) } \;.
\end{multline}
With this result available, it remains to repeat the same chain of bounds as for $ \wh{\mc{M}}_{T;\a}^{(1)}[f,g](\la) $ so as to infer that 
\beq
 \big| \wh{\mc{M}}_{T;\a}^{\, (2)}[f,g](\la) \big| \, \leq     \, C T^2 \cdot \norm{f-g }_{ L^{\infty}\big( \cup_{\a\in \{L,R\}} \mc{W}_{\a} \big)  }  \;.
\enq
Hence, provided that $0<T<T_0$ and $\eta > \tf{1}{(NT^4)}$, one concludes that
\beq
\norm{\, \wh{\mc{Q}}_{T;\a}^{\, (2)} \big[ f\big]   \, - \, \wh{\mc{Q}}_{T;\a}^{\, (2)} \big[ g \big]  }_{L^{\infty}(\mc{M}) } \; \leq  \;  C T^2
\cdot \underset{\a \in \{L,R\}}{\e{max}}\Big\{ \norm{f-g}_{L^{\infty}(\mc{W}_{\a}) } \Big\}
\label{ecriture borne contractivite Q2}
\enq
for some $C>0$.

We finally turn towards estimating $\wh{\mc{Q}}_{T;\a}^{\,(3)}$ introduced in
\eqref{definition Q T sigma 3}. Upon using the 
parameterisation \eqref{ecriture rep param en terme angle pour pts y sg alpha} 
for the endpoints $\wh{y}^{\, (\sg)}_{\a}[f]$ of the arcs $\wh{\ga}^{\, (\sg)} [f]$ given
in terms of the angles $\wh{\vth}^{\,(\sg)}_{\a}[f]$ introduced in Proposition
\ref{Proposition DA des angles vth et Lipschitz pour angles et t sg alpha}, one gets that 
\beq
\wh{\mc{Q}}_{T;\sg}^{\, (3)} \big[ f\big](\la)  \, - \, \wh{\mc{Q}}_{T;\sg}^{\, (3)} \big[ g \big](\la)  \; = \;  \wh{\mc{N}}_{T;\sg}^{\, (1)} \big[ f, g\big](\la) \, + \, \wh{\mc{N}}_{T;\sg}^{\, (2)} \big[ f, g\big](\la)
\enq
where, upon setting $z_{\vth}=-\i\tfrac{\zeta}{2}\, + \, \mf{c}_{\e{d}} T \ex{\i \vth }$,  
\begin{align}
 \wh{\mc{N}}_{T;\sg}^{\, (1)} \big[ f, g\big](\la) & = -\i T^2 \mf{c}_{\e{d}} \sg \sul{ \a \in \{L, R\} }{} \ups_{\a} \hspace{-3mm}   \Int{ \wh{\vth}^{\,(\sg)}_{\a}[g]  }{  \wh{\vth}^{\,(\sg)}_{\a}[f] }
\hspace{-3mm}  \dd \vth \; \ex{\i\vth}  R_{\e{c}}(\la,  z_{\vth} ) \ln\Big[ 1 + \ex{-\frac{\sg}{T}\wh{f}\, (z_{\vth} \,|\, \wh{\mathbb{X}}[f] ) } \Big]   \; , \\
 \wh{\mc{N}}_{T;\sg}^{\, (2)} \big[ f, g\big](\la) &= \i T  \mf{c}_{\e{d}}   \hspace{-3mm}   \Int{ \wh{\vth}^{\,(\sg)}_{L}[g]  }{  \wh{\vth}^{\,(\sg)}_{R}[g] }
\hspace{-3mm}  \dd \vth \; \ex{\i\vth}  R_{\e{c}}(\la,  z_{\vth} ) \Int{0}{1} \dd s 
\f{ \wh{f}\,\big(z_{\vth}\,|\, \wh{\mathbb{X}}[f] \big) \, - \, \wh{g}\, \big(z_{\vth}\,|\,\wh{\mathbb{X}}[g] \big)  }
{  1 + \ex{\frac{\sg}{T} \, \big[ s \wh{f}\,(z_{\vth} \,|\, \wh{\mathbb{X}}[f] )  + (1-s ) \wh{g}\,(z_{\vth} \,|\, \wh{\mathbb{X}}[g] ) \big] }  }  \;.
\end{align}

A direct application of
Proposition~\ref{Proposition bornage f et f hat sur arc circulaires rayon ordre T},
followed by implementing the Lipschitz bounds given in 
Proposition~\ref{Proposition DA des angles vth et Lipschitz pour angles et t sg alpha},
leads to the chain of bound for $\la \in \mc{M}$
\bem
     \, \Big| \, \wh{\mc{N}}_{T;\sg}^{\, (1)} \big[ f, g\big](\la) \Big| \; \leq  \;
        C T^2 \norm{ R_{\e{c}} }_{L^{\infty}(\mc{M}\times
	   \op{D}_{-\i\tf{\zeta}{2}, \, \varrho} ) } \cdot T^{M-|\op{Y}|} \cdot
     \sul{\a \in \{L,R\} }{} \Big| \,  \wh{\vth}^{\,(\sg)}_{\a}[g]  \, - \, \wh{\vth}^{\,(\sg)}_{\a}[f]  \Big| \\
     \; \leq \; C T^{2+M-|\op{Y}|} \cdot \underset{\a \in \{L,R\}}{\e{max}}\Big\{ \norm{f-g}_{L^{\infty}(\mc{W}_{\a}) } \Big\}  \;.
\end{multline}
%
%
Here, one picks $\varrho >0$ and small enough. 
Very similarly,
Proposition~\ref{Proposition bornage f et f hat sur arc circulaires rayon ordre T} and \eqref{ecriture bornes sur difference fnelle des u1}
lead to the estimate
%
%
%
\beq
     \, \Big| \, \wh{\mc{N}}_{T;\sg}^{\, (2)} \big[ f, g\big](\la) \Big| \; \leq  \; C T^2
        \norm{R_{\e{c}} }_{L^{\infty}(\mc{M}\times \op{D}_{-\i\tf{\zeta}{2}, \, \varrho})}
	\cdot \norm{f-g}_{L^{\infty}\big( \Dp{}\op{D}_{-\i\tf{\zeta}{2},\mf{c}_{\e{d}} T }  \cup_{\a\in \{L,R\}} \mc{W}_{\a} \big) } \;.
\enq

Thus, it follows that  for $0<T<T_0$ and $\eta > \tf{1}{(NT^4)}$
\beq
     \norm{\, \wh{\mc{Q}}_{T;\sg}^{\, (3)} \big[ f\big] \, - \,
        \wh{\mc{Q}}_{T;\sg}^{\, (3)} \big[ g \big]  }_{L^{\infty}(\mc{M}) } \; \leq \;
	C T^2 \cdot \norm{f-g}_{L^{\infty}\big( \mc{M} \setminus \op{D}_{-\i\tf{\zeta}{2},\mf{c}_{\e{d}} T }  \big) }
\label{ecriture borne contractivite Q3}
\enq
for some $C>0$ if $M \ge |\op{Y}| + 1$.


Ultimately, by putting all the upper bounds \eqref{ecriture borne contractivite Q1},
\eqref{ecriture borne contractivite Q2}, \eqref{ecriture borne contractivite Q3}
together, it follows that, for $0<T<T_0$ and $\eta > \tf{1}{(NT^4)}$ with $T_0,
\eta>0$ and small enough,
\beq
     \norm{ \,  \wh{\mc{L}}_{T}^{\, (2)} \big[ f \big]  \, - \,
        \wh{\mc{L}}_{T}^{\,(2)} \big[ g \big]  \, }_{ L^{\infty}(\mc{M})} \, \leq \,
	C  T \,  \norm{ f-g \, }_{ L^{\infty}\big( \mc{M} \setminus \op{D}_{-\i\tf{\zeta}{2},\mf{c}_{\e{d}} T }  \big) }
\label{ecriture contractivite L2}
\enq
for some $C>0$ if $M \ge |\op{Y}| + 1$. Hence, by putting \eqref{ecriture contractivite L1}
and \eqref{ecriture contractivite L2} together, the claim follows in the case of 
finite Trotter numbers. The details for the infinite Trotter number case are left to
the reader. \qed

\subsection{Existence and uniqueness of solutions }
\begin{theorem}
\label{Theorem existence et unicite sols NLIE}
Let  $C_{\mc{M}}>C_{\mc{M}}^{(0)}$ with $C_{\mc{M}}^{(0)}$ as given by
Proposition~\ref{Proposition stabilite operateur O} and assume that points $\mathrm{a)-d)}$ of
Hypothesis~\ref{Hypotheses solubilite NLIE} hold for $\op{X}, \op{Y}$.
Then, there exist $\eta_1, T_1>0$
small enough such that, for every $0<T<T_1$ and $\eta_1 > \tf{1}{(NT^4)}>0$ the operator $\wh{\mc{L}}_{T}$
admits a unique fixed point $\wh{u}$ on $\mc{E}_{\mc{M}}$. Likewise, for every $0<T<T_1$,
the operator $\mc{L}_{T}$ admits a unique
fixed point $u$ on $\mc{E}_{\mc{M}}$. Moreover, for this range of temperatures and Trotter numbers,  it holds that
\beq
\norm{u - \wh{u} \,  }_{L^{\infty}(\mc{M}) } \; \leq \;  \f{ C }{ N T^3} \;.
\label{ecriture deviation entre point fixe u et hat u}
\enq

Finally, denote by $\wh{u}_{X,Y,\zeta}$ the unique fixed point of $\wh{\mc{L}}_{T}$ associated with the collections of parameters $\op{X}$, $\op{Y}$, this for the anisotropy parameter $\zeta$.

Then, given two collections of parameters $\op{X}_0=\big\{ \op{x}_a\big\}_{a=1}^{|\op{X}_0|}$ and $\op{Y}_0=\big\{ \op{y}_a\big\}_{a=1}^{|\op{Y}_0|}$ satisfying
Hypothesis~\ref{Hypotheses solubilite NLIE} at anisotropy $\zeta_0$, there exists an open neighbourhood $\mc{V}_{\op{X}_0; \op{Y}_0}$ of
\beq
\bs{z}_0\,= \, \big( \bs{x}_0 , \bs{y}_0  \big)^{\e{t}} \qquad with \qquad
\left\{ \ba{ccc}  \bs{x}_0 & = & \big( \op{x}_1, \dots, \op{x}_{ |\op{X}_0| } \big)^{\e{t}} \vspace{1mm}  \\
 \bs{y}_0 & = & \big( \op{y}_1, \dots, \op{y}_{ |\op{Y}_0| } \big)^{\e{t}}  \ea \right. \;,
\enq
an open neighbourhood $\mc{S}_0 \subset \Cx$ of $\zeta_0$ such that the map $(\la,\bs{z},\zeta) \mapsto \wh{u}_{X,Y,\zeta}(\la)$ is holomorphic on $\mc{M} \times \mc{V}_{\op{X}_0; \op{Y}_0} \times \mc{S}_0$.
Here we agree upon
\beq
\bs{z}\,= \, \big( \bs{x} , \bs{y}  \big)^{\e{t}} \qquad with \qquad
\left\{ \ba{ccc}  \bs{x} & = & \big( x_1, \dots, x_{ |\op{X}_0| } \big)^{\e{t}} \vspace{1mm}  \\
 \bs{y} & = & \big( y_1, \dots, y_{ |\op{Y}_0| } \big)^{\e{t}}  \ea \right.
\label{definition coordonnee z et points partiels x et y}
\enq
and $\op{X} \, = \, \big\{x_a\big\}_{a=1}^{|\op{X}_0|}$ and $\op{Y} \, = \, \big\{ y_a \big\}_{a=1}^{|\op{Y}_0|}$.

\end{theorem}

\Proof 
The space $\mc{E}_{\mc{M}}$ is a complete metric space. By virtue of
Proposition~\ref{Proposition stabilite operateur O}, $\wh{\mc{L}}_{T}$ stabilises $\mc{E}_{\mc{M}}$
provided that $T<T_0$, $\tf{1}{(NT^4)}<\eta_0$ for some $\eta_{0}, T_{0}>0$. Then, by
Proposition~\ref{Proposition contractivite operateur}, there exists $\eta_0^{\prime}, T_0^{\prime}>0$ such
that, for any $0<T<T_{0}^{\prime}$ and $0 < \tf{1}{(NT^4)}< \eta_{0}^{\prime}$, $\wh{\mc{L}}_{T}$ is strictly contractive on $\mc{E}_{\mc{M}}$.
Thus, by the Banach fixed point theorem, $\wh{\mc{L}}_{T}$ admits a unique fixed point on $\mc{E}_{M}$
provided that $0<T<T_{1}=\e{min}\{ T_0,T_0^{\prime}\}$ and $0 < \tf{1}{(NT^4)}< \eta_1=\e{min}\{ \eta_{0}, \eta_{0}^{\prime}\}$.

\vspace{2mm}

Exactly the same reasoning can be repeated relatively to the operator $\mc{L}_{T}$.

\vspace{2mm}

Finally, to get the estimate \eqref{ecriture deviation entre point fixe u et hat u},
one observes that for any $f \in \mc{E}_{\mc{M}}$ it holds that 
$\mc{L}_{T}[f]=\wh{\mc{L}}_{T}[f+\veps_{\e{c}}-\mc{W}_N]$. Then, it is enough to
apply the bounds \eqref{ecriture contractivite L1 avec un control precis sur les domaines}
and \eqref{ecriture contractivite L2} which solely use the boundedness of the evaluated
functions on $\mc{M}\setminus \op{D}_{-\i\tf{\zeta}{2},\mf{c}_{\e{d}} T }$
to get that, for any $f \in \mc{E}_{\mc{M}}$,
\beq
\norm{  \wh{\mc{L}}_{T}[f]- \mc{L}_{T}[f] }_{L^{\infty}(\mc{M}) } \; \leq \; C \cdot
\norm{ \veps_{\e{c}}-\mc{W}_N }_{L^{\infty}\big(\mc{M} \setminus \op{D}_{-\i\tf{\zeta}{2},\mf{c}_{\e{d}} T }\big) }
\; \leq \; \f{ C^{\prime} }{ N T^3} \;.
\enq
Eventually one obtains the estimate
\beq
\e{d}_{\mc{E}_{\mc{M}}}\big( \, \wh{u}, u  \big) \; = \; \e{d}_{\mc{E}_{\mc{M}}}\Big( \, \wh{\mc{L}}_{T}[\, \wh{u} \, ], \mc{L}_{T}[u]  \Big) \; \leq \;
\e{d}_{\mc{E}_{\mc{M}}}\Big( \, \wh{\mc{L}}_{T}[\, \wh{u} \, ], \wh{\mc{L}}_{T} [u]  \Big) \,  +\,
\e{d}_{\mc{E}_{\mc{M}}}\Big( \, \wh{\mc{L}}_{T}[u], \mc{L}_{T}[u]  \Big) \;.
\enq
The previous estimate along with the contractivity of $\wh{\mc{L}}_{T}$
implies \eqref{ecriture deviation entre point fixe u et hat u}.

\vspace{2mm}

Consider the sequence $u_{n+1}\, = \, \wh{\mc{L}}_{T}[u_n]$ and $u_0=0$.
By construction, it holds that $\norm{ u_n-\wh{u}\,  }_{ L^{\infty}(\mc{M}) } \tend 0$
as $n \tend + \infty$, where $\wh{u}$ is the unique fixed point of
$\wh{\mc{L}}_{T}$ on $\mc{E}_{\mc{M}}$. Let $\bs{z}$ be as in
\eqref{definition coordonnee z et points partiels x et y} and associate with
it the collections of parameters $\op{X}$ and $\op{Y}$ as given below
\eqref{definition coordonnee z et points partiels x et y}. Further, consider
an  open neighbourhood $\wt{\mc{V}}_{\bs{\op{X}}_0;\bs{\op{Y}}_0 }
\subset \Cx^{ |\op{X}_0|+ |\op{Y}_0|}$ such that any
$\bs{z}\in \wt{\mc{V}}_{\bs{\op{X}}_0;\bs{\op{Y}}_0 }$ gives rise to sets
$\op{X},\op{Y}$ which enjoy Hypotheses \ref{Hypotheses solubilite NLIE},
this uniformly in $\zeta \in \wt{\mc{S}}_0$, with $\wt{\mc{S}}_0$ some
sufficiently small open neighbourhood of $\zeta_0$ in $\Cx$.  Let
$(\la, \bs{z},\zeta) \mapsto v_{\bs{z},\zeta}(\la)$ be a holomorphic map
on $\mc{M}\times \wt{\mc{V}}_{\bs{\op{X}}_0;\bs{\op{Y}}_0 } \times \wt{\mc{S}}_0$
such that $v_{\bs{z},\zeta} \in \mc{E}_{\mc{M}}$ pointwise in
$(\bs{z},\zeta) \in \wt{\mc{V}}_{\bs{\op{X}}_0;\bs{\op{Y}}_0 } \times \wt{\mc{S}}_0$.
Then, by virtue of the estimates obtained in Sections~\ref{SousSection Stabilite operateur},
\ref{SousSection Contractivite Operateur},
Propositions~\ref{Proposition stabilite operateur O},
\ref {Proposition contractivite operateur},
it is direct to see that $(\la, \bs{z},\zeta) \mapsto \wh{\mc{L}}_T\big[v_{\bs{z},\zeta}(*)\big]$
is holomorphic on $\mc{M}\times \wt{\mc{V}}_{\bs{\op{X}}_0;\bs{\op{Y}}_0 } \times \wt{\mc{S}}_0$.

\vspace{2mm}

Applying the latter to the sequence $u_n$, we infer that $u_n$ is a sequence
of holomorphic functions on $\mc{M}\times \wt{\mc{V}}_{\bs{\op{X}}_0;\bs{\op{Y}}_0 }
\times \wt{\mc{S}}_0$, such that $\la \mapsto u_n(\la; \bs{z},\zeta)$ converges
pointwise in $\bs{z},\zeta$ to $u(\la; \bs{z},\zeta)$ and
\beq
 \norm{u_n(*; \bs{z},\zeta)}_{ L^{\infty}(\mc{M}) } \, \leq \,  \mc{C}_{\mc{M}} T^2   \qquad \e{uniformly}\, \e{in} \quad   z \in  \wt{\mc{V}}_{\bs{\op{X}}_0;\bs{\op{Y}}_0 } \times \wt{\mc{S}}_0 \;.
\enq
Indeed, the estimates in Proposition~\ref{Proposition contractivite operateur} are independent on the choice of collection of parameters
$\op{X}, \op{Y}$ satisfying Hypothesis~\ref{Hypotheses solubilite NLIE} and these estimates depend continuously on $\zeta$. Then, taking $\rho>0$ such that
\beq
\Big\{ \times_{a=1}^{|\op{X}_0|} \ov{\op{D}}_{\op{x}_a,\rho} \Big\} \times  \Big\{ \times_{a=1}^{|\op{Y}_0|} \ov{\op{D}}_{\op{y}_a,\rho} \Big\}  \; \subset \; \wt{\mc{V}}_{\bs{\op{X}}_0;\bs{\op{Y}}_0 }
\qquad \e{and} \qquad \ov{\op{D}}_{\zeta_0,\rho}  \subset \wt{\mc{S}}_0 \;,
\enq
one has for all $n$
\beq
 u_n(\la; \bs{z},\zeta) \; = \; \Oint{ \Ga_{\rho} }{} \pl{a=1}{|\op{X}_0| + |\op{Y}_0| }\f{ \dd z_a^{\prime} }{ 2\i\pi (z_a^{\prime}-z_a) } \cdot
 \Oint{ \Dp{} \op{D}_{\zeta_0,\rho} }{} \f{ \dd \zeta^{\prime} }{ 2\i\pi (\zeta^{\prime} - \zeta) }    u_n(\la; \bs{z}^{\prime} ,\zeta^{\prime} )
\enq
with
\beq
\Ga_{\rho} \; = \; \Big\{ \times_{a=1}^{|\op{X}_0|} \Dp{}\op{D}_{\op{x}_a,\rho} \Big\} \times  \Big\{ \times_{a=1}^{|\op{Y}_0|} \Dp{}\op{D}_{\op{y}_a,\rho} \Big\} \;,
\enq
which entails by dominated convergence that  $(\la,\bs{z},\zeta)\mapsto u(\la; \bs{z},\zeta)$ is holomorphic on $\mc{M} \times \mc{V}_{\bs{\op{X}}_0;\bs{\op{Y}}_0 } \times \mc{S}_0$
for open neighbourhoods of $\bs{z}_0$ and $\zeta_0$ such that
\beq
\mc{V}_{\bs{\op{X}}_0;\bs{\op{Y}}_0 }  \subset \Big\{ \times_{a=1}^{|\op{X}_0|} \op{D}_{\op{x}_a,\rho} \Big\} \times  \Big\{ \times_{a=1}^{|\op{Y}_0|} \op{D}_{\op{y}_a,\rho} \Big\}
\qquad \e{and} \qquad
 \mc{S}_0 \; \subset \; \op{D}_{\zeta_0,\rho}  \;.
\enq

\qed

\section{Analysis of the quantisation conditions}
\label{Section conditions de quantification}
Now that we have established the existence of solutions to the non-linear problems at 
finite \eqref{ecriture NLIE a Trotter fini}, \eqref{ecriture monodromie Trotter fini}
and infinite \eqref{ecriture NLIE a Trotter infini}, \eqref{ecriture monodromie Trotter infini}
Trotter numbers, this uniformly for all
choices of collections of auxiliary parameters $\op{X}$ and $\op{Y}$ satisfying
Hypothesis~\ref{Hypotheses solubilite NLIE}, provided that $T$ and $\tf{1}{(NT^4)}$
are small enough, it remains to determine the possible collections of parameters
$\wh{\mf{X}}$ and $\wh{\mc{Y}}$, resp.\ $\mf{X}$ and $\mc{Y}$, which give rise to solutions
that also satisfy the quantisation
conditions at finite \eqref{ecriture conditions de quantifications Trotter fini}, resp.\
infinite \eqref{ecriture condition de quantification Trotter infini}, Trotter number.
This part of the problem will be dealt with in the present section, and the analysis
will be carried out under Hypotheses \ref{Hypotheses solubilite NLIE} and
\ref{Hypotheses eqns de quantification}. Namely, we will characterise all solution
sets that are compatible with both the quantisation conditions \textit{and} Hypotheses
\ref{Hypotheses solubilite NLIE} and \ref{Hypotheses eqns de quantification}. 
 
\subsection{Convenient factorisation formulae}
From now on, we shall assume that $\wh{u}\, (\la\,|\,\wh{\mathbb{Y}})$, resp.\
$u(\la\,|\,\mathbb{Y} )$, is the solution to the non-linear integral equation
\eqref{ecriture NLIE a Trotter fini}, resp.\ \eqref{ecriture NLIE a Trotter infini},
subject to the monodromy condition \eqref{ecriture monodromie Trotter fini},
resp.\ \eqref{ecriture monodromie Trotter infini}, and associated to the collection
of auxiliary parameters $\wh{\mathbb{Y}}$, resp.\ $\mathbb{Y}$, which, moreover, enjoy
Hypotheses~\ref{Hypotheses solubilite NLIE} and \ref{Hypotheses eqns de quantification}.
These auxiliary parameters are furthermore assumed to be such that the quantisation
conditions~\eqref{ecriture conditions de quantifications Trotter fini}, resp.\
\eqref{ecriture condition de quantification Trotter infini}, hold. In order to
distinguish them from the case of generic parameters, we shall henceforth adopt a specific
notation for the collections of parameters which do solve these conditions,
\beq
\wh{\op{X}} \oplus \wh{\op{X}}\;\!^{\prime}\; \hookrightarrow \; \wh{\mf{X}} \; = \; \big\{ \, \wh{x}_a\, \big\}_{1}^{| \wh{\mf{X}} |} \qquad \e{and} \qquad
\wh{\op{Y}} \oplus \wh{\op{Y}}\;\!^{\prime}\; \hookrightarrow \; \wh{\mc{Y}} \; = \; \big\{ \, \wh{y}_a \, \big\}_{1}^{| \wh{\mc{Y}} |}
\enq
in the case of finite Trotter number and without the $\wh{\,\,\,\,}$ in the
case of infinite Trotter number. Again, we stress that when we say that
$\wh{\mf{X}}$, $\wh{\mc{Y}}$ satisfy Hypotheses ~\ref{Hypotheses solubilite NLIE}-\ref{Hypotheses eqns de quantification}, we mean
that $\wh{\mf{X}}$, $\wh{\mc{Y}}$ satisfy Hypothesis \ref{Hypotheses eqns de quantification} and that
$\wh{\op{X}}, \wh{\op{Y}}$ satisfy Hypothesis ~\ref{Hypotheses solubilite NLIE}.

Below, most of the analysis will be carried out in the
finite Trotter number setting, since, owing to the continuity in $N$ of
$\wh{u}\, (\la \,|\, \wh{\mathbb{Y}})$, \textit{c.f.}\
Theorem~\ref{Theorem existence et unicite sols NLIE}, it is direct to convince
oneself that the infinite Trotter number results are obtained simply by taking
$N\tend + \infty$ on the level of the final formulae, which have a clear control
in $N$. We also stress that, owing to requirement $\mathrm{f)}$ of
Hypothesis~\ref{Hypotheses eqns de quantification}, $\wh{\mf{X}}$ and
$\wh{\mc{Y}}$ may simply be considered as sets -- and not collections of parameters --
since one does not have to focus on the multiplicity of the roots anymore --
the latter being always equal to $1$.

Further, analogously to the partitioning \eqref{definition racines singulieres et reguliere} of the generic roots $\op{Y}$
into the regular ones $\op{Y}_r$ and the shifted singular ones $\wt{\op{Y}}_{\e{sg}}$,
we introduce the sets 
\beq
\wh{\wt{\mc{Y}}}_{\e{sg}} \, = \, \Big\{ y \in \wh{\mc{Y}} \; : \; y - \i \zeta \in \wh{\mc{Y}}_{\e{sg}}  \Big\}\, = \, \Big\{ y + \i \zeta \; : \; y   \in \wh{\mc{Y}}_{\e{sg}}  \Big\} \;,
\enq
with
\beq
\wh{\mc{Y}}_{\e{sg}} \; = \; \Big\{ y - \i \zeta  \; : \; y \in \wh{\mc{Y}} \; \e{and} \; y-\i\zeta \in \e{Int}\big( \msc{C}_{\e{ref}} \big)  \Big\}
\enq
what leads to a partitioning of $\wh{\mc{Y}}$ into two disjoint sets 
\beq
\wh{\mc{Y}}\, = \, \wh{\mc{Y}}_{\e{r}} \sqcup \wh{\wt{\mc{Y}}}_{\e{sg}} \;. 
\enq
We will denote the associated collection of sets by $\wh{\mathbb{Y}}$, exactly as
it was introduced in  \eqref{definition ensemble Y hat et Y hat q}.

We shall now recast the solution $\wh{u}\, (\la\,|\,\wh{\mathbb{Y}} )$,
\textit{viz.}\ one solving the non-linear integral equation \textit{and}
the quantisation conditions, in a form that is most convenient for the low-$T$,
low-$\tf{1}{(NT^4)}$ analysis of the quantisation conditions
for the elements building up $\wh{\mf{X}}$ and $\wh{\mc{Y}}$. 
For that purpose, it is convenient to introduce the auxiliary function
\beq
\wh{\Phi}(\la \,|\, \wh{\mathbb{Y}}) \; = \;\f{1}{T}\Big( \mc{W}_N(\la)-\veps_{\e{c}}(\la) \Big) \, + \,  u_{1;\e{reg}}\big( \la\,|\, \wh{\mathbb{Y}} \, \big)
\, + \, \f{1}{T}   \wh{\mc{R}}_{T}\big[ \, \wh{u}( * \,|\, \wh{\mathbb{Y}} ) \big](\la)%
_{ \left| \substack{ q^{(\pm)}_{\wh{u}} \hookrightarrow
   \wh{q}^{\, (\pm)}_{\mathbb{Y}}[\wh{\,u\,}] \\[.5ex]
\msc{C}_{\wh{u}} \hookrightarrow  \wh{\msc{C}}_{\mathbb{Y}}[\wh{\,u\,}] } \right. } \;.
\enq
There $u_{1;\e{reg}}$ has been introduced in \eqref{definition u1 reg},
while $\wh{\mc{R}}_{T} \, = \, \wh{\mc{R}}_{T}^{\, (1)} \, +\, \wh{\mc{R}}^{\,(2)}_T$
is defined through \eqref{definition operateur reste de type 1},
\eqref{definition operateur reste de type 2}. The previous analysis ensures that 
\beq
\wh{\Phi}(\la\,|\, \wh{\mathbb{Y}}\, ) \; = \; u_{1;\e{reg}}(\la\,|\, \wh{\mathbb{Y}}\, )\, + \, \e{O}\big(T \big) \;,
\enq
and the control on the remainder is uniform
\begin{itemize}
\item in $\tf{1}{ (NT^4) }<\eta$,
\item in $\la$ uniformly away from $\pm \i\tf{\zeta}{2}+\i\pi \mathbb{Z}$,
\item with respect to the parameters forming $\wh{\mf{X}}$ and $\wh{\mc{Y}}$
as long as the set $\wh{\mathbb{Y}}$ fulfils the requirements $\mathrm{a)-g)}$
listed in Hypothesis~\ref{Hypotheses solubilite NLIE} and
Hypothesis~\ref{Hypotheses eqns de quantification}.
\end{itemize}

We also stress that the operators involved in  $\wh{\mc{R}}_{T}$ act on portions
of the curve  $\wh{\msc{C}}_{\mathbb{Y} }\big[ \, \wh{u} \, \big]$
where  $\wh{u}\in \mc{E}_{\mc{M}}$ is the fixed point of $\wh{\mc{L}}_T$
giving rise to $\wh{u}\, \big(\la\,|\,\wh{\mathbb{Y}} \big)$.

Upon using \eqref{ecriture eqn NL pour u forme asymptotique} and \eqref{ecriture DA uniforme u- sur tout le plan cplx}, this entails that
\beq
\ex{- \f{1}{T}  \wh{u}\,  ( \la\,|\, \wh{\mathbb{Y}}\,  ) } \; = \; \ex{ -\f{1}{T} \veps_{\e{c}}(\la) } \pl{y \in \wh{\mathbb{Y}}  }{} \f{ \sinh(\i\zeta+y-\la) }{ \sinh(\i\zeta - y+\la)  }
\cdot \ex{- \wh{\Phi}\,(\la\,|\, \wh{\mathbb{Y}}\, ) }  \cdot
\f{ \Big( 1+\ex{-\f{1}{T} \wh{u}\, (\la-\i   \zeta \,|\, \wh{\mathbb{Y}}\, )} \Big)^{\bs{1}_{ \msc{D}_{\mathbb{Y}; \i\pi }   }(\la-\i   \zeta)}  }
{ \Big( 1+\ex{-\f{1}{T} \wh{u}\, (\la + \i   \zeta \,|\, \wh{\mathbb{Y}}\, )}  \Big)^{\bs{1}_{ \msc{D}_{\mathbb{Y}; \i \pi } }(\la+\i  \zeta)}  } \;.  
\label{ecriture expression wh u apres NLIE}
\enq
Here we have introduced the domains 
\beq
\msc{D}_{\mathbb{Y}} \, = \, \e{Int}\Big( \, \wh{\msc{C}}_{\mathbb{Y}}\big[ \, \wh{u} \, \big] \Big)  \qquad \e{and} \qquad 
\msc{D}_{\mathbb{Y}; \i\pi } \, = \,\msc{D}_{\mathbb{Y}} +\i \pi \mathbb{Z}\; . 
\enq
Thus, for any $\la$ such that $\la \pm \i\zeta \not\in \msc{D}_{\mathbb{Y}; \i\pi}$, one has  the explicit factorisation 
\bem
\ex{- \f{1}{T}  \wh{u}\, (\la \,|\, \wh{\mathbb{Y}}\, ) } \; = \; \ex{ -\f{1}{T} \veps_{\e{c}}(\la) }
\pl{y \in \wh{\mc{Y}}_{\e{r}} }{} \bigg\{  \f{ \sinh(\i\zeta+y-\la) }{ \sinh(\i\zeta + \la- y )  }   \bigg\}
\cdot  \pl{y \in  \wh{\wt{\mc{Y}}}_{\e{sg}} }{} \bigg\{ \f{ \sinh(\i\zeta+y-\la)  \sinh( y-\la) }{ \sinh(\i\zeta +  \la- y )   \sinh(2 \i\zeta + \la- y )  }    \bigg\}  \\
\times  \pl{x \in \wh{\mf{X}} }{} \bigg\{  \f{ \sinh(\i\zeta + \la- x )  }{ \sinh(\i\zeta+x-\la) }   \bigg\}
\cdot  \ex{- \wh{\Phi}(\la \,|\, \wh{\mathbb{Y}}\, ) }   \;. 
\label{ecriture form eqn subsidiaire pour x1}
\end{multline}

If $\la - \i\zeta \in \msc{D}_{\mathbb{Y};\i\pi}$, it holds that
\bem
\ex{- \f{1}{T} \wh{u}\, (\la \,|\, \wh{\mathbb{Y}}\, ) } \; = \; \ex{ -\f{1}{T} \veps^{(-)}_{\e{c};2}(\la) }
\pl{y \in \wh{\mc{Y}}_{\e{r}} }{} \bigg\{  \f{ \sinh( \i \zeta + y-\la)  \sinh( 2 \i \zeta + y-\la)  }{ \sinh( \i \zeta + \la- y )    \sinh(  \la- y ) }   \bigg\}
   \pl{y \in  \wh{\wt{\mc{Y}}}_{\e{sg}} }{}
\bigg\{ - \f{ \sinh^2(\i\zeta+y-\la)  \sinh( 2\i\zeta + y - \la) }{ \sinh^2(\i\zeta +  \la- y )   \sinh(2 \i\zeta + \la- y )  }    \bigg\}   \\
\times \pl{x \in \wh{\mf{X}} }{} \bigg\{  \f{ \sinh(\i\zeta + \la- x )  \sinh(  \la- x )  }{ \sinh(\i\zeta+x-\la) \sinh(2 \i\zeta+x-\la) }   \bigg\}
\cdot  \ex{- \wh{\Phi}_2^{\, (-)}(\la\,|\, \wh{\mathbb{Y}}) }   \cdot \Big( 1+\ex{\f{1}{T} \wh{u}\, (\la-\i   \zeta \,|\, \wh{\mathbb{Y}}\, )} \Big)  \;.
\label{ecriture equation subsidiaire racine singulière}
\end{multline}
In the above expression, we agree upon 
\beq
\veps^{(-)}_{c;2}(\la) \,= \, \veps_{\e{c}}(\la) + \veps_{\e{c}}(\la-\i\zeta) \qquad \e{and} \qquad 
\wh{\Phi}_2^{\, (-)}(\la\,|\, \wh{\mathbb{Y}} ) \, = \, \wh{\Phi}(\la\,|\, \wh{\mathbb{Y}} \, ) \, + \, \wh{\Phi}(\la-\i\zeta\,|\,\wh{\mathbb{Y}}\, ) \;. 
\enq
Finally, if $\la + \i\zeta \in \msc{D}_{\mathbb{Y};\i\pi}$, then
\bem
\ex{- \f{1}{T}  \wh{u}\, (\la \,|\, \wh{\mathbb{Y}}\, )  } \; = \; \ex{ -\f{1}{T} \veps_{\e{c}}(\la) } \pl{y \in \wh{\mc{Y}}_{\e{r}} }{} \bigg\{  \f{ \sinh(\i\zeta+y-\la) }{ \sinh(\i\zeta + \la- y )  }   \bigg\}
\cdot  \pl{y \in  \wh{\wt{\mc{Y}}}_{\e{sg}} }{} \bigg\{ \f{ \sinh(\i\zeta+y-\la)  \sinh( y-\la) }{ \sinh(\i\zeta +  \la- y )   \sinh(2 \i\zeta + \la- y )  }    \bigg\}  \\
\times   \pl{x \in \wh{\mf{X}} }{} \bigg\{  \f{ \sinh(\i\zeta + \la- x )  }{ \sinh(\i\zeta+x-\la) }   \bigg\}
\cdot \f{  \ex{- \wh{\Phi}(\la\,|\, \wh{\mathbb{Y}}\, ) }  }{  1+\ex{-\f{1}{T} \wh{u}\, (\la+\i   \zeta\,|\, \wh{\mathbb{Y}}\, )}   }  \;
\label{ecriture exp de u hat sur D Y moins i zeta}
\end{multline}
holds, in which 
\bem
\ex{- \f{1}{T}  \wh{u}\, (\la+\i\zeta \,|\, \wh{\mathbb{Y}}\, ) } \; = \; \ex{ -\f{1}{T} \veps_{\e{c}}(\la+\i\zeta) } 
\pl{y \in \wh{\mc{Y}}_{\e{r}} }{} \bigg\{  \f{ \sinh( y - \la) }{ \sinh( 2 \i\zeta + \la- y )  }   \bigg\}
\cdot  \pl{y \in  \wh{\wt{\mc{Y}}}_{\e{sg}} }{} \bigg\{ \f{ \sinh(y-\la)  \sinh( y-\la- \i\zeta ) }{ \sinh( 2 \i\zeta +  \la- y )   \sinh(3 \i\zeta + \la- y )  }    \bigg\}  \\
\times   \pl{x \in \wh{\mf{X}} }{} \bigg\{  \f{ \sinh(2 \i\zeta + \la- x )  }{ \sinh( x - \la) }   \bigg\}
\cdot  \ex{- \wh{\Phi}\, (\la+ \i \zeta \,|\, \wh{\mathbb{Y}} ) }   \;. 
\end{multline}

At this stage already, one may make the following observation. The function
$\la \mapsto \ex{- \f{1}{T}  \wh{u}\, (\la+\i\zeta \,|\, \wh{\mathbb{Y}}\, ) }$ 
admits a zero at any regular root  $y \in \msc{D}_{\mathbb{Y};\i\pi}-\i\zeta$.
Hence, the extra part in \eqref{ecriture exp de u hat sur D Y moins i zeta},
involving $\ex{- \f{1}{T}  \wh{u}\, (\la+\i\zeta \,|\, \wh{\mathbb{Y}}\, ) }$, 
will not appear in the quantisation conditions for such regular roots. Therefore
the quantisation conditions only take a functionally different form for shifted
singular roots. 

Also, in the following we will denote
\beq
\wh{\mc{E}}\,(\la \,|\, \wh{\mathbb{Y}}\, )   \, = \,  \veps_{\e{c}}(\la) \, + \, T  \wh{\Phi}\,(\la \,|\, \wh{\mathbb{Y}}\, )  \qquad \e{and} \qquad 
\wh{\mc{E}}_2^{\, (-)}\,(\la \,|\, \wh{\mathbb{Y}}\, )   \, = \,  \veps_{\e{c};2}^{(-)}(\la) \, + \, T  \wh{\Phi}_2^{\, (-)}(\la \,|\, \wh{\mathbb{Y}}\, ) \;. 
\label{definition fcts composites des energies et phase habilees}
\enq

\subsection{Existence, uniqueness and structure of solutions to the quantisation equations}
Throughout the remainder of this section, we shall assume that  $\tf{\zeta}{\pi}$
is irrational. This hypothesis is necessary so as to ensure appropriate non-intersections
of domains and clearly spell out the concept of a maximal root. However, the existence
result for solutions, as stated in
Theorem~\ref{Theorem existence solutions racines particules et trous} to come,
is totally independent of this assumption.

Having fixed $P \in \mathbb{N}$, by virtue of the irrationality of $\tf{\zeta}{\pi}$,
it is easy to see that one may take $T$ to be small enough so that the domain
$\msc{D}_{\mathbb{Y}}$ satisfies
\beq
\msc{D}_{\mathbb{Y};\i\pi}\cap \big\{ \msc{D}_{\mathbb{Y};\i\pi} + \i p \zeta \big\} = \emptyset  \qquad \e{for}\; \e{any} \quad  p \in \intn{-P}{P}\setminus \{0\}\;.
\enq
This property will play an important role in the analysis to come. In fact, for $T,
\tf{1}{(NT^4)}$ small enough, one has the stronger property
\beq
\e{d}\Big(\msc{D}_{\mathbb{Y};\i\pi},  \msc{D}_{\mathbb{Y};\i\pi} + \i  p \zeta \Big)  \geq c \, > \, 0 \; , \quad p \in \intn{1}{P}
\label{ecriture propriete espacement fini pour domaines DY shiftes}
\enq
with $c$ uniform in $T,N$. It appears important to stress at this point that the lower bound
result \eqref{ecriture propriete espacement fini pour domaines DY shiftes} is also uniform
with respect to solution sets $\wh{\mf{X}}$ and $\wh{\mc{Y}}$ solving the quantisation
conditions, provided their cardinalities $|\wh{\mf{X}}\, |, |\wh{\mc{Y}}|$ are bounded.
Indeed, their influence on the shape of the boundary $\wh{\msc{C}}_{\mathbb{Y}}[\, \wh{u}\,]$
of $\msc{D}_{\mathbb{Y}}$ only appears on the level of small $\e{O}(T)$ modifications
which, moreover, are uniformly bounded with respect to the \textit{locii} of the
parameters composing the sets $\op{X}$ and $\op{Y}$, and \textit{a fortiori} with
respect to solution sets  $\wh{\mf{X}}$ and $\wh{\mc{Y}}$, fulfilling the
Hypotheses \ref{Hypotheses solubilite NLIE}, \ref{Hypotheses eqns de quantification},
provided that $|\wh{\mf{X}}\, |, |\wh{\mc{Y}}|$ are bounded.

The central result of this section is that sets $\wh{\mf{X}}$, $\wh{\mc{Y}}$, solving
the quantisation conditions \eqref{ecriture conditions de quantifications Trotter fini}
and fulfilling Hypotheses \ref{Hypotheses eqns de quantification}, exhibit a very
regular structure. The hole roots $\wh{x}_a$ are located in some neighbourhood
of the curve $\msc{C}_{\veps}$, whose size is controlled by the magnitude of $T$
and $\tf{1}{(NT^4)}$, while the roots $\wh{y}_a$ are located in some neighbourhood
of the curve
\beq
\Bigl\{\la \; : \; \Re\big[ \veps_{\e{c}}(\la) \big]=0\ \text{and}\ 0 \, \leq \, \Im (\la) \, \leq \, \tf{\zeta }{2} \Bigr\} \;.
\enq
The establishing of this result goes in two steps, the first one consists in showing
that, in full generality, the roots $\wh{x}_a$ are located in some neighbourhood
of $\msc{C}_{\veps}$, while the roots $\wh{y}_a$ form, up to $\e{o}(1)$ corrections
as $T, \tf{1}{N T^4} \tend 0$, complexes called thermal $r$-strings, whose top
element belongs to certain curves in the complex plane. Then, in
Theorem~\ref{Theorem classification forme racines particules et trous},
we show that, in the regime $0<\zeta <\tf{\pi}{2}$ of interest to us, only $r=1$
thermal strings can exist. Still, we stress that for generic
$\zeta\in \intoo{ \tf{\pi}{2} }{ \pi }$ more general $r$-strings may occur as
corroborated by our numerics, see
Conjecture~\ref{Conjecture classification forme racines particules et trous zeta general}.
To state the first result, we need to provide the definition of a thermal $r$-string.

\begin{defin}
\label{Defintion thermal r string} 
 A point $y\in \Cx$ is said to be the top of a thermal $r$ string, $r \in \mathbb{N}^{*}$, if the following set of conditions is fulfilled:
\beq
\Re\big[ \veps_{\e{c}; k}^{(-)}(y) \big] \, < \, 0 \quad for \quad k=1,\dots, r-1 \qquad and  \qquad \Re\big[ \veps_{\e{c}; r}^{(-)}(y) \big] \, = \, 0
\label{ecriture ensemble inequations pour r cordes thermales}
\enq
in which 
\beq
\veps_{\e{c}; k}^{(-)}(\la)\, =\; \sul{s=0}{k-1} \veps_{\e{c}}(\la-\i s \zeta) \;, 
\label{definition veps c k}
\enq
and $\veps_{\e{c}}$ is the dressed energy defined in \eqref{ecriture LIE pour veps c}.
Moreover, when $r=1$, the condition that $0 \leq \Im(y) \leq \tf{\zeta}{2}$ should be added.
\end{defin}

For the time being, we denote by $\{ r_k \}_{ k\in \mathbb{N}^{*} }$, $r_1<r_2<\cdots$
the subset of $\mathbb{N}^{*}$ corresponding to the values of $r$ where such solutions
exists. Note that it always holds that $r_1=1$, this independently of the value of $\zeta$.

The main result of the present analysis is the following

\begin{theorem}
\label{Theorem classification forme racines particules et trous}
Let $\wh{\mathbb{Y}}$ as in \eqref{definition ensemble Y hat et Y hat q} be a set solving
the quantisation conditions \eqref{ecriture conditions de quantifications Trotter fini}
such that $|\wh{\mathbb{Y}}| \, = \,  |\wh{\mc{Y}}|+|\wh{\wt{\mc{Y}}}_{\e{sg}}|
+|\wh{\mf{X}}|$ is bounded in $T$, $\tf{1}{(NT^4)}$ small and such that
Hypotheses~\ref{Hypotheses solubilite NLIE},
\ref{Hypotheses eqns de quantification} are satisfied.

Then there exist $T_0$, $\eta $ small enough, such that, for all $T_0>T>0$ and
$\eta> \tf{1}{(NT^4)}$, the sets $\wh{\mf{X}}$ and $\wh{\mc{Y}}$ may be represented as 
\beq
\wh{\mf{X}} \, = \, \Big\{ \, \wh{x}_1,\dots, \wh{x}_{n} \,  \Big\} \qquad and \qquad \wh{\mc{Y}} \, = \,   \Big\{ \,   \wh{y}_{1}, \dots, \wh{y}_{m} \, \Big\} \;,
\label{ecriture forme asymptotique ensembles solution trous et particules}
\enq
where the points $\wh{x}_k, \wh{y}_k$ solve the equations
\begin{align}
\Re\big[ \veps_{\e{c}}\big( \, \wh{x}_k \, \big) \big] & =  \e{o}(1) \;, \quad and \quad \wh{x}_k \in \msc{D}_{\mathbb{Y}}   \;,  \vspace{2mm}  \label{ecriture appartenanece Trotter fini courbe trous}\\[1ex]
\Re\big[ \veps_{\e{c}}\big( \, \wh{y}_k \, \big) \big] & =  \e{o}(1) \;, \quad and \quad
\wh{y}_k \in  \biggl\{ z\; : \; -\frac{\pi}{2} < \Im(z) \, \leq \frac{\pi}{2}\biggr\}
\setminus \msc{D}_{\mathbb{Y}} \;, 
\label{ecriture appartenanece Trotter fini courbe particules}
\end{align}
this when  $T$, $\tf{1}{(NT^4)} \tend 0^+$. In particular, there are no singular
roots, $\wh{\mc{Y}}_{\e{sg}}=\emptyset$, and no thermal $r$ strings with $r > 1$. 

Analogous results hold in the case of infinite Trotter number, where the sets  
\beq
\mf{X} \, = \, \bigl\{ x_1, \dots, x_n \bigr\} \qquad and \qquad
\mc{Y} \, = \, \bigl\{ y_1, \dots, y_m  \bigr\} 
\enq
are such that they satisfy Hypotheses~\ref{Hypotheses solubilite NLIE},
\ref{Hypotheses eqns de quantification} as well as
the quantisation conditions~\eqref{ecriture condition de quantification Trotter infini}
associated with the solution $u(\la\,|\,\mathbb{Y})$ to
\eqref{ecriture NLIE a Trotter infini} and \eqref{ecriture monodromie Trotter infini}.   
\end{theorem}

The idea of the proof is, as mentioned above, to first show that, up to
$\e{o}(1)$ corrections, the particle roots $\wh{y}_k$ are necessarily grouped
in thermal strings, and that, for $\zeta \in ]0, \pi/2[$ only 1 strings can exist.
Prior to presenting this proof of the theorem we have to establish the concepts
of maximal and weakly maximal roots in Definition~\ref{definition racine maximale},
which are the types of roots that may occur as tops of thermal strings. We
shall further establish a useful auxiliary lemma 
(Lemma~\ref{Lemme partition Y selon racines maximales} below) which states the
existence of a (weakly) maximal root and lifts the repulsion of roots, point e) of
Hypothesis~\ref{Hypotheses eqns de quantification}, to the level of thermal $r$ strings.

\begin{defin}
\label{definition racine maximale}
Let $\wh{\mc{Y}}$ satisfy Hypotheses \ref{Hypotheses solubilite NLIE},
\ref{Hypotheses eqns de quantification} and let $|\wh{\mc{Y}}|$ be uniformly bounded
in $T, \tf{1}{(NT^4)} \tend 0^+$. Pick some $\vsg>0$. A root $y \in \wh{\mc{Y}}$
is said to be maximal if
\begin{enumerate}

\item
$y \in \wh{\mc{Y}}_{\e{r}}$ is such that
\beq
\e{d}_{\i\pi}\big( y+\i\zeta, y^{\prime} \big) \, > \, \vsg T \; , \quad \forall \; y^{\prime} \in \wh{\mc{Y}}_{\e{r}}
\qquad and \qquad
\e{d}_{\i\pi}\big( y + 2 \i\zeta, y^{\prime} \big) \, > \, \vsg T \; , \quad \forall \; y^{\prime} \in \wh{\wt{\mc{Y}}}_{\e{sg}} \; ,
\enq
\item
$y \in \wh{\wt{\mc{Y}}}_{\e{sg}}$ is such that
\beq
\e{d}_{\i\pi}\big( y+\i\zeta, y^{\prime} \big) \, > \, \vsg T \; , \quad \forall \; y^{\prime} \in \wh{\mc{Y}}_{\e{r}} \;.
\enq
\end{enumerate}
Above, $\e{d}_{\i\pi}$ is as it has been introduced
in \eqref{definition i pi periodic distance}.

Furthermore, a root $y \in \wh{\mc{Y}}$ is said to be weakly maximal if $i)-ii)$
above hold with the exception of a single root $y^{\prime}\in \wh{\mc{Y}}$.
For that root one has the a priori lower bounds
\begin{enumerate}
\item
in case $y \in \wh{\mc{Y}}_{\e{r}}$
\beq
\e{d}_{\i\pi}\big( y+\i\zeta, y^{\prime} \big) \, > \, \ex{- \f{c_T}{T}  } \; , \quad  if \,  \; y^{\prime} \in \wh{\mc{Y}}_{\e{r}}
\qquad and \qquad
\e{d}_{\i\pi}\big( y + 2 \i\zeta, y^{\prime} \big) \, > \, \ex{- \f{c_T}{T}  }  \; , \quad if \; y^{\prime} \in \wh{\wt{\mc{Y}}}_{\e{sg}} \; ,
\enq
\item
in case $y \in \wh{\wt{\mc{Y}}}_{\e{sg}}$
\beq
\e{d}_{\i\pi}\big( y+\i\zeta, y^{\prime} \big) \, > \,  \ex{- \f{c_T}{T}  } \; , \quad  \; y^{\prime} \in \wh{\mc{Y}}_{\e{r}} \;,
\enq
\end{enumerate}
where $c_T=\e{o}(1)$ as $T\tend 0^+$.
\end{defin}


\begin{lemme}
\label{Lemme partition Y selon racines maximales}
Let $\wh{\mc{Y}}$ satisfy Hypotheses~\ref{Hypotheses solubilite NLIE},
\ref{Hypotheses eqns de quantification} and let $|\wh{\mc{Y}}|$ be bounded uniformly
in $T, \tf{1}{(NT^4)} \tend 0^+$. Pick $\vsg>0$ and small enough  such that
\beq
\vsg \, < \, \e{min} \bigg\{ \tfrac{ \mf{c}_{\e{rep}} }{ 4 } \; , \;  \tfrac{ C_{|\wh{\mc{Y}}| } }{ (1+|\wh{\mc{Y}}|) T }   \bigg\}
\qquad with \qquad
 C_{|\wh{\mc{Y}}| } \; = \; \e{min} \Big\{  \e{d}_{\i   \pi}(\i p \zeta, 0) \, ; \, p=1 , \dots,  1+|\wh{\mc{Y}}| \Big\}
\label{definition borne sup pour parametre controle vsg}
\enq
and with  $\mf{c}_{\e{rep}}$ as introduced in point e) of
Hypothesis~\ref{Hypotheses eqns de quantification}.

Then the set $\wh{\mc{Y}}$ of particle roots admits the decomposition
\beq
\wh{\mc{Y}} \; = \; \bigcup\limits_{ s=1}^{ p } \wh{\mc{Y}}^{(s)} \quad with \quad
\wh{\mc{Y}}^{(s)} \; = \; \big\{ \mf{y}_{ \ell , s } \big\}_{\ell=0}^{k_s} \quad and \quad  \wh{\mc{Y}}^{(s)}\cap \wh{\mc{Y}}^{(s^{\prime})} \, = \, \emptyset
\quad if \; s\not= s^{\prime}
\label{ecriture decomposition de hat Y en classes equivalence selon zeta shifts}
\enq
in which each  $\wh{\mc{Y}}^{(s)}$ contains at most one shifted
singular root from $\wh{\wt{\mc{Y}}}_{\e{sg}}$. The roots in $\wh{\mc{Y}}^{(s)}$ are
such that
\begin{enumerate}
\item
$\mf{y}_{0,s}$ is maximal in the sense of Definition~\ref{definition racine maximale};
\item
for $\ell= 0,\dots, k_s-1$ and uniformly in $T, \tf{1}{(NT^4)} \tend 0^+$, it holds that
\beq
\e{d}_{\i\pi}\big( \mf{y}_{\ell,s} \, , \, \mf{y}_{\ell+1,s}+\i n_{\ell,s} \zeta \big) \; \leq \; \vsg T \;,
\enq
where $n_{\ell,s}=1$ if $\mf{y}_{\ell,s} \in \wh{\mc{Y}}_{\e{r}}$ and $n_{\ell,s}=2$
if  $\mf{y}_{\ell,s} \in \wh{\wt{\mc{Y}}}_{\e{sg}}$;
\item
for any $y \in \wh{\mc{Y}}^{(s)}$ and
$y^{\prime} \in \wh{\mc{Y}} \setminus \wh{\mc{Y}}^{(s)}$ it holds that
\beq
\e{d}_{\i\pi}\big(y, y^{\prime} + \i m_{y} \zeta  \big) \, > \,   \vsg T \;,
\label{ecriture repulsivite racines entre chaines}
\enq
where $m_{y}=1$ if $y \in \wh{\mc{Y}}_{\e{r}}$ and $m_{y}=2$ if
$y \in\wh{\wt{\mc{Y}}}_{\e{sg}}$.
\end{enumerate}
\end{lemme}

Note  that the range of the indices $\ell,s$ may, in principle, vary with $T$ and $N$.

\Proof
Pick $\vsg>0$ satisfying the bounds
\eqref{definition borne sup pour parametre controle vsg}. We initiate the proof
by showing the existence of maximal roots. This will be done by contradiction.

First of all, we focus on the simpler case where $\wh{\wt{\mc{Y}}}_{\e{sg}}\, =\,
\emptyset$. Assume that there are no maximal roots. Then for any
$y \in \wh{\mc{Y}}_{\e{r}}=\wh{\mc{Y}}$ there exists $y^{\prime} \in  \wh{\mc{Y}}_{\e{r}}$
such that $\e{d}_{\i\pi}\big( y + \i \zeta , y^{\prime} \big) \leq \vsg T$. Thus, pick
some $y_0 \in \wh{\mc{Y}}$ and then $y_1\in \wh{\mc{Y}}$,
$\e{d}_{\i\pi}\big( y_0 + \i \zeta , y_1 \big) \leq \vsg T$. Necessarily,
$y_1$ is unique, for if $y_1, y_1^{\prime}$ both satisfy the upper bound on the
distance to $y_0+\i\zeta$, then one gets
\beq
\e{d}_{\i\pi} \big(y_1, y_1^{\prime} \big) \; \leq \; \e{d}_{\i\pi} \big(y_1, y_0 + \i \zeta \big) \; + \; \e{d}_{\i\pi} \big( y_0 + \i \zeta ,  y_1^{\prime} \big)  \; \leq \; 2 \vsg T
\, < \, \mf{c}_{\e{rep}} T
\enq
and this contradicts the repulsion of roots, point e) of
Hypothesis~\ref{Hypotheses eqns de quantification}. Then one may continue so until one builds
a chain of regular roots $y_0, \dots, y_{|\wh{\mc{Y}}|}$ such that
$\e{d}_{\i\pi}\big( y_s + \i \zeta , y_{s+1} \big) \leq \vsg T$
for $s=0,\dots, |\wh{\mc{Y}}| -1$. The roots are pairwise distinct,
for if one picks $a,b\in \intn{ 0 }{ |\wh{\mc{Y}}| }$, $a\not=b$, one may always
assume by symmetry that $a<b$ and set $b=a+k$ for some $k>0$. Then, one has the bound
\beq
\e{d}_{\i\pi}\big( y_a , y_{a+k} \big) \; \geq \; \big| \e{d}_{\i\pi}\big( y_a , y_{a} + \i k \zeta \big) \, - \, \e{d}_{\i\pi}\big( y_{a} + \i k \zeta , y_{a+k} \big)  \big| \;.
\enq
Further, one gets
\bem
\e{d}_{\i\pi}\big( y_{a} + \i k \zeta , y_{a+k} \big)  \; \leq \; \sul{s=1}{k} \e{d}_{\i\pi}\Big( y_{a+s-1} + \i (k+1-s) \zeta , y_{a+s} + \i (k-s) \zeta \Big) \\
\; = \; \sul{s=1}{k} \e{d}_{\i\pi}\big( y_{a+s-1} + \i  \zeta , y_{a+s}  \big)  \; \le \; k \vsg T \; \leq \;   |\wh{\mc{Y}}| \vsg T \;.
\end{multline}
Finally, one has that
\beq
 \e{d}_{\i\pi}\big( y_a , y_{a} + \i k \zeta \big)  \; \geq  \; \e{min}\Big\{ \e{d}_{\i\pi}\big( 0 ,  \i p \zeta \big)  \; , \; p=1,\dots,   |\wh{\mc{Y}}| \Big\}
 \; \geq \; C_{  |\wh{\mc{Y}}| } \;.
\enq
Thus, all-in-all,
\beq
\e{d}_{\i\pi}\big( y_a , y_{a+k} \big) \; \geq \; C_{  |\wh{\mc{Y}}| }  -    |\wh{\mc{Y}}| \vsg T \; > \;  \vsg T \;,
\enq
hence ensuring that the roots are indeed pairwise distinct. As a consequence, we have managed to build a sequence of $1+|\wh{\mc{Y}}|$ pairwise
distinct elements of $\wh{\mc{Y}}$, what is in contradiction with the cardinality $|\wh{\mc{Y}}|$ of $\wh{\mc{Y}}$.

\vspace{2mm}

We now assume that $\wh{\wt{\mc{Y}}}_{\e{sg}}\, \not=\, \emptyset$ and that there are no maximal roots.
In that case, picking $y_0 \in \wh{\wt{\mc{Y}}}_{\e{sg}}$ one has by construction
the existence of  $y_1\in \wh{\mc{Y}}_{\e{r}}$ such that $\e{d}_{\i\pi}\big( y_0 + \i \zeta , y_1 \big) \leq \vsg T$.
Exactly as before, it holds that $y_1$ is unique. Then, one continues to build the chain. Suppose thus,
that one has built $y_0 \in \wh{\wt{\mc{Y}}}_{\e{sg}}$, $y_1, \dots, y_k \in \wh{\mc{Y}}_{\e{r}}$ such that $\e{d}_{\i\pi}\big( y_s + \i \zeta , y_{s+1} \big) \leq \vsg T$ for
$s=0,\dots, k-1$. Then, by the non-existence of maximal roots, it holds that either there exists $y  \in \wh{\mc{Y}}_{\e{r}}$ such that
$\e{d}_{\i\pi}\big( y_k + \i \zeta , y \big) \leq \vsg T$ or $y^{\prime}  \in \wh{\wt{\mc{Y}}}_{\e{sg}}$ such that
$\e{d}_{\i\pi}\big( y_k + 2 \i \zeta , y^{\prime} \big) \leq \vsg T$. If the second case scenario were to happen, then one would have that
\beq
\e{d}_{\i\pi}\big( y_0 + (2+k) \i \zeta , y^{\prime} \big)  \; \leq  \;
 \e{d}_{\i\pi}\big( y_k + 2 \i \zeta , y^{\prime} \big) \, + \,
 \sul{s=0}{k-1} \e{d}_{\i\pi}\big( y_s + (2+k-s) \i \zeta , y_{s+1} + (1+k-s) \i \zeta  \big)   \; \leq \; (k+1)\vsg T \; .
\enq
However, since $y_0, y^{\prime} \in \wh{\wt{\mc{Y}}}_{\e{sg}}$ both belong to $\msc{D}_{\mathbb{Y};\i\pi}+\i\zeta$, one has the lower bound
\beq
\e{d}_{\i\pi}\big( y_0 + (2+k) \i \zeta , y^{\prime} \big)  \; \geq  \; \e{d}_{\i\pi}\Big( \msc{D}_{\mathbb{Y};\i\pi} + (3+k) \i \zeta ,  \msc{D}_{\mathbb{Y};\i\pi} +\i\zeta  \Big)
\; \geq  \; c
\enq
by virtue of \eqref{ecriture propriete espacement fini pour domaines DY shiftes}. Put together, these two bounds lead to a contradiction.

This entails that, except for $y_0$,  only regular roots may be present in the chain,
and so one may pick $y_{k+1} \in \wh{\mc{Y}}_{\e{r}}$ such that
$\e{d}_{\i\pi}\big( y_k + \i \zeta , y_{k+1} \big)  \; \leq  \; \vsg T$,
hence extending the chain by one more regular root. Eventually, in this way, one
constructs a chain of roots $y_0, \dots, y_{|\wh{\mc{Y}}|}$, all regular except $y_0$,
such that $\e{d}_{\i\pi}\big( y_s + \i \zeta , y_{s+1} \big) \leq \vsg T$.
One shows as before that the roots are pairwise distinct, which leads to a
contradiction with the cardinality of $\wh{\mc{Y}}$.

\vspace{3mm}

We have established the existence of maximal roots. Now, we pass on to the second
part of the proof. To start with we discuss the case of a maximal root $y_0$
belonging to $\wh{\wt{\mc{Y}}}_{\e{sg}}$. Then there are two possible sub-cases
to be distinguished. Either
\begin{itemize}

 \item[$\mathrm{i)}$] there exists
$y^{\prime} \in \wh{\mc{Y}}_{\e{r}}$ such that $\e{d}_{\i\pi}\big( y_0 , y^{\prime}+2\i\zeta \big) \, \leq \, \vsg T$;

 \item[$\mathrm{ii)}$] for all $y^{\prime}  \in \wh{\mc{Y}}_{\e{r}}$, $\e{d}_{\i\pi}\big( y_0 , y^{\prime}+2\i\zeta \big) \, > \, \vsg T$.

 \end{itemize}
 Recall that $y_0-\i\zeta \in \msc{D}_{\mathbb{Y};\i\pi}$, so that for shifted singular roots one has to focus on $2\i\zeta$ shifts.

In the second case scenario, one sets $\wh{\mc{Y}}^{(1)}=\wh{\mc{Y}}\setminus \{ y_0\}$.
Then, it holds that $\e{d}_{\i\pi}\big( y_0 , y+2\i\zeta \big)\, >\, \vsg T$ for any
$y \in \wh{\mc{Y}}^{(1)}$. This is so by $\mathrm{ii)}$ above for the regular roots
and by virtue of \eqref{ecriture propriete espacement fini pour domaines DY shiftes}
for shifted singular roots. One may then repeat the whole reasoning for $\wh{\mc{Y}}^{(1)}$
as long as there are maximal shifted singular roots present.

In the first case scenario, one sets $y_1 \, = \, y^{\prime}$
so that $\e{d}_{\i\pi}\big( y_0 , y_1 + 2\i\zeta \big) \, \leq \, \vsg T$.
As earlier on, one shows that the root $y_1$ is unique and then one continues.
Thus, suppose that one has built a chain $y_0 \in \wh{\wt{\mc{Y}}}_{\e{sg}}$,
$y_1,\dots, y_k \in \wh{\mc{Y}}_{\e{r}}$ of pairwise distinct roots such that
\beq
\e{d}_{\i\pi}\big( y_0 , y_1+2\i\zeta \big) \, \leq \, \vsg T \quad \e{and} \quad \e{d}_{\i\pi}\big( y_s , y_{s+1}+\i\zeta \big) \, \leq \, \vsg T \quad \e{for}
\;\; s=1,\dots k-1 \;.
\label{ecriture spacing properties for k chain starting with a singular root}
\enq
Then, there are three cases possible:
\begin{itemize}

 \item[$\mathrm{i)}$] there exists
$y^{\prime} \in \wh{\mc{Y}}_{\e{r}}$ such that $\e{d}_{\i\pi}\big( y_k , y^{\prime}+\i\zeta \big) \, \leq \, \vsg T$;

 \item[$\mathrm{ii)}$] there exists
$y^{\prime \prime} \in \wh{\wt{\mc{Y}}}_{\e{sg}}$ such that $\e{d}_{\i\pi}\big( y_k , y^{\prime\prime}+\i\zeta \big) \, \leq \, \vsg T$;

\item[$\mathrm{iii)}$] for all $y  \in \wh{\mc{Y}} $, $\e{d}_{\i\pi}\big( y_k , y + \i\zeta \big) \, > \, \vsg T$.

\end{itemize}

In the third case scenario, one sets $\wh{\mc{Y}}^{(1)}=\wh{\mc{Y}}\setminus \{ y_0, \dots, y_k\}$. The chain of roots $y_0, \dots, y_k$  is such that
$y_0$ is a maximal shifted singular root, the roots enjoy the spacing properties \eqref{ecriture spacing properties for k chain starting with a singular root}
and that
\beq
\forall y \in\wh{\mc{Y}}^{(1)} \quad  \e{d}_{\i\pi}\big( y_0 , y + 2\i\zeta \big) \, > \, \vsg T \qquad \e{and} \qquad
\forall y \in\wh{\mc{Y}}^{(1)} \quad  \e{d}_{\i\pi}\big( y_s , y + \i\zeta \big) \, > \, \vsg T \quad s=1,\dots, k\;.
\enq
The first inequality follows from the uniqueness of the root $y_1$, the one for $s=k$ follows from $\mathrm{iii)}$ above, and the
one for $s=1,\dots, k-1$ follows from the chain of bounds
\beq
 \e{d}_{\i\pi}\big( y_s , y + \i\zeta \big)\; \geq \;  \e{d}_{\i\pi}\big( y_{s+1}+\i\zeta , y + \i\zeta \big) \; - \;  \e{d}_{\i\pi}\big( y_s , y_{s+1} + \i\zeta \big)
 \; \geq \; \mf{c}_{\e{rep}} T \, - \, \vsg T \; \geq \; \vsg T \;,
\enq
where we used point $\mathrm{e)}$ of Hypotheses \ref{Hypotheses eqns de quantification}. Then one may repeat the whole reasoning for $\wh{\mc{Y}}^{(1)}$.


The whole process terminates at some point since $|\wh{\mc{Y}}|$ is finite.
Eventually, one obtains  chains
$ \Big\{ \big\{ \mf{y}_{\ell,s} \big\}_{\ell=0}^{k_s}\Big\}_{s=1}^{p^{\prime}} $ such that $\mf{y}_{0,s}$ is maximal,
$\mf{y}_{0,s} \in  \wh{\wt{\mc{Y}}}_{\e{sg}}$,  $\mf{y}_{\ell,s}\in \wh{\mc{Y}}_{\e{r}}$ for $\ell=1,\dots, k_s$ as well as
\beq
\e{d}_{\i\pi} \big( \mf{y}_{0,s} , \mf{y}_{1,s} + 2\i\zeta \big) \, \leq \, \vsg T \;\; , \quad \e{d}_{\i\pi} \big( \mf{y}_{\ell,s} , \mf{y}_{\ell+1,s} + \i\zeta \big) \, \leq \,  \vsg T
\quad \e{for} \quad s=1,\dots, k_s \;.
\enq
Further, one has that for any $y \in \wh{\mc{Y}}\setminus \big\{ \mf{y}_{\ell,s} \big\}_{\ell=0}^{k_s}$
\beq
\e{d}_{\i\pi} \big( \mf{y}_{0,s} , y + 2\i\zeta \big) \, >\, \vsg T \quad \e{and} \quad
\e{d}_{\i\pi} \big( \mf{y}_{\ell,s} ,y + \i\zeta \big) \, > \,  \vsg T \quad \e{for} \quad s=1,\dots, k_s \;.
\enq

It remains to focus on chains starting with a regular maximal root, if these exist in
$\wh{\mc{Y}}^{\prime} \; = \;  \wh{\mc{Y}} \setminus \Big\{ \big\{ \mf{y}_{\ell,s} \big\}_{\ell=0}^{k_s}\Big\}_{s=1}^{p^{\prime}}$.
By similar arguments as in the case of maximal shifted singular roots, picking
$y_0\in \wh{\mc{Y}}_{\e{r}}$, one shows that one may build a chain of pairwise distinct
roots $y_0, \dots, y_k$ containing at most one shifted singular root such that
\beq
\e{d}_{\i\pi} \big( y_{s} , y_{s+1} + \varkappa_{s} \i\zeta \big) \, \leq \, \vsg T  \quad
\text{for}\ s=0,\dots, k-1
\quad \e{and} \quad
\e{d}_{\i\pi} \big( y_{s} , y + \varkappa_{s} \i\zeta \big) \, > \, \vsg T  \quad \forall y \in \wh{\mc{Y}}^{\prime}\setminus \{y_r\}_{r=0}^{k}
\enq
for $s=0,\dots, k$,
where $\varkappa_s=1$ if $y_{s} \in \wh{\mc{Y}}_{\e{r}}$ and  $\varkappa_s=2$ if $y_{s} \in \wh{\wt{\mc{Y}}}_{\e{sg}}$.

One may then iterate the procedure until all maximal roots are exhausted in
$\wh{\mc{Y}}$, leading to the claim. \qed

We now proceed with the proof of
Theorem \ref{Theorem classification forme racines particules et trous}. The proof
is based on Lemma~\ref{Lemme partition Y selon racines maximales} and  on a number
of further technical lemmata gathered and proved in
Section~\ref{The auxiliary lemmas}. Those lemmata explore the possibilities
of having thermal $r$ strings, possibly including singular roots, as solutions
of the quantisation conditions for $T \tend 0^+$.

 
\Proof of Theorem \ref{Theorem classification forme racines particules et trous}.

Equation
\eqref{ecriture appartenanece Trotter fini courbe trous} is a direct consequence of
the definition of the domain $\msc{D}_{\mathbb{Y}}$. It remains to prove
\eqref{ecriture appartenanece Trotter fini courbe particules}.

Assume that one is given a solution $\big(\, \wh{u}\,(\la\,|\,\wh{\mathbb{Y}}\,)
\, , \wh{\mf{X}} \, ,  \wh{\mc{Y}}\, \big)$ to the non-linear problem such that the
solution sets $\wh{\mf{X}}$, $\wh{\mc{Y}}$ satisfy
Hypotheses~\ref{Hypotheses solubilite NLIE}, \ref{Hypotheses eqns de quantification}. 
Recall that $|\wh{\mf{X}}|+|\wh{\mc{Y}}|$ are assumed to be bounded in $T$ and $N$.
We first show that the roots in $\wh{\mc{Y}}$ organize themself in thermal $r$ strings
on an extracted subsequence in $T\tend 0^+$ -- and hence $N\tend +\infty$, since
$\eta \, > \,  \tf{1}{(NT^4)}$.


By virtue of Lemma \ref{Lemme partition Y selon racines maximales} the collection
of parameters $\wh{\mc{Y}}$ decomposes as
\eqref{ecriture decomposition de hat Y en classes equivalence selon zeta shifts}.
First of all, we focus on the maximal root $\mf{y}_{0,p}$ in the notation of that
lemma and call it $y_0$ from now on.
By Lemma~\ref{Lemme incompatibilite racine singuliere et racine maximale}, provided
that $T<T_0$ and $\eta > \tf{1}{(NT^4)}$ with $T_0$, $\eta$ small enough, $y_0$ must
be a regular root. Then, by virtue of Lemma~\ref{Lemme chaine de racines reguliere}, either 
\begin{enumerate}
\item
$\Re\big[ \veps_{\e{c}} (y_0) \big] = \e{o}(1)$ as $T\tend 0^+$, and hence $N\tend +\infty$
owing to the condition involving $\eta$; or
\item
one may extract a subsequence, as $T\tend 0^+$, such that for this subsequence,
there exists $y_1 \in \wh{\mc{Y}}\setminus \{y_0\}$ such that $y_0=y_1+\i\zeta \, 
+ \, \e{O}\Big( \ex{-\f{c}{T}} \Big)$ for some $c>0$. 
\end{enumerate}
In case $\mathrm{i)}$ $y_0$ converges to a thermal $1$ string. One may then set
$\wh{y}_{1;0}^{\, (1)}=y_0$ and iterate the analysis for the remaining roots
$\wh{\mc{Y}}\setminus \{y_0\}$. If $k_p=0$ in the notation of
Lemma~\ref{Lemme partition Y selon racines maximales}, then $\wh{\mc{Y}}^{(p)}
=\{y_0\}$, so that one may start over with $\mf{y}_{0, p-1}$ which is a maximal root.
If $k_p>0$, then there exists a unique $y^{\prime} \in \wh{\mc{Y}}$ such that
$\e{d}_{\i\pi}(y_0,y^{\prime}+\i\zeta)<\vsg T$. In fact, one has $y^{\prime}=\mf{y}_{1,p}$
in the notation of Lemma~\ref{Lemme partition Y selon racines maximales}.
By using  the properties of the roots arising in
Lemma~\ref{Lemme partition Y selon racines maximales} as well as point g)
of Hypothesis~\ref{Hypotheses eqns de quantification} along with
$\Re\big[ \veps_{\e{c}} (y_0) \big] = \e{o}(1)$, which imply that
$-\e{o}(1) \leq \Im( y_0) \leq \tf{\zeta}{2}-\eps$ for some $\eps>0$,
one infers from the quantisation condition for $y_0$ that, for some constants $C, d >0$
\beq
C T^d  \; \leq \; \Big| \ex{-\f{1}{T}  \veps_{\e{c}} (y_0)  } \sinh(\i\zeta + y^{\prime}-y_0 ) \Big|  \; \leq \; C^{-1} T^{-d} \;.
\enq
Then, since $\Re\big[ \veps_{\e{c}} (y_0) \big] = \e{o}(1)$, one gets that $ \big|
\sinh(\i\zeta + y^{\prime}-y_0 ) \big|  \geq \ex{- \f{1}{T} c_T}$ with $c_T=\e{o}(1)$.
Finally, by using the spacing bounds obtained in
Lemma~\ref{Lemme partition Y selon racines maximales}, it is easy to check  that
$y^{\prime}$ is a weakly maximal root. Using the
Lemmata \ref{Lemme chaine de racines reguliere},
\ref{Lemme prolongation chaine racines qui se termine sur une racine singuliere},
\ref{Lemme prolongation chaine de racines qui contiennent un racine singuliere dans le bulk}
and the repulsion of roots, point e) of Hypotheses \ref{Hypotheses eqns de quantification},
one may then continue to partition the set $\wh{\mc{Y}}$ into subsets that
will finally become thermal r strings on an extracted subsequence in $T\tend 0^+$
-- and hence $N\tend +\infty$.

Else, one is in case $\mathrm{ii)}$ above. Then, two sub-cases are possible, either
$y_1\in \wh{\mc{Y}}_{\e{r}}$ or $y_1\in \wh{\wt{\mc{Y}}}_{\e{sg}}$. In the first
case, one proceeds again with Lemma~\ref{Lemme chaine de racines reguliere}, in
the second case
Lemma~\ref{Lemme prolongation chaine racines qui se termine sur une racine singuliere}
applies. Pursuing the first sub-case one falls on another subdivision, either  
\begin{enumerate}
\item
$\Re\big[ \veps_{\e{c}} (y_0) + \veps_{\e{c}} (y_1)\big] = \e{o}(1)$ as $T\tend 0^+$,
and hence $N\tend +\infty$ owing to the condition involving $\eta$; or
\item
one may extract a subsequence, as $T\tend 0^+$, such that for this subsequence,
there exists $y_2 \in \wh{\mc{Y}}\setminus \{y_0, y_1\}$
such that $y_{k-1}=y_{k}+\i\zeta \,  + \, \e{O}\Big( \ex{-\f{c}{T}} \Big)$ for
some $c>0$ and $k=1,2$. 
\end{enumerate}
In case $\mathrm{i)}$ of this subdivision $(y_0,y_1)$ converges to a thermal $2$ string.
Thus, one may set $\wh{y}_{1;0}^{\,(2)} = y_0$,  $\wh{y}_{1;1}^{\,(2)} = y_1$. Then,
as before, if $k_p=1$ in the notations of Lemma~\ref{Lemme partition Y selon racines maximales},
then $\wh{\mc{Y}}^{(p)}=\{y_0, y_1\}$, so that one may start again with $\mf{y}_{0,p-1}$
which is a maximal root. If $k_p>2$, then there exists a unique $y^{\prime} \in \wh{\mc{Y}}$
such that $\e{d}_{\i\pi}(y_1,y^{\prime}+\i\zeta)<\vsg T$, and one shows as above that
$y^{\prime}$ is a weakly maximal root. Whatever the situation, one is able to repeat
the analysis for the remaining roots $\wh{\mc{Y}}\setminus \{y_0, y_1\}$ by focusing
on the maximal root $\mf{y}_{0,p-1}$ or on the weakly maximal root $\mf{y}_{2,p}=y^{\prime}$.
In case $\mathrm{ii)}$ again two situations are possible, either
$y_2\in \wh{\mc{Y}}_{\e{r}}$ or $y_2\in \wh{\wt{\mc{Y}}}_{\e{sg}}$, and we may iterate
our arguments by either applying Lemma~\ref{Lemme chaine de racines reguliere} again
or by switching to
Lemma~\ref{Lemme prolongation chaine racines qui se termine sur une racine singuliere}.

All-in-all, it has become clear that,
by repeated use of Lemma~\ref{Lemme chaine de racines reguliere},
one can continue to build a chain of regular roots $y_0,\dots, y_{k-1}$ of a certain maximal
length $k \le k_p + 1$ such that, for an extracted subsequence in $T\tend 0^+$, 
\begin{itemize}
\item[i)] $y_p \, = \, y_{p+1} \, + \, \i\zeta \, + \, \de_{p}$ with $\de_{p} = \e{O}\Big( \ex{-\f{1}{T} d_p }  \Big)$  for some $d_p>0$ and $p=0,\dots k-2$\;;
\item[ii)] $\Re\Big[ \sul{p=0}{s} \veps_{\e{c}}(y_p)  \Big] \, \leq \, - c_s <0$  for  $s=0,\dots, k-2$\;; 
\end{itemize}
and such that it either terminates, if $\Re\Big[ \sum_{p=0}^{k-1} \veps_{\e{c}}(y_p)
\Big]=\e{o}(1)$ as $T\tend 0^+$, or continues with a shifted singular root
$y_{k} \in \wh{\wt{\mc{Y}}}_{\e{sg}}$. If the chain terminates, then $(y_0,y_1,\dots, y_{k-1})$
converges to a thermal $k$ string and one may set $\wh{y}_{1;0}^{\,(k)} = y_0$, 
$\wh{y}_{1;1}^{\,(k)} = y_1$, $\dots$, $\wh{y}_{1;k-1}^{\,(k)} = y_{k-1}$. As before
there are two possiblities for the termination. If $k_p=k-1$ in the notations of
Lemma~\ref{Lemme partition Y selon racines maximales}, then $\wh{\mc{Y}}^{(p)}=
\{y_0, y_1, \dots, y_{k-1}\}$, so that one may start again with $\mf{y}_{0,p-1}$
which is a maximal root. If $k_p>k-1$, then there exists a unique $y^{\prime} \in \wh{\mc{Y}}$ such that $\e{d}_{\i\pi}(y_{k-1},y^{\prime}+\i\zeta)<\vsg T$,
and one shows as above that $y^{\prime}$ is a weakly maximal root. Whatever the situation, one is able to repeat the analysis
for the remaining roots $\wh{\mc{Y}}\setminus \{y_0,   \dots, y_{k-1} \}$ by focusing on the maximal root $\mf{y}_{0,p-1}$ or on the weakly maximal root
$\mf{y}_{k,p}=y^{\prime}$.

If $\Re\Big[ \sum_{p=0}^{k-1} \veps_{\e{c}}(y_p)  \Big] \, \not= \, \e{o}(1)$ as
$T\tend 0^+$, then, due to the maximality of $y_0$ there exists
$y_{k}\in \wh{\wt{\mc{Y}}}_{\e{sg}}$ such that one may extract a subsequence satisfying 
\beq
     y_{k-1} \, = \, y_{k } \, + \, \i\zeta \, + \, \de_{k-1} \quad \e{with} \quad  \de_{k-1} =\e{O}\Big( \ex{-\f{1}{T} d_{k-1} }  \Big) \qquad \e{and} \qquad 
\Re\Big[ \sul{p=0}{k-1} \veps_{\e{c}}(y_p)  \Big] \, \leq \, - c_{k-1} < 0 \;, 
\enq
for some constants $c_{k-1}, d_{k-1} >0$, and one is exactly in the setting
of Lemma~\ref{Lemme prolongation chaine racines qui se termine sur une racine singuliere}. 
One then sets $y_{k+1}=y_k-\i\zeta$ and two options remain: either 
\beq
 \Re\Big[ \sul{p=0}{k+1} \veps_{\e{c}}(y_p)  \Big]=\e{o}(1) \quad  \e{as} \quad T\tend 0^+ \; , 
\enq
in which case the chain terminates. Then $(y_0,y_1,\dots, y_{k+1})$ converges to a thermal
$k+2$ string and one may set $\wh{y}_{1;p}^{\,(k+2)} = y_p$. As before, if $k_p=k+1$ in
the notations of Lemma \ref{Lemme partition Y selon racines maximales}, then
$\wh{\mc{Y}}^{(p)}=\{y_0, \dots, y_{k+1}\}$, so that one may start over with $\mf{y}_{0,p-1}$
which is a maximal root. If $k_p>k+1$, then there exists a unique $y^{\prime} \in \wh{\mc{Y}}$
such that $\e{d}_{\i\pi}(y_{k+1},y^{\prime}+\i\zeta)<\vsg T$, and one shows as above that
$y^{\prime}$ is a weakly maximal root. Whatever the situation, one is able to repeat the analysis
for the remaining roots $\wh{\mc{Y}}\setminus \{y_0,   \dots, y_{k+1} \}$ by focusing on
the maximal root $\mf{y}_{0,p-1}$ or on the weakly maximal root $\mf{y}_{k+2,p}=y^{\prime}$.

Else, one has an extracted subsequence and a regular root $y_{k+2} \in \wh{\mc{Y}}_{\e{r}}$
such that
\beq
y_{k + 2} \, = \, y_{k + 1} \, - \, \i\zeta \, + \, \de_{k+1} \quad \e{with} \quad  \de_{k+1} =  \e{O}\Big( \ex{-\f{1}{T} d_{k+1} } \Big) \qquad \e{and} \qquad 
\Re\Big[ \sul{p=0}{k+1} \veps_{\e{c}}(y_p)  \Big] \, \leq \, - c_{k+1} < 0 \;. 
\enq
Then, with the sequence $y_0,\dots, y_{k+2}$, one is in the setting of Lemma~\ref{Lemme prolongation chaine de racines qui contiennent un racine singuliere dans le bulk}. 
One may continue applying this lemma until this produces a chain of roots
$y_0, \dots, y_{\ell} $  which satisfy, in a subsequential limit $T\tend 0^+$, the
constraints
\begin{enumerate}
\item
$y_p \, = \, y_{p+1} \, + \, \i\zeta \, + \, \de_{p}$ with $\de_{p} = \e{O}\Big( \ex{-\f{1}{T} d_p }  \Big)$  for some $d_p>0$ and $p=0,\dots \ell-1$;
\item
$\Re\Big[ \sul{p=0}{s} \veps_{\e{c}}(y_p)  \Big] \, \leq \, - c_s <0$  for  $s=0,\dots, \ell-1$; 
\item
$\Re\Big[ \sul{p=0}{\ell} \veps_{\e{c}}(y_p)  \Big]=\e{o}(1)$ as $T\tend 0^+$\;. 
\end{enumerate}
The fact that the chain has to terminate at some point follows from the finiteness,
uniformly in $N, T$ of the set $\wh{\mc{Y}}$. Thus, one concludes that
$(y_0,y_1,\dots, y_{\ell})$ converges to a thermal $\ell+1$ string and one may set
$\wh{y}_{1;p}^{\,(\ell+1)} = y_p$. Then, as before, if $k_p=\ell$ in the notations of
Lemma \ref{Lemme partition Y selon racines maximales}, $\wh{\mc{Y}}^{(p)}=
\{y_0,   \dots, y_{\ell}\}$, so that one may start again with $\mf{y}_{0,p-1}$
which is a maximal root. If $k_p>\ell$, then there exists a unique $y^{\prime}
\in \wh{\mc{Y}}$ such that $\e{d}_{\i\pi}(y_{\ell},y^{\prime}+\i\zeta)<\vsg T$,
and one shows as above that $y^{\prime}$ is a weakly maximal root. Whatever the
situation, one is able to repeat the analysis for the remaining roots
$\wh{\mc{Y}}\setminus \{y_0,   \dots, y_{\ell} \}$ by focusing on the maximal root
$\mf{y}_{0,p-1}$ or on the weakly maximal root $\mf{y}_{\ell+1,p}=y^{\prime}$.


Finally, by using the repulsion of chains \eqref{ecriture repulsivite racines entre chaines},
one may just focus on the roots $\wh{\mc{Y}} \setminus \{y_a \}_{a=0}^{\ell}$ and repeat
the above reasoning for these. By finiteness of $\wh{\mc{Y}}$ one eventually arrives to the
conclusion that an extracted subsequence of solution sets $\wh{\mf{X}}$ and $\wh{\mc{Y}}$
to the quantisation conditions  \eqref{ecriture conditions de quantifications Trotter fini}
may be represented as 
\beq
\wh{\mf{X}} \, = \, \big\{ \, \wh{x}_1,\dots, \wh{x}_n \,  \big\} \qquad \e{and} \qquad \wh{\mc{Y}} \, = \, \bigg\{ \Big\{ \big\{ \, \wh{y}_{a;\ell}^{\; (r_k)} \big\}_{\ell=0}^{r_k-1} \Big\}_{a=1}^{n_{r_k} } \bigg\}_{k=1}^{p}
\enq
in which 
\beq
 \Re\big[ \veps_{\e{c}}\big( \wh{x}_k \big) \big] = \e{o}(1) \;, \quad \e{and} \quad \wh{x}_k \in \msc{D}_{\mathbb{Y}}\;, 
\enq
this when  $T, \tf{1}{(NT^4)} \tend 0^+$, while for some $c>0$
\beq
\wh{y}_{a;\ell}^{\; (r_k)}  \; = \; \wh{y}_{a;0}^{\; (r_k)} \, - \, \ell \i\zeta \, + \, \e{O}( \ex{- \f{c}{T}} ) \; , \; \ell = 1,\dots, r_k-1 \;, 
\enq
and $\wh{y}_{a;0}^{\; (r_k)}$ satisfies 
\beq
\Re\Big[\veps_{\e{c}; \ell }^{(-)}\big(  \,  \wh{y}_{a;0}^{\; (r_k)} \,  \big)\Big] \, < \, -c_{\ell, r_k} \;, \quad \e{for} \quad \; \ell=1,\dots, r_k-1 \qquad \e{and} \qquad 
\Re\Big[\veps_{\e{c}; r_k-1 }^{(-)}\big(   \, \wh{y}_{a;0}^{\; (r_k)} \,  \big)\Big] \, = \, \e{o}(1)
\enq
for some constants $c_{\ell, r_k} > 0$, this when  $T$, $\tf{1}{(NT^4)} \tend 0^+$. Recall that
$\veps_{\e{c}; \ell }^{(-)}$ appearing above has been defined in \eqref{definition veps c k}.

Now assume that, among the roots of $\mc{\wh{Y}}$, one may produce a subsequence
in $T\tend 0^+$ which does not satisfy the classification scheme described above. 
Then, for this extracted subsequence, by repeating the above reasoning, one obtains
an extracted sub-subsequence which falls into the above classification scheme. 
This leads to a contradiction. Therefore, the result holds in the strong sense,
\textit{viz}.\ not only for an extracted subsequence.

Let $\mc{D}_{\veps}=\big\{ \la \in \Cx \, : \, |\Im(\la)| \, \leq \, \tf{\zeta}{2} \;
\e{and} \; \Re\big[\veps(\la) \big] \; < \; 0  \big\}$ be the bounded domain in the
strip of width $\pi$ centred around $\R$ encapsulated by the curve $\Re[\veps(\la)]=0$
and introduce $\mc{D}_{\veps;\i\pi} = \mc{D}_{\veps} + \i\pi \mathbb{Z}$.  Then, observe
that, by virtue of Lemma~\ref{Lemme positivite dans plan cplx de real vepsc}, one has that if 
\beq
 \la \not\in  \ov{\mc{D}}_{\veps;\i\pi} \qquad \e{then} \qquad
 \Re\big[\veps_{\e{c}}(\la) \big] \;> \; 0  \;. 
\enq
The latter entails that, for the inequalities
\beq
\Re\Big[\veps_{\e{c}; \ell }^{(-)}\big(  \,  \wh{y}_{a;0}^{\; (r_k)} \,  \big)\Big]
\, < \, -c_{\ell, r_k} \quad \text{for}\quad \; \ell=1,\dots, r_k-1 
\enq
to hold with $c_{\ell, r_k} > 0 $ for the top part $\wh{y}_{a;0}^{\; (r_k)} $ of a string
$\big\{ \wh{y}_{a;\ell}^{\; (r_k)}  \big\}_{\ell=0}^{r_k-1}$, one necessarily needs to have
that $\wh{y}_{a;0}^{\; (r_k)} \in \mc{D}_{\veps;\i\pi}$. However, by
Lemma~\ref{Lemme ptes positivite pour veps 2 c}, one has that
\beq
\Re\Big[  \veps_{\e{c};2}^{(-)} \big( \wh{y}_{a;0}^{\; (r_k)}  \big)  \Big] \, > \,0\;.
\enq
for $\wh{y}_{a;0}^{\; (r_k)} \in \ov{\mc{D}}_{\veps;\i\pi}$.
This implies two things:
\begin{enumerate}
\item
if $r_k \, >\, 2$, the second condition at $\ell=1$, $\Re\Big[
\veps_{\e{c};2}^{(-)} \big( \, \wh{y}_{a;0}^{\; (r_k)} \,  \big)  \Big] \, < \, - c_{2,r_k}$,
cannot be met, so that no string of length greater than 2 may exist; 
\item
if $r_k\,  = \, 2$, then the root ending condition, $\Re\Big[
\veps_{\e{c};2}^{(-)} \big( \, \wh{y}_{a;0}^{\; (2)} \,  \big)  \Big] \, = \, \e{o}(1) $,
cannot be met for $T$ and $\tf{1}{(NT^4)}$ small enough, so that no string of length 2 may exist.
\end{enumerate}
Thus, we have shown that any solution set to the quantisation conditions which enjoys
Hypotheses~\ref{Hypotheses solubilite NLIE}, \ref{Hypotheses eqns de quantification}
necessarily takes the form outlined in equations
\eqref{ecriture forme asymptotique ensembles solution trous et particules}-%
\eqref{ecriture appartenanece Trotter fini courbe particules},
which also implies that $\wh{\mc{Y}}_{\e{sg}}=\emptyset$. \qed

\vspace{2mm}

We now show the reciprocal of
Theorem~\ref{Theorem classification forme racines particules et trous}, namely the
existence of solutions to the non-linear problem.

\begin{theorem}
\label{Theorem existence solutions racines particules et trous}
Assume one is given
\begin{itemize}
\item
$\mf{s}  \in \mathbb{Z}$ and  four fixed integers $|\mf{X}^{(R)}|, |\mf{X}^{(L)}|,
|\mc{Y}^{(R)}|, |\mc{Y}^{(L)}|$ such that
\beq
\mf{s} \, + \, |\mc{Y}^{(R)}| \, + \,  |\mc{Y}^{(L)}| \, - \, |\mf{X}^{(R)}| \, - \, |\mf{X}^{(L)}| \, = \, 0 \; ;
\enq
\item
a collection of integers possibly depending on $T$
\beq
0 \, \leq  \, p_{1}^{(\a)} \, < \, \dots \, < \, p_{  |\mc{Y}^{(\a)}|  }^{(\a)} \qquad and \qquad
0 \, \leq  \, h_{1}^{(\a)} \, < \, \dots \, < \, h_{  |\mf{X}^{(\a)}|  }^{(\a)}
\label{ecriture suite croissante entiers pour theoreme existence solutions}
\enq
such that $T p_{a}^{(\a)}$ and $Th_a^{(\a)}$ admit a limit when $T\tend 0^+$.
\end{itemize}
\begin{enumerate}
\item
Then, there exist $T_0, \eta>0$ which only depend on the integers $\mf{s}, |\mc{Y}^{(\a)}|,
|\mf{X}^{(\a)}|$ such that for any $0< \tf{1}{(NT^4)}<\eta $ and $0<T<T_0$, there exists
a unique solution  $\Bigl(\, \wh{u}\, \big(\la \,|\,\wh{\mathbb{Y}}\,  \big)
\, , \wh{\mf{X}} \, ,  \wh{\mc{Y}}  \Bigr)$ to the non-linear problem
\eqref{ecriture NLIE a Trotter fini}, \eqref{ecriture monodromie Trotter fini}
and \eqref{ecriture conditions de quantifications Trotter fini} which enjoys
Hypotheses \ref{Hypotheses solubilite NLIE}, \ref{Hypotheses eqns de quantification}.
\item
For this solution, it holds that
\beq
\wh{\mf{X}} \; = \; \bigcup\limits_{\a\in \{L,R\}}^{}
\Big\{ \, \wh{x}_{a}^{\; (\a)}  \Big\}_{a=1}^{ |\mf{X}^{(\a)}| }
\quad and \quad
\wh{\mc{Y}} \; = \; \bigcup\limits_{\a\in \{L,R\}}^{}
\Big\{  \, \wh{y}_{a}^{\; (\a)}  \Big\}_{a=1}^{ |\mc{Y}^{(\a)}| }
\label{ecriture ensembles X et Y solutions a}
\enq
in which
\beq
\left\{ \ba{ccc} \wh{u}\, \Big( \, \wh{x}_{a}^{\; (\a)}\,\big|\,\wh{\mathbb{Y}}\,  \Big)
                         &  =  &  - \,  2\i\pi T \ups_{\a} \big(  h_{a}^{(\a)} \, + \, \tfrac{1}{2} \big)   \vspace{2mm} \\
\wh{u}\, \Big(\,  \wh{y}_{a}^{\; (\a)}\,\big|\, \wh{\mathbb{Y}}\,  \Big)
                         &  = & 2\i\pi T  \ups_{\a} \big(  p_{a}^{(\a)} \, + \, \tfrac{1}{2} \big)  \ea \right.
\qquad and \qquad
\left\{ \ba{ccc}  \wh{u}^{\,\prime}  \Big( \, \wh{x}_{a}^{\; (\a)}\,\big|\, \wh{\mathbb{Y}}\,  \Big) &\not=& 0  \vspace{2mm} \\
\wh{u}^{\, \prime} \Big(\,  \wh{y}_{a}^{\; (\a)}\,\big|\, \wh{\mathbb{Y}}\,  \Big)  &\not=& 0  \ea \right. \;.
\label{ecriture conditions de quantification logarithmiques Trotter fini}
\enq
\item
Any such solution gives rise to a Bethe Eigenstate of the quantum transfer matrix.
\item
Solutions subordinate to different choices of integers $p_{a}^{(\a)}$ and $h_a^{(\a)}$
are different.
\item
An analogous statement holds relatively to the unique existence of a solution
$\Big(u \big(\la \,|\, \mathbb{Y}\,  \big) \, , \mf{X} \, ,  \mc{Y}  \Big)$ with
\beq
 \mf{X} \; = \; \bigcup\limits_{\a\in \{L,R\}}^{} \Big\{  x_{a}^{\; (\a)}  \Big\}_{a=0}^{ |\mf{X}^{(\a)}| }
\quad and \quad
 \mc{Y} \; = \; \bigcup\limits_{\a\in \{L,R\}}^{} \Big\{  y_{a}^{\; (\a)}  \Big\}_{a=0}^{ |\mc{Y}^{(\a)}| }
\enq
to the non-linear problem \eqref{ecriture NLIE a Trotter infini},
\eqref{ecriture monodromie Trotter infini}
and \eqref{ecriture condition de quantification Trotter infini} which enjoys
Hypotheses \ref{Hypotheses solubilite NLIE}, \ref{Hypotheses eqns de quantification}
and the infinite Trotter number equivalent of the quantisation conditions
\eqref{ecriture conditions de quantification logarithmiques Trotter fini}.
\item
In addition, the solution points admit the asymptotic expansion
\beq
\wh{x}_{a}^{\; (\a)} \, = \,  x_{a}^{(\a)}  \; + \; \e{O}\biggl( \f{1}{ N T^3 }\biggr) \qquad and \qquad
 \wh{y}_{a}^{\; (\a)}    \, = \,   y_{a}^{(\a)}   \; + \; \e{O}\biggl( \f{1}{NT^3 }\biggr)   \;,
\label{ecriture DA grand N des pts sol de NLIE pblm}
\enq
where the infinite Trotter number particle-hole parameters admit the low-$T$ expansion
\beq
x_{a}^{\, (\a)} \, = \, \lim_{T\tend 0^+} \Big\{ \veps_{\a}^{-1}\Big(  -2\i\pi T  \ups_{\a} \big(  h_a^{(\a)} \, + \, \tfrac{1}{2} \big)  \Big) \Big\}
          \, + \, \e{o}\big(1 \big)  \;
\quad  and  \quad
y_{a}^{\, (\a)} \, = \, \lim_{T\tend 0^+} \Big\{ \veps_{\a}^{-1}\Big(  2\i\pi T  \ups_{\a} \big(  p_a^{(\a)} \, + \, \tfrac{1}{2} \big)  \Big) \Big\}
\, + \, \e{o}\big(1 \big)  \;.
\label{ecriture convergence low T des racines a trotter infini}
\enq
\end{enumerate}
\end{theorem}


\Proof
First of all, we partition the sets built up from the integers $p_a^{(\a)}$ and $h_a^{(\a)}$ into finer
subsets, enjoying the property that elements in the same subset admit the same
leading low-$T$ behaviour. Namely
\beq
\Big\{ h_a^{(\a)} \Big\}_{a=1}^{|\mf{X}^{(\a)}|} \; = \; \Big\{ \big\{ h_{a;\ell}^{(\a)} \big\}_{\ell=1}^{\varkappa_a^{(\a)} } \Big\}_{a=0}^{n_{\a}} \; , \quad  \e{resp}. \quad
\Big\{ p_a^{(\a)} \Big\}_{a=1}^{|\mc{Y}^{(\a)}|} \; = \; \Big\{ \big\{ p_{a;s}^{(\a)} \big\}_{s=1}^{\mf{y}_a^{(\a)} } \Big\}_{a=0}^{m_{\a}} \;,
\label{ecriture partition entiers originaix en groupes}
\enq
so that for any $\ell , \ell^{\prime}=1,\dots, \varkappa_a^{(\a)}$, resp.\
$s, s^{\prime}=1,\dots, \mf{y}_a^{(\a)}$,
\beq
T \Big( h_{a;\ell}^{(\a)} \, - \, h_{a;\ell^{\prime}}^{(\a)} \Big) \limit{T}{0^+}0 \;,
\quad  \e{resp}. \quad
T \Big( p_{a;\ell}^{(\a)} \, - \, p_{a;\ell^{\prime}}^{(\a)} \Big) \limit{T}{0^+}0 \;,
\label{ecriture entiers convergant vers le meme point}
\enq
but if $a\not=b$
\beq
T \Big| h_{a;\ell}^{(\a)} \, - \, h_{b;\ell^{\prime}}^{(\a)} \Big| \; > \; C \;,
\quad  \e{resp}. \quad
T \Big| p_{a;\ell}^{(\a)} \, - \, p_{b;\ell^{\prime}}^{(\a)} \Big| \; > \; C
\label{ecriture entiers ayant une limite differente}
\enq
for some $C>0$ as $T\tend 0^+$. Moreover, we single out the integers $h_{0,\ell}^{(\a)}$,
resp.\ $p_{0,s}^{(\a)}$, by demanding that these specifically satisfy
\beq
T  h_{0;\ell}^{(\a)}  \limit{T}{0^+}0 \; ,
\quad  \e{resp}. \quad
T  p_{0;\ell}^{(\a)}  \limit{T}{0^+}0 \;.
\enq
Note that such integers may, of course, be non-existing among the original integers
$p_a^{(\a)}$ and $h_a^{(\a)}$. In that case, one should simply understand that
$\varkappa_0^{(\a)}=0$ or $\mf{y}_0^{(\a)}=0$.

Further, we introduce
\begin{itemize}
\item
the collection of points $\op{x}^{(\a)}_0=\ups_{\a} q, \op{x}^{(\a)}_1,
\dots, \op{x}^{(\a)}_{n_{\a}}$ with $-\tf{\zeta}{2} < \Im\big[ \op{x}^{(\a)}_a \big] \leq 0$
and  $\ups_{\a} \Re\big[ \op{x}^{(\a)}_a \big] \geq 0$,
\item
the collection of points $\op{y}^{(\a)}_0=\ups_{\a} q, \op{y}^{(\a)}_1,
\dots, \op{y}^{(\a)}_{m_{\a}}$, with $ 0\leq  \Im\big[ \op{y}^{(\a)}_a \big] < \tf{\zeta}{2}$
and $\ups_{\a} \Re\big[ \op{y}^{(\a)}_a \big] \geq 0$;
\end{itemize}
such that, for any $\ell=1,\dots, \varkappa_a^{(\a)}$, resp.\ $s=1,\dots, \mf{y}_a^{(\a)}$,
\beq
\veps_{\e{c}}\big( \op{x}_a^{(\a)} \big) \; + \; 2\i\pi T  \ups_{\a} \big(  h_{a;\ell}^{(\a)} \, + \, \tfrac{1}{2} \big)  \limit{T}{0^+}0 \; ,
\qquad  \e{resp}. \qquad
\veps_{\e{c}}\big( \op{y}_a^{(\a)} \big) \; - \; 2\i\pi T   \ups_{\a} \big(  p_{a;s}^{(\a)} \, + \, \tfrac{1}{2} \big)   \limit{T}{0^+}0  \;.
\label{ecriture contrainte sur entiers ha ell et p a s}
\enq
We stress that the above defines the points $\op{x}_a^{(\a)}$  and $\op{y}_a^{(\a)}$
uniquely and that $\op{x}_a^{(\a)}\not=\op{x}_b^{(\a^{\prime})}$, resp.\
$\op{y}_a^{(\a)}\not=\op{y}_b^{(\a^{\prime})}$, if $\a \not=\a^{\prime}$
or $\a=\a^{\prime}$ and $a\not=b$.

Next, introduce the vectors
\beq
\bs{z} \, = \, \big( \bs{z}^{(L)} , \bs{z}^{(R)} \big)^{\op{t}} \qquad \e{and} \qquad
\bs{z}^{(\a)} \, = \, \big( \bs{x}^{(\a)} , \bs{y}^{(\a)} \big)^{\op{t}}
\enq
with $\op{t}$ being the transpose, while
\beq
\bs{x}^{(\a)}\; = \; \Big( x_1^{(\a)},\dots, x_{\mc{N}_{\a}}^{(\a)}  \Big)^{\op{t}} \qquad \e{and} \qquad
\bs{y}^{(\a)}\; = \; \Big( y_1^{(\a)},\dots, y_{\mc{M}_{\a}}^{(\a)}  \Big)^{\op{t}} \;,
\enq
where we have set $\mc{N}_{\a} \, = \, \sul{a=1}{ n_{\a} } \varkappa_{a}^{(\a)}$
and $\mc{M}_{\a} \, = \, \sul{a=1}{ m_{\a} } \mf{y}_{a}^{(\a)}$. We also define
the corresponding sets
\beq
\op{Y} \, = \,  \Big\{ \big\{ y_a^{(\a)} \big\}_{a=1}^{\mc{M}_{\a}} \Big\}_{\a \in \{L,R\}}   \quad \e{and} \quad
                              \op{X} \, = \, \Big\{   \big\{ x_a^{(\a)} \big\}_{a=1}^{ \mc{N}_{\a} } \Big\}_{\a \in \{L,R\}} \;.
\label{definition des ensembles Y et X pour resoudre les contraintes}
\enq
Finally, we shall need the vectors $\bs{z}_{\e{ref}} \, = \, 
\big( \bs{z}^{(L)}_{\e{ref}} , \bs{z}^{(R)}_{\e{ref}} \big)^{\op{t}}$ and
$\bs{z}_{\e{ref}}^{(\a)} \, = \, \big( \bs{x}^{(\a)}_{\e{ref}} , \bs{y}^{(\a)}_{\e{ref}} \big)^{\op{t}}$ with
\beq
\bs{x}^{(\a)}_{\e{ref}}\; = \; \Big( \underbrace{\op{x}_1^{(\a)},\dots,\op{x}_1^{(\a)}}_{ \varkappa_1^{(\a)} \, \e{times} } \, , \, \dots  \, , \,
 \underbrace{\op{x}_{n_{\a} }^{(\a)},\dots,\op{x}_{n_{\a}}^{(\a)}}_{ \varkappa_{n_{\a} }^{(\a)} \, \e{times} }  \Big)^{\op{t}} \qquad \e{and} \qquad
\bs{y}^{(\a)}_{\e{ref}}\; = \; \Big( \underbrace{\op{y}_1^{(\a)},\dots,\op{y}_1^{(\a)}}_{ \mf{y}_1^{(\a)} \, \e{times} } \, , \, \dots  \, , \,
 \underbrace{\op{y}_{m_{\a} }^{(\a)},\dots,\op{y}_{m_{\a}}^{(\a)}}_{ \mf{y}_{m_{\a} }^{(\a)} \, \e{times} }  \Big)^{\op{t}} \;.
\enq

Now let $U_{\a}$ be a sufficiently small open neighbourhood of $\bs{z}_{\e{ref}}^{(\a)}$
in $\Cx^{\mc{N}_{\a}+\mc{M}_{\a}}$. Since $\op{x}_a^{(\a)}\not= \pm q$,
$\op{y}_a^{(\a)}\not= \pm q$ for $a\geq 1$ by hypothesis, it holds that
the coordinates of $\bs{z}_{\e{ref}}^{(\a)}$ are all uniformly away from $\pm q$.
Reducing $U_{\a}$ if need be, one concludes that the sets $\op{Y}, \op{X}$ defined
in terms of $\bs{z}\in U_{\a}$ as introduced in
\eqref{definition des ensembles Y et X pour resoudre les contraintes} satisfy
\beq
\e{d}_{\i\pi}\Big(\op{Y}, \pm q \Big) \, > \, \mf{c}_{\e{loc}}  \qquad \e{and} \qquad \e{d}_{\i\pi}\Big(\op{X}, \pm q \Big) \, > \, \mf{c}_{\e{loc}}
\enq
for some $\mf{c}_{\e{loc}}>0$.

Then, by construction, given $\ups \in \{\pm\}$,  the sets $\op{Y}, \op{X}$
defined in terms of $\bs{z}_{\e{ref}}^{(\a)}\in U_{\a}$ as introduced in
\eqref{definition des ensembles Y et X pour resoudre les contraintes} satisfy
\beq
\e{d}_{\i\pi}\Big(\op{Y}-\i\ups \zeta, \pm q \Big) \, > \, \mf{c} \;, \quad
\e{d}_{\i\pi}\Big(\op{Y}-\i\ups \zeta, \msc{C}_{\e{ref}} \Big) \, > \, \mf{c}_{\e{ref}} \cdot T
\enq
as well as
\beq
\op{Y}\cap \ov{\op{D}}_{-\i\tf{\zeta}{2}, \mf{c}_{\e{sep}} T } \, = \,
\op{Y}\cap \ov{\op{D}}_{-3\i\tf{\zeta}{2}, \mf{c}_{\e{sep}} T } \, = \,
\emptyset \qquad \e{and} \qquad
\op{X}\cap \ov{\op{D}}_{\pm \i\tf{\zeta}{2}, \mf{c}_{\e{sep}} T }  \, = \, \emptyset
\enq
for some $\mf{c}, \mf{c}_{\e{ref}}, \mf{c}_{\e{sep}}>0$ and $T>0$ small enough.
Thus, we have just shown that the collections of parameters $\op{X}, \op{Y}$
satisfy points $\mathrm{a)-d)}$ of Hypothesis~\ref{Hypotheses solubilite NLIE}.

Hence, taken the integers $h_{0;\ell}^{(\a)}$, $\ell=1,\dots, \varkappa_{0}^{(\a)}$,
and $p_{0;s}^{(\a)}$, $s=1,\dots, \mf{y}_{0}^{(\a)}$, which satisfy
$Th_{0;\ell}^{(\a)} +  T p_{0;\ell}^{(\a)} = \e{o}(1)$ by virtue of
\eqref{ecriture contrainte sur entiers ha ell et p a s}, we may introduce the function
$\wh{u}\,\big(\la\,|\,\wh{\mathbb{X}} \big)$ associated with $\wh{u}$, the unique
fixed point of the operator $\wh{\mc{L}}_T$, as ensured by
Theorem~\ref{Theorem existence et unicite sols NLIE}.
Here
%
%
%
\beq
\wh{\mathbb{X}} \; = \; \wh{\op{Y}}_{\e{tot}} \ominus \wh{\op{X}}_{\e{tot}} \quad \e{with} \quad
\begin{cases}
\; \wh{\op{Y}}_{\e{tot}} \; = \; \op{Y} \oplus \wh{\op{Y}}^{\prime} \\[1ex]
\; \wh{\op{X}}_{\e{tot}} \; = \; \op{X} \oplus \wh{\op{X}}^{\prime} ,
\end{cases}
\label{ecriture definition Yq en plusieurs variables}
\enq
in which $\op{Y}$, $\op{X}$ are defined by means of
$\bs{z}\in U_L\times U_R$ as in
\eqref{definition des ensembles Y et X pour resoudre les contraintes}, and
\beq
 \wh{\op{Y}}^{\prime}  \; = \;  \bigcup\limits_{\a \in \{L,R\} }  \Big\{ \, \wh{\op{y}}_{0;s}^{\, (\a)} \Big\}_{s=1}^{\mf{y}_0^{(\a)}}  \qquad \e{and} \qquad
\wh{\op{X}}^{\prime}  \; = \; \bigcup\limits_{\a \in \{L,R\} }  \Big\{ \, \wh{\op{x}}_{0;\ell}^{\, (\a)}  \Big\}_{\ell=1}^{\varkappa_0^{(\a)}}
\enq
are defined in terms of the unique solutions, belonging to a small open neighbourhood
of $\ups_{\a} q$, to the coupled system of equations
\beq
\wh{u}\, \Big( \wh{\op{x}}_{0;\ell}^{\, (\a)}\,\big|\,\wh{\mathbb{X}} \, \Big)
\; = \; -2\i\pi T \ups_{\a} \Big(h_{0;\ell}^{(\a)} \, + \, \tfrac{1}{2} \Big)
\qquad \e{and} \qquad
\wh{u}\, \Big( \wh{\op{y}}_{0;s}^{\, (\a)}\,\big|\, \wh{\mathbb{X}}\, \Big)
\; = \; 2\i\pi T \ups_{\a} \Big(p_{0;s}^{(\a)} \, + \, \tfrac{1}{2} \Big) \;.
\label{definition x 0 ell et y 0 ell pour variable z generique}
\enq
Recall that
Proposition~\ref{Proposition existence et continuite parameters particule trou partiels}
ensures that $\wh{\op{x}}_{0;\ell}^{\, (\a)}$ and $\wh{\op{y}}_{0;s}^{\, (\a)}$ are
all holomorphic with respect to $\bs{z} \in U_{L}\times U_{R}$, provided that these
neighbourhoods of $\bs{z}_{\e{ref}}^{(L/R)}$ are taken sufficiently small. Likewise,
Theorem~\ref{Theorem existence et unicite sols NLIE} ensures that
$\bs{z} \mapsto \wh{u}\, \big( z_a\,|\,\wh{\mathbb{X}}\big)$ is holomorphic with respect to
$\bs{z} \in U_{L}\times U_{R}$.

We are now in position to introduce the holomorphic map
\beq
\wh{\Phi}_T : U_L \times U_R \;   \tend  \;   \Big( \Cx^{\mc{N}_{L}} \times   \Cx^{\mc{M}_{L}} \Big) \times \Big( \Cx^{\mc{N}_{R}} \times   \Cx^{\mc{M}_{R}} \Big)
\quad \e{given}\; \e{by} \quad
\wh{\Phi}_T \big( \bs{z} \big) \; = \;
\begin{pmatrix}
   \wh{\Phi}_T^{(L)} \big( \bs{z} \big) \\[1ex]
   \wh{\Phi}_T^{(R)} \big( \bs{z} \big)
\end{pmatrix} \;,
\label{definition map Phi}
\enq
where
\beq
\wh{\Phi}_T^{(\a)} \big( \bs{z} \big) \; = \;
\Big( \wh{u}\, \big( \, x_1^{(\a)} \,|\, \wh{\mathbb{X}} \, \big)   \; , \dots, \; \wh{u}\, \big( \, x^{(\a)}_{\mc{N}_{\a}} \,|\,  \wh{\mathbb{X}} \, \big)  , \;
\wh{u}\, \big( \, y_1^{(\a)} \,|\,  \wh{\mathbb{X}} \, \big)   \; , \dots, \;\wh{u}\, \big( \, y^{(\a)}_{\mc{M}_{\a}} \,|\,  \wh{\mathbb{X}} \, \big) \Big)^{\op{t}} \;.
\label{definition fct Phi}
\enq
In order to compute the differential of $\wh{\Phi}_T$, it is convenient to
represent $\wh{u}$ in the form
\beq
\wh{u}\, \big(\la\,|\,\wh{\mathbb{X}} \big)  \; = \; \veps_{\e{c}} \big(  \la   \big) \, + \,
T \wh{\mc{V}}\big(\la\,|\,\bs{z}  \big) \;,
\enq
where we have introduced
\beq
     \wh{\mc{V}}\bigl(\la\,|\,\bs{z} \bigr) \; = \;
     u_{1}\big( \la\,|\,\wh{\mathbb{X}}  \, \big) \, + \,
        \f{1}{T}\wh{\mc{R}}_{T}\big[ \, \wh{u}( *  \,|\,\wh{\mathbb{X}} ) \big](\la)
\, +\, \f{1}{T}\Big( \mc{W}_N(\la)-\veps_{\e{c}}(\la) \Big) \;.
\enq
Then, one readily gets to the representation
\beq
\op{D}_{ \bs{z}  }\wh{\Phi}_T \; = \;
\begin{pmatrix}
\op{D}_{ \bs{z}^{(L)}  }\wh{\Phi}_T^{(L)}  &
\op{D}_{ \bs{z}^{(R)}  }\wh{\Phi}_T^{(L)} \\[1ex]
\op{D}_{ \bs{z}^{(L)}  }\wh{\Phi}_T^{(R)}  &
\op{D}_{ \bs{z}^{(R)}  }\wh{\Phi}_T^{(R)}
\end{pmatrix}
\enq
in which  the partial differentials take the form
\beq
\op{D}_{ \bs{z}^{(\a)}  }\wh{\Phi}_T^{(\be)} \; = \;
\begin{pmatrix}
\veps_c' \bigl(x_a^{(\be)}\bigr) \,  \de_{ab} \de_{\a \be}
\, + \, T  \Dp{x_b^{(\a)}} \wh{\mc{V}}\big( \la\,|\,\bs{z} \big)\bigr|_{\la=x_a^{(\be)}}
& T \Dp{ y_b^{(\a)} } \wh{\mc{V}}\big(\la\,|\,\bs{z} \big)\bigr|_{\la=x_a^{(\be)}} \\[2ex]
T  \Dp{ x_b^{(\a)} } \wh{\mc{V}}\big(\la\,|\,\bs{z} \big)\bigr|_{ \la=y_a^{(\be)}}^{}
& \veps_c' \bigl(y_a^{(\be)}\bigr)\, \de_{ab} \de_{\a \be}
\, + \, T  \Dp{y_b^{(\a)} } \wh{\mc{V}}\big(\la\,|\,\bs{z} \big)\bigr|_{\la=y_a^{(\be)}}
\end{pmatrix} \;.
\label{ecriture differentielle de Phi}
\enq
%
%
%

Observe that
\beq
T \wh{\mc{V}}\bigl(\la\,|\,\bs{z} \bigr) \; = \;
\e{O}\big( T, 1/(NT) \big) \qquad \e{when} \qquad   | \Im (\la) | \leq \tf{\zeta}{2}
\enq
with $\la$  uniformly away from $\pm \i\tf{\zeta}{2}$, and recall the estimates for
the partial derivatives of $\wh{\op{x}}_{0;\ell}^{\, (\a)}, \wh{\op{y}}_{0;s}^{\, (\a)}$
as ensured by \eqref{ecriture estimees en T sur les trous et particules proches de pm q}
given in
Proposition~\ref{Proposition existence et continuite parameters particule trou partiels}.
This then allows one to infer that
\beq
\bigl| \det\big[ \op{D}_{ \bs{z}  }\wh{\Phi}_T  \big] \bigr| \; \geq  \; c >0
\label{ecriture module det hat phi non nul}
\enq
uniformly in $T, \tf{1}{NT^4}$ small enough.

The latter ensures that there exists an open, $T, N$ independent, neighbourhood
$\wt{U}_{L}\times \wt{U}_{R}$, $\wt{U}_{\a}\subset U_{\a}$, of the point
$\bs{z}_{\e{ref}}$ such that $\wh{\Phi}_T: \wt{U}_{L}\times \wt{U}_{R}
\tend \wh{\Phi}_T \big( \wt{U}_L\times \wt{U}_{R} \big)$ is a biholomorphism
\cite{RangeHolomorphicFctsInManyVariables}. Moreover, observe that
\beq
\wh{\Phi}_T \, = \, \Phi_{0} \, + \, T \de \wh{\Phi}_T \qquad \e{with} \qquad
\Phi_0 \, = \, \begin{pmatrix} \Phi_0^{(L)} \\[1ex]  \Phi_0^{(R)}  \end{pmatrix} \;,
\label{ecriture introduction Phi_0 map}
\enq
where
\beq
\Phi_0^{(\a)}\big( \bs{z} \big) \, = \, \Big( \veps_{\e{c}}\big( x_1^{(\a)} \big) \, , \,  \dots \, , \, \veps_{\e{c}}\big( x_{\mc{N}_{\a}}^{(\a)} \big) \, , \,
 \veps_{\e{c}}\big( y_1^{(\a)} \big) \, , \,  \dots \, , \, \veps_{\e{c}}\big( y_{\mc{M}_{\a}}^{(\a)} \big) \Big)^{\op{t}} \;.
\label{definition coordonnee Phi0 map}
\enq
The map $\Phi_0$ is open, thus, since $\wh{\Phi}_T$ is a biholomorphism and
$\de \wh{\Phi}_T$ a norm bounded holomorphism, the open set
$\wh{\Phi}_T \big( \wt{U}_L\times \wt{U}_R \big)$ contains an open, $N,T$ independent,
neighbourhood of the point $\wh{\Phi}_T\big( \bs{z}_{\e{ref}} \big)$
and, in particular, the point $\Phi_0 \big( \bs{z}_{\e{ref}} \big)$.
One may arrive to the same conclusion by applying the multidimensional variant of the
Rouch\'{e} theorem, Theorem \ref{Theoreme Rouche generalise poru plusieurs vars complexes},
as it was done in the proof of
Proposition~\ref{Proposition existence et continuite parameters particule trou partiels}.

Now introduce
\begin{equation}
\bs{h}^{(\a)}_{\mc{N}_{\a}} =    -   2\i\pi T   \ups_{\a}\Big(     h_{1;1}^{(\a)} \, + \, \tfrac{1}{2}   \, , \, \dots \, , \,
   h_{n_{\a} ; \varkappa_{n_{\a}}^{(\a)} }^{(\a)} \, + \, \tfrac{1}{2} \Big) \;,  \quad
\bs{p}^{(\a)}_{\mc{M}_{\a}} =  2\i\pi T \ups_{\a} \Big(  p_{1;1}^{(\a)} \, + \, \tfrac{1}{2}   \, , \, \dots \, , \,
 p_{m_{\a} ; \mf{y}_{m_{\a}}^{(\a)} }^{(\a)} \, + \, \tfrac{1}{2}   \Big) \;.
\end{equation}
Since,
\beq
\Phi_0\big( \bs{z}_{\e{ref}}    \big)\, - \,
\Big( \bs{h}^{(L)}_{\mc{N}_{L}} \, , \,   \bs{p}^{(L)}_{\mc{M}_{L}} \, , \,  \bs{h}^{(R)}_{\mc{N}_{R}} \, , \,   \bs{p}^{(R)}_{\mc{M}_{R}}  \Big)^{\op{t}} \;  \limit{T}{0^+}  \; 0 \;,
\enq
it holds by the above that for $T$ small enough
\beq
\Big( \bs{h}^{(L)}_{\mc{N}_{L}} \, , \,   \bs{p}^{(L)}_{\mc{M}_{L}} \, , \,  \bs{h}^{(R)}_{\mc{N}_{R}} \, , \,   \bs{p}^{(R)}_{\mc{M}_{R}}  \Big)^{\op{t}}
\in \wh{\Phi}_T\big( \,  \wt{U}_{R}\times \wt{U}_L \,    \big) \;.
\enq
As a consequence, there exists a unique $  \wh{\bs{z}}   \in \wt{U}_{R}\times \wt{U}_L$ such that
\beq
\wh{\Phi}_{T}\big(\, \wh{ \bs{z}}\,  \big) \, = \,
\Big( \bs{h}^{(L)}_{\mc{N}_{L}} \, , \,   \bs{p}^{(L)}_{\mc{M}_{L}} \, , \,  \bs{h}^{(R)}_{\mc{N}_{R}} \, , \,   \bs{p}^{(R)}_{\mc{M}_{R}}  \Big)^{\op{t}}
\enq
upon choosing the parameterisation $\wh{\bs{z}} \, = \,
\big( \, \wh{\bs{z}}^{\, (L)} \, , \,   \wh{\bs{z}}^{\, (R)}  \, \big)^{\op{t}}$,
with $\wh{\bs{z}}^{\,(\a)}  \, = \,
\big( \, \wh{\bs{x}}^{\, (\a)} \, , \,   \wh{\bs{y}}^{\, (\a)}  \, \big)^{\op{t}}$, where
\beq
\wh{ \bs{x} }^{\, (\a)} \; = \; \Big( \wh{\op{x}}^{\, (\a)}_{1;1}, \dots, \wh{\op{x}}^{\, (\a)}_{ 1;\varkappa_1^{(\a)}} , \dots,
\wh{\op{x}}^{\, (\a)}_{n_{\a};\varkappa_{n_{\a}}^{(\a)}} \Big) \quad \e{and} \quad
\wh{\bs{y}}^{\, (\a)} \; = \; \Big( \wh{\op{y}}^{\, (\a)}_{1;1}, \dots, \wh{\op{y}}^{\, (\a)}_{1;\mf{y}_1^{(\a)}}, \dots,
\wh{\op{y}}^{\, (\a)}_{m_{\a};\mf{y}_{m_{\a}}^{(\a)}} \Big) \;.
\enq
With these notations we can also define the corresponding sets
\beq
 \wh{\mf{X}} \; = \; \bigcup\limits_{\a\in \{L,R\}}^{} \Big\{ \big\{ \, \wh{\op{x}}_{a;\ell}^{\, (\a)} \big\}_{\ell=1}^{\varkappa_a^{(\a)} } \Big\}_{a=0}^{n_{\a}}
\qquad \e{and} \qquad
 \wh{\mc{Y}} \; = \; \bigcup\limits_{\a\in \{L,R\}}^{} \Big\{ \big\{ \,  \wh{\op{y}}_{a;s}^{\, (\a)} \big\}_{s=1}^{\mf{y}_a^{(\a)} } \Big\}_{a=0}^{m_{\a}}
\label{ecriture ensembles X et Y solutions avec parametrisation speciale}
\enq
in which it is understood that $ \wh{\op{x}}_{0;\ell}^{\, (\a)} $ and $ \wh{\op{y}}_{0;s}^{\, (\a)} $ are
solving \eqref{definition x 0 ell et y 0 ell pour variable z generique} subordinate to the vector $\wh{\bs{z}}$.

We now establish that $ \wh{\op{x}}_{a;\ell}^{\, (\a)} $ and $ \wh{\op{y}}_{a;s}^{\, (\a)}$
converge to their infinite Trotter number counterparts and that the latter are well defined.
Introduce $\Phi_{T}$, the infinite Trotter number $N$ counterpart of $\wh{\Phi}_{T}$.
Similarly as above one shows the existence and uniqueness of solutions on
$ \bs{z}  \in \wt{U}_L\times \wt{U}_R$ to
\beq
\Phi_{T}\big( \bs{z}   \big) \, = \,
\Big( \bs{h}^{(L)}_{\mc{N}_{L}} \, , \,   \bs{p}^{(L)}_{\mc{M}_{L}} \, , \,  \bs{h}^{(R)}_{\mc{N}_{R}} \, , \,   \bs{p}^{(R)}_{\mc{M}_{R}}  \Big)^{\op{t}}  \;.
\label{definition map Psi T infinite Trotter number limit}
\enq
Observe that Theorem \ref{Theorem ppl existance solution NLIE} implies the bound
\beq
\norm{ \Phi_{T} \, - \, \wh{\Phi}_{T} }_{L^{\infty}( U_L \times U_R) } \; \leq \; \f{C}{N T^3} \;.
\enq
Then, upon inserting the above estimate into the relation $\wh{\Phi}_{T}\big(\, \wh{ \bs{z}}\,  \big) \, = \, \Phi_{T}\big( \bs{z}\,  \big)$,
one readily infers that, as $N \tend + \infty$ and $T$ is small enough,
\beq
 \wh{\bs{z}}   \, - \,   \bs{z}   \; = \; \e{O}\biggl( \f{1}{N T^3} \biggr) \;.
\label{label ecriture convergence vers pt solution hat z vers pt sol z}
\enq
The latter implies \eqref{ecriture DA grand N des pts sol de NLIE pblm} for the roots which correspond to
$ \wh{\op{x}}_{a;\ell}^{\, (\a)} $ and $ \wh{\op{y}}_{a;s}^{\, (\a)} $ with $a \geq 1$. One obtains the estimates
relative to the roots $ \wh{\op{x}}_{0;\ell}^{\, (\a)} $ and $ \wh{\op{y}}_{0;s}^{\, (\a)} $, should these be present,
by similar handlings involving the mapping $\Psi_{T}$, subordinate to the solution point $\bs{z}$ given in
\eqref{label ecriture convergence vers pt solution hat z vers pt sol z}, that was introduced in Proposition
\ref{Proposition existence et continuite parameters particule trou partiels}. This ensures the validity of
\eqref{ecriture DA grand N des pts sol de NLIE pblm} for any root.

Finally, it follows from
\beq
 \Phi_{T}\big( \bs{z}   \big) \; - \;  \Phi_{0}\big( \bs{z}_{\e{ref}}  \big) \; \limit{T}{0^+} \; 0
\enq
and a similar property of the map $\Psi_{T}$ introduced in Proposition \ref{Proposition existence et continuite parameters particule trou partiels}
that $  \bs{z}   \, \limit{T}{0^+} \,   \bs{z}_{\e{ref}} $, namely that \eqref{ecriture convergence low T des racines a trotter infini} holds.

It has already been established that the solution parameters $\wh{\mf{X}}$ and $\wh{\mc{Y}}$
as introduced in \eqref{ecriture ensembles X et Y solutions avec parametrisation speciale}
satisfy Hypothesis~\ref{Hypotheses solubilite NLIE}. Thus, it remains to prove that they
also satisfy Hypothesis~\ref{Hypotheses eqns de quantification}.

First of all, observe that one has the decomposition
\bem
\veps_{\e{c}}\big( \,  \wh{\op{x}}_{s;t}^{\, (\a)} \big) \, - \, \veps_{\e{c}}\big( \, \wh{\op{x}}_{s^{\prime};t^{\prime} }^{\,(\a)} \big) \; = \;
\veps_{\e{c}}\big( \op{x}_{s}^{(\a)} \big) \, - \, \veps_{\e{c}}\big( \op{x}_{s^{\prime} }^{(\a)} \big)  \\
\; - \; T \Big( \wh{\mc{V}}\big(\,  \wh{\op{x}}_{s;t}^{\,(\a)}\,|\, \wh{\bs{z}} \,  \big)
\, - \, \wh{\mc{V}}\big(\,  \wh{\op{x}}_{s^{\prime};t^{\prime}}^{\,(\a)}\,|\, \wh{\bs{z}} \,  \big)   \Big)
\, + \, \Bigl[ -  2\i\pi T \ups_{\a}   \big(h_{s;t}^{(\a)} + \tfrac{1}{2} \big)   \, - \, \veps_{\e{c}}\big( \op{x}_{s}^{(\a)} \big) \Bigr]  \\
\, - \, \Bigl[ - 2\i\pi T   \ups_{\a} \big(h_{s^{\prime};t^{\prime}}^{(\a)} + \tfrac{1}{2} \big)
\, - \, \veps_{\e{c}}\big( \op{x}_{s^{\prime}}^{(\a)} \big)   \Bigr]\;.
\label{ecriture difference energies en divers point de type trous}
\end{multline}
The last two lines are $\e{O}(T)$. Moreover, for $s\not=s^{\prime}$, since
$\op{x}_{s}^{(\a)} \not= \op{x}_{s^{\prime} }^{(\a)} $ and both
have negative imaginary part, one has that $\big|\veps_{\e{c}}\big( \op{x}_{s}^{(\a)} \big)
\, - \, \veps_{\e{c}}\big( \op{x}_{s^{\prime} }^{(\a)} \big) \big| > c$ for some $N,T$
independent $c>0$. This entails that
\beq
\big|\veps_{\e{c}}\big( \,  \wh{\op{x}}_{s;t}^{\, (\a)} \big) \, - \, \veps_{\e{c}}\big( \,  \wh{\op{x}}_{s^{\prime};t^{\prime} }^{\,(\a)} \big) \big| > \f{c}{2} \;,
\enq
which, in its turn, guarantees that $\big| \,  \wh{\op{x}}_{s;t}^{\,(\a)} \, - \,
\wh{\op{x}}_{s^{\prime};t^{\prime} }^{\, (\a)} \big|>c^{\prime}>0$.

We now focus on the $s=s^{\prime}$ case. Then, one may recast
\eqref{ecriture difference energies en divers point de type trous} in the form
\beq
\veps_{\e{c}}\big( \,  \wh{\op{x}}_{s;t}^{\, (\a)} \big) \, - \, \veps_{\e{c}}\big( \,  \wh{\op{x}}_{s;t^{\prime} }^{\, (\a)} \big) \; = \;
 T \Big( \wh{\mc{V}}\big(\,  \wh{\op{x}}_{s;t^{\prime}}^{\,(\a)}\,|\, \wh{\bs{z}} \,  \big)
\, - \, \wh{\mc{V}}\big(\,  \wh{\op{x}}_{s;t}^{\, (\a)}\,|\, \wh{\bs{z}} \,  \big)   \Big)
\, + \,2\i\pi T   \ups_{\a} \Big( h_{s;t^{\prime}}^{(\a)} \, - \, h_{s;t}^{(\a)} \Big) \;.
\label{ecriture difference dressed energies}
\enq
Further, denote
\beq
    \wh{\op{x}}_{s;t}^{\, (\a)} \,-\,   \wh{\op{x}}_{s;t^{\prime} }^{\, (\a)} \, = \,  \ex{\i \vth}
    \big| \,   \wh{\op{x}}_{s;t}^{\, (\a)} \,-\,   \wh{\op{x}}_{s;t^{\prime} }^{\, (\a)}  \big|
\qquad \e{and} \qquad
\veps^{\prime}_{\e{c}}\Big(\, \wh{\op{x}}_{s;t'}^{(\a)} \, + \, u \big( \, \wh{\op{x}}_{s;t}^{\, (\a)} \,-\, \wh{\op{x}}_{s;t^{\prime} }^{\, (\a)} \big) \Big)
\; = \; \om_u \ex{\i\psi_u} \;,
\enq
where $u \in \intff{0}{1}$. Then, one observes that
\bem
\biggl| \Int{\wh{\op{x}}_{s;t'}^{\, (\a)}}{\wh{\op{x}}_{s;t}^{\, (\a)}} \hspace{-1mm} \dd u \:
\veps^{\prime}_{\e{c}}(u) \biggr| \; = \;
\biggl| \Int{0}{1} \dd u \:
\veps^{\prime}_{\e{c}}\Big(\, \wh{\op{x}}_{s;t'}^{(\a)} \, + \,
u \big( \, \wh{\op{x}}_{s;t}^{\, (\a)} \,-\, \wh{\op{x}}_{s;t^{\prime} }^{\, (\a)} \big) \Big)
\cdot  \big( \, \wh{\op{x}}_{s;t}^{\, (\a)} \,-\,
\wh{\op{x}}_{s;t^{\prime} }^{\, (\a)}  \big)  \biggr| \\[-1ex]
\; = \; \big| \,   \wh{\op{x}}_{s;t}^{\, (\a)} \,-\,   \wh{\op{x}}_{s;t^{\prime} }^{\, (\a)}  \big| \cdot
\biggl| \Int{0}{1} \dd u \: \om_u \ex{\i(\vth+\psi_u)} \biggr|
\; \geq \; \big| \,   \wh{\op{x}}_{s;t}^{\, (\a)} \,-\,   \wh{\op{x}}_{s;t^{\prime} }^{\, (\a)}  \big| \cdot
\e{inf}\Big\{ |\veps^{\prime}_{\e{c}}(z)| \; : \; z \in U_s^{(\a)} \Big\}  \\[-1ex]
\times \max \bigg\{ \e{inf}\Big\{ |\cos\big[\vth +\e{arg}(\veps^{\prime}_{\e{c}}(z))\big]| \; : \; z \in U_s^{(\a)} \Big\},
\e{inf}\Big\{ |\sin\big[\vth +\e{arg}(\veps^{\prime}_{\e{c}}(z))\big]| \; : \; z \in U_s^{(\a)} \Big\} \bigg\} \\[-1ex]
\; \geq \; c \cdot \big| \,   \wh{\op{x}}_{s;t}^{\, (\a)} \,-\,   \wh{\op{x}}_{s;t^{\prime} }^{\, (\a)}  \big| \;.
\end{multline}
Above, $U_s^{(\a)}$ is a sufficiently small open and convex neighbourhood of $\op{x}^{(\a)}_{s}$ such that
$\wh{\op{x}}_{s;t}^{\, (\a)}, \wh{\op{x}}_{s;t^{\prime}}^{\, (\a)} \in U_s^{(\a)}$. In the last lower
bound, we used that $\veps^{\prime}\big( \op{x}_{s}^{(\a)} \big)\not= 0$ and the continuity of $z \mapsto \e{arg}(\veps^{\prime}(z))$ on $U_s^{(\a)}$.
Thus, taken that $\wh{\mc{V}}$ is bounded on $\wt{U}_L\times \wt{U}_R$, one has that
\beq
c \cdot \big| \,   \wh{\op{x}}_{s;t}^{\, (\a)} \,-\,   \wh{\op{x}}_{s;t^{\prime} }^{\, (\a)}  \big|  \; \leq \;
C T \, + \, 2\pi T \big| h_{s;t^{\prime}}^{(\a)} \, - \, h_{s;t}^{(\a)} \big| \;.
\enq
This ensures that $\wh{\op{x}}_{s;t}^{\, (\a)} \,-\,   \wh{\op{x}}_{s;t^{\prime} }^{\, (\a)} =\e{o}(1)$ as $T, \tf{1}{(NT^3)} \tend 0^+$, with $0< \tf{1}{(NT^4)}< \eta$.
In particular, one gets that in this limit
\beq
 \wh{\mc{V}}\big(\,  \wh{\op{x}}_{s;t^{\prime}}^{\,(\a)}\,|\, \wh{\bs{z}} \,  \big)
\, - \, \wh{\mc{V}}\big(\,  \wh{\op{x}}_{s;t}^{\, (\a)}\,|\, \wh{\bs{z}} \,  \big)  \; = \;\e{o}(1) \;.
\enq
Thus, upper bounding the difference of dressed energies in \eqref{ecriture difference dressed energies}, one gets that
\beq
 \big| \,   \wh{\op{x}}_{s;t}^{\, (\a)} \,-\,   \wh{\op{x}}_{s;t^{\prime} }^{\, (\a)}  \big|
 \cdot \e{sup}\Big\{ \big| \veps^{\prime}(u) \big| \; : \; u \in U_s^{(\a)} \Big\} \; \geq \;
- T \e{o}(1) \, + \, 2\pi T \big| h_{s;t^{\prime}}^{(\a)} \, - \, h_{s;t}^{(\a)} \big| \;.
\label{estimee borne inf sur distance entre points solutiuon}
\enq
Since the integers $h_{s;t}^{(\a)}$ are pairwise distinct, this
ensures that for $T, \tf{1}{(NT^4)}$ small enough, point $\mathrm{e)}$ of
Hypothesis~\ref{Hypotheses eqns de quantification} holds for $\wh{\mf{X}}$.
A very similar reasoning leads to the same conclusion for $\wh{\mc{Y}}$.

The remaining points of Hypothesis~\ref{Hypotheses eqns de quantification}
are then trivially satisfied. The fact that the roots are each of multiplicity $1$
follows from the estimates \eqref{estimee borne inf sur distance entre points solutiuon}.
Since $\op{x}^{(\a)}_s \not \in \op{D}_{-\i\tf{\zeta}{2}, 2 \eps}$
and  $\op{y}^{(\a)}_s \not \in \op{D}_{\i\tf{\zeta}{2}, 2 \eps}$ for some $\eps>0$,
it follows directly from the convergence of $\wh{\op{x}}_{s;t}^{\, (\a)} $, resp.\
$\wh{\op{y}}_{s;t}^{\, (\a)}$, to $\op{x}^{(\a)}_s$, resp.\ $\op{y}^{(\a)}_s$ that
$\wh{\op{x}}_{s;t}^{\, (\a)} \not \in \op{D}_{-\i\tf{\zeta}{2}, \eps}$ and
$\wh{\op{y}}_{s;t}^{\, (\a)}\not \in \op{D}_{\i\tf{\zeta}{2}, \eps}$.

Furthermore, since $\wh{\op{x}}_{s;t}^{\, (\a)} \tend \op{x}_{s}^{\, (\a)}$ and $\wh{\op{y}}_{s;t}^{\, (\a)} \tend \op{y}_{s}^{\, (\a)}$
when $T, \tf{1}{(NT^3)} \tend 0^+$, one gets that
\beq
\wh{u}^{\, \prime}\Big( \, \wh{\op{x}}_{s;t}^{\, (\a)}\,\big|\, \wh{\mathbb{Y}} \; \Big) \; = \; \veps_{\e{c}}^{\prime}\big( \op{x}_{s;t}^{\, (\a)}  \big) \; + \; \e{O}(T) \not= 0
\qquad \e{and} \qquad
\wh{u}^{\, \prime}\Big( \, \wh{\op{y}}_{s;t}^{\, (\a)}\,\big|\, \wh{\mathbb{Y}} \; \Big) \; = \; \veps_{\e{c}}^{\prime}\big(\op{y}_{s;t}^{\, (\a)}  \big) \; + \; \e{O}(T) \not= 0 \;.
\enq

Finally, it remains to establish that different choices of the original integers, say
\beq
 \Big( \Big\{ p_{a}^{(L)} \Big\}_{a=1}^{|\mc{Y}^{(L)}|}  ;   \Big\{ p_{a}^{(R)} \Big\}_{a=1}^{|\mc{Y}^{(R)}|}  ;
               \Big\{ h_{a}^{(L)} \Big\}_{a=1}^{|\mf{X}^{(L)}|}  ;   \Big\{ h_{a}^{(R)} \Big\}_{a=1}^{|\mf{X}^{(R)}|}   \Big)  \; \not=  \;
 \Big( \Big\{ \wt{p}_{a}^{(L)} \Big\}_{a=1}^{|\wt{\mc{Y}}^{(L)}|}  ;   \Big\{ \wt{p}_{a}^{(R)} \Big\}_{a=1}^{|\wt{\mc{Y}}^{(R)}|}  ;
               \Big\{ \wt{h}_{a}^{(L)} \Big\}_{a=1}^{|\wt{\mf{X}}^{(L)}|}  ;   \Big\{ \wt{h}_{a}^{(R)} \Big\}_{a=1}^{|\wt{\mf{X}}^{(R)}|}   \Big) \;,
\enq
give rise to distinct solutions $\Big( \wh{u}\,\big(\la \,|\, \wh{\mathbb{Y}} \big),
\wh{\mf{X}} ,  \wh{\mc{Y}}\Big)$ and
$\Big( \wh{\wt{u}}\,\big(\la \,|\, \wh{\wt{\mathbb{Y}}} \big), \wh{\wt{\mf{X}}},
\wh{\wt{\mc{Y}}}\Big)$. Assume that this is not so, \textit{viz}.\ that there exists
two distinct, not necessarily equal in cardinality, collections of integers which
give rise to equal solutions, \textit{viz}.\ that $\wh{u}\,\big(\la \,|\, \wh{\mathbb{Y}} \big)
\, =  \, \wh{\wt{u}}\,\big(\la\,|\,\wh{\wt{\mathbb{Y}}} \big)$. Note that here the parameters
$\wh{\mf{X}}$, resp.\  $\wh{\mc{Y}}$, associated with
$\wh{u}\,\big(\la\,|\, \wh{\mathbb{Y}} \big)$ are to be determined from the $h_a^{(\a)}$,
resp.\ $p_a^{(\a)}$, with the help of $\wh{u}\,\big(\la \,|\, \wh{\mathbb{Y}} \big)$,
while the parameters $\wh{\wt{\mf{X}}}$, resp.\ $\wh{\wt{\mc{Y}}}$, associated with
$\wh{ \wt{u}}\,\big(\la \,|\, \wh{\wt{\mathbb{Y}}} \big)$ are to be determined from the
$\wt{h}_a^{(\a)}$, resp.\ $\wt{p}_a^{(\a)}$, with the help of
$\wh{ \wt{u}}\,\big(\la \,|\, \wh{\wt{\mathbb{Y}}} \big)$. Of course, since there
is equality between these two functions, one may use both functions to determine
either of the parameters by using the appropriate set of integers. Owing to Hypotheses
\ref{Hypotheses solubilite NLIE}, \ref{Hypotheses eqns de quantification} and the
content of Section~\ref{Section NLIE equivalente et contour Cu}, one may go back to the
original non-linear integral equation \eqref{ecriture NLIE a Trotter fini}. One obtains that
\begin{align*}
\wh{u}\,\big( \xi\,|\, \wh{\mathbb{Y}}\big) & =  h \, -\, T\mf{w}_N(\xi)  \; - \;
\i \pi \mf{s} T  \; - \; \i T \sul{y \in \wh{\mathbb{Y}}_{ \e{ref} ; \varkappa}  }{}\th_{+}( \xi-y)
\; - \; T\Oint{ \msc{C}_{\e{ref}}  }{} \dd \la \: K(\xi-\la) \cdot
\msc{L}\mathrm{n}_{  \msc{C}_{\e{ref}} }\Big[ 1+  \ex{ - \f{1}{T}\wh{u} } \, \Big]\big( \la\,|\, \wh{\mathbb{Y}}\big) \;, \\
\wh{\wt{u}}\,\big( \xi\,|\, \wh{\wt{\mathbb{Y}}}\big) & =  h \, -\, T\mf{w}_N(\xi)  \; - \; \i \pi \wt{\mf{s}} T  \; - \; \i T
\sul{y \in \wh{\wt{\mathbb{Y}}}_{ \e{ref} ; \varkappa}  }{}\th_{+}( \xi-y)
\; - \; T\Oint{ \msc{C}_{\e{ref}}  }{} \dd \la \: K(\xi-\la) \cdot
\msc{L}\mathrm{n}_{  \msc{C}_{\e{ref}} }\Big[ 1+  \ex{ - \f{1}{T}\wh{\wt{u}} } \, \Big]\big( \la\,|\, \wh{\wt{\mathbb{Y}}}\big)
\end{align*}
in which $\wh{\mathbb{Y}}_{ \e{ref} ; \varkappa}$, resp.\
$\wh{\wt{\mathbb{Y}}}_{ \e{ref} ; \varkappa}$, is defined in terms of $\wh{\mathbb{Y}}$,
resp.\ $\wh{\wt{\mathbb{Y}}}$, by means of
\eqref{defintion mathbb X ref et son hat},
\eqref{defintion mathbb X ref et son hat avec un monodromie varkappa}. Upon using that
$\wh{\wt{u}}\,\big( \xi\,|\, \wh{\wt{\mathbb{Y}}}\big)=
\wh{u}\,\big( \xi\,|\, \wh{\mathbb{Y}}\big)$, one infers that
\beq
(-1)^{\mf{s}}\pl{y \in \wh{\mathbb{Y}}_{ \e{ref} ; \varkappa}  }{} \ex{-\i\th( \xi-y) } \; = \;
(-1)^{ \wt{\mf{s}} }\pl{y \in \wh{\wt{\mathbb{Y}}}_{ \e{ref} ; \varkappa}  }{} \ex{-\i\th( \xi-y) } \;,
\enq
on $\msc{C}_{\e{ref}}$ and thus on $\Cx$ by meromorphic continuation. Then, comparing
the poles and zeroes of both sides of this expression implies that
$\wh{\mathbb{Y}}_{ \e{ref} ; \varkappa}=
\wh{\wt{\mathbb{Y}}}_{ \e{ref} ; \varkappa}$ and $\mf{m}=\wt{\mf{m}}$.
Going back to the non-linear integral equations satisfied by the two functions,
one gets that $\mf{s}=\wt{\mf{s}}$. In order to deduce $ \wh{\mathbb{Y}}$ from
$\wh{\mathbb{Y}}_{ \e{ref} ; \varkappa}$, resp. $\wh{\wt{\mathbb{Y}}}$ from
$\wh{\wt{\mathbb{Y}}}_{ \e{ref} ; \varkappa}$, one should add or subtract the zeroes of
$1+\ex{- \f{1}{T} \wh{u}(\la\,|\,\wh{\mathbb{Y}}) }$, resp.\
$1+\ex{- \f{1}{T} \wh{\wt{u}}(\la\,|\, \wh{\wt{\mathbb{Y}}}) }$,
that are located between $\msc{C}_{\e{ref}}$ and $\wh{\msc{C}}_{\mathbb{Y}}[\,\wh{u}\,]$,
resp.\ $\wh{\msc{C}}_{\wt{\mathbb{Y}}}[\, \wh{\wt{u}}\, ]$. Since the functions are
equal, their associated contours are also equal and so are the associated zeroes
located between these contours and $\msc{C}_{\e{ref}}$. This entails that
$\wh{\mathbb{Y}} = \wh{\wt{\mathbb{Y}}}$ and, as a consequence, that
$ \wh{\mf{X}} \, = \, \wh{\wt{\mf{X}}}$ and  $\wh{\mc{Y}}=\wh{\wt{\mc{Y}}}$. Therefore,
with obvious notations, one gets
\beq
2\i\pi T \ups_{\a} \Big( p_a^{(\a)} \, + \, \tfrac{1}{2} \Big) \, = \,
\wh{u}\,\big( \, \wh{y}_a^{\, (\a)}\,|\, \wh{\mathbb{Y}} \big) \, = \,
\wh{\wt{u}}\,\big( \, \wh{y}_a^{\, (\a)}\,|\, \wh{\mathbb{Y}} \big) \, = \, \wh{\wt{u}}\,\big( \, \wh{\wt{y}}_a^{\, (\a)}\,|\, \wh{\mathbb{Y}} \big)
\; = \;  2\i\pi T \ups_{\a} \Big( \wt{p}_a^{(\a)} \, + \, \tfrac{1}{2} \Big)
\enq
and similarly for $p\hookrightarrow h$, $y \hookrightarrow x$, and
$\ups_\a \hookrightarrow - \ups_\a$. This is in contradiction to the distinctiveness
of the sets of original integers.

Finally, the fact that the solution associated with solving the quantisation conditions
\eqref{ecriture conditions de quantification logarithmiques Trotter fini} gives rise
to a Bethe Eigenstate of the quantum transfer matrix follows from
Proposition~\ref{Proposition non nullite vecteur de Bethe via formule des normes}.

Obviously, the same reasoning holds as well at infinite Trotter number, from which
one infers the statements present in the last of the theorem.
\qed

\vspace{3mm}


The analysis in Section~\ref{The auxiliary lemmas} below inspires a conjecture on
the form of the solutions sets $\wh{\mf{X}}$ and $\wh{\mc{Y}}$ in the large-$N$
and low-$T$ limits, for general $\zeta \in \intoo{0}{\pi}$.
\begin{conj}
\label{Conjecture classification forme racines particules et trous zeta general} 
Let $0<\zeta <\pi$ be generic, $\mf{s}\in \mathbb{Z}$ and let $1=r_1<\dots < r_p$ the
allowed thermal string lengths at this value of $\zeta$. Let $\wh{u}\,
\big(\la \,|\, \wh{ \mathbb{Y} } \,  \big)$ be a solution to the non-linear integral equation
\eqref{ecriture NLIE a Trotter fini} and monodromy condition
\eqref{ecriture monodromie Trotter fini} such that the collections of parameters
$\big(\, \wh{\mf{X}},\, \wh{\mc{Y}}\big)$ subject to
Hypotheses~\ref{Hypotheses solubilite NLIE}, \ref{Hypotheses eqns de quantification},
solve the quantisation conditions \eqref{ecriture conditions de quantifications Trotter fini}, 
fulfil $\mf{s} \, +\,  |\wh{\mc{Y}}|\,+\,|\wh{\mc{Y}}_{\e{sg}}|\,=\,|\wh{\mf{X}}|$ with
$|\wh{\mc{Y}}|, |\wh{\wt{\mc{Y}}}_{\e{sg}}|, |\wh{\mf{X}}|$ all bounded in
$T$, $\tf{1}{(NT^4)}$ small.

Then, there exist $T_0$, $\eta $ small enough, such that, for all $T_0>T>0$ and
$\eta> \tf{1}{(NT^4)}$, the sets $\wh{\mf{X}}$ and $\wh{\mc{Y}}$ may be represented as 
\beq
\wh{\mf{X}} \, = \, \big\{ \, \wh{x}_1,\dots, \wh{x}_n \,  \big\} \qquad and \qquad
\wh{\mc{Y}} \, = \, \bigg\{ \Big\{ \big\{ \, \wh{y}_{a;\ell}^{\; (r_k)} \big\}_{\ell=0}^{r_k-1} \Big\}_{a=1}^{n_{r_k} } \bigg\}_{k=1}^{p}
\enq
in which 
\beq
\Re\big[ \veps_{\e{c}}\big( \wh{x}_a \big) \big] = \e{o}(1) \quad and \quad \wh{x}_a \in \msc{D}_{\mathbb{Y}}\;, 
\enq
this when  $T, \tf{1}{(NT^4)} \tend 0^+$, while 
\beq
\wh{y}_{a;\ell}^{\; (r_k)}  \; = \; \wh{y}_{a;0}^{\; (r_k)} \, - \, \ell \i\zeta \, + \, \e{O}\big( \ex{- \f{c}{T}} \big) \; , \quad \ell = 1,\dots, r_k-1 \;,
\enq
and $\,\wh{y}_{a;0}^{\; (r_k)}$ satisfies for $\ell = 1,\dots, r_k-1$
\beq
\Re\Big[\veps_{\e{c}; \ell }^{(-)}\big(  \,  \wh{y}_{a;0}^{\; (r_k)} \,  \big)\Big] \, < \, -c_{\ell, r_k} \qquad and \qquad 
\Re\big[ \veps_{\e{c}; r_k }^{(-)}\big(   \, \wh{y}_{a;0}^{\; (r_k)} \,  \big) \big] \, = \, \e{o}(1)
\enq
for some constants $c_{\ell, r_k} > 0$, this when  $T, \tf{1}{(NT^4)} \tend 0^+$. Above
$\veps_{\e{c}; \ell }^{(-)}$ has been defined in \eqref{definition veps c k}. 

Reciprocally, assume one is given collections $\Big\{ \Big\{ p_{a}^{(r_k)}
\Big\}_{a=1}^{n_{r_k}} \Big\}_{k=1}^{p}$ and $\Big\{ h_a \Big\}_{a=1}^{n}$
of pairwise distinct relative integers $p_{a}^{(r_k)}, h_a\in \mathbb{Z}$ such
that  $T p_{a}^{(r_k)}$ and $Th_a$ admit a limit when $T\tend 0^+$
and so that $2\pi \lim_{T \tend 0^+}\big\{ T p_{a}^{(r_k)} \big\} $
belongs to the interior of the range of $\Im\big[ \veps_{\e{c}; r_k }^{(-)} \big]$
when restricted to the curve $\Re\big[ \veps_{\e{c}; r_k }^{(-)} \big]$.

Then there exists a solution $\wh{u}\, \big(\la \,|\, \wh{\mathbb{Y}} \, \big)$
to the non-linear integral equation \eqref{ecriture NLIE a Trotter fini}, the
monodromy conditions  \eqref{ecriture monodromie Trotter fini}
and the quantisation conditions \eqref{ecriture conditions de quantifications Trotter fini}
satisfying Hypotheses~\ref{Hypotheses solubilite NLIE},
\ref{Hypotheses eqns de quantification} such that
\beq
\wh{\mf{X}} \, = \, \big\{ \, \wh{x}_1,\dots, \wh{x}_n \,  \big\} \qquad and \qquad
\wh{\mc{Y}} \, = \, \bigg\{ \Big\{ \big\{ \, \wh{y}_{a;\ell}^{\; (r_k)} \big\}_{\ell=0}^{r_k-1} \Big\}_{a=1}^{n_{r_k} } \bigg\}_{k=1}^{p} \;.
\enq
The parameters $\wh{x}_a \in \msc{D}_{\mathbb{Y}}$ and $\,\wh{y}_{a;\ell}^{\; (r_k)}
\in \Cx \setminus\msc{D}_{\mathbb{Y};\i\pi}$ are such that
\beq
\wh{u}\, \Big( \, \wh{x}_{a}   \,\big|\, \wh{\mathbb{Y}}\,  \Big)
                         \;  =  \;   \,  2\i\pi T   \big(  h_{a}  \, + \, \tfrac{1}{2} \big)   \qquad and \qquad
\sul{\ell=1}{r_k}\wh{u}\, \Big(\,  \wh{y}_{a;\ell}^{\; (r_k)} \,\big|\, \wh{\mathbb{Y}}\,  \Big)
                         \;  = \; 2\i\pi T   \big(p_{a}^{(r_k)}  \, + \, \tfrac{ r_k }{2} \big)   \;,
\enq
while $\wh{y}_{a;\ell}^{\; (r_k)}  \; = \;
\wh{y}_{a;0}^{\; (r_k)} \, - \, \ell \i\zeta \, + \, \e{O}\big( \ex{- \f{c}{T}} \big)$.
An analogous statement holds for the case of the infinite Trotter number non-linear
integral equation \eqref{ecriture NLIE a Trotter infini}
and quantisation conditions \eqref{ecriture condition de quantification Trotter infini}.
The solution points $\wh{x}_{a}$ and $\,\wh{y}_{a;\ell}^{\; (r_k)}$ converge, with
$\e{O}\Big( \tfrac{1}{NT^3 }\Big)$ speed, to their infinite Trotter number
counterparts $x_{a}$ and $y_{a;\ell}^{\; (r_k)}$.
\end{conj}

Note that, while some elements of a proof of the above conjecture are available in
the present material, in particular the classification of the string structure
of the particle roots, others are still missing -- mainly the holomorphic in the
auxiliary parameters solvability theory for the non-linear integral equation
\eqref{ecriture NLIE a Trotter fini}. Here, in fact, the main ingredient missing
pertains to a full characterisation of the map $\veps$, in particular, to the question
of whether or not it remains a double-cover map in the regime $\tf{\pi}{2} < \zeta < \pi$.

We would like to close this subsection by establishing a no-go theorem which ensures
that only the choices of integers given in
\eqref{ecriture suite croissante entiers pour theoreme existence solutions}
do give rise to solutions to the non-linear problem which, in its turn, gives
rise to solutions to the Bethe Ansatz equations.

\begin{prop}
\label{Proposition non existence solutions NLIE avec entier coincidant qui donne une solution pblm non lineaire}
Assume one is given
\begin{itemize}
\item
$\mf{s}  \in \mathbb{Z}$ and  four fixed integers $|\mf{X}^{(R)}|, |\mf{X}^{(L)}|,
|\mc{Y}^{(R)}|, |\mc{Y}^{(L)}|$ such that
\beq
\mf{s} \, + \, |\mc{Y}^{(R)}| \, + \,  |\mc{Y}^{(L)}| \, - \, |\mf{X}^{(R)}| \, - \, |\mf{X}^{(L)}| \, = \, 0 \; ;
\enq
\item
a collection of integers $\big\{  p_{a}^{(\a)}  \big\}_{a=1}^{ |\mc{Y}^{(\a)}| }$,
$\big\{  h_{a}^{(\a)}  \big\}_{a=1}^{ |\mf{X}^{(\a)}| }$ with $\a \in \{L,R\}$,
possibly depending on $T$, such that at least two integers in the same group coincide,
\textit{viz}.\ there exists $a\not=b$ and $\a$ such that either $p_a^{(\a)}=p_b^{(\a)}$
or $h_a^{(\a)}=h_b^{(\a)}$.
\end{itemize}

Then, there exist $T_0, \eta>0$ which only depend on the integers
$\mf{s},  |\mc{Y}^{(\a)}| , |\mf{X}^{(\a)}|$ such that for any $0< \tf{1}{(NT^4)}<\eta $ and
$0<T<T_0$, there exists a unique solution
$\Big(\,\wh{u}\,\big(\la\,|\,\wh{\mathbb{Y}}\, \big)\,, \wh{\mf{X}} \,, \wh{\mc{Y}}\Big)$
to the non-linear problem \eqref{ecriture NLIE a Trotter fini},
\eqref{ecriture monodromie Trotter fini} and
\eqref{ecriture conditions de quantifications Trotter fini} which enjoys
Hypotheses \ref{Hypotheses solubilite NLIE}, \ref{Hypotheses eqns de quantification}
with the exception of point $\mathrm{f)}$. For this solution, it holds that
\beq
 \wh{\mf{X}} \; = \; \bigcup\limits_{\a\in \{L,R\}}^{} \Big\{ \, \wh{x}_{a}^{\; (\a)}  \Big\}_{a=1}^{ |\mf{X}^{(\a)}| }
\quad and \quad
 \wh{\mc{Y}} \; = \; \bigcup\limits_{\a\in \{L,R\}}^{} \Big\{  \, \wh{y}_{a}^{\; (\a)}  \Big\}_{a=1}^{ |\mc{Y}^{(\a)}| }
\label{ecriture ensembles X et Y solutions b}
\enq
in which
\beq
\left\{ \ba{ccc} \wh{u}\, \Big( \, \wh{x}_{a}^{\; (\a)}\,\big|\, \wh{\mathbb{Y}}\,  \Big)
                         &  =  &  - \,  2\i\pi T \ups_{\a} \big(  h_{a}^{(\a)} \, + \, \tfrac{1}{2} \big)   \vspace{2mm} \\
\wh{u}\, \Big(\,  \wh{y}_{a}^{\; (\a)}\,\big|\, \wh{\mathbb{Y}}\,  \Big)
                         &  = & 2\i\pi T  \ups_{\a} \big(  p_{a}^{(\a)} \, + \, \tfrac{1}{2} \big)  \ea \right.
\qquad and \qquad
\left\{ \ba{ccc}  \wh{u}^{\,\prime}  \Big( \, \wh{x}_{a}^{\; (\a)}\,\big|\, \wh{\mathbb{Y}}\,  \Big) &\not=& 0  \vspace{2mm} \\
\wh{u}^{\, \prime} \Big(\, \wh{y}_{a}^{\; (\a)}\,\big|\, \wh{\mathbb{Y}}\,  \Big)  &\not=& 0  \ea \right. \;.
\label{ecriture equations quantifications avec egalite entre entiers}
\enq
Solution points subordinate to equal integers belonging to the same group are equal,
namely if there exists $a\not=b$ and $\a$ such that $p_a^{(\a)}=p_b^{(\a)}$, resp.\
$h_a^{(\a)}=h_b^{(\a)}$, then $\wh{y}_{a}^{\; (\a)}=\wh{y}_{b}^{\; (\a)}$,
resp.\ $\wh{x}_{a}^{\; (\a)}=\wh{x}_{b}^{\; (\a)}$. In particular, owing to the
existence of larger than one multiplicities, such solutions fail to satisfy
\eqref{equation quantification particles et tous cas Trotter fini}.
\end{prop}

\Proof
One may proceed exactly as in the proof of
Theorem~\ref{Theorem existence solutions racines particules et trous} so as to show
the existence and uniqueness of solutions to the non-linear problem
\eqref{ecriture NLIE a Trotter fini}, \eqref{ecriture monodromie Trotter fini}
under Hypothesis~\ref{Hypotheses solubilite NLIE}. Moreover, one also shows in this
way the unique solvability of the quantisation conditions
\eqref{ecriture conditions de quantifications Trotter fini} written in the logarithmic form
\eqref{ecriture equations quantifications avec egalite entre entiers} even when some
of the integers with different labels are equal. This leads to a solution enjoying
Hypothesis~\ref{Hypotheses eqns de quantification}, with the exception of point $\mathrm{f)}$.

We now show that such a solution contains roots of multiplicities larger than one,
as explained in the statement of the Proposition. Recall the partitioning of roots
\eqref{ecriture partition entiers originaix en groupes} introduced in the
proof of Theorem~\ref{Theorem existence solutions racines particules et trous}. Such
roots enjoy the properties \eqref{ecriture entiers convergant vers le meme point}--\eqref{ecriture entiers ayant une limite differente}.
Then, we partition the integers
$h_{a;\ell}^{(\a)}$, $p_{a;s}^{(\a)}$ into groups built up of pairwise distinct integers, while
keeping track of their multiplicities:
\beq
\big\{ h_{a;\ell}^{(\a)} \big\}_{\ell=1}^{\varkappa_a^{(\a)} } \; = \;
\Big\{ \big\{ \,  \wt{h}_{a;\ell}^{(\a)} \big\}^{\oplus \mf{u}_{a,\ell} } \Big\}_{\ell=1}^{ \wt{\varkappa}_a^{(\a)} } \;, \quad a = 0, \dots, m_\a \;, \qquad 
 \big\{ p_{a;s}^{(\a)} \big\}_{s=1}^{\mf{y}_a^{(\a)} } \; = \;
\Big\{ \big\{ \,  \wt{p}_{a;s}^{(\a)} \big\}^{\oplus \mf{v}_{a,s}} \Big\}_{s=1}^{ \wt{\mf{y}}_a^{(\a)} }  \;,
\quad a = 0, \dots, n_\a \;.
\enq
Then we define the limiting points $\op{x}_{a}^{(\a)}$ and $\op{y}_a^{(\a)}$ as
in \eqref{ecriture contrainte sur entiers ha ell et p a s} and also introduce the vectors
\beq
\wt{\bs{z}} \, = \, \big( \, \wt{\bs{z}}^{\, (L)} , \wt{\bs{z}}^{\, (R)} \big)^{\op{t}} \qquad \e{and} \qquad
\wt{\bs{z}}^{\, (\a)} \, = \, \big( \, \wt{\bs{x}}^{\, (\a)} , \wt{\bs{y}}^{\, (\a)} \big)^{\op{t}}
\enq
with $\op{t}$ being the transpose, while
\begin{align}
\wt{\bs{x}}^{\, (\a)}  &  = \Big( x_1^{(\a)},\dots, x_{ \wt{\mc{N}}_{\a} }^{(\a)}  \Big)^{\op{t}} \; \equiv \;
\Big( x_{1;1}^{(\a)}, \dots, x_{1;\wt{\varkappa}_1^{\, (\a)}}^{\,(\a)}, \dots, x_{n_{\a};\wt{\varkappa}_{n_{\a}}^{\, (\a)}}^{\,(\a)}  \Big)^{\op{t}} \;,  \\[1ex]
\wt{\bs{y}}^{\, (\a)}  &  =  \Big( y_1^{(\a)},\dots, y_{ \wt{\mc{M}}_{\a}}^{(\a)}  \Big)^{\op{t}} \; \equiv \;
\Big( y_{1;1}^{(\a)}, \dots, y_{1;\wt{\mf{y}}_1^{\, (\a)}}^{\,(\a)}, \dots, y_{m_{\a};\wt{\mf{y}}_{m_{\a}}^{\, (\a)}}^{\,(\a)}  \Big)^{\op{t}} \;.
\end{align}
There we have set $\wt{\mc{N}}_{\a} \, = \, \sul{a=1}{ n_{\a} } \wt{\varkappa}_{a}^{\, (\a)}$ and $\wt{\mc{M}}_{\a} \, = \, \sul{a=1}{ m_{\a} } \wt{\mf{y}}_{a}^{\,(\a)}$.
We further introduce the vectors
\beq
\wt{\bs{z}}_{\e{ref}} \, = \,  \big( \, \wt{\bs{z}}^{\, (L)}_{\e{ref}} , \wt{\bs{z}}^{\,(R)}_{\e{ref}} \big)^{\op{t}} \qquad \e{and} \qquad
\wt{\bs{z}}_{\e{ref}}^{\,(\a)} \, = \, \big( \, \wt{\bs{x}}^{\,(\a)}_{\e{ref}} , \, \wt{\bs{y}}^{\, (\a)}_{\e{ref}} \big)^{\op{t}}
\enq
with
\beq
\wt{\bs{x}}^{\, (\a)}_{\e{ref}}\; = \; \Big( \underbrace{\op{x}_1^{(\a)},\dots,\op{x}_1^{(\a)}}_{ \wt{\varkappa}_1^{\, (\a)} \, \e{times} } \, , \, \dots  \, , \,
 \underbrace{\op{x}_{n_{\a} }^{(\a)},\dots,\op{x}_{n_{\a}}^{(\a)}}_{ \wt{\varkappa}_{n_{\a} }^{\,(\a)} \, \e{times} }  \Big)^{\op{t}} \quad \text{and} \quad
\wt{\bs{y}}^{\,(\a)}_{\e{ref}}\; = \; \Big( \underbrace{\op{y}_1^{(\a)},\dots,\op{y}_1^{(\a)}}_{ \wt{\mf{y}}_1^{\,(\a)} \, \e{times} } \, , \, \dots  \, , \,
 \underbrace{\op{y}_{m_{\a} }^{(\a)},\dots,\op{y}_{m_{\a}}^{(\a)}}_{ \wt{\mf{y}}_{m_{\a} }^{\, (\a)} \, \e{times} }  \Big)^{\op{t}}
\enq
and take $\wt{V}_{\a}$ to be a sufficiently small open neighbourhood of
$\wt{\bs{z}}_{\e{ref}}^{\,(\a)}$ in $\Cx^{\wt{\mc{N}}_{\a} + \wt{\mc{M}}_{\a}}$.

Next, we introduce the vectors
\beq
\wt{\bs{\mf{z}}} \, = \, \big( \, \wt{\bs{\mf{z}}}^{\, (L)} , \wt{\bs{\mf{z}}}^{\, (R)} \big)^{\op{t}} \qquad \e{and} \qquad
\wt{\bs{\mf{z}}}^{\, (\a)} \, = \, \big( \, \wt{\bs{u}}^{\, (\a)} , \wt{\bs{v}}^{\, (\a)} \big)^{\op{t}}
\enq
with $\op{t}$ being the transpose, while
\beq
\wt{\bs{u}}^{\, (\a)}  \;  =  \; \Big( x_{0;1}^{(\a)}, \dots, x_{0;\wt{\varkappa}_0^{\, (\a)}}^{\,(\a)}\Big)^{\op{t}} \qquad \e{and} \qquad
\wt{\bs{v}}^{\, (\a)}  \;   = \;   \Big( y_{0;1}^{(\a)}, \dots, y_{0;\wt{\mf{y}}_0^{\, (\a)}}^{\,(\a)} \Big)^{\op{t}} \;.
\enq

We are now in position to set up the main steps of our analysis. First of all, for any
$f \in \mc{E}_{\mc{M}}$, we introduce the map
\beq
\wt{f}\big(\la \,|\, \wt{\mathbb{X}}_{\e{gen}} \big)  \;  = \;
        \mc{W}_N(\la) \, + \, T \, u_1\big( \la \,|\, \wt{\mathbb{X}}_{\e{gen}} \big)
	\, + \, f(\la) \;,
\enq
where $u_1$ is as given in \eqref{definition fonction u 1}, while
\beq
  \wt{\mathbb{X}}_{\e{gen}}   \; = \;\wt{\op{Y}}_{\e{gen}} \ominus \wt{\op{X}}_{\e{gen}}
\qquad \text{and} \qquad
\begin{cases}
\; \wt{\op{Y}}_{\e{gen}} \; = \;  \wt{\op{Y}} \oplus \wt{\op{Y}}\; \! ^{\prime}_{\e{var}} \\[1ex]
\; \wt{\op{X}}_{\e{gen}} \; = \;   \wt{\op{X}} \oplus \wt{\op{X}}\; \! ^{\prime}_{\e{var}}.
\end{cases}
\enq
The collections of parameters appearing above are built from the components of the
vectors $\wt{\bs{x}}^{\,(\a)}, \wt{\bs{y}}^{\,(\a)}, \wt{\bs{u}}^{\,(\a)}$ and~$\wt{\bs{v}}^{\,(\a)}$:
\beq
\begin{cases}\; \wt{\op{Y}} \, =  \,
\Bigl\{\Big\{ \big\{ \,  y_{a;s}^{(\a)} \big\}^{\oplus \mf{v}_{a,s} }
       \Big\}_{s=1}^{ \wt{\mf{y}}_a^{(\a)}}\Bigr\}_{a=1}^{m_\a} \\[2ex]
\; \wt{\op{Y}}\; \! ^{\prime}_{\e{var}} \, = \,
\Big\{ \big\{ \,  y_{0;s}^{(\a)} \big\}^{\oplus \mf{v}_{0,s} }
       \Big\}_{s=1}^{ \wt{\mf{y}}_0^{(\a)}}
\end{cases}
\qquad  \e{and} \qquad
\begin{cases}
\; \wt{\op{X}} \, = \,
\Bigl\{\Big\{ \big\{ \,  x_{a;\ell}^{(\a)} \big\}^{\oplus \mf{u}_{a,\ell}}
       \Big\}_{\ell=1}^{ \wt{\mf{y}}_a^{(\a)}}\Bigr\}_{a=1}^{n_\a} \\[2ex]
\; \wt{\op{X}}\; \! ^{\prime}_{\e{var}} \, = \,
\Big\{ \big\{ \,  x_{0;\ell}^{(\a)} \big\}^{\oplus \mf{u}_{0,\ell} }
       \Big\}_{\ell=1}^{ \wt{\varkappa}_0^{(\a)} } \;.
\end{cases}
\label{defintion des ensembles X,Y tile et X,Y var}
\enq
Similarly as in the proof of
Proposition~\ref{Proposition existence et continuite parameters particule trou partiels},
one sets $\wt{\mf{h}}_{\a} \, = \, \wt{\mf{y}}^{\, (\a)}_0  + \wt{\varkappa}^{\, (\a)}_0 $,
$\wt{\mf{h}}=\wt{\mf{h}}_{L} + \wt{\mf{h}}_{R}$,
$\wt{\mc{W}}= \op{D}_{-q, \eps}^{\wt{\mf{h}}_{L}}\times \op{D}_{q, \eps}^{\wt{\mf{h}}_{R}}
\subset \Cx^{ \wt{\mf{h}} }$ with $\eps>0$ and small enough
and introduces the holomorphic map
\beq
\wt{\Psi}_T : \wt{\mc{W}} \;   \tend  \;  \Cx^{\mf{h}}
\quad \text{given by} \quad
\wt{\Psi}_T \big( \, \wt{\bs{\mf{z}}} \, \big) \; = \;
\begin{pmatrix}
\wt{\Psi}_T^{(L)} \big( \, \wt{\bs{\mf{z}}} \, \big) \\[1ex]
\wt{\Psi}_T^{(R)}  \big( \, \wt{\bs{\mf{z}}} \, \big)
\end{pmatrix}
\enq
and where
\beq
\wt{\Psi}_T^{(\a)} \big( \, \wt{\bs{\mf{z}}} \, \big) \; = \;
\Big( \wt{f}\, \big( \, x_{0;1}^{(\a)} \,|\, \wt{\mathbb{X}}_{\e{gen}} \big)   \; , \dots, \;
          \wt{f}\, \big( \, x^{(\a)}_{ 0;\, \wt{\varkappa}^{\, (\a)}_0 } \,|\, \wt{\mathbb{X}}_{\e{gen}} \big)  , \;
\wt{f}\, \big( \, y_{0;1}^{(\a)} \,|\, \wt{\mathbb{X}}_{\e{gen}} \big)   \; , \dots,
                         \;\wt{f}\, \big( \, y^{(\a)}_{ 0; \wt{\mf{y}}^{\, (\a)}_0 } \,|\, \wt{\mathbb{X}}_{\e{gen}} \big) \Big)^{\op{t}} \;.
\enq
Exactly as in the proof of
Proposition~\ref{Proposition existence et continuite parameters particule trou partiels},
one shows that $\wt{\Psi}_T$ is a bi-holomorphism onto its image, from which follows
the unique solvability of the equation
\beq
\wt{\Psi}_T\big( \, \wt{\bs{\mf{z}}} \, \big)\, = \,
 \Big( \bs{h}^{(L)}_{  \wt{\varkappa}^{\, (L)}_0 } \, , \,   \bs{p}^{(L)}_{  \wt{\mf{y}}^{\, (L)}_0 } \, ,
                              \,  \bs{h}^{(R)}_{ \wt{\varkappa}^{\,(R)}_0 } \, , \,   \bs{p}^{(R)}_{ \wt{\mf{y}}^{\, (R)}_0 }  \Big)^{\op{t}}
\enq
in which
\beq
\bs{h}^{(\a)}_{ \wt{\varkappa}^{\, (\a)}_0 } =  -  \ups_{\a} 2\i\pi T \Big( \,  \wt{h}_{0;1}^{\, (\a)} \, + \, \tfrac{1}{2} \, , \, \dots \, , \,
\wt{h}_{ 0; \wt{\varkappa}^{\,(\a)}_0 }^{(\a)} \, + \, \tfrac{1}{2}  \Big)^{\op{t}}  \;, \quad
\bs{p}^{(\a)}_{ \wt{\mf{y}}^{\, (\a)}_0 } = \ups_{\a} 2\i\pi T \Big(  \, \wt{p}_{0;1}^{\, (\a)} \, + \, \tfrac{1}{2} \, , \, \dots \, , \,
 \wt{p}_{ 0; \wt{\mf{y}}^{\,(\a)}_0 }^{(\a)} \, + \, \tfrac{1}{2}  \Big)^{\op{t}} \;.
\enq
We denote the coordinates of that solution $ \wt{\op{y}}_{0;s}^{\, (\a)},
\wt{\op{x}}_{0;\ell}^{\, (\a)}$, introduce
\beq
 \wt{\op{Y}}\;\!^{\prime}  \; = \;  \bigcup\limits_{\a \in \{L,R\} }
 \Big\{ \big\{ \, \wt{\op{y}}_{0;s}^{\, (\a)} \big\}^{\oplus \mf{v}_{0,s} }  \Big\}_{s=1}^{ \wt{\mf{y}}_0^{\, (\a)}}
\qquad \e{and} \qquad
\wt{\op{X}}\;\!^{\prime}  \; = \; \bigcup\limits_{\a \in \{L,R\} }
\Big\{  \big\{ \, \wt{\op{x}}_{0;\ell}^{\, (\a)}  \big\}^{\oplus \mf{u}_{0,\ell} } \Big\}_{\ell=1}^{ \wt{\varkappa}_0^{\, (\a)}}
\enq
and further denote
\beq
\wt{\mathbb{X}} \; = \; \wt{\op{Y}}_{\e{tot}} \ominus \wt{\op{X}}_{\e{tot}} \quad \e{with} \quad
\begin{cases}
\; \wt{\op{Y}}_{\e{tot}} \, = \, \wt{\op{Y}} \oplus \wt{\op{Y}}\;\!^{\prime}   \\[1ex]
\; \wt{\op{X}}_{\e{tot}} \, = \, \wt{\op{X}} \oplus \wt{\op{X}}\;\!^{\prime},
\end{cases}
\enq
in which $\wt{\op{Y}}$, $\wt{\op{X}}$ have been defined by means of
$\,\wt{\bs{z}}\in \wt{V}_L\times \wt{V}_R$ as given in
\eqref{defintion des ensembles X,Y tile et X,Y var}.

Exactly as in the proof of Theorem~\ref{Theorem existence solutions racines particules et trous},
one justifies that one may introduce the solution $\wh{u}\,\big(\la \,|\, \wt{\mathbb{X}} \big)$
associated with $\wh{u}$, the unique fixed point of the operator $\wh{\mc{L}}_T$, as ensured
by Theorem~\ref{Theorem existence et unicite sols NLIE}. Exactly as in the proof of
Proposition~\ref{Proposition existence et continuite parameters particule trou partiels}
one justifies that $\wt{\op{x}}_{0;\ell}^{\, (\a)}$ and $\wt{\op{y}}_{0;s}^{\, (\a)}$
are all holomorphic with respect to $\wt{\bs{z}} \in \wt{V}_{L}\times \wt{V}_{R}$, provided
that these neighbourhoods of $\bs{z}_{\e{ref}}^{(L/R)}$ are taken sufficiently small.
Likewise, Theorem \ref{Theorem existence et unicite sols NLIE} ensures that
$\,\wt{\bs{z}} \mapsto \wh{u}\, \big( z_a \,|\, \wt{\mathbb{X}} \big)$ is holomorphic with
respect to $\,\wt{\bs{z}} \in \wt{V}_{L}\times \wt{V}_{R}$.

Analogously to our previous considerations we define the holomorphic map
\beq
\wt{\Phi}_T : \wt{V}_L \times \wt{V}_R \;   \tend  \;   \Big( \Cx^{ \wt{\mc{N}}_{L} } \times   \Cx^{ \wt{\mc{M}}_{L} } \Big) \times \Big( \Cx^{ \wt{\mc{N}}_{R} }
\times   \Cx^{ \wt{\mc{M}}_{R}} \Big)
\quad \e{given}\; \e{by} \quad
\wt{\Phi}_T \big( \, \wt{\bs{z}} \, \big) \; = \;
\begin{pmatrix}
\wt{\Phi}_T^{(L)} \big( \,  \wt{\bs{z}} \, \big) \\[1ex]
\wt{\Phi}_T^{(R)} \big( \, \wt{\bs{z}} \, \big)
\end{pmatrix}
\enq
and where
\beq
\wt{\Phi}_T^{(\a)} \big( \, \wt{\bs{z}} \big) \; = \;
\Big( \wh{u}\, \big( \, x_1^{(\a)} \,|\, \wt{\mathbb{X}} \, \big)   \; , \dots, \; \wh{u}\, \big( \, x^{(\a)}_{ \wt{\mc{N}}_{\a}} \,|\,  \wt{\mathbb{X}} \, \big)  , \;
\wh{u}\, \big( \, y_1^{(\a)} \,|\,  \wt{\mathbb{X}} \, \big)   \; , \dots, \;\wh{u}\, \big( \, y^{(\a)}_{  \wt{\mc{M}}_{\a}} \,|\,  \wt{\mathbb{X}} \, \big) \Big)^{\op{t}} \;.
\enq
As in the proof of Theorem \ref{Theorem existence et unicite sols NLIE} one shows that
$\wt{\Phi}_T : \wt{V}_L \times \wt{V}_R \mapsto \wt{\Phi}_T\big( \wt{V}_L \times \wt{V}_R  \big)$
is a biholomorphism and that $\wt{\Phi}_T\big( \wt{V}_L \times \wt{V}_R  \big)$ contains
a fixed $T,N$-independent open neighbourhood of $\Phi_0\big( \wt{V}_L \times \wt{V}_R \big)$
with $\Phi_0$ as defined through \eqref{ecriture introduction Phi_0 map}--\eqref{definition coordonnee Phi0 map}.

Now, by introducing
\beq
\bs{h}^{(\a)}_{ \wt{\mc{N}}_{\a}} =  - 2\i\pi T   \ups_{\a}\, \Big(  \,   \wt{h}_{1;1}^{\, (\a)} \, + \, \tfrac{1}{2}   \, , \, \dots \, , \,
\wt{h}_{n_{\a} ; \wt{\varkappa}_{n_{\a}}^{(\a)} }^{\, (\a)} \, + \, \tfrac{1}{2}    \Big)^{\op{t}}
\;, \quad
\bs{p}^{(\a)}_{ \wt{\mc{M}}_{\a} } =  2\i\pi T \ups_{\a} \,  \Big(  \, \wt{p}_{1;1}^{\, (\a)} \, + \, \tfrac{1}{2}   \, , \, \dots \, , \,
 \wt{p}_{ m_{\a} ; \wt{\mf{y}}_{m_{\a}}^{(\a)} }^{\, (\a)} \, + \, \tfrac{1}{2}   \Big)^{\op{t}} \;,
\enq
we see that there exists a unique $\wt{\bs{z}}   \in \wt{U}_{R}\times \wt{U}_L$ such that
\beq
\wt{\Phi}_{T}\big(\, \wt{ \bs{z}}\,  \big) \, = \,
\Big( \bs{h}^{(L)}_{ \wt{\mc{N}}_{L} } \, , \,   \bs{p}^{(L)}_{ \wt{\mc{M}}_{L} } \, , \,  \bs{h}^{(R)}_{ \wt{\mc{N}}_{R} } \, , \,   \bs{p}^{(R)}_{ \wt{\mc{M}}_{R} }  \Big)^{\op{t}}  \;.
\enq
Upon parameterising the coordinates of that solution by $ \wt{\op{x}}^{\, (\a)}_{ a; \ell} $
and $ \wt{\op{y}}^{\, (\a)}_{ a; s} $
 we set
\beq
 \wt{\mf{X}} \; = \; \bigcup\limits_{\a\in \{L,R\}}^{} \bigg\{ \Big\{ \big\{ \, \wt{\op{x}}_{a;\ell}^{\, (\a)} \big\}^{\oplus \mf{u}_{a;\ell} }
 \Big\}_{\ell=1}^{ \wt{\varkappa}_a^{(\a)} } \bigg\}_{a=0}^{n_{\a}}
\qquad \e{and} \qquad
 \wt{\mc{Y}} \; = \; \bigcup\limits_{\a\in \{L,R\}}^{} \bigg\{ \Big\{ \big\{ \,  \wt{\op{y}}_{a;s}^{\, (\a)} \big\}^{\oplus \mf{v}_{a;s} }
 \Big\}_{s=1}^{ \wt{\mf{y}}_a^{(\a)} } \bigg\}_{a=0}^{m_{\a}} \;.
\enq
All-in-all, agreeing upon $\wt{\mathbb{Y}}\, = \,  \wt{\mc{Y}} \ominus  \wt{\mf{X}} $
we have built the solution to the system of equations
\beq
\wh{u}\, \Big( \,  \wt{\op{x}}_{a;\ell}^{\, (\a)}  \,\big|\, \wt{\mathbb{Y}} \, \Big) \; = \; -2\i\pi T \ups_{\a} \Big(\, \wt{h}_{a;\ell}^{(\a)} \, + \, \tfrac{1}{2} \Big)
\qquad \e{and} \qquad
\wh{u}\, \Big(\,  \wt{\op{y}}_{a;s}^{\, (\a)}  \,\big|\, \wt{\mathbb{Y}}\, \Big) \; = \; 2\i\pi T \ups_{\a} \Big( \, \wt{p}_{a;s}^{(\a)} \, + \, \tfrac{1}{2} \Big) \;.
\enq
We now introduce the vectors
\beq
\bs{\mf{z}}_{\e{sol}} \, = \, \big( \, \bs{\mf{z}}^{\, (L)} ,  \bs{\mf{z}}^{\, (R)} \big)^{\op{t}} \qquad \e{and} \qquad
\bs{\mf{z}}^{\, (\a)} \, = \, \big(  \bs{u}^{\, (\a)} , \bs{v}^{\, (\a)} \big)^{\op{t}}
\;,
\enq
whose coordinates are given by
\beq
\bs{u}^{\, (\a)} \, = \, \Big( \underbrace{\wt{\op{x}}_{0;1}^{(\a)}, \dots, \wt{\op{x}}_{0;1}^{(\a)}}_{ \mf{u}_{0,1}^{(\a)} },
\dots, \underbrace{ \wt{\op{x}}_{0;\wt{\varkappa}_0^{\, (\a)}}^{\,(\a)} , \dots,   \wt{\op{x}}_{0;\wt{\varkappa}_0^{\, (\a)}}^{\,(\a)} }
_{ \mf{u}_{ 0,\wt{\varkappa}_0^{\, (\a)}}^{\,(\a)} }  \Big)^{\op{t}}
\;, \quad
%
%
%
%
%
%
%
%
\bs{v}^{\, (\a)} \,   = \, \Big( \underbrace{\wt{\op{y}}_{0;1}^{(\a)}, \dots, \wt{\op{y}}_{0;1}^{(\a)}}_{ \mf{v}_{0,1}^{(\a)} },
\dots, \underbrace{ \wt{\op{y}}_{0;\wt{\varkappa}_0^{\, (\a)}}^{\,(\a)} , \dots,   \wt{\op{y}}_{0;\wt{\varkappa}_0^{\, (\a)}}^{\,(\a)} }
_{ \mf{v}_{ 0,\wt{\varkappa}_0^{\, (\a)}}^{\,(\a)} }  \Big)^{\op{t}} \;.
\enq
Analogously, we introduce the vectors
\beq
\bs{z}_{\e{sol}} \, = \, \big( \, \bs{z}^{\, (L)} ,  \bs{z}^{\, (R)} \big)^{\op{t}} \qquad \e{and} \qquad
\bs{z}^{\, (\a)} \, = \, \big(  \bs{x}^{\, (\a)} , \bs{y}^{\, (\a)} \big)^{\op{t}} \;,
\enq
%
%
whose building blocks read
\beq
\bs{x}^{\, (\a)}  \,  =  \, \Big( \underbrace{\wt{\op{x}}_{1;1}^{(\a)}, \dots, \wt{\op{x}}_{1;1}^{(\a)}}_{ \mf{u}_{1,1}^{(\a)} },
\dots, \underbrace{ \wt{\op{x}}_{n_{\a};\wt{\varkappa}_{n_{\a}}^{\, (\a)}}^{\,(\a)} , \dots,   \wt{\op{x}}_{n_{\a};\wt{\varkappa}_{n_{\a}}^{\, (\a)}}^{\,(\a)} }
_{ \mf{u}_{ n_{\a} ,\wt{\varkappa}_{n_{\a}}^{\, (\a)}}^{\,(\a)} }  \Big)^{\op{t}}, \
%
%
%
%
%
%
%
%
\bs{y}^{\, (\a)}  \,   = \,    \Big( \underbrace{\wt{\op{y}}_{1;1}^{(\a)}, \dots, \wt{\op{y}}_{1;1}^{(\a)}}_{ \mf{v}_{1,1}^{(\a)} },
\dots, \underbrace{ \wt{\op{y}}_{m_{\a};\wt{\varkappa}_{m_{\a}}^{\, (\a)}}^{\,(\a)} , \dots,   \wt{\op{y}}_{ m_{\a} ;\wt{\varkappa}_{ m_{\a} }^{\, (\a)}}^{\,(\a)} }
_{ \mf{v}_{ m_{\a},\wt{\varkappa}_{ m_{\a} }^{\, (\a)}}^{\,(\a)} }  \Big)^{\op{t}}.
\enq
We continue on by introducing the vectors of integers
\begin{align}
\bs{h}^{(\a)}_{ \varkappa^{\, (\a)}_0 }\; & = -  \ups_{\a} 2\i\pi T \Big( \,
\underbrace{ \wt{h}_{0;1}^{\, (\a)} \, + \, \tfrac{1}{2} \, , \, \dots \, , \,   \wt{h}_{0;1}^{\, (\a)} \, + \, \tfrac{1}{2} }_{ \mf{u}_{0,1}^{(\a)} }
\, , \, \dots \, , \,
\underbrace{ \wt{h}_{ 0; \wt{\varkappa}^{\,(\a)}_0 }^{(\a)} \, + \, \tfrac{1}{2}  \, ,
\, \dots \, , \,  \wt{h}_{ 0; \wt{\varkappa}^{\,(\a)}_0 }^{(\a)} \, + \, \tfrac{1}{2} }
_{ \mf{u}_{0, \wt{\varkappa}^{\,(\a)}_0 }^{(\a)} } \Big)^{\op{t}}  \; ,  \\[-1ex]
\bs{p}^{(\a)}_{ \mf{y}^{\, (\a)}_0 }\; & = \;\;  \ups_{\a} 2\i\pi T \Big(  \,
\underbrace{ \wt{p}_{0;1}^{\, (\a)} \, + \, \tfrac{1}{2} \, , \, \dots \, , \,\wt{p}_{0;1}^{\, (\a)} \, + \, \tfrac{1}{2} }_{ \mf{v}_{0,1}^{(\a)} },  \, \dots \, , \,
 \underbrace{   \wt{p}_{ 0; \wt{\mf{y}}^{\,(\a)}_0 }^{(\a)} \, + \, \tfrac{1}{2} , \, \dots \, , \,   \wt{p}_{ 0; \wt{\mf{y}}^{\,(\a)}_0 }^{(\a)} \, + \, \tfrac{1}{2} }
_{ \mf{v}_{0,\wt{\mf{y}}^{\,(\a)}_0 }^{(\a)} }\Big)^{\op{t}} \;.
\end{align}
as well as
\begin{align}
\bs{h}^{(\a)}_{ \mc{N}_{\a}} \; & =  -   2\i\pi T   \ups_{\a}\, \Big(  \,
\underbrace{ \wt{h}_{1;1}^{\, (\a)} \, + \, \tfrac{1}{2}   \, , \, \dots \, , \, \wt{h}_{1;1}^{\, (\a)} \, + \, \tfrac{1}{2}  }_{ \mf{u}_{1,1}^{(\a)} } ,   \, \dots \, , \,
  \underbrace{   \wt{h}_{n_{\a} ; \wt{\varkappa}_{n_{\a}}^{(\a)} }^{\, (\a)} \, + \, \tfrac{1}{2}  \, , \, \dots \, , \,
     \wt{h}_{n_{\a} ; \wt{\varkappa}_{n_{\a}}^{(\a)} }^{\, (\a)} \, + \, \tfrac{1}{2}  }_{ \mf{u}_{n_{\a} ; \wt{\varkappa}_{n_{\a}}^{(\a)} }^{(\a)} }  \Big)^{\op{t}} \;,  \\[-1ex]
\bs{p}^{(\a)}_{ \mc{M}_{\a} } \; & = \;\; 2\i\pi T \ups_{\a} \,  \Big(  \,
\underbrace{ \wt{p}_{1;1}^{\, (\a)} \, + \, \tfrac{1}{2}   \, , \, \cdots \, , \, \wt{p}_{1;1}^{\, (\a)} \, + \, \tfrac{1}{2}  }_{ \mf{v}_{1,1}^{(\a)} }
\, , \, \cdots \, , \,
\underbrace{  \wt{p}_{ m_{\a} ; \wt{\mf{y}}_{m_{\a}}^{(\a)} }^{\, (\a)} \, + \, \tfrac{1}{2} \, , \, \cdots \, , \,
 \wt{p}_{ m_{\a} ; \wt{\mf{y}}_{m_{\a}}^{(\a)} }^{\, (\a)} \, + \, \tfrac{1}{2}    }_{ \mf{v}_{ m_{\a} ; \wt{\mf{y}}_{m_{\a}}^{(\a)} }^{(\a)} } \Big)^{\op{t}} \;.
\end{align}

Thus, recalling the biholomorphisms $\wh{\Psi}_{T}$, see
\eqref{definition composantes left and right de la map Psi T}, \eqref{definition fct Psi}
from Proposition~\ref{Proposition existence et continuite parameters particule trou partiels}
and $\wh{\Phi}_{T}$, see \eqref{definition fct Phi}, \eqref{definition map Phi}, from
Theorem~\ref{Theorem existence solutions racines particules et trous}, we get that the
unique solution to the quantisation conditions written in logarithmic form
\eqref{ecriture equations quantifications avec egalite entre entiers} that was discussed
in the first part of this proof may be expressed in terms of the vectors written above as
\begin{align}
\wh{\Psi}_{T; \wh{\bs{z}}} \, \big(\, \wh{\bs{\mf{z}}}  \, \big) \; & = \;
\Big( \bs{h}^{(L)}_{ \varkappa^{\, (L)}_0 } \, , \,   \bs{p}^{(L)}_{ \mf{y}^{\, (L)}_0  }
\, , \,  \bs{h}^{(R)}_{  \varkappa^{\, (R)}_0  } \, , \,   \bs{p}^{(R)}_{ \mf{y}^{\, (R)}_0  }  \Big)^{\op{t}}  \;,  \\[1ex]
\wh{\Phi}_{T} \big( \,  \wh{\bs{z}} \,  \big) \; & = \;
\Big( \bs{h}^{(L)}_{ \mc{N}_{L} } \, , \,   \bs{p}^{(L)}_{ \mc{M}_{L} } \, , \,  \bs{h}^{(R)}_{ \mc{N}_{R} }
\, , \,   \bs{p}^{(R)}_{ \mc{M}_{R} }  \Big)^{\op{t}}  \;.
\end{align}
Note that above we have made the dependence of $\wh{\Psi}_{T}$ on $\bs{z}$ explicit. Also, one has that $\wh{\bs{\mf{z}}} \in \mc{S}$,
with $\mc{S}$ as introduced just above of \eqref{ecriture biholomorphisme hat psi T avec domaines},
while   $\wh{\bs{z}} \in \wt{U}_{L} \times \wt{U}_{R} $, with $\wt{U}_{\a}$ as introduced just below of \eqref{ecriture module det hat phi non nul}.
Upon writing down the equations characterising $\wh{\bs{\mf{z}}}$ and  $\wh{\bs{z}} $
coordinate wise and observing that the set $\wh{\mathbb{Y}}$ canonically associated with the vectors $\bs{\mf{z}}_{\e{sol}}, \bs{z}_{\e{sol}}$
just coincides with $\wt{\mathbb{Y}}$, we get that  $\bs{\mf{z}}_{\e{sol}}, \bs{z}_{\e{sol}}$ also satisfy the above equations. Since
$\bs{z}_{\e{sol}} \in \wt{U}_{L} \times \wt{U}_{R}$ while $\bs{\mf{z}}_{\e{sol}}\in \mc{S}$, the solutions have to coincide, \textit{viz}.
\beq
\wh{\bs{\mf{z}}} \; = \; \bs{\mf{z}}_{\e{sol}} \qquad \e{and} \qquad \wh{\bs{z}} \; =\; \bs{z}_{\e{sol}} \;.
\enq
This entails the statement relative to the properties of coinciding roots.
Finally, one readily checks by using that $\veps_{\e{c}}^{\prime}$ does not vanish on the curve
$\Big\{ \la  \; : \;  \Re\big[\veps_{\e{c}}(\la)\big]=0 \; , \;  \big| \Im(\la) \big| \leq \tf{\pi}{2}\Big\} $ that
for the constructed solution it holds that
\beq
\wh{u}^{\,\prime}  \Big( \, \wh{x}_{a}^{\; (\a)}\,\big|\, \wh{\mathbb{Y}}\, \Big) \; \not= \; 0
\qquad \e{and} \qquad
\wh{u}^{\, \prime} \Big(\,  \wh{y}_{a}^{\; (\a)}\,\big|\, \wh{\mathbb{Y}}\, \Big) \;\not= \; 0 \;,
\enq
which contradicts the existence of higher-than-one multiplicities for the roots,
\textit{c.f.}\ \eqref{equation quantification particles et tous cas Trotter fini}.

This entails the claim. \qed

\subsection{Auxiliary lemmata used in the proof of Theorem
\ref{Theorem existence solutions racines particules et trous}}
\label{The auxiliary lemmas}

\subsubsection{Preliminary remarks}

Lemma~\ref{Lemme positivite dans plan cplx de real vepsc} establishes that 
\beq
\Re\big[ \veps_{\e{c}}(\la) \big]>0 \qquad  \e{for} \qquad \la \in     \Cx \setminus    \, \ov{\mc{D}}_{\veps;\i\pi}  \;,
\enq
where, as we remind the reader, $\mc{D}_{\veps}=
\Big\{ z \in \Cx \, : \,  |\Im(z)|\leq \tf{\zeta}{2} \;\;
\e{and} \; \; \Re[\veps_c(z)] < 0 \Big\}$ and
$\mc{D}_{\veps;\i\pi} = \mc{D}_{\veps} + \i\pi\mathbb{Z}$.
Observe that $\veps_{\e{c}}(\la) \tend  h$ as $\Re(\la) \tend \infty$ and that 
\beq
\e{d}\Big( \big\{ \msc{D}_{\mathbb{Y}} + \i \zeta \big\}
\setminus \op{D}_{ \i\frac{\zeta}{2}, \eps } + \i\pi \mathbb{Z}\, ,
 \ov{\mc{D}}_{\veps ; \i\pi} \Big) \geq C >0
\label{ecriture borne inf sur distance DY shifte par rapport D veps}
\enq
with $\eps>0$ and small enough, and for some constant $C$ uniform in $T$ and
$\tf{1}{ (NT^4) }$ small enough. From the above it follows that for $T_0>0$
there exists a $T$-independent constant $c > 0$ such that for all $T<T_0$
\beq
\Re\big[ \veps_{\e{c}}( y  ) \big] \, \geq \, c \, > \,  0 \quad \e{with} \quad y \in \big\{ \msc{D}_{\mathbb{Y}} + \i \zeta \big\} \setminus \op{D}_{ \i\frac{\zeta}{2}, \eps } \, . 
\label{ecriture borne inf sur reelle de veps c sur certains domaines}
\enq

Prior to going into the technicalities of the proofs, we would also like to observe that,
given $\eps^{\prime}>0$ and small enough, it holds that
\beq
\e{d}\Big( \,  \wh{\mf{X}}, \msc{C}_{\e{ref}} \setminus {\textstyle \bigcup_{\sg=\pm}}
\op{D}_{\sg q, \eps^{\prime} }   \Big) \geq c T 
\label{ecriture estimation distance de trous a la frontiere C ref}
\enq
for some $c>0$ small enough and $N, T$-independent, provided that $\tf{1}{ (N T^4) }$
is small enough. Indeed, by construction, one has that $\wh{\mf{X}} \subset
\e{Int}( \msc{C}_{\e{ref}} ) \bigcup_{\sg=\pm} \op{D}_{\sg q, \eps^{\prime} }$.
This may be justified upon closer inspection of the contour deformation
procedure $\msc{C}_{\e{ref}} \hookrightarrow \wh{\msc{C}}_{\mathbb{Y}}[\,\wh{u}\,]$
in the non-linear integral equation explored in the proof of
Proposition~\ref{Proposition domaine local holomorphie de F}. First of all, the
hole solution set $\wh{\mf{X}}_{\e{ref}}$, see \eqref{definition Y ref}, subordinate
to the contour $ \msc{C}_{\e{ref}}$, satisfies by construction
$\wh{\mf{X}}_{\e{ref}} \subset \ov{ \e{Int}( \msc{C}_{\e{ref}} ) }$. The deformation
procedure $ \msc{C}_{\e{ref}} \hookrightarrow \wh{\msc{C}}_{\mathbb{Y}}[\,\wh{u}\,]$
may pick some zeroes of $\la \mapsto 1\, + \, \ex{-\f{1}{T} \wh{u}(\la\,|\,\wh{\mathbb{Y}})}$
that are located in a uniformly $\e{O}(T)$-sized ball around $\pm q$. Thus $\wh{\mf{X}}
\subset \ov{\e{Int}( \msc{C}_{\e{ref}} ) } \bigcup_{\sg=\pm} \op{D}_{\sg q, \eps^{\prime}}$.
Further, by construction, \textit{c.f.}~\eqref{definition contour Cref},
$\msc{C}_{\e{ref}}\setminus \bigcup_{\sg=\pm} \op{D}_{\sg q, \eps^{\prime} }$
is contained in the domain
\beq
\bigcup\limits_{\a \in \{L,R\} }^{} \!  \veps_{\a}^{-1}\Big( \mc{G}_{\a}  \Big)  \,
\cup \op{D}_{-\i\f{\zeta}{2}, \mf{c}_{\e{d}} T} \;,
\enq
where $\mc{G}_{\a}$ is as introduced in
Proposition~\ref{Proposition borne sup sur exponentielle en tempe de la fct f lambda mathbbY}.
That Proposition also ensures that
\beq
\ex{\mp \f{1}{T} \wh{u}\, (\la\,|\,\wh{\mathbb{Y}}) }
\;  = \; \e{O}\Big( T^{M-|\wh{\mc{Y}}|} \Big)  \quad
\e{on} \quad \veps_{\a}^{-1}\Big( \mc{G}_{\a}^{(\pm)}  \Big) \;,
\enq
see \eqref{ecriture domain bornage exponentielle de proche de C ref} for the
definition of  $\mc{G}_{\a}^{(\pm)}$. This upper bound ensures that there cannot
exist hole roots in $\veps_{\a}^{-1}\Big(  \mc{G}_{\a}^{(+)} \cup \mc{G}_{\a}^{(-)}  \Big)$.
Moreover, by point g) of Hypothesis~\ref{Hypotheses eqns de quantification}, one
has that $\wh{\mf{X}} \cap \ov{\op{D}}_{-\i\tf{\zeta}{2}, \eps} = \emptyset $ for some $\eps >0$.
By using the explicit form of  $\mc{G}_{\a}^{(\pm)}$, the facts that the size of the domain
\beq
\veps\bigg(   \veps_{\a}^{-1}\Big(  \mc{G}_{\a}^{(+)} \cup \mc{G}_{\a}^{(-)}  \Big) \cup \op{D}_{-\i\f{\zeta}{2}, \eps} \bigg)
\label{ecrtuire image par veps de veps inverse ou on enleve voisinage T indpt de moins i zeta sur 2}
\enq
is bounded in $T$, the points  $0$ and $\i \tf{\pi}{2}$ are uniformly away from it, and
that $\big(\veps^{-1}_{\a}\big)^{\prime}$ is bounded there, one then obtains
the bound \eqref{ecriture estimation distance de trous a la frontiere C ref}.

An analogous statement holds with regard to the roots $\wh{\mc{Y}}$. In particular,
\beq
\wh{\mc{Y}} \subset \Big\{ z \in \Cx \; : \; |\Im(z)|\leq \tf{\pi}{2} \; \e{and} \; z \in \e{Ext}( \msc{C}_{\e{ref}} )   \Big\} \, {\textstyle \bigcup_{\sg=\pm}} \op{D}_{\sg q, \eps^{\prime} }
\enq
modulo $\i\pi$.

The proof of the structure of the possible sets $\wh{\mc{Y}}$ and $\wh{\mf{X}}$ 
in the low-$T$ limit relies on several technical lemmata. We first show that neither
maximal roots nor weakly maximal roots can be singular. Then we characterise the
possibilities to construct strings of roots attached to a maximal or weakly
maximal particle root. 

Also, from now on, all relations should be understood mod $\i\pi$.

\subsubsection{Maximal or weakly maximal roots cannot be singular}
\begin{lemme}
\label{Lemme incompatibilite racine singuliere et racine maximale}  
There exist $T_0, \eta$ small enough such that, for $0<T<T_0$ and $\eta> \tf{1}{(NT^4)}$,
if $y_0\in \wh{\mc{Y}}$ is a maximal or weakly maximal root, then $y_0$ is necessarily
a regular root, \textit{viz}.\ $y_0 \in \wh{\mc{Y}}_{\e{r}}$.
\end{lemme}

\Proof 
Assume that $y_0 \in \wh{\wt{\mc{Y}}}_{\e{sg}}$ is a shifted singular root.
Then, upon using \eqref{ecriture equation subsidiaire racine singulière}
in order to write down the quantisation condition
\eqref{ecriture conditions de quantifications Trotter fini} for $y_0$, one sees that
the latter takes the form
\bem
-1 \; = \; \ex{ -\f{1}{T} \wh{\mc{E}}^{\,(-)}_2(y_0 \,|\, \wh{\mathbb{Y}}\, ) } 
\pl{y \in \wh{\mc{Y}}_{\e{r}} }{} \bigg\{  \f{ \sinh( \i \zeta + y-y_0)  \sinh( 2 \i \zeta + y-y_0)  }{ \sinh( \i \zeta + y_0- y )    \sinh(  y_0- y ) }   \bigg\}\\
\times
\pl{y \in  \wh{\wt{\mc{Y}}}_{\e{sg}} }{} \bigg\{ - \f{ \sinh^2(\i\zeta + y - y_0)  \sinh( 2\i\zeta + y - y_0) }{ \sinh^2(\i\zeta + y_0 - y )   \sinh(2 \i\zeta + y_0- y )  }    \bigg\}
\cdot   \pl{x \in \wh{\mf{X}} }{} \bigg\{  \f{ \sinh(\i\zeta + y_0 - x )  \sinh(  y_0 -  x )  }{ \sinh(\i\zeta + x - y_0 ) \sinh(2 \i\zeta  + x - y_0) }   \bigg\}  \;.
\label{ecriture condition subsidiaires racine singuliere}
\end{multline}
We remind the reader that  $\wh{\mc{E}}^{\,(-)}_2(* \,|\,\wh{\mathbb{Y}}\,)$ was introduced
in \eqref{definition fcts composites des energies et phase habilees}.

Recall that according to point $\mathrm{g)}$ of
Hypothesis~\ref{Hypotheses eqns de quantification} $y_0 \not \in \op{D}_{\i\tf{\zeta}{2},\eps}$
mod $\i\pi$, meaning that $y_0 \in \big\{ \msc{D}_{\mathbb{Y}} + \i \zeta \big\}
\setminus \op{D}_{ \i\tf{\zeta}{2}, \eps }$ mod $\i\pi$. Hence, it follows by
\eqref{ecriture borne inf sur distance DY shifte par rapport D veps} that
$\e{d}\big( y_0, \ov{\mc{D}}_{\veps;\i\pi} \big) \, > \,  C $ uniformly in $T$
and $\tf{1}{(NT^4)}$ small enough. Then,
Lemma~\ref{Lemme positivite dans plan cplx de real vepsc} ensures
that there exists a $T$-independent constant $c_0>0$ such that
$\Re\big[ \veps_{\e{c}}( y_0 ) \big] \, \geq \, c_0 > 0$. 
Thus, since $y_0-\i\zeta \in \msc{D}_{\mathbb{Y}}\setminus \op{D}_{-\i\tf{\zeta}{2},\eps}$
mod $\i\pi$, one has that $\Re\big[ \veps_{\e{c}}( y_0 -\i \zeta) \big] =\e{O}(-T \ln T)$.
Hence, for $T$  and $\tf{1}{(NT^4)}$ small enough, one concludes that
$\Re\big[ \veps_{\e{c};2}^{(-)}( y_0) \big] \, \geq \, \tf{3 c_0}{4} > 0$. Taken that
uniformly on
\beq
\Bigl\{ \big\{ \msc{D}_{\mathbb{Y}} + \i \zeta \big\} \setminus \op{D}_{ \i\tf{\zeta}{2}, \eps }  \Bigr\} \cup
\Bigl\{  \msc{D}_{\mathbb{Y}}\setminus \op{D}_{-\i\tf{\zeta}{2},\eps} \Bigr\}
\enq
one has the estimate $\mc{W}_N-\veps_{\e{c}}=\e{O}\Big( \tfrac{1}{NT}\Big)$,
one infers the lower bound
\beq
\Re\big[ \wh{\mc{E}}^{\,(-)}_{2}( y_0 \,|\, \wh{\mathbb{Y}}\, ) \big] \, \geq \, \tf{ c_0}{2} > 0 \; .
\enq
This lower bound entails that \eqref{ecriture condition subsidiaires racine singuliere}
contains, when $T$, $\tf{1}{(NT^4)}$ are small, an exponentially small pre-factor
$\ex{ -\f{1}{T} \wh{\mc{E}}^{\, (-)}_2(y_0 \,|\, \wh{\mathbb{Y}}\, ) }$. Thus,
the only way for equation \eqref{ecriture condition subsidiaires racine singuliere}
to have solutions is that $y_0$ is located, with at least the same exponential
precision, in the vicinity of a pole of the expression. We now discuss the possible scenarios,
discarding them one-by-one. 
\begin{itemize}
\item
Assume  there exists $y\in \wh{\wt{\mc{Y}}}_{\e{sg}} $ such that, for some $c>0$,  
\beq
y_0= y- \i\zeta +\e{O}\Big(  \ex{- \f{c}{T} } \Big)   \qquad \e{or} \qquad   y_0 =  y- 2 \i\zeta  +\e{O}\Big(  \ex{- \f{c}{T} } \Big) \;.
\enq
Both situations  cannot take place because of the spacing properties of $\msc{D}_{\mathbb{Y}}$,
\eqref{ecriture propriete espacement fini pour domaines DY shiftes}. Indeed, since
$y_0, y \in  \wh{\wt{\mc{Y}}}_{\e{sg}}$ \textit{i.e.}\
$y_0, y \in \msc{D}_{\mathbb{Y};\i\pi}+\i\zeta$,
the above would imply that
\beq
\e{O}\Big(  \ex{- \f{c}{T} } \Big) =
\e{d}_{\i \pi} (y_0, y-\i\zeta) \geq
\e{d}\Bigl(\msc{D}_{\mathbb{Y};\i\pi} + \i \zeta \,,\, \msc{D}_{\mathbb{Y};\i\pi} \Bigr)
\enq
or
\beq
\e{O}\Big(  \ex{- \f{c}{T} } \Big) = \e{d}_{\i \pi}(y_0, y - 2\i\zeta) \geq
\e{d}\Bigl( \msc{D}_{\mathbb{Y};\i\pi} + \i \zeta \,,\,
\msc{D}_{\mathbb{Y};\i\pi} - \i \zeta \Bigr)
\enq
which contradicts \eqref{ecriture propriete espacement fini pour domaines DY shiftes}
for $T$ small enough.
 
 \item Assume   there exists $y\in \wh{\mc{Y}}_{\e{r}} $ such that, for some $c>0$,  
\beq
y_0= y   +\e{O}\Big(  \ex{- \f{c}{T} } \Big)  \qquad \e{or} \qquad   y_0 =  y- \i\zeta  +\e{O}\Big(  \ex{- \f{c}{T} } \Big) \;.
\enq
The first equation cannot hold as it would contradict the repulsion principle, point e) of Hypothesis \ref{Hypotheses eqns de quantification},
while the second case contradicts the maximality or weak maximality of $y_0$, namely that $\e{d}_{\i\pi}\big( y_0+\i\zeta, y \big) \, > \, \vsg T$ for some $\vsg>0$
and small enough in the maximal case, and likewise in the weakly maximal case for all but at most one $y\in \wh{\mc{Y}}_{\e{r}} $ for which one has
$\e{d}_{\i\pi}\big( y_0+\i\zeta, y \big) \, > \, \ex{- \f{c_T}{T}}$ for some $c_T=\e{o}(1)$.

 \item Assume there exists $x \in \wh{\mf{X}} $ such that, for some $c>0$,  
\beq
y_0 =  x + 2 \i\zeta  + \e{O}\Big(  \ex{- \f{c}{T} } \Big)  \qquad \e{or} \qquad    y_0= x + \i \zeta   + \e{O}\Big(  \ex{- \f{c}{T} } \Big)  \;.
\enq
 Since $x \in \msc{D}_{\mathbb{Y}}$ and $y_0 \in \msc{D}_{\mathbb{Y}}+\i\zeta$, the
 first case would imply that
$\e{d}\Big(  \msc{D}_{\mathbb{Y};\i\pi}  \,,\, \msc{D}_{\mathbb{Y};\i\pi}+\i\zeta \Big)
=  \e{O}\Big(  \ex{- \f{c}{T} } \Big) $, which contradicts the estimates
\eqref{ecriture propriete espacement fini pour domaines DY shiftes} for $T$ small enough.
More analysis is, however, needed so as to rule out the second possibility.
\end{itemize}

Assume that there exists $x_1 \in \wh{\mf{X}}$ such that
$y_0= x_1 + \i \zeta   - \vth_0$, with $ \vth_0 = \e{O}\big( \ex{- \f{c}{T} } \big)$
for some $c>0$. Starting from \eqref{ecriture form eqn subsidiaire pour x1}, one
readily gets the equation satisfied by $x_1$
\bem
-1 \; = \; \ex{ -\f{1}{T} \wh{\mc{E}}(x_1 \,|\, \wh{\mathbb{Y}}\, ) } \pl{y \in \wh{\mc{Y}}_{\e{r}} }{} \bigg\{  \f{ \sinh(\i\zeta+y-x_1) }{ \sinh(\i\zeta +x_1- y )  }   \bigg\}
\cdot  \pl{y \in  \wh{\wt{\mc{Y}}}_{\e{sg}}\setminus \{y_0 \} }{} \bigg\{ \f{ \sinh(\i\zeta+y-x_1)  \sinh( y-x_1) }{ \sinh(\i\zeta +  x_1 - y )   \sinh(2 \i\zeta + x_1 - y )  }    \bigg\}  \\
\times \f{ \sinh(\i\zeta + y_0 -x_1)  \sinh( y_0 - x_1) }{ \sinh(\i\zeta +  x_1 - y_0 )   \sinh(2 \i\zeta + x_1 - y_0 )  }
\cdot \pl{x \in \wh{\mf{X}} }{} \bigg\{  \f{ \sinh(\i\zeta + x_1 - x )  }{ \sinh(\i\zeta + x - x_1) }   \bigg\}  \;. 
\label{ecriture equation racine trou avec particule singuliere} 
\end{multline}
Due to the relation linking $y_0$ and $x_1$, the above equation contains an exponentially small term appearing in the denominator of the second line.
We first establish that this is the only exponentially small quantity present in the denominator of \eqref{ecriture equation racine trou avec particule singuliere}. 
Since  $\Dp{} \msc{D}_{\mathbb{Y}}$ is a curve of width at most $\e{O}(-T \ln T)$ around $\msc{C}_{\veps}$ with the exception of a neighbourhood of $-\i\tf{\zeta}{2}$ which is avoided at distance $\mf{c}_{\e{d}} T$, the product over the hole roots appearing above
is bounded from below and above by constants, this uniformly in $T, \tf{1}{(NT^4)}$.
Moreover, 
\begin{itemize}
 
 \item  assume that there exists $y\in \wh{\wt{\mc{Y}}}_{\e{sg}} \setminus \{ y_0 \} $ such that 
\beq
x_1= y - \i\zeta  + \e{O}\Big(  \ex{- \f{c}{T} } \Big)  \qquad \e{or} \qquad   x_1 =  y - 2 \i\zeta  + \e{O}\Big(  \ex{- \f{c}{T} } \Big) \;.
\enq
In the first case one infers that  $y=y_0 \, + \,  \e{O}\big(  \ex{- \f{c}{T} } \big)$,
hence contradicting the repulsion principle, item $\mathrm{e)}$ of
Hypothesis~\ref{Hypotheses eqns de quantification}. In the second case one infers
$y_0 = y - \i\zeta + \e{O}\big(  \ex{- \f{c}{T} } \big)$. Since $y, y_0
\in \msc{D}_{\mathbb{Y};\i\pi}+\i\zeta$, one sees that in this case scenario
\beq
\e{O}\Big(  \ex{- \f{c}{T} } \Big) \; = \; \e{d}_{\i \pi} \big(y_0,y-\i\zeta \big) \, \geq \, \e{d}\Big(\msc{D}_{\mathbb{Y};\i\pi}+\i\zeta, \msc{D}_{\mathbb{Y};\i\pi} \Big) \;,
\enq
hence, contradicting \eqref{ecriture propriete espacement fini pour domaines DY shiftes} for
$T$ low enough.
\item 
Assume that there exists $y\in \wh{\mc{Y}}_{\e{r}} $ such that 
\beq
x_1= y - \i\zeta \, + \, \e{O}\Big(  \ex{- \f{c}{T} } \Big)  \;.
\enq
Then,  one infers that $y=y_0 \, +\, \e{O}\Big(  \ex{- \f{c}{T} } \Big) $, which
contradicts the repulsion principle, item $\mathrm{e)}$ of
Hypothesis~\ref{Hypotheses eqns de quantification}.
\end{itemize}

The properties of $\msc{D}_{\mathbb{Y}}$ and the fact that
$x_1 \not \in \op{D}_{-\i\tf{\zeta}{2},\eps}$, \textit{c.f.}\ point $\mathrm{g)}$ of
in Hypothesis \ref{Hypotheses eqns de quantification},
ensure that one has the upper bound 
\beq
\big| \Re\big[ \, \wh{\mc{E}}\,(x_1\,|\,\wh{\mathbb{Y}}\, ) \big] \big| \,
   \leq  \, - C T \ln T \;, 
\label{ecriture bornes sur terme energetique dans eqn pour trou 1}
\enq
for some $T$-independent constant $C > 0$ that only depends on $|\wh{\mc{Y}}|$ and
$|\wh{\mf{X}}|$, provided  also that $\eta > \tf{1}{(NT^4)}$ with $\eta$ small enough.
Therefore, the pre-factor $\ex{ - \f{1}{T} \wh{\mc{E}}\,(x_1\,|\,\wh{\mathbb{Y}}\, ) }$
cannot compensate for the exponential blow up begot by the second line. Still, in
principle, the exponentially small factor in the denominator could be 
compensated by some exponentially small factor appearing in the products. 

\begin{itemize}
\item  Assume that there exists $y\in \wh{ \wt{\mc{Y}} }_{\e{sg}} \setminus \{y_0 \} $ such that 
\beq
x_1 \, = \,  y \, + \, \e{O}\Big(  \ex{- \f{c}{T} } \Big)  \qquad \e{or} \qquad   x_1 \, =\,   y + \i\zeta  \, + \, \e{O}\Big(  \ex{- \f{c}{T} } \Big) \;.
\enq
The first case implies that $y_0\, =\, y  \, + \, \i\zeta 
\, + \, \e{O}\big(  \ex{- \f{c}{T} } \big)$. Since $y_0$ and $y$ are two shifted singular roots,
they both belong to $\msc{D}_{\mathbb{Y};\i\pi} + \i \zeta$. Hence, the first case implies that
$\e{d}\Big( \msc{D}_{\mathbb{Y};\i\pi} + \i \zeta, \msc{D}_{\mathbb{Y};\i\pi} + 2\i \zeta \Big)
\, = \,  \e{O}\big(  \ex{- \f{c}{T} } \big)$, thus contradicting
\eqref{ecriture propriete espacement fini pour domaines DY shiftes}.
The second case leads to  $y_0  \, =\,  y \, + \, 2 \i \zeta  \, + \,
\e{O}\big(  \ex{- \f{c}{T} } \big)$, and thus, taken that $y, y_0$ are shifted singular
roots, implies that $\e{d}\Big( \msc{D}_{\mathbb{Y};\i\pi} + \i \zeta, \msc{D}_{\mathbb{Y};\i\pi}
+ 3\i \zeta \Big) \, = \,  \e{O}\big(  \ex{- \f{c}{T} } \big)$, hence contradicting
\eqref{ecriture propriete espacement fini pour domaines DY shiftes} for $T$ low enough.
\item
Assume that there exists $y\in \wh{\mc{Y}}_{\e{r}} $ such that 
\beq
x_1 \, = \, y \, + \, \i\zeta \, + \, \e{O}\Big(  \ex{- \f{c}{T} } \Big) \;.
\label{ecriture cas de figure ou un trou egale une raguliere shifte par i zeta}
\enq
%
%
Again, there are no direct contradictions to such a situation and it requires a deeper analysis.
\end{itemize}

Assume that there exists $y_1 \in \wh{\mc{Y}}_{\e{r}}  $ such that $x_1 = y_1 +
\i\zeta - \vth_1$ with $\vth_1 \, = \, \e{O}\big(  \ex{- \f{c}{T} }\big)$.
We recall that $y_0=x_1+\i\zeta-\vth_0$. Thus, $y_0 = y_1 + 2 \i\zeta - \vth_2$, where
$\vth_2\, =\, \vth_1 \, +\, \vth_0 \, = \, \e{O}\big(\ex{- \f{c}{T} }\big)$.
Note that, by virtue of the repulsion property, \textit{c.f.}\ point $\mathrm{e)}$
of Hypothesis~\ref{Hypotheses eqns de quantification}, there may exist at most one
such $y_1$.


By virtue of Lemma~\ref{Lemme impossibilite presence trou dans chaine de racines eg terminant sur une racine sing}, there exist $C, \wt{C} > 0$ and $d \in \mathbb{N}$,
only depending on $|\wh{\mc{Y}}|$, $|\wh{\mf{X}}|$, such that 
\beq
C^{-1} T^{d} \, \leq \,  \biggl| \f{  \sinh(\i\zeta +  y_1 - x_1 )  }{  \sinh(\i\zeta +  x_1 - y_0 ) }  \biggr|   \, \leq  \, C T^{-d} \quad viz.\quad
\wt{C}^{-1} T^{d} \, \leq \,  \biggl| \f{ \vth_1  }{  \vth_0 }  \biggr|
\, \leq  \, \wt{C} T^{-d}
\label{ecriture inegalite secteur trou sur les deviations}
\enq
and, for some $c_T=\e{o}(1)$,
\beq
C^{-1} T^{d}  \ex{-\f{1}{T} c_T } \, \leq \,  \bigg| \ex{ -\f{1}{T} [ \veps_{\e{c}}(y_0  ) +  \veps_{\e{c}}(y_0 -\i\zeta ) ]  }
\f{  \sinh(2 \i\zeta +  y_1 - y_0 )  }{  \sinh(\i\zeta +  x_1 - y_0 ) }  \bigg|   \, \leq  \, C\,  T^{-d}
\enq
\textit{viz.},
\beq
\wt{C}^{-1} T^{ \tilde{d} }  \ex{-\f{1}{T} c_T } \, \leq \, 
\biggr| \ex{ -\f{1}{T} \veps_{\e{c}}(y_0)  } \f{ \vth_2  }{  \vth_0 }  \biggl|
\, \leq  \, \wt{C} T^{-\tilde{d}} \,,
\label{ecriture inegalite secteur particules sur les deviations}
\enq
where $\tilde d > 0$ as well.
Above, we used that, since $y_0\not\in \op{D}_{\i \tf{\zeta}{2},\eps}$ for some
$\eps>0$, \textit{c.f.}\ point g) of Hypotheses \ref{Hypotheses eqns de quantification},
\textit{viz}.\ $y_0-\i\zeta  \in \msc{D}_{\mathbb{Y}}\setminus \op{D}_{-\i \tf{\zeta}{2},\eps}$,
one has that $\veps_{\e{c}}(y_0-\i\zeta)=\e{O}(-T\ln T)$.

The inequality in \eqref{ecriture inegalite secteur trou sur les deviations} and
the fact that $\vth_2=\vth_1+\vth_0$ imply that
\beq
%
\biggl| \f{ \vth_2  }{  \vth_0 } \biggr| \, \leq \, \check{C} T^{-d}
\label{ecriture borne sup et inf pour delta 2 sur delta 1}
\enq
for some constant $\check{C}>0$. However, from
\eqref{ecriture inegalite secteur particules sur les deviations} one infers that 
\beq
c \ex{ \f{ c^{\prime} }{T} } \le \;
\biggr| \f{ \vth_2  }{  \vth_0 }  \biggr|
\quad \e{with} \quad c, c^{\prime}>0 \;.
\label{ecriture borne contradictoire sur ration delta 2 a delta 0}
\enq
This is clearly in contradiction to \eqref{ecriture borne sup et inf pour delta 2 sur delta 1}
as $T\tend 0^+$, thus entailing the claim. \qed




\subsubsection{The pure chain of regular roots}

We now establish the first pattern allowing one for an inductive construction of a
string of regular roots. 

\begin{lemme}
\label{Lemme chaine de racines reguliere}  
There exists $T_0, \eta$ small enough such that, uniformly in $0<T<T_0$ and
$\eta > \tf{1}{(NT^4)}$ the following holds. 
 
Let $y_0, \dots , y_{k} \in \wh{\mc{Y}}_{\e{r}}$ be such that $y_0$ is a maximal
or weakly maximal root and, if $k \geq 1 $, that
\begin{enumerate}
\item
$y_p \, = \, y_{p+1} \, + \, \i\zeta \, + \, \de_{p}$ with
$\de_{p} = \e{O}\Big( \ex{-\f{1}{T} d_p }  \Big)$  for some $d_p>0$ and $p=0,\dots k-1$;
\item
$\Re\Big[ \sul{p=0}{s} \veps_{\e{c}}(y_p)  \Big] \, \leq \, - c_s <0$ for $s=0, \dots, k-1$.
\end{enumerate}
Then, either for some $C>0$ 
\beq
\Big| \Re\Big[ \sul{p=0}{k} \veps_{\e{c}}(y_p)  \Big] \Big|
\, \leq   \, C \eta_T  \qquad with \qquad \eta_T \tend 0^+ \;\; as \;\; T\tend 0^+\;, 
\label{ecriture condition arret de la chain inductive pour corde type racine reguliere pures}
\enq
or one may extract a subsequence in $T\tend 0^+$ such that, for that subsequence,
there exists a particle root $y_{k+1}\in \wh{\mc{Y}}_{\e{r}} \sqcup \wh{\wt{\mc{Y}}}_{\e{sg}}$
such that 
\beq
y_k \, = \, y_{k + 1} \, + \, \i\zeta \, + \, \de_{k} \quad with \quad  \de_{k} =\e{O}\Big( \ex{-\f{1}{T} d_k }  \Big) \qquad and \qquad 
\Re\Big[ \sul{p=0}{k} \veps_{\e{c}}(y_p)  \Big] \, \leq \, - c_k < 0
\label{ecriture etape k+1 de conctruction de chaine de racines}
\enq
for certain constants $c_k, d_k >0$. 
\end{lemme}

Note that owing to the condition $\eta> \tf{1}{(NT^4)}$, having $T\tend 0^+$ necessarily
implies that $N\tend +\infty$ fast enough.

\Proof 
If $ \Re\Big[ \sum_{p=0}^{k} \veps_{\e{c}}(y_p)  \Big] \tend 0$ as $T\tend 0^+$, then there
is nothing else to prove. 

Otherwise, one may write down an equation for $y_k$ having convenient properties by
taking the product of the subsidiary conditions for $y_0,\dots, y_k$. This yields
\bem
(-1)^{k+1} \; = \; \pl{s=0}{k} \bigg\{ \ex{ -\f{1}{T} \wh{\mc{E}}\, ( y_s \,|\, \wh{\mathbb{Y}}\, ) } \bigg\} \cdot 
\pl{s=0}{k} \Bigg\{  \pl{y \in \wh{\mc{Y}}_{\e{r};k} }{} \bigg\{  \f{ \sinh(\i\zeta + y- y_s) }{ \sinh(\i\zeta + y_s- y )  } \bigg\}
\cdot \pl{x \in \wh{\mf{X}} }{} \bigg\{  \f{ \sinh(\i\zeta + y_s - x )  }{ \sinh(\i\zeta + x - y_s) }   \Bigg\}
  \bigg\} \\
\times
\pl{s=0}{k}  \pl{y \in  \wh{\wt{\mc{Y}}}_{\e{sg}} }{} \bigg\{ \f{ \sinh(\i\zeta + y- y_s )  \sinh( y- y_s ) }{ \sinh(\i\zeta +  y_s - y )   \sinh(2 \i\zeta + y_s - y )  }    \bigg\}  \;. 
\label{ecriture produit sur k racines regulieres}
\end{multline}
Above, we have introduced 
\beq
\wh{\mc{Y}}_{\e{r};k} \, = \, \wh{\mc{Y}}_{\e{r}} \setminus \big\{ y_0, \cdots, y_k \big\} \;. 
\enq

Since we are in the situation when $ \Re\Big[ \sum_{p=0}^{k} \veps_{\e{c}}(y_p)  \Big]
\not\tend 0$ as $T\tend 0^+$, one may extract a subsequence such that
$\Re\Big[ \sum_{p=0}^{k} \veps_{\e{c}}(y_p)  \Big]  \tend \pm  2 c_k$ as $T\tend 0^+$,
this for some constant $c_k>0$.

For this subsequence, assume that $ \Re\Big[ \sum_{p=0}^{k} \veps_{\e{c}}(y_p)  \Big]
\, \geq \, c_k >0$, this provided that $T$ is small enough. Then, for the equation to
be satisfied, one needs to compensate the exponentially small term appearing in the \textit{rhs}
of \eqref{ecriture produit sur k racines regulieres}. This would mean that any
of the following cases would hold true.
\begin{itemize}
\item
There exists $y \in \wh{\mc{Y}}_{\e{r};k}$ and $s \in \intn{0}{k}$ such that
$y-\i\zeta  \,= \, y_s \, +\,  \e{O}\big( \ex{-\f{c}{T}} \big)$. 
However, for $s=0$, that would contradict the (weak) maximality of $y_0$,
\textit{viz}.\ that for some $\vsg>0$ one has
$\e{d}_{\i\pi}\big( y_0 + \i\zeta , y^{\prime} \big)>\vsg T$
for all but at most one $y^{\prime} \in \wh{\mc{Y}}$ and for that one
$\e{d}_{\i\pi}\big( y_0 + \i\zeta , y^{\prime} \big)> \ex{-\f{1}{T}c_T}$.
In its turn, for $s \in \intn{1}{k}$, this would lead to $y \, =\,  y_{s-1}
\, +\,  \e{O}\big( \ex{-\f{c}{T}} \big)$, hence contradicting the repulsion
property $\mathrm{e)}$ of Hypothesis~\ref{Hypotheses eqns de quantification}.
\item
There exists $x \in \wh{\mf{X}}$ and $s \in \intn{0}{k}$ such that $x+\i\zeta
\, =\,  y_s   \, +\,  \e{O}\big( \ex{-\f{c}{T}} \big)$. Recall that by point
$\mathrm{a)}$ of Hypothesis~\ref{Hypotheses solubilite NLIE}, shifted
singular roots belong to $\e{Int}(\msc{C}_{\e{ref}}) \setminus
\bigcup_{\sg=\pm}\op{D}_{\sg q, \mf{c}}+ \i \zeta + \i\pi \mathbb{Z}$ and so,
regular roots belong to the exterior of
\beq
\Big\{ \e{Int}(\msc{C}_{\e{ref}}) \setminus {\textstyle \bigcup_{\sg=\pm}}
\op{D}_{\sg q, \mf{c}} + \i \zeta + \i\pi \mathbb{Z}  \Big\} \cup
\Big\{ \e{Int}(\msc{C}_{\e{ref}})\setminus {\textstyle \bigcup_{\sg=\pm}}
\op{D}_{\sg q, \mf{c}} + \i\pi \mathbb{Z}  \Big\}  \;.
\enq
Taken that  $x \in   \e{Int}(\msc{C}_{\e{ref}}) \bigcup_{\sg=\pm}\op{D}_{\sg q, \mf{c}} $,
this leads to the chain of upper bounds
\beq
\label{estilem9}
 \e{O}\big( \ex{-\f{c}{T}} \big) \, = \, \e{d}_{\i\pi}\big( y_s-\i\zeta, x\big) \, \geq  \,
\e{min} \Big\{ \e{d}_{\i \pi} \big( y_s-\i\zeta,  \msc{C}_{\e{ref}} \setminus
{\textstyle \bigcup_{\sg=\pm}} \op{D}_{\sg q, \mf{c}}  \big) \, , \,
\e{d}_{\i \pi} \big( y_s-\i\zeta, {\textstyle \bigcup_{\sg=\pm}}
\op{D}_{\sg q, \mf{c}}  \big)  \Big\}  \; \geq  \; \mf{c}_{\e{ref}} T \, ,
\enq
where, in the last bound, we used points a) and b) of
Hypothesis~\ref{Hypotheses solubilite NLIE}, leading to a contradiction.
\item
There exists $y \in \wh{\wt{\mc{Y}}}_{\e{sg}}$ and $s \in \intn{0}{k}$ such that
$y-\i\zeta = y_s  \, +\,  \e{O}\big( \ex{-\f{c}{T}} \big)$
or $y-2\i\zeta = y_s  \, +\,  \e{O}\big( \ex{-\f{c}{T}} \big)$.
The first case scenario, for $s \in \intn{1}{k}$, would lead to $y = y_{s-1}
\, +\,  \e{O}\big( \ex{-\f{c}{T}} \big)$, hence contradicting the repulsion property,
\textit{c.f.}\ $\mathrm{e)}$ of Hypothesis~\ref{Hypotheses eqns de quantification}.
When $s=0$, one observes that by point $\mathrm{a)-b)}$ of
Hypothesis~\ref{Hypotheses solubilite NLIE} $\e{d}_{\i\pi}\big( y-\i\zeta , \pm q\big) > \mf{c}$
and $\e{d}_{\i\pi}\big( y-\i\zeta , \msc{C}_{\e{ref}} \big) > \mf{c}_{\e{ref}} T$.
Thus, since it also holds that $y-\i\zeta \in \e{Int}\big(  \msc{C}_{\e{ref}} \big)$
and $\e{d}_{\i\pi}\big( y-\i\zeta, y_0\big) \; = \;  \e{O}\big( \ex{-\f{c}{T}} \big)$,
we get that $y_0 \in \e{Int}\big( \msc{C}_{\e{ref}} \big)$. However, since
$\wh{\msc{C}}_{\mathbb{Y}}[\,\wh{u}\,]$ is displaced from $\msc{C}_{\e{ref}}$
on a $\e{O}(T)$ scale close to $\pm q$ and by an order $\e{O}(c T)$ with $c>0$ small
enough, $c \ll \mf{c}_{\e{ref}}$, away from some $T$-independent neighbourhood of $\pm q$,
it is direct to see that, since $\e{d}_{\i\pi}\big( y_0 , \pm q\big) > \tf{\mf{c}}{2}$,
$y_0\in \msc{D}_{\mathbb{Y};\i\pi}$, which cannot be by construction.


The second case scenario, for $s=0$ would contradict the (weak) maximality of $y_0$,
since the latter implies that $\e{d}_{\i\pi}\big(y_0+2\i\zeta, y^{\prime} \big)> \vsg T$
for all but at most one $y^{\prime} \in \wh{\wt{\mc{Y}}}_{\e{sg}}$, for which it
rather holds $\e{d}_{\i\pi}\big(y_0+2\i\zeta, y^{\prime} \big)>  \ex{- c_T \f{1}{T}}$
with $c_T = \e{o} (T)$.
For $s=1$, the relation between $y_1$ and $y_0$ would yield that $y = y_1 + 2\i\zeta 
\, +\,  \e{O}\big( \ex{-\f{c}{T}} \big)= y_0 + \i\zeta  \, +\,  \e{O}\big( \ex{-\f{c}{T}} \big)$.
Then one argues as for $s=0$ in the first case scenario. Finally, for
$s \in \intn{2}{k}$, it  would lead to $y = y_{s-2} \, +\,  \e{O}\big( \ex{-\f{c}{T}} \big)$,
hence contradicting the repulsion property, \textit{c.f.} point $\mathrm{e)}$ of
Hypothesis~\ref{Hypotheses eqns de quantification}.
\end{itemize}
Therefore, it cannot be that $ \Re\Big[ \sum_{p=0}^{k} \veps_{\e{c}}(y_p)  \Big]
\, \geq \, c_k > 0 $ for $T$ small enough.

It remains to consider the case of a negative real part.
Given the subsequence as above, we shall assume that
$\Re\Big[ \sum_{p=0}^{k} \veps_{\e{c}}(y_p)  \Big] \, \leq \, -c_k <0$,
this provided that $T$ is small enough. 
Then, for equation \eqref{ecriture produit sur k racines regulieres} to be satisfied,
the only possibility is to compensate for the exponentially large term appearing on
its \textit{rhs} by approaching exponentially fast to zero  with  some of the factors
in the numerator.
It turns out that one should treat the cases of a one-root-long chain ($k=0$) and
the one of longer chains ($k>0$) separately.

\vspace{2mm}

{ $\bullet$} $k=0$ case

\vspace{2mm}

For $k=0$ one way to compensate for the exponentially large driving term is to have
$x \in \wh{\mf{X}}$ such that $x-\i\zeta \,  =\,  y_0  \, +\,  \e{O}\big( \ex{-\f{c}{T}} \big)$.
Then one looks at the equation for $x$ which is directly inferred from
\eqref{ecriture form eqn subsidiaire pour x1}. It has the exponentially small factor
$\sinh(\i\zeta + y_0 - x)$ present in the numerator. The latter cannot be compensated
by $\ex{-\f{1}{T} \veps_c(x) }$ which decays to zero algebraically in $T$.
A cancellation of the exponentially small contribution can thus only be achieved
if $x$ approaches with exponential precision some pole of the expression.
However, if there exists $y\in \wh{\mc{Y}}_{\e{r}}$ such that $x \, = \,
y \, - \, \i\zeta  \, +\,  \e{O}\big( \ex{-\f{c}{T}} \big)$, then, since
$y-\i\zeta \not\in \op{D}_{\pm q, \mf{c}}$ by virtue of point a) of
Hypothesis~\ref{Hypotheses solubilite NLIE} and the fact that $1+\ex{- \f{1}{T}
\wh{u}(\la \,|\, \wh{\mathbb{Y}} \, ) }$ has no roots in
$\bigcup_{\a \in \{L, R\}} \bigl(\wh{\msc{D}}_{II; \a}\cup \wh{\msc{D}}_{III; \a}\bigr)$,
\textit{c.f.}\ \eqref{definition domain Df pour la deformation de contours}
of Proposition~\ref{Proposition domaine local holomorphie de F}, it follows that
$x \in \e{Int}\big( \msc{C}_{\e{ref}}\big) \setminus
\bigcup_{\sg=\pm}\op{D}_{\sg q, \tf{\mf{c}}{2}}$. Hence, taken that
$y-\i\zeta \in \Cx \setminus \Big\{ \big\{ \e{Int}(\msc{C}_{\e{ref}}) +\i\pi \mathbb{Z}\big\}
\bigcup_{\sg=\pm} \op{D}_{\sg q , \mf{c}} \Big\}$, one obtains
\beq
\e{O}\big( \ex{-\f{c}{T}} \big) \, = \, \e{d}_{\i\pi}\big( y-\i\zeta , x \big) \, \geq \,
\e{d}\Big(x, \msc{C}_{\e{ref}} \setminus {\textstyle \bigcup_{\sg=\pm}}
\op{D}_{\sg q ,   \tf{\mf{c}}{2} } \Big)   \;.
\label{ecriture discussion x moins i zeta egal y0 est impossible}
\enq
This is however in contradiction for $T$, $\tf{1}{(NT^4)}$ low enough with
\eqref{ecriture estimation distance de trous a la frontiere C ref}.

An exponentially large driving term can also be compensated if there exists
$y\in \wh{\wt{\mc{Y}}}_{\e{sg}}$ such that $ x \, = \,  y \, - \, \i\zeta
\, +\,  \e{O}\big( \ex{-\f{c}{T}} \big)$ or $ x \,  = \,  y  \, - \,  2\i\zeta
\, +\,  \e{O}\big( \ex{-\f{c}{T}} \big)$. The first case leads to 
$y_0 = y \, - \, 2\i\zeta \, +\,  \e{O}\big( \ex{-\f{c}{T}} \big)$, which contradicts the (weak) maximality of $y_0$, since
the latter implies that $\e{d}_{\i\pi}\big(y_0+2\i\zeta, y^{\prime} \big)> \vsg T$
for all but at most one $y^{\prime} \in \wh{\wt{\mc{Y}}}_{\e{sg}}$, for which
it rather holds that $\e{d}_{\i\pi}\big(y_0+2\i\zeta, y^{\prime} \big)> \ex{-\f{1}{T}c_T}$,
where $c_T = \e{o} (T)$. The second case leads to the estimate
\beq
\e{O}\big( \ex{-\f{c}{T}} \big) \, = \, \e{d}_{\i\pi}\big( y-2\i\zeta , x \big) \, \geq \,
\e{d}\Big(\msc{D}_{\mathbb{Y};\i\pi}  \, , \, \msc{D}_{\mathbb{Y};\i\pi} -\i\zeta \Big)   \;.
\enq
This contradicts
\eqref{ecriture propriete espacement fini pour domaines DY shiftes} for $T$ low enough.

All-in-all, one concludes that the equation for the hole root cannot be satisfied in
presence of the root configuration $x \, -\, \i\zeta \, =  \,  y_0   \, +\,
\e{O}\big( \ex{-\f{c}{T}} \big)$.

Another way to compensate for the exponentially large driving term is that
$y_0 \, = \, y   \, +\,  \e{O}\big( \ex{-\f{c}{T}} \big)$ for some
$y \in \wh{\wt{\mc{Y}}}_{\e{sg}}$. However, this cannot happen as it would
contradict the repulsion principle, point e) of
Hypothesis~\ref{Hypotheses eqns de quantification}.

All of this means that, in order to compensate for the exponentially large
driving term, there exists $y \in \wh{\wt{\mc{Y}}}_{\e{sg}} \sqcup \wh{\mc{Y}}_{\e{r};k}$
such that $y_0 \, = \, y \, + \, \i\zeta  \, +\,  \e{O}\big( \ex{-\f{c}{T}} \big)$,
which is exactly the statement given in
\eqref{ecriture etape k+1 de conctruction de chaine de racines}.

\vspace{2mm}

{ $\bullet$} $k>0$ case

\vspace{2mm}

Now, we continue with the general case $k>0$ and show that the statement of the
lemma holds for $k+1$. We have to consider the following possibilities.
\begin{itemize}
\item
There exists $x \in \wh{\mf{X}}$ and $s \in \intn{0}{k}$ such that
$x\, - \, \i\zeta \, = \,  y_s  \, +\,  \e{O}\big( \ex{-\f{c}{T}} \big)$. 
If $s=0$, the situation has already been discarded by the discussion proper to the $k=0$ case.
For $s \in \intn{1}{k}$ it would imply that $x \, = \,  y_{s-1}  \,+\,
\e{O}\big( \ex{-\f{c}{T}} \big)$. By point a) of Hypothesis~\ref{Hypotheses solubilite NLIE},
one has that that $\e{d}_{\i\pi}(y_{s-1}, \pm q) \geq \mf{c}$.
Thus, one has that $x \not\in \op{D}_{\pm q , \tf{\mf{c}}{2} }$.

Since $y_{s-1} \in \e{Ext}(\msc{C}_{\e{ref}}) \bigcup_{\sg=\pm}\op{D}_{\sg q, \mf{c}}$
and $x \in \e{Int}(\msc{C}_{\e{ref}})$  one gets the lower bound
\beq
\e{O}\big( \ex{-\f{c}{T}} \big) \, = \, \e{d}_{\i\pi}\big(x,y_{s-1}\big) \; \geq  \;
\e{d}_{\i\pi}\Big(x, \msc{C}_{\e{ref}} \setminus
{\textstyle \bigcup_{\sg=\pm}} \op{D}_{\sg q ,   \tf{\mf{c}}{2} } \Big)   \;.
\label{ecriture estimee distance racine tour et racine particule}
\enq
This is however in contradiction for $T$, $\tf{1}{(NT^4)}$ low enough
with  \eqref{ecriture estimation distance de trous a la frontiere C ref}.
 
\item
There exists $y \in \wh{\mc{Y}}_{\e{r};k}$ and $s \in \intn{0}{k}$ such that
$y \, + \, \i\zeta \, = \,  y_s  \, +\,  \e{O}\big( \ex{-\f{c}{T}} \big) $. 
If  $s\in \intn{0}{k-1}$, this would imply that $y \, = \, y_{s+1}  \, +\,
\e{O}\big( \ex{-\f{c}{T}} \big)$, hence contradicting the repulsion property of
the roots, point e) of Hypothesis~\ref{Hypotheses eqns de quantification}.
However, for $s=k$ such a configuration is possible, and this corresponds exactly
to the statement given in \eqref{ecriture etape k+1 de conctruction de chaine de racines}.
\item
There exists $y \in \wh{\wt{\mc{Y}}}_{\e{sg}}$ and $s \in \intn{0}{k}$ such that
$y \, +\, \i\zeta \, = \, y_s  \, +\,  \e{O}\big( \ex{-\f{c}{T}} \big)$
or $y \, = \,  y_s  \, +\,  \e{O}\big( \ex{-\f{c}{T}} \big)$. Clearly, the
second scenario cannot take place owing to the repulsion property of the roots,
point e) of Hypothesis~\ref{Hypotheses eqns de quantification}. As for the first
case scenario,  when $s \in \intn{0}{k-1}$, it would lead to $y \,  =  \,
y_{s+1}  \, +\,  \e{O}\big( \ex{-\f{c}{T}} \big)$, which again cannot be
owing to the repulsion property. However, for $s=k$, such a configuration is
possible, and this corresponds exactly to the statement given in
\eqref{ecriture etape k+1 de conctruction de chaine de racines}.
\end{itemize}

Thus, we have established the claim for $k+1$. \qed

\subsubsection{ A chain of regular roots ending on a singular root}
The previous result indicates that one may go down a chain of particle
roots either by picking a regular or a singular root. Below, we discuss the
continuation mechanism of a chain in the case when the lowermost root is singular. 

\begin{lemme}
\label{Lemme prolongation chaine racines qui se termine sur une racine singuliere}
There exists $T_0,$ $\eta$ small enough such that, uniformly in $0<T<T_0$ and
$\eta > \tf{1}{(NT^4)}$ the below holds. 
 
Let $k\geq 1$ and assume that there exist $y_0, \dots , y_{k-1} \in \wh{\mc{Y}}_{\e{r}}$
such that $y_0$ is a maximal or weakly maximal root, $y_k\in \wh{\wt{\mc{Y}}}_{\e{sg}}$, and
\begin{enumerate}
\item
$y_p \, = \, y_{p+1} \, + \, \i\zeta \, + \, \de_{p}$ with $\de_{p} =   \e{O}\big( \ex{-\f{1}{T} d_p} \big)$  for some $d_p>0$ and  $p=0,\dots k-1$;
\item
$\Re\Big[ \sul{p=0}{s} \veps_{\e{c}}(y_p)  \Big] \, \leq \, - c_s <0$  for  $s=0,\dots, k-1$.
\end{enumerate}
Let $y_{k+1}=y_k-\i\zeta$.

Then, either 
\beq
\Big| \Re\Big[ \sul{p=0}{k+1} \veps_{\e{c}}(y_p)  \Big] \big|  \, \leq   \, C \eta_T   \qquad with \qquad \eta_T \tend 0^+ \;\; as \;\; T\tend 0^+\;, 
\label{ecriture condition arret de la chain inductive pour corde type racine reguliere pures et une racine singuliere}
\enq
or one may extract a subsequence in $T\tend 0^+$ such that, for that subsequence of
temperatures going to $0$, there exists a regular root $y_{k+2}\in \wh{\mc{Y}}_{\e{r}}$
such that 
\beq
y_{k + 2} \, = \, y_{k + 1} \, - \, \i\zeta \, - \, \de_{k+1} \quad with \quad  \de_{k+1} =  \e{O}\big( \ex{-\f{1}{T} d_{k+1} } \big) \qquad and \qquad
\Re\Big[ \sul{p=0}{k+1} \veps_{\e{c}}(y_p)  \Big] \, \leq \, - c_{k+1} < 0 \;. 
\label{ecriture condition de passage au rang suivant pour chaine de racines terminant sur racine singuliere}
\enq

\end{lemme}

\Proof 
By taking the product over the subsidiary conditions defining the particle roots
$y_0,\dots, y_{k-1}$ and the singular root $y_{k}$, one gets upon using the 
convention that $y_{k+1}=y_k-\i\zeta$,
\bem
(-1)^k \; = \; \pl{s=0}{k+1} \bigg\{ \ex{ -\f{1}{T} \wh{\mc{E}}\, ( y_s \,|\, \wh{\mathbb{Y}}\, ) } \bigg\} \cdot
\pl{s=0}{k+1} \Bigg\{  \pl{y \in \wh{\mc{Y}}_{\e{r};k-1} }{} \bigg\{  \f{ \sinh(\i\zeta + y- y_s) }{ \sinh(\i\zeta + y_s- y )  } \bigg\} \cdot
\pl{x \in \wh{\mf{X}} }{} \bigg\{  \f{ \sinh(\i\zeta + y_s - x )  }{ \sinh(\i\zeta + x - y_s) }   \bigg\}
  \Bigg\} \\
\times \pl{y \in  \wh{\wt{\mc{Y}}}_{\e{sg};k} }{} \Bigg[ - \f{ \sinh^2(\i\zeta + y- y_k )  \sinh(2\i\zeta + y- y_k ) }{ \sinh^2(\i\zeta +  y_k - y )   \sinh(2 \i\zeta + y_k - y )  } 
\cdot \pl{s=0}{k-1}  \bigg\{ \f{ \sinh(\i\zeta + y- y_s )  \sinh( y- y_s ) }{ \sinh(\i\zeta +  y_s - y )   \sinh(2 \i\zeta + y_s - y )  }    \bigg\}  \Bigg]\;. 
\label{ecriture produit sur racines regulieres et une signuliere}
\end{multline}
Here we agree upon 
\beq
\wh{\mc{Y}}_{\e{r};k-1} \, = \, \wh{\mc{Y}}_{\e{r}}\setminus \{y_0,\dots, y_{k-1} \} \qquad \e{and} \qquad 
\wh{\wt{\mc{Y}}}_{\e{sg};k} \, = \, \wh{\wt{\mc{Y}}}_{\e{sg}}\setminus \{y_k \}  \;. 
\label{definition de har Y r k-1 et Y sh tilde k}
\enq

If $ \Re\Big[ \sum_{p=0}^{k+1} \veps_{\e{c}}(y_p)\Big] \tend 0$ as $T\tend 0^+$, then
there is nothing else to prove. 
Otherwise, one is in the situation when $\Re\Big[ \sum_{p=0}^{k+1} \veps_{\e{c}}(y_p) \Big]
\not\tend 0$ as $T\tend 0^+$. Hence, one may extract a subsequence such that 
$\Re\Big[ \sum_{p=0}^{k+1} \veps_{\e{c}}(y_p)  \Big]  \tend \pm 2 c_{k+1}$ as $T\tend 0^+$,
this for some constant $c_{k+1}>0$.
 
First assume that for this subsequence
\beq
 \Re\Big[ \sul{p=0}{k+1} \veps_{\e{c}}(y_p)  \Big] \, \geq \, c_{k+1} > 0
\label{ecriture propriete de parte reelle somme energies habillee borne inf positive}
\enq
provided that $T, \tf{1}{(NT^4)}$ are small enough. Then, for the equation to be
satisfied, one needs to compensate the exponentially small term appearing in the \textit{rhs}
of \eqref{ecriture produit sur racines regulieres et une signuliere}. This would mean
that either of the following holds.
\begin{itemize}
\item
There exists $y \in \wh{\mc{Y}}_{\e{r};k-1}$ and $s \in \intn{0}{k+1}$ such
that $y-\i\zeta \, = \, y_s \,  + \,   \e{O}\big( \ex{-\f{c}{T}} \big) $. 
For $s=0$, this would contradict the (weak) maximality of $y_0$, the latter ensuring that
for some $\vsg>0$ small enough $\e{d}_{\i\pi}\big( y_0+\i\zeta, y^{\prime} \big) \geq \vsg T$
for all but at most one $y^{\prime} \in \wh{\mc{Y}}_{\e{r}}$ for which it rather holds
$\e{d}_{\i\pi}\big( y_0+\i\zeta, y^{\prime} \big) \geq \ex{-\f{1}{T}c_T}$ with $c_T=\e{o}(1)$.
For $s \in \intn{1}{k+1}$ this would lead to $y = y_{s-1} \, + \,
\e{O}\big( \ex{-\f{c}{T}} \big) $, hence contradicting the repulsion property of the roots,
point e) of Hypothesis~\ref{Hypotheses eqns de quantification}.
\item
There exists $y \in \wh{\wt{\mc{Y}}}_{\e{sg};k}$ and $s \in \intn{0}{k}$ such that,
in the first case scenario, $y \, - \, \i\zeta \, = \, y_s  \,  + \,
\e{O}\big( \ex{-\f{c}{T}} \big) $ or, in the second case scenario,
$y \, - \, 2\i\zeta \, = \,  y_s  \,  + \,   \e{O}\big( \ex{-\f{c}{T}} \big)$. 
 
The first case scenario, for any $s$ one would have $y-\i\zeta=
y_k \, + \, \i(k-s) \zeta \,  + \,   \e{O}\big( \ex{-\f{c}{T}} \big) $ and, since
$y-\i\zeta, y_k-\i\zeta \in \msc{D}_{\mathbb{Y};\i\pi}$ by definition, we would get that
\beq
\e{O}\big( \ex{-\f{c}{T}} \big) \geq  \e{d}_{\i\pi}\big(y_k-\i\zeta +\i (k+1-s)\zeta ,   y-\i\zeta \big) \, \geq  \,
\e{d}_{\i\pi}\big(\msc{D}_{\mathbb{Y};\i\pi} +\i (k+1-s)\zeta  ,  \msc{D}_{\mathbb{Y};\i\pi} \big)  \, ,
\enq
hence contradicting \eqref{ecriture propriete espacement fini pour domaines DY shiftes},
which ensures that the latter distance is uniformly bounded from below.

The reasoning in the second case scenario is much similar and leads to the chain of upper bounds
\beq
 \e{O}\big( \ex{-\f{c}{T}} \big) \geq  \e{d}_{\i\pi}\big(y_k-\i\zeta +\i (k+2-s)\zeta ,   y-\i\zeta \big) \, \geq  \,
\e{d}_{\i\pi}\big(\msc{D}_{\mathbb{Y};\i\pi} +\i (k+2-s)\zeta  ,  \msc{D}_{\mathbb{Y};\i\pi} \big)  \, ,
\enq
hence is once more contradicting
\eqref{ecriture propriete espacement fini pour domaines DY shiftes}.
\item
There exists $x \in \wh{\mf{X}}$ and $s \in \intn{0}{k+1}$ such that
$x \, + \, \i\zeta \, =  \, y_s \,  + \,   \e{O}\big( \ex{-\f{c}{T}} \big)$. 
 
For $s\in \intn{0}{k-1}$ this would imply that $y_{s+1} \, = \, x  \,  + \,
\e{O}\big( \ex{-\f{c}{T}} \big)$ hence leading to the estimates
\eqref{ecriture estimee distance racine tour et racine particule}, which
contradict \eqref{ecriture estimation distance de trous a la frontiere C ref}.
For $s=k+1$ this would imply that $x \, + \,  2 \i\zeta \, =\,
y_k  \,  + \,   \e{O}\big( \ex{-\f{c}{T}} \big)$. Thus,
$\e{d}\big( y_k, 2\i\zeta +  \msc{D}_{\mathbb{Y};\i\pi} \big)=
\e{O}\big( \ex{-\f{c}{T}} \big)$. Since, by construction,
$y_k-\i\zeta \in \msc{D}_{\mathbb{Y}}$ one infers that
$\e{d}\Big(\msc{D}_{\mathbb{Y}} + \i \zeta\, , \, \msc{D}_{\mathbb{Y}}  +  2\i\zeta  \Big)
\leq \e{O}\big( \ex{-\f{c}{T}} \big)$, which contradicts
\eqref{ecriture propriete espacement fini pour domaines DY shiftes}.

Finally, one cannot immediately exclude the possibility of having
$x \, + \, \i\zeta \, = \,  y_k  \,  + \,  \e{O}\big( \ex{-\f{c}{T}} \big)$. For this,
one has to inspect the equation satisfied by the associated hole root $x$.
Then, the equation for $x$ contains the exponentially small term
$\sinh(\i\zeta + x - y_k)$ in its denominator. One can check by the above reasoning
that it can only be compensated by the contribution of regular roots present in
the numerator of that equation. Thus, there exists a root $y \in \wh{\mc{Y}}_{\e{r}}$
such that $\i\zeta + y - x =  \vth_{k+1} = \e{O}\big( \ex{-\f{c}{T}} \big)$, implying
that $2\i\zeta + y - y_k =  \vth_{k+2} = \e{O}\big( \ex{-\f{c}{T}} \big)$.
It follows that $\i\zeta + x - y_k = \vth_{k+1} + \vth_{k+2} = \vth_k$. Then, invoking
Lemma~\ref{Lemme impossibilite presence trou dans chaine de racines eg terminant sur une racine sing},
one sees that there exists $C>0$ and $n\in \mathbb{N}$ such that
\beq
\label{lem11useslem10}
C^{-1}T^{n} \, < \, \bigg| \f{\sinh(\i\zeta + y - x ) }{ \sinh(\i\zeta + x - y_k) }  \bigg|  \, < \, C T^{-n}
\enq
and
\beq
C^{-1}T^{n} \ex{- \f{1}{T} c_T} \, < \, \bigg| \ex{-\f{1}{T} \sum_{p=0}^{k+1}\veps_c(y_p)  }  \f{\sinh(2 \i\zeta + y - y_k ) }{ \sinh(\i\zeta + x - y_k) }  \bigg|  \, < \, C T^{-n}  \;.
\enq
Thus, for some other constant $\wt{C}>0$, 
\beq
\wt{C}^{-1}T^{n} \, < \, \bigg|  \f{ \vth_{k+1} }{ \vth_k }  \bigg|  \, < \, \wt{C} T^{-n}  \qquad \e{and} \qquad 
\wt{C}^{-1}T^{n} \ex{- \f{1}{T} c_T} \, < \, \bigg| \ex{-\f{1}{T}\sum_{p=0}^{k+1}\veps_c(y_p)  } \f{ \vth_{k+2} }{ \vth_k }  \bigg|  \, < \, \wt{C} T^{-n}  \;.
\label{ecriture bornes sup et inf sur delta k sur k+1 et k+2}
\enq
The first inequality along with $\vth_{k}+\vth_{k+1}=\vth_{k+2}$ implies that 
\beq
\bigg|  \f{ \vth_{k+2} }{ \vth_k }  \bigg|  \, < \, L T^{-n}
\enq
for some constant $L>0$. However, this is in contradiction, when
$T, \tf{1}{(NT^4)}\tend 0^+$, with the second inequality
in \eqref{ecriture bornes sup et inf sur delta k sur k+1 et k+2},
owing to \eqref{ecriture propriete de parte reelle somme energies habillee borne inf positive}.
\end{itemize}

Therefore, the only remaining possibility to investigate is when for this subsequence
\beq
\Re\Big[ \sul{p=0}{k+1} \veps_{\e{c}}(y_p)  \Big] \, \leq \, -c_{k+1} <0 \;,
\enq
provided that $T$ is small enough.
In this last case scenario, for the quantisation equation to be satisfied, the only
possibility is to compensate for the exponentially large term appearing in the \textit{rhs}
of \eqref{ecriture produit sur racines regulieres et une signuliere} by approaching
exponentially fast to zero  with  some of the factors present in the numerator. 
Thus, we are in one of the following situations
\begin{itemize}
\item
There exists $x \in \wh{\mf{X}}$ and $s \in \intn{0}{k+1}$ such that
$x \,  - \, \i\zeta \, =  \, y_s \, +  \, \e{O}\big( \ex{-\f{c}{T}} \big)$. 
When $s=0$, one infers, by repeating the reasoning outlined in the paragraph
around \eqref{ecriture discussion x moins i zeta egal y0 est impossible},
that such a situation cannot happen. For $s \in \intn{1}{k+1}$ it would imply
that $x \, = \,y_{s-1} \, +  \, \e{O}\big( \ex{-\f{c}{T}} \big)$ and one arrives
at a contradiction by repeating the reasoning in the paragraph around
\eqref{ecriture estimee distance racine tour et racine particule}.
\item
There exists $y \in \wh{\wt{\mc{Y}}}_{\e{sg};k}$ and $s \in \intn{0}{k-1}$ such
that $y \, + \, \i\zeta \, = \, y_s \, +  \, \e{O}\big( \ex{-\f{c}{T}} \big)$
or $y \, = \, y_s \, +  \, \e{O}\big( \ex{-\f{c}{T}} \big)$. 
In the former case one infers that $y \, = \, y_{s+1} \, +  \,
\e{O}\big( \ex{-\f{c}{T}} \big)$. Thus, in both cases, we have a contradiction
with the repulsion property of the roots, point e) of
Hypothesis~\ref{Hypotheses eqns de quantification}.
\item
There exists $y \in \wh{\wt{\mc{Y}}}_{\e{sg};k}$ such that $y \, + \, \i\zeta \, = \,
y_k \, +  \, \e{O}\big( \ex{-\f{c}{T}} \big) $ or $y \, + \,  2\i\zeta \, = \,
y_k \, +  \, \e{O}\big( \ex{-\f{c}{T}} \big)$. Such a relation between two singular
roots cannot exist owing to $\e{d}\big(\msc{D}_{\mathbb{Y};\i\pi} + \i\zeta  \, , \,
\msc{D}_{\mathbb{Y};\i\pi} \big) \geq c $ and $\e{d}\big(  \msc{D}_{\mathbb{Y};\i\pi}
+ 2 \i\zeta  \, , \,  \msc{D}_{\mathbb{Y};\i\pi}  \big) \geq c$,
see \eqref{ecriture propriete espacement fini pour domaines DY shiftes}.

item
There exists $y \in \wh{\mc{Y}}_{\e{r};k-1}$ and $s \in \intn{0}{k+1}$ such that
$y \, + \, \i\zeta \, = \,  y_s \, +  \, \e{O}\big( \ex{-\f{c}{T}} \big)$.  
If $s\in \intn{0}{k-1}$, that would imply that $y \, = \,  y_{s+1} \, +  \,
\e{O}\big( \ex{-\f{c}{T}} \big) $, hence contradicting the repulsion property of the roots,
point e) of Hypotheses \ref{Hypotheses eqns de quantification}.
For $s=k$, one would get that $y \, = \,  y_{k+1} \, +  \, \e{O}\big( \ex{-\f{c}{T}} \big)$.
By point b) of Hypothesis~\ref{Hypotheses solubilite NLIE}, one has that
$\e{d}\big( y_{k} - \i\zeta, \msc{C}_{\e{ref}}\big) \geq \mf{c}_{\e{ref}} T $
and, since
\beq
\e{d}\big( \msc{D}_{\mathbb{Y}}, \msc{C}_{\e{ref}} \big) \leq c T
\label{ecriture borne sup sur distance de msc D Y et C ref}
\enq
for some absolute constant $c$ only depending on $|\wh{\mf{X}}|, |\wh{\mc{Y}}|$,
for  $\mf{c}_{\e{ref}}$ large enough as assumed, one gets
that $y\in  \msc{D}_{\mathbb{Y}}$, a contradiction.

However, for $s=k+1$, one gets that $y  \, + \,  2 \i\zeta \, = \,
y_k \, +  \, \e{O}\big( \ex{-\f{c}{T}} \big)$, which in principle may happen.
This case scenario exactly corresponds to
\eqref{ecriture condition de passage au rang suivant pour chaine de racines terminant sur racine singuliere}.
\end{itemize}

This allows one to conclude. \qed

\begin{lemme}
\label{Lemme impossibilite presence trou dans chaine de racines eg terminant sur une racine sing}
Let $k\geq 0$ and, if $k > 0$,  $y_0, \dots,  y_{k-1} \in \wh{\wt{\mc{Y}}}_{\e{r}}$ with
$y_0$ being a maximal or weakly maximal root. Let $y_k \in \wh{\wt{\mc{Y}}}_{\e{sg}}$,
$y_{k+1}=y_k-\i\zeta$, and let, for $p=0,\dots, k-1$, the roots be arranged as
\beq
y_p \, = \, y_{p+1} \,  + \,  \i\zeta  \, + \, \de_{p} \quad
\text{with} \quad \de_p \, = \, \e{O}\Big( \ex{-\f{1}{T}d_p} \Big) \quad and \quad d_p>0 \;.
\enq
%
%
%
Assume that there exists $x_1\in \wh{\mf{X}}$ and $y_{k+2}\in \wh{\mc{Y}}_{\e{r}}$ such that
\beq
\i\zeta + x_1-y_{k}\; = \; \vth_k \; , \quad  \i\zeta + y_{k+2}-x_1\; = \; \vth_{k+1} \; ,  \quad and \quad
2\i\zeta+y_{k+2}-y_k \; = \; \vth_{k+2} \;,
\enq
where $\vth_k, \vth_{k+1}, \vth_{k+2}=\e{O}\big( \ex{-\f{c}{T} } \big)$.

Then there exist
constants $C_x, C_y>0$ and integers $d_x, d_y$, all uniform in $T, \tf{1}{(NT^4)} \tend 0^+$
and only depending on $|\wh{\mc{Y}}|$ and $|\wh{\mf{X}}|$, such that one has the bounds
\beq
\begin{split}
C^{-1}_x T^{d_x} \, & \leq \,  \bigg| \f{  \sinh(\i\zeta +  y_{k+2} - x_1 )  }{  \sinh(\i\zeta +  x_1 - y_k ) }  \bigg|   \, \leq  \, C_x T^{-d_x} \;,
%
\\[1ex]
C_y^{-1} T^{d_y} \ex{- \f{1}{T} c_T} \, & \leq \,
\bigg| \ex{ -\f{1}{T} \sum_{s=0}^{k+1} \veps_{\e{c}}(y_s ) }
\f{  \sinh(2 \i\zeta +  y_{k+2} - y_k )  }{  \sinh(\i\zeta +  x_1 - y_k ) }  \bigg|   \, \leq  \, C_y\,  T^{-d_y} \;,
\end{split}
\label{ecriture bornes sur yk+2 x1 et yk pour exclision de config avec une paire racine sing et particule}
\enq
where $c_T=\e{o}(1)$ as $T\tend 0^+$.
\end{lemme}

\Proof
The estimates can be obtained by looking separately at the quantisation conditions
\eqref{ecriture conditions de quantifications Trotter fini} for the roots $y_k$ and
$x_1$. The quantisation conditions for $y_k$, \textit{c.f.}\
\eqref{ecriture produit sur racines regulieres et une signuliere}, may be recast in the form
\beq
(-1)^k \; = \; \pl{s=0}{k+1}\ex{ -\f{1}{T} \wh{\mc{E}}(y_s \,|\, \wh{\mathbb{Y}}\, ) } \cdot  \f{  \sinh( \i\zeta +  y_{k+2} - y_{k+1} )  }{  \sinh(\i\zeta +  x_1 - y_k ) } \cdot
\pl{\ell=1}{4} \mc{P}_{\ell} \;,
\label{ecriture equation concatenee pour la racine maximale y0}
\enq
where, making use of the notation \eqref{definition de har Y r k-1 et Y sh tilde k}
and upon using the relation between the roots $y_k, y_{k+2}$ and $x_1$, and further
introducing for convenience $\de_k=0$ and $\de_{k+1}=-\vth_{k+2}$, one has
\bem
\mc{P}_1\; =  \pl{s=0}{k} \pl{y \in \wh{\mc{Y}}_{\e{r};k-1} }{} \! \!
\bigg\{  \f{ \sinh( \i \zeta + y-y_s)     }{ \sinh( \i \zeta + y_s- y )    }   \bigg\} \cdot
\pl{ \substack{ y \in \wh{\mc{Y}}_{\e{r};k-1}  \\ \setminus \{ y_{k+2} \} } }{} \! \!
\bigg\{  \f{ \sinh( \i \zeta + y-y_{k+1} )     }{ \sinh( \i \zeta + y_{k+1}- y )    }   \bigg\}  \\
\; =  \pl{y \in \wh{\mc{Y}}_{\e{r};k-1} }{}
\!\! \bigg\{ \f{ \sinh(  y-x_1+\vth_k)}{\sinh(  \i\zeta + y_0 -  y)}
\prod_{s=1}^{k}  \f{\sinh(y-y_{s}-\de_{s-1})}{\sinh(  y_{s-1}- y-\de_{s-1} )}
 \bigg\} \cdot
\pl{ \substack{ y \in \wh{\mc{Y}}_{\e{r};k-1}  \\ \setminus \{ y_{k+2} \} } }{} \! \!
\bigg\{  \f{ \sinh(  y-y_{k+2}-\de_{k+1} )     }{  \sinh( y_{k}- y- \de_k )    }   \bigg\}
\end{multline}
and
\beq
\mc{P}_2 \; = \; \f{ \sinh( \i \zeta + y_k-x_1)   }{  \sinh( \i \zeta + y_{k+1}-y_{k+2})  }
 \; =  \; \f{   \sinh( 2\i \zeta - \vth_k )  }
 {   \sinh( 2 \i \zeta - \vth_{k+2 }) } \;.
\enq
Further, upon using the structure of the $\i\zeta$- shifts between the roots $y_p$, one obtains
\bem
\mc{P}_3 \; = \; \pl{x \in \wh{\mf{X}} \setminus \{x_1\} }{} \bigg\{  \f{ \sinh(\i\zeta + y_k - x )   }{ \sinh(\i\zeta + x - y_k ) }   \bigg\}
\cdot \pl{ \substack{ s=0 \\ s \ne k } }{ k+1 }  \pl{x \in \wh{\mf{X}}  }{} \bigg\{  \f{ \sinh(\i\zeta + y_s - x )   }{ \sinh(\i\zeta + x - y_s ) }   \bigg\} \\
\; = \;  \pl{x \in \wh{\mf{X}} \setminus \{x_1\} }{} \bigg\{  \f{ \sinh(2 \i\zeta + x_1 - x + \vth_k )  }
{ \sinh(x - x_1-\vth_k )  }   \bigg\}  \cdot
\pl{ \substack{ s=0 \\ s \ne k } }{ k+1 }  \pl{x \in \wh{\mf{X}}  }{} \bigg\{  \f{ \sinh(\i\zeta(2+k-s) + x_1 - x + \ga_s )   }{ \sinh(\i\zeta(s-k) + x - x_1-\ga_s ) }   \bigg\}
\end{multline}
for some $\ga_s = \e{O}\big( \ex{- \f{c}{T} } \big)$. Finally,  for some $\ga_s^{\prime} = \e{O}\big( \ex{- \f{c}{T} } \big)$, one has
\bem
\mc{P}_4 \; = \; \pl{y \in  \wh{\wt{\mc{Y}}}_{\e{sg};k} }{} \Bigg[
\bigg\{ - \f{ \sinh^2(\i\zeta + y - y_k)  \sinh( 2\i\zeta + y - y_k ) }{ \sinh^2(\i\zeta + y_k - y )   \sinh(2 \i\zeta + y_k- y )  }    \bigg\}
\pl{s=0}{k-1}  \bigg\{  \f{  \sinh(\i\zeta + y -y_s) \sinh(y-y_s)  }{  \sinh(\i\zeta + y_s -y) \sinh(2\i\zeta + y_s-y)    } \bigg\} \Bigg] \\ \mspace{136.mu}
\; = \;  \pl{y \in  \wh{\wt{\mc{Y}}}_{\e{sg};k} }{} \Bigg[
\bigg\{ - \f{ \sinh^2(\i\zeta + y - y_k)  \sinh( 2\i\zeta + y - y_k ) }{ \sinh^2(\i\zeta + y_k - y )   \sinh(2 \i\zeta + y_k- y )  }    \bigg\}  \\
\times \pl{s=0}{k-1}  \bigg\{  \f{  \sinh(\i\zeta(1+s-k) + y -y_k+\ga_s^{\prime}) \sinh(\i\zeta(s-k) + y -y_k+\ga_s^{\prime})  }
{  \sinh(\i\zeta(1-s+k) + y_k -y-\ga_s^{\prime}) \sinh(\i\zeta(2+k-s) + y_k-y - \ga_s^{\prime})    } \bigg\} \Bigg] \;.
\end{multline}

We now  upper and lower bound the $\mc{P}_{\ell}$ one after the other. First of all,
since $\Re(x_1)=\Re(y_s)+\e{O}\big( \ex{-\f{c}{T¨}}\big)$ for $s = 0, \dots, k + 2$ and
since $x_1 \in \e{Int}\big( \msc{C}_{\e{ref}}\big) \bigcup_{\sg=\pm}\op{D}_{\sg q , c T }$
for some $c > 0$ that only depends on $|\wh{\mf{X}}|, |\wh{\mc{Y}}|$, one concludes
that $\Re(y_{s})$ is bounded in $T$ for $s=0,\dots,k+2$. Therefore, for any
$y\in \wh{\mc{Y}}_{\e{r};k-1}$ such that  $| \Re(y) | \geq K$, with $K>0$ and large enough,
one infers that
\beq
C \; \leq \; \bigg|  \f{ \sinh(\i\zeta + y-y_{s})   }{ \sinh(  \i \zeta + y_s- y )   }  \bigg| \; \leq \; C^{-1}
\enq
for some constant $C>0$ and $s=0,\dots, k$. Thus, it remains to focus on the case of
regular roots such that $| \Re(y)|  \leq K$.

In that case, the repulsion property, item e) of
Hypothesis \ref{Hypotheses eqns de quantification}, and obvious upper bounds imply that 
\beq
 C^{\prime} \geq | \sinh(  y_{s-1}- y -\de_{s-1}) | \geq c^{\prime} T \quad \e{and} \quad   C^{\prime} \geq | \sinh(  y-y_{s}-\de_{s-1}) | \geq c^{\prime} T \;.
\enq
Here $C^{\prime}, c^{\prime} >0$ and the bounds hold for
for $s=1,\dots, k$ and $y\in \wh{\mc{Y}}_{\e{r};k-1}$. We stress that for $s=k$
one has to invoke that $y_k \not\in \wh{\mc{Y}}_{\e{r};k-1}$. Quite similarly, one obtains 
\beq
 C^{\prime} \geq | \sinh(  y_{k}- y -\de_{k}) | \geq c^{\prime} T \quad \e{and} \quad   C^{\prime} \geq | \sinh(  y-y_{k+2}-\de_{k+1}) | \geq c^{\prime} T
\enq
for all $y\in \wh{\mc{Y}}_{\e{r};k-1}\setminus \{y_{k+2}\}$.

The (weak) maximality property of $y_0$ and the boundedness of $\Re(y), \Re(y_0)$
ensure that there exist $C^{\prime}, c^{\prime}>0$ such that
\beq
C^{\prime} \geq |\sinh(  \i \zeta + y_0- y )| \geq c^{\prime} T  \qquad \e{or} \qquad
C^{\prime} \geq |\sinh(  \i \zeta + y_0- y )| \geq \ex{-\f{1}{T} c_T}
\enq
for all $y\in \wh{\mc{Y}}_{\e{r};k-1}$. Here the second bounds holds for at most
a single $y\in \wh{\mc{Y}}_{\e{r};k-1}$. Finally, observe that the relation
$x_1=y_k-\i\zeta + \vth_k$ and the relation  $\e{d}_{\i\pi}(y_k-\i\zeta, \pm q)
\geq \mf{c}$ issuing from point a) of
Hypothesis~\ref{Hypotheses solubilite NLIE} both ensure that $\e{d}(x_1, \pm q)
\geq \tf{ \mf{c} }{ 2}$  provided that $T$ is small enough. Then, since
\beq
x_1 \in \Big\{ \e{Int}\big( \msc{C}_{\e{ref}} \big)\setminus
{\textstyle \bigcup_{\sg=\pm}} \op{D}_{\sg q , \tf{\mf{c}}{2}} + \i \pi \mathbb{Z}\Big\} 
\qquad \e{and} \qquad
y \in \Cx \setminus \bigg\{ \Big\{ \e{Int}\big( \msc{C}_{\e{ref}} \big) + \i \pi \mathbb{Z} \Big\}\cup \msc{D}_{\mathbb{Y}} \bigg\}
\quad \e{if} \quad y\in \wh{\mc{Y}}_{\e{r};k-1} \; ,
\enq
one has that
\beq
\e{d}_{\i\pi}(y,x_1) \, \geq \,  \e{d}_{\i\pi} \Big(\msc{C}_{\e{ref}} \setminus
{\textstyle \bigcup_{\sg=\pm}} \op{D}_{\sg q, \f{\mf{c}}{2}} \, , \wh{\mf{X}} \Big) \geq cT
\quad \e{for} \; \e{any} \quad y\in \wh{\mc{Y}}_{\e{r};k-1} \;.
\enq
Here we invoked \eqref{ecriture estimation distance de trous a la frontiere C ref}.
Thus, taken that $\vth_k$ is exponentially small and assuming that $|\Re(y)|\leq K$,
one arrives to the estimates
\beq
C^{\prime} \geq | \sinh(y-x_1+\vth_k)|  \geq c^{\prime} T
\enq
with $C', c' > 0$. As a consequence, all-in-all, there exist $C_1 > 0$ and
$d_1 \in {\mathbb N}$ such that
\beq
C_1 T^{d_1}  \; \leq  \; \big| \mc{P}_1 \big| \; \leq \; C_1^{-1} T^{-d_1} \ex{\f{1}{T} c_T} \;.
\enq
%
%
%

The estimation of $\mc{P}_2$ is immediate, and one obtains
$C_2\leq |\mc{P}_2| \leq C_2^{-1}$ for some $C_2 > 0$.

In order to get the bounds for $\mc{P}_3$, one first observes that owing to
the boundedness of $\Re(x_1)$, $\Re(y_s)$ is bounded from above in $T$ for
any $s$ and thus all arguments appearing in the product formula for $\mc{P}_3$
are so. Then one recalls the estimates
\eqref{ecriture propriete espacement fini pour domaines DY shiftes}
which readily allow one to infer that, given $\ga = \e{O}\big(\ex{-\f{c}{T}} \big)$
and for any $p\not=0$ and $x \in \wh{\mf{X}}$, it holds that
\beq
%
\e{d}_{\i\pi}\big(x \pm \i p \zeta , x_1  \big) \, -  \, \e{O}\big(\ex{-\f{c}{T}} \big)
\, \geq \, \e{d}\big( \msc{D}_{\mathbb{Y};\i\pi}, \msc{D}_{\mathbb{Y};\i\pi} + \i p \zeta \big) \, -  \, \e{O}\big(\ex{-\f{c}{T}} \big) \geq c \;,
\enq
for some $c>0$ which implies, in turns, that $|\sinh\big( \i p \zeta \pm (x-x_1) + \ga \big) |
\, \geq \, c'$ for a certain $c' > 0$. Finally, when $p=0$, one invokes the repulsion property,
point e) of Hypothesis~\ref{Hypotheses eqns de quantification}, to see that
$|\sinh(x - x_1-\ga_k )| \, \geq \, c T$, for some $c>0$. Thus, since upper
bounds hold trivially for each factor, one infers that there exists $d_3 \in {\mathbb N}$
such that
\beq
C_3\leq |\mc{P}_3| \leq C_3^{-1} T^{-d_3}
\enq
for some $C_3 > 0$.
%
%

Finally, in order to bound $\mc{P}_{4}$, one uses that, given
$\ga = \e{O}\big(\ex{-\f{c}{T}} \big)$, for any $p\not=0$ and
$y \in  \wh{\wt{\mc{Y}}}_{\e{sg};k}$, one has for $T$ low enough
\beq
%
%
\e{d}\big( \msc{D}_{\mathbb{Y};\i\pi} + (1+p)\i\zeta , \msc{D}_{\mathbb{Y};\i\pi} + \i \zeta \big)   \, -  \, \e{O}\big(\ex{-\f{c}{T}} \big) \geq c
\enq
and therefore $|\sinh\big( \i p \zeta + y-y_k + \ga \big) | \, \geq \, c'$ for
some $c' > 0$.
Furthermore, by the repulsion property, point e) of
Hypothesis~\ref{Hypotheses eqns de quantification}, one gets that, given
$\ga = \e{O}\big(\ex{-\f{c}{T}} \big)$, for any
$y \in \wh{\wt{\mc{Y}}}_{\e{sg};k}$ one also has that $|\sinh\big(  y-y_k + \ga \big)| \geq c T$.
The upper bounds being trivial, all-in-all this ensures that
\beq
C_4 T^{d_4} \leq |\mc{P}_4| \leq C_4^{-1}
\enq
for some $C_4 > 0$ and some $d_4 \in {\mathbb N}$.
%
%

At this stage, it remains to recast the quantisation conditions for $y_k$ in the form
\beq
\f{ \sinh(\i\zeta + y_{k+2} - y_{k+1})   }{ \sinh(\i\zeta +  x_1 - y_k )}
\pl{s=0}{k+1} \ex{-\f{1}{T} \veps_{\e{c}}(y_s)  } \; = \;
(-1)^k \pl{s=0}{k+1}
\ex{\wh{\Phi}(y_s \,|\, \wh{\mathbb{Y}}\,)} \cdot \pl{\ell=1}{4}\mc{P}_{\ell}^{-1} \;.
\enq
The boundedness of $\wh{\Phi}(\la\,|\,\wh{\mathbb{Y}}\,)$ along with the previous
estimates then leads to the second bound in
\eqref{ecriture bornes sur yk+2 x1 et yk pour exclision de config avec une paire racine sing et particule}.


The derivation of the first bound in
\eqref{ecriture bornes sur yk+2 x1 et yk pour exclision de config avec une paire racine sing et particule}
is obtained by carrying out precise bounds in the quantisation condition for
$x_1 \in \wh{\mf{X}}$. The latter may be recast in the form
\beq
-1 \; = \;  \ex{ -\f{1}{T} \wh{\mc{E}}(x_1 \,|\, \wh{\mathbb{Y}}\, ) } \cdot \f{ \sinh(\i\zeta + y_{k+2} - x_1)   }{ \sinh(\i\zeta +  x_1 - y_k )  } \cdot
\pl{\ell=1}{4}\wt{\mc{P}}_{\ell} \;,
\enq
where we have set
\beq
\wt{\mc{P}}_{1} \; = \; \pl{y \in \wh{\mc{Y}}_{\e{r}}\setminus \{y_{k+2} \} }{} \bigg\{  \f{ \sinh(\i\zeta+y-x_1) }{ \sinh(\i\zeta +x_1- y )  }   \bigg\}
\; = \;  \pl{y \in \wh{\mc{Y}}_{\e{r}}\setminus \{y_{k+2}\} }{} \bigg\{  \f{ \sinh(y-y_{k+2} + \vth_{k+1}) }{ \sinh(y_k- y +\vth_k )  }   \bigg\}
\enq
and
\beq
\wt{\mc{P}}_2 \; = \; \f{ \sinh(\i\zeta + y_k-x_1)  \sinh( y_k-x_1) }{ \sinh(\i\zeta +  x_1 - y_{k+2} )   \sinh(2 \i\zeta + x_1 - y_k )  }
\; = \; \f{ \sinh( 2 \i\zeta - \vth_k)  \sinh( \i\zeta - \vth_k) }{ \sinh(2\i\zeta - \vth_{k+1} )   \sinh( \i\zeta + \vth_k )  } \;.
\enq
Further, making again use of convention \eqref{definition de har Y r k-1 et Y sh tilde k}, one has

\beq
\wt{\mc{P}}_{3} \; =  \hspace{-2mm}  \pl{y \in  \wh{\wt{\mc{Y}}}_{\e{sg};k} }{}\hspace{-2mm}
\bigg\{ \f{ \sinh(\i\zeta+y-x_1)  \sinh( y-x_1) }{ \sinh(\i\zeta +  x_1 - y )   \sinh(2 \i\zeta + x_1 - y )  }    \bigg\}
 \; =  \hspace{-2mm}  \pl{y \in  \wh{\wt{\mc{Y}}}_{\e{sg};k} }{} \hspace{-2mm}
\bigg\{ \f{ \sinh(y-y_{k+2}+\vth_{k+1} )  \sinh( y-x_1) }{ \sinh(y_k - y + \vth_k)   \sinh(2 \i\zeta + x_1 - y )  }    \bigg\}   \;.
\enq
Finally,
\beq
\wt{\mc{P}}_4 \; = \;   \pl{x \in \wh{\mf{X}} \setminus \{x_1\} }{} \bigg\{  \f{ \sinh(\i\zeta + x_1 - x )  }{ \sinh(\i\zeta + x - x_1) }   \bigg\} \;.
\enq
%
%
%

$\wt{\mc{P}}_1$ may be bounded as before by splitting the bounds according to
$|\Re(y)|\geq K$ for some $K>0$ large enough and $|\Re(y)|\leq K$ and invoking the
repulsion property of the particle roots, point e) of
Hypothesis~\ref{Hypotheses eqns de quantification}. One gets
\beq
\wt{C}_1 T^{\wt{d}_1}  \; \leq  \; \big| \wt{\mc{P}}_1 \big| \; \leq \; \wt{C}_1^{-1} T^{-\wt{d}_1}
\enq
for $\wt{d}_1 \in {\mathbb N}$ and $\wt{C}_1 > 0$.
%
%
Furthermore, it is direct to estimate that, for some $\wt{C}_2 > 0$,
\beq
\wt{C}_2   \; \leq  \; \big| \wt{\mc{P}}_2 \big| \; \leq \; \wt{C}_2^{-1}  \; .
\enq
%
%
%

The bounds for $\wt{\mc{P}}_3$ may be obtained by invoking the repulsion property
of the roots, point e) of Hypothesis~\ref{Hypotheses eqns de quantification},
as well as the lower bounds
\beq
\big| \sinh( y- x_1) \big| \, \geq \,
c \quad \e{and} \quad
\big| \sinh( 2\i\zeta + x_1-y) \big| \, \geq \,
c \;,
\enq
which hold for a $c > 0$ and all $y \in  \wh{\wt{\mc{Y}}}_{\e{sg};k}$, since
$\e{d}\big( \msc{D}_{\mathbb{Y};\i\pi}, \msc{D}_{\mathbb{Y};\i\pi} + \i   \zeta \big) > c'$ and
$\e{d}\big( \msc{D}_{\mathbb{Y};\i\pi} + 2\i \zeta, \msc{D}_{\mathbb{Y};\i\pi} + \i \zeta \big)
> c'$ for some $c' > 0$. Altogether, this yields that, for $\wt{C}_3 > 0$ and
$\wt{d}_3 \in {\mathbb N}$,
\beq
\wt{C}_3 T^{\wt{d}_3}  \; \leq  \; \big| \wt{\mc{P}}_3 \big| \; \leq \; \wt{C}_3^{-1} T^{-\wt{d}_3} \, .
\enq
%
%
%
Finally, in the estimates relative to $\wt{\mc{P}}_4$, one uses that
$\tf{\zeta }{2} + \eps^{\prime}-\eps  \geq |\Im(x-x_1)| \geq  0$ which leads to
\beq
\wt{C}_4 \; \leq  \; \big| \wt{\mc{P}}_4 \big| \; \leq \; \wt{C}_4^{-1} \;, \quad
\wt{C}_4 > 0 \;.
\enq
%
%
%


One now recasts the quantisation conditions for $x_1$ in the form
\beq
\f{ \sinh(\i\zeta + y_{k+2} - x_1)   }{ \sinh(\i\zeta +  x_1 - y_k )  }  \; = \;
- \ex{ \f{1}{T} \wh{\mc{E}}(x_1 \,|\, \wh{\mathbb{Y}}\, ) }  \cdot \pl{\ell=1}{4}\wt{\mc{P}}_{\ell}^{-1}
\enq
and observes that $x_1 \in \msc{D}_{\mathbb{Y}} \setminus \op{D}_{ -\i \tf{\zeta}{2},\eps}$
as follows from point g) of Hypothesis~\ref{Hypotheses eqns de quantification}.
Hence, one has the estimate that $\Re[\veps_{\e{c}}(x_1) ] = \e{O}(-T \ln T)$. This last
estimate, along with the boundedness of $\wh{\Phi}(\la\,|\,\wh{\mathbb{Y}}\,)$, leads to
the first bound in \eqref{ecriture bornes sur yk+2 x1 et yk pour exclision de config avec une paire racine sing et particule}. \qed

\subsubsection{A chain of regular roots containing one singular root in the inside}
We now establish the continuation property allowing one to characterise the 
possibilities of extending a chain of particle roots containing a singular root
somewhere in its inside. 

\begin{lemme}
\label{Lemme prolongation chaine de racines qui contiennent un racine singuliere dans le bulk}
There exists $T_0, \eta$ small enough such that, uniformly in $0<T<T_0$ and
$\eta > \tf{1}{(NT^4)}$ the below holds. 
 
Let $\ell \geq 2$ and assume that there exist $y_0, \dots , y_{k-1}, y_{k+2}, \dots, y_{\ell}
\in \wh{\mc{Y}}_{\e{r}}$ such that $y_0$ is a maximal or weakly maximal root,
$y_k\in \wh{\wt{\mc{Y}}}_{\e{sg}}$ and
\begin{enumerate}
\item
$y_p \, = \, y_{p+1} \, + \, \i\zeta \, + \, \de_{p}$ with
$\de_{p} =  \e{O}\big( \ex{-\f{1}{T}d_{p}  } \big)$  for some $d_p>0$ and $p=0,\dots \ell-1$;
\item
$\Re\Big[ \sul{p=0}{s} \veps_{\e{c}}(y_p)  \Big]
\, \leq \, - c_s <0$  for  $s=0,\dots, \ell-1$.
\end{enumerate}
Above, it should be understood that $y_{k+1}=y_k-\i\zeta$ and $\de_{k}=0$.  
 
Then, either  
\beq
\Re\Big[ \sul{p=0}{\ell} \veps_{\e{c}}(y_p)  \Big] \, \leq   \, C \eta_T   \qquad with \qquad \eta_T \tend 0^+ \;\; as \;\; T\tend 0^+\;, 
\enq
or one may extract a subsequence in $T\tend 0^+$ such that, for that subsequence
of temperatures going to $0$, there exists a particle root $y_{\ell+1}\in
\wh{\mc{Y}}_{\e{r}}$ such that 
\beq
y_{\ell + 1} \, = \, y_{\ell} \, - \, \i\zeta \, + \, \de_{ \ell } \quad  with  \quad  \de_{\ell} =  \e{O}\big( \ex{-\f{c}{T}} \big) \qquad  and  \qquad 
\Re\Big[ \sul{p=0}{\ell} \veps_{\e{c}}(y_p)  \Big] \, \leq \, - c_{\ell} < 0 \;. 
\enq
\end{lemme}

\Proof 
If $ \Re\Big[ \sum_{p=0}^{\ell} \veps_{\e{c}}(y_p)  \Big] \tend 0$ as $T\tend 0^+$,
then there is nothing else to do. 

Otherwise, one is in the situation when $ \Re\Big[ \sum_{p=0}^{\ell} \veps_{\e{c}}(y_p) \Big]
\not\tend 0$ as $T\tend 0^+$. Hence, one may extract a subsequence such that
$\Re\Big[ \sum_{p=0}^{\ell} \veps_{\e{c}}(y_p) \Big] \tend \pm 2 c_{\ell}$ as
$T\tend 0^+$, this for some constant $c_{\ell}>0$.
 
For this subsequence, one starts by taking the product over all the subsidiary
conditions defining the regular particle roots $y_0,\dots, y_{k-1}, y_{k+2},\dots,
y_{\ell}$ and the singular root $y_{k}$. This yields
\bem
(-1)^{ \ell - 1 } \; = \; \pl{ s = 0 }{ \ell } \bigg\{ \ex{ -\f{1}{T} \wh{\mc{E}}( y_s \,|\, \wh{\mathbb{Y}}\, ) } \bigg\} \cdot 
\pl{ s=0 }{ \ell } \Bigg\{  \pl{y \in \wh{\mc{Y}}_{\e{r};\ell}^{(k)} }{} \bigg\{  \f{ \sinh(\i\zeta + y- y_s) }{ \sinh(\i\zeta + y_s- y )  } \bigg\} \cdot 
\pl{x \in \wh{\mf{X}} }{} \bigg\{  \f{ \sinh(\i\zeta + y_s - x )  }{ \sinh(\i\zeta + x - y_s) }   \bigg\}
\Bigg\} \\
\times \pl{y \in  \wh{\wt{\mc{Y}}}_{\e{sg};k} }{} \Bigg[ - \f{ \sinh^2(\i\zeta + y- y_k )  \sinh(2\i\zeta + y- y_k ) }{ \sinh^2(\i\zeta +  y_k - y )   \sinh(2 \i\zeta + y_k - y )  } 
\cdot \pl{ \substack{ s=0 \\ \not= k, k+1} }{\ell}  \bigg\{ \f{ \sinh(\i\zeta + y- y_s )  \sinh( y- y_s ) }{ \sinh(\i\zeta +  y_s - y )   \sinh(2 \i\zeta + y_s - y )  }    \bigg\}  \Bigg]\;, 
\label{ecriture produit sur k racines regulieres etc}
\end{multline}
where we agree upon 
\beq
\label{rangeregwithouttwo}
\wh{\mc{Y}}_{\e{r};\ell}^{(k)} \, = \, \wh{\mc{Y}}_{\e{r}}\setminus \{y_0,\dots, y_{k-1} , y_{k+2},\dots, y_{\ell} \} \qquad \e{and} \qquad 
\wh{\wt{\mc{Y}}}_{\e{sg};k} \, = \, \wh{\wt{\mc{Y}}}_{\e{sg}}\setminus \{y_k \}  \;. 
\enq

First assume that, for the selected subsequence, $\Re\Big[ \sum_{p=0}^{\ell}
\veps_{\e{c}}(y_p)  \Big] \, \geq \, c_{\ell} >0$, when $T$ is small enough.
Then, for the equation to be satisfied, one needs to compensate the exponentially
small term appearing on the \textit{rhs}
of \eqref{ecriture produit sur k racines regulieres etc} by approaching a
pole of the prefactors. There are the following possibilities.
\begin{itemize}
\item
There exists $y \in \wh{\mc{Y}}_{\e{r};\ell}^{(k)}$ and $s \in \intn{0}{\ell}$
such that $y \, - \,  \i\zeta \, = \, y_s \, + \,  \e{O}\big( \ex{-\f{c}{T}} \big)$. 
For $s=0$ this would contradict the (weak) maximality of $y_0$, for $s \in \intn{1}{\ell}$
this would lead to $y \, = \, y_{s-1}\, + \,  \e{O}\big( \ex{-\f{c}{T}} \big)$.
For $s \in \intn{1}{\ell} \setminus \{ k+2\}$ the latter contradicts the repulsion
property of the roots, point e) of Hypothesis~\ref{Hypotheses eqns de quantification}.
For $s=k+2$ one obtains that $y \, = \, y_{k+1}\, + \,  \e{O}\big( \ex{-\f{c}{T}} \big)$.
Since point b) of Hypothesis~\ref{Hypotheses solubilite NLIE} ensures that
$\e{d}\big(y_k-\i\zeta, \msc{C}_{\e{ref}} \big)\geq \mf{c}_{\e{ref}} T$ and since
$\e{d}\bigl(z, \msc{C}_{\e{ref}}\bigr) \le cT$ for all $z \in \wh{\msc{C}}_{\mathbb Y}
[\,\wh{u}\,]$ and a constant
$c>0$ that only depends on $ |\wh{\mf{X}}|$ and $|\wh{\mc{Y}}|$, taken that $\mf{c}_{\e{ref}}$
is large enough, one would get that $y \in \msc{D}_{\mathbb{Y};\i\pi}$, a contradiction.
\item
There exists $x \in \wh{\mf{X}}$ and $s \in \intn{0}{\ell}$ such that
$x \, + \, \i\zeta \, = \, y_s \, + \,  \e{O}\big( \ex{-\f{c}{T}} \big)$. 
Then, by using the explicit form of the roots $y_{s}$, one has that
$x \, = \, y_{k+1} \, + \, \i(k-s) \zeta \, + \,  \e{O}\big( \ex{-\f{c}{T}} \big)$.
If $s \not= k$, taken that $x, y_{k+1} \in \msc{D}_{\mathbb{Y};\i\pi}$, one would
be led to $\e{d}\Big(\msc{D}_{\mathbb{Y}; \i \pi },
\msc{D}_{\mathbb{Y}; \i \pi } + \i (k-s) \zeta \Big) \, = \,  \e{O}\big( \ex{-\f{c}{T}} \big)$
which contradicts \eqref{ecriture propriete espacement fini pour domaines DY shiftes}.
If $s=k$, then upon repeating the arguments of
Lemma~\ref{Lemme impossibilite presence trou dans chaine de racines eg terminant sur une racine sing}, one would obtain the estimates
\beq
C_x T^{d_x} \; \leq \;
\biggl|\f{ \sinh(\i\zeta + y_{k+2}-x) }{ \sinh(\i\zeta +x-y_k)} \biggl| \,
\leq \, C_x^{-1} T^{-d_x}
\enq
and
\beq
C_y T^{d_y} \ex{-\f{1}{T}c_T}\; \leq \;
\Biggl|  \pl{s=0}{\ell} \ex{-\f{1}{T}\veps_{\e{c}}(y_s) } \cdot
\f{ \sinh(2\i\zeta + y_{k+2}-y_k)  }{  \sinh(\i\zeta +x-y_k)  } \Biggr| \,
\leq \, C_y^{-1} T^{-d_y}
\enq
for some constants $C_x, C_y>0$ and integers $d_x, d_y \in \mathbb{N}$. Pursuing a
similar reasoning as in the proof of Lemma~%
\ref{Lemme prolongation chaine racines qui se termine sur une racine singuliere}
below equation \eqref{lem11useslem10}, one sees that the latter bounds cannot hold
simultaneously for $T, \tf{1}{(NT^4)}$ small enough,
since $\Re\Big[ \sum_{p=0}^{\ell} \veps_{\e{c}}(y_p)  \Big] \, \geq \, c_{\ell} >0$
by hypothesis, hence leading to a contradiction.
\item
There exists $y \in \wh{\wt{\mc{Y}}}_{\e{sg};k}$ and $s \in \intn{0}{\ell}\setminus \{ k, k+1\}$
such that one of the following two scenarios holds $y\, -\, \i\zeta \, =\,
y_s \, + \,  \e{O}\big( \ex{-\f{c}{T}} \big)$ or $y\, -\, 2\i\zeta \, = \,
y_s \, + \,  \e{O}\big( \ex{-\f{c}{T}} \big)$. 
Focusing on the first case scenario, $y\, -\, \i\zeta \, =\,  y_s \, + \, 
\e{O}\big( \ex{-\f{c}{T}} \big)$ for some $s \in \intn{0}{\ell}\setminus \{ k, k+1\}$,
and supposing that $s=0$, one may argue as in the proof of
Lemma~\ref{Lemme chaine de racines reguliere} below equation \eqref{estilem9}
that $y_0\in \msc{D}_{\mathbb{Y};\i\pi}$ which cannot be.
Else, when $s \in \intn{1}{\ell}\setminus \{ k, k+1, k+2 \}$, one has $y \, =\, 
y_{s-1} \, + \,  \e{O}\big( \ex{-\f{c}{T}} \big)$ which contradicts the repulsion
property between the roots, point e) of Hypothesis~\ref{Hypotheses eqns de quantification}.
Finally, for $s=k+2$, one gets $y \, =\,  y_{k+1} \, + \, \e{O}\big( \ex{-\f{c}{T}} \big)$.
Since $y \in \msc{D}_{\mathbb{Y};\i\pi}+\i\zeta$ and $y_{k+1} \in \msc{D}_{\mathbb{Y};\i\pi}$
that would lead to
\beq
\e{O}\big( \ex{-\f{c}{T}} \big) \; = \; \e{d}_{\i\pi}(y,y_k) \; \geq  \; \e{d}\Big( \msc{D}_{\mathbb{Y};\i\pi}+\i\zeta , \msc{D}_{\mathbb{Y};\i\pi} \Big) \;,
\enq
hence contradicting \eqref{ecriture propriete espacement fini pour domaines DY shiftes}.

Now turn to the second case scenario, $y\, -\, 2 \i\zeta \, =\,  y_s \, + \,
\e{O}\big( \ex{-\f{c}{T}} \big)$ for some $s \in \intn{0}{\ell}\setminus \{ k, k+1\}$.
For $s=0$ one obtains a contradiction to the (weak) maximality of $y_0$ if $T$ is small enough.
For $s=1$ one ends up with $y \, =\,  y_0 +\i\zeta \, + \,  \e{O}\big( \ex{-\f{c}{T}} \big)$
and one may argue as above that $y_0\in \msc{D}_{\mathbb{Y};\i\pi}$ which cannot be.
For $s \in \intn{2}{\ell}\setminus \{ k, k+1, k+3\}$, one has $y \, =\,
y_{s-2} \, + \,  \e{O}\big( \ex{-\f{c}{T}} \big)$ which contradicts the repulsion
property between the roots, point e) of Hypothesis~\ref{Hypotheses eqns de quantification}.
Finally, for $s=k+3$, one gets $y \, =\,  y_{k+1} \, + \,  \e{O}\big( \ex{-\f{c}{T}} \big)$.
Since $y \in \msc{D}_{\mathbb{Y};\i\pi}+\i\zeta$ and $y_{k+1} \in \msc{D}_{\mathbb{Y};\i\pi}$
that would lead to
\beq
\e{O}\big( \ex{-\f{c}{T}} \big) \; = \; \e{d}_{\i\pi}(y,y_k) \; \geq  \; \e{d}\Big( \msc{D}_{\mathbb{Y};\i\pi}+\i\zeta , \msc{D}_{\mathbb{Y};\i\pi} \Big) \;,
\enq
hence contradicting \eqref{ecriture propriete espacement fini pour domaines DY shiftes}.
\item
There exists $y \in \wh{\wt{\mc{Y}}}_{\e{sg};k}$ such that $y \, = \, y_k \, + \,
\i\zeta \, + \,  \e{O}\big( \ex{-\f{c}{T}} \big)$ or $y \, =  \, y_k \, + \, 2\i\zeta
\, + \,  \e{O}\big( \ex{-\f{c}{T}} \big)$. Since
$y, y_k\in \msc{D}_{\mathbb{Y};\i\pi}+\i\zeta$, these estimates would lead to
\beq
\e{O}\big( \ex{-\f{c}{T}} \big) \; = \; \e{d}_{\i\pi}(y,y_k) \; \geq  \; \e{d}\Big( \msc{D}_{\mathbb{Y};\i\pi}+\i(1+p)\zeta , \msc{D}_{\mathbb{Y};\i\pi} + \i \zeta \Big)
\enq
for either $p=1$ or $p=2$, hence contradicting \eqref{ecriture propriete espacement fini pour domaines DY shiftes}.
\end{itemize} 
 
It remains to consider the case, when for the subsequence of interest
$\Re\Big[ \sum_{p=0}^{ \ell } \veps_{\e{c}}(y_p)  \Big] \, \leq \, -c_{\ell} <0$, provided
that $T$ is small enough.
Then, for equation \eqref{ecriture produit sur k racines regulieres etc}
to be satisfied, the only possibility is to compensate for the exponentially large
term appearing in its \textit{rhs} by approaching  exponentially fast to zero  with
some of the factors present in the numerator. Thus, one of the following considerations
applies.
\begin{itemize}
\item
There exists $x \in \wh{\mf{X}}$ and $s \in \intn{0}{\ell}$ such that
$x\, - \, \i\zeta \, =  \, y_s \, + \,  \e{O}\big( \ex{-\f{c}{T}} \big)$.
For $s=0$ it is enough to repeat the reasoning outlined in the paragraph around
\eqref{ecriture discussion x moins i zeta egal y0 est impossible}. For
$s \in \intn{1}{\ell}\setminus \{k+2\}$ one would get $x  \, = \,
y_{s-1} \, + \,  \e{O}\big( \ex{-\f{c}{T}} \big)$. By point a) of
Hypothesis~\ref{Hypotheses solubilite NLIE} one has that
$\e{d}_{\i\pi}(y_{s-1},\pm q)\geq \mf{c}$ so that $\e{d}_{\i\pi}(x,\pm q) \geq \mf{c}/2$.
Hence, $x \in \e{Int}(\msc{C}_{\e{ref}})\setminus
\bigcup_{\sg=\pm} \op{D}_{\sg q, \tf{\mf{c}}{2}}$.
Since for $s \not=k+2$ the root $y_{s-1}$ is not contained in $\e{Int}(\msc{C}_{\e{ref}})$,
one has
\beq
\e{O}\big( \ex{-\f{c}{T} } \big) \, = \, \e{d}_{\i\pi}( x, y_{s-1} ) \, \geq  \,
\e{d}_{\i\pi}\Big(\, \wh{\mf{X}},  \msc{C}_{\e{ref}}\setminus
{\textstyle \bigcup_{\sg=\pm}} \op{D}_{\sg q, \tf{\mf{c}}{2} } \Big) \;,
\enq
hence contradicting \eqref{ecriture estimation distance de trous a la frontiere C ref}.
Finally, for $s=k+2$, such a situation would imply that $x \, + \, \i\zeta \,- \, y_{k} = \vth_k$,
$y_{k+2}\,-\, x \, + \, \i\zeta \, = \, \vth_{k+1}$ and $2\i\zeta + y_{k+2} - y_{k} \,
= \, \vth_{k+2}$ where $\vth_k, \vth_{k+1}, \vth_{k+2} \, = \,  \e{O}\big( \ex{-\f{c}{T}} \big)$.
Thus, by virtue of Lemma~%
\ref{Lemme impossibilite presence trou dans chaine de racines eg terminant sur une racine sing},
there exist constants $C_x, C_y>0$ and integers $d_x, d_y$ such that
\beq
C_x T^{d_x} \; \leq \; \biggl|
\f{ \sinh(\i\zeta + y_{k+2}-x) }{ \sinh(\i\zeta +x-y_k)} \biggr| \,
\leq \, C_x^{-1} T^{-d_x}
\enq
and
\beq
C_y T^{d_y} \ex{-\f{1}{T} c_T} \; \leq \;
\Biggl|  \pl{s=0}{k+1} \ex{-\f{1}{T}\veps_{\e{c}}(y_s) }
\f{ \sinh(2\i\zeta + y_{k+2}-y_k)  }{  \sinh(\i\zeta +x-y_k)  } \Biggr| \,
\leq \, C_y^{-1} T^{-d_y} \;.
\enq
Since $\vth_{k+2}=\vth_{k+1}+\vth_{k}$ and $\sum_{s=0}^{k+1} \Re[\veps_{\e{c}}(y_s)]
\leq -c_{k+1}<0$ by hypothesis, this leads to a contradiction.
\item
There exists $y \in \wh{\wt{\mc{Y}}}_{\e{sg};k}$ and $s \in \intn{0}{\ell}\setminus \{k\}$
such that $y \, + \, \i\zeta \, = \,  y_s \, + \,  \e{O}\big( \ex{-\f{c}{T}} \big)$ or
$y \, = \,  y_s \, + \,  \e{O}\big( \ex{-\f{c}{T}} \big)$. For $s \ne \ell$ the first
case reduces to $y \, = \,  y_{s+1} \, + \, \e{O}\big( \ex{-\f{c}{T}} \big)$. Thus, if
$s \ne \ell$ in the first case and $s \ne k + 1$ in the second case, the repulsion
property of the roots, point e) of Hypothesis~\ref{Hypotheses eqns de quantification},
is violated, which cannot be. If $s = \ell$ in the first case, then $y \, = \, y_{k}
\, - \, \i (\ell - k + 1)\zeta \, + \,  \e{O}\big( \ex{-\f{c}{T}} \big)$, leading to
\beq
\e{O}\big( \ex{-\f{c}{T}} \big) \; = \;
\e{d}\big(y, y_k - \i (\ell - k + 1) \zeta \big) \; \geq \;
\e{d}_{\i\pi}\Big(\msc{D}_{\mathbb{Y};\i\pi} + \i \zeta,
\msc{D}_{\mathbb{Y};\i\pi} - \i (\ell - k) \zeta   \Big)
\enq
which contradicts \eqref{ecriture propriete espacement fini pour domaines DY shiftes}.
Similarly, in the remaining exceptional case, $y = y_k - \i \zeta +
\e{O}\bigl(\ex{-\f{c}{T}}\bigr)$.
Hence,
\beq
\e{O}\big( \ex{-\f{c}{T}} \big) \; = \;
\e{d}\big(y, y_k - \i \zeta \big) \; \geq \;
\e{d}_{\i\pi}\Big(\msc{D}_{\mathbb{Y};\i\pi} + \i \zeta,
\msc{D}_{\mathbb{Y};\i\pi} \Big)
\enq
which again contradicts \eqref{ecriture propriete espacement fini pour domaines DY shiftes}.
\item
There exists $y \in \wh{\mc{Y}}_{\e{r};\ell}^{(k)}$ and $s \in \intn{0}{\ell}$ such that
$y \, = \,  y_s \, - \, \i\zeta \, + \,  \e{O}\big( \ex{-\f{c}{T}} \big)$.
If $s = k$ one argues as in the first item below equation \eqref{rangeregwithouttwo}
that then $y \in \msc{D}_{\mathbb{Y};\i\pi}$, which is impossible for regular roots.
If $s\in \intn{0}{\ell-1} \setminus \{k\}$, one obtains the relation $y \, = \,
y_{s+1} \, + \,  \e{O}\big( \ex{-\f{c}{T}} \big)$ which contradicts the repulsion property
of the roots, point e) of Hypothesis~\ref{Hypotheses eqns de quantification}. However,
there is no obstruction here in having such a relation for $s=\ell$.
\end{itemize}

This allows one to conclude. \qed









\section{\boldmath The low-$T$ behaviour of simple spectral observables}
\label{Section Low T expansion of spectral observables}

\subsection{The low-$T$ behaviour of the particle-hole roots}
In this sub-section we establish Proposition
\ref{Proposition caracterisation dvpt basse T pour part et trous}. Namely, we determine 
the first few terms in the low-$T$ expansion of the hole $x_a\in \mf{X}$ and particle
 $y_a\in \mc{Y}$ roots, this when the infinite Trotter limit has been taken. As it will be
shown, both types of roots  are located in an $\e{O}(T)$ tubular neighbourhood of the curve 
\beq
\msc{C}^{(\e{tot})}_{\veps}=\Big\{ \la \, : \, \Re\big[ \veps(\la) \big]=0 \;  , \;  |\Im[\la]| \leq \tf{\zeta}{2} \Big\} \; .
\enq
Since
\beq
     \Im\big[ \veps  \big] \; : \; \msc{C}^{(\e{tot})}_{\veps} \tend \R
\enq
is a double cover map (see Appendix~\ref{Properties_of_the_dressed_energies}),
it is convenient to split this curve in two pieces.

$\Im\big[ \veps  \big] $ becomes a diffeomorphism along the curves
$\Ga^{(\a)}_{\veps}$, $\a\in \{L, R\}$, where 
\beq
\Ga^{(\a)}_{\veps} \; = \; \Big\{\la \in   \msc{C}^{(\e{tot})}_{\veps} \; : \; \ups_{\a}\Re[\la]>0  \Big\} \;. 
\enq
Note that, when endowed with a counterclockwise orientation,
$\la \mapsto \Im\big[ \veps(\la)  \big]$ is increasing along $\Ga^{(R)}_{\veps}$ and
$\Ga^{(L)}_{\veps}$. 

In this section, we continue to denote by $u(\la\,|\, \mathbb{Y})$ the solution to the
infinite Trotter number non-linear integral equation associated with the fixed point
$u\in \mc{E}_{\mc{M}}$, where $\mathbb{Y}=\mc{Y}\ominus \mf{X}$ is the collection of
particle-hole parameters solving the quantisation
conditions~\eqref{ecriture condition de quantification Trotter infini}.

To be specific about the solution we focus on, we first pick positive pairwise distinct integers
\beq
 h_a^{(\a)}  \; , \quad a=1, \dots, |\mf{X}^{(\a)}|  \qquad \e{and}\qquad  p_a^{(\a)} \; , \quad  a=1, \dots,  |\mc{Y}^{(\a)}|
\enq
with $\a \in \{L,R\}$ and $|\mf{X}^{(\a)}|$, $ |\mc{Y}^{(\a)}|$ uniformly bounded in $T$.
Further, these positive integers are such that $T h_a^{(\a)}$ and
$T p_{a}^{(\a)}$ converge as $T\tend 0^+$ and we define points
\beq
     \wt{\op{x}}^{\, (\a)}_a \; = \;
        \veps^{-1}_{\a}\Bigl(- \lim_{T\tend 0^+} \ups_\a 2\i\pi T
	   \bigl(h_a^{(\a)} + {\textstyle \frac{1}{2}}\bigr) \Bigr) \qquad \e{and} \qquad
     \wt{\op{y}}^{\, (\a)}_a \; = \;
        \veps^{-1}_{\a}\Bigl(\lim_{T\tend 0^+} \ups_\a 2\i\pi T
	   \bigl(p_a^{(\a)} + {\textstyle \frac{1}{2}}\bigr) \Bigr) \;.
\enq
Should some of these points coincide, we keep track of their multiplicities.  This operation gives rise to collections of points
$\op{x}^{(\a)}_{a}$, $\op{y}^{(\a)}_{a}$, and associated multiplicities $\varkappa_{a}^{(\a)}$, $\mf{y}_{a}^{(\a)}$ as in
Theorem \ref{Theorem existence solutions racines particules et trous}.  Then, one has a solution
to the infinite Trotter number quantisation conditions as ensured by Theorem \ref{Theorem existence solutions racines particules et trous}.

\begin{figure}
\begin{center}
\includegraphics[width=0.75\textwidth]{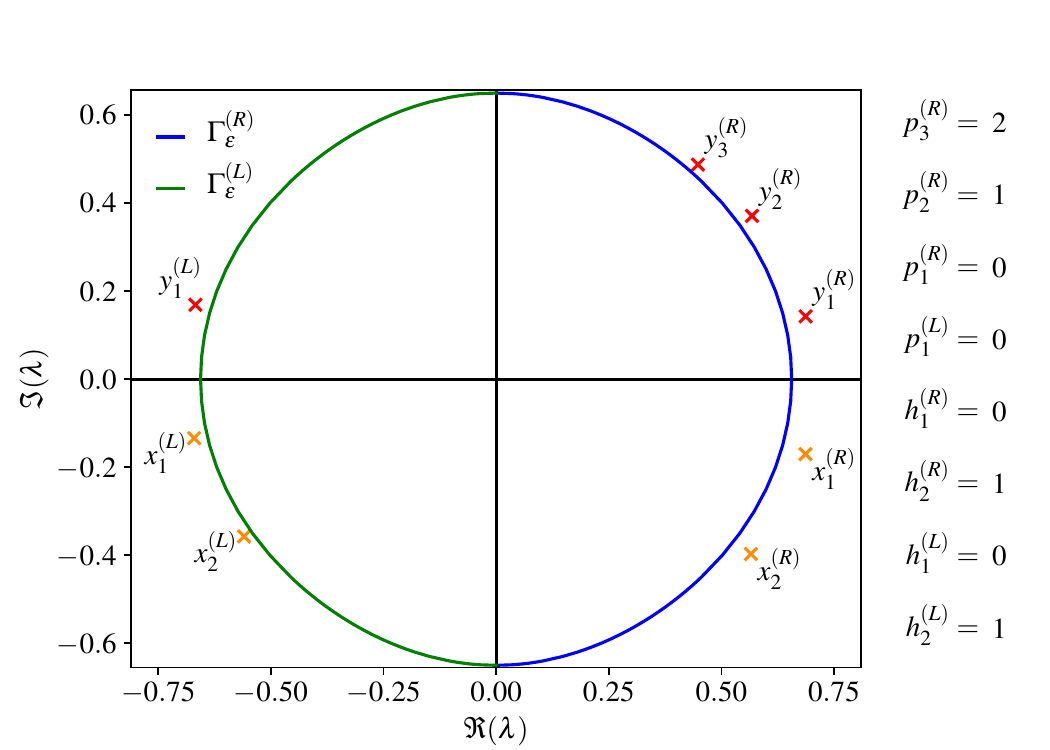}
\caption{Example of a particle-hole configuration with particle roots $y_a^{(\alpha)}$,
hole roots $x_a^{(\alpha)}$ and parameters $p_a^{(\alpha)}$, $h_a^{(\alpha)}$ obtained
by solving \eqref{equation pour part trous type +} and \eqref{equation pour part trous type -}
numerically for the choice of parameters $J=1$,  $h=2$, $\zeta = 1.3$
($\Rightarrow \Delta \approx 0.27)$, $T = 0.2$.}
\label{ph_example_configuration}
\end{center}
\end{figure}

The particle/hole roots solving these quantisation conditions and located close to
the curve $\Ga^{(R)}_{\veps}$ are denoted by $y_a^{(R)}/x_a^{(R)}$ and are solutions to 
\beq
u\big( y_a^{(R)} \,|\, \mathbb{Y} \big) \, = \, 2\i\pi T \, \big(   p_a^{(R)} \, + \,  \tfrac{1}{2}  \big)  \qquad \e{and} \qquad
u\big( x_a^{(R)} \,|\, \mathbb{Y} \big) \, = \, - 2\i\pi T \, \big(    h_a^{(R)} \, + \,  \tfrac{1}{2} \big)
\label{equation pour part trous type +}
\enq
with $p_a^{(R)}, h_a^{(R)} \in \mathbb{N}$. In their turn, the particle/hole roots located 
close to the curve $\Ga^{(L)}_{\veps}$ are denoted by $y_a^{(L)}/x_a^{(L)}$ and solve
the equations
\beq
u\big( y_a^{(L)} \,|\, \mathbb{Y} \big) \, = \, - 2\i\pi T\,  \big(   p_a^{(L)} \, + \,  \tfrac{1}{2}  \big)  \qquad \e{and} \qquad
u\big( x_a^{(L)} \,|\, \mathbb{Y} \big) \, = \, 2\i\pi T \, \big(    h_a^{(L)} \, + \,  \tfrac{1}{2} \big)
\label{equation pour part trous type -}
\enq
with $p_a^{(L)}, h_a^{(L)} \in \mathbb{N}$. 
Note that the change of signs from \eqref{equation pour part trous type +}
to \eqref{equation pour part trous type -} is in accordance with the behaviour
of $\Im[\veps (\la)]$ along the oriented curves $\Ga^{(\a)}_{\veps}$ (see
Fig.~\ref{ph_example_configuration} for an example).
Accordingly, the particle $\mc{Y}$  or hole $\mf{X}$ solution sets partition as $\mf{X}\, =\, \mf{X}^{(L)}\cup \mf{X}^{(R)}$ and 
 $\mc{Y}\, = \, \mc{Y}^{(L)}\cup \mc{Y}^{(R)}$, where 
\beq
 \mf{X}^{(\a)} \; = \; \Big\{ x_a^{(\a)} \Big\}_{1}^{ |\mf{X}^{(\a)}| } \qquad \e{and} \qquad 
 \mc{Y}^{(\a)} \; = \; \Big\{ y_a^{(\a)} \Big\}_{1}^{ |\mc{Y}^{(\a)}| } \;. 
\enq

In order to obtain the results relative to the existence of a low-$T$ asymptotic
expansion for the roots, we need to partition the roots further
according to
\beq
\left\{ \ba{ccc}  \mc{Y}^{(R)} & = & \mc{Y}^{(\e{fR})} \oplus  \mc{Y}^{(+)}   \\
                \mc{Y}^{(L)} & = &  \mc{Y}^{(\e{fL})}  \oplus  \mc{Y}^{(-)}  \ea \right.
\qquad \e{and} \qquad
\left\{ \ba{ccc}  \mf{X}^{(R)} & = & \mf{X}^{(\e{fR})} \oplus  \mf{X}^{(+)}   \\
                \mf{X}^{(L)} & = &  \mf{X}^{(\e{fL})}   \oplus  \mf{X}^{(-)}  \ea \right.  \;.
\label{ecriture partition close far roots}
\enq
Moreover, we agree to parameterise the various roots as $\mc{Y}^{(\a)} =
\big\{  y_a^{(\a)} \big\}_{a=1}^{n_p^{(\a)} }$ and $\mf{X}^{(\a)}  =
\big\{  x_a^{(\a)} \big\}_{a=1}^{n_h^{(\a)} }$  for  $ \a\in \big\{ +, -, \e{fR}, \e{fL} \big\}$.
Under such a reparameterisation, $\mc{Y}$ is recast as in
\eqref{definition ensemble Y partitiones divers types particules},
$\mf{X}$ as in \eqref{definition ensemble X partitiones divers types trous}.
We split the sets of integers \eqref{ecriture entiers pa et ha}
describing these roots into four families
\beq
\Big\{  p_a^{(\a)} \Big\}_{a=1}^{n_p^{(\a)} } \; , \quad \e{resp}. \quad  \Big\{  h_a^{(\a)} \Big\}_{a=1}^{n_h^{(\a)} } \;, \quad  \a\in \Big\{ +, -, \e{fR}, \e{fL} \Big\} \;.
\label{ecriture partition close far integers}
\enq
The integers may vary with $T$ but in such a way that
\beq
\limsup_{T\tend 0^+} T p_a^{(\pm)} \, = \, \limsup_{T\tend 0^+} T h_a^{(\pm)} \; = \; 0 \qquad \e{and} \qquad
\liminf_{T\tend 0^+} T p_a^{(\a)},\ \liminf_{T\tend 0^+} T h_a^{(\a)} \; > \; 0 \; , \quad \a \in \big\{ \e{fR}, \e{fL} \big\} \;.
\label{ecriture limite des close far integers en echelle T}
\enq
To each collection of these integers, we associate the vectors
\begin{align}
     \bs{h}_{  |\mf{X}^{(\a)}| }^{(\a)} & = -  2\i\pi T \ups_{\a} \, \Big(    h_1^{(\a)} \, + \, \tfrac{1}{2} \;, \dots ,
\,  h_{  |\mf{X}^{(\a)}| }^{(\a)}        \, + \, \tfrac{1}{2}   \Big)^{\op{t}}  \;,   \label{ecriture parametres h alpha X alpha} \\[1ex]
     \bs{p}_{  |\mc{Y}^{(\a)}| }^{(\a)} & =  \; \;  2\i\pi T  \ups_{\a} \, \Big( p_1^{(\a)} \, + \, \tfrac{1}{2} \;, \dots ,  \, p_{  |\mc{Y}^{(\a)}| }^{(\a)}
     \, + \,  \tfrac{1}{2}   \Big)^{\op{t}}    \; .
\label{ecriture parametres p alpha Y alpha}
\end{align}
Finally, we introduce the vectors
\beq
\bs{z}_{\e{loc}}\;=\;\big( \bs{z}^{(-)} , \bs{z}^{(+)} \big)^{\op{t}} \; , \quad \bs{z}_{\e{far}}\;=\;\big( \bs{z}^{(fL)} , \bs{z}^{(fR)} \big)^{\op{t}} \quad \e{with} \quad
\bs{z}^{(\a)} \; = \; \big( \bs{x}^{(\a)} , \bs{y}^{(\a)} \big)^{\op{t}}
\enq
where
\beq
\bs{x}^{(\a)} \; = \; \big( x_1^{(\a)}, \dots, x_{|\mf{X}^{(\a)}|}^{(\a)} \big)^{\op{t}} \qquad \e{and} \qquad
\bs{y}^{(\a)} \; = \; \big( y_1^{(\a)}, \dots, y_{|\mc{Y}^{(\a)}|}^{(\a)} \big)^{\op{t}} \;.
\enq

Now, recall the map $\Phi_{T}$ introduced in the proof of Theorem
\ref{Theorem existence solutions racines particules et trous},
\textit{c.f.}\ \eqref{definition map Psi T infinite Trotter number limit}. Further,
let $\Psi_{T ; \bs{z}_{\e{far}}}$ be the infinite Trotter number equivalent of the map
$\wh{\Psi}_{T}$ introduced in the proof of
Proposition~\ref{Proposition existence et continuite parameters particule trou partiels}, see
\eqref{definition composantes left and right de la map Psi T}, \eqref{definition fct Psi}.
Here, we made it explicit that $\Psi_{T ; \bs{z}_{\e{far}}}$ depends parametrically on the
position of the far-type roots. We refer to the proof of
Proposition~\ref{Proposition existence et continuite parameters particule trou partiels}
for more details on the matter. Then, it follows from the construction given in
Proposition~\ref{Proposition existence et continuite parameters particule trou partiels}
and Theorem~\ref{Theorem existence solutions racines particules et trous} that
$\bs{z}_{\e{loc}}$ and $\bs{z}_{\e{far}}$ are given by
\begin{align}
  \bs{z}_{\e{far}}   & = \Phi^{-1}_{T}\bigg(
\Big(   \bs{h}_{  |\mf{X}^{(fL)}| }^{(fL)} \, , \,  \bs{p}_{  |\mc{Y}^{(fL)}| }^{(fL)} \, , \,
\bs{h}_{   |\mf{X}^{(fR)}| }^{(fR)} \, , \,  \bs{p}_{  |\mc{Y}^{(fR)}| }^{(fR)} \Big)^{\op{t}} \bigg) \;, \\[1ex]
  \bs{z}_{\e{loc}}   & = \Psi^{-1}_{T;  \bs{z}_{\e{far}} }\bigg(
\Big(   \bs{h}_{  |\mf{X}^{(-)}| }^{(-)} \, , \,  \bs{p}_{  |\mc{Y}^{(-)}| }^{(-)} \, , \,
\bs{h}_{   |\mf{X}^{(+)}| }^{(+)} \, , \,  \bs{p}_{  |\mc{Y}^{(+)}| }^{(+)} \Big)^{\op{t}} \bigg) \;.
\end{align}
The invertibility of the above maps is ensured by
Proposition~\ref{Proposition existence et continuite parameters particule trou partiels}
and Theorem~\ref{Theorem existence solutions racines particules et trous}.

Since $T\mapsto \Phi^{-1}_{T}(u)$ and $(T, \bs{z}_{\e{loc}} ) \mapsto
\Psi^{-1}_{T;  \bs{z}_{\e{far}} }(u) $ are smooth, it follows that the particle and
hole roots admit the low-$T$ asymptotic expansions
\beq
y_a^{(\a)} \, \simeq  \, \sul{k \geq 0 }{} y_{a;k}^{(\a)} \cdot T^{k} \qquad \e{and} \qquad 
x_a^{(\a)} \, \simeq  \, \sul{k \geq 0 }{} x_{a;k}^{(\a)} \cdot T^{k},
\qquad \a \in \{L, R\} \;. 
\enq
The coefficients in the expansions can then be computed order by order by inserting
these expansions into the logarithmic quantisation conditions
\eqref{equation pour part trous type +}, \eqref{equation pour part trous type -}.
Note that in this way of writing the expansion, it is understood that the coefficients
are still taken to be functions of the variables given in
\eqref{ecriture parametres h alpha X alpha}, \eqref{ecriture parametres p alpha Y alpha}.
This way of understanding the low-$T$ expansion is quite convenient in that the
variables gathered in \eqref{ecriture parametres h alpha X alpha},
\eqref{ecriture parametres p alpha Y alpha} may have, in principle, a low-$T$ expansion
of their own -- possibly involving non-integer powers of $T$ \textit{etc}.\ -- so that
mixing the two would have just obscured the picture. The above expansion may in fact
be seen as a way to dissociate the model-dependent part of the expansion induced by
the $T$ dependence of $\Psi_{T}$ and the one originating from the dependence of the image
points on $T$.

The leading terms of the expansions are given by 
\beq
y_{a;0}^{(\a)} \, = \, \veps^{-1}_{\a}\Big( \ups_{\a}  2\i\pi T \big( p_a^{(\a)} \, + \,  \tfrac{1}{2}  \big)   \Big) \qquad \e{and} \qquad 
x_{a;0}^{(\a)} \, = \, \veps^{-1}_{\a}\Big( - \ups_{\a}  2\i\pi T \big( h_a^{(\a)} \, + \,  \tfrac{1}{2}  \big)   \Big) \;.
\enq
Here, we recall that $\veps_{\a}^{-1}$ is the inverse of $\veps$ taken along the
$\Ga^{(\a)}_{\veps}$ curve. We also stress that for \textit{fixed} integers
$p_a^{(\a)}$ or $h_a^{(\a)}$ -- \textit{viz}.\ those \textit{not} scaling with
$T$ -- this leading term is, in fact, $\ups_{\a} q + \e{O}(T)$.

A straightforward Taylor expansion shows that
\beq
     y_{a;1}^{(\a)} \, = \, - \f{1}{\veps^{\prime}\big(y_{a;0}^{(\a)} \big)}
        \cdot   u_1\big( y_{a;0}^{(\a)}  \,|\, \mathbb{Y}_0 \big)
	              \;, \quad
     x_{a;1}^{(\a)} \, = \, - \f{1}{\veps^{\prime}\big(x_{a;0}^{(\a)} \big)}
        \cdot    u_1\big( x_{a;0}^{(\a)}  \,|\, \mathbb{Y}_0 \big)
	             \;,
\enq
where $\mathbb{Y}_0$ is obtained from $\mathbb{Y}$ by replacing
$x_a^{(\a)}$, resp.  $y_a^{(\a)}$, with $x_{a;0}^{(\a)}$, resp. $y_{a;0}^{(\a)}$, namely
\beq
\mathbb{Y}_0 \; = \; \bigg\{ \Big\{ y_{a;0}^{(L)} \Big\} \cup \Big\{ y_{a;0}^{(R)} \Big\} \bigg\}
\ominus \bigg\{ \Big\{ x_{a;0}^{(L)} \Big\} \cup \Big\{ x_{a;0}^{(R)} \Big\} \bigg\} \;.
\enq
Similarly, proceeding to the second order,
%
%
%
%
%
%
\begin{multline}
     y_{a;2}^{(\a)} =
        - \frac{\veps''(y_{a;0}^{(\a)})}{2 \veps'(y_{a;0}^{(\a)})}\bigl(y_{a;1}^{(\a)}\bigr)^2
        - \frac{u_1'(y_{a;0}^{(\a)}\,|\,\mathbb{Y}_0)}{\veps'(y_{a;0}^{(\a)})} y_{a;1}^{(\a)}
        - \frac{u_2(y_{a;0}^{(\a)}\,|\,\mathbb{Y}_0)}{\veps'(y_{a;0}^{(\a)})} \\
        - \frac{1}{\veps'(y_{a;0}^{(\a)})}
        \sum_{\beta\in\{L,R\}}\Biggl\{
	  \sum_{b=1}^{|\mf{X}^{(\beta)}|}
	  \bigl(\partial_{x_b^{(\beta)}} u_1(z\,|\,\mathbb{Y})\bigr) \, x_{b;1}^{(\beta)} +
	  \sum_{b=1}^{|\mc{Y}^{(\beta)}|}
	  \bigl(\partial_{y_b^{(\beta)}} u_1(z\,|\,\mathbb{Y})\bigr) \, y_{b;1}^{(\beta)}
	  \Biggr\}_{ \,|\, \mathbb{Y} = \mathbb{Y}_0,\ z = y_{a;0}^{(\a)}} \;.
\end{multline}
The expression for the second term in the corresponding expansion of the hole rapidities is 
obtained under the substitution $y \hookrightarrow x$ in the above formula. 

Note that for fixed integers $p_a^{(\a)}$ or $h_a^{(\a)}$ -- \textit{viz}.\
those \textit{not} scaling with $T$ -- one may make the particle/hole-root expansions
more explicit by further expanding in $T$ the leading contributions $x_{a;0}^{(\a)}$
and $y_{a;0}^{(\a)}$. Upon assuming that $|\mf{X}^{(\beta)}|, |\mc{Y}^{(\beta)}|$ are uniformly bounded in T,  this yields to first order in $T$
\begin{align}
     x_a^{(\a)} & =  \ups_{\a} q -
        \f{ 2\i\pi  T }{\veps^{\prime}(q)}
	\bigg\{h_a^{(\a)}+\f{1}{2}
        \, + \, \f{\ups_\a}{2\i\pi}    u_1( \ups_{\a} q \,|\, \mathbb{Y}_0)  \bigg\}  \,  + \, \e{O}\Big( T^2 \e{max}^{2}\big\{h_b^{(\a^{\prime})},  p_b^{(\a^{\prime})} \big\}  \Big) \;,
      \label{DA basse T des trous CFT b} \\[1ex]
      y_a^{(\a)} & = \ups_{\a} q +
         \f{ 2\i\pi  T }{\veps^{\prime}(q)}
	 \bigg\{p_a^{(\a)}+\f{1}{2}
         \, - \, \f{\ups_\a}{2\i\pi}   u_1( \ups_{\a} q \,|\, \mathbb{Y}_0) \bigg\}    \,  + \, \e{O}\Big( T^2 \e{max}^{2}\big\{h_b^{(\a^{\prime})},  p_b^{(\a^{\prime})} \big\}  \Big)  \;.
     \label{DA basse T des particules CFT b} 
\end{align}

Due to their very different interpretations and structure, one should partition the
sets of particles and holes into those containing particle, resp.\ hole,
roots that are uniformly away from the endpoints of the Fermi zone and those
that collapse on either of the endpoints of the Fermi zone, as it was done in 
\eqref{definition ensemble Y partitiones divers types particules},
\eqref{definition ensemble X partitiones divers types trous}. Also, it is useful to 
introduce the associated partitioning of the integers
\eqref{ecriture partition des ensembles d entiers selon far et close left and right}. 
 
Since the particle, resp.\ hole, roots associated with fixed integers $p_a^{(\sg)}$,
resp.\ $h_a^{(\sg)}$, $\sg\in \{\pm\}$ will collapse to an endpoint of the Fermi
zone, on a macroscopic level it suffices to distinguish the overall difference
between the numbers of particles $n_p^{(\sg)}$ and holes $n_h^{(\sg)}$ collapsing on $\pm q$:
\beq
\ell^{(\sg)} \; = \; \sg \Big(n_p^{(\sg)} \, - \, n_h^{(\sg)} \Big)  \;. 
\enq

In particular, it is easy to see that, taken a set $\mathbb{Y}=  \mc{Y}  \ominus \mf{X}$
built from the sets introduced in
\eqref{definition ensemble Y partitiones divers types particules},
\eqref{definition ensemble X partitiones divers types trous}
it holds that 
\beq
u_1(\la\,|\, \mathbb{Y}) \, = \, u_1(\la\,|\, \mathbb{Y}^{(\e{far})})  \;  + \, \e{O}\big( T  \big) \;,
\label{ecriture DA de u1 pour particles collapsant sur zone fermi} 
\enq
with $\mathbb{Y}^{(\e{far})}$ as given by \eqref{definition ensemble reduit YM mathbb}.

In the sub-sections to come, we will establish the low-$T$ expansion for the infinite
Trotter number limit of the Eigenvalues of the quantum transfer matrix starting from their
integral representation given in Corollary~\ref{Corolaire limit Trotter vp QTM},
\textit{c.f.}~\eqref{ecriture forme explicite basse T pour  vp QTM},
\eqref{definition impulsion thermale}. Likewise, we will establish the low-$T$ expansion
for the energetic part $\mc{E}\big( \mathbb{Y} \big)$ of the Eigenvalues \textit{c.f.}~%
\eqref{definition eto E de Y}, \eqref{definition mc E de Y}.
However, first of all, we need to establish a technical result that will allow us to
carry out the low-$T$ expansions of interest.

\subsection{Low-$T$ expansion of an auxiliary quantity}
Let $g^{\prime}$ be a function meromorphic in an open neighbourhood of the curve
$\msc{C}_{ \e{ref}}$ and define, first, the function $\gammaup^{\prime}_{\e{c}}$ as the
unique solution to the linear integral equation
\beq
\Big(\e{id} \, + \, \op{K}_{\msc{C}_{\veps}}\Big)\big[ \gammaup^{\prime}_{\e{c};-} \big] (\la) \, = \, g^{\prime}(\la)\;,
\enq
where the $-$ regularisation should be understood as it was introduced in \eqref{ecriture LIE pour veps c}. This regularisation is essential should $g^{\prime}$
have poles on $\msc{C}_{\veps}$.

Then define its antiderivative by the integral representation 
\beq
     \gammaup_{\e{c}}(\la) \, - \, \lim_{\a \tend 0^-} \Int{ \msc{C}_{\veps} }{ } \f{ \dd \mu }{ 2\pi } \:
        \th(\mu-\la)  \gammaup^{\prime}_{\e{c}}(\mu+\i\a) \; = \; g(\la) \;.
\enq
Assume that $g$ extends into an $\i\pi$-periodic, meromorphic function on
$\Cx\setminus\Ga_{g}$, where $\Ga_{g}$ is the curve describing the system of cuts for $g$. 
Since $\th$ admits an $\i\pi$ periodic, odd extension to $\Cx\setminus\Ga_{\th}$,
the above integral representation allows one to continue $\gammaup_{\e{c}}$ to
an $\i\pi$ periodic meromorphic function on $\Cx\setminus\Ga_{\gammaup}$ for some
curve $\Ga_{\gammaup}$ which describes the system of cuts of the analytic continuation.

\begin{prop}
\label{Proposition derssing of quantities in low-T limit}
Let $\mf{X}$, $\mc{Y}$ be collections of parameters solving the quantisation conditions
\eqref{equation pour part trous type +}-\eqref{equation pour part trous type -}
and subject to Hypotheses~\ref{Hypotheses solubilite NLIE} and
\ref{Hypotheses eqns de quantification}. Let $u(\la \,|\, \mathbb{Y})$ be the associated
solution to the non-linear integral equation \eqref{ecriture NLIE a Trotter infini}
subject to the index condition \eqref{ecriture monodromie Trotter infini}. Let $\mathbb{Y}_{\e{ref}}$
denote the set of particle-hole roots relatively to the contour $\msc{C}_{\e{ref}}$ as introduced in
\eqref{definition Y ref}.

Then, given $g$ as above such that $g^{\prime}$ has at most a single pole in an open neighbourhood of $\msc{C}_{\e{ref}}$
which lies at $-\i\tf{\zeta}{2}$, the quantity
\beq
     \mc{G}(\mathbb{Y}) \, = \,
        \sul{ y \in \mathbb{Y}_{\e{ref};\varkappa} }{} g( y  )
	\ - \,
        \Oint{ \msc{C}_{ \e{ref}}  }{} \f{ \dd \la }{ 2 \i \pi } \:
	g^{\prime}(\la ) \,
        \msc{L}\mathrm{n}_{\msc{C}_{ \e{ref}}}
	\big[ 1+  \ex{ -\f{1}{T}u   } \, \big](\la \,|\, \mathbb{Y})
\label{definition fonction auxiliaire G}
\enq
admits the low-$T$ asymptotic expansion 
\beq
 \mc{G}(\mathbb{Y}) \, = \,  \f{1}{T}\mc{G}_{-1}  \, + \, \mc{G}_{0}(\mathbb{Y}) \,  + \,  T \mc{G}_{1}(\mathbb{Y}) \, + \, \e{O}\big( T^{2} \big) \;,
\label{ecriture DA basse T pour G}
\enq
where 
\beq
\mc{G}_{-1}  \; = \;  -  \lim_{\a\tend 0^-}\Int{ \msc{C}_{\veps}  }{ }  \f{\dd \mu }{ 2\i\pi} \veps_{\e{c}}(\mu + \i\a ) g^{\prime}(\mu + \i\a )
\enq
as well as 
\beq
\mc{G}_{0}(\mathbb{Y})\; = \; \sul{ y \in \mathbb{Y} }{} \gammaup_{\e{c}}( y  ) \; + \; \f{ \mf{s} }{2} \big[ \gammaup_{\e{c}}(q)-\gammaup_{\e{c}}(-q) \big]
\enq
and 
\beq
\mc{G}_{1}(\mathbb{Y})\; = \; \sul{ \sg = \pm }{}  \f{  \sg \, \gammaup^{\prime}_{\e{c}}( \sg q ) }{ 4\i\pi \veps^{\prime}_{\e{c}}(\sg q) }
\bigg\{  \Big( u_1(\sg q \,|\, \mathbb{Y}) \Big)^2 \, + \, \f{\pi^2}{3}  \bigg\} \;.
\enq
\end{prop}

\Proof 
By Proposition~\ref{Proposition domaine local holomorphie de F} we may deform
the integration contour to $\msc{C}_{\mathbb{Y}}[u]$. Then it follows from
equation \eqref{decomposition fondamentale integrale contre log} given in
Lemma~\ref{Lemme DA fonction holomorphe versis log}, that one may recast
$\mc{G}(\mathbb{Y})$, given in \eqref{definition fonction auxiliaire G}, in
the form below, more adapted for computing the low-$T$ expansion:
\bem
     \mc{G}(\mathbb{Y}) \; = \;  \sul{ y \in \mathbb{Y} }{} g( y  )
 \ - \, \lim_{\a \tend 0^-} \Int{ \msc{C}_{\veps}  }{}  \f{\dd \mu }{ 2\i\pi T} \: 	u(\mu+\i\a\,|\, \mathbb{Y}) g^{\prime}(\mu +\i\a )  \\
     - \, \Biggl\{ \Int{ q }{ q^{(+)}_{\mathbb{Y}}[u]}\! \!   \f{\dd \mu}{2\i\pi T} \:
        u( \mu  \,|\, \mathbb{Y} )
       \, + \, \Int{q^{(-)}_{\mathbb{Y}}[u]}{ - q }  \! \! \f{\dd \mu}{2\i\pi T} \:
       u( \mu  \,|\, \mathbb{Y} )  \bigg\} \cdot g^{\prime}(\mu)
     \, - \, \Oint{\msc{C}_{ \mathbb{Y}}[u]}{ } \f{ \dd \mu }{ 2 \i \pi } \:
     g^{\prime}(\mu ) \cdot
     \ln\Big[ 1+  \ex{ -\f{1}{T}|u|(\mu \,|\, \mathbb{Y})   } \Big]
     \;. 
\label{ecriture G sous une forme adaptee au DA}
\end{multline}
Using \eqref{expression racines q pm mathbbY} one obtains that
%
%
%
\bem
     \Biggl\{ \Int{q}{ q^{(+)}_{\mathbb{Y}}[u] }\! \! \f{\dd \mu}{2\i\pi} \:
        u( \mu  \,|\, \mathbb{Y} )  \, + \,
	\Int{ q^{(-)}_{\mathbb{Y}}[u] }{ - q } \! \! \f{\dd \mu}{2\i\pi} \:
	u( \mu  \,|\, \mathbb{Y} )  \Biggr\} \cdot g^{\prime}(\mu) \\
     \; = \; 
     \f{1}{4\i\pi} g^{\prime}(  q^{(-)}_{\mathbb{Y}}[u] )
        u^{\prime}(  q^{(-)}_{\mathbb{Y}}[u]  \,|\, \mathbb{Y} ) \,
	\Big( q^{(-)}_{\mathbb{Y}}[u] + q \Big)^2 \; - \; 
     \f{1}{4\i\pi} g^{\prime}(  q^{(+)}_{\mathbb{Y}}[u] )
        u^{\prime}(  q^{(+)}_{\mathbb{Y}}[u]  \,|\, \mathbb{Y} ) \,
	\Big( q^{(+)}_{\mathbb{Y}}[u] - q \Big)^2 \; + \; \e{O}\bigl(T^3 \bigr) \\[1ex]
     \; = \;-   T^2 \sul{\sg=\pm}{}
        \f{ \sg g^{\prime}(  \sg q )   }{4\i\pi  \veps^{\prime}_{\e{c}}(\sg q) }
	\cdot  \Big( u_1 \big( \sg q \,|\, \mathbb{Y} \big)\Big)^2  \; + \; \e{O} \bigl(T^3 \bigr) \;.
\end{multline}
Moreover, by \eqref{ecriture DA integrale avec u lisse},
\beq
     \, - \, \Oint{\msc{C}_{ \mathbb{Y}} [u]}{ } \f{ \dd \mu }{ 2 \i \pi } \:
        g^{\prime}(\mu ) \cdot
        \ln\Big[ 1+  \ex{ -\f{1}{T}|u|(\mu \,|\, \mathbb{Y})   } \Big]
     \; = \; T \sul{\sg=\pm}{}
        \f{ \sg g^{\prime}(  \sg q )   }{4\i\pi  \veps^{\prime}_{\e{c}}(\sg q) }
	\cdot  \f{\pi^2 }{3}
	\; + \; \e{O} \bigl(T^2 \bigr) \;. 
\enq
Finally, by inserting the low-$T$ expansion
\eqref{ecriture DA de u Trotter fini low T ordre 2} of $u(\la\,|\, \mathbb{Y})$ into the
remaining integral in \eqref{ecriture G sous une forme adaptee au DA}, one arrives at
the expression \eqref{ecriture DA basse T pour G}, with $\mc{G}_{-1}$ already taking
the claimed form, while 
\beq
\mc{G}_{0}(\mathbb{Y}) \; = \;  \sul{ y \in \mathbb{Y} }{} g( y  )  \ - \,
\lim_{\a \tend 0^-} \Int{ \msc{C}_{\veps}  }{ }  \f{\dd \mu }{ 2\i\pi} u_1(\mu\,|\, \mathbb{Y}) g^{\prime}(\mu+\i\a )
\enq
and 
\beq
     \mc{G}_{1}(\mathbb{Y}) \; = \;
        \sul{\sg=\pm}{}  \f{ \sg  }{4\i\pi  \veps^{\prime}_{\e{c}}(\sg q) }  \cdot  
     \bigg\{ \Big( u_1 \big( \sg q \,|\, \mathbb{Y} \big) \Big)^2
     \,  + \, \f{\pi^2 }{3} \bigg\} \cdot \Big( \e{id}
     \, - \, \op{R}_{\e{c}} \Big)[g^{\prime}](  \sg q ) \;,
\enq
where \eqref{function_u2_explicitly} has been inserted. It follows immediately from
the definition of the function $\gammaup^{\prime}_{\e{c}}$ that
$\mc{G}_{1}(\mathbb{Y})$ already takes the desired form. 

Further, going back to the definition of the dressed phase and charge, it is
readily seen that 
\bem
     - \Int{\msc{C}_{\veps}  }{ } \f{\dd \mu }{ 2\i\pi} \: u_1(\mu\,|\, \mathbb{Y}) g^{\prime}(\mu+\i\a ) \; = \; \f{ \mf{s} }{2}
\Int{ \msc{C}_{\veps}  }{ }  \dd \mu \: g^{\prime}(\mu+\i\a )  \cdot \Big(\e{id}-\op{R}_{\e{c}} \Big)[1](\mu)  \\
        \, + \, \sul{ y \in \mathbb{Y}   }{} \Int{ \msc{C}_{\veps}  }{ }  \dd \mu \:
g^{\prime}(\mu+\i\a ) \cdot	\Big(\e{id}-\op{R}_{\e{c}} \Big)[\tfrac{1}{2\pi} \th(*-y) ](\mu) \;.
\end{multline}
Then, using that the resolvent kernel is a symmetric function, one may move
the action of the resolvent onto $g^{\prime}$, leading to 
\bem
     - \lim_{\a \tend 0^-} \Int{ \msc{C}_{\veps}  }{ }  \f{\dd \mu }{ 2\i\pi} \: u_1(\mu\,|\, \mathbb{Y}) g^{\prime}(\mu +\i\a  )
\; = \; \f{ \mf{s}  }{2}	\lim_{\a \tend 0^-} \Int{ \msc{C}_{\veps}  }{ }  \dd \mu \: \gammaup_{\e{c}}^\prime(\mu+\i\a )
     \, + \, \sul{ y \in \mathbb{Y}   }{} \lim_{\a \tend 0^-} \Int{ \msc{C}_{\veps}  }{ } \frac{ \dd \mu }{2\pi} \: \gammaup^{\prime}_{\e{c}}(\mu+\i\a )  \cdot  \th(\mu - y) \\
     \; = \;  \f{s }{2} \big[\gammaup_{\e{c}}(q) -\gammaup_{\e{c}}(-q) \big]
     \; - \;  \sul{ y \in \mathbb{Y}   }{}
        \Big\{ g(y) \, - \,  \gammaup_{\e{c}}(y) \Big\} \;. 
\end{multline}
This entails the claim. \qed




\subsection{The low-$T$ expansion of $\mc{P}(\mathbb{Y})$}

We now establish Proposition~\ref{Propostion DA fin pour la vp de la QTM} and
Proposition~\ref{Propostion infimum pour les vp de la QTM}. To start with, observe
that $\mc{P}(\mathbb{Y})$ introduced in \eqref{definition impulsion thermale} is
exactly of the form \eqref{definition fonction auxiliaire G} with $g=p_0$ as
defined in \eqref{definition de p0}. Then, it follows that $\gammaup_{\e{c}}=p_{\e{c}}$
as defined through \eqref{definition dressed momentum c deforme}. 

Hence, by Proposition \ref{Proposition derssing of quantities in low-T limit}, one
has the low-$T$ expansion
\bem
     \mc{P}\big( \mathbb{Y} \big) \; = \;
        - \f{1}{T} \lim_{\a\tend 0^-} \Int{ \msc{C}_{\veps} }{ } \f{\dd \mu }{ 2\i\pi} \:
	  \veps_{\e{c}}(\mu+\i\a) p_0^{\prime}(\mu+\i\a)
     \; + \; \sul{ y \in \mathbb{Y} }{} p_{\e{c}}( y  )
     \; + \; \f{ \mf{s}}{2} \big[ p_{\e{c}}(q)-p_{\e{c}}(-q) \big] \\
     \; + \; T \sul{\sg = \pm}{} 
     \f{  \sg \, p^{\prime}_{\e{c}}( \sg q ) }{ 4\i\pi \veps^{\prime}_{\e{c}}(\sg q) }
     \bigg\{ \Big( u_1(\sg q \,|\, \mathbb{Y})  \Big)^2
     \, + \, \f{\pi^2}{3}  \bigg\}
     \; + \; \e{O}(T^2) \;. 
\end{multline}
Upon deforming the integration contour to $\intff{-q}{q}$, the first term of the expansion
yields $\mc{P}_{-1}$ as given in \eqref{definition P moins 1}. Then, by using the partition
of roots \eqref{definition ensemble Y partitiones divers types particules},
\eqref{definition ensemble X partitiones divers types trous}, the expansion of 
$u_1$ \eqref{ecriture DA de u1 pour particles collapsant sur zone fermi} as well as
the low-$T$ expansion \eqref{DA basse T des trous CFT},
\eqref{DA basse T des particules CFT} for the roots collapsing at $\pm q$, one
arrives at \eqref{ecriture dvpt basse tempe pour P de Y}.

We now pass on to proving Proposition \ref{Propostion infimum pour les vp de la QTM}.
In order to make the minimisation easier to solve, it appears convenient to slightly
reorganise \eqref{ecriture dvpt basse tempe pour P de Y}. 

First of all, one observes that the lowest possible value for $\Im[\varpi_{1}(\mathbb{H})]$, with $\varpi_{1}$ given by \eqref{definition fct varpi1 de H},
is attained at a fully packed configuration which corresponds to 
\beq
     \mathbb{H}_{\e{min}} \;=\;
        \bigcup_{ \sg = \pm }\mathbb{H}_{\e{min}}^{(\sg)}
	   \qquad \e{with} \qquad \mathbb{H}_{\e{min}}^{(\sg)}
     \, = \, \Big\{ \{a-1\}_{1}^{n_p^{(\sg)} } \; ; \;  \{ a-1\}_{1}^{n_h^{(\sg)} }
                   \Big\} \;. 
\label{definition dist part trous CFT minimale}
\enq
Moreover, a direct calculation gives 
\beq
     \varpi_{1}(\mathbb{H}_{\e{min}}) \; = \;
        \f{   \i \pi }{ \op{v}_F } \sul{ \sg = \pm }{}
	   \Big\{ (n_p^{(\sg)})^2 + (n_h^{(\sg)})^2 \Big\} \;. 
\enq
Starting from the explicit expression 
\beq
     u_1\big( \la\,|\, \mathbb{Y}^{(\e{far})} \big) \, = \,
        -\i\pi \mf{s}  Z_{\e{c}}(\la) \, - \,
	2\i\pi \ell^{(+)}  \phi_{\e{c}}(\la,q) \, + \,
	2\i\pi  \ell^{(-)} \phi_{\e{c}}(\la,-q) 
     \,  - \, 2\i\pi \hspace{-1.2cm}
     \sul{y \in \mc{Y}^{(\e{f}R)} \oplus \mc{Y}^{(\e{f}L)} \ominus \mf{X}^{(\e{f}R)}
          \ominus \mf{X}^{(\e{f}L)} }{} \hspace{-1.2cm} \phi_{\e{c}}(\la,y) 
\enq
and using the zero-monodromy condition in the form
\beq
      \ell^{(+)}-\ell^{(-)}  \, = \,
        | \mf{X}^{(\e{f}R)} | \, + \, 
	| \mf{X}^{(\e{f}L)} | \, - \, | \mc{Y}^{(\e{f}R)} | \, - \, 
	| \mc{Y}^{(\e{f}L)} |  - \mf{s}
\enq
as well as the
identities \cite{KorepinSlavnovNonlinearIdentityScattPhase,SlavnovNonlinearIdentityScattPhase}
\beq
 1 + \phi_\e{c} (q,q) = \frac{1}{2Z_\e{c}(q)} + \frac{Z_\e{c}(q)}{2} \;, \quad
     \phi_\e{c} (q,-q) = \frac{1}{2Z_\e{c}(q)} - \frac{Z_\e{c}(q)}{2}
     \label{Slavnov_Smatrix_identiy}
\enq
and the symmetry $\phi_\e{c} (\la, \mu) = - \phi_\e{c} (- \la, - \mu)$, it is
straightforward to verify that
\beq
     u_1\big(  \sg q \,|\, \mathbb{Y}^{(\e{far})} \big)   \, = \,
     u^{(\sg)}\big( \mathbb{Y}^{(\e{far})} \big) \, + \,
     2\i\pi \biggl( \ell^{(\sg)} \, - \, \ell^{(-)}Z_{\e{c}}(q)
                    \, + \, \f{ \sg \mf{s} }{ 2 Z_{\e{c}}(q) } \biggr) \;,
\label{ecriture dvpmt simplifie de u1 avec YM  non vide}
\enq
where 
\beq
     u^{(+)}\big( \mathbb{Y}^{(\e{far})} \big) =
        -2\i\pi  \hspace{-1.2cm}
	   \sul{y \in \mc{Y}^{(\e{f}R)} \oplus \mc{Y}^{(\e{f}L)} \ominus
	        \mf{X}^{(\e{f}R)} \ominus \mf{X}^{(\e{f}L)} }{}  
     \hspace{-1.2cm} \Big\{ \phi_{\e{c}}(q,y) -  \phi_{\e{c}}(q,q) -1 \Big\} \;, \quad
     u^{(-)}\big( \mathbb{Y}^{(\e{far})} \big) =
        -2\i\pi  \hspace{-1.2cm}
	   \sul{y \in \mc{Y}^{(\e{f}R)} \oplus \mc{Y}^{(\e{f}L)} \ominus
	        \mf{X}^{(\e{f}R)} \ominus \mf{X}^{(\e{f}L)} }{} 
     \hspace{-1.2cm} \Big\{ \phi_{\e{c}}(-q,y) -  \phi_{\e{c}}(-q,q)  \Big\} \;. 
\enq
This allows one to reorganise the quantities as 
\beq
     \mc{P}_{1}( \mathbb{Y}^{(\e{far})} ) \, + \,
        \varpi_{1}(\mathbb{H})  \; = \;  \msc{P}_1( \mathbb{Y}^{(\e{far})} )
	\, + \,  \Pi_{1}(\mathbb{H}) \;, 
\enq
where 
\beq
     \msc{P}_1( \mathbb{Y}^{(\e{far})} ) \, = \,
        \f{ \i\pi }{\op{v}_F}
	   \sul{\sg=\pm}{}
	      \Biggl(\frac{u^{(\sg)}\big( \mathbb{Y}^{(\e{far})} \big)}{ 2\i\pi }
	      \, - \, 
           2 \ell^{(-)}   Z_{\e{c}}(q)
	         \, + \,  \f{\sg \mf{s}}{Z_{\e{c}}(q)}\Biggr)
           \frac{u^{(\sg)}\big( \mathbb{Y}^{(\e{far})} \big)}{ 2\i\pi } \;,
\enq
while 
\beq
     \Pi_{1}(\mathbb{H}) \; = \; \f{2\i\pi}{\op{v}_F}
        \bigg\{- \frac{1}{12} + Z_{\e{c}}^{2}(q) \big(  \ell^{(-)} \big)^2
	       + \f{\mf{s}^2}{4   Z_{\e{c}}^{2}(q)  }
	       + \sul{\sg=\pm }{} n_p^{(\sg)} n_h^{(\sg)}  \bigg\} \; + \; 
     \varpi_{1}(\mathbb{H}) \, - \, \varpi_{1}(\mathbb{H}_{\e{min}}) \;.
\enq
Taken altogether, the low-$T$ expansion of $\mc{P}\big( \mathbb{Y} \big)$
is now rewritten in the form
\beq
     \mc{P}( \mathbb{Y} )\, - \,\f{1}{T} \mc{P}_{-1} \, = \,
        \msc{P}_0( \mathbb{Y}^{(\e{far})} ) \; + \;
	T  \Big(   \msc{P}_1 ( \mathbb{Y}^{(\e{far})}) \, + \,
	\Pi_{1}(\mathbb{H}) \Big)  \; + \; \e{O}(T^2) \;.
\label{ecriture dvpt basse tempe ordonne pour P de Y moins dominant state a}
\enq
This rewriting is appropriate for studying the minimising root configuration. $\msc{P}_0( \mathbb{Y}^{(\e{far})} )$
appearing above has been introduced in \eqref{definition mas P0}

Now $p_{\e{c}}(\la)=p(\la)$ if $|\Im(\la)| < \tf{\zeta}{2}$, and one can show that 
\beq
     \Im[p(\la)]>0 \quad \e{for} \quad  0\, <\, \Im(\la) \, < \,
        \f{\pi}{2} \;, \qquad \e{while} \qquad 
     \Im[p(\la)]<0 \quad \e{for} \quad  -\f{\pi}{2}  \, <\, \Im(\la) \, < \, 0 \;. 
\label{ecriture estimee sur partie imaginaie impuslion dans plan complexe}
\enq
Thus, indeed, as claimed in Proposition~\ref{Propostion infimum pour les vp de la QTM},
for $T$ low enough, $\Im\Big[\msc{P}_0( \mathbb{Y}^{(\e{far})}) \Big]$ attains its minimum for a
configuration of roots which satisfies
\beq
\mc{Y}^{(\e{fR})}\, = \, \mc{Y}^{(\e{fL})} \, = \, \mf{X}^{(\e{fR})}\, = \, \mf{X}^{(\e{fL})} \, = \, \emptyset \; .
\enq
On this class of solutions, the expression for $\mathbb{Y}^{(\e{far})}$
reduces to $\mathbb{Y}^{(\e{far})} \, = \,
\oplus_{\sg= \pm} \{ \sg q \}^{ \oplus \sg \ell^{(\sg)} }$ and
$\msc{P}_1(\mathbb{Y}^{(\e{far})} ) = 0$, so that, for $T$ low enough,
\begin{multline}
     \Im  \Big( \mc{P}( \mathbb{Y} )\, - \,\f{1}{T} \mc{P}_{-1} \Big) \; \geq \;
        \Im  \Big( \mc{P}( \mathbb{Y}_{\e{min}} )\, - \,\f{1}{T} \mc{P}_{-1} \Big) \\[1ex]
	\; = \; -  \f{\pi T }{ 6  \op{v}_F} 
     \; + \; \f{2 \pi T}{ \op{v}_F} \bigg\{Z_\e{c}^2(q) \big(\ell^{(-)} \big)^2   \, + \,   \f{ \mf{s}^2 }{4 Z_\e{c}^2(q)} \, + \,
        \sul{\sg=\pm }{} n_p^{(\sg)} n_h^{(\sg)} \bigg\} \; + \; \e{O}(T^2) \;.
\label{ecriture premiere borne inf sur impulsion}
\end{multline}
Here $ \mathbb{Y}_{\e{min}}$ corresponds to the configuration of roots $\mf{X}, \mc{Y}$
subordinate to the choice of particle-hole integers in
\eqref{equation pour part trous type +}, \eqref{equation pour part trous type -} as
encoded through \eqref{definition dist part trous CFT minimale}. It is the root
configuration that minimises the right hand side of
\eqref{ecriture premiere borne inf sur impulsion} for fixed given numbers $n_p^{(\sg)}$,
$n_h^{(\sg)}$ of particle and hole roots.

The terms in curly brackets on the right hand side of
\eqref{ecriture premiere borne inf sur impulsion} are all non-negative. The parameters
$\mf{s}$, $n_p^{(\sg)}$, $n_h^{(\sg)}$ are still free as long as they fulfil the
zero-monodromy constraint
$0 \, = \, - \mf{s} \, - \,  \big( n_p^{(+)} + n_p^{(-)} \big) \, + \,
\big( n_h^{(+)} + n_h^{(-)} \big)$, \textit{viz.}
$-\mf{s} \, = \, \ell^{(+)} \, - \, \ell^{(-)}$.

For the right hand side of \eqref{ecriture premiere borne inf sur impulsion} to be
minimal we must necessarily set $\mf{s} = 0$ and $\ell^{(-)} = 0$ and $n_p^{(\sg)} n_h^{(\sg)}=0$. The latter immediately imposes
that $\ell^{(+)}=0$ as well and thus $n_p^{(+)} = n_h^{(+)} = n_p^{(-)} = n_h^{(-)} = 0$. Hence, we
have shown that the unique admissible root configuration that minimises $\Im
\bigl(\mc{P}(\mathbb{Y})\bigr)$ in the low-$T$ limit is $\mf{X}=\mc{Y}=\emptyset$,
$\mf{s}=0$ for which
\beq
     \mc{P}( \emptyset ) \, = \, \f{1}{T} \mc{P}_{-1}
	-  \f{\i \pi T }{ 6  \op{v}_F} \; + \; \e{O}(T^2)
\enq
as stated in Proposition~\ref{Propostion infimum pour les vp de la QTM}.

%
%
%
%
%
%
%
%
%
%
%
%
%
%
%
%

\subsection{The low-$T$ expansion of $\mc{E}(\mathbb{Y})$}
Recall that  $\mc{E}(\mathbb{Y})$ has been defined in \eqref{definition mc E de Y}.
A direct application of Proposition~\ref{Proposition derssing of quantities in low-T limit}
along with the expansions \eqref{DA basse T des trous CFT}, \eqref{DA basse T des particules CFT}
leads to the expansion given in \eqref{ecriture dvpt basse tempe pour E de Y}.

We shall now slightly rewrite \eqref{ecriture dvpt basse tempe pour E de Y}
in the spirit of the previous subsection. This will allow us to identify
the contribution from  the root configuration $\mf{X}=\mc{Y}=\emptyset$
and $\mf{s}=0$ that is believed to characterize the dominant Eigenvalue.
Observe that, for the fully packed configuration introduced
in \eqref{definition dist part trous CFT minimale}, one obtains
\beq
     \vsg_{1}(\mathbb{H}_{\e{min}}) \; = \;
        \i \pi \sul{ \sg = \pm }{} \sg  \Big\{ (n_p^{(\sg)})^2 + (n_h^{(\sg)})^2 \Big\} \;. 
\enq
Then, upon using \eqref{ecriture dvpmt simplifie de u1 avec YM  non vide}, one
arrives at
\beq
     \mc{E}_{1}( \mathbb{Y}^{(\e{far})} ) \, + \,
        \vsg_{1}(\mathbb{H}) \; = \;
	\msc{E}_1( \mathbb{Y}^{(\e{far})} )  \, + \,  \Ups_{1}(\mathbb{H}) \;, 
\enq
where 
\begin{align}
     & \msc{E}_1(\mathbb{Y}^{(\e{far})} ) \, = \,
        \i\pi \sul{ \sg=\pm}{} \sg \,
	\Biggl(\frac{u^{(\sg)}\big(\mathbb{Y}^{(\e{far})} \big)}{2\i\pi} \, - \, 
     2   \ell^{(-)}  Z_\e{c}(q)  \, + \, \f{\sg \mf{s}}{Z_\e{c}(q)}\Biggr)
        \frac{  u^{(\sg)}\big( \mathbb{Y}^{(\e{far})} \big)}{2\i\pi} \;, \\[1ex]
      & \Ups_{1}(\mathbb{H}) \; = \;
         2 \i\pi \bigg\{- \mf{s} \ell^{(-)}  \, + \,
	    \sul{\sg=\pm }{} \sg \, n_p^{(\sg)} n_h^{(\sg)}  \bigg\} \; + \; 
      \vsg_{1}(\mathbb{H}) \, - \, \vsg_{1}(\mathbb{H}_{\e{min}}) \;.
\end{align}

Taken altogether, this yields the expansion 
\beq
     \mc{E}( \mathbb{Y} )\, - \, \mc{E}( \emptyset ) \, = \,
        \msc{E}_0( \mathbb{Y}^{(\e{far})} ) \; + \;
	T \Big( \msc{E}_1 ( \mathbb{Y}^{(\e{far})} ) + \Ups_{1}(\mathbb{H}) \Big)
	\; + \; \e{O}(T^2) \;,
\label{ecriture dvpt basse tempe ordonne pour P de Y moins dominant state b}
\enq
in which  
\beq
     \mc{E}( \emptyset ) \, = \, \f{1}{T} \mc{E}_{-1} \; + \; \e{O}(T^2)
\enq
is associated with the root configuration $\mf{X}=\mc{Y}=\emptyset$ and $\mf{s}=0$.

\subsection{The large-$N$ low-$T$ expansion of the ``norms'' and existence of Eigenvalues}
So far, we have produced a large class of solutions to the non-linear problem presented
in Proposition~\ref{Proposition description pblm non lineaire}. As stated in that Proposition,
each such solution naturally gives rise to a collection of admissible roots
$\la_1,\dots, \la_{N^{\prime}}$ solving the system of Bethe Ansatz equations.
This then allows one to build candidates for the Eigenvalues of the quantum transfer
matrix, in the sense that there exists a `Bethe vector'
$\bs{\Psi}\big( \la_1,\dots, \la_{N^{\prime}} \big)$, parameterised by the given set
of Bethe roots, such that
\beq
\op{t}_{\mf{q}}(\xi) \cdot \bs{\Psi}\big( \la_1,\dots, \la_{N^{\prime}} \big) \; = \;  \wh{\La}\,\big( \xi \,|\, \{\la_k\}_1^{N^{\prime}} \big) \cdot
\bs{\Psi}\big( \la_1,\dots, \la_{N^{\prime}} \big)  \;,
\enq
where $ \wh{\La}$ is given by \eqref{ecriture valeur propre qtm}. For
$\wh{\La}\,\big( \xi \,|\, \{\la_k\}_1^{N^{\prime}} \big)$ produced in this
way to be indeed an Eigenvalue of $\op{t}_{\mf{q}}(\xi)$, it is necessary that
$\bs{\Psi}\big( \la_1,\dots, \la_{N^{\prime}} \big) \not=0$. The non-vanishing
of the Bethe vector may be proved by constructing a vector $\bs{\Phi} \in \mf{h}_{\mf{q}}$,
with $\mf{h}_{\mf{q}}$ being the Hilbert space on which the quantum transfer
matrix acts, such that
\beq
\Big( \bs{\Phi} \, , \, \bs{\Psi}\big( \la_1,\dots, \la_{N^{\prime}} \big)  \Big) \; \not=\; 0 \;.
\label{equation PS entre etat Bethe et un autre etat avec recouvrement non nul}
\enq
The existence of such a vector will be the main result of this subsection.

\begin{prop}
\label{Proposition non nullite vecteur de Bethe via formule des normes}

There exists $T_0, \eta>0$ which only depend on the integers $\mf{s},  |\wh{\mc{Y}}^{(\a)}| , |\wh{\mf{X}}^{(\a)}|$ such that for any $0< \tf{1}{(NT^4)}<\eta $ and
$0<T<T_0$, the unique solution  $\Big( \, \wh{u}\, \big(\la \,|\, \wh{\mathbb{Y}}\,  \big) \, , \wh{\mf{X}} \, ,  \wh{\mc{Y}}  \Big)$
to the non-linear problem \eqref{ecriture NLIE a Trotter fini},  \eqref{ecriture monodromie Trotter fini}
and \eqref{ecriture conditions de quantifications Trotter fini} which enjoys  Hypotheses \ref{Hypotheses solubilite NLIE}-\ref{Hypotheses eqns de quantification}
and which solves the logarithmic quantisation conditions \eqref{ecriture conditions de quantification logarithmiques Trotter fini}
gives rise to a non-zero Bethe Eigenstate $ \bs{\Psi}\big( \la_1,\dots, \la_{N^{\prime}} \big)$ with $N^{\prime}=N-\mf{s}$.

\end{prop}

\Proof
By using the quantum inverse scattering method
\cite{FaddeevSklyaninTakhtajanSineGordonFieldModel}, given a set of roots
$\la_1,\dots, \la_{N^{\prime}}$ solving the Bethe Ansatz equations, one may
produce a vector  $\wt{\bs{\Psi}}\big( \la_1,\dots, \la_{N^{\prime}} \big)$
such that its scalar product with $ \bs{\Psi}\big( \la_1,\dots, \la_{N^{\prime}} \big)$
reduces to an expectation value of entries of the quantum monodromy matrix
\beq
  \Big( \wt{\bs{\Psi}}\big( \la_1,\dots, \la_{N^{\prime}} \big) \, , \, \bs{\Psi}\big( \la_1,\dots, \la_{N^{\prime}} \big) \Big) \; = \;
\Biggl( \bs{v}_{\e{ref}} \, , \, \pl{a=1}{N^{\prime} } \op{C}(\mu_a) \cdot  \pl{a=1}{N^{\prime} } \op{B}(\mu_a) \bs{v}_{\e{ref}} \Biggr) \;.
\label{ecriture PS dual et vecteur de Bethe comme Norme}
\enq
Here, we recall (see, \textit{e.g.},
\cite{GohmannKlumperSeelFinieTemperatureCorrelationFunctionsXXZ})
that the quantum monodromy matrix $\op{T}_{0,\mf{q}}(\xi)$ is an operator valued matrix
on an auxiliary space $\mf{h}_0\simeq \Cx^2$ whose entries are operators on the
Hilbert space $\mf{h}_{\mf{q}}$ associated with the quantum transfer matrix
\beq
     \op{T}_{0,\mf{q}}(\xi) \; = \;
        \begin{pmatrix}
	   \op{A}(\xi)  &  \op{B}(\xi)  \\
	   \op{C}(\xi)  &  \op{D}(\xi)
        \end{pmatrix}
	   \qquad \e{with} \qquad
     \e{tr}_{\mf{h}_0}\big[ \op{T}_{0,\mf{q}}(\xi)  \big] \; = \; \op{t}_{\mf{q}}(\xi) \;.
\enq
We further recall that $\big[\op{C}(\la), \op{C}(\mu)]=0=\big[\op{B}(\la), \op{B}(\mu)]$
so that the order in the products does not matter. Finally, $\bs{v}_{\e{ref}}$ is a
fixed reference vector which, in the case of the quantum transfer matrix, takes the form
\beq
     \bs{v}_{\e{ref}} \; = \;
        \bs{v}_{+}\otimes \bs{v}_{-} \otimes \bs{v}_{+} \otimes \cdots
	\otimes  \bs{v}_{-} \qquad \e{with} \qquad
     \bs{v}_{+} \; = \; \big(1 \,, 0 \big)^{\op{t}} \quad \e{and}
     \quad  \bs{v}_{-} \; = \; \big( 0 \,, 1 \big)^{\op{t}}  \;.
\enq

The main advantage of \eqref{ecriture PS dual et vecteur de Bethe comme Norme} is
that the expression appearing in the \textit{rhs} of the equality corresponds to the
so-called\symbolfootnote[3]{There exist quantum integrable models for which the formula gives the
actual norm of the Bethe state, but that is not the case for the  model under study.}
``norm'' of the Bethe state $\bs{\Psi}\big( \la_1,\dots, \la_{N^{\prime}} \big)$
for which there exist a closed expression. It is given in terms of a determinant
and some product prefactors. The closed formula was first conjectured in
\cite{GaudinMcCoyWuNormXXZ} and the conjecture was backed up by more convincing,
yet still heuristic, calculations in \cite{KorepinNormBetheStates6-Vertex}. The norm
formula was rigorously proven for the XXZ spin-$1/2$ transfer matrix in
\cite{KitanineMailletTerrasFormfactorsperiodicXXZ} and for all integrable
models subordinate to a six-vertex $R$ matrix, as is the case of the quantum
transfer matrix, in \cite{SlavnovScalarProducts6VertexProofBasedOnDualFields}.
See also \cite{KozKitMailSlaTer6VertexRMatrixMasterEquation} for another proof,
this time based on explicit algebraic resummations of combinatorial formulae.

The ``norm'' formula takes the explicit form
\bem
\mc{N}\big( \{\la_a\}_{ 1 }^{ N^{\prime} }\big) \; = \;
\Biggl( \bs{v}_{\e{ref}} \, , \, \pl{a=1}{N^{\prime} } \op{C}(\la_a) \cdot  \pl{a=1}{N^{\prime} } \op{B}(\la_a) \bs{v}_{\e{ref}} \Biggr) \\
\; = \; \pl{ a = 1 }{ N^{\prime} } \Biggl\{\f{ \wh{u}^{\, \prime}(\la_a\,|\,\wh{\mathbb{Y}})}{T} a(\la_a) d(\la_a)  \Biggr\} \cdot
\f{ \pl{a,b=1}{ N^{\prime} }  \sinh(\la_a-\la_b - \i\zeta) }{ \pl{a \not= b }{ N^{\prime} }  \sinh(\la_a-\la_b )  }
\cdot \e{det}_{N^{\prime} } \Biggl[ \de_{ab} \, + \, 2\pi \i T  \f{ K(\la_a-\la_b) }{  \wh{u}^{\, \prime}(\la_b\,|\,\wh{\mathbb{Y}}) } \Biggr]  \;.
\label{ecriture formule des normes vecteur de Bethe}
\end{multline}
Note that in case that there are roots with multiplicities higher than one, the
above formula should be understood as a limit when some $\la_a$ coalesce. The
well-definiteness of the limit follows from standard handlings with determinants,
see \textit{e.g.} \cite{DaleVeinDeterminantsAndAppsinMathPhys}.
Also, above,  $\wh{u}$ is defined in terms of the Bethe roots $\la_1,\dots, \la_{N^{\prime}}$
through \eqref{definiton hat u en terme racines de Bethe} while the functions $a$ and
$d$ take the explicit form
\begin{align}
a(\xi) & = \ex{ \f{h}{2T} } \Biggl\{ \f{ \sinh\big[\xi - \tf{\aleph}{N}-\i\tf{\zeta}{2} \big]  \sinh\big[\xi + \tf{\aleph}{N}+\i\tf{\zeta}{2} \big]  }
{\sinh^{2}(-\i\zeta)} \Biggr\}^N  \; , \\[1ex]
d(\xi) & = \ex{ -\f{h}{2T} } \Biggl\{ \f{ \sinh\big[\xi + \tf{\aleph}{N} + 3\i\tf{\zeta}{2} \big]  \sinh\big[\xi - \tf{\aleph}{N}+\i\tf{\zeta}{2} \big]  }
{\sinh^{2}(-\i\zeta)} \Biggr\}^N \; .
\end{align}
We remind that $\aleph$ is as introduced in \eqref{definition aleph}.
We now provide a rewriting of the formula which allows one to check that
\eqref{ecriture formule des normes vecteur de Bethe} is indeed non-zero.
First of all, a direct residue calculation, \textit{e.g.} on the level of
the Fredholm series representation for the discrete determinant, leads to the
representation
\beq
\e{det}_{N^{\prime} } \Biggl[ \de_{ab} \, + \, 2\pi \i T \f{ K(\la_a-\la_b) }{  \wh{u}^{\, \prime}(\la_b\,|\,\wh{\mathbb{Y}} ) } \Biggr] \; = \;
\underset{L^2\big( \wh{\msc{C}}_{\e{mod}} \big)}{\e{det}}\Big[ \e{id} \,  + \, \wh{\mc{K}}  \Big] \quad \e{with} \quad
\wh{\msc{C}}_{\e{mod}} \; = \; \wh{\msc{C}} \cup\Ga_{+}\big( \, \wh{\mc{Y}} \, \big) \cup \Ga_{-}\big( \, \wh{\mf{X}} \, \big) \;.
\label{ecriture determinant norme comme determinant de Fredholm}
\enq
Here $\wh{\msc{C}} = \wh{\msc{C}}_{\mathbb{Y}}\bigl[ \, \wh{u} \, \bigr]$,
$\Ga_{+}\big( \, \wh{\mc{Y}} \, \big)$ consists of tiny counterclockwise loops
around the pairwise distinct elements of $\wh{\mc{Y}}$, while
$\Ga_{-}\big( \, \wh{\mf{X}} \, \big)$ consists of tiny clockwise loops around
the pairwise distinct elements of $\wh{\mf{X}}$. These loops are such that they
do not encircle any other zeroes of $ 1\, + \, \ex{ \f{1}{T}\wh{u}(\mu\,|\,\wh{\mathbb{Y}})}$.
Finally, $\wh{\mc{K}}$ is the integral operator on $L^2\big( \wh{\msc{C}}_{\e{mod}} \big)$
acting with the integral kernel
\beq
     \mc{K}(\la,\mu) \; = \;
        \f{-K(\la-\mu)}{1\, + \, \ex{ \f{1}{T}\wh{u}(\mu\,|\,\wh{\mathbb{Y}})}  }  \;.
\enq

The Fredholm determinant appearing in
\eqref{ecriture determinant norme comme determinant de Fredholm} may be
recast in a form more convenient for our purposes. First of all, one observes that
$\e{id} \,  + \, \wh{\mc{K}}$ is invertible on $L^2\big( \wh{\msc{C}}\; \big)$.
Indeed, the operator has a smooth kernel and the contour $\wh{\msc{C}}$ is compact
so that \cite{DudleyGonzalesBarriosMetricConditionForOpToBeTraceClass}
ensures that $ \wh{\mc{K}}$ is trace class. The invertibility of the operator will
then follow from the non-vanishing of its Fredholm determinant which is well
defined.

Observe that, for $\mu \in \wh{\msc{C}}$, one may decompose the operator's
integral kernel as
\beq
\mc{K}(\la,\mu) \; = \; \mc{K}_0(\la,\mu) \, + \,  \de\mc{K}(\la,\mu) \quad \e{where} \quad
\mc{K}_0(\la,\mu)  \; = \; - K(\la-\mu) \bs{1}_{ \wh{\msc{C}}^{\, (-)} }(\mu)\;,
\label{ecriture decomposition asymptotique noyau hat cal K}
\enq
while the perturbing kernel takes the explicit form
\beq
\de\mc{K}(\la,\mu)  \; = \;  -   K(\la-\mu)  \sul{ \sg= \pm  }{}   \f{ - \sg  \bs{1}_{ \wh{\msc{C}}^{(\sg)} }(\mu)  }
 {1\, + \, \ex{ \f{1}{T} |\; \! \wh{u}\; \!|( \mu  \,|\, \wh{\mathbb{Y}})}} \;.
\enq
Above, $\wh{\msc{C}}^{\, (-)}$ is the ``upper'' part of the contour $\wh{\msc{C}}$
joining $\wh{q}_{\mathbb{Y}}^{\, (+)}[\, \wh{u} \, ]$ to
$\wh{q}_{\mathbb{Y}}^{\, (-)}[\, \wh{u} \, ]$, which are the unique solutions to
\beq
\wh{u}\Big(  \, \wh{q}_{\mathbb{Y}}^{\, (\pm)}[\, \wh{u} \, ] \,|\, \wh{\mathbb{Y}} \Big) \; = \; 0
\enq
located in the neighbourhood of $\pm q$. Similarly, $\wh{\msc{C}}^{\, (+)}$ is the ``lower''
part of the contour $\wh{\msc{C}}$ joining $\wh{q}_{\mathbb{Y}}^{\, (-)}[\, \wh{u} \, ]$
to $\wh{q}_{\mathbb{Y}}^{\, (+)}[\, \wh{u} \, ]$.

It is direct to see, by using estimates analogous to those leading to
\eqref{ecriture DA integrale avec u lisse} of
Lemma~\ref{Lemme DA fonction holomorphe versis log}, that
\beq
     \e{tr}\Big[\wh{\de\mc{K}}\Big] \; = \;
        \Int{ \wh{\msc{C}} }{} \hspace{-1mm} \dd \la \:
	   \de\mc{K}(\la,\la)  \limit{T}{0^+} 0 \qquad \e{and} \qquad
     \norm{\, \wh{\de\mc{K}} \, }_{\e{HS}} \; = \;
        \norm{\, \de\mc{K}\,}_{L^2 \big( \wh{\msc{C}} \times \wh{\msc{C}} \big) }
	\limit{T}{0^+} 0 \;,
\enq
where one keeps $0 < \tf{1}{(NT^4)} <\eta$ with $\eta>0$ and small enough. Thus, since
\beq
\underset{L^2\big( \wh{\msc{C}}\,  \big)}{\e{det}}\Big[ \e{id} \,  + \, \wh{\mc{K}}  \Big] \; = \; \ex{ \e{tr} [  \wh{ \mc{K}}  ]  } \cdot
\underset{L^2\big( \wh{\msc{C}}  \, \big)}{\e{det}_2}\Big[ \e{id} \,  + \, \wh{\mc{K}}  \Big] \;,
\enq
where $\det_{2}$ is the $2$-determinant defined for Hilbert-Schmidt
operators \cite{GohbergGoldbargKrupnikTracesAndDeterminants}, and
owing to the Lipschitz continuity of 2-determinants of Hilbert-Schmidt
operators \cite{GohbergGoldbargKrupnikTracesAndDeterminants},
\beq
\Big| \e{det}_2\big[\e{id} \, + \, \op{A} \big] \, - \, \e{det}_2\big[\e{id} \, + \, \op{B} \big] \Big| \; \leq \;
\norm{\op{A}-\op{B}}_{\e{HS}} \cdot \ex{ C \big( \norm{\op{A} }_{\e{HS}} + \norm{\op{B} }_{\e{HS}} + 1 \big)^2  } \;,
\enq
one obtains that
\beq
\underset{L^2\big( \wh{\msc{C}}\, \big)}{\e{det}}\Big[ \e{id} \,  + \, \wh{\mc{K}}  \Big]  \, - \,
\underset{L^2\big( \wh{\msc{C}} \, \big)}{\e{det}}\Big[ \e{id} \,  + \, \wh{\mc{K}}_0  \Big]  \; = \; \e{o}(1)
\enq
as $T\tend 0^+$, while $0 < \tf{1}{(NT^4)} <\eta$ with $\eta>0$.

Direct contour deformations in the Fredholm series representation
for $\e{det}\big[ \e{id}+\wh{\mc{K}}_0\big] $ allow one to infer that
\beq
\underset{L^2\big( \wh{\msc{C}}  \, \big)}{\e{det}}\Big[ \e{id} \,  + \, \wh{\mc{K}}_0  \Big]  \; = \;
\underset{L^2\big( \wh{\mc{I}}  \big)}{\e{det}}\Big[ \e{id} \,  + \, \op{K}  \Big] \;,
\quad \e{where} \quad  \wh{\mc{I}} \; = \; \big[ \, \wh{q}_{\mathbb{Y}}^{\, (-)}[\, \wh{u} \, ] \, ; \, \wh{q}_{\mathbb{Y}}^{\, (+)}[\, \wh{u} \, ] \, \big]\; ,
\label{ecriture egalite determinants et definition intervalle hat I}
\enq
while
$\op{K}$ acts on $L^2\big( \wh{\mc{I}}  \big)$ with integral kernel $K(\la-\mu)$. Recall that
\beq
\wh{q}_{\mathbb{Y}}^{\, (\pm)}[\, \wh{u} \, ] \; = \;  \pm q \; + \; \e{O}\Big(T \, + \, \f{ 1 }{  NT^4  } \Big) \;,
\enq
so that these endpoints remain bounded when $T, \tf{ 1 }{  (NT^4)  } $ are small enough.
Moreover, it is well-known that $\underset{ L^2( J ) }{ \e{det} }\Big[ \e{id} \,  +
\, \op{K}  \Big] \not= 0$ for any segment $J$ of $\R$, see \textit{e.g.}\
\cite{KozProofOfDensityOfBetheRoots}, and that it is a continuous function of the
endpoints of $J$. This entails that
$\underset{L^2\big( \wh{\msc{C}}\, \big)}{\e{det}}\Big[ \e{id} \,  + \, \wh{\mc{K}}  \Big]$
is uniformly away from $0$ and hence that $ \e{id} \,  + \, \wh{\mc{K}} $ is invertible
provided that $T$ and $\tf{ 1 }{  (NT^4)  }$ are small enough.

Upon computing the contour integrals, one conludes that the determinant of
$ \e{id} \,  + \, \wh{\mc{K}}$ on $ L^2\big( \wh{\msc{C}}_{\e{mod}} \big)$
is equal to the determinant of the block matrix operator on
$L^2\big( \wh{\msc{C}}\; \big) \oplus \Cx^{|\wh{\mc{Y}}|} \oplus \Cx^{|\wh{\mf{X}}|}$
with integral kernel
\beq
\begin{pmatrix} \de(\la-\mu) \,  + \, \mc{K}(\la,\mu) & \f{ 2\pi \i T K(\la-y_b) }{ \wh{u}^{\, \prime}\big( y_b \,|\, \wh{\mathbb{Y}} \big)  }
                                                                                          & -  \f{2\pi \i T K(\la-x_b) }{ \wh{u}^{\, \prime}\big( x_b \,|\, \wh{\mathbb{Y}} \big)  } \\[3ex]
                    \mc{K}(y_a,\mu)      & \de_{ab} \, + \,   \f{2\pi \i T K(y_a - y_b) }{ \wh{u}^{\, \prime}\big( y_b \,|\, \wh{\mathbb{Y}} \big)  }
                                                                                       & -  \f{2\pi \i T K(y_a-x_b) }{ \wh{u}^{\, \prime}\big( x_b \,|\, \wh{\mathbb{Y}} \big)  } \\[3ex]
                    \mc{K}(x_a,\mu)   &    \f{2\pi \i T K(x_a - y_b) }{ \wh{u}^{\, \prime}\big( y_b \,|\, \wh{\mathbb{Y}} \big) }
                                                                              &    \de_{ab}   -  \f{ 2\pi \i T K(x_a-x_b) }{ \wh{u}^{\, \prime}\big( x_b \,|\, \wh{\mathbb{Y}} \big)  }
\end{pmatrix} \;.
\enq
Here we made use of the formal $\de$-function representation for the integral kernel of
the identity operator. $\la, \mu \in \wh{\msc{C}}$, the indices associated with particle
roots run through $\intn{1}{|\wh{\mc{Y}}|}$, while those associated with hole roots
run through $\intn{1}{|\wh{\mf{X}}|}$. Then, recalling the formula for block determinants
\beq
\det \begin{bmatrix} A & B \\ C & D \end{bmatrix}
\; = \; \det\big[ A \big] \cdot \det\big[ D \, - \, C A^{-1} B \big]
\enq
and using that the inverse operator $\e{id}-\wh{\mc{R}}$ to $\e{id}+\wh{\mc{K}}$
has its resolvent satisfying $\wh{\mc{R}} \cdot \wh{\mc{K}} \, = \,
\wh{\mc{K}} - \wh{\mc{R}}$, one gets the identity
\beq
\underset{ L^2\big( \wh{\msc{C}}_{\e{mod}} \big)}{\e{det}}\Big[ \e{id} \,  + \, \wh{\mc{K}}  \Big]
\; = \; \underset{L^2\big( \wh{\msc{C}}\, \big)}{\e{det}}\Big[ \e{id} \,  + \, \wh{\mc{K}}  \Big] \cdot
\det\big[ \mc{M} \big] \;,
\label{ecriture factorisation Fdet sur Cmod en F det et det fini}
\enq
where $\mc{M}$ is a $\big( |\wh{\mc{Y}}| + |\wh{\mf{X}}| \big) \times \big( |\wh{\mc{Y}}| + |\wh{\mf{X}}| \big)$ matrix given by
\beq
\mc{M} \; = \; \begin{pmatrix} \de_{ab} \, + \,   \f{2\pi \i T \mc{R}(y_a , y_b) }{ \wh{u}^{\, \prime}\big( y_b \,|\, \wh{\mathbb{Y}} \big)  }
                                                                           & -  \f{2\pi \i T \mc{R}(y_a , x_b) }{ \wh{u}^{\, \prime}\big( x_b \,|\, \wh{\mathbb{Y}} \big)  }  \\[3ex]
                \f{2\pi \i T \mc{R}(x_a , y_b) }{ \wh{u}^{\, \prime}\big( y_b \,|\, \wh{\mathbb{Y}} \big)  }
                                   & \de_{ab} \, - \,   \f{2\pi \i T \mc{R}(x_a , x_b) }{ \wh{u}^{\, \prime}\big( x_b \,|\, \wh{\mathbb{Y}} \big)  }     \end{pmatrix} \;.
\enq

We already established that the Fredholm determinant appearing in the \textit{rhs}
of \eqref{ecriture factorisation Fdet sur Cmod en F det et det fini} does not
vanish. Thus, in order to get the non-vanishing of the original determinant appearing
in the \textit{lhs} of \eqref{ecriture factorisation Fdet sur Cmod en F det et det fini},
it remains to show that $\mc{M}$ is invertible. Taken that the particle and hole roots
are assumed to be uniformly away from $\pm \i\tf{\zeta}{2}$, and that for
$T, \tf{1}{(NT^4)}$ small $\wh{u}^{\,  \prime}(\la\,|\,\wh{\mathbb{Y}} \big)$
is uniformly away from $0$ as long as one stays in some sufficiently small
open neighbourhood of the curve $\Re[\veps_{\e{c}}]=0$, the invertibility of $\mc{M}$ will follow,
for $T$ small enough, as soon as we establish that the resolvent kernel $\wh{\mc{R}}(\la , \mu)$
is bounded on a set containing the particle and hole roots, \textit{e.g.}
\beq
\Big\{ \la \; : \; | \Im (\la) | \leq \tf{\zeta}{2}-\eps \, \Big\} \times \Big\{ \mu \; : \; | \Im (\mu) | \leq \tf{\zeta}{2}-\eps \, \Big\}
\label{ecriture ensemble de bornage pour le resolvent hat cal R}
\enq
for some $\eps>0$ but small enough, this uniformly in $T, \tf{1}{(NT^4)}$ small enough.

Upon inserting the expansion \eqref{ecriture decomposition asymptotique noyau hat cal K}
into the linear integral equation for the resolvent kernel $\mc{R}$
and then  deforming the integration along $\wh{\msc{C}}^{\, (-)}$ in the action of
$\wh{\mc{K}}_0$ to $\wh{\mc{I}}$,
\textit{c.f.}\ \eqref{ecriture egalite determinants et definition intervalle hat I},
while taking into account the change of orientation, one obtains 
\beq
K(\la-\mu) \; = \; \mc{R}(\la,\mu) \; +  \hspace{-2mm}  
\Int{ \wh{q}_{\mathbb{Y}}^{\, (-)}[\, \wh{u} \, ] }{ \wh{q}_{\mathbb{Y}}^{\, (+)}[\, \wh{u} \, ] } \hspace{-3mm} \dd \nu \: K(\la-\nu) \mc{R}(\nu,\mu)
\, + \, \Int{ \wh{\msc{C}} }{}   \dd \nu \, \de \mc{K}(\la,\nu)  \mc{R}(\nu,\mu) \;.
\enq
Upon inverting the operator $\e{id} + \op{K}$ acting on $L^2\big(\, \wh{\mc{I}} \, \big)$
and denoting its resolvent kernel by $\mc{L}$, one gets a linear integral
equation satisfied by $\mc{R}$,
\beq
\mc{L}(\la,\mu) \; = \; \mc{R}(\la,\mu)
\, + \, \Int{ \wh{\msc{C}} }{}   \dd \nu \: {\de \mc{K}}_{\e{mod}}(\la,\nu)  \mc{R}(\nu,\mu) \;,
\enq
in which
\beq
{\de\mc{K}}_{\e{mod}}(\la,\mu)  \; = \;  -   \mc{L}(\la,\mu)  \sul{ \sg= \pm  }{}   \f{ - \sg  \bs{1}_{ \wh{\msc{C}}^{(\sg)} }(\mu)  }
 {1\, + \, \ex{ \f{1}{T} |\;\!\wh{u} \;\! |( \mu \,|\, \wh{\mathbb{Y}})}} \;.
\enq

Recall that $\mc{L}$ admits the series representation
\cite{GohbergGoldbargKrupnikTracesAndDeterminants}
\beq
\mc{L}(\la,\mu) \; = \; \f{ 1 }{ \underset{L^2\big( \wh{\mc{I}}  \big)}{\e{det}}\Big[ \e{id} \,  + \, \op{K}  \Big] }
\cdot \sul{n \geq 0  }{} \f{1}{(n+1)!} \Int{  \wh{\mc{I}}   }{} \dd^n \nu \:
\det_{n+1}
\begin{bmatrix}  K(\la-\mu)& K(\la-\nu_1)&\cdots & K(\la-\nu_n) \\
                    K(\nu_1-\mu) &  K(\nu_1-\nu_1) & \cdots & K(\nu_1-\nu_n) \\
                    \vdots &  \vdots &      &  \vdots  \\
                      K(\nu_n-\mu) &  K(\nu_n-\nu_1) & \cdots & K(\nu_n-\nu_n)
		      \end{bmatrix} \;.
\label{ecriture serie de Fredholm pour resolvent hat L}
\enq
The latter entails that $\mc{L}$ is bounded and holomorphic on
\eqref{ecriture ensemble de bornage pour le resolvent hat cal R}.
Finally, denoting $\de \mc{L}_{\e{mod}}(\la,\nu)$ the resolvent kernel of
$\e{id} + \wh{\de \mc{K}}_{\e{mod}}$ on $L^2\big(\, \wh{\msc{C}}\; \big)$,
one obtains an integral representation for $\mc{R}$ in the form
\beq
\mc{R}(\la,\mu) \; = \; \mc{L}(\la,\mu)
\, - \, \Int{ \wh{\msc{C}} }{}   \dd \nu \: {\de \mc{L}}_{\e{mod}}(\la,\nu) \mc{L}(\nu,\mu) \;.
\enq
Now ${\de \mc{L}}_{\e{mod}}(\la,\mu)$ admits a series expansion similar
to \eqref{ecriture serie de Fredholm pour resolvent hat L}, with integrals over
$\wh{\msc{C}}$, which allows one to infer that
\beq
{\de \mc{L}}_{\e{mod}}(\la,\mu) \; = \;
- \mc{H}(\la,\mu) \sul{\sg = \pm}{} \f{\sg \bs{1}_{ \wh{\msc{C}}^{(\sg)} }(\mu)}
{1\, + \, \ex{ \f{1}{T} | \; \! \wh{u} \; \! |( \mu \,|\, \wh{\mathbb{Y}})}}
\enq
for some function $\mc{H}$, bounded and holomorphic
on \eqref{ecriture ensemble de bornage pour le resolvent hat cal R}.

By repeating the reasoning leading to the expansion \eqref{ecriture DA integrale avec u lisse}
given in Lemma~\ref{Lemme DA fonction holomorphe versis log}, one infers that
$\mc{R}(\la,\mu) \; = \; \mc{L}(\la,\mu) \, + \, \e{O}(T^2)$ uniformly
on \eqref{ecriture ensemble de bornage pour le resolvent hat cal R} and
in $T, \tf{1}{(NT^4)}$ small. It thus holds that $|\det[\mc{M}]|>c$ uniformly in
$T, \tf{1}{(NT^4)}$ small. All-in-all we have established that in this regime of $N$ and $T$
\beq
\e{det}_{ N^{\prime} } \Biggl[ \de_{ab} \, + \, 2\pi \i T
\f{ K(\la_a-\la_b) }{  \wh{u}^{\, \prime}(\la_b\,|\,\wh{\mathbb{Y}}) } \Biggr] \; = \;
\underset{ L^2\big( \wh{\msc{C}}\, \big) }{ \e{det} }\Big[ \e{id} \,  + \, \wh{\mc{K}}  \Big] \cdot
\det\big[ \mc{M} \big]
\enq
with each of the two terms appearing in the \textit{rhs} of this formula
being uniformly away from zero.

By following the techniques described \textit{e.g.}\
in \cite{KozDugaveGohmannThermaxFormFactorsXXZ}, one may establish that
\beq
\pl{ a = 1 }{ N^{\prime} } \Biggl\{\f{ \wh{u}^{\, \prime}\big( \la_a\,|\,\mathbb{Y} \big) }{ T } a(\la_a) d(\la_a) \Biggr\} \cdot
 \f{ \pl{a,b=1}{ N^{\prime} }  \sinh(\la_a-\la_b - \i\zeta) }{ \pl{a \not= b }{ N^{\prime} }  \sinh(\la_a-\la_b )  } \; = \; \op{c}_N \cdot
\mc{W}\big( \wh{\mf{X}}, \wh{\mc{Y}} \big) \cdot \ex{  \wh{\vp}_2} \cdot \pl{ y \in \wh{\mathbb{Y}} }{} \ex{\wh{\vp}_1(y)} \;.
\label{ecriture produit singulier comme contributions integrales}
\enq
Above,
\bem
\mc{W}\big( \wh{\mf{X}}, \wh{\mc{Y}} \big) \; = \;
(-1)^{|\wh{\mf{X}}|} \big[ \sinh(-\i\zeta) \big]^{ |\wh{\mf{X}}| + |\wh{\mc{Y}}|} \cdot
\pl{y \in \wh{\mc{Y}}}{}
\f{ \wh{u}^{\, \prime}\bigl(y\,|\,\mathbb{Y} \bigr) }{ T } \cdot \pl{x\in\wh{\mf{X}}}{}
\f{ \wh{u}^{\, \prime}\bigl(x\,|\,\mathbb{Y} \bigr) }{ T } \cdot \\
\times
\pl{ \substack{ y \not= y^{\prime}  \\ y, y^{\prime}\in \wh{\mc{Y}} } }{} \f{  \sinh(y-y^{\prime}-\i\zeta)  }{  \sinh(y-y^{\prime})  }
\cdot \pl{ \substack{ x \not= x^{\prime}  \\ x, x^{\prime}\in \wh{\mf{X}} } }{} \f{  \sinh(x-x^{\prime}-\i\zeta)  }{  \sinh(x-x^{\prime})  }
\cdot \pl{ \substack{ y \in \wh{\mc{Y}} \\ x \in \wh{\mf{X}}  } }{} \f{  \sinh(y-x) \sinh(x-y)  }{  \sinh(y-x-\i\zeta) \sinh(x-y-\i\zeta)  } \;.
\end{multline}
The remaining building blocks are defined in terms of the auxiliary function
\beq
\wh{\msc{L}}(\la) \; = \;
\f{1}{2\i\pi} \msc{L}\e{n}_{ \wh{\msc{C}} } \;
\Big[ 1\, + \, \ex{-\f{1}{T}\wh{u}(* \,|\, \wh{\mathbb{Y}})} \Big](\la)
\qquad  \e{with}  \qquad \wh{\msc{C}} = \wh{\msc{C}}_{\mathbb{Y}}\big[ \, \wh{u} \, \big]\;.
\label{definition compacte pour le logarithme et pour le ctr integration}
\enq
One has
\begin{align}
\wh{\vp}_1(y) & = \Int{  \wh{\msc{C}}  }{} \dd \la \: \wh{\msc{L}}(\la)\sul{\sg=\pm}{} \Big\{  \coth(y-\la-\i\sg \zeta) \, - \, \coth (y-\la)  \Big\} \;, \\
 \wh{\vp}_2 & = - \Int{  \wh{\msc{C}}  }{} \!  \dd \la \: \wh{\msc{L}}(\la)  \hspace{-2mm}
\Int{\wh{\msc{C}}^{\,\! \prime} \subset \wh{\msc{C}}}{} \hspace{-3mm} \dd \mu \:
\wh{\msc{L}}(\mu) \,
 \Big\{  \coth^{\prime}(\mu-\la-\i \zeta) \, - \, \coth^{\prime} (\mu - \la)  \Big\} \;.
\end{align}
In the second integral $ \wh{\msc{C}}^{\, \!\prime} $ corresponds to a slight
deformation of the contour $\wh{\msc{C}}$ in such a way that no poles of 
$\Dp{\la}\wh{\msc{L}}(\la)$ are crossed and such that the deformed contour
stays completely inside of the original contour. Finally,
\bem
\op{c}_N  \; = \; \pl{a=1}{N^{\prime}}  \bigg\{ \f{ a(\la_a) d(\la_a) }{ \big[ \sinh(\la_a + \tf{\aleph}{N} + \i\tf{\zeta}{2}) \big]^N   } \bigg\} \cdot
\pl{a=1}{ N} \Big\{ \sinh(\mu_a + \tf{\aleph}{N} -\i\tf{\zeta}{2})    \Big\}^{N}\\
\times \pl{ y \in \wh{\mathbb{Y}} }{} \bigg\{ \f{ \sinh(y + \tf{\aleph}{N} +3\i\tf{\zeta}{2}) \sinh(y + \tf{\aleph}{N} - \i\tf{\zeta}{2})  }
                                                                      { \sinh(y + \tf{\aleph}{N} + \i\tf{\zeta}{2})  }  \bigg\}^{N} \\
\times \exp\bigg\{- N \Int{  \wh{\msc{C}}  }{} \dd \la \: \wh{\msc{L}}(\la) \,
\Big\{  \coth(\la +\tf{\aleph}{N} + \tf{3 \i \zeta}{2}) \, - \,   \coth(\la +\tf{\aleph}{N} + \tf{ \i \zeta}{2}) \Big\}  \bigg\} \;.
\end{multline}
Above, we have made use of the set $\{\mu_a\}_1^N$ of zeroes of $1+\ex{ - \f{1}{T} \wh{u}(s \,|\, \wh{\mathbb{Y}} )  }$
located inside of $\wh{\msc{C}}$, \textit{viz}. $\{\mu_a\}_1^N \, = \, \{\la_a\}_1^{N^{\prime}}\ominus \wh{\mathbb{Y}} $.

We now discuss the finiteness and boundedness away from zero of the
building blocks appearing in the \textit{rhs} of
\eqref{ecriture produit singulier comme contributions integrales}.
According to Proposition~\ref{Proposition description pblm non lineaire},
$\wh{u}(\mu\,|\,\wh{\mathbb{Y}}) $ gives rise to an admissible solution
of the Bethe Ansatz equations. In particular, we have
\beq
\la_a\not \in \Biggl\{  \pm \f{\aleph}{N }- \i \f{\zeta}{2} \, , \,\f{\aleph}{N } +  \i \f{\zeta}{2} \, , \, -\f{\aleph}{N }-  3\i \f{\zeta}{2} \Biggr\} \:.
\enq
Hence, the first product over the Bethe roots in $\op{c}_N$ does not vanish.
The second product involving the roots $\mu_a$ does not vanish since, by
construction of $\wh{\msc{C}}$, for any $\eps>0$,
there exist $\eta>0$  and $T_0>0$ small enough such that for $\tf{1}{NT^4} < \eta$ and $0<T<T_0$
and any $a\in \intn{1}{N}$, it holds that $-\tf{\zeta}{2}-\eps \leq \Im(\mu_a) \leq \eps$.
Likewise, since, for any $s \in \wh{\mc{Y}}\cup \wh{\mf{X}}$,
it holds that $-\tf{\zeta}{2}+\eps \leq \Im(s) \leq \tf{\zeta}{2}-\eps$, the
product over $y \in  \wh{\mathbb{Y}}$ is also non-zero and finite.
Finally, $\wh{\msc{L}} \in L^1\cap L^2\big(\, \wh{\msc{C}} \, \big)$ as well as
\beq
s \mapsto \coth\Bigl( s + \tfrac{\aleph}{N}  \, + \,  \i \tfrac{\zeta}{2} \, + \, \i \tfrac{\zeta}{2}(1\pm 1) \Bigr) \;.
\enq
In total, this entails the finiteness of the integral term entering the definition
of $\op{c}_N$. The term $\wh{\mc{W}}\big( \wh{\mf{X}} \, , \, \wh{\mc{Y}} \big)$ as well
is finite and away from zero as can be seen from the allowed range of the imaginary
parts of the particle and hole roots, from the repulsion principle for the roots,
point e) of Hypotheses \ref{Hypotheses eqns de quantification}, and from the fact that all roots
have multiplicity one.  The finiteness of the integrals defining $\wh{\vp}_1(y)$
and $\wh{\vp}_2$ follows from similar arguments.

To sum-up, we have established that
\beq
\mc{N}\big( \{\la_a\}_{ 1 }^{ N^{\prime} }\big) \; = \; \op{c}_N \cdot
\mc{W}\big( \wh{\mf{X}}, \wh{\mc{Y}} \big) \cdot
\ex{ \wh{\vp}_2} \cdot \pl{ y \in \wh{\mathbb{Y}} }{} \ex{\wh{\vp}_1(y)}  \cdot
\underset{L^2\big(\wh{\msc{C}}\, \big)}{\e{det}}\Big[ \e{id} \,  + \, \wh{\mc{K}}  \Big]
\cdot \det\big[ \mc{M} \big]
\enq
with each term being finite and bounded away from zero. This entails the claim. \qed

\subsection{The dominant Eigenvalue}

In this subsection we prove the two Theorems stated in the first part of the introduction.
We start by proving the explicit characterisation of $\wh{\La}_{\e{max};\e{BA}}$,
and then we prove the conditional Theorem giving a closed expression for the
free energy of the XXZ chain.

\subsubsection{Proof of Theorem \ref{Theorem form close La max BA}}

First of all, by virtue of Theorem \ref{Theorem ppl existance solution NLIE},
specialised to the case $\bs{\op{X}}=\bs{\op{Y}}=\emptyset$ and $\mf{s}=0$,
one gets that there exist  $\eta>0$, $T_0>0$ and $C_{\mc{M}}^{(0)}>0$, such
that for any
\beq
T_0 > T > 0 \;  , \quad  \eta > \f{1}{N T^4} \quad \e{and} \quad  C_{\mc{M}} \, >\,  C_{\mc{M}}^{(0)}
\enq
the non-linear integral equation \eqref{ecriture NLIE Trotter fini VP dominante}
is well defined on the space $\wh{\mc{E}}_{\mc{M}}$ introduced in \eqref{definition espace EM hat}
and admits a unique solution $\wh{u}_{\e{max}}$ subject to the null-index condition
\eqref{ecriture null index condition}. Moreover, Theorem~\ref{Theorem existence et unicite sols NLIE}
ensures that the solution is analytic in $\zeta \in \intof{0}{\tf{\pi}{2}}$, pointwise in $\xi$.
The fact that $\wh{u}_{\e{max}}$ allows one to construct a non-vanishing Bethe Eigenvector
of the quantum transfer matrix $\op{t}_{\mf{q}}(\xi)$ is then ensured by
Proposition~\ref{Proposition non nullite vecteur de Bethe via formule des normes}.
The form of the associated Eigenvalue \eqref{ecriture Trotter fini VP dominantes BA}
follows from \eqref{ecriture forme vp QTM}, specialised to the case
$\wh{\mc{Y}}\, =\, \wh{\mf{X}} \, = \, \emptyset$. The obtained Eigenvalue is
also analytic in $\zeta \in \intof{0}{\tf{\pi}{2}}$, as readily inferred from
the $\zeta$-analyticity of $\wh{u}_{\e{max}}$ and the very form of the integral
representation \eqref{ecriture Trotter fini VP dominantes BA}.

\vspace{2mm}

We now turn to the maximality statement. Thus, we consider integers  $h_a^{(\a)}, p_a^{(\a)}$
satisfying the constraints \eqref{ecriture suite croissante entiers p et h bornee en T}-\eqref{ecriture limitation sur nombre entiers particle et trou}.
Pick $T_0, \eta>0$ small enough as in
Proposition~\ref{Proposition non nullite vecteur de Bethe via formule des normes}. Then,
for any $0< \tf{1}{(NT^4)}<\eta $ and $0<T<T_0$, consider the unique solution
$\Big( \, \wh{u}\, \big(\la \,|\, \wh{\mathbb{Y}}\,  \big) \, , \wh{\mf{X}} \, ,  \wh{\mc{Y}}  \Big)$
to the non-linear problem \eqref{ecriture NLIE a Trotter fini},
\eqref{ecriture monodromie Trotter fini}, associated with the choice
of integers $h_a^{(\a)}, p_a^{(\a)}$. The parameters building up $\wh{\mf{X}}$
and $\wh{\mc{Y}}$ solve the logarithmic quantisation conditions
\eqref{ecriture conditions de quantification logarithmiques Trotter fini}.
The solution of interest produces a non-zero Bethe Eigenstate by virtue of
Proposition~\ref{Proposition non nullite vecteur de Bethe via formule des normes}.
We remind the reader that, by construction, the associated roots do enjoy
Hypotheses \ref{Hypotheses solubilite NLIE}-\ref{Hypotheses eqns de quantification}.
Analogously, consider the unique solution
$\Big(u \big(\la \,|\, \mathbb{Y}\,  \big) \, , \mf{X} \, ,  \mc{Y}  \Big)$
to the non-linear problem \eqref{ecriture NLIE a Trotter infini},
\eqref{ecriture monodromie Trotter infini} and
\eqref{ecriture condition de quantification Trotter infini}
subordinate to the same choice of integers $h_a^{(\a)}, p_a^{(\a)}$.

Then, by virtue of \eqref{ecriture DA grand N des pts sol de NLIE pblm} in
Theorem~\ref{Theorem existence solutions racines particules et trous}, and
the estimate \eqref{ecriture deviation entre point fixe u et hat u} of
Theorem~\ref{Theorem existence et unicite sols NLIE}, it holds that
\beq
\norm{ \wh{u}\, \big(* \,|\, \wh{\mathbb{Y}}\,  \big) - u \big(* \,|\, \mathbb{Y}\,  \big)   }_{L^{\infty}(\mc{M}) } \; \leq \;  \f{ C }{ N T^3} \;.
\enq
Above, $C$ is uniform with respect to the choice of integers $h_a^{(\a)}, p_a^{(\a)}$
satisfying the constraints \eqref{ecriture suite croissante entiers p et h bornee en T}-\eqref{ecriture limitation sur nombre entiers particle et trou}.

Now, one goes back to the set of particle-hole parameters relative to
$\msc{C}_{\e{ref}}$, \textit{c.f.}\ Section~\ref{Section NLIE equivalente et contour Cu}.
Note that the cardinalities of the new sets $\wh{\mc{Y}}_{\e{ref}}, \wh{\mf{X}}_{\e{ref}}$
differ from those of $\wh{\mf{X}}, \wh{\mc{Y}}$ only by a finite
amount, this uniformly in the choices of integers $h_a^{(\a)}, p_a^{(\a)}$
satisfying the constraints \eqref{ecriture suite croissante entiers p et h bornee en T}-%
\eqref{ecriture limitation sur nombre entiers particle et trou}.
Then, by putting all of the above together, and upon using the explicit form of the integral
representation \eqref{ecriture forme vp QTM}, one obtains
\beq
 \ln \Big\{ \wh{\La}\, \Big( 0 \big|\, \wh{u}(*\,|\, \wh{\mathbb{Y}}_{\e{ref}; \varkappa} ), \wh{\mc{Y}}_{\e{ref}}, \wh{\mf{X}}_{\e{ref}} \Big)  \Big\} \, = \,
 -\i\pi \mf{s}  \, + \, \f{h}{2T}   \, - \, \f{2J}{T}\cos(\zeta)  \, + \,  \i \mc{P}(\mathbb{Y})  \, + \, \e{O}\bigg( \f{1 }{ N T^4 } \bigg)  \;.
\enq
Here, the control on the remainder only depends on
$|\wh{\mc{Y}}_{\e{ref}}|+|\wh{\mf{X}}_{\e{ref}}|+|\mf{s}|$ and is thus uniform
over the choices of integers as in \eqref{ecriture suite croissante entiers p et h bornee en T}-%
\eqref{ecriture limitation sur nombre entiers particle et trou}.
We recall that $\mathbb{Y}_{\e{ref}; \varkappa}$ is as in
\eqref{defintion mathbb X ref et son hat avec un monodromie varkappa}.
Its definition involves the index of $u$ with respect to $\msc{C}_{\e{ref}}$, \textit{c.f.}\
\eqref{ecriture monodromie Trotter infini}. Finally,  $\mc{P}(\mathbb{Y})$
is as introduced in \eqref{definition impulsion thermale}.

Further, Proposition~\ref{Propostion DA fin pour la vp de la QTM}
provides the form of the low-$T$ expansion for $\mc{P}(\mathbb{Y})$,
\textit{c.f.}\ \eqref{ecriture dvpt basse tempe pour P de Y}.
By virtue of \eqref{ecriture estimee sur partie imaginaie impuslion dans plan complexe},
$\e{Im}\big[ \msc{P}_{0}( \mathbb{Y}^{(\e{far})} ) \big] \, \geq  \, 0$
and attains its minimum for $\mathbb{Y}^{(\e{far})} = \emptyset$. Here,
we recall that $\mathbb{Y}^{(\e{far})}$ has been defined in
\eqref{definition ensemble reduit YM mathbb} and $\msc{P}_{0}( \mathbb{Y}^{(\e{far})} )$
is given in \eqref{definition mas P0}. Moreover,
\beq
\Im[\varpi_{1}(\mathbb{H})] \, \geq \, \Im[\varpi_{1}(\mathbb{H}_{\e{min}})] \,= \,  \f{  \pi }{ \op{v}_F } \sul{ \sg = \pm }{}	   \Big\{ (n_p^{(\sg)})^2 + (n_h^{(\sg)})^2 \Big\} \; ,
\enq
with $\varpi_{1}$ given by \eqref{definition fct varpi1 de H} and
$\mathbb{H}_{\e{min}}$ as defined in \eqref{definition dist part trous CFT minimale}.
From here it follows that, if
\beq
\Big( \wh{u}(*\,|\, \wh{\mathbb{Y}}), \wh{\mc{Y}}_{\e{ref}}, \wh{\mf{X}}_{\e{ref}} \Big) \not= \Big( \wh{u}_{\e{max}}, \emptyset, \emptyset \Big) \;,
\enq
then
\beq
 \ln \Big| \wh{\La}\, \Big( 0 \big|\, \wh{u}(*\,|\, \wh{\mathbb{Y}}), \wh{\mc{Y}}_{\e{ref}}, \wh{\mf{X}}_{\e{ref}} \Big) \Big| \, \leq \,
\ln \Big\{ \wh{\La}_{\e{max}; \e{BA}} \Big\} \, - \, \e{O}\Big(T \e{max}\{h_a^{(\pm)}, p_a^{(\pm)} \} \Big) \, + \, \e{O}(T^2) \, + \,  \e{O}\Big( \f{1 }{ N T^4 } \Big)  \;.
\enq
The two remainders $ \e{O}(T^2) \, + \,  \e{O}\Big( \f{1 }{ N T^4 } \Big) $
only depend on $|\wh{\mc{Y}}_{\e{ref}}| + |\wh{\mf{X}}_{\e{ref}}| + |\mf{s}|$.
In its turn, the first remainder only involves the integers $h_a^{(\pm)}, p_a^{(\pm)}$
such that $Th_a^{(\pm)} \tend 0$, $p_a^{(\pm)} \tend 0$ as $T \tend 0^+$,
see \eqref{ecriture partition close far roots}, \eqref{ecriture partition close far integers},
\eqref{ecriture limite des close far integers en echelle T}. This remainder is strictly
positive, hence ensuring that the claim about maximality follows.

\vspace{2mm}

We finish the proof by establishing the complex conjugation property.
This will be achieved by first establishing the invariance of the
integration contour $\msc{C}_{\e{ref}}$, defined in
\eqref{definition contour Cref}, with respect to the involutive
transformation $\Ups:z \mapsto \Ups(z) \, = \, -z^*$. For this
purpose it is sufficient to show that
\begin{equation} \label{to_show_for_cref_invariance}
    \Ups(\gamma^{(\pm)}\bigr) = \gamma^{(\pm)} \;, \quad
    \veps_L^{-1} \bigl(\Gamma_{\delta T}^{(L)}\bigr) =
    \Ups\Bigl(\veps_R^{-1} \bigl(\Gamma_{\delta T}^{(R)}\bigr)\Bigr) \;,
\end{equation}
\textit{c.f.}\ \eqref{definition contour Cref}.

First of all, observe that the dressed energy satisfies
$\veps\bigl(\Ups(\la)\bigr) \, = \, - \Ups\bigl(\veps(\la)\bigr)$
as can be seen by taking the complex conjugate of
\eqref{definition energie habille et energie nue} at $Q=q$, invoking
that $\veps$ is even on $\R$, and using the symmetry property
$K\bigl(\Ups(\la)\bigr) \, = \, - \Ups\bigl(K(\la)\bigr)$ of the
integral kernel.

Now $\Ups\Bigl(\op{D}_{ -\i \f{\zeta}{2}, \mf{c}_{\e{d}} T}\Bigr) =
\op{D}_{ -\i \f{\zeta}{2}, \mf{c}_{\e{d}} T}$ and $y_L^{(\pm)} \in
\Bigl(\op{D}_{ -\i \f{\zeta}{2}, \mf{c}_{\e{d}} T} \cap {\rm Int} \,(\msc{U}_{L;\veps})\Bigr)$ by construction, implying that
$\Ups\bigl(y_L^{(\pm)}\bigr) \in \Bigl(\op{D}_{ -\i \f{\zeta}{2}, \mf{c}_{\e{d}} T} \cap {\rm Int} (\msc{U}_{R;\veps})\Bigr)$. We have as well that
$\Ups\bigl(y_L^{(\pm)}\bigr) \in \veps_R^{-1}(\pm \de_T - \i {\mathbb R})$
which follows from
\begin{equation} \label{map_reflect_yL}
     \veps\bigl(\Ups(y_L^{(\pm)})\bigr) =
        - \Ups\bigl(\veps(y_L^{(\pm)})\bigr) = \pm \delta_T - \i \mf{t}_L^{(\pm)}
\end{equation}
and $\mf{t}_L^{(\pm)} > 0$. Hence, by \eqref{definition points finaux sur petit disque},
\begin{equation} \label{reflect_yL_is_yR}
     \Ups (y_L^{(\pm)}) = y_R^{(\pm)} \;.
\end{equation}
The continuity of $\Ups$ then implies the first equation
\eqref{to_show_for_cref_invariance}.

By reinserting \eqref{reflect_yL_is_yR} back into \eqref{map_reflect_yL}
we see that $\mf{t}_L^{(\pm)} = - \mf{t}_R^{(\pm)}$ and
therefore $- \Ups (\Gamma_{\de_T}^{(R)}) = \Gamma_{\de_T}^{(L)}$,
implying in turn that
\begin{equation}
     \veps\,\Bigl(\Ups\bigl(\veps_R^{-1} (\Gamma_{\de_T}^{(R)})\bigr)\Bigr) =
     - \Ups\Bigl(\veps\,\bigl(\veps_R^{-1} (\Gamma_{\de_T}^{(R)})\bigr)\Bigr) =
     - \Ups\bigl(\Gamma_{\de_T}^{(R)})\bigr) = \Gamma_{\de_T}^{(L)} \;.
\end{equation}
Acting with $\veps_L^{-1}$ on this equation we obtain the second equation
\eqref{to_show_for_cref_invariance}.

We continue with the invariance of the nonlinear integral equation
under $\Ups$. Pick $\varkappa= -\i\frac{\zeta}{2} + \i \mf{c}_{\e{d}} T$
in \eqref{definition logarithm de 1+a} and set
$\mf{g}(\mu) \, = \, \Big(\wh{u}(-\mu^*) \Big)^*$,
with $\wh{u}$ being the unique solution to
\eqref{ecriture NLIE Trotter fini VP dominante}. It is obvious
that $\mf{g}\in \wh{\mc{E}}_{\mc{M}}$. Further, it is direct
to check that
\beq
\Big( \msc{L}\mathrm{n}_{ \msc{C}_{\e{ref}} } \Big[ 1+  \ex{ - \f{1}{T}\wh{u} } \, \Big](-\nu^*) \Big)^* \, = \,   \msc{L}\mathrm{n}_{ \msc{C}_{\e{ref}} }
        \Big[ 1+  \ex{ - \f{1}{T}\mf{g} } \, \Big](\nu)  \;.
\enq
Likewise, a direct calculation gives that
\beq
\Big( \mf{w}_N(-\xi^*) \Big)^*  \; = \; \mf{w}_{N}(\xi)
\enq
with $\mf{w}_{N}$ as given through \eqref{definition fct mathfrak w N}. Thus, one gets
\begin{align}
\mf{g}(\xi)  & \; = \; \Bigl(\, \wh{u}\,(-\xi^*) \Bigr)^* 
               \, = \, h \, -\, T\mf{w}_N(\xi)
	       \; - \; T\Oint{ \msc{C}_{\e{ref}}  }{} \dd \la \: \Bigl( K(-\xi^*-\la) \Bigr)^*
               \cdot \Bigl( \msc{L}\mathrm{n}_{ \msc{C}_{\e{ref}} }
	             \Bigl[ 1+  \ex{ - \f{1}{T}\wh{u} } \, \Bigr](\la) \Bigr)^* \\
             & \; = \; h \, -\, T\mf{w}_N(\xi)
	       \; - \; T\Oint{ \msc{C}_{\e{ref}}  }{} \dd \la \:   K( \xi -\la)
               \cdot \msc{L}\mathrm{n}_{ \msc{C}_{\e{ref}} }
	       \Bigl[ 1+  \ex{ - \f{1}{T}\mf{g} } \, \Bigr](\la) \;,
\end{align}
where, in the second line, we changed variables $\la \hookrightarrow -\la^*$.
Similar handlings and the invariance of $\msc{C}_{\e{ref}}$ under $\Ups$
ensure that $\mf{g}$ satisfies the zero index condition. Hence, by virtue
of the uniqueness of the solutions to the non-linear integral equation subject
to the zero-index condition on $\wh{\mc{E}}_{\mc{M}}$, one has that
$\mf{g}=\wh{u}$.

The very same handlings on the level of the integral representation
\eqref{ecriture Trotter fini VP dominantes BA} lead to
$\Bigl( \wh{\La}_{\e{max};\e{BA}} \Bigr)^* \, = \,\wh{\La}_{\e{max};\e{BA}}$. \qed

\subsubsection{Proof of Theorem \ref{Theorem identification vp dominante et energie libre XXZ}}
\label{SousousSection preuve energie libre et identification vp dominante}

In order to establish the result, we need to go a bit deeper into the construction
of the quantum transfer matrix which is an operator on $\mf{h}_{\mf{q}} \, = \,
\bigotimes_{a=1}^{2N} \wt{\mf{h}}_a$, with $\wt{\mf{h}}_a \simeq \Cx^2$.
It is defined as the trace over the auxiliary space $\mf{h}_{0}$ of the
quantum monodromy matrix,
\beq
 \op{t}_{ \mf{q} }(\xi) \, = \, \e{tr}_{\mf{h}_0} \Big[ \op{T}_{\mf{q};0}(\xi) \Big] \;.
\enq
In its turn, $ \op{T}_{\mf{q};0}$ is defined as an ordered product of $\op{R}$-matrices,
\beq
\op{T}_{\mf{q};0} (\xi) \; = \; \op{R}_{2N,0}^{\mf{t}_{2N}}\Big( -\tfrac{ \aleph }{N}-\xi \Big) \, \op{R}_{0,2N-1}\Big( \xi -\tfrac{  \aleph  }{N}\Big) \cdots
 \op{R}^{\mf{t}_{2}}_{2, 0}\Big( -\tfrac{  \aleph  }{N}-\xi \Big)\,
\op{R}_{0, 1}\Big( \xi -\tfrac{ \aleph }{N}\Big) \cdot   \ex{ \f{h}{2T} \sg^z_0} \;,
\enq
with $\aleph$ as introduced in \eqref{definition aleph}.
Above, $\op{R}$ represents the $\op{R}$-matrix of the six-vertex model
\beq
\op{R}(\la)\; = \;  \f{1}{ \sinh(\eta) } \left( \ba{cccc} \sinh\big( \eta + \la \big) 	& 0 	& 0	 & 0  \\
							    0   &  \sinh\big( \la \big)   &   \sinh(\eta)   &  0  \\
								0   & \sinh(\eta)  & \sinh\big(  \la \big)   &    0  \\
							      0 	& 0		 &    0  			&   \sinh\big( \eta + \la \big)  \ea \right) \; ,
\enq
while the notation $\op{R}_{ab}$ stands for the embedding of $\op{R}$ into
${\rm End}\, \bigl(\mf{h}_0\otimes \mf{h}_{\mf{q}}\bigr)$, $\mf{h}_0\simeq \Cx^2$,
which acts as the identity operator on all spaces $\mf{h}_{k}$,
$k \not= a,b$ and as $\op{R}(\la)$ on the spaces $\mf{h}_{a} \otimes \mf{h}_b$.
Finally, the superscript $\mf{t}_a$ stands for the partial transposition
with respect to the space $\mf{h}_a$.

It is a matter of direct calculations to check that $\op{T}_{\mf{q};0} (\xi)
\, = \, \big( \op{T}_{\mf{q};0} (-\xi^*)  \big)^*$, where $*$ refers to the
complex conjugation. In particular, this ensures that
$\op{t}_{\mf{q}}( \xi ) \, = \, \big( \,\op{t}_{\mf{q}}( -\xi^* ) \big)^*$.
From this relation, one infers that if $\bs{\Psi}$ is an Eigenvector of
$\op{t}_{\mf{q}}( \xi )$ associated with the Eigenvalue $\wh{\La}(\xi)$,
\textit{i.e.}\ $\op{t}_{\mf{q}}( \xi ) \bs{\Psi} \, = \, \wh{\La}(\xi) \bs{\Psi}$,
then
\beq
%
%
%
\op{t}_{\mf{q}}( \xi ) \bs{\Psi}^* \, = \, \Big( \wh{\La}(-\xi^*) \Big)^* \bs{\Psi} ^* \;.
\enq
This implies that if $\wh{\La}_k \, = \, \wh{\La}_k(0)$ is an
Eigenvalue of $\op{t}_{\mf{q}}( 0 )$ so is $ \wh{\La}_k^{\,*}$.

Thus, assume that $\wh{\La}_{\e{max}} \not \in \R$. Then,
$\wh{\La}_{\e{max}}^{\,*}\not=\wh{\La}_{\e{max}} $ is also an
Eigenvalue of $\op{t}_{\mf{q}}( 0 )$ and both are equal in modulus.
This is not possible owing to Conjecture~\ref{Conjecture VP dominante matrice transfer}.

Because  $\wh{\La}_{\e{max}}$ is non-degenerate, the holomorphic implicit function
theorem ensures that $\zeta \mapsto  \wh{\La}_{\e{max}}$ is real analytic
on $\intof{0}{\tf{\pi}{2}}$. So is $\wh{\La}_{\e{max};\e{BA}}$ by virtue of
Theorem \ref{Theorem form close La max BA}.

It follows from a direct calculation that $\wh{\La}_{\e{max}}=\wh{\La}_{\e{max};\e{BA}}$
when $\zeta=\tf{\pi}{2}$. Indeed, at $\zeta=\tf{\pi}{2}$, the Bethe Ansatz
equations trivialise, so that they can be solved explicitly, hence
producing very explicit non-zero Bethe Eigenvectors which can be shown
to be complete and all Eigenvalues can be explicitly compared.
Because of the continuity in $\zeta \in \intof{0}{\tf{\pi}{2}}$
of the Eigenvalues of $\op{t}_{\mf{q}}( 0 )$ as zeroes of the
characteristic polynomial, having smooth coefficients in $\zeta$, and
Conjecture~\ref{Conjecture VP dominante matrice transfer}, ensuring the
existence of a finite spectral gap, one must necessarily have
\beq
\wh{\La}_{\e{max}}=\wh{\La}_{\e{max};\e{BA}} \qquad \e{for} \qquad \zeta \in \intof{ \tfrac{\pi}{2} - \eps_N }{  \tfrac{\pi}{2}}
\enq
for some $\eps_N>0$ and small enough. Since the two real analytic functions
coincide on a set containing an accumulation point, they are equal everywhere.

Conjecture \ref{Conjecture commutativite vp QTM}, then ensures that
\beq
f = -T \lim_{N\tend + \infty} \ln \wh{\La}_{\e{max};\e{BA}} \;.
\enq
The infinite Trotter limit may then be taken by using the estimate
\eqref{ecriture deviation entre point fixe u et hat u} given in
Theorem~\ref{Theorem existence et unicite sols NLIE}, thus leading to
\eqref{ecriture forme close energie libre XXZ}. The form of the
low-$T$ expansion given in \eqref{ecriture DA basse T de energie libre}
then follows from Proposition \ref{ecriture DA P emptyset}. \qed

\section*{Conclusion}

This work developed a detailed and rigorous study  of the spectrum of the quantum
transfer matrix that can be obtained from a large subset of solutions to the
Bethe-Ansatz equations subordinate to it. More precisely, in this paper we have
set up a rigorous framework allowing one to take effectively the infinite
Trotter-number limit on the level of a non-linear problem -- solving a non-linear
integral equation and the associated quantisation conditions -- which is equivalent
to the system of Bethe-Ansatz equations subordinate to the model. Our technique
allows us to effectively construct a rather large class of solutions
to the Bethe-Ansatz equations in the large Trotter number and small temperature regime.
In particular, we are able to exhibit one solution giving rise to a particular
Eigenvalue $\wh{\La}_{\e{max};\e{BA}}$ of the quantum transfer matrix.
This Eigenvalue is such that the low-temperature expansion of
$\lim_{N\tend +\infty}\wh{\La}_{\e{max};\e{BA}}$
takes the precise form conjectured for the leading low-$T$ asymptotics of the
infinite Trotter limit of the  dominant Eigenvalue of the quantum transfer matrix,
as predicted with the help of conformal field theory methods. We were able to check
that, among the class of solutions we considered,  $\wh{\La}_{\e{max};\e{BA}}$
is indeed dominant with respect to all other Eigenvalues which are computable within
our scheme. It is reasonable to expect that, in fact, we have constructed the
true dominant Eigenvalue of the quantum transfer matrix. Further, we were able
to construct a class of Eigenvalues $\wh{\La}_{k;\e{low}}$ whose low-$T$ asymptotics
reproduce, in the infinite Trotter number limit, the spectrum of the $c=1$
free Boson conformal field theory. This allowed us to prove a long-standing
conjecture issuing from the physics of such models. Also, the low-$T$ asymptotic expansions
we obtained indicate the appearance of two decoupled conformal field theories at order $T$
which already couple at order $T^2$: this is some form of manifestation of the
breaking of conformal invariance for finite $T$.

We believe that our techniques have a wide scope of applicability and that they will
allow to finally address, at least within the regimes discussed in our work, the
rigorous complete analysis of the Bethe-Ansatz equations. Furthermore, we would like to
underline some potential fall-offs of our work. We have established that there are no strings in the Bethe root patterns for $0 < \Delta < 1$
and have provided a very explicit description of the excitations, among the states we considered.
These properties open the perspective to obtain very explicit form factor series, in the low-$T$ regime, and
to study them analytically. Next, our results provide a precise \textit{a posteriori} knowledge of the function u.
In particular, we have provided a precise description of the locations of the singularities and zeroes of $1 + \ex{-\f{u}{T} }$.
This allows us to justify the use of straight contours in numerical studies of the NLIE, which appeared as
a empirical observation so far. Finally, the main ingredient of our analysis consists in a precise characterisation of
the dressed energy function $\veps_{\e{c}}$ which catches the leading asymptotics of the root
distribution. Our analysis will thus be applicable to other systems of rank $1$ Bethe Ansatz equation
as long as one has a similar understanding of the special function driving the leading asymptotics.

\section*{Acknowledgements}
SF and FG acknowledge financial support by the DFG in the framework
of the research unit FOR 2316. The work of KKK is supported by the
CNRS, by the ``Projet international de coop\'eration scientifique
No.~PICS07877'': \textit{Fonctions de corr\'elations dynamiques
dans la cha\^ine XXZ \`a temp\'erature finie}, Allemagne,
2018-2020 and by the ERC Project LDRAM: ERC-2019-ADG Project 884584.  
Part of this material is based upon work supported by the National
Science Foundation under Grant No.~1440140, while KKK visited the
Mathematical Sciences Research Institute in Berkeley, California,
during the ``Universality and Integrability in Random Matrix Theory
and Interacting Particle Systems'' semester held in the Fall of 2021. 
KKK is grateful for the unique working conditions at the PMU de la Poste,
where part of this manuscript was written down. FG and KKK thank Andreas
Kl\"umper and Junji Suzuki for numerous stimulating discussions
related to the topics tackled in this paper.

\appendix

\section{Algebraic sums of sets}
\label{Appendix Section algebraic sum of sets}
Consider $n$ complex numbers $z_1,\dots, z_n \in \Cx$, distinct or not. If some
of these coincide, the total number thereof equal to a given complex number
$z$ is called their multiplicity and denoted $k_{z}$. To the collection of these
numbers one associates the set
\beq
\{ z_a \}_{1}^{n} \; = \; \Big\{ (\la, k_{\la} ) \; : \; \la \in \{z_1,\dots, z_n \}  \Big\}
\enq
in which  $\{z_1,\dots, z_n \} $ stands for the usual set build up from the numbers  $z_1,\dots, z_n$.

Further, given a finite set $\Om$ and a map  $n : \Om \tend \mathbb{Z}$, the weighted
cardinality $|A|$ of the set $A=\big\{ (x, n_x) \, : \, x \in \Om \big\}$ is defined
by $|A|=\sul{ x \in \Om }{} n_x$. Given a function $f$ on the set $\Om$ and $A$ defined
as above, we agree upon the shorthand notation
\beq
\sul{\la \in A}{} f(\la) \, \equiv \,\sul{x \in \Om }{} n_x \,  f(x)  \qquad \e{and} \qquad \pl{\la \in A}{} f(\la) \, \equiv \, \pl{x \in \Om }{} \big\{ f(x)\big\}^{n_x}  \;.
\enq

There is a natural way to define the algebraic sum $\oplus$ and difference $\ominus$ of two sets
$A=\big\{ (x, n_x) \, : \, x \in \Om_{A} \big\}$ and $B=\big\{ (y, n_y) \, : \, x \in \Om_{B} \big\}$, namely
\beq
A \oplus B \; = \; \Big\{ ( x, n_x+m_x ) \; : \; x \in \Om_A\cup \Om_B \Big\} \;, \qquad A \ominus B \; = \; \Big\{ ( x, n_x-m_x ) \; : \; x \in \Om_A\cup \Om_B \Big\}
\label{definition difference algebrique ensemble}
\enq
in which one understands that the maps $n$ and $m$ are extended as $n_x=0$, resp. $m_x=0$, on $\Om_{B}\setminus \Om_{A}$, resp. $\Om_{A}\setminus \Om_{B}$.
Furthermore, if $\Om_A$, $\Om_B$ are two sets, then $\Om_A \ominus \Om_B \; \equiv  \; A\ominus B$, where $A,B=\big\{ (x, 1) \, : \, x \in \Om_{A,B} \big\}$.

We now arrive to the main advantage of this construction in that this setting allow one to introduce
compact notations for sums and products. Given a function $f$ on $\Om_A\cup\Om_B$ parameterising sets $A,B$ as above, our conventions imply
\beq
\sul{\la \in A\ominus B}{}   f(\la)   \; = \;
\sul{x \in \Om_A}{} n_x  f(x) \, - \,  \sul{y \in \Om_B}{} m_y f(y) \;  ,   \quad
\pl{\la \in A\ominus B}{} f(\la) \, = \,  \f{ \pl{ x \in \Om_A }{} \big\{ f(x) \big\}^{n_x}  }{  \pl{y \in \Om_B }{}  \big\{ f(y) \big\}^{m_y}  }   \;.
\label{defintion convention somme produit et cardinalite ensembles}
\enq
Finally, given a point $x \in \Cx$, $\{x\}^{\oplus n }$ denotes the set $\{(x,n)\}$, meaning that one should understand
\beq
\sul{t \in \{x\}^{\oplus n} }{} f(t) = n f(x) \, .
\label{definition ensemble repetee}
\enq

\section{Useful integrals and integration lemmas}
\label{Appendix Section ptes integrales et espaces fnels}

\begin{lemme}
\label{lemme Natte integrale elementaire}

Let $g$ be $\mc{C}^{k+1}$ on $\intff{-\de}{\de}$ with $\de\geq -MT \ln T$, and $M>0$ large enough. Then
\beq
     \Int{-\de}{\de} \dd t \: g(t) \, \ln\big[ 1+\ex{-\f{\abs{t}}{T}} \big] =
\sul{r=0}{ \lfloor \tf{k}{2} \rfloor }  2 T^{2r +1}  \big(1-2^{-1-2r}\big) \zeta(2+2r) g^{(2r) }(0)
 \; +\; \e{O}\big(T^{k+2}\big) \;,
\enq
where $\lfloor x \rfloor$ is the integer part of $x$.

\end{lemme}

\Proof
By the Taylor-integral formula, given any function $g$ that is $\mc{C}^{k+1}$ on
$\intff{-\de}{\de}$, there exists a continuous function $u_k\pa{t}$ on $\intff{-\de}{\de}$
such that
\beq
g\pa{t} = \sul{r=0}{ k } \f{t^r}{r!}g^{\pa{r}}\pa{0} + u_k\pa{t} \quad \e{with} \quad \abs{u_k\pa{t}} \, \leq\,  C_k \cdot  |t|^{k+1} \; , \quad
\e{uniformly} \; \e{on} \; \intff{-\de}{\de} \;.
\enq
Thence,
\beq
     \Int{-\de}{\de} \dd t \: g(t) \ln\big[ 1 + \ex{-\f{\abs{t}}{T}} \big] \;  =\;
\sul{r=0}{  \lfloor \tf{k}{2} \rfloor } \f{g^{(2r)}(0) T^{2r +1}  }{ (2r)! } \Int{-\f{\de}{T}}{ \f{\de}{T} } \dd t \: t^{2r} \ln\big[ 1 + \ex{-\abs{t}} \big]
\, +\,  \Int{-\de}{ \de } \dd t \: u_k(t) \ln\big[ 1 + \ex{-\f{\abs{t}}{T}} \big] \;.
\label{equation DA local avec fonction g}
\enq
The remainder is a $\e{O}\big( T^{k+2} \big)$ as
\beq
\abs{\Int{-\de}{ \de } \dd z \: u_k\pa{z} \ln\pac{1+\ex{-\f{\abs{z}}{T}}}}  \leq
2 C_k T^{k+2} \Int{0}{+\infty} \dd t \: t^{k+1} \ln\pac{1+\ex{- t } } = \e{O}\pa{T^{k+2}}
\;.
\enq
Moreover, up to $\e{O}\pa{T^{M}}$ corrections, one can extend the integration in the sum in \eqref{equation DA local avec fonction g} from $\intff{-\tf{\de}{T}}{\tf{\de}{T}}$
to $\R$. The integral representation
\beq
\big( 1-2^{-1-s} \big)\zeta(s+2) \,  = \,  \f{1}{\Ga(s+1)}  \Int{0}{+\infty} \dd t \:
t^{s} \ln\big[ 1+\ex{-t} \big] \; ,
\enq
allows one to conclude. \qed
%
%
%




\subsection{Integral equations on $L^2\big( \msc{C}_{\veps} \big)$}

\begin{lemme}
\label{Lemme inversibilite de Id + K sur courbe C eps}

 The operators $\e{id}+\op{K}_{\msc{C}_{\veps}}$ and $\e{id}+\op{K}_{ \wh{\msc{C}}_{\veps}}$, \textit{c.f.} \eqref{ecriture LIE pour veps c}, are invertible.

\end{lemme}

\Proof
We only discuss the proof in the case of $\msc{C}_{\veps}$, as both cases are similar.

Obviously, the operator $\op{K}_{\msc{C}_{\veps}}$ is compact. Since $\msc{C}_{\veps}$ is smooth and compact while the integral kernel $(\la,\mu)\mapsto K(\la-\mu)$
is smooth on $\msc{C}_{\veps}^2$, by the metric entropy condition of \cite{DudleyGonzalesBarriosMetricConditionForOpToBeTraceClass}, $\op{K}_{\msc{C}_{\veps}}$ is also trace class.
Hence, its Fredholm determinant is well-defined. It is given by the absolutely convergent series
\bem
\underset{ L^2(\msc{C}_{\veps} ) }{ \det } \Big[ \e{id}+\op{K}_{\msc{C}_{\veps}} \Big]\; = \; \sul{n \geq 0}{} \f{1}{n!} \Int{ \msc{C}_{\veps}^n }{} \dd^n\la \det_n\Big[ K(\la_a-\la_b) \Big] \\[-1ex]
\; = \;  \sul{n \geq 0}{} \f{1}{n!} \Int{-q}{q} \dd^n\la \det_n\Big[ K(\la_a-\la_b) \Big]
\; = \; \underset{ L^2(\intff{-q}{q} ) }{ \det } \Big[ \e{id}+\op{K} \Big] \; \not=\; 0 \;,
\end{multline}
where the second equality follows from Morera's theorem and the fact that $(\la,\mu)\mapsto K(\la-\mu)$ is analytic on $\e{Int}\Big(  \msc{C}_{\veps}\cup\intff{q}{-q}  \Big)^2$.
In the third equality, we used that $\e{id}+\op{K}$ is invertible on $ L^2(\intff{-q}{q} )$, see \cite{KozProofOfDensityOfBetheRoots,KozDugaveGohmannThermoFunctionsZeroTXXZMassless}.
Since $\underset{ L^2(\msc{C}_{\veps} ) }{ \det } \Big[ \e{id}+\op{K}_{\msc{C}_{\veps}} \Big]
\ne 0$, it then follows that the operator
$\e{id}+\op{K}_{\msc{C}_{\veps}}$ is invertible. \qed




\subsection{Completeness of the metric space $\mc{E}_{\mc{M}}$}

\begin{prop}
\label{Proposition caractere espace metrique complet pour EM}
The metric space $(\mc{E}_{\mc{M}}, \op{d}_{\mc{E}_{\mc{M}}} )$ defined through
\eqref{definition espace fonctionnel principal}, \eqref{ecriture metrique pour espace EM} is complete.
\end{prop}

\Proof
Let $f_n$ be a Cauchy sequence in $\mc{E}_{\mc{M}}$, then $f_n$ is a Cauchy sequence of continuous functions from $\mc{M}$, as defined through Fig.~\ref{Domaine pour point fixe NLIE},
to the closed disc $\ov{\op{D}}_{0, C_{\mc{M}} T^2}$ of radius $C_{\mc{M}} T^2$ centred at $0$ as defined in \eqref{definition disque ouvert generique}.
Hence, since that space is complete, it converges to some $f\in \mc{C}^{0}\big(\mc{M}, \ov{\op{D}}_{0, C_{\mc{M}} T^2} \big)$. Then, for any $\eps>0$, pick $n$
so that one has $\norm{f-f_n}_{L^{\infty}(\mc{M})}<\eps$. Then,
\beq
\big| f(\la) \big| \, \leq \, \norm{f-f_n}_{L^{\infty}(\mc{M})} \, + \, \big| f_n(\la) \big| \quad \e{so}\; \e{that} \quad
\limsup_{ \substack{ \la \tend \infty \\ \la \in \mc{M} }  } \big| f(\la) \big| \, \leq \,\eps
\enq
namely, $f(\la) \limit{ \la \in \mc{M} }{ \infty } 0$. Finally, given a compact $K\Subset\mc{M}$, one has
\beq
f_n(z) \; = \; \Oint{ \Dp{} K }{}  \f{\dd s }{2\i\pi} \cdot \f{ f_n(s) }{ s -  z } \qquad \e{for} \; \e{any} \qquad z \in \e{Int}(K) \;.
\enq
One may apply the dominated convergence theorem for any $z$ such that
$\op{d}(z,\Dp{}K)>0$ leading to a Cauchy integral representation for $f(z)$,
$z \in \e{Int}(K)$. This entails that $f \in \mc{O}(\mc{M})$.  \qed




\subsection{The multidimensional Rouch\'{e} theorem}

\begin{theorem} {\bf Aizenberg and Yuzhakov}
\cite{AizenbergYuzhakovIntRepAndMultidimensionalResidues}.
\label{Theoreme Rouche generalise poru plusieurs vars complexes}
Let $U\subset \Cx^n$ be open and $f: U \tend \Cx^n$ be a biholomorphic map. Pick
$\varrho_{a}>0$ such that  the special analytic polyhedron
\beq
\mc{D}_{\bs{\varrho}} \; = \; \Big\{ \bs{z} \in U \; :
   \; |f_{a}(\bs{z})| \,<\, \varrho_a\; , \; a=1,\dots, n  \Big\}
\enq
and its associated skeleton $\Ga_{\bs{\varrho}}$,
\beq
\Ga_{\bs{\varrho}} \; = \; \Big\{ \bs{z} \in \ov{\mc{D}}_{\bs{\varrho}} \; :
   \; |f_{a}(\bs{z})| \, = \, \varrho_a\; , \; a=1,\dots, n  \Big\} \;,
\enq
satisfy $\ov{\mc{D}}_{\bs{\varrho}}\subset U$. Assume that $g\, : \,  U \tend \Cx^n$
is holomorphic and such that
\beq
\big|  g_a(\bs{z})\big| \, \leq \, \big|  f_a(\bs{z})\big|  \quad for \; all \;\;  \bs{z}\in  \Ga_{\bs{\varrho}} \;\; and \;\;  a=1,\dots, n \;.
\enq
Then
\begin{enumerate}
\item
$f$ and $f+g$ have the same number of zeroes in $\mc{D}_{\bs{\varrho}}$;
\item
for any holomorphic $\vp \, : \, U \tend \Cx$
\beq
     \Int{ \Ga_{\bs{\varrho}}}{}
	\f{ \dd^n z }{ (2\i\pi)^n } \:
	\f{\vp(\bs{z}) \,
	   \e{det}\big[ \op{D}_{\bs{z}} (f+g) \big]}{\prod_{a=1}^{n} \big( f_a+g_a)(\bs{z})}
	\; = \;
     \sul{ \bs{w} \in \op{E}_{f+g} }{} \vp(\bs{w}) \;,
\enq
where $\op{E}_{f+g} \; = \; \Big\{ \bs{w} \in \mc{D}_{\bs{\varrho}} \; : \;
(f+g)(\bs{w})=\bs{0} \Big\}$.
\end{enumerate}
\end{theorem}




\section{Properties of the dressed energies $\veps$,
$\veps_{\e{c}}$ and $\veps_{\e{c};2}^{(-)}$}
\label{Properties_of_the_dressed_energies}

Throughout this section, we assume that $0<\zeta<\tf{\pi}{2}$.

\subsection{Several properties of the dressed energy $\veps$}

Recall that the dressed energy $\veps$ is defined as the solution $\veps(\la\mid q)$ to \eqref{definition energie habille et energie nue} with $q$ being the Fermi point.
Clearly it is meromorphic on $\Cx\setminus  \bigcup_{\ups=\pm} \Big\{\intff{-q}{q} + \i \ups \zeta + \i\pi \mathbb{Z} \Big\}$ and admits smooth $\pm$ boundary values on
$\intoo{-q}{q} + \i \ups \zeta + \i\pi \mathbb{Z}$,  with $\ups \in \{\pm \}$. These boundary values exhibit
$\e{O}\Big( (\la+\i\ups\zeta  \pm q +\i\pi n) \ln (\la+\i\ups\zeta  \pm q +\i\pi n) \Big) $ singularities
at the endpoints $-\i\ups \zeta \pm q  - \i\pi n$, $\ups\in \{\pm \}$, of these intervals. This can be seen by recasting the original integral equation  \eqref{definition energie habille et energie nue} at $Q=q$
in the form
\bem
\veps(\la) \; = \; \veps_0(\la) \, + \, \Int{-q}{q} \f{ \dd \mu }{2\i\pi} \veps(\mu) \coth(\la-\mu+\i\zeta) \; - \; \Int{-q}{q} \f{\dd \mu }{2 \i\pi } \Big[ \veps(\mu)- \veps(\la-\i\zeta) \Big] \coth(\la-\mu-\i\zeta) \\
 \, + \, \f{ \veps(\la-\i\zeta)  }{ 2\i\pi } \ln \bigg( \f{ \sinh(\la-q- \i\zeta) }{ \sinh(\la+q-\i\zeta) } \bigg)
\end{multline}
from which the mentioned singularity structure when $\la \tend \pm q +\i\zeta$ immediately follows since $\veps(\pm q)=0$.
The remaining cases follow from $\big[ \veps(\la^*) \big]^* \, = \, \veps(\la)$ and $\i\pi$ periodicity.

\begin{lemme}
\label{Lemme caracterisation des courbes des valeurs limites de veps}

 The maps $s \mapsto \Im\Big[ \veps_{\pm}\big( s + \i\zeta \big)\Big]$ are strictly increasing on $\intff{-q}{q}$. Moreover, the oriented curves
\beq
\ga_{\pm}= \Big\{  \veps_{\pm}\big( s + \i\zeta \big) \, : \, s \in \intff{\pm q }{ \mp q }   \Big\}
\label{definition des courbes gamma pm}
\enq
form a Jordan curve $\ga_+\cup \ga_-$ in $\Cx$. In particular the bounded domain $\e{Int}\big( \ga_+\cup \ga_- \big)$ in $\Cx$ is connected.

\end{lemme}

\Proof
We start by establishing the strict-increase property. By using that $\veps$ is
meromorphic in the strip $\Big\{ z \in \Cx \; : \; |\Im(z)|< \zeta\Big\}$
and that its sole singularities there are simple poles at $\pm \i \tf{\zeta}{2}$ such that
\beq
\e{Res}\Big( \veps(\la) \dd \la , \la = \i \f{\zeta}{2}\Big) \, = \, 2\i J \sin(\zeta) \;,
\enq
one may put the integral equation \eqref{definition energie habille et energie nue} for $\veps$ in the form
\beq
\veps(\la) \; = \; \veps_0(\la) \, + \, 4 \pi  J \sin(\zeta) K\Big( \la - \i \tfrac{\zeta}{2}\Big) \, +
\hspace{-2mm} \Int{ \intff{-q}{q}^{\e{c}} }{} \hspace{-3mm} \dd \mu \, K(\la-\mu) \veps(\mu)
\; - \; \Int{ \R }{}  \! \dd \mu \,   K(\la-\mu-\i\zeta)\veps_-(\mu + \i \zeta) \;,
\enq
provided that $0<\Im (\la) < \zeta$ and where we agree upon $ \intff{-q}{q}^{\e{c}}=\R \setminus  \intff{-q}{q}$.
This thus yields that $\eta(\la) = \Im\Big[ \veps_{-}\big( \la + \i\zeta \big)\Big]$ solves the linear integral equation
\beq
\big( \e{id} \, + \, \op{K}_{\R} \big)\big[ \eta \big](\la) \, = \,
\Fint{ \intff{-q}{q}^{\e{c}} }{} \hspace{-1mm} \f{\dd \mu}{4\pi} \Big\{ \coth(\la-\mu+2\i\zeta) + \coth(\la-\mu - 2\i\zeta) - 2 \coth(\la-\mu)   \Big\} \veps(\mu)  \;.
\enq
One may readily compute the distributional Fourier transform of
\beq
G(\la) \; = \;  \f{1}{4\pi} \Big\{ \coth(\la +2\i\zeta) + \coth(\la - 2\i\zeta) - 2 \op{P}\coth(\la)   \Big\}
\enq
with $\op{P}$ being the principal value:
\beq
\mc{F}[G](k) \; = \; \lim_{\a \tend 0^+} \bigg\{ -\i  \f{ \sinh\big[ k (\tfrac{\pi-\a}{2}-\zeta)\big] \sinh\big[ k (\zeta-\tfrac{\a}{2})\big] }{  \sinh\big[ k \tfrac{\pi}{2} \big]   } \bigg\}
\; = \; \lim_{\a \tend 0^+} \bigg\{ -\i  \f{ \sinh\big[ k (\tfrac{\pi}{2}-\zeta-\a)\big] \sinh\big[ k  \zeta \big] }{  \sinh\big[ k \tfrac{\pi}{2} \big]   } \bigg\} \;.
\enq
There we used the convention that
\beq
\mc{F}[f](k)  \, = \, \Int{ \R }{} \dd \la f(\la) \ex{\i k \la} \;  ,  \quad viz. \quad \mc{F}^{-1}[f](\la)  \, = \, \Int{ \R }{} \f{\dd k}{2\pi} f(k) \ex{ - \i k \la}
\enq
for any $f \in L^1(\R)$.

Since
\beq
\mc{F}[K](k) \; = \; \f{ \sinh\big[k(\tfrac{\pi}{2}-\zeta ) \big] }{ \sinh\big[k\tfrac{\pi}{2} \big]  } \quad \e{and} \quad
1 + \mc{F}[K](k) \; = \; 2 \f{ \cosh\big[k\tfrac{\zeta}{2} \big]   \sinh\big[k\tfrac{\pi-\zeta}{2}  \big] }{ \sinh\big[k\tfrac{\pi}{2} \big]  }
\enq
one gets that
\beq
\f{ \mc{F}[G](k) }{1 + \mc{F}[K](k) }\; = \; \lim_{\a \tend 0^+} \bigg\{ -\i  \f{ \sinh\big[ k (\tfrac{\pi}{2}-\zeta-\a)\big] \sinh\big[ k \tfrac{\zeta}{2}\big] }
{     \sinh\big[k\tfrac{\pi-\zeta}{2}  \big]  } \bigg\} \;.
\enq
As a consequence, one gets that
\beq
\eta(\la) \; = \; \lim_{\a \tend 0^+} \Fint{ \intff{-q}{q}^{\e{c}} }{} \hspace{-2mm} \dd \mu \, \veps(\mu)  \, \psi_{\a}(\la-\mu) \;,
\enq
in which
\beq
\psi_{\a}(\la) \; = \; \Int{ \R }{} \f{ \dd k }{2 \i \pi} \ex{-\i k \la } \f{ \sinh\big[ k (\tfrac{\pi}{2}-\zeta-\a )\big] \sinh\big[ k \tfrac{ \zeta }{2} \big] }
{    \sinh\big[k\tfrac{\pi-\zeta}{2}  \big]  } \;.
\enq

The above Fourier transform can be evaluated in closed form by using the identity
\beq
\Int{\R-\i0^+}{} \f{\dd k }{8\i\pi} \f{ \ex{-\i k \la } }{ \sinh\big[ k \tfrac{\pi-\zeta}{2} \big] } \; = \; \f{ 1  }{ 2(\pi-\zeta) \Big( 1 + \ex{ \f{2\pi \la }{\pi-\zeta} } \Big)  }  \;.
\enq
Upon sending $\a\tend 0^+$, one gets
\beq
\psi_{0}(\la) \; = \; \f{ \ex{ \f{-\pi \la }{\pi-\zeta} } }{ 4(\pi-\zeta) } \Bigg\{ \f{ \ex{\i \frac{\pi \zeta}{\pi-\zeta } } }{  \sinh\big[ \frac{\pi}{\pi-\zeta} (\la-\i \zeta ) \big] }
\, + \, \f{ \ex{-\i \frac{\pi \zeta}{\pi-\zeta } } }{  \sinh\big[ \frac{\pi}{\pi-\zeta} (\la+\i\zeta) \big] }
\, - \, \f{ 2 }{  \sinh\big[ \frac{\pi \la }{\pi-\zeta} \big] }\Bigg\} \;.
\enq
Then, reducing to a common denominator yields
\beq
\psi_{0}(\la) \, = \, f(\la)\cdot g(\la) \quad \text{with} \quad
   \begin{cases}
      f(\la)  =  -
      \f{ 1-\cos \big[ \frac{ 2 \pi \zeta}{ \pi-\zeta  }\big]   }{  \sinh\big[ \frac{\pi}{\pi-\zeta} (\la+\i \zeta ) \big] \sinh\big[ \frac{\pi}{\pi-\zeta} (\la - \i \zeta ) \big]  } \\[1ex]
      g(\la) = \f{    1 }{   4(\pi-\zeta) }  \coth\big[ \frac{\pi \la }{\pi-\zeta} \big]
      \;.
   \end{cases}
\enq
It follows that $\psi_{0}$ is odd. Moreover, for $\la \in \R^+$, since $f(\la)<0$ and $f^{\prime}(\la)>0$ while $g(\la)>0$ and $g^{\prime}(\la)<0$, one gets that
$\psi_0^{\prime}(\la)=f(\la) g^{\prime}(\la)+g(\la) f^{\prime}(\la)>0$ for $\la\in \R^*$.

As a consequence, for $\la \in \intoo{-q}{q}$
\beq
\eta^{\prime}(\la) \; =  \hspace{-2mm}  \Int{ \intff{-q}{q}^{\e{c}} }{} \hspace{-2mm} \dd \mu \, \veps(\mu)  \, \psi^{\prime}_0(\la-\mu) \; > \; 0 \, ,
\enq
since $\veps|_{\intff{-q}{q}^{\e{c}} }>0$. Here, we stress that the integral
representation for $\eta$ is well-defined at $\la=\pm q$ since $\veps(\pm q)=0$.
That concludes the proof of the first part of the statement.

\vspace{2mm}

Now observe that $\veps_+(\la+\i \zeta) - \veps_-(\la+\i\zeta) = \veps(\la)$. The strict increase of $\la \mapsto \Im\big[\veps_+(\la+\i \zeta)\big] = \Im\big[\veps_-(\la+\i \zeta)\big]$
on $\intff{-q}{q}$ ensures that each of the curves does not intersect with itself. Moreover, since $\veps(\pm q)=0$ the two curves $\ga_{\pm}$ meet at the values $\la=\pm q$ of their parameterisations.
Finally, away from these two points, the two curves never intersect. Else, there would exist $\la_1, \la_2 \in \intoo{-q}{q}$ such that $\veps_+(\la_1+\i \zeta) = \veps_-(\la_2+\i \zeta)$.
The monotonicity properties of the imaginary parts imply that one necessarily has $\la_1=\la_2$. However, then  $\Re\big[ \veps_+(\la_1+\i \zeta) - \veps_-(\la_1+\i\zeta) \big]= \veps(\la_1)< 0$
since $\la_1 \in \intoo{-q}{q}$ by hypothesis, hence leading to a contradiction. This establishes that the oriented curve $\ga_+\cup \ga_-$ is a Jordan curve. \qed



\begin{prop}
\label{Proposition double recouvrement de veps}

Let
\beq
\msc{U}_{\veps} \, = \, \Big\{  z \in \Cx \; : \; -\f{\pi}{2} \, <  \, \Im(z)  \, \leq \, \f{\pi}{2}   \Big\}  \setminus \bigcup_{\ups=\pm} \Big\{\intff{-q}{q} + \i \ups \zeta \Big\} \;,
\enq
and let  $\ga_{\pm}$ be as introduced in \eqref{definition des courbes gamma pm}. For every $z \in \Cx\setminus \ov{ \e{Int}\big(\ga_+\cup\ga_-\big) }$,
$\la \mapsto \veps(\la) - z$ has two zeroes, counted with multiplicities, in $\msc{U}_{\veps}$. These zeroes are double if $z \in \{\veps(0), \veps(\i\tfrac{\pi}{2}) \}$
and they are simple otherwise.

In particular, the dressed energy
\beq
\veps \, : \,  \msc{U}_{\veps} \setminus \big\{0, \i\tfrac{\pi}{2}\big\}
\mapsto  \veps\big( \msc{U}_{\veps} \setminus \big\{0, \i\tfrac{\pi}{2}\big\} \big) \, = \,
\Cx\setminus \Big\{ \ov{ \e{Int}\big(\ga_+\cup\ga_-\big) }
\cup \{\veps(0), \veps(\i\tfrac{\pi}{2}) \} \Big\}
\enq
is  a double covering map.

Moreover, the maps $\veps_{L/R}=\veps|_{\msc{U}_{L/R;\veps} }: \msc{U}_{L/R;\veps} \tend  \veps\big( \msc{U}_{\veps} \big) $ are biholomorphisms. Here we agree upon
\begin{align*}
      \msc{U}_{L;\veps} & = \Big\{  z \in \msc{U}_{\veps} \; :   \; \Re(z)<0 \;\; or \; \; \Re(z)=0 \; \; and \; \; 0 < \Im(z) < \f{\pi}{2} \Big\} \;, \\[1ex]
      \msc{U}_{R;\veps} & = \Big\{   z \in \msc{U}_{\veps} \; :   \; \Re(z)>0 \;\; or \; \; \Re(z)=0 \; \; and \; \; -\f{\pi}{2} < \Im(z) < 0  \Big\} \; .
\end{align*}
\end{prop}

\begin{figure}[h]
\begin{center}
\includegraphics{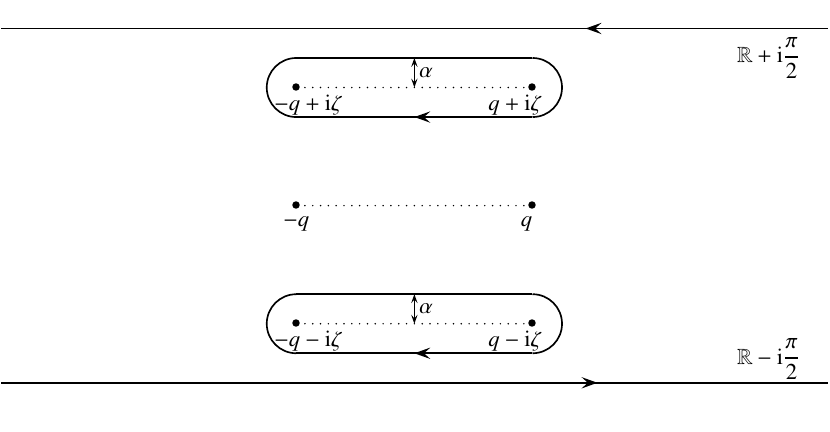}
\caption{Definition of the integration contour $\Ga_{\a}$ for $\mc{J}_{\a}(z)$.}
\label{contour integration pour nombre de zeros de veps}
\end{center}
\end{figure}

\Proof
Introduce the contour $\Ga_{\a}$ as in
Fig.~\ref{contour integration pour nombre de zeros de veps}
and consider the integral
\beq
\mc{J}_{\a}(z) \, = \, \Int{ \Ga_{\a} }{} \vspace{-3mm} \f{ \dd \mu }{2 \i \pi } \f{ \veps^{\prime}(\mu) }{ \veps(\mu) - z }
\enq
which is well-defined for $\a$ small enough provided that $|\Re(z)|$ is large enough. Indeed, it is easy to see that  $\veps\big(\R+\i\tf{\pi}{2}\big)\subset \intff{-M}{M}$ for some $M>0$
while given that $\veps$ has continuous boundary values on $\intff{-q}{q} + \i \zeta$, the values of $\veps$ on the curves surrounding
$\intff{-q}{q} + \i \zeta$ stay close to the set
\beq
\Big\{ \veps_+(\mu+\i\zeta) \, , \; \mu \in \intff{-q}{q} \Big\} \cup \Big\{ \veps_-(\mu+\i\zeta) \, , \; \mu \in \intff{-q}{q} \Big\}
\enq
which is bounded.
The same holds for the curves surrounding $\intff{-q}{q} - \i \zeta$ since $\veps$ is even.
By construction, one can compute $\mc{J}_{\a}(z)$ by the residues which gives
\beq
\mc{J}_{\a}(z) \, = \, \# \Big\{  \la \in \e{Int}(\Ga_{\a}) \; : \; \veps(\la) =z \Big\} \, - \, 2 \;,
\enq
where we used that $\veps$ has two simple poles inside the integration domain at $\pm \i \tf{\zeta}{2}$. The zeroes are counted with their multiplicities.

Since $\veps$ has differentiable\symbolfootnote[2]{By this we mean that it holds as well that $\veps^{\prime}(\la)=\e{O}\Big(   \ln \big( \la \mp q +\i\ups \zeta\big)  \Big)$. }
$\e{O}\Big( \big( \la \mp q +\i\ups \zeta\big)  \ln \big( \la \mp q +\i\ups \zeta\big)  \Big)$ behaviour at the endpoints $\pm q -\i\ups \zeta$, $\ups \in \{\pm 1\}$,
and smooth boundary values on $\intoo{-q}{q} \pm \i \zeta$, by virtue of it being $\i\pi$-periodic, one may recast the $\a \tend 0^+$ limit of $\mc{J}_{\a}(z)$ as
\beq
\mc{J}_{0^+}(z) \, = \, \Int{ -q }{q} \vspace{-2mm} \f{ \dd \mu }{2 \i \pi }
\bigg\{ \f{ \veps^{\prime}_+(\mu+\i\zeta) }{ \veps_+(\mu+\i\zeta) - z } \, - \, \f{ \veps^{\prime}_-(\mu+\i\zeta) }{ \veps_-(\mu+\i\zeta) - z }
\, + \,  \f{ \veps^{\prime}_+(\mu-\i\zeta) }{ \veps_+(\mu-\i\zeta) - z } \, - \, \f{ \veps^{\prime}_-(\mu-\i\zeta) }{ \veps_-(\mu-\i\zeta) - z }  \bigg\} \;.
\enq
This integral may be reduced further by observing that, due to $\veps$ being even, $\veps_{\pm}(s+\i\zeta) \, = \, \veps_{\mp}(-s-\i\zeta) $, what yields
\beq
\mc{J}_{0^+}(z) \, = \, \Int{ -q }{q} \vspace{-2mm} \f{ \dd \mu }{ \i \pi }
\bigg\{ \f{ \veps^{\prime}_+(\mu+\i\zeta) }{ \veps_+(\mu+\i\zeta) - z } \, - \, \f{ \veps^{\prime}_-(\mu+\i\zeta) }{ \veps_-(\mu+\i\zeta) - z }  \bigg\} \;.
\enq
For $\Re(z)<-\aleph$ with $\aleph$ large enough, one may take the integral explicitly which gives
\beq
\mc{J}_{0^+}(z) \, = \, \Bigg[  \f{1}{\i\pi} \ln \bigg( \f{ \veps_+(\mu+\i\zeta) - z }{ \veps_-(\mu+\i\zeta) - z }   \bigg) \Bigg]^{q}_{-q} \;= \; 0 \;.
\enq
The vanishing stems from the fact that $\veps_{+}(\pm q + \i\zeta) - \veps_-(\pm q + \i\zeta) = \veps(\pm q) = 0$. Since $z \mapsto \mc{J}_{0^+}(z)$ is analytic
on the connected domain $\Cx \setminus \ov{ \e{Int}\big( \ga_+\cup \ga_- \big) }$, it follows that it vanishes there.
Therefore, for any $z\in \Cx \setminus \ov{ \e{Int}\big( \ga_+\cup \ga_- \big) }$, $\veps - z$ has exactly two zeroes in $\msc{U}_{\veps}$
counted with multiplicities.  Moreover, if $\la \in \msc{U}_{\veps}$ is such that $\veps(\la)=z$, then
since $\veps$ is even and $\i\pi$ periodic, it follows that $-\la$ and $\i\pi-\la$ are also solutions to this equation. Thus, for $z \not \in \big\{ \veps(0), \veps(\i\tfrac{\pi}{2})\big\}$,
the solutions to $\veps(\la)=z$ appear in two distinct pairs
\beq
\{ \la_0, -\la_0 \} \quad \e{if} \quad  \Im(\la_0)\not= \tfrac{\pi}{2} \quad \e{and} \quad  \{\la_0, \i\pi-\la_0\} \quad \e{if} \quad  \Im(\la_0) = \tfrac{\pi}{2} \; .
\enq
When $z \in \big\{ \veps(0), \veps(\i\tfrac{\pi}{2})\big\}$, then parity and $\i\pi$-periodicity readily allow one to conclude that $0$ or $\tf{\i\pi}{2}$ are double
zeroes of the corresponding maps.
\qed

\vspace{3mm}




Recall \cite{KozFaulmannGohmannDressedEnergyInComplexPlane} that the closed curve
\beq
\msc{C}_{\veps; \e{tot}} = \Big\{ \la \in \Cx \; : \; -\tfrac{\zeta}{2} \leq \Im(\la) \leq \tfrac{\zeta}{2} \; \e{with} \; \Re[\veps(\la)]=0 \Big\}
\enq
starts at $\i \tf{\zeta}{2}$, passes through the point $- q$, joins  $-\i\tf{\zeta}{2}$
and then goes back to $\i\tf{\zeta}{2}$ by passing through $q$.
The intersection of this curve with the lower Poincar\'{e} half-plane $\mathbb{H}^{-}$: $\msc{C}_{\veps}=\msc{C}_{\veps; \e{tot}} \cap \ov{\mathbb{H}^{-}}$
may be partitioned in two branches:
\begin{itemize}
\item[i)] One starting from $-q$ and going to $-\i\tf{\zeta}{2}$ which corresponds to $\Im[\veps(\la)]$ monotonically increasing from $0$ to $+\infty$.
This curve is explicitly given as $\veps_{L}^{-1}(\intff{0}{+\i\infty})$, with $\veps_{L}^{-1}$ as introduced in Proposition \ref{Proposition double recouvrement de veps}.

\item[ii)] One going from $-\i\tf{\zeta}{2}$ to $q$ which corresponds to $\Im[\veps(\la)]$ monotonically increasing from $-\infty$ to $0$.
 This curve is explicitly given as $\veps_{R}^{-1}(\intff{-\i\infty}{0})$, with $\veps_{R}^{-1}$ as introduced in Proposition \ref{Proposition double recouvrement de veps}.
\end{itemize}

We now describe the local behaviour of $\veps^{-1}_{L/R}(s)$ when $\Im(s) \tend \pm \infty$ with $\Re(s)$ staying bounded for the preimage point $-\i\tf{\zeta}{2}$.
This will then allow us to provide a complete description of the solutions to the intersection problem introduced in
\eqref{definition probleme intersection Trotter fini et infini}, \eqref{definition des points limites des arcs ga sg et hat ga sg}.

\begin{lemme}
 \label{Lemme comportement local autour de la singularite pour inverse de veps}
 Let $\tau$ be as in \eqref{definition tau} and assume that $\Im(s) \tend  \ups_{\a} \infty$, $\a \in \{L,R\}$, with $\ups_L=+$ and $\ups_{R}=-$. Then, it holds that
\beq
\veps_{\a}^{-1}(s) \, = \, - \i\f{\zeta}{2} \, - \, \f{ 2\i J \sin \zeta  }{ s -\tau } \bigg( 1+\e{O}\Big( \f{1}{s-\tau }\Big)^2 \bigg) \;.
\label{ecriture comportement inverse veps proche de -i zeta sur 2}
\enq
In particular, there exists $C \in {\mathbb C}$ such that
\beq
\f{1}{\veps^{\prime}\Big( \veps_{\a}^{-1}(s) \Big) } \; \widesim{ \Im(s) \tend \ups_{\a} \infty } \; \f{ C }{ s^2  } \;.
\label{ecriture comportement asymptotique de veps prime le long de veps alpha}
\enq
\end{lemme}

\Proof
It is direct to establish that
\beq
\veps_0\big( t -\i \tfrac{\zeta}{2} \big) \; = \;  - \f{  2\i J \sin (\zeta) }{ t } \, + \, h  \,   - \,  2 J \cos(\zeta) \, + \, \e{O}(t)
\enq
so that, by using the linear integral equation \eqref{definition energie habille et energie nue} satisfied by $\veps$,  one gets
\beq
\veps \big( t -\i \tfrac{\zeta}{2} \big) \; = \;  - \f{  2\i J \sin (\zeta) }{ t } \, +  \, \tau \, + \, \e{O}(t) \;.
\label{ecriture dvpm petit t pour veps}
\enq
Thence, by solving $\veps \big( t -\i \tfrac{\zeta}{2} \big) \, = \, s$ in the limit where $\Im(s) \tend \pm \infty$, one obtains
\eqref{ecriture comportement inverse veps proche de -i zeta sur 2}. The rest of the claim follows directly. \qed

\begin{prop}
\label{Proposition DA des angles vth et Lipschitz pour angles et t sg alpha}

 Let $T_0, \eta>0$ be small enough. Then, for any $0<T<T_0$ and $\eta >\tf{1}{(NT^4)}$, given $f \in \mc{E}_{\mc{M}}$ the intersection problem
\eqref{definition probleme intersection Trotter fini et infini}, \eqref{definition des points limites des arcs ga sg et hat ga sg}
for $\Big( \, \wh{\mf{t}}^{\, (\sg)}_{\a}[f], \wh{\vth}_{\a}^{\,(\sg)}[f]\Big)$, $\a\in \{L,R\}$ and $\sg\in \{\pm\}$, expressed as
\beq
\veps_{\a}^{-1}\Big( \, \wh{\mf{z}}^{\, (\sg)}_{\a}[f] + \i \wh{\mf{t}}^{\, (\sg)}_{\a}[f] \Big) \, = \, - \i \f{\zeta}{2} \; + \; \mf{c}_{\e{d}} T \ex{ \i \wh{\vth}_{\a}^{\,(\sg)}[f] } \qquad with \qquad
   \begin{cases}
      \wh{\vth}_{L}^{\, (\sg)}[f] \in \intff{-\pi}{-\tfrac{\pi}{2}}\cup \intff{\tfrac{\pi}{2}}{\pi} \; \e{mod} \;  2\pi \\[1ex]
   \wh{\vth}_{R}^{\, (\sg)}[f] \in \intff{-\tfrac{\pi}{2}}{\tfrac{\pi}{2}}   \; \e{mod} \; 2\pi
   \end{cases}
\label{equation donnant forme solution angles intersection avec D - i zeta sur 2 cd T}
\enq
and constrained so that
\beq
\wh{\vth}_{L}^{\, (\sg)}[f] \, = \, -\sg \pi\, +\, \e{o}(1) \qquad and  \qquad
\wh{\vth}_{R}^{\, (\sg)}[f]  \, = \,   \e{o}(1) \quad  as \quad T +\tf{1}{(NT^4)} \tend 0^+ \;,
\enq
admits  a unique solution. This solution is such that one has the low-$T$ expansion for the intersection angle $\wh{\vth}_{\a}^{\,(\sg)}[f]$ and
the intersection coordinate $\wh{\mf{t}}^{\, (\sg)}_{\a}[f]$:
\begin{align}
  \wh{\vth}_{\a}^{\, (\sg)}[f]  & = -\sg \pi \de_{\a , L}\, + \,   \f{  T \mf{c}_{\e{d}} \ups_{\a} }{ 2  J \sin \zeta } \Re \bigl(  \tau \, - \,  \sg  \de_T  \bigr)   \, + \, \e{O}(T^2) \;,
\label{ecriture DA angle vth a basse T} \\
    \wh{\mf{t}}_{\a}^{\, (\sg)}[f]  & = - \f{ 2  \ups_{\a} J \sin \zeta }{  T \mf{c}_{\e{d}}  } \, + \, \Im \bigl(\tau \, - \,  \sg  \de_T\bigr) \, + \, \e{O}(T)
\label{ecriture DA mf t a basse T}
\end{align}
with a remainder that is uniform in $f \in \mc{E}_{\mc{M}}$.  Moreover, there exists
$\rho>0$ and small enough such that one has the Lipschitz continuity
\beq
\Big| \, \wh{\mf{t}}^{\, (\sg)}_{\a}[f]  \, - \, \wh{\mf{t}}^{\, (\sg)}_{\a}[g] \Big| \, + \, \Big| \, \wh{\vth}_{\a}^{\, (\sg)}[f]  \, - \,\wh{\vth}_{\a}^{\, (\sg)}[g] \Big| \; \leq \;
C \cdot \norm{ f\,-\, g }_{L^{\infty}\big(  \cup_{\a\in \{L,R\}} \veps_{\a}^{-1}( \op{D}_{0,2\rho} ) \big)  } \quad \e{for} \quad f\,, \,  g \in \mc{E}_{\mc{M}} \;.
\enq
Exactly the same conclusions hold for the quantities  $\Big( \, \mf{t}^{\, (\sg)}_{\a}[f], \vth_{\a}^{(\sg)}[f]\Big)$ with the only difference that the constraint involving $\eta$
is absent.

\end{prop}

\Proof
Observe that one has the representation
\beq
\veps \big( t -\i \tfrac{\zeta}{2} \big) \; = \;  - \f{  2\i J \sin (\zeta) }{ t } \, +  \, \tau \, + \, \de\veps (t)
\enq
with
\beq
\de\veps (t) \, = \, \veps_{0} \big( t -\i \tfrac{\zeta}{2} \big) \, + \, \f{  2\i J \sin (\zeta) }{ t }  -h +2J \cos(\zeta)
\, - \, \Int{-q}{q} \dd \mu \: \veps(\mu) \Big[ K\big(  t -\mu -\i \tfrac{\zeta}{2}  \big) \, - \, K\big(  \mu + \i \tfrac{\zeta}{2}  \big) \Big] \;.
\enq
$t\mapsto \de\veps (t)$ is smooth in a neighbourhood of $0$, and it follows from Lemma~\ref{Lemme comportement local autour de la singularite pour inverse de veps} that $\de\veps(t)= \e{O}\big( t \big)$.
Thus, the intersection equation can be recast in the form
\beq
 \, - \, \f{  2\i J \sin (\zeta) }{ \mf{c}_{\e{d}} T } \ex{ - \i  \wh{\vth}_{\a}^{\, (\sg)}[f]   } \, + \, \tau \, + \, \de \veps\Big(  \mf{c}_{\e{d}} T  \ex{  \i  \wh{\vth}_{\a}^{\, (\sg)}[f]   } \Big) \; = \;
 \wh{\mf{z}}^{\, (\sg)}_{\a}[f] + \i \wh{\mf{t}}^{\, (\sg)}_{\a}[f] \;.
\enq

Taking the real and imaginary parts and recentering on the part going to $0$: $ \wh{\vth}_{\a}^{\, (\sg)}[f] = -\sg \pi \de_{\a, L} +  \wh{\vp}_{\a}^{\, (\sg)}[f] $, one gets the system of equations
\begin{align}
\sin\Big( \wh{\vp}_{\a}^{\, (\sg)}[f] \Big) & = - \f{ (-1)^{\de_{\a,L} } \mf{c}_{\e{d}} T }{  2  J \sin (\zeta) }
\Re\bigg\{ \, \wh{\mf{z}}^{\, (\sg)}_{\a}[f] \, - \,  \tau  \, - \,  \de \veps\Big(  \mf{c}_{\e{d}} T  \ex{  \i  \wh{\vth}_{\a}^{\, (\sg)}[f]   } \Big)  \bigg\} \;,  \\[1ex]
 \wh{\mf{t}}^{\, (\sg)}_{\a}[f] &=  - \f{  2  J \sin (\zeta) }{  (-1)^{\de_{\a,L} }  \mf{c}_{\e{d}} T }  \cos\Big( \wh{\vp}_{\a}^{\, (\sg)}[f] \Big)
\, + \, \Im\bigg\{ \,   \tau \, -\, \wh{\mf{z}}^{\, (\sg)}_{\a}[f] \, + \,  \de \veps\Big(  \mf{c}_{\e{d}} T  \ex{  \i  \wh{\vth}_{\a}^{\, (\sg)}[f]   } \Big)  \bigg\} \;.
\label{ecriture expression explicite pour hat t sg alpha}
\end{align}
The first equation allows one to determine $\wh{\vp}_{\a}^{\, (\sg)}[f]$. Upon introducing the map
\beq
\Xi_{T}(\vp) \; = \; \sin(\vp) \, - \, \f{ (-1)^{\de_{\a,L} } \mf{c}_{\e{d}} T }{  2  J \sin (\zeta) }
\Re\bigg\{ \, \tau  \, +  \,  \de \veps\Big(  \ups_{\a} \mf{c}_{\e{d}} T  \ex{  \i  \vp    } \Big)  \bigg\}
\enq
one readily gets that $\Xi$ is smooth provided that $T$ is small enough and $\Xi_{T}^{\prime}(\vp) \,  = \, \cos(\vp)\,+\,\e{O}(T^2)$ uniformly in $\vp$.
In particular,
\beq
\Xi_{T}: \intff{ - \tfrac{\pi}{4} }{  \tfrac{\pi}{4} } \tend \Big[ \Xi_{T}\big(- \tfrac{\pi}{4} \big)  \, ; \,     \Xi_{T}\big( \tfrac{\pi}{4} \big)   \Big]
\enq
is an
increasing smooth diffeomorphism. Since  $\Xi_{T}(\vp) \,  = \, \sin(\vp)\,+\,\e{O}(T)$, it follows that there exists $\kappa>0$ small enough such that
$\intff{-\kappa}{\kappa}\subset \big[ \Xi_{T}\big(- \tfrac{\pi}{4} \big)  \, ; \,     \Xi_{T}\big( \tfrac{\pi}{4} \big)   \big] $.
Thus, the low-$T$ large $NT$ expansion \eqref{ecriture DA basse tempe de tau et eta f} ensures that, uniformly in $0<T<T_0$ and $\eta > \tf{1}{ (NT^4) }$, one has
\beq
 \wh{\vp}_{\a}^{\, (\sg)}[f] \; = \; \Xi_{T}^{-1}\bigg( - \f{ (-1)^{\de_{\a,L} } \mf{c}_{\e{d}} T }{  2  J \sin (\zeta) }
\Re\big[\, \wh{\mf{z}}^{\, (\sg)}_{\a}[f] \big]  \bigg) \;.
\enq
This implies, in particular, that
\beq
 \wh{\vp}_{\a}^{\, (\sg)}[f] \; = \; \Xi_{T}^{-1}(0) \, - \, \Big(\Xi_{T}^{-1}\Big)^{\prime}(0)  \f{ (-1)^{\de_{\a,L} } \mf{c}_{\e{d}} T }{  2  J \sin (\zeta) }
\Re\big[\, \wh{\mf{z}}^{\, (\sg)}_{\a}[f] \big] \; + \; \e{O}(T^2) \;.
\enq
Since $\Big(\Xi_{T}^{-1}\Big)^{\prime}(0) =1 + \e{O}(T^2)$,
$\Xi_{T}^{-1}(0)\, = \, \f{ (-1)^{\de_{\a,L} } \mf{c}_{\e{d}} T }{  2  J \sin (\zeta) } \Re(\tau) \, + \, \e{O}(T^2)$, upon
inserting \eqref{ecriture DA basse tempe de tau et eta f}, this yields
\eqref{ecriture DA angle vth a basse T}, while \eqref{ecriture DA mf t a basse T}
follows by means of \eqref{ecriture expression explicite pour hat t sg alpha}.

By Lipschitz continuity of the functions $\Xi^{-1}$ and $\veps$ and by Proposition
\ref{Proposition continuite inverse f et sa comparaison a veps inverse} one gets
\beq
\Big|  \wh{\vp}_{\a}^{\, (\sg)}[f]  \, - \,  \wh{\vp}_{\a}^{\, (\sg)}[g]  \Big| \, \leq \, C \big|\, \wh{\mf{z}}^{\, (\sg)}_{\a}[f]  \, - \, \wh{\mf{z}}^{\, (\sg)}_{\a}[g]  \big|
\; \leq \; C^{\prime} \norm{f-g}_{L^{\infty}\big(  \cup_{\a\in \{L,R\}} \veps_{\a}^{-1}( \op{D}_{0,2\rho} ) \big)  } \;,
\enq
in which $\rho$ is as given by Proposition \ref{Proposition continuite inverse f et sa comparaison a veps inverse}.
Finally, starting from the representation
\eqref{ecriture expression explicite pour hat t sg alpha} for
$\wh{\mf{t}}^{\, (\sg)}_{\a}[f]$, by using the above bounds
and the smoothness of $\de \veps$ in a neighbourhood of the origin, one readily gets that
\beq
\Big| \,  \wh{\mf{t}}_{\a}^{\, (\sg)}[f]  \, - \,  \wh{\mf{t}}_{\a}^{\, (\sg)}[g]  \Big| \; \leq \; C \,
                         \norm{f-g}_{L^{\infty}\big(  \cup_{\a\in \{L,R\}} \veps_{\a}^{-1}( \op{D}_{0,2\rho} ) \big) } \;.
\enq
This concludes the proof. \qed

\subsection{The dressed energy on curved contours and string sums}

Given a function $f$ meromorphic on $\Cx$ having cuts on a union of curves $\Ga_{f}$, we define
\beq
f^{(-)}_{k}(\la) \; = \; \sul{r=0}{k-1} f(\la - \i r \zeta)
\enq
whenever it makes sense.


It is easy to check that
\begin{align}
K_k^{(-)}(\la) & = \f{1}{2\i\pi} \Big\{  \coth(\la-\i k \zeta) \, + \, \coth(\la-\i (k-1) \zeta)    \, - \, \coth(\la + \i \zeta) \, - \, \coth( \la )   \Big\} \;, \\[1ex]
\veps_{0;k}^{(-)}(\la) & = k h \, - \, 2\i J \sin(\zeta) \Big\{  \coth(\la + \i \tfrac{\zeta}{2})  \, - \, \coth(\la + \i (\tfrac{1}{2}-k)\zeta )  \Big\} \;.
\end{align}
The above along with \eqref{ecriture LIE pour veps c},
\eqref{ecriture relation entre veps c et veps} entails that
\beq
\veps_{\e{c};k}^{(-)}(\la) \,= \, \veps_{0;k}^{(-)}(\la) \, - \, \Int{ \Dp{} \mc{D}_{\veps}^{(\da)} }{} \hspace{-3mm} \dd \mu \, K_k^{(-)}(\la-\mu) \veps(\mu)
\label{ecriture expression pour vepsck en terme de vepsk etc}
\enq
in which we agree upon
\beq
 \mc{D}_{\veps}^{(\da)} \; = \; \mc{D}_{\veps} \cap \ov{\mathbb{H}^{-}} \qquad \e{and} \qquad  \mc{D}_{\veps}^{(\ua)} \; = \; \mc{D}_{\veps} \cap \ov{\mathbb{H}^{+}} \;,
\label{definition D veps up and down}
\enq
where we defined
\beq
\mc{D}_{\veps} \; = \;  \Big\{ \la \in \Cx\; : \; |\Im(\la)| \leq  \tfrac{\zeta}{2} \quad \e{and} \quad \Re[\veps(\la)] <0 \Big\} \;.
\enq
The last integral in \eqref{ecriture expression pour vepsck en terme de vepsk etc} can be taken by residues, what ultimately yields
\beq
\veps_{\e{c};k}^{(-)}(\la) \,= \, \veps_{k}^{(-)}(\la) \, + \, \Big( \veps \bs{1}_{  \mc{D}_{\veps; \i\pi }^{(\da)}   }   \Big)(\la-\i k \zeta) \,  +\,
\Big( \veps \bs{1}_{  \mc{D}_{\veps; \i\pi }^{(\da)}   }   \Big)(\la-\i (k-1) \zeta) \,  - \, \Big( \veps \bs{1}_{  \mc{D}_{\veps; \i\pi }^{(\da)}   }   \Big)(\la + \i  \zeta)
\,  - \,\Big( \veps \bs{1}_{  \mc{D}_{\veps; \i\pi }^{(\da)}   }   \Big)(\la  )  \;.
\label{expression explicite pour vepsc k}
\enq
Above, we agree upon the shorthand notation
\beq
\mc{D}_{\veps; \i\pi }^{(\da)} \, = \, \mc{D}_{\veps }^{(\da)} + \i \pi \mathbb{Z} \;.
\enq

We now establish a few properties of the functions $\veps_{\e{c}}$ and
$\veps^{(-)}_{\e{c};2}$ which play an important role in the analysis in the main
body of the text.

\begin{lemme}
\label{Lemme positivite dans plan cplx de real vepsc}
  If $\la \not\in  \ov{ \mc{D}_{\veps; \i\pi } }$  with $\mc{D}_{\veps; \i\pi } \, = \, \mc{D}_{\veps } + \i \pi \mathbb{Z} $ , then  $\Re\big[\veps_{\e{c}}(\la) \big]>0$ \;.
\end{lemme}

\Proof
First of all, observe that for $k=1$, \eqref{expression explicite pour vepsc k} reduces to
\beq
\veps_{\e{c}}(\la) \,= \, \veps(\la) \, + \, \Big( \veps \bs{1}_{  \mc{D}_{\veps; \i\pi }^{(\da)}   }   \Big)(\la-\i \zeta)
\,  - \, \Big( \veps \bs{1}_{  \mc{D}_{\veps; \i\pi }^{(\da)}   }   \Big)(\la  + \i \zeta) \;.
\enq
In order to check the validity of the statement, one should discuss depending on whether $\la  \pm \i \zeta \in \mc{D}_{\veps; \i\pi }^{(\da)} $ or not.
It was established in \cite{KozFaulmannGohmannDressedEnergyInComplexPlane} that, when   $\zeta \in \intoo{0}{ \tf{\pi}{2} }$, it holds that
\beq
\Re\big[\veps(\la) \big]>0 \quad \e{if}  \quad \la \not\in  \ov{ \mc{D}_{\veps; \i\pi }  }   \qquad \e{while} \qquad
\Re\big[\veps(\la) \big]<0 \quad \e{if}  \quad \la \in  \mc{D}_{\veps; \i\pi }   \;.
\label{ecriture proprietes positivite de veps dans C}
\enq
Since the curve $\Dp{}\mc{D}_{\veps}$ intersects $\R$ at the points $\pm q$, the above entails that
\beq
\veps(\la)>0 \quad \e{on} \quad \R\setminus \intff{-q}{q}  \quad \e{while} \quad  \veps(\la)<0 \quad \e{on} \quad  \intoo{-q}{q} \;.
\label{ecriture positivite de veps sur R}
\enq

\subsubsection*{$\bullet$ $\la \not\in \overline{\mc{D}_{\veps; \i\pi }}$ and
$\la  \pm \i \zeta \not\in \mc{D}_{\veps; \i\pi }^{(\da)}$ }
In this case, one simply has $\veps_{\e{c}}(\la) \,= \, \veps(\la) $ and the claim follows
from the mentioned previous results.

\subsubsection*{$\bullet$ $\la \not\in \overline{\mc{D}_{\veps; \i\pi }}$ and
$\la  - \i \zeta \not\in \mc{D}_{\veps; \i\pi }^{(\da)}$
\normalfont{but}  $\la  + \i \zeta \in \mc{D}_{\veps; \i\pi }^{(\da)}$ }
In this case, one has $\veps_{\e{c}}(\la) \,= \, \veps(\la)  \,  - \,  \veps (\la  + \i \zeta)$ and, again, on directly concludes by virtue of \eqref{ecriture proprietes positivite de veps dans C}.

\subsubsection*{$\bullet$ $\la \not\in \overline{\mc{D}_{\veps; \i\pi }}$ and
$\la  +  \i \zeta \not\in \mc{D}_{\veps; \i\pi }^{(\da)}$
\normalfont{but}  $\la  - \i \zeta \in \mc{D}_{\veps; \i\pi }^{(\da)}$ }
In this case, one has  $\veps_{\e{c}}(\la) \,= \, \veps(\la)  \,  + \,  \veps (\la  - \i \zeta)$ so that the signs of the real parts do not obviously combine as earlier on.

First of all, observe that in the case of interest, there exists $p\in \mathbb{Z}$ such that $\la_p -\i\zeta = \la-\i p \pi- \i \zeta \in\mc{D}_{\veps  }^{(\da)}$. In particular, one has that
 $\Im\big( \la_p \big) \in \intoo{ \tf{\zeta}{2} }{ \zeta }$. In order to estimate the
 sign of $\Re\big[\veps_{\e{c}}(\la) \big]$ in this case, one should rely on an appropriate integral representation for $\veps$. It is well know  \cite{KozDugaveGohmannThermoFunctionsZeroTXXZMassless,YangYangXXZStructureofGS}
 that one may represent $\veps$ as the unique solution to the linear integral equation
\beq
\veps(\la) \, = \, \veps_{\infty}(\la) \; + \; \Int{\R \setminus I_q}{} \hspace{-2mm } \dd \mu \, R(\la-\mu) \veps(\mu) \quad \e{with} \quad \veps_{\infty}(\la) \; = \; \f{ h \pi }{2 (\pi- \zeta) } \, - \,
\f{ 2\pi J \sin(\zeta) }{ \zeta \cosh\big(  \frac{\pi \la }{ \zeta } \big) } \;,
\label{ecriture rep int pour veps dans la rep duale}
\enq
where $R$ corresponds to the resolvent kernel of the operator $\e{id}+\op{K}_{\R}$ acting on $L^2(\R)$ with integral kernel $ K(\la-\mu)$. One has $R(\la) \; = \; R_{I}(\la)$
for $|\Im(\la)|< \tf{\zeta}{2}$, where $R_{I}$ is given by the integral representation
\beq
R_{I}(\la) \, = \, \f{ \pi  }{ 2 \zeta (\pi-\zeta) } \Int{ \R }{}   \dd \mu  \:
\f{1}{ \cosh\Big(  \frac{\pi (\la-\mu) }{ \zeta } \Big)  }
K\biggl( \f{ \pi \mu}{ \pi - \zeta } \bigg| \, \wt{\zeta} \, \biggr)
\quad \e{where} \quad \wt{\zeta} \, = \,  \f{ \pi \zeta }{ 2(\pi - \zeta) } \;,
\enq
while
\beq
K(\la \mid \a ) \; = \; \f{1}{2\i\pi} \Big\{ \coth(\la - \i \a) \, - \, \coth(\la+\i\a)  \Big\}  \;.
\enq
 When $\tf{\zeta}{2} < | \Im(\la) | < \zeta$, one may obtain the expression
 for $R$ by analytically continuing the representation $R(\la) \; = \; R_{I}(\la)$  from the strip $|\Im(\la)|< \tf{\zeta}{2}$.
This yields
\beq
R(\la) \, = \,  R_{I}(\la) \; + \;
   \begin{cases}
      \f{\pi}{\pi - \zeta } \,
      K\biggl( \f{ \pi (\la-\i \tf{\zeta}{2}) }{ \pi - \zeta } \bigg| \, \wt{\zeta} \, \biggr)
      \quad \text{for}\ \f{\zeta}{2} <  \Im(\la)  < \zeta \\[2ex]
      \f{\pi}{\pi - \zeta } \,
      K\biggl( \f{ \pi (\la+\i \tf{\zeta}{2}) }{ \pi - \zeta } \bigg| \, \wt{\zeta} \, \biggr)
      \quad \text{for}\ - \zeta <  \Im(\la)  < -  \f{\zeta}{2} \;.
   \end{cases}
\label{ecriture continuation analytique R dans bande II}
\enq
Hence, by using the $\i\zeta$ anti-periodicity of $R_I$ as well as the $\i\pi$ periodicity of the resolvent $R$ one gets that
\beq
\veps_{\e{c}}(\la) \; = \; \veps(\la_p) \, + \,  \veps(\la_p-\i\zeta)
\; = \; \f{ h \pi }{ \pi - \zeta } \, + \, \f{\pi}{\pi-\zeta }
\Int{\R \setminus I_q}{} \hspace{-2mm} \dd \mu \:
K\biggl( \f{\la_p-\mu-\i \tf{\zeta}{2}}{1 - \tf{\zeta}{\pi} } \bigg| \, \wt{\zeta} \, \biggl)
\veps(\mu) \;.
\label{ecriture repr int pour veps c}
\enq
In order to take the real part of this expression it is useful to apply the identity
\beq
2 \Re\Big[  K(x+i y \mid s) \Big] \, = \, K(x\mid s-y) \, + \, K(x\mid s+y) \;,
\enq
which yields
\beq
2 \Re\bigg[ K\biggl( \f{  \la_p-\mu-\i \tf{\zeta}{2} }{ 1 - \tf{\zeta}{\pi} } \bigg| \, \wt{\zeta} \, \biggr) \bigg] \; = \;
K\biggl( \f{  \Re(\la)-\mu }{ 1 - \tf{\zeta}{\pi} } \bigg| \, \de_1  \biggr)  \, + \, K\biggl( \f{  \Re(\la)-\mu }{ 1 - \tf{\zeta}{\pi} } \bigg| \, \de_2  \biggr) \;,
\enq
where
\beq
\de_1 = \f{ \Im(\la_p) }{ 1 - \tf{\zeta}{\pi}  } \qquad \e{and} \qquad \de_2 = \f{ \zeta - \Im(\la_p) }{ 1 - \tf{\zeta}{\pi}  } \;.
\enq
It is easy to check that $\de_2 \in \intoo{0}{\tf{\pi}{2}}$ so that $K(s\mid \de_2)>0$ on $\R$ while $\de_1$ may take values in $\intoo{0}{\pi}$, depending on the value of $\zeta$ and $\Im(\la_p)$
and thus the associated kernel may be either positive or negative.
If $\de_1, \de_2 \in \intoo{0}{\tf{\pi}{2}}$, then since $\veps|_{\R\setminus I_q}>0$, taking the real part of \eqref{ecriture repr int pour veps c} yields that
\beq
\Re\big[\veps_{\e{c}}(\la) \big] \, >  \, \f{ \pi h }{ \pi  - \zeta} \;.
\enq
It thus remains to consider the case when $\de_{1}\in \intoo{ \tf{\pi}{2} }{ \pi }$. In that case, upon using the bounds $h> \veps(\la) >0$ for $\la \in \R \setminus I_q$ and
$K(s\mid \de_1)<0$ on $\R$, one gets the chain of lower bounds
\beq
\Re\big[ \veps_{\e{c}}(\la) \big]  \, >  \, \f{ \pi h }{ \pi  - \zeta} \, + \,
\f{\pi h }{2 (\pi-\zeta) } \Int{\R \setminus I_q}{} \hspace{-2mm} \dd \mu \,
K\biggl( \f{  \Re(\la)-\mu }{ 1 - \tf{\zeta}{\pi} } \bigg| \, \de_1 \biggr)
\, > \,  \f{  3 h }{ 2 } \, .
\enq
where we made profit of $\Int{\R}{} \dd \mu \: K(\mu\mid \de) = 1- \tf{2\de}{\pi}$.

\subsubsection*{ $\bullet$ $\la \not\in \mc{D}_{\veps; \i\pi }$
  \normalfont{but}  $\la  \pm  \i \zeta \in \mc{D}_{\veps; \i\pi }^{(\da)}$ }


 Taken the previous two discussions, one readily concludes that in that last case it
 also holds that $\Re\big[ \veps_{\e{c}}(\la) \big]>0$. \qed

 \vspace{5mm}

\begin{lemme}
 \label{Lemme ptes positivite pour veps 2 c}

 Assume that $\zeta \in \intoo{0}{ \tf{\pi}{2} }$.  If $\la \in  \ov{\mc{D}_{\veps; \i\pi } } $, then  $\Re\big[\veps_{\e{c};2}^{(-)}(\la) \big]>0$ \;.
\end{lemme}

\Proof
For $k=2$, the expression \eqref{expression explicite pour vepsc k} reduces to
\beq
\veps_{\e{c};2}^{(-)}(\la) \,= \, \veps(\la) \, + \, \veps(\la-\i\zeta)  \, + \, \Big( \veps \bs{1}_{  \mc{D}_{\veps; \i\pi }^{(\da)}   }   \Big)(\la- 2\i  \zeta) \,  +\,
\Big( \veps \bs{1}_{  \mc{D}_{\veps; \i\pi }^{(\da)}   }   \Big)(\la-\i   \zeta) \,  - \, \Big( \veps \bs{1}_{  \mc{D}_{\veps; \i\pi }^{(\da)}   }  \Big)(\la + \i  \zeta)
\,  - \,\Big( \veps \bs{1}_{  \mc{D}_{\veps; \i\pi }^{(\da)}   }   \Big)(\la  )  \;.
\label{expression explicite pour vepsc 2}
\enq
By $\i\pi$-periodicity of the involved functions, it is enough to focus on the case when $\la \in  \ov{\mc{D}_{\veps } }$.
In such a case, since $\zeta \in \intoo{0}{\tf{\pi}{2}}$, one can check that $\la \pm \i \zeta \not\in  \ov{\mc{D}_{\veps; \i\pi } }$,  hence
slightly simplifying  \eqref{expression explicite pour vepsc 2}. To push further this simplification, one needs to distinguish cases.

\subsubsection*{$\bullet$ $\la \in \ov{\mc{D}_{\veps }^{(\ua)} }$ }

In the considered range of $\zeta$s, it is direct to check that then
$\la - 2\i\zeta \not\in \ov{\mc{D}_{\veps}^{(\da)}}$ so that one has
\beq
\veps_{\e{c};2}^{(-)}(\la) \,= \, \veps(\la) \, + \, \veps(\la-\i\zeta)  \;.
\enq
Upon using the integral representation \eqref{ecriture rep int pour veps dans la rep duale} along with \eqref{ecriture continuation analytique R dans bande II},
one concludes by means of analogous handlings to those outlined in the proof of Lemma~\ref{Lemme positivite dans plan cplx de real vepsc} that then,
\beq
\Re\big[ \veps_{\e{c};2}^{(-)}(\la) \big] \, > \, \f{3 h }{2} >0 \;.
\enq

\subsubsection*{ $\bullet$ $\la \in \ov{\mc{D}_{\veps }^{(\da)} }$ }

In that case, one gets that $-\tf{5 \zeta}{2} \leq \Im(\la-2\i\zeta)  \leq -2\zeta$.
One may check that when $0 < \zeta< \tf{ 2\pi }{ 5 }$, $\la -2\i\zeta \not\in  \ov{\mc{D}_{\veps;\i\pi }^{(\da)} }$ so that
\beq
\veps_{\e{c};2}^{(-)}(\la) \, = \, \veps(\la-\i\zeta)
\enq
and, since $\la-\i\zeta \not\in \ov{\mc{D}_{\veps;\i\pi } } $, one has by virtue of Lemma \ref{Lemme positivite dans plan cplx de real vepsc}
that $\Re\big[ \veps_{\e{c};2}^{(-)}(\la) \big] \, > \, 0 $.

When $\tf{ 2\pi }{ 5 } < \zeta< \tf{\pi}{2}$, there are two possible cases:
\beq
\veps_{\e{c};2}^{(-)}(\la) \, = \, \left\{ \ba{ccc}  \veps(\la-\i\zeta)      & \e{if} & \la -2\i\zeta \not\in  \ov{\mc{D}_{\veps;\i\pi }^{(\da)} }  \vspace{2mm} \\
                                                \veps(\la-\i\zeta) \, + \, \veps(\la-2\i\zeta)    & \e{if} & \la -2\i\zeta  \in  \ov{\mc{D}_{\veps;\i\pi }^{(\da)} } \ea \right.
\enq
In the first case, one concludes as above. In the second one, one necessarily has that $\la-2\i\zeta \in \ov{\mc{D}_{\veps }^{(\da)} }-\i\pi$,
\textit{viz}. $\check{\la}=\la+\i\pi-2\i\zeta \in  \ov{\mc{D}_{\veps }^{(\da)} }$.
Then, by repeating the same reasoning as above, one infers that $\Re\big[ \veps_{\e{c};2}^{(-)}(\la) \big] \, > \, \f{3 h }{2} >0$ which entails the claim. \qed

\end{document}